\documentclass [12pt,oneside]{report}
\usepackage{setspace}

\input{amssymb.sty}
\usepackage[dvips]{graphicx}
\usepackage{rotate}
\usepackage{longtable}
\usepackage[dvips]{lscape}
\usepackage{supertabular}
\usepackage{rotate}

\usepackage{fancyhdr}
\fancypagestyle{plain}{%
\fancyhf{}
\fancyhead[R]{\thepage}

}

\newcommand{\pipi}{\pi^{+}\pi^{-}}
\newcommand{\KK}{K^{+}K^{-}}
\newcommand{\ppbar}{p\overline{p}}
\newcommand{\hadpair}{h^{+}h^{-}}
\newcommand{\hhbar}{h\overline{h}}
\newcommand{\ee}{e^{+}e^{-}}
\newcommand{\mumu}{\mu^{+}\mu^{-}}
\newcommand{\leppair}{l^{+}l^{-}}

\newcommand{\pipiJpsi}{\pi^{+}\pi^{-}J/\psi}
\newcommand{\CpipiJpsi}{C^{\pipiJpsi}}
\newcommand{\JPll}{J/\psi \rightarrow \leppair}

\newcommand{\eetomm}{\ee \rightarrow m^{+}m^{-}}
\newcommand{\eetopipi}{\ee \rightarrow \pipi}
\newcommand{\eetoKK}{\ee \rightarrow \KK}
\newcommand{\eetoppbar}{\ee \rightarrow \ppbar}
\newcommand{\eetohh}{\ee \rightarrow \hadpair}
\newcommand{\eetohhbar}{\ee \rightarrow \hhbar}

\newcommand{\eetoll}{\ee \rightarrow \leppair}
\newcommand{\Bhabha}{\ee \rightarrow \ee}
\newcommand{\dimuon}{\ee \rightarrow \mumu}
\newcommand{\gamgamprod}{\ee \rightarrow \gamma\gamma}

\newcommand{\eetoISRJPhhbar}{\ee \rightarrow \gamma J/\psi, J/\psi \rightarrow \hhbar}

\newcommand{\eetoISRJPKK}{\ee \rightarrow \gamma J/\psi, J/\psi \rightarrow \KK}

\newcommand{\egam}{E_{\gamma}}

\newcommand{\psip}{\psi(2S)}

\newcommand{\psiptohhbar}{\psip \rightarrow \hhbar}
\newcommand{\psiptopipi}{\psip \rightarrow \pipi}
\newcommand{\psiptoKK}{\psip \rightarrow \KK}
\newcommand{\psiptoppbar}{\psip \rightarrow \ppbar}
\newcommand{\psiptomumu}{\psip \rightarrow \mumu}
\newcommand{\psiptoee}{\psip \rightarrow \ee}

\newcommand{\DtoKpi}{D^{0} \rightarrow K^{-}\pi^{+}}

\newcommand{\mff}{|F_{m}(s)|}

\newcommand{\piff}{|F_{\pi}(s)|}

\newcommand{\piffsq}{|F_{\pi}(s)|^{2}}
\newcommand{\kff}{|F_{K}(s)|}

\newcommand{\kffsq}{|F_{K}(s)|^2}

\newcommand{\prelecffq}{G^P_{E}(Q^2)}
\newcommand{\prmagffq}{G^P_{M}(Q^2)}
\newcommand{\prelecffqsqtl}{G^{P,tl}_{E}(Q^2)}
\newcommand{\prmagffqsqtl}{G^{P,tl}_{M}(Q^2)}

\newcommand{\prelecff}{|G^{P}_{E}(s)|}
\newcommand{\prmagff}{|G^{P}_{M}(s)|}

\newcommand{\etal}{\it{et~al.}}
\newcommand{\msbar}{\overline{\mathrm{MS}}}
\newcommand{\qqbar}{q\overline{q}}
\newcommand{\ppbartoee}{\ppbar \rightarrow \ee }

\begin{document}
\baselineskip=18pt

\onehalfspacing

\renewcommand{\thefootnote}{\fnsymbol{footnote}}
\bibliographystyle{unsrt}

\pagestyle{myheadings}
\pagenumbering{arabic}

\setcounter{page}{0}
\def\baselinestretch{1.5}
\begin{titlepage}         
\hoffset 0.in
\begin{center}
\large NORTHWESTERN UNIVERSITY\\
\vspace{0.45in}
{\Large\bf Precision Measurements of the Timelike Electromagnetic Form Factors 
of the Pion, Kaon, and Proton}\\
\vspace{.35in}
\large A DISSERTATION\\
\vspace{.1in}
\large SUBMITTED TO THE GRADUATE SCHOOL\\
\mbox{IN~PARTIAL~FULFILMENT~OF~THE~REQUIREMENTS}\\
\vspace{.2in}
\large for the degree\\
\vspace{.05in}
\large DOCTOR OF PHILOSOPHY\\
\vspace{.1in}
\large Field of Physics and Astronomy\\
\vspace{.25in}
\large By\\
\vspace{.15in}
\large Peter Karl Zweber\\
\vspace{.15in}
\large EVANSTON, ILLINOIS\\                        
\vspace{.1in}
\large June 2006\\   
\end{center}
\end{titlepage}
\def\baselinestretch{2.0}

\setcounter{page}{2}
\newpage

\vspace*{\fill}
\begin{center}
\copyright  $\;$Copyright by Peter Karl Zweber 2006\\ 
\vspace{7 mm}
All Rights Reserved
\end{center}

\newpage

\chapter*{\centering \ \large ABSTRACT}
\addcontentsline{toc}{chapter}{ABSTRACT}

\vspace{-1cm}
{\centering {\Large\bf
 Precision Measurements of the Timelike Electromagnetic Form Factors 
of the Pion, Kaon, and Proton}\\
}
\vspace{20pt}
\begin{center}
Peter Karl Zweber
\end{center}	
\vspace{20pt}

\doublespacing

Using 20.7 pb$^{-1}$ of $e^+e^-$ annihilation data taken at $\sqrt{s}=3.671$ GeV 
with the CLEO--c detector, precision measurements of the electromagnetic form 
factors of the charged pion, charged kaon, and proton have been made for timelike 
momentum transfer of $|Q^2|=13.48$ GeV$^2$ by the reaction $e^+e^-\to h^+h^-$.  
The measurements are the first ever with identified pions and kaons of $|Q^2|>4$ 
GeV$^2$, with the results 
$|F_\pi(13.48\;\mathrm{GeV}^2)|=0.075\pm0.008(\mathrm{stat})
\pm0.005(\mathrm{syst})$ 
and 
$|F_K(13.48\;\mathrm{GeV}^2)|=0.063\pm0.004(\mathrm{stat})
\pm0.001(\mathrm{syst})$.  
The result for the proton, assuming $|G^P_E(Q^2)|=|G^P_M(Q^2)|$, is 
$|G^P_M(13.48\;\mathrm{GeV}^2)|=0.0139^{+0.0024}_{-0.0018}(\mathrm{stat})
\pm0.0006(\mathrm{syst})$, 
which is in agreement with earlier results.

\newpage

\vspace*{2.0 in}
\begin{center}
To my parents, brothers, and Adie, for all of their support and patience.
\end{center}

\newpage
\chapter*{\Large Acknowledgements}
\addcontentsline{toc}{chapter}{Acknowledgements}

\vspace{-1.0cm}

I would first like to thank my thesis advisor, Professor Kamal K. Seth.  He was 
truly instrumental in my development of becoming an experimental scientist.  
Our discussions, sometimes heated, taught me to be well prepared and well spoken 
in the defense of my positions.

I would also like to thank the other members of his group.  They are Sean Dobbs, 
Dave Joffe, Zaza Metreveli, Willi Roethel, Amiran Tomaradze, and Ismail Uman.  In 
particular, the numerous conversations with Zaza allowed me the opportunity to expand 
my understanding of particle physics concepts.

I would also like to thank all of the people I had the privilege to encounter 
on the CLEO experiment and in Ithaca in general.  I would like to particularly acknowledge 
Stefan Anderson, Basit Athar, Karl Berkelman, Dave Besson, 
V\'{e}ronique Boisvert, Devin Bougie, Matt Chasse, 
Christine Crane, Istvan Danko, Jean Duboscq,  Richard Galik, 
Justin Hietala, Lauren Hsu, Curtis Jastremsky, 
Tim and Lynde Klein, Brian Lang, Norm Lowrey, 
Hanna Mahlke-Kr\"{u}ger, Alan Magerkurth, Paras Naik, Jim Napolitano, 
Mark Palmer, Rukshana Patel, Todd Pedlar, 
Jon Rosner, Isaac Robinovitz, Batbold Sanghi, Master Shake for asking 
``Who is the Drizzle?'', Matt and Katie Shepherd, Alex Smith, Chris Stepaniak, 
Gocha Tatishvili, Gregg and Jana Thayer, David Urner, Mike Watkins, 
Mike and Tammie Weinberger, and Alexis Wynne.  These people, and numerous others, 
made my graduate student days in Ithaca a bearable experience.  

I would like to acknowledge the support of my family.  My parents, Vince and Edrie, 
and brothers, Jeff and Eric, were very supportive and occasionally provoking during 
my days as a graduate student.  I would also like to acknowledge my relatives 
in the Chicago area: Gladys Cowman, Marge Cowman, and Eva Mary Cowman, in which two of them 
(Gladys and Marge) are no longer with us.  I would especially like to thank Mary for 
her hospitality during the months of my dissertation writing.  

I would like to acknowledge my ``children'': the bitches, Stoica and Rapscallion, and 
the feline predator, Odin.

Finally, I would like to acknowledge my fianc\'{e} Adrienne Gloor.  
Thanks for all of the support you provided along the way.  
I love you very much, and I will be coming home soon.

\tableofcontents
\listoftables
\addcontentsline{toc}{chapter}{List of Tables}
\listoffigures
\addcontentsline{toc}{chapter}{List of Figures}

\baselineskip=24pt
\chapter{Introduction}

This dissertation is devoted to the study of the structure of the three lightest 
strongly interacting hadrons, the two lightest mesons, the pion and 
the kaon, and the lightest baryon, the proton, by measuring their electromagnetic 
form factors.  In order to put this study in perspective, it is useful to briefly review 
particle physics.  

The study of particle physics is the study of fundamental particles and 
the interactions between them.  The modern framework which incorporates the 
fundamental particles and interactions is called the Standard Model.  
There are four fundamental interactions, and they are, in decreasing order of strength, 
the strong (often called nuclear or hadronic), 
electromagnetic, weak, and gravitational.  With respect to the strong interaction, 
the relative strength of the electromagnetic, weak, and gravitational interactions are 
$\sim 1/137$, $10^{-5}$, and $10^{-39}$ \cite{perkinsbook}, respectively.  
The gravitational interaction is not yet well understood and is not included in 
the Standard Model.  

Particles are called fundamental when they are structureless and pointlike.  While 
being structureless, fundamental particles have an intrinsic property called spin.  
The spin of particles, fundamental or composite, have either integer or 
half-integer values.  Particles with integer values of spin are governed by 
Bose-Einstein 
statistics and are hence called bosons, and particles with half-integer values are 
governed by Fermi-Dirac statistics and are called fermions.  

The strong, electromagnetic, and weak interactions are all mediated 
by spin-1 vector bosons.  The strong interaction is mediated by 
gluons (denoted by $g$), electromagnetism by photons ($\gamma$), and 
the weak by $W$ and $Z$ bosons.  Table \ref{tab:forces} lists the 
various properties of the mediators.  

\begin{table}[h]
\caption[Fundamental interaction mediators in the Standard Model.]
{Fundamental interaction mediators in the Standard Model.  
The numerical values are taken from Ref. \cite{PDG2004}.  
The electric charge $|e|$ is the charge of an electron, $|e|$ = 1.602$\times10^{-19}$ 
Coulombs.}
\begin{center}
\begin{tabular}{|c|c|c|c|}
\hline
Force & Mediator & Electric & Mass (GeV) \\
      &          & Charge   &            \\
\hline
\hline
Strong & $g$ & 0 & 0 \\
\hline
Electromagnetic & $\gamma$ & 0 & 0 \\
\hline
Weak & $W^{\pm}$ & $\pm$1$|e|$ & 80.425(38) \\
     & $Z^{0}$   & 0      & 91.1876(21) \\
\hline
\end{tabular}
\label{tab:forces}
\end{center}
\end{table}

The fundamental particles which comprise all 
of the known matter in the universe are spin-1/2 fermions.  The particles 
which can interact strongly are called quarks and the ones which cannot are called leptons.  
For every charged lepton, there is a corresponding weakly interacting partner, the 
neutrino.  For example, the lightest charged lepton is the electron ($e$), and 
its corresponding neutrino partner is called the electron neutrino ($\nu_e$).  
The charged-and-neutrino lepton combination forms a family or generation.  
Two other families of leptons exist.  They are the muon ($\mu$) and tau ($\tau$) and 
the corresponding muon neutrino ($\nu_{\mu}$) and tau neutrino ($\nu_{\tau}$).  
The $\mu$ and $\tau$ have the same general 
properties as the electron except with larger masses.  
Just as the leptons can be formed into generations, 
the quarks are also grouped into generations.  The first generation of quarks consists 
of the up ($u$) and down ($d$) quarks, the second consists of the strange ($s$) and 
charm ($c$) quarks, and the third consists of the bottom ($b$) and top ($t$) quarks.  
The type of quark is also called the flavor of the quark.  
Table \ref{tab:particles} lists the properties of the fundamental fermions.  Each 
fundamental fermion has a corresponding antiparticle, which has the opposite charge.   

\begin{table}[h]
\caption[Fundamental particles in the Standard Model.]
{Fundamental particles in the Standard Model.  
The numerical values are taken from Ref. \cite{PDG2004}.  
The electric charge $|e|$ is the charge of an electron, 
$|e|$ = 1.602$\times10^{-19}$ Coulombs.  
The masses of the quarks are the so-called 'current quark masses'.  
The upper limits on the neutrino masses are at 90$\%$ confidence level.}
\begin{center}
\begin{tabular}{|c|c|c|c|c|c|c|}
\hline
 & \multicolumn{3}{|c|}{Leptons} & \multicolumn{3}{|c|}{Quarks} \\ 
\hline
Generation & Name & Electric & Mass & Name & Electric & Mass \\
or Family  &      & Charge   &      &      & Charge   &      \\
\hline
\hline
I & $e^-$     & $-1|e|$ & 511 keV  & $u$ & $+\frac{2}{3}|e|$ & 1.5$-$4 MeV \\
  & $\nu_{e}$ &  0  & $<$ 3 eV & $d$ & $-\frac{1}{3}|e|$ & 4$-$8 MeV \\
\hline
II & $\mu^-$     & $-1|e|$ & 106 MeV  & $c$ & $+\frac{2}{3}|e|$ & 1.15$-$1.35 GeV \\
   & $\nu_{\mu}$ &  0 & $<$ 0.19 keV & $s$ & $-\frac{1}{3}|e|$ & 80$-$130 MeV \\
\hline
III & $\tau^-$     & $-1|e|$ & 1.78 GeV & $t$ & $+\frac{2}{3}|e|$ & 174 GeV \\
    & $\nu_{\tau}$ &  0 & $<$ 18.2 eV & $b$ & $-\frac{1}{3}|e|$ & 4.1$-$4.4 GeV \\
\hline
\end{tabular}
\label{tab:particles}
\end{center}
\end{table}

As mentioned above, not all of the fundamental fermions participate in all three 
interactions.  
The quarks interact through all three interactions, 
the charged leptons only through the electromagnetic 
and weak, and neutrinos only via the weak.
The vector bosons mediating the interaction couple to the 'charges' of the particles.  
The most familiar type of charge is electric charge.  The propagator of the 
electromagnetic interaction, the photon, couples to the electric charge of the particle.  
The propagators of the weak interactions, the $W$ and $Z$, couple to the fermions 
through the so-called the 'weak' charge.  The strong force couples 
through the so-called the 'color' charge, 
first described by Greenberg \cite{greenbergcolor} in 
1964.  While there is only one type of electric charge (positive ($+$) and negative ($-$)), 
color has three charges denoted by red, blue, and green ($r,b,g$, and 
$\overline{r},\overline{b},\overline{g}$).  The strong interaction is described by 
the SU(3) symmetry group called color SU(3).  One possible representation of this 
symmetry group is
\begin{equation}
r\overline{g},~~r\overline{b},~~g\overline{b},~~ g\overline{r},~~ b\overline{r}, 
~~b\overline{g},~~ \sqrt{\frac{1}{2}}(r\overline{r}-g\overline{g}), 
~~\sqrt{\frac{1}{6}}(r\overline{r}+g\overline{g}-b\overline{b}), 
\label{eq:gluoncolor}
\end{equation}
where $r$, $\overline{r}$, $g$, $\overline{g}$, $b$, and $\overline{b}$, denote the red, 
antired, green, antigreen, blue, and antiblue color charges, respectively.  
Each color combination 
listed in Eqn. \ref{eq:gluoncolor} is ascribed to a gluon.

Another facet of the strong interaction is that free quarks do not exist in nature.  
Quarks bind together into particles 
called hadrons.  Hadrons have only been observed in two configurations; 
quark-antiquark pairs ($q\overline{q}$) and quark triplets ($qqq$).  The former 
are called mesons and have integer spin values, while the latter are called baryons 
and have half-integer spins.  Hadrons are color neutral.  For a given meson, 
the quark possesses one type of color charge while the antiquark possesses its 
anticolor.  The convention for baryons to be color neutral is for each quark to have a 
different color.  The mechanism for the absence of free quarks is called 
confinement, whose origin is related to the fact that gluons themselves carry color.  
Since gluons carry color, they can bind to each other.  This self-coupling 
phenomena is not present in electromagnetism because the photons are 
electrically neutral.  

The quantum theory describing the strong interaction is called Quantum Chromodynamics 
(QCD).  It is described by the QCD Lagrangian \cite{PDG2004} 
\begin{equation}
{\cal L} = -\frac{1}{4}F^a_{\mu\nu}F^{a\mu\nu} + 
\sum_{q}\overline{\psi_q}(i{\not\!D}-m_q)\psi_q,
\end{equation}
where $F^a_{\mu\nu}$ is the strength of the gluon field, 
$\psi_q$ is the quark wave function, 
$\not\!D$ is the covariant derivative, 
$m_q$ is the quark mass, 
and $a$ is an index for the three color charges.  
The strength of the gluon field is given by
\begin{equation}
F^a_{\mu\nu} = \partial_{\mu}A^a_{\nu} - \partial_{\nu}A^a_{\mu} 
+ g f^{abc}A^b_{\mu}A^c_{\nu},
\end{equation}
where $A^i_{\nu}$ $(i = a, b, c)$ are the gauge potentials of the gluon fields, 
$g$ is coupling constant for the gluon, 
and $f^{abc}$ are the structure constants of the gauge group.  The covariant derivative 
is given by 
\begin{equation}
\not\! D_{\mu} = \partial_{\mu} - i g\sum_{a}A^a_{\mu} t^a, 
\end{equation}
where $t^i$ are the gauge representation matrices.  The coupling 
constant $g$ is normally rewritten in terms of the strong coupling constant 
$\alpha_s$ = $g^2/4\pi$.  

The strong interaction can be characterized by an empirical 
potential.  A commonly used potential is the so-called Cornell potential 
\cite{Cornellpotential},
\begin{equation}
V_{strong} = \frac{\kappa}{r} + br,
\end{equation}
where $r$ is the interquark distance and $\kappa$ and $b$ are coefficients with units of 
\\(energy)$\cdot$(length) and (energy)/(length), respectively.  The $1/r$ 
part describes the standard Coulombic potential of the electromagnetic 
interaction.  The part proportional to $r$ describes the confinement aspect of the 
potential; the larger the distance between the quarks, 
the larger the force binding them together.  
Potentials are only a simple approximation to the theory of strong interactions which 
is described in terms of quantum field theory.

In 1974, Wilson \cite{OriginalLattice} showed how to quantize a gauge field theory 
on a discrete lattice in Euclidean space-time preserving exact gauge invariance.  
He applied this calculational technique to the strong coupling regime of QCD.  
In these Lattice Gauge calculations (Lattice QCD), space-time is replaced by a four 
dimensional hypercubic lattice of size $L^3T$.  The sites are separated by the lattice 
spacing $a$.  The quarks and gluons fields are defined at discrete points.  Physical 
problems are solved numerically by Monte Carlo simulations requiring only the quarks 
masses as input parameters. 

\begin{figure}[!tb]
\begin{center}
\includegraphics[width=13cm]{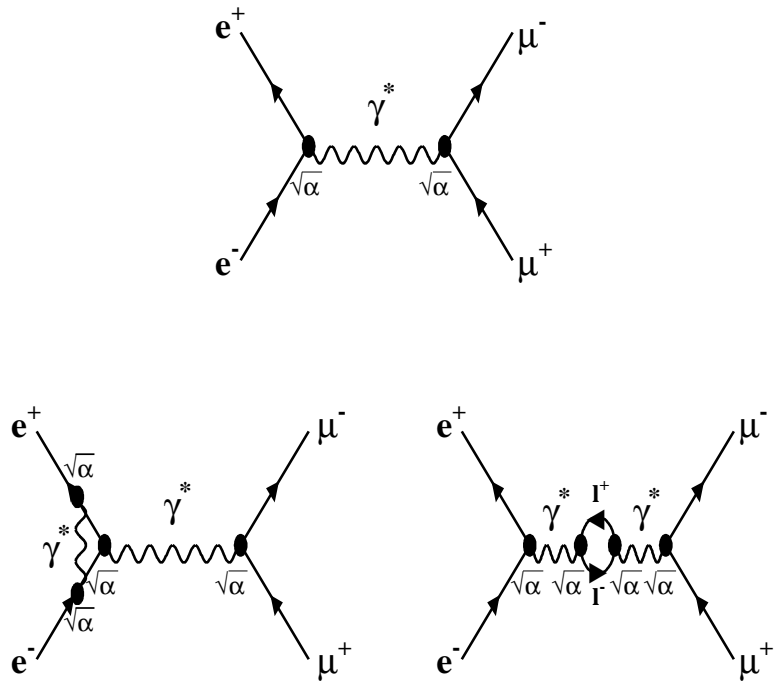}
\caption[Feymann diagrams for $\dimuon$.]
{Feymann diagrams for $\dimuon$.  The top figure is the lowest order process.  
The bottom figures are higher order corrections.}
\label{fig:feyperbex}
\end{center}
\end{figure}

An important simplification used with quantum field theories is the perturbative 
expansion.  Experimentally, only the initial and final states 
of an interaction are observed 
while the internal action is not.  Theoretical models and predictions are 
made to describe the nature of the unobserved interaction.  The simplest 
interaction is when a single mediating boson interacts between the initial and final 
states.  At each point that the mediator couples to a particle, a 
coupling constant is added the overall process.  
There are also processes which include more than one internal interaction.  The 
more the internal interactions, the larger the number of coupling constants that 
are included in the final process.  If the coupling constant is small, 
the more complicated internal processes (higher order processes) have a lower 
significance in the overall process.  
For example, the quantum theory describing the electromagnetic interaction, 
Quantum Electrodynamics (QED), has a coupling constant which is given by the 
fine-structure constant $\alpha$ = $e^2/(4\pi\epsilon_{0}\hbar c)$ = $1/137$, 
where $e$ is the charge of an electron, $\epsilon_{0}$ is the permittivity of free space, 
$\hbar$ is the Planck constant,and c is the speed of light.  
An example QED process is dimuon production from $\ee$ annihilations, i.e.,
\begin{equation}
\dimuon.
\end{equation}
Feymann diagrams for this process are shown in Figure \ref{fig:feyperbex}.  The lowest 
order term is the $\ee$ annihilating to a virtual photon followed by that photon producing a 
$\mumu$ pair.  Figure \ref{fig:feyperbex} also shows some example higher order processes.  
At each point where a photon couples to a fermion line, 
one order of $\sqrt{\alpha}$ is added to the process.  
Each of the higher order processes have an extra $\alpha$ term.  Their contribution 
to the overall process is at the percent level.  The higher order electromagnetic 
processes can therefore be neglected 
because the electromagnetic coupling constant $\alpha$ is small.

\begin{figure}[!tb]
\begin{center}
\includegraphics[width=15cm]{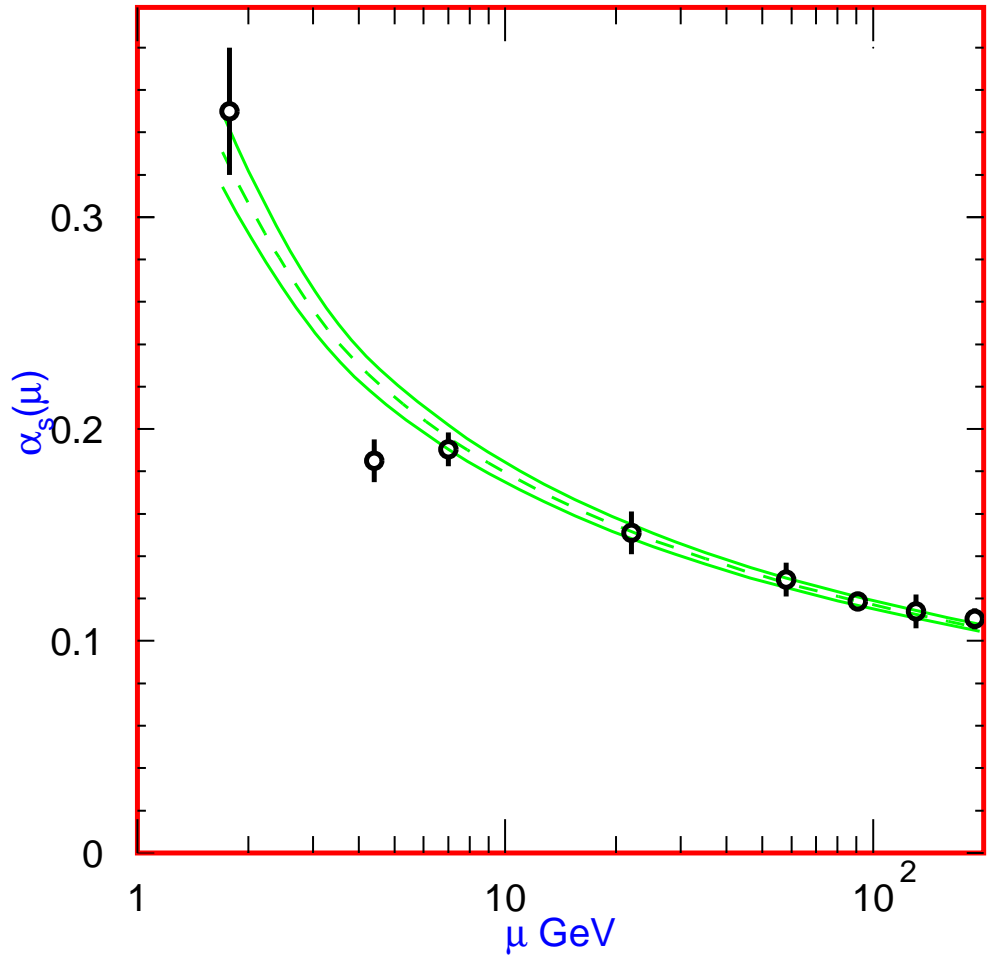}
\caption[Strong coupling constant as a function of energy.]
{Strong coupling constant as a function of energy \cite{PDG2004}.}
\label{fig:alphas}
\end{center}
\end{figure}

A major difficulty with QCD is the size of strong coupling constant.  
Gross and Wilczek \cite{asyfree1} and Politzer \cite{asyfree2} showed that the strong 
coupling constant is energy dependent ($\alpha_{s} \rightarrow \alpha_{s}(\mu)$, where 
$\mu$ is the energy scale), and it decreases with increasing energy.  
The one loop form of the strong coupling constant is 
\begin{equation}
\alpha_{s}(\mu) = \frac{4\pi}{\beta~\mathrm{ln}(\mu^2/\Lambda^2)},
\label{lab:ch1alphas}
\end{equation}
where $\beta =  11 - \frac{2}{3}n_{f}$, $n_f$ is the number of flavors, and 
$\Lambda = 0.2-0.3$ GeV is the QCD scale parameter.  
Figure \ref{fig:alphas} shows the variation of $\alpha_{s}(\mu)$ as a function 
of energy; it decreases from $\sim$0.25 at 3 GeV to $\sim$0.11 at 100 GeV.  
As $\mu \rightarrow \infty$, $\alpha_{s}(\mu) \rightarrow 0$.  
This behavior is called asymptotic freedom, 
and therefore QCD is said to be asymptotically free.  
For small $\alpha_{s}(\mu)$, perturbative calculations can be made, and the formalism 
is called Perturbative Quantum Chromodynamics (PQCD).  PQCD has been used to describe the 
electromagnetic form factors of hadrons, as described later.

Historically, the idea of form factors is related to the 'size' of subatomic particles, 
beginning with the determination of nuclear size by Rutherford 
in alpha scattering experiments.  
He found that the 'size' of the nuclei was of the order of 10 fermis.  
Once it was recognized that the 'size' should 
be measured by a probe which itself was 'sizeless', electron scattering became the 
means of choice.  The concept of form factors was formalized, with the form factor 
$F(\textbf{q}^2)$ defined as the multiplicative factor in 
\begin{equation}
\left(\frac{d\sigma}{d\Omega}\right) = 
\left(\frac{d\sigma}{d\Omega}\right)_{point}\times F(\textbf{q}^2),
\end{equation}
where $(d\sigma/d\Omega)_{point}$ is the differential cross section for 
scattering off a pointlike target and $\textbf{q}$ is the momentum transferred to the 
target.  It can be shown that with this definition the form factor is the Fourier 
transform of the charge density $\rho(\textbf{r})$
\begin{equation}
F(\textbf{q}^2) = \int {\mathrm d}^3{\mathrm r}~\rho(\textbf{r})
~{\mathrm{exp}}\left(i~\frac{\textbf{q}\cdot\textbf{x}}{\hbar}\right).
\end{equation}
For small momentum transfers, $F(\textbf{q}^2)$ only measures the 'size', or 
rms radius, of the charge distribution.  The classic experiments of Hofstadter 
and colleagues at Stanford \cite{Hofstadter1,Hofstadter2} showed that for large 
momentum transfers considerable more details of the charge and current distributions of 
the nuclei could be obtained.

The first measurements of electron scattering by nucleons were made by Hofstadter and 
colleagues at Stanford \cite{Hofstadter3,Hofstadter4} and by Wilson and colleagues at 
Cornell \cite{Wilson1}.  Since then, many more measurements, with much larger momentum 
transfers and higher precision, have been made.  Most of these are electron 
elastic scattering 
measurements in which the momentum transfer is spacelike.  Since the advent of the 
$\ee$ colliders, measurements in which a $\hhbar$ pair ($h$ = hadron) is produced 
in an $\ee$ annihilation have also been reported.  In these measurements, momentum 
transfer is timelike.  However, these measurements have 
been generally confined to small momentum transfers and have poorer precision.  
We describe these in detail in the following.

In the commonly used metric, momentum transfers are defined as 
\begin{displaymath}
Q^2 = -q^2 = t = (p_{1} - p_{2})^2,~~~~~~\mathrm{spacelike}~Q^2
\end{displaymath}
\begin{equation}
-Q^2 = q^2 = s = (p_{1} + p_{2})^2,~~~~~\mathrm{timelike}~Q^2
\end{equation}
where the subscripts 1 and 2 refer to the colliding particles and $q$ and $p_i$ are 
four-momenta.  For protons, the differential cross section for the timelike momentum 
transfer, $|Q^2| = s$, is described in terms of two form factors, $F^P_1(s)$ and 
$F^P_2(s)$, by \cite{cabibbogatto}
\begin{displaymath}
\frac{d\sigma}{d\Omega} = 
\frac{\alpha^2\beta_p}{4s}~[~|F^P_{1}(s) + \kappa_{p}F^P_{2}(s)|^{2}
(1 + \mathrm{cos}^{2}\theta)~~~~~~~~~~~~~~~~~~~~~ 
\end{displaymath}
\begin{equation}
~~~~~~~~~~~~~~~+ \left(\frac{4m^2_{p}}{s}\right)\left|F^P_{1}(s) + 
\left(\frac{s}{4m^2_{p}}\right)\kappa_{p}F^P_{2}(s)\right|^{2}\mathrm{sin}^{2}\theta~],
\label{eq:prf1f2diffcs}
\end{equation}
where $\beta_{p}$ is the proton velocity measured in the center-of-mass system of the 
annihilation, 
$\kappa_{p}$ is the anomalous magnetic moment of the proton, 
$m_p$ is the mass of the proton, and $\theta$ is the angle between the incident positron 
and the produced proton.  The form factor $F^P_1(s)$ is called the Dirac form factor 
and relates to both the electric and magnetic scattering from a spin-1/2 Dirac particle, 
and $F^P_2(s)$, called the Pauli form factor, is related to the additional magnetic 
scattering contribution arising from the anomalous part of the proton magnetic moment.  
It has become conventional to use the so-called Sachs form factors $G^P_E(s)$ and 
$G^P_M(s)$ instead of $F^P_1(s)$ and $F^P_2(s)$, with 
\begin{equation}
G^{P}_{E}(s) = F^{P}_{1}(s) + \left(\frac{s}{4m^{2}_{p}}\right) \kappa_{p}F^{P}_{2}(s) 
~~~~~~~~G^{P}_{M}(s) = F^{P}_{1}(s) + \kappa_{p}F^{P}_{2}(s), 
\label{eq:prffdefs}
\end{equation}
so that 
\begin{equation}
\frac{d\sigma}{d\Omega} = 
\frac{\alpha^2\beta_p}{4s}[~\prmagff^{2}~(1 + \mathrm{cos}^{2}\theta) + 
\left(\frac{4m^{2}_{p}}{s}\right)\prelecff^{2}~\mathrm{sin}^{2}\theta~].
\label{eq:prrldiffcs}
\end{equation}
The normalization is done at $Q^2 = 0$, with 
\begin{displaymath}
F^{P}_{1}(0) = 1,~~~~~F^{P}_{2}(0) = 1,~~~~~~~~~~~~~~~~~~\mathrm{and} 
\end{displaymath}
\begin{equation}
G^{P}_{E}(0) = 1,~~~~~G^{P}_{M}(0) = 1+\kappa_{p} = \mu_p = 2.79, 
\end{equation}
where $\mu_p$ is the magnetic moment of the proton in units of the nuclear magneton.  

For spin-0 charged mesons, e.g., the pion and kaon, there is no magnetic scattering 
and only the electric form factor survives, with \cite{cabibbogatto}
\begin{equation}
\frac{d\sigma}{d\Omega}(\eetomm) = 
\frac{\alpha^2}{8s}~\beta^3_m~|F_m(s)|^2~\mathrm{sin}^{2}\theta, 
\label{eq:mffdiffcs}
\end{equation}
where the $\beta^3_m~\mathrm{sin}^{2}\theta$ dependence is a direct consequence of the 
fact that the meson pair must be produced in a p-wave state.  The normalization is 
\begin{equation}
F_m(0) = 1.
\end{equation}

The earliest attempts to understand electromagnetic form factors were in terms of the 
Vector Dominance Model (VDM).  In VDM it is assumed that the photon (from $\ee$ 
annihilation, for example) 'converts' into a vector meson, and the vector meson 
interact hadronically with the hadron whose electromagnetic structure is 
being probed.  The model was originally invented to understand the relation between 
$\rho \rightarrow \ee$ and $\rho \rightarrow \pipi$, and therefore the pion form factor 
near the mass of the $\rho$ meson \cite{GounarisSakurai}.  It was later extended to larger 
energies by including known and hypothesized recurrences of the $\rho$ and $\phi$ mesons 
\cite{Dubnickaetal1,Dubnickaetal2}.  Examples of recent extensions of VDM are 
Ref. \cite{deMeloetal} and Ref. \cite{Bruchetal}.  The VDM calculations of the form 
factors cannot be called predictions; they are fits to the existing experimental 
data at relatively low momentum transfers, 
and involve a large number of parameters (32 and 15 for the pion 
form factor predictions in Ref. \cite{deMeloetal} and Ref. \cite{Bruchetal}, 
respectively, and 26 for the kaon form factor prediction in Ref. \cite{Bruchetal}).  
The available data in the large momentum transfer region has been either 
non-existent or of very poor quality, as discussed later.  
It is worth noting that in the limit of flavor SU(3) invariance, $F_\pi(s) = F_K(s)$, 
and attempts to take account of SU(3) breaking do not lead to any large deviations from 
this \cite{Bruchetal,ktlff_VDM}.  In this dissertation we will not discuss VDM 
predictions any further, and will concentrate on QCD-based models for form factors.

The earliest attempts to describe the $Q^2$ variation of the electromagnetic form factors 
in terms of QCD were made by Brodsky and Farrar \cite{ffscaling1,ffscaling2} and 
Matveev, Muradyan, and Tavkhelidze \cite{ffscaling3}.  Their 'dimensional scaling' 
considerations lead to the prediction that 
exclusive scattering 
scales as $s^{2-n}$, where $n$ is the total number of leptons, photons, and 
quark components, i.e., elementary fields, in the initial and final states.  This directly 
leads to the so-called 'quark counting rule' prediction that 
\begin{equation}
F_{h}(Q^2) \sim (Q^2)^{1-n_q},
\end{equation}
where $n_q$ is the number of quarks contained in the hadron.  Thus, the 
form factor for pions and kaons ($n_q$ = 2) scales as 
\begin{equation}
F_{\pi,K}(Q^2) \sim (Q^2)^{-1}
\end{equation}
and the proton ($n_q$ = 3) scales as
\begin{equation}
G^{P}_{E}(Q^2) = G^{P}_{M}(Q^2) \sim (Q^2)^{-2}.
\end{equation}
It is important to remember that 'dimensional scaling' or the 'quark counting rule' is 
strictly valid only for $|Q^2| = s \rightarrow \infty$.

Farrar and Jackson \cite{FarrarJackson_PionPQCD} obtained the same $Q^2$ behavior 
for the pion form factor by solving the light-cone pion Bethe-Salpeter equation 
in QCD, with the additional result relating $F_\pi(Q^2)$ to the pion decay constant.  
Lepage and Brodsky \cite{LepageBrodsky_PQCDFF} obtained the same result in a more 
systematic analysis of PQCD with the 'factorization' hypothesis.  In this model, 
exclusive processes can be described in terms of two factorizable parts - a necessarily 
nonperturbative part involving the wave functions of the 
initial and final states of the hadron 
and a hard scattering part which can be described 
perturbatively.  The latter leads to both the 'quark counting rule' behavior and the 
connection of the $\pi$ and $K$ form factors to their decay constants.  The result 
is that asymptotically ($Q^2 \rightarrow \infty$) the pion and kaon form factor are 
\begin{equation}
Q^2~F_{\pi,K}(Q^2) \rightarrow 8\pi~\alpha_s(Q^2)~f^2_{\pi,K},
\end{equation}
where $f_{\pi}$ = 130.7 $\pm$ 0.4 MeV \cite{PDG2004} is the pion decay constant, and 
$f_{K}$ =  159.8 $\pm$ 1.5 MeV \cite{PDG2004} is the kaon decay constant.  This 
prediction gives an absolute normalization to the pion and kaon form factor.  
For the proton, the asymptotic behavior of the magnetic form factor, neglecting leading 
logarithms, is 
\begin{equation}
Q^4~G^P_M(Q^2) \rightarrow C~\alpha^2_s(Q^2),
\end{equation}
where $C$ is an arbitrary constant.  
This prediction is not absolutely normalized but it does contain 
the $\sim$ $Q^{-4}$ behavior predicted by the 'quark counting rule'.
It was argued \cite{LepageBrodsky_PQCDFF} that these predictions were consistent 
with the then existing form factor data for $Q^2 > 5$ GeV$^2$.

Objections have been raised that the above predictions can not applied to the existing 
data because, at the available momentum transfers, the asymptotic regime has not 
been reached.  Isgur and Llewellyn Smith \cite{Isgur_LS_argue1,Isgur_LS_argue2} and 
Radyushkin \cite{Rad_argue} have argued that the perturbative part of the form factor 
can only describe $\sim$10$\%$ 
of the cross sections and that the other 90$\%$ 
is dominated by nonperturbative, or 'soft', processes.  A more detailed description of 
theoretical considerations of the electromagnetic form factors are given in Chapter 2.  
 
\begin{figure}[htbp]
\begin{center}
\includegraphics[width=14.0cm]{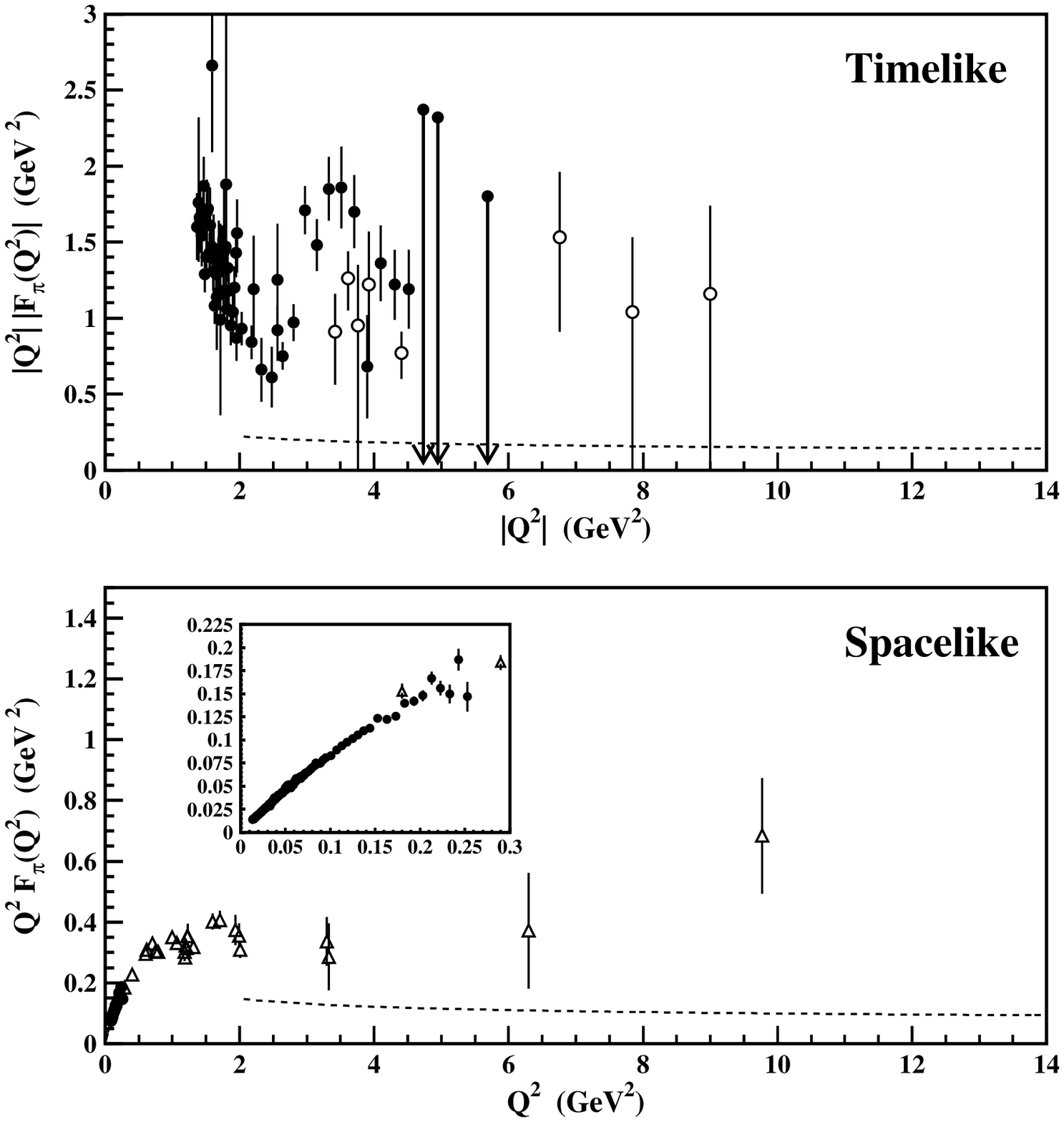}
\caption[Experimental status of the charged pion electromagnetic form factor.]
{Experimental status of the charged pion electromagnetic form factor for 
timelike (top) and spacelike (bottom) momentum transfers.  
The solid points in the top plot are measurements where $\pipi$ events 
are positively observed \cite{kpicommontlff_1}-\cite{pitlff_3}.  
The open points in the top plot are measurements where $\pipi$ and $\KK$ events 
are observed and VDM predictions are used to determine the number 
of $\pipi$ events \cite{pitlff_VDM}.  Results are only shown in the 
region $|Q^2| >$ 1.37 GeV$^2$ in the top plot as to exclude the $\rho$ resonance. 
The solid points in the bottom plot are from $\pi-e$ scattering experiments 
\cite{pislff_sc1}-\cite{pislff_sc4} and 
the open triangles are from experiments using electroproduction of charged pions 
\cite{pislff_elprod1}-\cite{pislff_elprod6}.  
The dashed lines in the top and bottom plot are the PQCD predictions by 
Brodsky $\etal$ \cite{Brodskyetal_tlPQCD} and Lepage and 
Brodsky \cite{LepageBrodsky_PQCDFF}, respectively.}
\label{fig:pi_tlsl}
\end{center}
\end{figure}

\begin{figure}[!tb]
\begin{center}
\includegraphics[width=7cm]{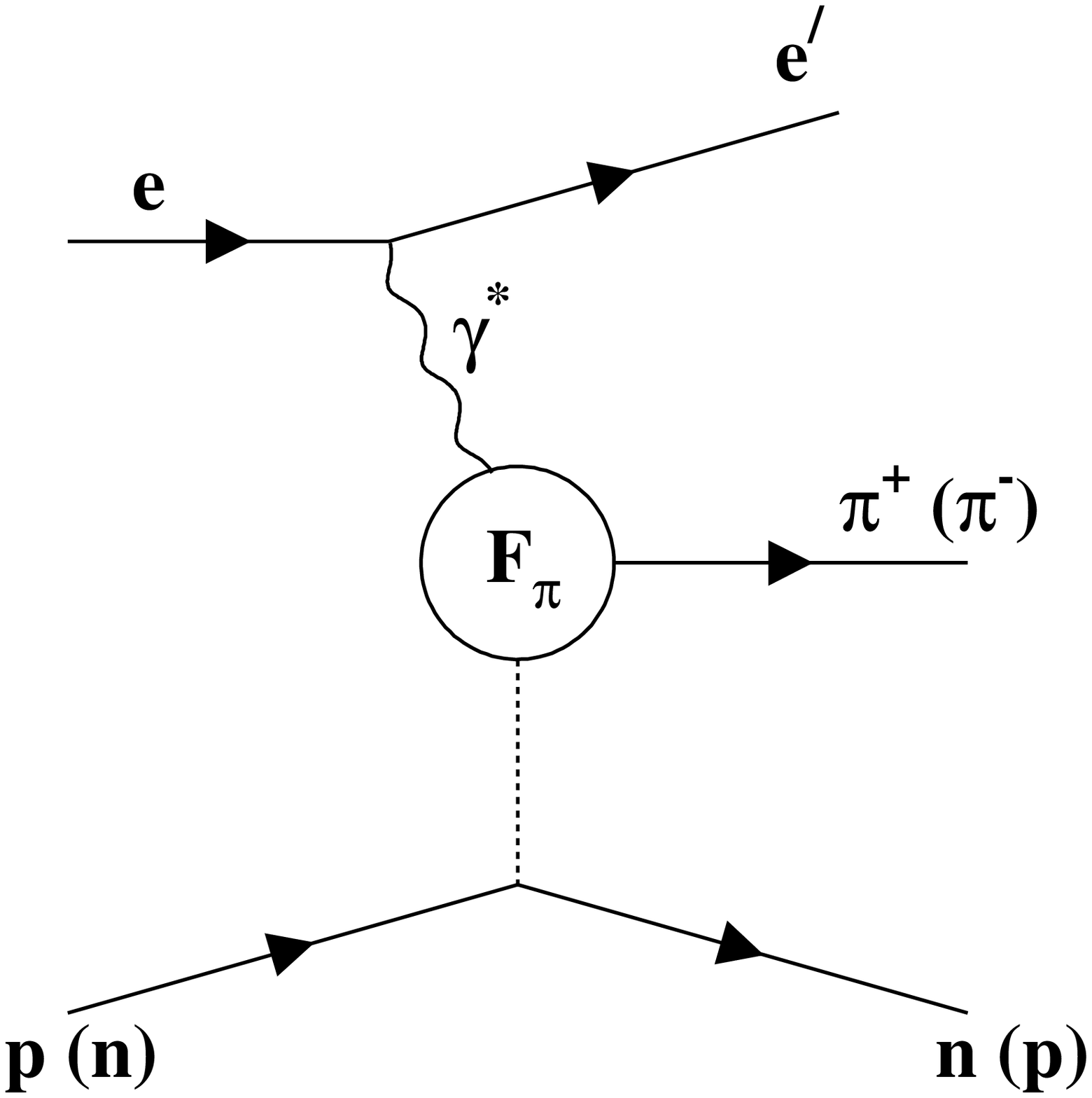}
\caption[Feymann diagram for electroproduction of charged pions.]
{Feymann diagram for electroproduction of charged pions.}
\label{fig:feyelpiff}
\end{center}
\end{figure}

The experimental data for the charged pion electromagnetic form factor with 
timelike and spacelike momentum transfers are shown in Figure \ref{fig:pi_tlsl}.   
The spacelike pion form factor is measured using two different techniques.  The first, 
and cleaner, method is by elastically scattering charged pions 
off electrons bound in atomic targets through the reaction
\begin{equation}
\pi^{-}~e^{-} \rightarrow \pi^{-}~e^{-}.
\end{equation}
This method is limited to the determination of the charged radius of the pion 
because of the very small 
momentum transfers which can be achieved.  For example, even with a 300 GeV beam, 
$Q^2$(max) = 0.12 GeV$^2$ was realized \cite{pislff_sc4}.
The second method is by pion production off nucleons in the reactions 
\begin{displaymath}
e^{-}~p \rightarrow ~e^{-}~\pi^{+}~n,
\end{displaymath}
\vspace*{-1.2cm}
\begin{equation}
e^{-}~n \rightarrow ~e^{-}~\pi^{-}~p.
\end{equation}
The incoming electron in these reactions emits 
a virtual photon and the virtual photon interacts with the 'pion cloud' surrounding 
the nucleon.  The Feynman diagram governing the lowest order process 
is shown in Figure \ref{fig:feyelpiff}.  Pion electroproduction for the form factor 
determination, first proposed by Frazer \cite{Frazer}, has recently come under 
considerable criticism.  Carlson and Milana  \cite{carlsonmilana_elprod} 
and others \cite{StermanStoler} have pointed out that the electromagnetic 
form factor of the pion is not well determined from electroproduction experiments.  
The concerns arise due to the magnitude of the pion-nucleon form factor, the other 
competing uncalculated QCD processes, and the inability of many measurements to 
separate the 
transverse and longitudinal parts of the measured cross section.  
The uncertainties in the 
pion form factor measured from electroproduction experiments range from 
$\pm23\%$ to $\pm51\%$ 
for $Q^2 >$ 3 GeV$^2$ \cite{pislff_elprod4}.

The determination of the pion form factor in the timelike region has 
different problems.  The timelike form factor is measured in the reaction 
\begin{equation}
\ee \rightarrow \pipi,
\end{equation}
and it is theoretically a well defined reaction \cite{cabibbogatto}.  
However, because of the experimental problems, 
no direct measurements of the pion form factor 
are available with $|Q^2|$ $>$ 4.5 GeV$^2$.  
In the only existing measurements for $|Q^2|$ $>$ 4.5 GeV$^2$ 
meson pairs ($\pipi$ and $\KK$) were observed, but it was not possible 
to individually identify the $\pipi$ events \cite{pitlff_VDM}. 
The pion form factors were determined 
using VDM predictions, as described earlier, to divide the number of observed 
meson pair events into an approximately equal number of $\pipi$ and $\KK$ events.  
Without considering the effect of using a theoretical model to 
determine form factors, the uncertainties in the measurements 
are by themselves quite large, $\pm$41$\%$ at $|Q^2|$ = 6.76 GeV$^2$ and 
$\pm$100$\%$ with $|Q^2| >$ 7 GeV$^2$.

\begin{figure}[htbp]
\begin{center}
\includegraphics[width=14.5cm]{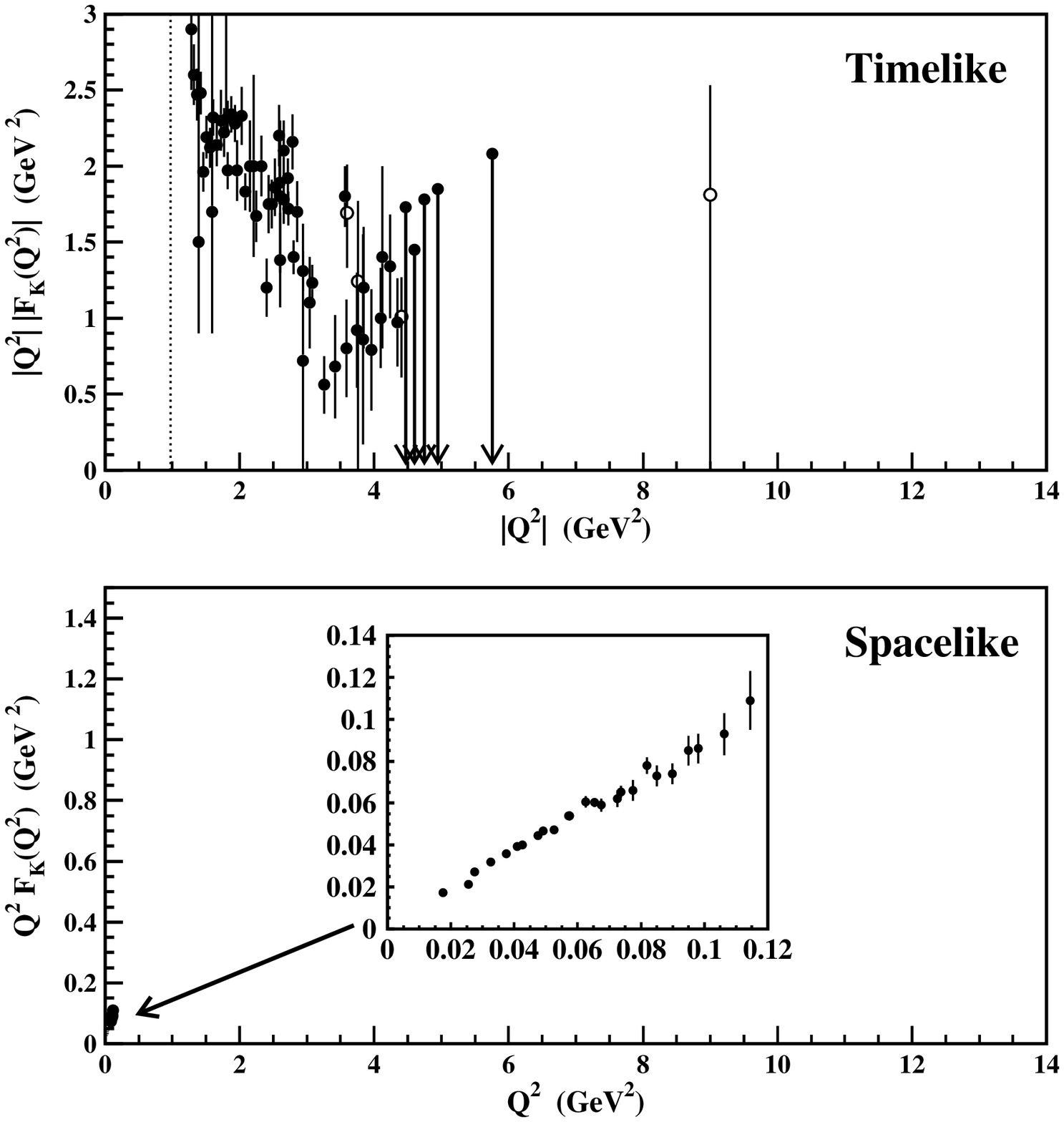}
\caption[Experimental status of the charged kaon electromagnetic form factor.]
{Experimental status of the charged kaon electromagnetic form factor for 
timelike (top) and spacelike (bottom) momentum transfers.  
The solid points in the top plot are measurements where $\KK$ events are 
positively observed \cite{kpicommontlff_1}-\cite{kpicommontlff_3}, 
\cite{ktlff_1}-\cite{ktlff_4}.  
The open points in the top plot are measurements where $\pipi$ and $\KK$ events 
are observed and VDM predictions were used 
to determine the number of $\KK$ events \cite{ktlff_VDM}.  
Results are only shown in the region $|Q^2| >$ 1.28 GeV$^2$ in the top plot 
as to exclude the $\phi$ resonance.  
The solid points in the bottom plot are data from $K-e$ scattering experiments 
\cite{kslff_sc1,kslff_sc2}.  
The vertical dotted line in the top plot represents the $\KK$ production threshold 
($|Q^2| = (2m_{K})^2$ = 0.975 GeV$^2$).}
\label{fig:k_tlsl}
\end{center}
\end{figure}

The experimental data for the charged kaon electromagnetic form factor 
with timelike and spacelike momentum transfers are shown in Figure \ref{fig:k_tlsl}.  
As for the pion form factor, the kaon form factor in the spacelike region is measured 
by elastically scattering charged kaons from electrons bound in atomic targets 
through the reaction
\begin{equation}
K^{-}~e^{-} \rightarrow K^{-}~e^{-}.
\end{equation}
The highest $Q^2$ for spacelike momentum transfer is 0.115 GeV$^2$, 
and it is used only to determine the kaon charge radius.  The form factor in the timelike 
region is measured in the reaction 
\begin{equation}
\ee \rightarrow \KK.
\end{equation}
As with the pion form factor, no direct measurements of the 
kaon form factor are available with $|Q^2|$ $>$ 4.5 GeV$^2$.  Only one measurement 
exists above $|Q^2|$ $>$ 4.5 GeV$^2$, and once again, the experiment observed 
meson pairs ($\pipi$ and $\KK$) but was not able to individually identify $\KK$ 
events \cite{ktlff_VDM}.  The kaon form factors were determined 
using VDM predictions to divide the total number of observed hadronic pair 
events (2 to 8 events) into 1 and 4 events of $\KK$ \cite{ktlff_VDM}.

\begin{figure}[htbp]
\begin{center}
\includegraphics[width=15cm]{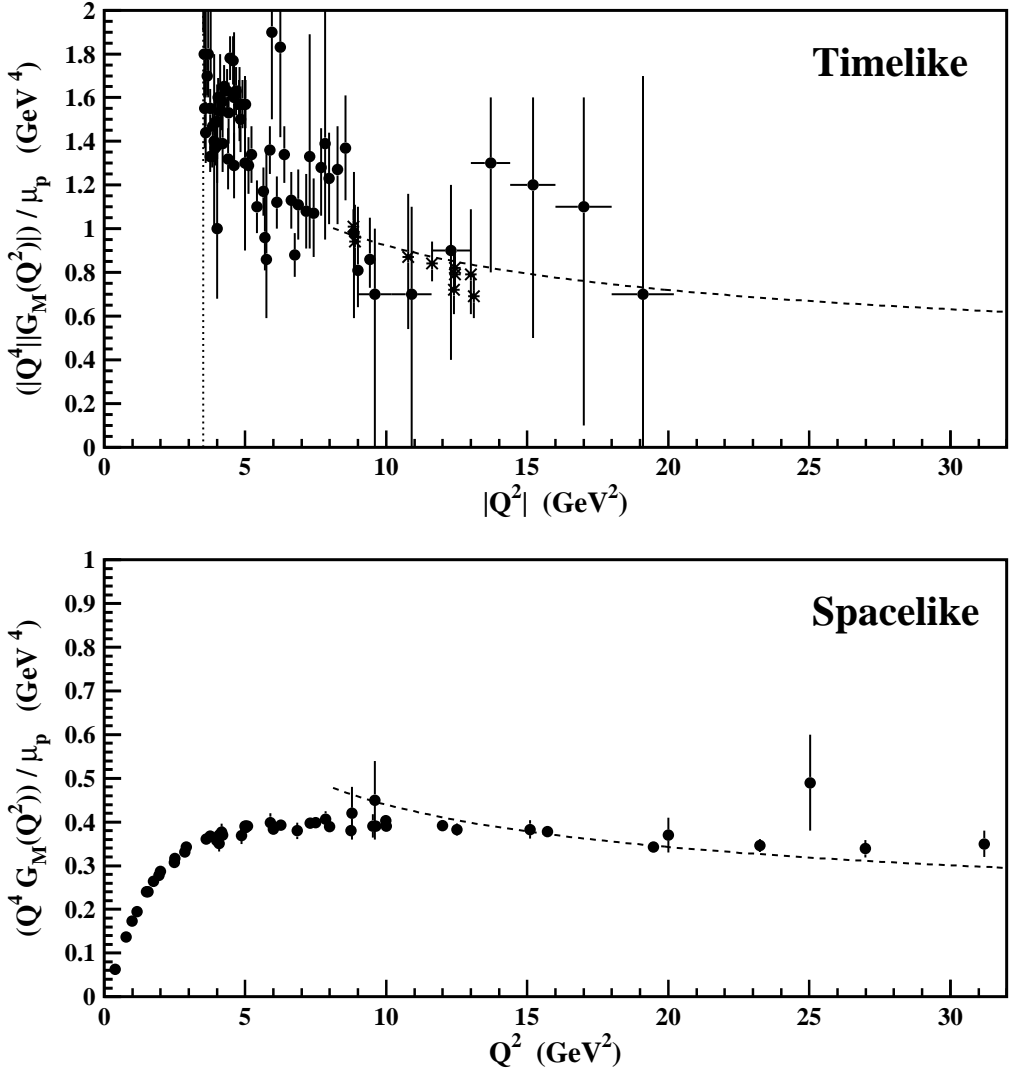}
\caption[Experimental status of the proton magnetic form factor.]
{Experimental status of the proton magnetic form factor for timelike (top) 
and spacelike (bottom) momentum transfers.  
The solid points for timelike momentum transfer are from Refs. 
\cite{prtlff_2}-\cite{BABARppbarISR}, while the stars are 
the E760/E835 measurements from $\ppbartoee$ events \cite{e835_1}-\cite{e835_3}.
The timelike data assumes $|G^P_E(Q^2)|$ = $|G^P_M(Q^2)|$.  
The data points for spacelike momentum transfer are from Refs. 
\cite{prslff_gmonly1}-\cite{prslff_gmonly6} assuming $G^P_E(Q^2)$ = $G^P_M(Q^2)/\mu_p$.
The vertical dotted line in the top plot represents the $\ppbar$ production 
threshold ($|Q^2| = (2m_{p})^2$ = 3.52 GeV$^2$).  The dashed lines are the 
arbitrarily-normalized PQCD prediction 
$|Q^4||G^P_M(Q^2)| \propto \alpha^2_s(|Q^2|)$ \cite{LepageBrodsky_PQCDFF}.}
\label{fig:pr_tlsl}
\end{center}
\end{figure}

The experimental data for the proton electromagnetic form 
factor is shown in Figure \ref{fig:pr_tlsl}.  
As shown in Eqn. \ref{eq:prrldiffcs}, for large $Q^2 = s$, $G^P_M(Q^2)$ overwhelms 
$G^P_E(Q^2)$, and it becomes impossible to separately measure 
the electric form factor $G^P_E(Q^2)$ and the magnetic form factor $G^P_M(Q^2)$.  
In the results shown in Figure \ref{fig:pr_tlsl}, 
$G^P_E(Q^2)$ = $G^P_M(Q^2)/\mu_p$ is assumed.  
The proton magnetic form factor 
in the spacelike region has been measured up to $Q^2$ = 31.2 GeV$^2$ with an uncertainty 
of $\sim$10$\%$ from elastically scattering electrons off proton 
targets \cite{prslff_gmonly6}
\begin{equation}
e^{-}~p \rightarrow e^{-}~p.
\end{equation}
$G^P_E(Q^2)$ and $G^P_M(Q^2)$ separation was done only for $Q^2 <$ 8.83 GeV$^2$ 
using the Rosenbluth technique.  
In the timelike region, the proton form factors are measured by $\ee$ annihilation
\begin{equation}
\ee \rightarrow \ppbar,
\end{equation}
and in the time-reversed $p\overline{p}$ annihilation to the $\ee$ final state
\begin{equation}
\ppbar \rightarrow \ee.
\end{equation}
The highest energy measurement from direct $\ee$ annihilations is 
$|Q^2|$ = 9.42 GeV$^2$ and has a 22$\%$ uncertainty \cite{prtlff_9}, while measurements 
from $\ppbar$ annihilations go up to $|Q^2|$ = 13.11 GeV$^2$ with a 
14$\%$ uncertainty \cite{e835_2}.  Recently, a BaBar experiment 
has reported measurement of the proton from factor from $\ee$ annihilation 
produced via initial state radiation \cite{BABARppbarISR} in the reaction
\begin{equation}
\ee \rightarrow \gamma (\ee) \rightarrow \gamma (\ppbar).
\end{equation}
They were able to measure the timelike form factor up to $|Q^2|$ $\sim$ 20 GeV$^2$ 
but with uncertainties between $\pm40\%$ and $\pm100\%$ in large $Q^2$ bins.  
Figure \ref{fig:pr_tlsl} also shows that, at the present momentum transfers, 
the timelike form factor is approximately twice as large as the spacelike form factor.

\begin{figure}[htbp]
\begin{center}
\includegraphics[width=15.0cm]{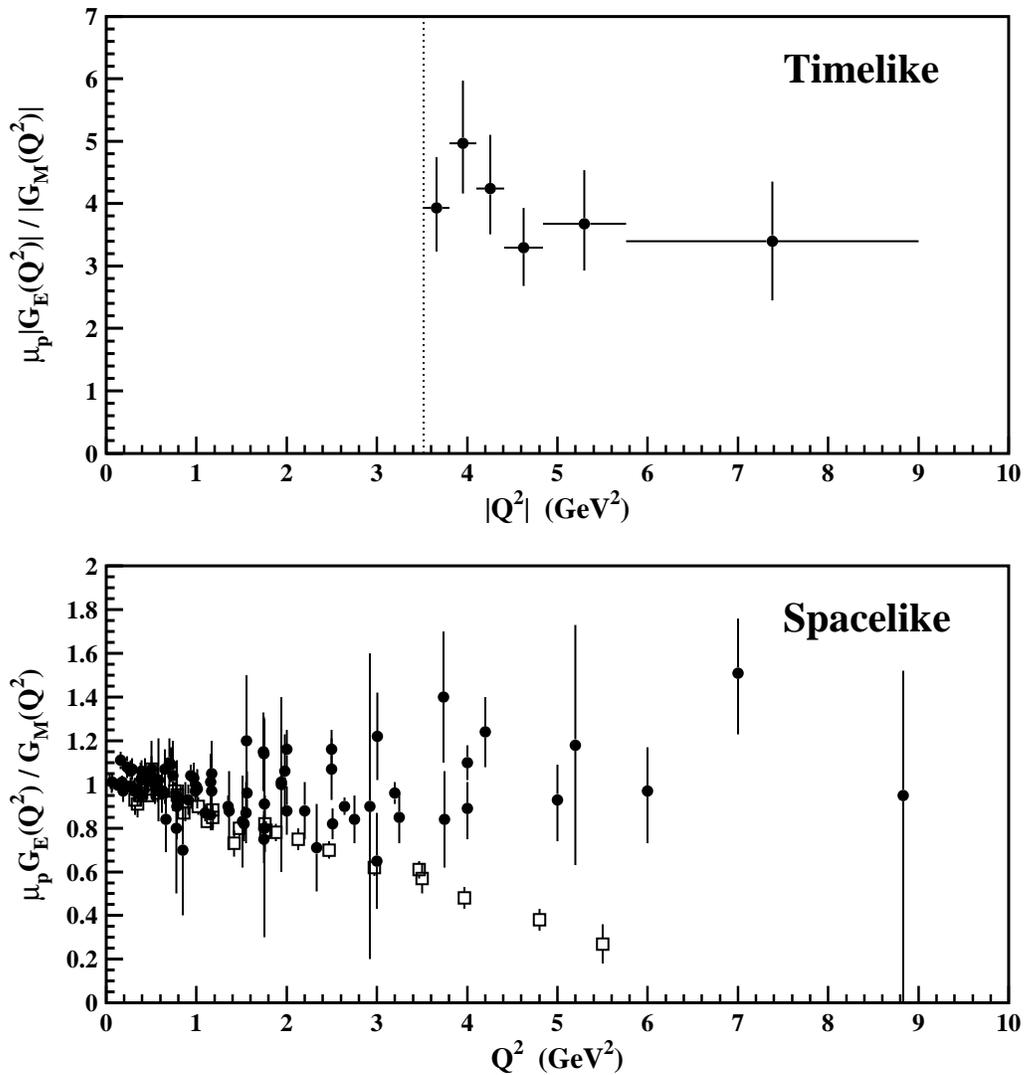}
\caption[Experimental results for the ratio $\mu_pG^P_E(Q^2)/G^P_M(Q^2)$ in the 
timelike and spacelike regions.]
{Experimental results for the ratio $\mu_pG^P_E(Q^2)/G^P_M(Q^2)$ in the timelike (top) 
and spacelike (bottom) regions.  
Top: The solid points are the experimental results from Ref. \cite{BABARppbarISR}.  
The vertical dotted line in the top plot represents the $\ppbar$ production 
threshold ($|Q^2| = (2m_{p})^2$ = 3.52 GeV$^2$).  
Bottom: The open squares are from the polarization 
transfer \cite{prslff_pol1}-\cite{prslff_pol4} measurements, and the solid 
points are from the Rosenbluth separation \cite{prslff_ros1}-\cite{prslff_ros13} 
measurements.  
Note that the polarization transfer results show a monotonic decrease in the ratio, 
while the Rosenbluth results are all consistent with the ratio being constant $\approx$ 1.} 
\label{fig:prtlslgegmintro}
\end{center}
\end{figure}

The $G^P_E(Q^2)/G^P_M(Q^2)$ ratio for the proton has been measured in the 
spacelike region with $Q^2 <$ 9 GeV$^2$ using two different methods, Rosenbluth 
separation and polarization transfer.  The differential cross section for Rosenbluth 
separation \cite{rosenbluthsep} is 
\begin{equation}
\left(\frac{d\sigma}{d\Omega}\right)_{exp} 
= \left(\frac{d\sigma}{d\Omega}\right)_{point} 
\times\left(~\frac{[G^P_E(Q^2)]^2 + \tau[G^P_M(Q^2)]^2}{1+\tau} 
+ 2\tau[G^P_M(Q^2)]^2~\mathrm{tan}^2\frac{\theta_e}{2}~\right),
\end{equation}
where $\tau$ = $\frac{Q^2}{4m^2_p}$ and $\theta_e$ is defined as the angle between the 
incident and scattered electron. Precision measurements of $G^P_E(Q^2)$ in 
Rosenbluth separation experiments are limited because the overall cross section becomes 
less sensitive to it with increasing momentum transfer.  
The measurements using polarization transfer consist of elastically scattering 
of polarized electrons off proton targets.  Scattering of longitudinally 
polarized electrons 
off unpolarized proton targets gives the recoiling proton either 
polarization transverse to ($P_{t}$), or in the same longitudinal direction ($P_{l}$), 
of the recoiling proton, with respect to the 
scattering plane defined by the incident electron and the recoiling proton.  The 
$G^P_E/G^P_M$ ratio at a given $Q^2$ is determined 
in the polarization transfer experiments as \cite{poltrans_th1,poltrans_th2}
\begin{equation}
\frac{G_E}{G_M} = 
\frac{P_{t}}{P_{l}}\frac{E+E^{\prime}}{2m_p}\mathrm{tan}\frac{\theta_e}{2},
\end{equation}
where $E$ and $E^{\prime}$ are the energies of the incident and scattered electron, 
respectively.
The results are shown in Figure \ref{fig:prtlslgegmintro}, 
where the ratio is plotted as $\mu_p\prelecffq/\prmagffq$.  

As shown in Figure \ref{fig:prtlslgegmintro}, $\mu_p\prelecffq/\prmagffq$ 
measured by the polarization transfer experiments differs from those 
using Rosenbluth separation. The ratio determined from Rosenbluth 
separation is consistent with the electric and magnetic form factors being equal, or 
$\prelecffq$ = $\prmagffq/\mu_p$, while the ratio from polarization transfer decreases 
linearly.  
A linear extrapolation of the polarization transfer measurements would lead to 
$G^P_E$ = 0 at $Q^2$ $\sim$ 7.5 GeV$^2$.  The only measurement of the electric-to-magnetic 
form factor 
ratio in the timelike region is from the BaBar experiment \cite{BABARppbarISR}, which 
measures the angular distribution of $\ppbar$ events produced in $\ee$ annihilations via 
initial state radiation.  As shown in Figure \ref{fig:prtlslgegmintro}, BaBar finds 
that the ratio in the $|Q^2|$ range 4.4$-$9.0 GeV$^2$ is consistent 
with $|G^P_E(Q^2)|/|G^P_M(Q^2)|$ = 1, which is predicted by Eqn. 
\ref{eq:prffdefs} for the $\ppbar$ threshold condition, $|Q^2| = s = 4m_{p}^2$.  
The form factor ratio in the timelike region is also given by Eqn. \ref{eq:prrldiffcs}, 
and therefore it contains the same sensitive issue 
for extracting $|\prelecffq|$ by the Rosenbluth separation method.

We have used the data collected with the CLEO detector at the 
Laboratory of Elementary Particle Physics in Ithaca, NY, to measure 
the electromagnetic form factors of the 
charged pion, charged kaon, and proton with timelike momentum transfer.  
The Cornell Electron Storage Ring (CESR) began colliding $\ee$ beams 
in 1979 at a center-of-mass energy of $\sqrt{s} \sim$ 10 GeV.  The original CLEO detector, 
CLEO I, was constructed to study quarkonium spectroscopy by means of $\ee$ annihilations.  
The CLEO I detector was upgraded to CLEO II, CLEO II.V, and CLEO III 
between 1979 and 2003.  In 2003, CESR was redesigned to run in the 
lower energy region of $\sqrt{s}$ = 3-5 GeV, and the CLEO III detector 
was modified to the present CLEO-c detector.

The CLEO-c detector collected 20.7 pb$^{-1}$ of $\ee$ annihilation data at a 
center-of-mass energy $\sqrt{s}$ = 3.671 GeV, or $|Q^2|$ = 13.48 GeV$^2$, between January 
and April of 2004.  This data sample is used in this dissertation to measure 
the timelike electromagnetic form factors of the charged pion, charged kaon, 
and proton at $|Q^2|$ = 13.48 GeV$^2$.

Before presenting the results of our measurements, 
in Chapter 2, we discuss the current theoretical interpretations 
of the electromagnetic form factors.  
In Chapter 3 we describe the Cornell Electron Storage Ring and the CLEO-c detector 
used to produce and observe the $\ee$ annihilations to the exclusive final 
states $\pipi$, $\KK$, and $\ppbar$.  
In Chapter 4 we describe our measurements and the analysis procedure for determining the 
respective electromagnetic form factors.  
Finally, in Chapter 5 we summarize the experimental results and 
compare them to existing experimental results and theoretical predictions.  
We conclude by discussing the impact of our measurements on the validity of applying 
PQCD at $|Q^2|$ $\sim$ 10 GeV$^2$.  
Appendix A lists the individual event properties for events which satisfy the 
$\eetopipi$ and $\eetoppbar$ selection criteria. 
Appendix B lists the existing experimental results of the pion, kaon, 
and proton form factors used throughout this dissertation.

\baselineskip=24pt
\chapter{Form Factors - Theoretical}

In this chapter I review the present status of the theoretical understanding of the 
electromagnetic form factors of hadrons.  As always, the development of theoretical 
models is closely related to the availability of experimental data.  Unfortunately the 
available experimental data for $\pi$, $K$, and $p$ are limited in two ways.  Very little 
data are available for timelike or spacelike form factors for large momentum 
transfers \\ ($>$ 5 GeV$^2$), and the few data which are available generally suffer from 
very large, up to $\pm$100$\%$, uncertainities.  Because of these limitations theoretical 
models have generally concentrated on $Q^2 \lesssim 2$ GeV$^2$ and for spacelike 
momentum transfers.  

In the following review, the theoretical developments are organized in the following 
manner.  As stated in Chapter 1, I confine myself here to the developments in the 
framework of Quantum Chromodynamics (QCD) and do not discuss pure Vector Dominance 
Models (VDM) related predictions.  Within the framework of QCD I discuss the basic 
ideas of the three prevalent approaches: (1) Factorization based Perturbative QCD, 
(2) QCD sum rules, and (3) Lattice QCD.  These considerations apply to all hadrons, 
including $\pi$, $K$, and $p$.  The presentation is then divided in two parts, 
pseudoscalar mesons ($\pi$ and $K$) and baryons ($p$).  I discuss theoretical predictions 
for pion form factors since nearly all meson form factor predictions 
relate to pions, but I note that, 
in general, the theoretical models can be extended to kaons, although explicit predictions 
are few.  I conclude with the proton form factor.

\section{Theoretical Formalisms}

\begin{figure}[!tb]
\begin{center}
\includegraphics[width=15cm]{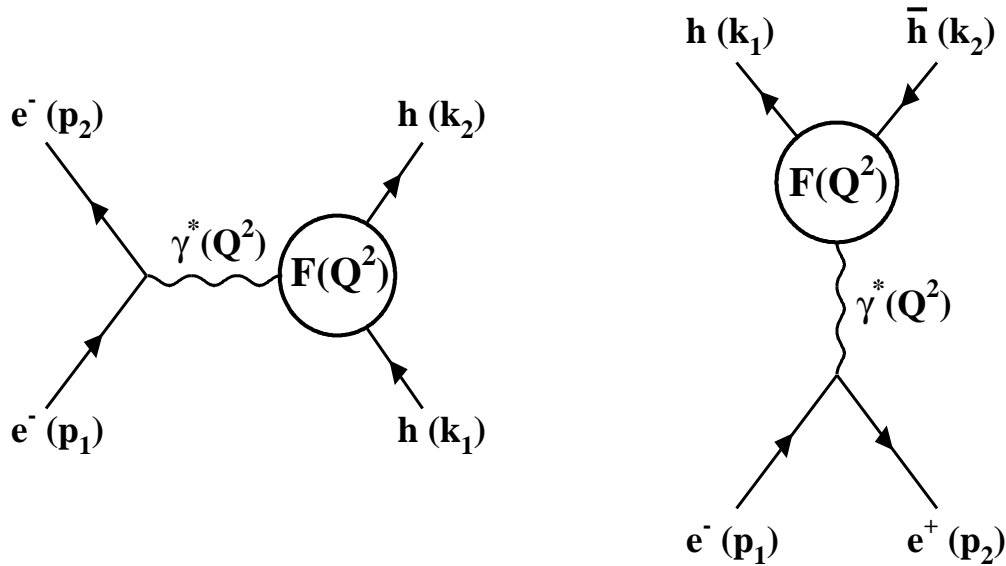}
\caption[Feymann diagrams for studying the electromagnetic form factors of a hadron.]
{Feymann diagrams for studying the electromagnetic form factors of a hadron. 
Left: Spacelike momentum transfer from electron scattering.  
The initial and final four-momenta of the electron are $p_1$ and $p_2$, respectively.  
The initial and final four-momenta of the hadron are $k_1$ and $k_2$, respectively.  
The four-momentum of the virtual photon is defined as $Q^2 = -q^2 = t = (p_{1} - p_{2})^2$.
Right: Timelike momentum transfer from $\ee$ annihilations.  
The initial four-momenta of the electron and positron are $p_1$ and $p_2$, respectively. 
The final four-momenta of the hadron and 'anti'hadron are $k_1$ and $k_2$, respectively.  
The four-momentum of the virtual photon is defined as $-Q^2 = q^2 = s = (p_{1} + p_{2})^2$}.
\label{fig:feyfh}
\end{center}
\end{figure}

Different theoretical predictions interpret the photon-hadron vertex in different ways.  
Figure \ref{fig:feyfh} shows the Feymann diagrams for the photon-hadron interactions 
for spacelike and timelike momentum transfers.  
The most fundamental theories are based on the modern quantum field theory describing the 
strong interaction, Quantum Chromodynamics (QCD).  They attempt to describe the 
behavior of the strong interaction between quarks within the hadron.  At large momentum 
transfers, the strong interaction can be described through a power series expansion 
in terms of the strong coupling constant, $\alpha_s(Q^2)$.  This procedure is known 
as Perturbative Quantum Chromodynamics (PQCD) and its complete formalism for hadronic 
form factors has been described by Lepage and Brodsky \cite{LepageBrodsky_PQCDFF} 
in 1980.  

\subsection{Factorization}

\begin{figure}[!tb]
\begin{center}
\includegraphics[width=12cm]{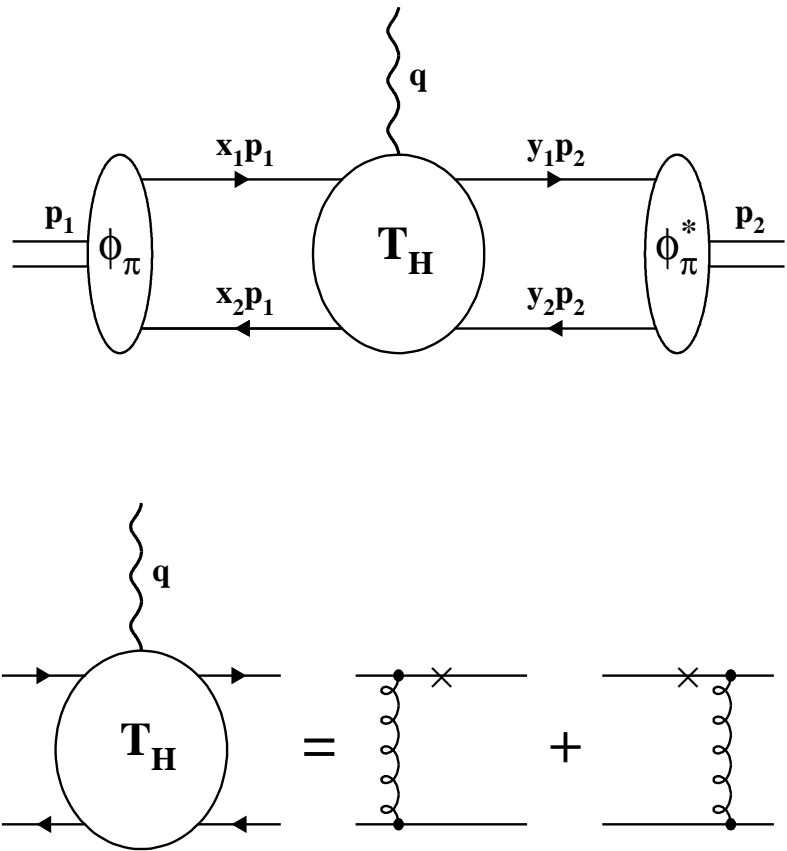}
\caption[PQCD factorization diagrams for the pion.]
{PQCD factorization diagrams for the pion.  
The top figure shows the pion approaching from the left in the form of the 
initial pion distribution amplitude ($\phi_{\pi}$), 
the virtual photon interacting with the pion (T$_{\mathrm{H}}$), 
and the final pion distribution amplitude ($\phi^{\ast}_{\pi}$) of the outgoing pion.  
The bottom figure shows the leading order terms for the hard scattering 
amplitude  (T$_{\mathrm{H}}$).  The crosses represent the quark-photon interaction.}
\label{fig:feypihard}
\end{center}
\end{figure}

Figure \ref{fig:feypihard} shows the PQCD factorization scheme diagrams associated with 
elastic pion scattering by a virtual photon.  The incoming pion has a momentum $p_1$.  
The probability that it will be in a state consisting 
of two collinear quarks carrying  momenta $x_1p_1$ and $x_2p_1$ 
(where $x_i$ is the momentum fraction of the $i$th quark, satisfying $\Sigma_i x_i = 1$) 
is given by distribution amplitude $\phi_\pi$.  
One of the quarks is struck by the photon carrying momentum $q$.  
A portion of the momentum absorbed by the struck quark must be distributed 
to the other quark in order for the two quarks to remain bound together.  
The momentum is exchanged through the emission of gluons.  To the lowest order, 
as shown on the bottom of Figure \ref{fig:feypihard}, this is done by through the 
exchange of a single hard gluon.  The transfer of momentum by the gluon redistributes the 
momenta carried by the quarks, which are represented by $y_1p_2$ and $y_2p_2$ (where 
$y_i$ is the momentum fraction carried by the final state quarks).  
The probability that the two valence quarks will come out collinear and reform 
a pion with momentum $p_2 = p_1 +q$ is given by the distribution amplitude $\phi^*_\pi$.

The central feature of applying QCD based perturbation theory to the description of the 
form factor is the separation of the process into the perturbative and the 
nonperturbative parts.
The photon-pion interaction, denoted by $T_H$, probes the short-distance aspect 
of the pion.  The scale of this interaction is set by the momentum transfer of 
the photon, $Q^2 = q^2$. If the momentum transfer is large enough, 
the strong coupling constant associated with the gluon transferring momentum 
(which is on the order of $Q^2$) to the spectator quark will be small 
(see Eqn. \ref{lab:ch1alphas} for the momentum dependence of $\alpha_s(Q^2)$).  
Hence, $T_H$ can be described within PQCD.  
The probability of finding the pion with two valence quarks is given 
by the distribution amplitude.  It is governed by long-distance QCD 
and has to be treated nonperturbatively.  

For the general case of a given hadron, the form factor is expressed in the 
factorized PQCD scheme by \cite{LepageBrodsky_PQCDFF}
\begin{equation}
F(Q^2) = \int^{1}_{0}\int^{1}_{0}[dx][dy]~\phi^{\ast}(y_i,Q^2)
~T_H(x_i,y_i,Q^2,\mu^2_R)~\phi(x_i,Q^2),
\label{eq:genericfact}
\end{equation}
where the hard scattering amplitude is denoted by $T_H(x_i,y_i,Q^2,\mu^2_R)$, 
the incoming and outgoing hadron distribution amplitudes 
are denoted by $\phi(x_i,Q^2)$ and $\phi^{\ast}(y_i,Q^2)$, respectively, and $\mu^2_R$ 
is the renormalization scale of the strong coupling constant, $\alpha_s(\mu^2_R)$.  
The subscript $i$ is defined by the number of constituent 
quarks (i.e., $i$ = 2 for mesons and $i$ = 3 for baryons).  
The integration variable of the quark momentum fraction is 
$[dx] \equiv (\sum_{i}dx_i)~\delta(1 - \Sigma_i x_i)$ and the same for $[dy]$.

Any prediction of a physical process using perturbation theory 
needs to be independent of the renormalization scale.  For a given 
process, the ideal procedure is to evaluate every term in the $\alpha_s(\mu^2_R)$ 
power series expansion.  This is nearly impossible because more complicated 
contributions arise at higher orders, and therefore the power series must be truncated.  
The general form of the power expansion of the form factor in $\alpha_s(\mu^2_R)$ is
\begin{equation} 
F(Q^2,\mu^2_R) = \alpha_s(\mu^2_R)F^{(0)}(Q^2,\mu^2_R)
[~1+\alpha_s(\mu^2_R)F^{(1)}(Q^2,\mu^2_R)+\cdots], 
\end{equation}
where $F^{(0)}(Q^2,\mu^2_R)$ is the leading order (LO) contribution, 
$F^{(1)}(Q^2,\mu^2_R)$ is the next-to-leading (NLO) contribution, 
and higher order contributions are represented by the dots.  The truncation of the series 
is determined by the renormalization scale, $\mu^2_R$.  
The renormalization scale setting is important for an exclusive process like 
the form factor of the hadron.  In the following discussions it will be set 
by the momentum transfer of the virtual photon, $\mu^2_R = Q^2$, unless 
otherwise specified.

The value of $\alpha_s(Q^2)$ determines the validity of applying perturbation theory.  
The behavior of $\alpha_s(Q^2)$ at low $Q^2$ is not only large but is undefined when 
$Q^2$ = $\Lambda^2$, a.k.a. the Landau pole.  One possible method to avoid this 
behavior is to 'freeze' the value of $Q^2$ by introducing an effective gluon mass.  
Originally proposed by Parisi and Petronzio \cite{ParisiPetronzio_FrozenAlphas} 
and Cornwall \cite{Cornwall_FrozenAlphas}, 
the 'frozen' version of $\alpha_s(Q^2)$ modifies the one loop form of $\alpha_s(Q^2)$ 
(given by Eqn. \ref{lab:ch1alphas}) into \cite{Cornwall_FrozenAlphas}
\begin{equation} 
\alpha_s(Q^2) = \frac{4\pi}{\beta_0~\mathrm{ln}
\left(\frac{Q^2+4m^2_g}{\Lambda^2}\right)},  
\label{eq:frozenalphas}
\end{equation}
where $m^2_g$ is the effective gluon mass.  This definition of $\alpha_s(Q^2)$ 
freezes its value in the range $Q^2 < 4m^2_g$.  
For larger values of $Q^2$, the gluon mass has a small and negligible effect 
on $\alpha_s(Q^2)$, and this modification is not relevant.

\subsection{QCD Sum Rules}

The QCD sum rule (QCDSR) approach is based on the pioneering work of Shifman, Vainshtein, 
and Zakharov \cite{QCDSRTheory}.  Its premise is that the properties of a hadron 
are dictated by its interactions with the QCD vacuum, composed of violent fluctuations 
of virtual gluons ($G^a_{\mu\nu}G^a_{\mu\nu}$) and quark-antiquark pairs ($\qqbar$).   
These interactions are governed by the nonperturbative aspects of QCD.  

The relationship between the hadron and the QCD vacuum is described 
in terms of the operator product expansion of hadronic currents, which consist of the 
constituent quarks of the hadron of interest.  
Its explicit form is expressed in terms of time ordered hadronic currents 
$j_1$ and $j_2$ by \cite{QCDSRTheory}
\begin{equation} 
i\int d^4x~\mathrm{e}^{iqx} \langle 0|~T[j_2(x)j_1(0)]~|0 \rangle = 
\sum_{n} C^{12}_n(q) O_n,
\label{eq:genQCDSR1}
\end{equation}  
where $q$ and $x$ are the momentum and position of the hadron current $j_2$ with 
respect to the current $j_1$.
The interactions between the QCD vacuum and the hadrons are described by local field 
operators, $O_n$, and their coefficients, $C^{12}_n(q)$.  The operators are 
defined by their twist, where twist is defined as the canonical dimension of the operator 
minus its spin.  Higher twist operators describe higher order interactions between 
the hadronic currents and the QCD vacuum.      

Each hadronic current in Eqn. \ref{eq:genQCDSR1} has a characteristic 
energy scale defined by $s_i$ , which has units of energy squared ($i = 1,2$ are with 
respect to the hadronic currents $j_1$ and $j_2$, respectively).  
There is a corresponding energy threshold, denoted as $s_0$, 
which is the maximum energy for which the quarks that comprise the current 
can be associated with the hadron of interest (e.g., $\pi$, $K$, $p$).  
For energies above $s_0$, the currents are contaminated by higher resonance states 
with the same quark structure.  
This energy threshold defines the region for the so-called quark-hadron duality 
\cite{QCDSRTheory}. 

The QCDSR is used in three different ways to determine the electromagnetic form factor 
of a hadron.  They are briefly reviewed below.  

The first way is the \textit{distribution amplitude moment method}.  
This formalism originates from the work by 
Chernyak and Zhitnitsky \cite{ChernyakZhitnitsky_QCDSR}.  
Here one determines distribution amplitudes based on their moments, 
which are then used in the PQCD factorization scheme.  

\begin{figure}[!tb]
\begin{center}
\includegraphics[width=12cm]{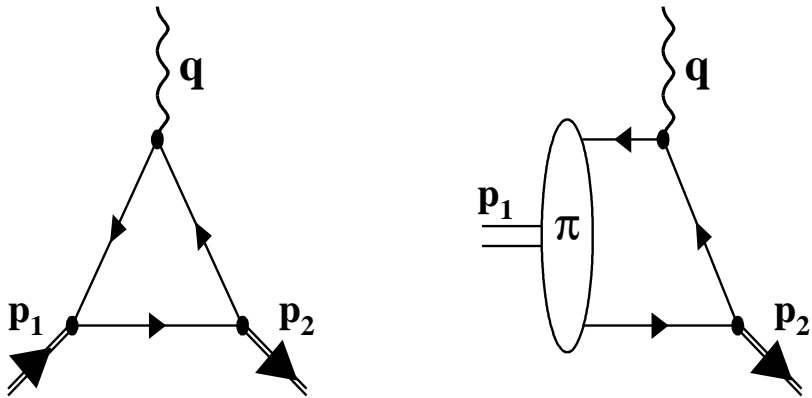}
\caption[The three-point amplitude and correlator function diagrams 
used by the QCD sum rules for the pion form factor.]
{The three-point amplitude and correlator function diagrams 
used with QCD sum rules for the pion form factor.  
Left: The three-point amplitude diagram.  The four-momenta of the incoming pion current, 
virtual photon, and outgoing pion current are $p_1$, $q$, and $p_2$, respectively.  
Right: The correlator function diagram.  The four-momenta of the incoming composite pion, 
virtual photon, and outgoing pion current are $p_1$, $q$, and $p_2$, respectively.  
The quarks are denoted by the single arrowed lines in both cases.}
\label{fig:feypisrcur}
\end{center}
\end{figure}

In the second way one determines the electromagnetic form factors by using the 
\textit{three-point amplitude method}.  The initial application 
of the three-point amplitude in describing of the pion form factor was made 
independently by Nesterenko and Radyushkin \cite{NesterenkoRadyushkin_QCDSR} 
and Ioffe and Smilga \cite{IoffeSmilga_QCDSR}. It consists of replacing the 
matrix element in the left hand side of Eqn. \ref{eq:genQCDSR1} by 
\cite{NesterenkoRadyushkin_QCDSR}
\begin{equation} 
\langle 0|~T[j_\alpha(y)j^{em}_\mu(0) j_\beta(x)]~|0 \rangle,
\label{eq:threepoint}
\end{equation}
where $j_\beta(x)$ and $j_\alpha(y)$ are the incident and final state hadron currents 
and $j^{em}_\mu(0)$ is the electromagnetic interaction with one of the quarks.  
The process is schematically illustrated for pion form factor 
in Figure \ref{fig:feypisrcur} (left).  
The incoming pion current with momentum $p_1$ breaks into its 
quark and antiquark representation, 
the photon interacts with one of the quarks, 
and the outgoing quarks recombine into a pion current with momentum $p_2$.  

Two different treatments of the energy scales of the hadronic currents 
are used with the three-point amplitude method.  
The first, called the \textit{square representation}, treats the scales 
independently, i.e., $0 < s_1 < s_0$ and $0 < s_2 < s_0$.  
The other, called the \textit{triangle representation},  replaces the energy scales 
in the square representation with a triangle of the same area, i.e., 
$0 < s_1 + s_2 < S_0 = \sqrt{2}~s_0$.  

The third way determines the electromagnetic form factors by using the 
\textit{correlator function method}.  
It consists of replacing the matrix element in the left hand side 
of Eqn. \ref{eq:genQCDSR1} by \cite{BraunHalperin_QCDSR}
\begin{equation} 
\langle 0|~T[j_\nu(0)j^{em}_\mu(x)]~|\pi^+(p_1)\rangle,
\label{eq:srcorfnct}
\end{equation}  
where the initial pion is described by $|\pi^+(p)\rangle$, 
the final pion by the current $j_\nu(x)$, and $j^{em}_\mu(0)$ is the electromagnetic 
interaction with one of the quarks. 
The process is schematically illustrated for pion form factor 
in Figure \ref{fig:feypisrcur} (right).  
The composite incoming pion with momentum $p_1$ breaks into its quark 
and antiquark representation, the photon interacts with one of the quarks, 
and the outgoing quarks recombine into its pion, described by the hadronic current, with 
momentum $p_2$.  The correlator function method can be used to determine 
distribution amplitudes \cite{BraunHalperin_QCDSR}.


\subsection{Lattice QCD}

In 1974, Wilson \cite{OriginalLattice} showed how to quantize a gauge field theory 
on a discrete lattice in Euclidean space-time preserving exact gauge invariance.  
He applied this calculational technique to the strong coupling regime of QCD.  
In these Lattice Gauge calculations (Lattice QCD), space-time is replaced by a four 
dimensional hypercubic lattice of size $L^3T$.  The sites are separated by the lattice 
spacing $a$.  The quarks and gluons fields are defined at discrete points.  Physical 
problems are solved numerically by Monte Carlo simulations requiring only the quarks 
masses as input parameters. 

Lattice QCD has been used to calculate the electromagnetic form 
factor of the hadron (e.g., for the pion form factor, see Ref. \cite{Lattice1}) 
using lattice correlation functions.  
The correlation functions connect a hadron creation operator at time $t_i$, 
a hadron annihilation operator at $t_f$, and a vector current insertion at 
$t$ ($t_i < t < t_f$).  
All of the calculations have so far been performed in the 'quenched' approximation, 
in which no virtual quark-antiquark pairs are allowed to be produced from the vacuum.

\section{Meson Form Factors in Theory}

The electromagnetic form factor of a spin-0 meson, studied with 
\textbf{spacelike} momentum transfers, is related to the following matrix element 
\begin{equation}
\langle m(p_2)|~j^{em}_\mu~|m(p_1)\rangle = (p_2+p_1)_\mu F(Q^2),
\label{eq:pionslmatrixel}
\end{equation}
where the electromagnetic current 
$j^{em}_\mu$ = $\Sigma_f~e_f~\overline{q}_f \gamma_\mu q_f$ is expressed in terms of 
quarks $q_f$ with flavor $f$ and electric charge $e_f$; the spacelike momentum 
transfer is defined as $Q^2$ = $-q^2$ = $(p_1-p_2)^2$, and $p_1$ and $p_2$ are the 
initial and final momenta of the meson, respectively.  The form factor $F(Q^2)$ 
measures the deviation of the meson from a Dirac point particle.  The matrix element 
for \textbf{timelike} momentum transfers is obtained by replacing 
$\langle m(p_2)|~j^{em}_\mu~|m(p_1)\rangle$ with 
$\langle m(p_1)\overline{m}(p_2)|~j^{em}_\mu~|0\rangle$.  
The timelike momentum transfer is defined as 
$-Q^2$ = $q^2$ = $s$ = $(p_1+p_2)^2$, where $s$ is the center of mass energy square of 
the system, and $p_1$ and $p_2$ are the momenta of the meson 
and the 'anti' meson, respectively.   

This section is organized as follows.  The pion form factor will be discussed, 
concentrating mostly on PQCD and QCDSR.  The description of the kaon form factor based 
on PQCD and QCDSR follows.  
The majority of the theoretical literature is devoted to the discussion of form factors 
in the spacelike region, and it will be reviewed.  Predictions of the form factors in the 
timelike region are described at the end of each subsection.

\subsection{Perturbative Quantum Chromodynamics}

The formalism for form factor predictions based on the PQCD factorization scheme 
has been described in Section 2.1.1. The process for the pion form factor is shown 
schematically in Figure \ref{fig:feypihard}.  The lowest order contribution to the 
meson form factor is the interaction of a single hard gluon between the two valence 
quarks.  The momentum dependence of the hard gluon propagator is proportional to 
$1/Q^2$.  The form factor is therefore $F(Q^2) \sim 1/Q^2$, 
consistent with the 'quark counting rules' \cite{ffscaling1,ffscaling2,ffscaling3}. 
Higher order corrections arise from higher Fock, i.e., non-valence, 
states and other nonperturbative effects, which are suppressed with respect to 
the single hard gluon exchange.

The formalism for describing the meson form factor by the PQCD factorization scheme 
was determined independently by 
Farrar and Jackson \cite{FarrarJackson_PionPQCD}, 
Efremov and Radyushkin \cite{EfremovRadyushkin_PionPQCD}, 
and Lepage and Brodsky \cite{LepageBrodsky_PionPQCD}.  
As described by Eqn. \ref{eq:genericfact} in Section 2.1.1., 
the meson form factor is expressed in the factorized PQCD scheme by 
\begin{equation}
F(Q^2) = \int^{1}_{0}\int^{1}_{0}[dx][dy]~\phi^{\ast}(y_i,Q^2)
~T_H(x_i,y_i,Q^2,\mu^2_R)~\phi(x_i,Q^2),
\label{eq:Fpifact}
\end{equation}
The hard scattering amplitude $T_H(x_i,y_i,Q^2,\mu^2_R)$ incorporates the 
short-distance interactions between the constitute quarks inside the meson.  The 
lowest order contribution from the emission of a single hard gluon is given as 
\cite{LepageBrodsky_PQCDFF}
\begin{equation} 
T_H(x_i,y_i,Q^2,\mu^2_R) = \frac{8\pi~C_F~\alpha_s(\mu^2_R)}{Q^2}
\left[~\frac{e_1}{x_2y_2}+\frac{e_2}{x_1y_1}\right],
\label{eq:piTH1storder}
\end{equation}
where $e_1$ and $e_2$ are the electric charges of the quarks and 
$C_F = (n^2_c-1)/2n_c = 4/3$ with $n_c = 3$ denoting the number of colors.

\begin{figure}[!tb]
\begin{center}
\includegraphics[width=12cm]{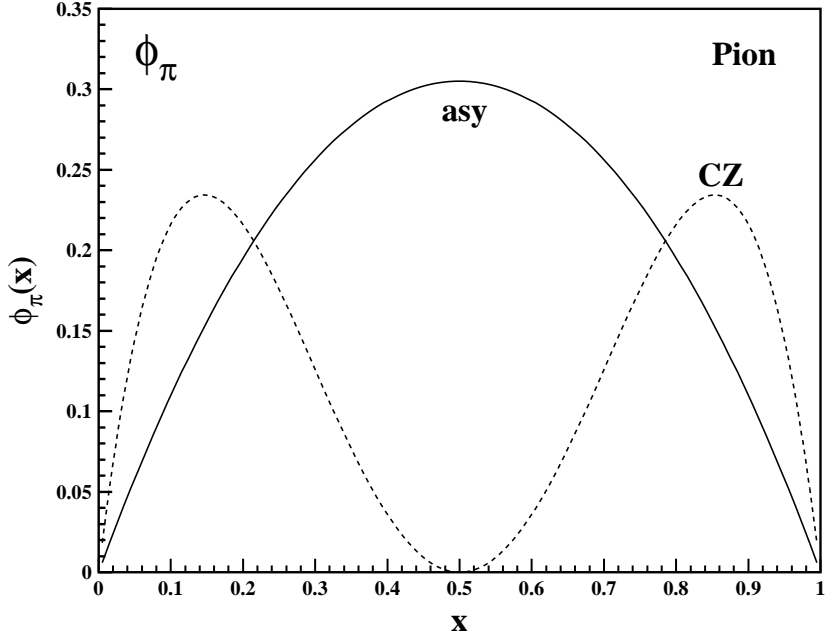}
\caption[Asymptotic and CZ pion distribution amplitudes as a function of quark 
momentum fraction.]
{Asymptotic and CZ distribution amplitudes for the pion as a function of quark 
momentum fraction $x$.  The solid line is the asymptotic form 
(asy: $\phi^{asy}_{\pi}(x) = (f_\pi\cdot \sqrt{3/2})x(1-x)$) 
and the dashed line is the Chernyak-Zhitnitsky form 
(CZ: $\phi^{CZ}_{~\pi}(x) = (f_\pi\cdot 15/4)x(1-x)(2x-1)^2$).  
The pion decay constant is taken to be unity in this figure.}
\label{fig:pida}
\end{center}
\end{figure}

The meson distribution amplitude contains all of the nonperturbative aspects of the 
interaction.  
The meson distribution amplitude, $\phi(x_i,Q^2)$, is related to the integral 
of the full meson wave function, $\psi(x_i,\mathbf{k}_{\perp,i})$, over the transverse 
momentum $\mathbf{k}_{\perp,i}$ of the $i$th quark by \cite{LepageBrodsky_PQCDFF}
\begin{equation}
\phi(x_i,Q^2) = \int^{Q^2}\frac{d^2\mathbf{k}_{\perp,i}}{16\pi^3}
~\psi(x_i,\mathbf{k}_{\perp,i}), 
\end{equation}
The general solution of the distribution amplitude is a series 
of Gegenbauer polynomials $C^{3/2}_n$ \cite{LepageBrodsky_PQCDFF}
\begin{equation}
\phi(x_i,Q^2) = x_1 x_2 \sum^{\infty}_{n=0} a_n~C^{3/2}_n(x_1-x_2)
[~1 + O(\alpha_s(Q^2))~], 
\label{eq:DAsoln}
\end{equation}
where $a_n$ are the coefficients of the polynomial. 
In the large $Q^2$, or asymptotic, limit ($Q^2 \to \infty$), 
the first term of the distribution amplitude (Eqn. \ref{eq:DAsoln}) dominates.  
The distribution amplitude is therefore $\phi^{asy}(x_i,Q^2) = a_0~x_1 x_2$, 
or $\phi^{asy}(x_i,Q^2) = a_0~x(1-x)$ with the replacements 
$x_1 = x$ and $x_2 = 1-x$, and is 
commonly referred to as the \textit{asymptotic} distribution amplitude.  
The quark momentum fraction dependence of the asymptotic distribution amplitude 
is shown in Figure \ref{fig:pida}.  

\begin{figure}[!tb]
\begin{center}
\includegraphics[width=14cm]{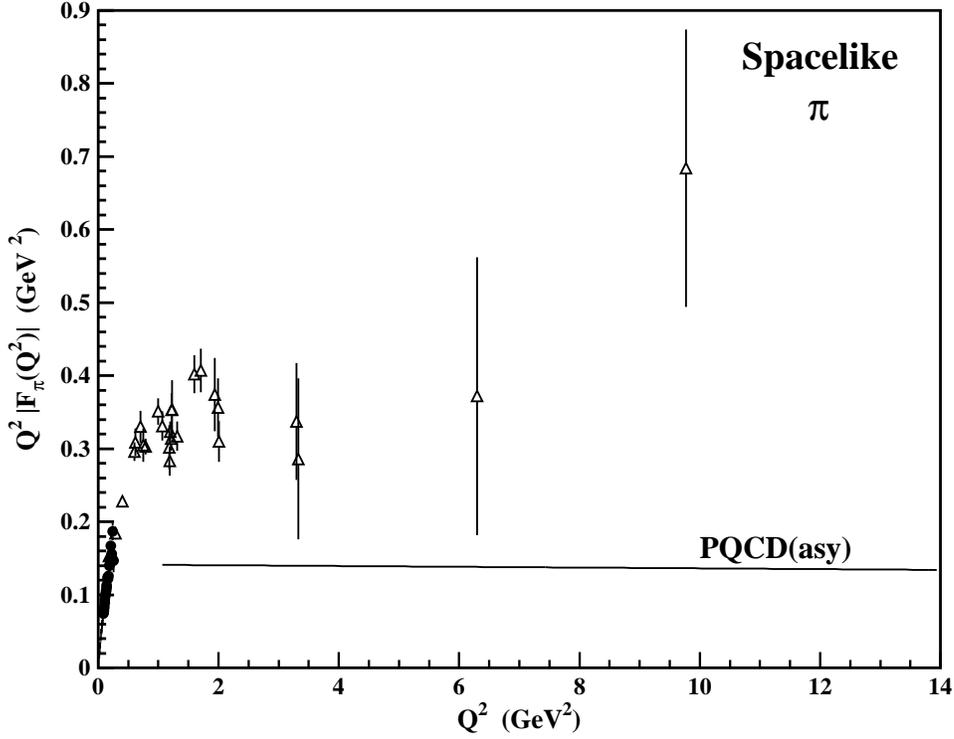}
\caption[Asymptotic PQCD prediction of the spacelike pion form factor.]
{Asymptotic PQCD prediction of the spacelike pion form factor.  
The line is the PQCD prediction by Lepage and Brodsky 
with $\Lambda = 0.316$ GeV \cite{LepageBrodsky_PQCDFF}.  
The solid points and open triangles are experimental data 
\cite{pislff_sc1}-\cite{pislff_elprod6}.}  
\label{fig:pislpqcd1}
\end{center}
\end{figure}

For the pion, 
the normalization of the distribution amplitude is determined from the matrix element 
between the quark-antiquark pair and the composite pion, i.e., 
$\langle 0|\overline{d}\gamma_5\gamma_\mu u|\pi^+(p)\rangle$.  The 
$a_0$ coefficient in the asymptotic distribution amplitude 
is fixed by relating it to the weak decay process 
$\pi^+ \to \mu^+\nu_\mu$.  This leads to 
$\langle 0|\overline{d}\gamma_5\gamma_\mu u|\pi^+(p)\rangle$ = $p_\mu \sqrt{2/3}~a_0$ 
= $p_\mu f_\pi$, or $a_0$ = $\sqrt{3/2} f_\pi$, 
where $f_{\pi}$ = 130.7 $\pm$ 0.4 MeV \cite{PDG2004} is the pion decay constant.  
The asymptotic pion distribution amplitude is therefore \cite{LepageBrodsky_PQCDFF} 
\begin{equation}
\phi^{asy}(x_i,Q^2) = \sqrt{3/2}~f_{\pi}~x_1 x_2
\label{eq:asypida}
\end{equation}
Substituting the asymptotic 
distribution amplitude (Eqn. \ref{eq:asypida}) and the hard scattering 
amplitude (Eqn. \ref{eq:piTH1storder}) into the factorization expression 
(Eqn. \ref{eq:Fpifact}), in the limit of large spacelike momentum transfer, 
the pion form factor is 
\cite{LepageBrodsky_PQCDFF}
\begin{equation}
F_{\pi}(Q^2) = \frac{8\pi~f^2_{\pi}~\alpha_s(Q^2)}{Q^2},
\label{eq:q2pqcd1}
\end{equation}
or
\begin{equation}
Q^2~F_{\pi}(Q^2) = 8\pi~f^2_{\pi}~\alpha_s(Q^2)~\mathrm{GeV}^2 
= 0.43~\alpha_s(Q^2)~\mathrm{GeV}^2.
\label{eq:pqcd1}
\end{equation}
The pion form factor in the large $Q^2$ limit is dominated by 
the hard gluon emission between the valence quarks 
and is absolutely normalized by the pion decay constant.  This result will be 
referred to as the \textit{asymptotic form factor prediction}. 
The spacelike form factor prediction \cite{LepageBrodsky_PQCDFF} 
for the pion, as a function of $Q^2$, with $\Lambda = 0.316$ GeV, 
is shown in Figure \ref{fig:pislpqcd1}.  
It is nearly factor three smaller than the data in the $Q^2 = 1-2$ GeV$^2$ range 
in which the data have reasonable errors.

\begin{figure}[!tb]
\begin{center}
\includegraphics[width=14cm]{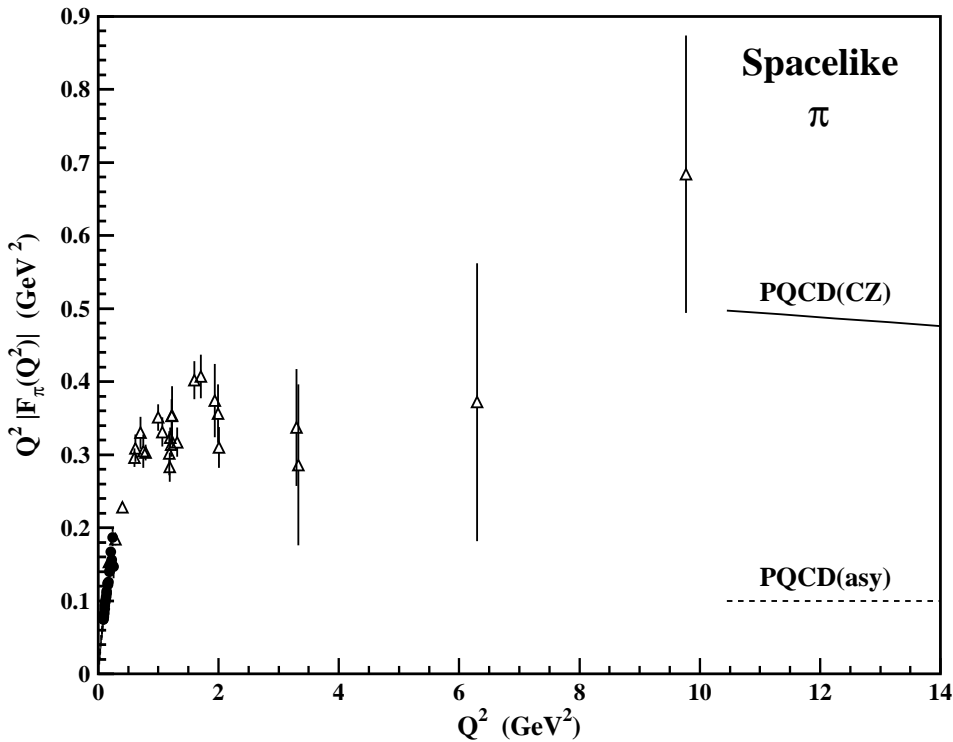}
\caption[PQCD factorization prediction of the spacelike pion form factor 
using the CZ pion distribution amplitude.]
{PQCD factorization prediction of the spacelike pion form factor using the CZ pion 
distribution amplitude.  
The predictions are by Chernyak and Zhitnitsky \cite{ChernyakZhitnitsky_QCDSR}.  
The solid line is the prediction using the CZ distribution amplitude, 
and the dashed is the prediction using the asymptotic distribution amplitude.  
The solid points and open triangles are experimental data 
\cite{pislff_sc1}-\cite{pislff_elprod6}.}
\label{fig:pislsrcz}
\end{center}
\end{figure}

Pion distribution amplitudes are also determined from the QCDSR using the 
distribution amplitude moment method.  The moments of the 
pion distribution amplitude are defined as \cite{ChernyakZhitnitsky_QCDSR}
\begin{equation} 
\langle\xi^n\rangle = \int^{1}_{-1} d\xi~\xi^n~\phi(\xi).
\label{eq:pidamoment}
\end{equation}
where $\xi = x_1-x_2 = 2x-1$ is the momentum fraction difference between the two valence 
quarks.  By using the QCDSR, the moments are found to be $\langle\xi^0\rangle =$ 1, 
$\langle\xi^2\rangle \simeq$ 0.46, and $\langle\xi^4\rangle \simeq$ 0.30 
\cite{ChernyakZhitnitsky_QCDSR}.
The distribution amplitude, derived by 
Chernyak and Zhitnitsky \cite{ChernyakZhitnitsky_QCDSR}, which reproduces 
these moments is 
\begin{equation} 
\phi^{CZ}_{~\pi}(x) = \frac{15}{4}~f_{\pi}~(1-\xi^2)\xi^2 
= 15~f_{\pi}~x(1-x)(2x-1)^2.
\label{eq:czpida}
\end{equation}
This asymmetric double-humped distribution amplitude forces one quark to carry \\
$\sim$$85\%$ of the pion momentum.  Figure \ref{fig:pida} 
compares the momentum fraction dependence of the Chernyak-Zhitnitsky (CZ) 
and asymptotic distribution amplitudes.  Using the CZ distribution amplitude in the PQCD 
factorization scheme results in a spacelike pion form factor prediction which 
is about five times larger than the prediction from the asymptotic distribution 
amplitude, as shown in Figure \ref{fig:pislsrcz}.  
   
\begin{figure}[!tb]
\begin{center}
\includegraphics[width=14cm]{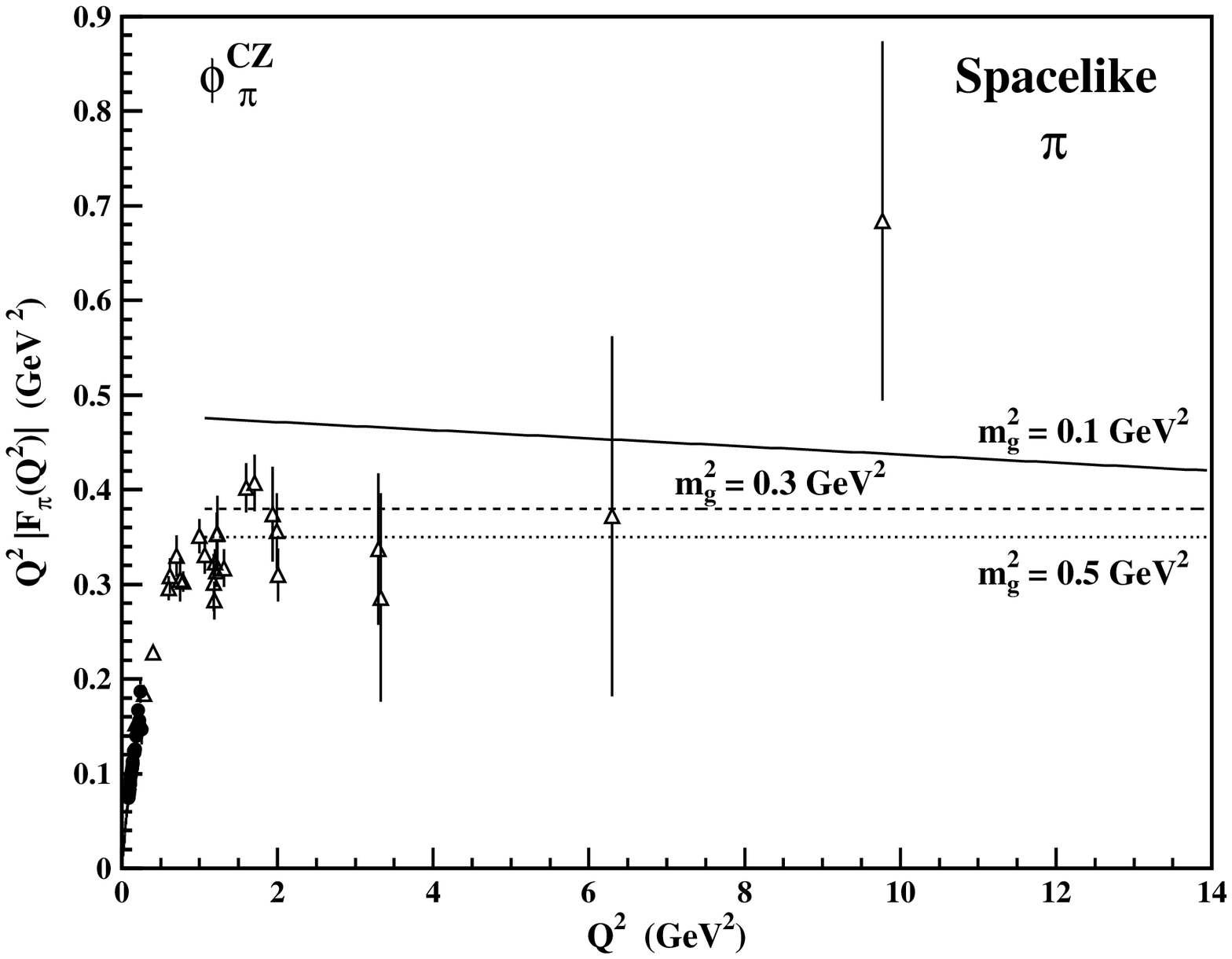}
\caption[PQCD factorization prediction of the spacelike pion form factor using the CZ pion 
distribution amplitude using the 'frozen' version of $\alpha_s(Q^2)$.]
{PQCD factorization prediction of the spacelike pion form factor using the CZ 
distribution amplitude and the 'frozen' version of $\alpha_s(Q^2)$.  
The predictions are by Ji and Amiri \cite{JiAmiri_QCDSR}.  
The solid, dashed, and dotted lines are 
with $m^2_g$ = 0.1, 0.3, and 0.5 GeV$^2$, respectively.  
The solid points and open triangles are experimental data 
\cite{pislff_sc1}-\cite{pislff_elprod6}.}
\label{fig:pislczfrozen}
\end{center}
\end{figure}

Ji and Amiri \cite{JiAmiri_QCDSR} have calculated the spacelike pion form factor using 
the CZ distribution amplitude and the 'frozen' version of $\alpha_s(Q^2)$ 
(see Eqn. \ref{eq:frozenalphas} for definition of frozen $\alpha_s(Q^2)$).  
The predictions are similar to those by 
Chernyak and Zhitnitsky \cite{ChernyakZhitnitsky_QCDSR}, and are 
in reasonable agreement with the data, as shown in Figure \ref{fig:pislczfrozen}.

Arguments have been raised about the validity of the PQCD predictions to 
describe the 
existing experimental data.  Isgur and Llewellyn Smith 
\cite{Isgur_LS_argue1,Isgur_LS_argue2} have argued that the spacelike 
form factor prediction with the asymptotic and CZ distribution amplitudes contain 
significant contributions from regions where most of the 
momentum of the pion is carried by one quark ($x \approx$ 0 or 1).  
Near $x \to 0,1$, the gluon virtuality $x_iy_iQ^2$ is small 
but $\alpha_s(x_iy_iQ^2$) is quite large.  Therefore PQCD should not be applied 
in these regions.  This issue is called the \textit{end-point problem}.  Isgur and 
Llewellyn Smith argue that the PQCD form factor prediction should be restricted 
to regions where the gluon virtuality is above some minimum value so 
that higher order effects can be appropriately neglected.  
Table \ref{tab:isgurlstable} lists the percentage of the form factor predictions 
which arise in the valid region above different momentum transfer cutoffs.  
As shown in Table \ref{tab:isgurlstable}, with a cutoff at $x_iy_iQ^2 =$ 1 GeV$^2$, 
the valid part of the PQCD form factor prediction 
with the asymptotic distribution amplitude 
is only 2$\%$ at $Q^2$ = 2 GeV$^2$ and 52$\%$  at $Q^2$ = 16 GeV$^2$.  
The situation is even worse for prediction using the CZ distribution amplitude.  
Isgur and Llewellyn Smith therefore conclude \cite{Isgur_LS_argue1,Isgur_LS_argue2} 
that the form factor at currently accessible energies gets most 
of its contribution from higher order nonperturbative effects.  

\begin{table}[h]
\caption[Valid contributions of the pion PQCD form factor predictions 
using the asymptotic and CZ distribution amplitudes.]
{Valid contributions of the pion PQCD form factor predictions 
using the asymptotic and CZ distribution amplitudes.  
The values are the percentage of the original PQCD prediction 
which remains after excluding the end-point regions, defined by the $x_iy_iQ^2$ cutoff. 
This table is reproduced from Ref. \cite{Isgur_LS_argue2}.}
\medskip
\begin{center}
\begin{tabular}{|c|c|c|c|c|c|c|}
\hline
 & \multicolumn{3}{|c|}{$\phi^{asy}_{\pi}(x)$} 
 & \multicolumn{3}{|c|}{$\phi^{CZ}_{~\pi}(x)$} \\
\hline
Cutoff (GeV$^2$) & 0.25 & 0.5 & 1.0 & 0.25 & 0.5 & 1.0 \\ 
\hline
Q$^2$ (GeV$^2$): & & & & & & \\ 
1  & 13$\%$ &  2$\%$ &  0$\%$ &  2$\%$ &  1$\%$ &  0$\%$ \\ 
2  & 32$\%$ & 13$\%$ &  2$\%$ &  8$\%$ &  2$\%$ &  1$\%$ \\ 
4  & 52$\%$ & 32$\%$ & 13$\%$ & 16$\%$ &  8$\%$ &  2$\%$ \\ 
8  & 68$\%$ & 52$\%$ & 32$\%$ & 27$\%$ & 16$\%$ &  8$\%$ \\ 
16 & 80$\%$ & 68$\%$ & 52$\%$ & 41$\%$ & 27$\%$ & 16$\%$ \\ 
\hline
\end{tabular}
\label{tab:isgurlstable}
\end{center}
\end{table}

Li and Sterman \cite{LiSterman_PionSud} extended the PQCD factorization scheme 
to include effects from the transverse momenta of the quarks.  
The transverse momenta are suppressed by QCD radiative corrections, 
the so-called \textit{Sudakov effects}.  Sudakov effects 
arise from the QCD radiative corrections to the quark propagator 
and the photon-quark vertex. 
By performing a Fourier transform between the transverse momenta of the quark 
$\mathbf{k}_{\perp,i}$ and the quark-antiquark impact parameter $\mathbf{b}$, 
the PQCD form factor expression becomes \cite{LiSterman_PionSud}
\begin{equation}
F(Q^2) = \int^{1}_{0}[dx][dy]\int 
\frac{d^2\mathbf{b}}{(4\pi^2)^2}~\Phi^{\ast}(y_i,\mathbf{b},Q^2)
~T_H(x_i,y_i,b,Q^2)~\Phi(x_i,\mathbf{b},Q^2),
\end{equation}
where the new wave function is given by \cite{LiSterman_PionSud}  
\begin{equation}
\Phi^{\ast}(x,\mathbf{b},Q^2) = \int d^2\mathbf{k}_{\perp}
~\mathrm{e}^{-i\mathbf{b}\cdot\mathbf{k}_{\perp}}~\psi(x,k_{\perp}).
\end{equation}
The impact parameter constraints the maximum allowed distance between the quarks. 
This allows $T_H(x_i,y_i,b,Q^2)$ to truly describe short distance processes by requiring 
$b < 1/\Lambda$ ($\sim$ 0.66 fm for $\Lambda$ = 300 MeV).  
The PQCD form factor prediction with the inclusion of the Sudakov effects 
is \cite{LiSterman_PionSud}
\begin{displaymath}
F(Q^2) = \int^{1}_{0}[dx][dy]~\phi^{\ast}(y_i,Q^2)\phi^{\ast}(x_i,Q^2)
~~~~~~~~~~~~~~~~~~~~~~~~~~~~~~~~~~
\end{displaymath}
\begin{equation}
\times ~\int^{\infty}_{0} b~db~T_H(x_i,y_i,b,Q^2,t)~\mathrm{exp}[-S(x_i,y_i,b,Q^2,t)].
\label{eq:resumSud}
\end{equation}
The exp$[-S(x_i,y_i,b,Q^2,t)]$ term is the Sudakov form factor containing the QCD 
radiative corrections.  
The variable $t$ is the largest mass scale in the hard scattering amplitude, 
i.e., $t = max(\sqrt{x_iy_i}Q,1/b)$.  
The large $Q^2$ PQCD form factor prediction with the inclusion of Sudakov suppression 
is \cite{LiSterman_PionSud}

\begin{figure}[!tb]
\begin{center}
\includegraphics[width=14cm]{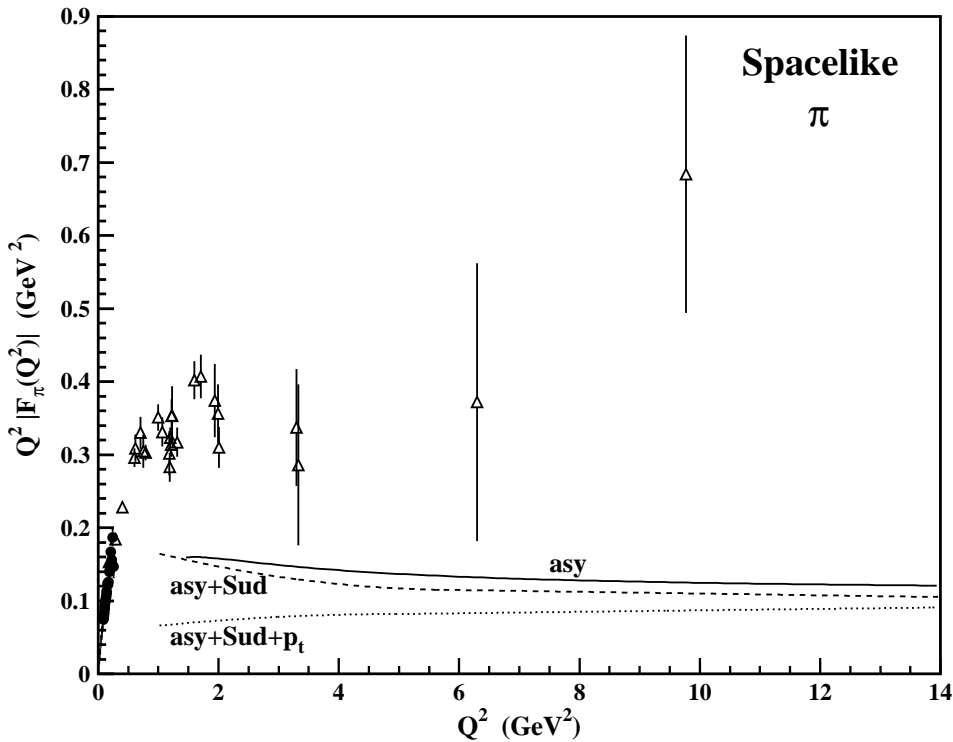}
\caption[Sudakov and intrinsic transverse momenta effects 
on the spacelike pion form factor.]
{Sudakov and intrinsic transverse momenta effects on the spacelike pion 
form factor.  
The predictions are by Jakob and Kroll \cite{JakobKroll_Ptrans}.  
The solid line is the asymptotic (asy) form factor prediction, 
the dashed line includes the Sudakov effect (asy+Sud), 
and the dotted line includes both the Sudakov and 
intrinsic transverse momenta effects (asy+Sud+p$_t$).  
The solid points and open triangles are experimental data 
\cite{pislff_sc1}-\cite{pislff_elprod6}. }
\label{fig:pislsudpt}
\end{center}
\end{figure}

\newpage
\begin{displaymath}
F_\pi(Q^2) = 8\pi~C_F~\int^{1}_{0}[dx][dy]~\phi^{\ast}(y_i,Q^2)\phi^{\ast}(x_i,Q^2)
~~~~~~~~~~~~~~~~~~~~~~~~~~~~~~~~~~~~~~~~~~~ 
\end{displaymath}
\begin{equation}
\times ~\int^{\infty}_{0} b~db~\alpha_s(t)~K_0(\sqrt{x_iy_i}Qb)
~\mathrm{exp}[-S(x_i,y_i,b,Q^2,t)],
\label{eq:resumSudresult}
\end{equation}
where $K_0$ is the modified Bessel function of order zero.  This expression is 
referred to as the \textit{resummed PQCD form factor}.  
With the resummed asymptotic form factor prediction, 
50$\%$ of the contribution to the form factor arises from 
the regions with $b/\Lambda \le 0.39$ and $\le 0.25$ for $Q/\Lambda$ = 10 and 20, 
respectively \cite{LiSterman_PionSud}.  The comparison between the asymptotic 
prediction for the spacelike pion form factor with and without 
the inclusion of the Sudakov effects is shown in Figure \ref{fig:pislsudpt}.  
Inclusion of the Sudakov effect decreases the spacelike pion form factor prediction.

Jakob and Kroll \cite{JakobKroll_Ptrans} have argued that the intrinsic transverse 
momentum in the pion should be included in the PQCD prediction of the 
pion form factor.  They define the following pion wave function \cite{JakobKroll_Ptrans}
\begin{equation}
\Psi(x_i,\mathbf{k}_{\perp}) = \frac{f_\pi}{2\sqrt{6}}~
\phi^{asy}(x_i)\chi(x_i,\mathbf{k}_{\perp}) = 
A~x(1-x)~\mathrm{exp}\left(-\frac{\beta^2\mathbf{k}_{\perp}}{x(1-x)}\right).
\label{eq:ptranswf}
\end{equation}
The variables $A = 10.07$ and $\beta^2 = 0.883$ GeV$^{-2}$ are chosen so the 
average transverse momentum is 350 MeV.  After inserting 
Eqn. \ref{eq:ptranswf} into the resummed PQCD form factor 
expression (Eqn. \ref{eq:resumSudresult}), the intrinsic transverse momentum produces 
further suppression of the spacelike form factor, as shown in Figure \ref{fig:pislsudpt}.  
Jakob and Kroll also argue \cite{JakobKroll_Ptrans} 
that the difference between the PQCD prediction and the experimental 
data originates from higher order and nonperturbative effects, 
but that their contributions 
become equal to the perturbative contribution 
at $Q^2 \approx 5$ GeV$^2$ \cite{JakobKroll_Ptrans}.  Unfortunately, even doubling the 
Jakob and Kroll prediction at $Q^2$ = 5 GeV$^2$ leaves it short of the experimental 
data by more of a factor two.

The NLO term of the hard scattering amplitude were derived independently by 
Field, Gupta, Otto, and Chang \cite{Fieldetal_NLOPQCD}, 
Dittes and Radyushkin \cite{DittesRadyushkin_NLOPQCD}, 
and Braaten and Tse \cite{BraatenTse_NLOPQCD}.  The NLO term contains ultraviolet 
(UV) and infrared (IR) divergences.  The ultraviolet divergences 
are removed through the choice of renormalization schemes, with the two most common being 
the modified minimal subtraction ($\msbar$) \cite{MSbar_RenormScheme} 
and momentum subtraction (MOM) \cite{MOM_RenormScheme} schemes.  
The IR divergences are absorbed into renormalized distribution amplitudes.  
The asymptotic pion form factor with NLO corrections is \cite{Fieldetal_NLOPQCD}
\begin{equation} 
Q^2~F_\pi(Q^2) = 8\pi~f^2_\pi~\alpha_s(Q^2)[~1+A~\alpha_s(Q^2)+\cdots~], 
\label{eq:nlopqcd}
\end{equation}
where $A$ = 2.1 in the $\msbar$ scheme with $\Lambda_{\msbar}$ = 0.5 GeV and 
$A$ = 0.72 in the MOM scheme 
with $\Lambda_{\mathrm{MOM}}$ = 1.3 GeV \cite{Fieldetal_NLOPQCD}.  
Inclusion of NLO corrections from the distribution amplitudes was 
determined by Meli\'{c}, Ni\v{z}i\'{c}, and Passek \cite{Melicetal_PQCD}, 
and the effect from the NLO asymptotic and CZ distribution amplitudes was found to be 
on the order of 1$\%$ and 6$\%$, respectively, 
with respect to the LO spacelike pion form factor.  

\begin{figure}[!tb]
\begin{center}
\includegraphics[width=14cm]{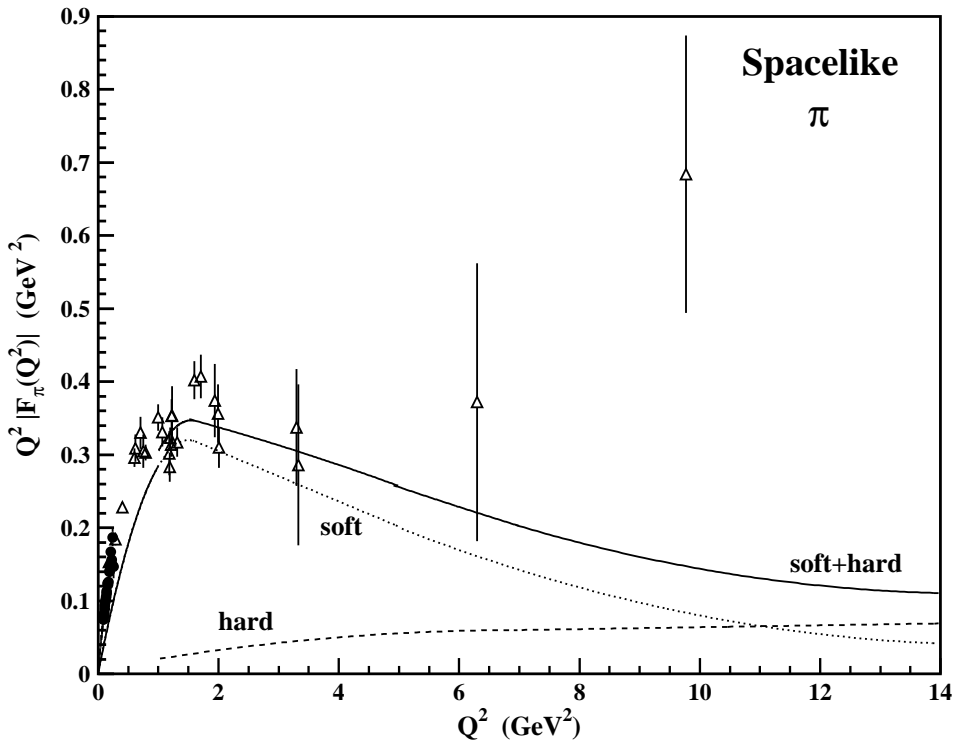}
\caption[PQCD $k_T$ factorization prediction of the spacelike pion form factor.]
{PQCD $k_T$ factorization prediction of the spacelike pion form factor.  
The predictions are by Huang $\etal$ \cite{Huangetal_PQCDkt}.  
The dotted, dashed, and solid lines are the soft, hard, and soft+hard total contribution 
to the form factor, respectively.  
The solid points and open triangles are experimental data 
\cite{pislff_sc1}-\cite{pislff_elprod6}.}
\label{fig:pislpqcdkt}
\end{center}
\end{figure}

The effect of higher helicity states on the spacelike pion form factor 
was studied by Huang, Wu, and Wu \cite{Huangetal_PQCDkt}.  
They found, by explicitly keeping the transverse momentum of the quark 
and gluon propagators ($k_T$ factorization formalism), 
that the higher helicity state ($\lambda_1 + \lambda_2$ = 1, 
where $\lambda_i$ is the helicity of the $i$th quark) 
slightly decreases the form factor as compared to considering only the usual helicity state 
($\lambda_1 + \lambda_2$ = 0).  
In addition, they also studied the effect of including a soft, nonperturbative 
contribution and found that the soft contribution is less than the hard 
contribution for $Q^2 > 11$ GeV$^2$, as shown in Figure \ref{fig:pislpqcdkt}.

\begin{figure}[!tb]
\begin{center}
\includegraphics[width=14cm]{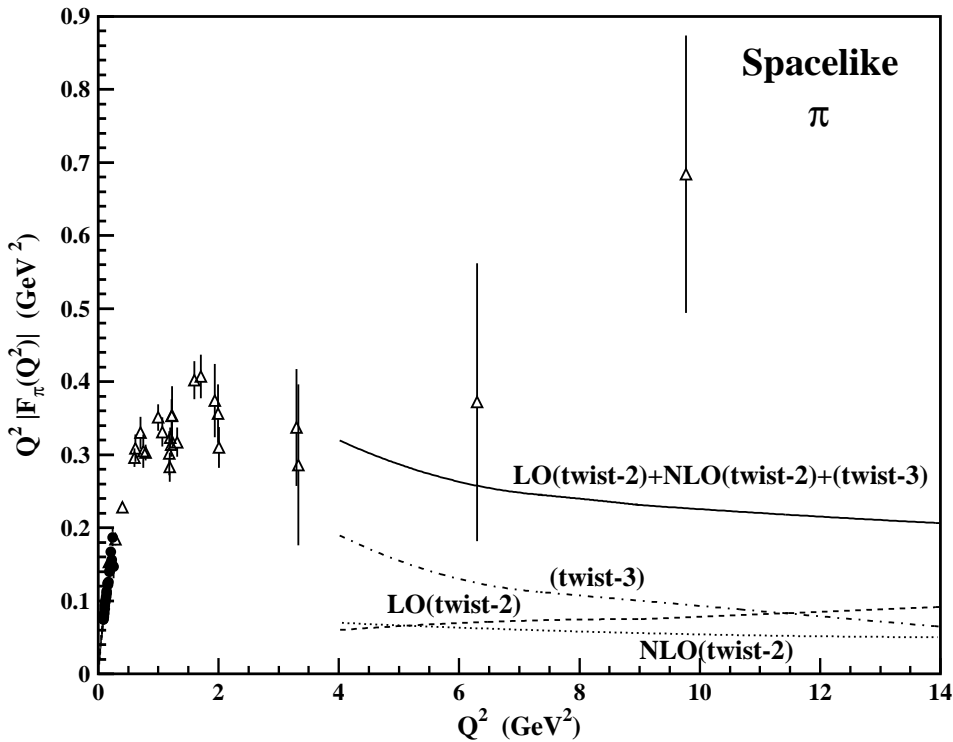}
\caption[PQCD prediction of the spacelike pion form factor using a modified wave function 
of twist-3 accuracy.]
{PQCD prediction of the spacelike pion form factor using a modified wave function 
of twist-3 accuracy.  The predictions are by Huang $\etal$ \cite{Huangetal_PQCDkt}.  
The dashed, dotted, dash-dotted, and solid lines are the LO twist-2, NLO twist-2, 
twist-3, and total contributions to the pion form factor.  
The solid points and open triangles are experimental data 
\cite{pislff_sc1}-\cite{pislff_elprod6}.}
\label{fig:pislpqcdt3}
\end{center}
\end{figure}

Huang and Wu \cite{HuangWu_QCDSRT3} used the QCDSR distribution amplitude moment method 
to determine a higher order, twist-3 wave function based on its moments 
and the inclusion of explicit transverse momentum dependence.  It should be 
noted that the asymptotic wave function is of twist-2 and the higher twist 
denotes contributions from higher Fock, i.e., non leading order, states.  
While the wave function is double-humped, it was found to have better end-point suppression 
than the asymptotic wave function \cite{HuangWu_QCDSRT3}.  
The wave function was used to predict the 
spacelike pion form factor, as shown in Figure \ref{fig:pislpqcdt3}.  
The fact that the twist-3 contribution is found to be more than twice the twist-2 
contribution for $Q^2 < 6$ GeV$^2$ is not a comfortable feature of these calculations.  
The twist-3 contribution becomes smaller than the LO twist-2 hard scattering 
contribution at $Q^2 \approx 10$ GeV$^2$.

\begin{figure}[!tb]
\begin{center}
\includegraphics[width=14cm]{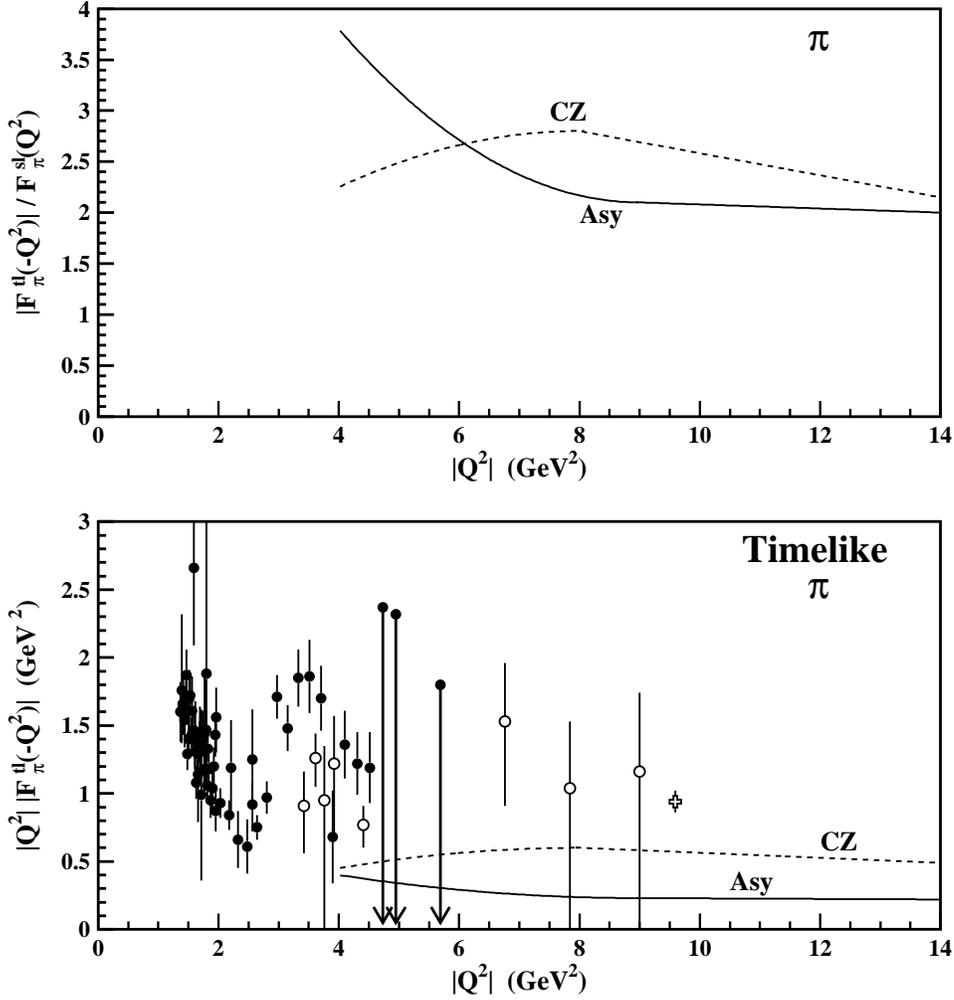}
\caption[PQCD predictions of the $|F^{tl}_{\pi}(Q^2)|/F^{sl}_{\pi}(Q^2)$ ratio and 
timelike pion form factor.]
{PQCD predictions of the $|F^{tl}_{\pi}(Q^2)|/F^{sl}_{\pi}(Q^2)$ ratio (top) and 
timelike pion form factor (bottom).
The predictions are by Gousset and Pire \cite{GoussetPire_tlPQCD}.  
The solid lines are the PQCD factorization prediction 
with the asymptotic distribution amplitude, 
and the dashed lines are the PQCD factorization prediction 
with the CZ distribution amplitude.  
The solid points are from $\eetopipi$ measurements with pions experimentally identified 
\cite{kpicommontlff_1}-\cite{pitlff_3}.  The open points are from $\eetohh$ 
measurements with the pion fraction of the observed $\hadpair$ determined according to 
a VDM prescription \cite{pitlff_VDM}.  
The value denoted with the plus symbol comes from interpreting the 
$J/\psi \rightarrow \pi^{+}\pi^{-}$ branching ratio as a pion form factor measurement 
as in Ref. \cite{Milanaetal_jpsipipi}.}
\label{fig:pitlgandp}
\end{center}
\end{figure}

So far, only the predictions for the spacelike form factor of the pion have 
been described.  The predictions for the timelike form factor of the pion are scarce.  
Actually, in the PQCD formalism there are only two.  
Gousset and Pire \cite{GoussetPire_tlPQCD} analytically continued the Sudakov form factor 
(discussed in Eqn. \ref{eq:resumSud}) from the spacelike region into the timelike 
region by the following replacement: $Q \rightarrow -iW$, where $W^2 = s$.  
They found \cite{GoussetPire_tlPQCD} 
that this causes an enhancement in the timelike-to-spacelike ratio of the pion form factor 
from both the asymptotic and CZ distribution amplitudes, as shown in the Figure 
\ref{fig:pitlgandp}(top).  The $Q^2 F_{\pi}$ prediction including this enhancement 
is shown for the timelike pion form factor in Figure \ref{fig:pitlgandp}(bottom). 
We note that even with this enhancement the timelike PQCD predictions for 
both the asymptotic and CZ distribution amplitudes are 
$\sim$1/4 and $\sim$1/2 the value determined at 
$Q^2 = M^2_{J/\psi} = 9.6$ GeV$^2$ \cite{Milanaetal_jpsipipi}. 


\begin{figure}[!tb]
\begin{center}
\includegraphics[width=14cm]{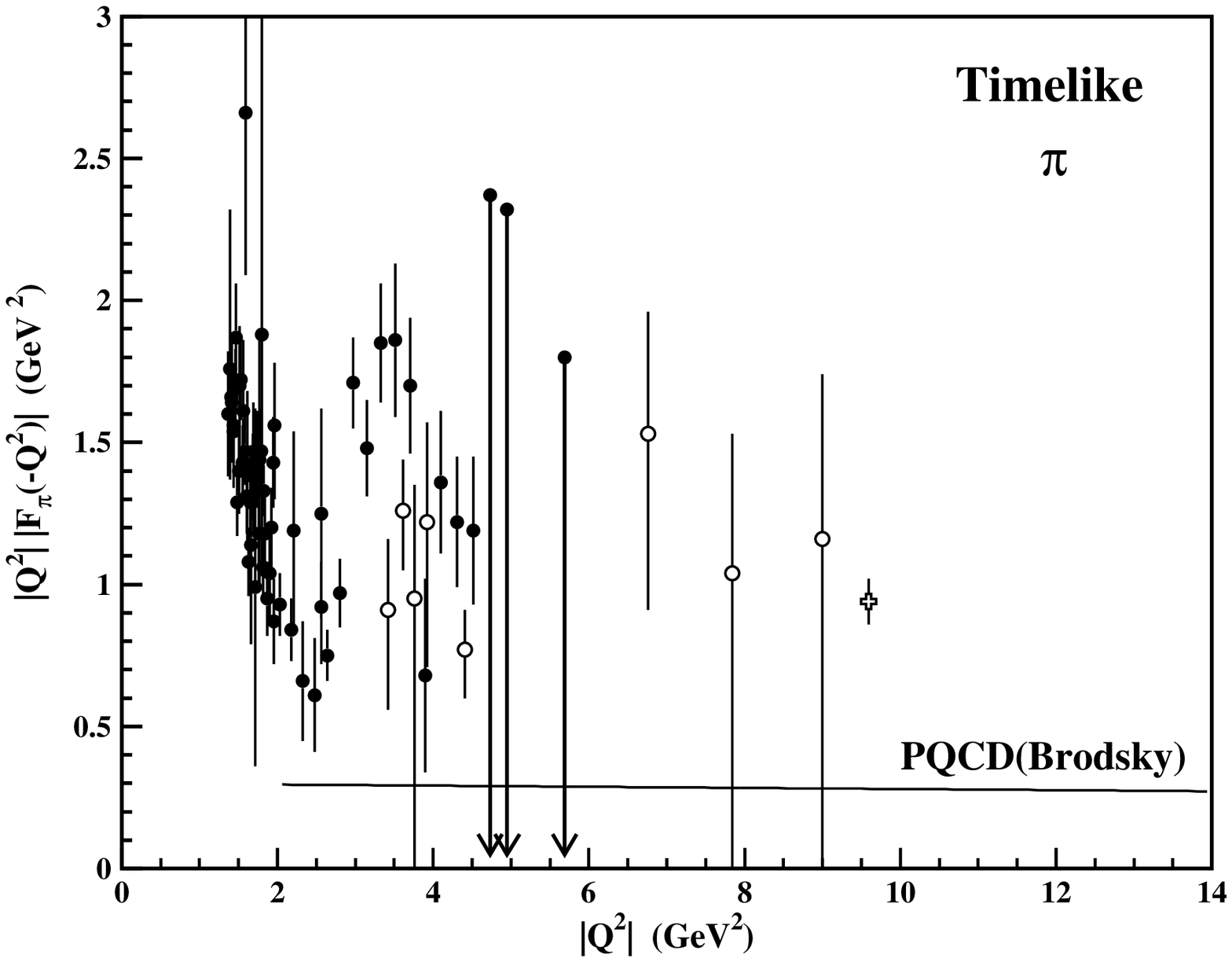}
\caption[PQCD prediction of the timelike pion form factor.]
{PQCD prediction of the timelike pion form factor.  
The solid line is the PQCD prediction by Brodsky $\etal$ \cite{Brodskyetal_tlPQCD}.  
The solid points are from $\eetopipi$ measurements with pions experimentally identified 
\cite{kpicommontlff_1}-\cite{pitlff_3}.  The open points are from $\eetohh$ 
measurements with the pion fraction of the observed $\hadpair$ determined according to 
a VDM prescription \cite{pitlff_VDM}.  
The value denoted with the plus symbol comes from interpreting the 
$J/\psi \rightarrow \pi^{+}\pi^{-}$ branching ratio as a pion form factor measurement 
as in Ref. \cite{Milanaetal_jpsipipi}.}
\label{fig:pitlpqcd}
\end{center}
\end{figure}

Brodsky $\etal$ \cite{Brodskyetal_tlPQCD} have studied the timelike form factor 
by analytically continuing the strong coupling constant from the spacelike 
region into the timelike region.  
The timelike-to-spacelike ratio of the pion form factor is \cite{Brodskyetal_tlPQCD}  
\begin{equation} 
\frac{|F^{tl}_\pi(-Q^2)|}{F^{sl}_\pi(Q^2)} = 
\frac{|\alpha_s(-Q^{2})|}{\alpha_s(Q^{2})}. 
\end{equation}
Using the asymptotic distribution amplitude and the 'frozen' version of $\alpha_s(Q^2)$ 
(for definition of frozen $\alpha_s(Q^2)$, see Eqn. \ref{eq:frozenalphas}) 
with an effective gluon mass of $m^2_g = 0.19$ GeV$^2$,  
the $|F^{tl}_\pi(-Q^2)|/F^{sl}_\pi(Q^2)$ ratio 
was found to be $\sim$1.5 for $Q^2 < 10$ GeV$^2$\cite{Brodskyetal_tlPQCD}.  
Figure \ref{fig:pitlpqcd} compares 
the $|Q^2||F^{tl}_\pi(-Q^2)|$ prediction to the existing timelike data.  
This timelike PQCD prediction is $\sim$1/3 the value determined from the $J/\psi$ decay, 
but is comparable to the Gousset and Pire prediction using the asymptotic 
distribution amplitude \cite{GoussetPire_tlPQCD}. 
Bakulev, Radyushkin, and Stefanis \cite{Bakulevetal_tlQCDSR}, 
in a contrary analysis of the analytic continuation of the strong coupling constant, 
found no enhancement in the timelike PQCD form factor due to a different parameterization 
of the strong coupling constant.

\newpage
\clearpage
\subsection{QCD Sum Rules}

\begin{figure}[!tb]
\begin{center}
\includegraphics[width=14cm]{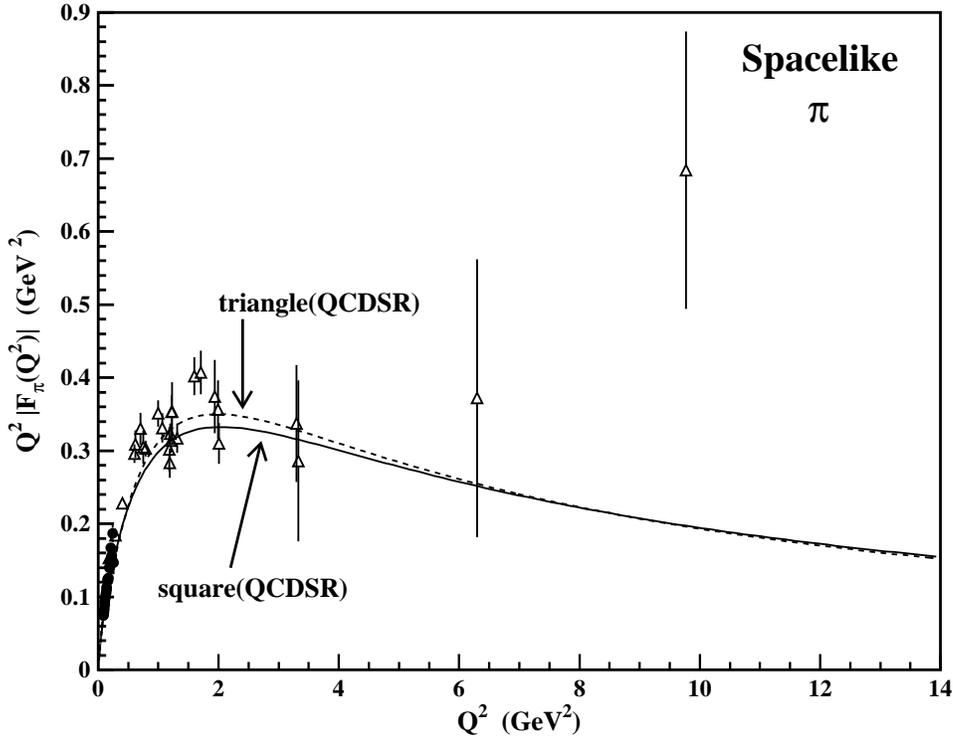}
\caption[Predictions of the spacelike pion form factor using the three-amplitude QCDSR 
method.]  
{Predictions of the spacelike pion form factor using the three-amplitude QCDSR 
method.  
The predictions are by Nesterenko and Radyushkin \cite{NesterenkoRadyushkin_QCDSR}.  
The solid line is the prediction using the square representation 
(Eqn. \ref{eq:sqlocalsr}).  
The dashed line is the prediction using the triangle representation 
(Eqn. \ref{eq:trilocalsr}) 
The solid points and open triangles are experimental data 
\cite{pislff_sc1}-\cite{pislff_elprod6}.}
\label{fig:pislqcdsr1}
\end{center}
\end{figure}

The prediction of the spacelike pion form factor using the QCDSR 
three-amplitude method was performed independently 
by Nesterenko and Radyushkin \cite{NesterenkoRadyushkin_QCDSR} 
and Ioffe and Smilga \cite{IoffeSmilga_QCDSR}. 
The spacelike pion form factor from the square representation 
(see Section 2.1.2. for definition of QCDSR variables and representations) 
is \cite{NesterenkoRadyushkin_QCDSR}
\begin{equation} 
F_\pi(Q^2) = \frac{s_0}{4\pi^2 f^2_\pi}
\left[1-\frac{1+6s_0/Q^2}{(1+4s_0/Q^2)^{3/2}}\right],
\label{eq:sqlocalsr}
\end{equation}
An alternative prediction for the spacelike pion form factor, 
from the triangle representation, is \cite{NesterenkoRadyushkin_QCDSR}
\begin{equation} 
F_\pi(Q^2) = \frac{S_0}{4\pi^2 f^2_\pi(1+Q^2/2S_0)^{2}},
\label{eq:trilocalsr}
\end{equation}
where $S_0 = \sqrt{2}~s_0$, as described in Section 2.1.2.  
Figure \ref{fig:pislqcdsr1} shows that the two spacelike pion form factor predictions 
using the QCDSR three-amplitude method are consistent with 
the experimental data with $s_0 = 4\pi^2f^2_\pi = 0.7$ GeV$^2$.

\begin{figure}[!tb]
\begin{center}
\includegraphics[width=14cm]{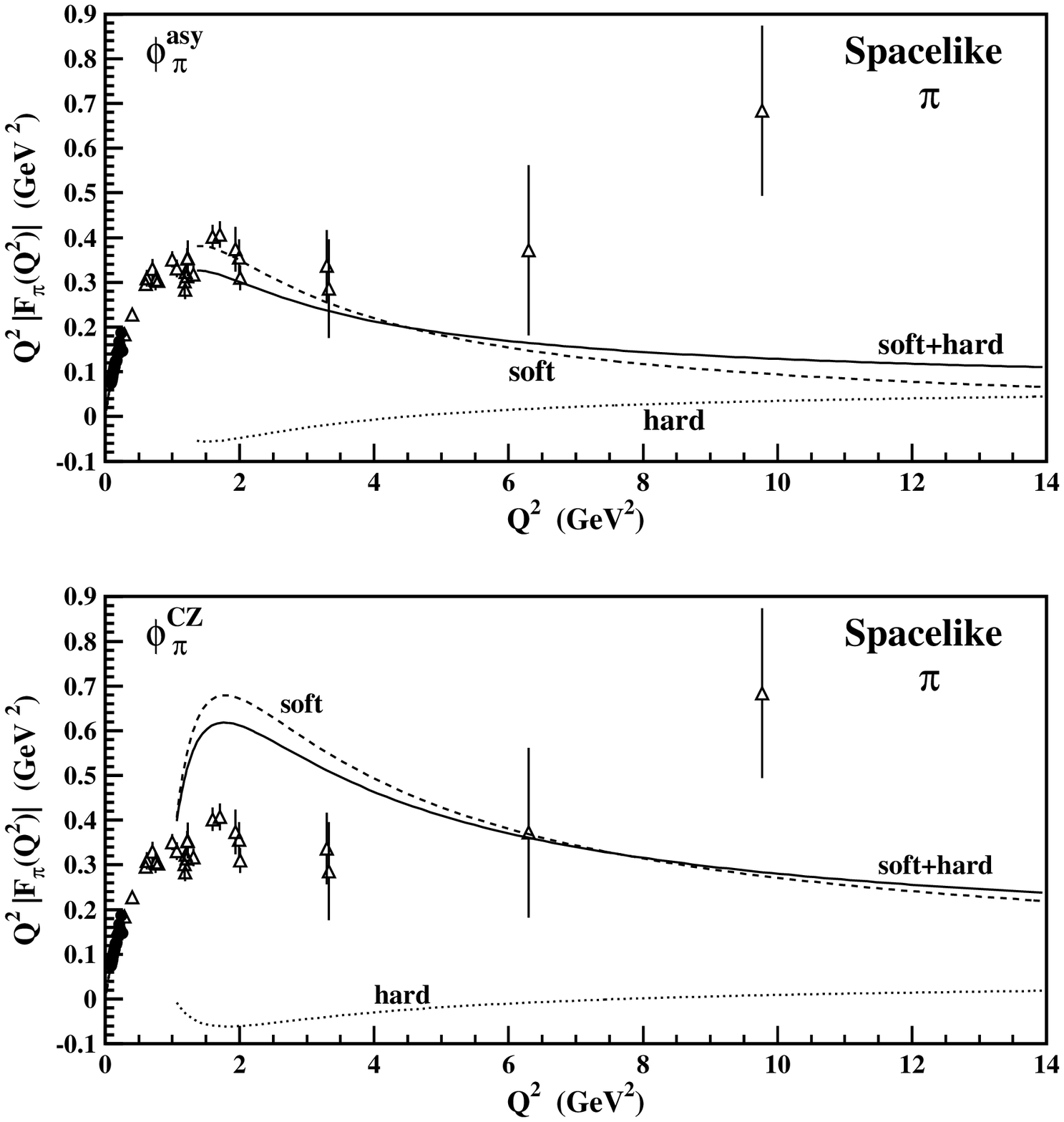}
\caption[Twist-2 contributions to the spacelike pion form factor based on the 
QCDSR correlator function method.]
{Twist-2 contributions to the spacelike pion form factor based on the QCDSR 
correlator function method.  
The predictions are by Braun $\etal$ \cite{Braunetal_QCDSR}.  
The dashed, dotted, and solid lines are the soft, hard, and soft+hard twist-2 
contributions, respectively.  
The predictions in the top and bottom plots are derived using the asymptotic and CZ pion 
distribution amplitudes, respectively.  
The solid points and open triangles are experimental data 
\cite{pislff_sc1}-\cite{pislff_elprod6}.}
\label{fig:pislcorsrt2}
\end{center}
\end{figure}

\begin{figure}[!tb]
\begin{center}
\includegraphics[width=14cm]{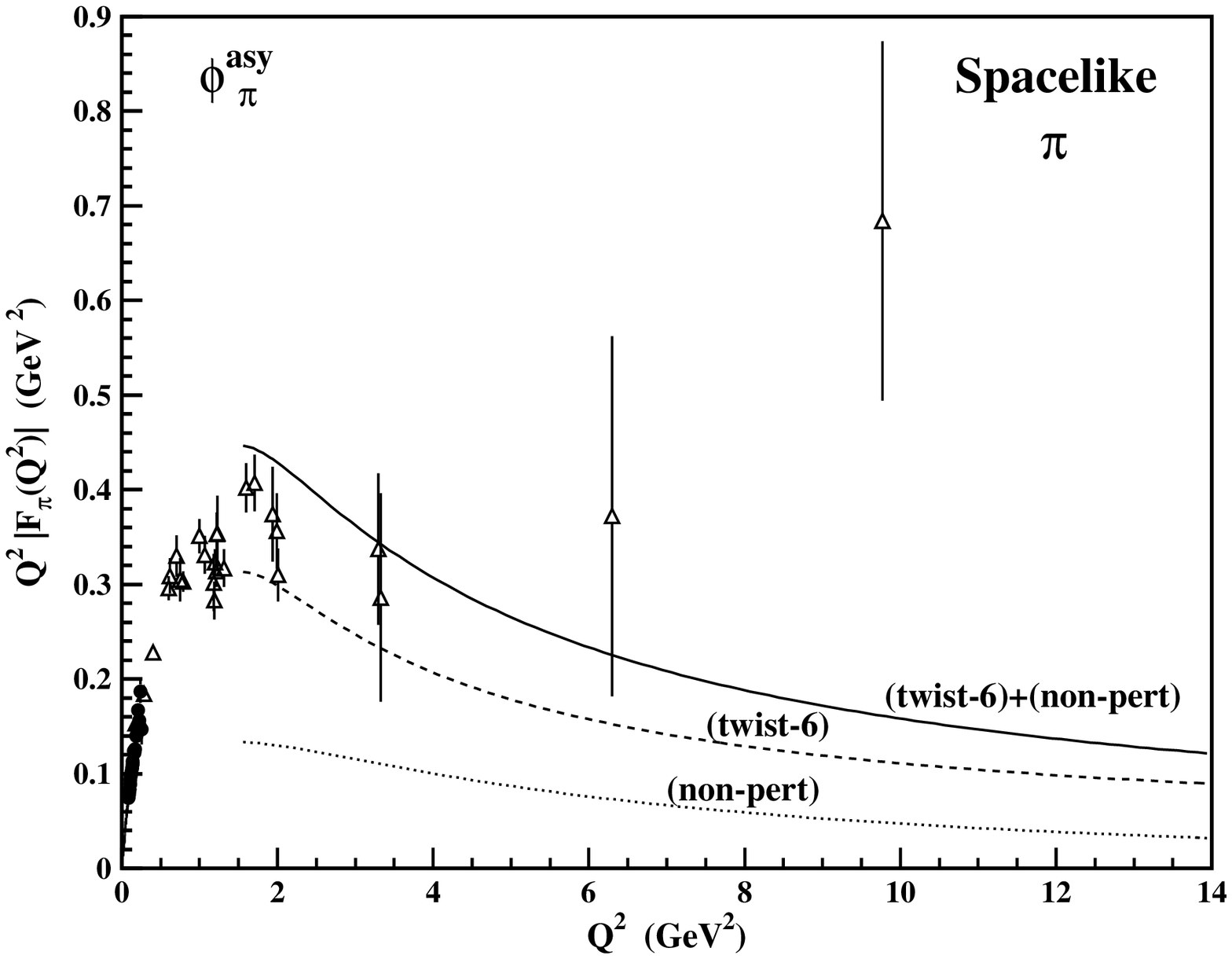}
\caption[Comparison of the correlator function QCDSR prediction 
at twist-6 accuracy to the nonperturbative 
contributions of the spacelike pion form factor.]
{Comparison of the correlator function QCDSR prediction 
at twist-6 accuracy to the nonperturbative 
contributions of the spacelike pion form factor.  
The predictions are by Braun $\etal$ \cite{Braunetal_QCDSR}.  
The dashed line is the QCDSR prediction determined to twist-6 accuracy, the dotted line 
is nonperturbative contribution, and the solid line is the sum of the contributions.  
Predictions were determined using the asymptotic distribution amplitude.  
The solid points and open triangles are experimental data 
\cite{pislff_sc1}-\cite{pislff_elprod6}.}
\label{fig:pislcorsrfull}
\end{center}
\end{figure}

Using the QCDSR correlator function method, 
Braun $\etal$ \cite{Braunetal_QCDSR} studied the effect of the twist-2 
hard and soft contributions to the spacelike pion form factor (note that the 
hard scattering process used in PQCD is a twist-2 effect).  
The soft contribution is found to dominate the hard contribution 
but leads to a slight cancellation to the overall form factor, as shown in 
Figure \ref{fig:pislcorsrt2} for both the asymptotic and CZ distribution amplitudes.  
Braun $\etal$ \cite{Braunetal_QCDSR} also studied the pion form factor 
to twist-6 accuracy.  
They also determined \cite{Braunetal_QCDSR} 
the twist-2 nonperturbative contribution to the form factor, 
defined as the difference between the total twist-2 form 
factor (hard+soft) and the LO and NLO perturbative contributions 
(Eqn. \ref{eq:nlopqcd}).  The predictions are shown in Figure \ref{fig:pislcorsrfull}.   
The nonperturbative contributions do not contribute more than 1/3 to 
the total form factor over the entire $Q^2$ range.

\begin{figure}[!tb]
\begin{center}
\includegraphics[width=14cm]{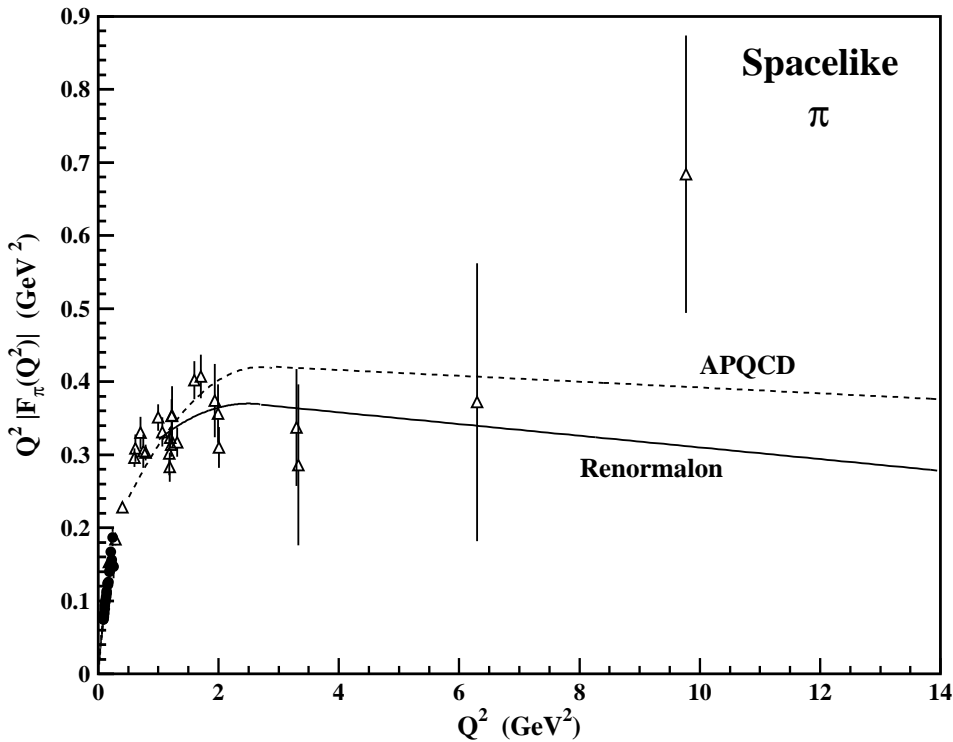}
\caption[Renormalon model and Analytic Perturbative QCD predictions 
of the spacelike pion form factor.]
{Renormalon model and Analytic Perturbative QCD predictions 
of the spacelike pion form factor.  
The solid line is the renormalon model prediction by Agaev \cite{Agaev5}.  
The dashed line is the Analytic Perturbative QCD (APQCD) prediction by 
Bakulev $\etal$ \cite{Bakulevetal_QCDSR1}.
The solid points and open triangles are experimental data 
\cite{pislff_sc1}-\cite{pislff_elprod6}.}
\label{fig:pislnewalpha}
\end{center}
\end{figure}

Two different predictions based on the QCDSR correlator function method 
were made for the spacelike pion form factor 
addressing the low $Q^2$ behavior of $\alpha_s(Q^2)$.  
Agaev \cite{Agaev5} redefined $\alpha_s(Q^2)$ using the renormalon model \cite{renormalon} 
and used the QCDSR correlator function method to determine a new distribution 
amplitude based on its moments to twist-4 accuracy.  
The prediction, shown in Figure \ref{fig:pislnewalpha}, 
is in agreement with the existing experimental data.  
Bakulev $\etal$ \cite{Bakulevetal_QCDSR1} replaced $\alpha_s(Q^2)$ with its analytic 
image based on Analytic Perturbative Theory \cite{APT1,APT2,APT3}.  They also determined 
a well behaved distribution amplitude worked to NLO, and used 
the three-amplitude QCDSR prediction in the square representation
at low $Q^2$ (Eqn. \ref{eq:sqlocalsr}).  They found 
their predictions to be consistent with the existing experimental data, as shown in 
Figure \ref{fig:pislnewalpha}.

\begin{figure}[!tb]
\begin{center}
\includegraphics[width=14cm]{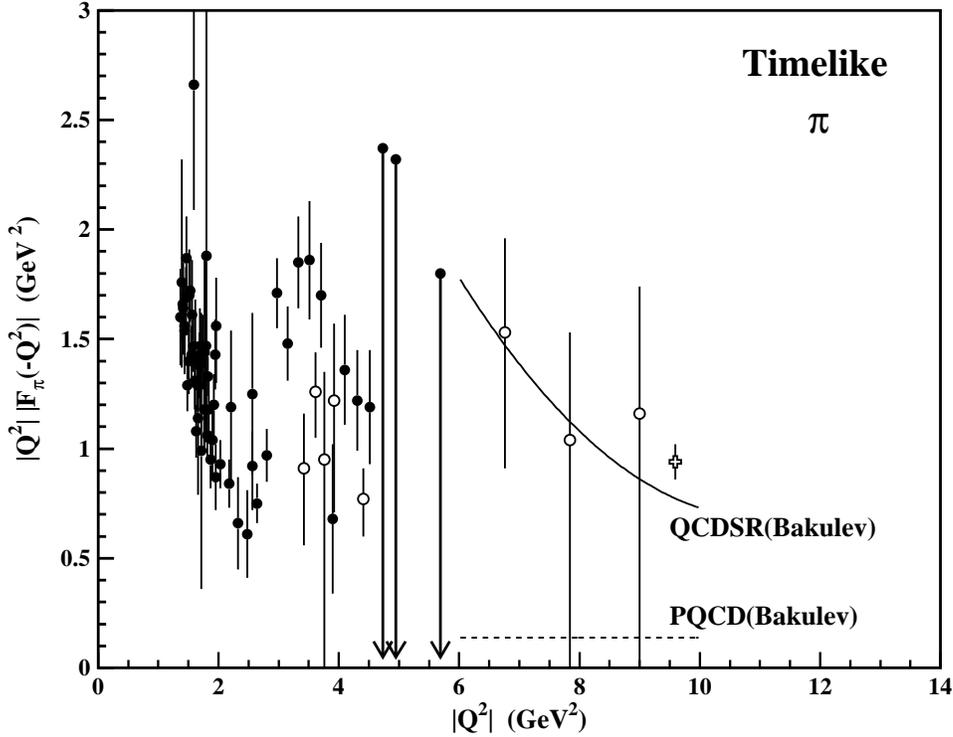}
\caption[Comparison of QCDSR and PQCD predictions of the timelike pion form factor.]
{Comparison of QCDSR and PQCD predictions of the timelike pion form factor.  
The predictions are by Bakulev $\etal$ \cite{Bakulevetal_tlQCDSR}.  
The solid line is the three-amplitude QCDSR using the triangle representation.  
The dashed line is the asymptotic PQCD prediction.  
The solid points are from $\eetopipi$ measurements with pions experimentally identified 
\cite{kpicommontlff_1}-\cite{pitlff_3}.  The open points are from $\eetohh$ 
measurements with the pion fraction of the observed $\hadpair$ determined according to 
a VDM prescription \cite{pitlff_VDM}.  
The value denoted with the plus symbol comes from interpreting the 
$J/\psi \rightarrow \pi^{+}\pi^{-}$ branching ratio as a pion form factor measurement 
as in Ref. \cite{Milanaetal_jpsipipi}.}
\label{fig:pitlqcdsr}
\end{center}
\end{figure}

Only one prediction exists for the timelike pion form factor based on the QCDSR.  
Bakulev $\etal$ \cite{Bakulevetal_tlQCDSR} analytically continued the $Q^2$ 
behavior of the spacelike pion form factor derived with the three-amplitude QCDSR method 
using the triangle representation (Eqn. \ref{eq:trilocalsr}).  
As discussed at the end of Section 2.2.1, Bakulev $\etal$ \cite{Bakulevetal_tlQCDSR} 
also determined a PQCD prediction of the timelike pion form factor by 
analytically continuing the strong coupling constant.  
They used the following parameterization for $\alpha_s(Q^2)$ \cite{Bakulevetal_tlQCDSR} 
\begin{equation} 
\alpha_s(q^2) = \frac{4\pi}{b_0}~\mathrm{arctan}
\left(\frac{\pi}{\mathrm{ln}(q^2/\Lambda^2)}\right),
\label{eq:alphastlsr}
\end{equation}
where $q^2$ = $-Q^2 > 0$ is the timelike momentum transfer.  
Bakulev $\etal$ do not find an enhancement in the timelike pion form factor 
from their asymptotic PQCD prediction \cite{Bakulevetal_tlQCDSR}.  
Their QCDSR prediction, along with their determination of the PQCD prediction 
with a fixed $\alpha_s(Q^2)$ = 0.3, is shown in Figure \ref{fig:pitlqcdsr}.  
The QCDSR prediction is consistent with the existing experimental data and $\sim$4 larger 
than the PQCD prediction at $Q^2$ = 10 GeV$^2$.

\subsection{Lattice QCD}

Lattice QCD predictions \cite{Lattice1}-\cite{Lattice7} are only available for the 
pion form factor in the spacelike region.  They are found to be consistent 
with the VDM monopole form of the form factor for spacelike momentum transfers 
\begin{equation}
Q^2 F_\pi(Q^2) = Q^2\left(\frac{m^2_\rho}{m^2_\rho+Q^2}\right),
\label{eq:monopole}
\end{equation}
where $m_\rho = 770$ MeV \cite{PDG2004} 
is the mass of the $\rho$ meson, but the predictions only exist in the limited 
momentum transfer range of $Q^2 < 3.5$ GeV$^2$.  If extended to $Q^2 = 13.5$ GeV$^2$, 
Eqn. \ref{eq:monopole} would lead to $Q^2 F_{\pi}(Q^2) = 0.57$ GeV$^2$.

\subsection{Other Models}

Other models have been proposed to explain the observed behavior in the 
experimental data for the spacelike form factor of the pion.  
They include instanton-induced contributions, 
meson cloud corrections, and predictions based on the effect of a gluon string 
tube connecting the valence quarks. 

\begin{figure}[!tb]
\begin{center}
\includegraphics[width=14cm]{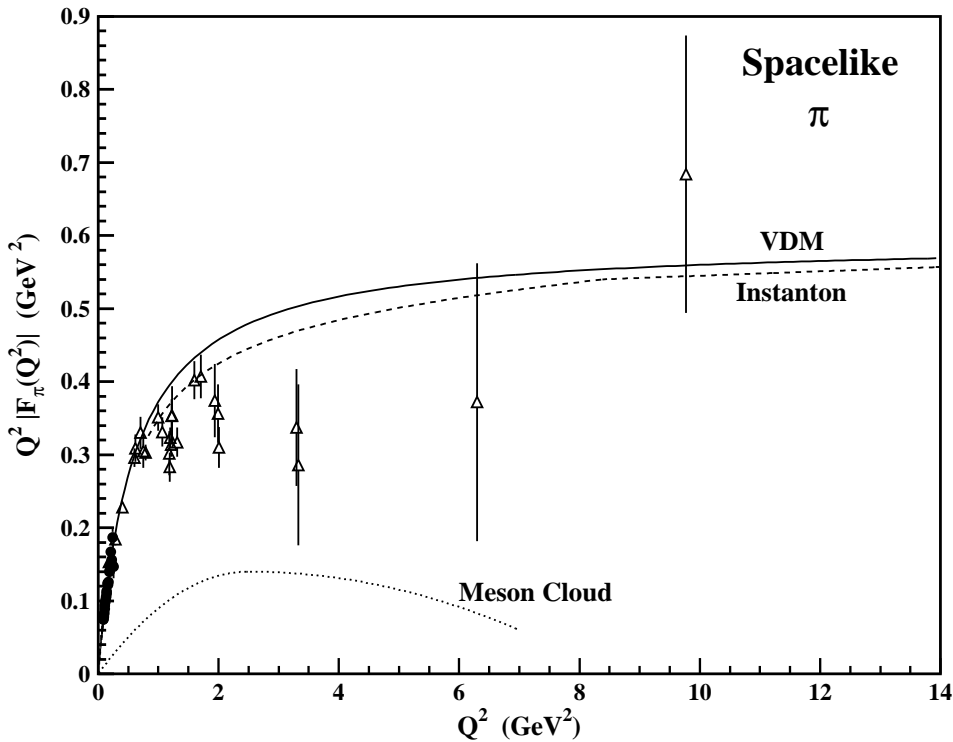}
\caption[Instanton and meson cloud model predictions of the spacelike pion form factor.]
{Instanton and meson cloud model predictions of the spacelike pion form factor.  
The dashed line is the instanton model prediction 
by Faccioli $\etal$ \cite{Facciolietal_Instanton}.  
The dotted line is the meson cloud model prediction 
by Carvalho $\etal$ \cite{Carvalhoetal_MesonCloud}.  
The solid line is the VDM monopole prediction.  
The solid points and open triangles are experimental data 
\cite{pislff_sc1}-\cite{pislff_elprod6}.}
\label{fig:pislother}
\end{center}
\end{figure}

In the instanton model the spacelike form factor of the pion 
was calculated by Faccioli $\etal$ \cite{Facciolietal_Instanton}.  They considered 
the effect of the interaction between the valence quarks in the 
pion with a single instanton, an intense classical vacuum field with the same quantum 
numbers as the pion.  The prediction is shown in Figure \ref{fig:pislother} and found 
to agree with the VDM monopole form of the form factor (Eqn. \ref{eq:monopole}).  
The authors also state \cite{Facciolietal_Instanton} that the theory will break down for 
$Q^2 > 20$ GeV$^2$ without the addition of multi-instanton effects.  

In the meson cloud model it assumed that the pion can occasionally fluctuate 
to higher Fock states consisting of a vector and pseudoscalar meson pair.  
While the contribution to the form factor from these higher Fock states are 
expected to decrease faster than the perturbative contributions, 
Carvalho $\etal$ \cite{Carvalhoetal_MesonCloud} considered the interaction 
of the virtual photon on a pion which fluctuates into $K,K^\ast$ and $\pi,\rho$ pairs.  
The accuracy of the model is drawn into question because the pion decay constant 
is determined to be an order of magnitude smaller than the experimental value.  
The prediction is shown in Figure \ref{fig:pislother}.  It is found that 
its maximum contribution is at $Q^2 \sim 2$ GeV$^2$, 
where it accounts for $\sim$$40\%$ of the experimental value, and has a 
decreasing contribution at higher values of $Q^2$.

\begin{figure}[!tbp]
\begin{center}
\includegraphics[width=15cm]{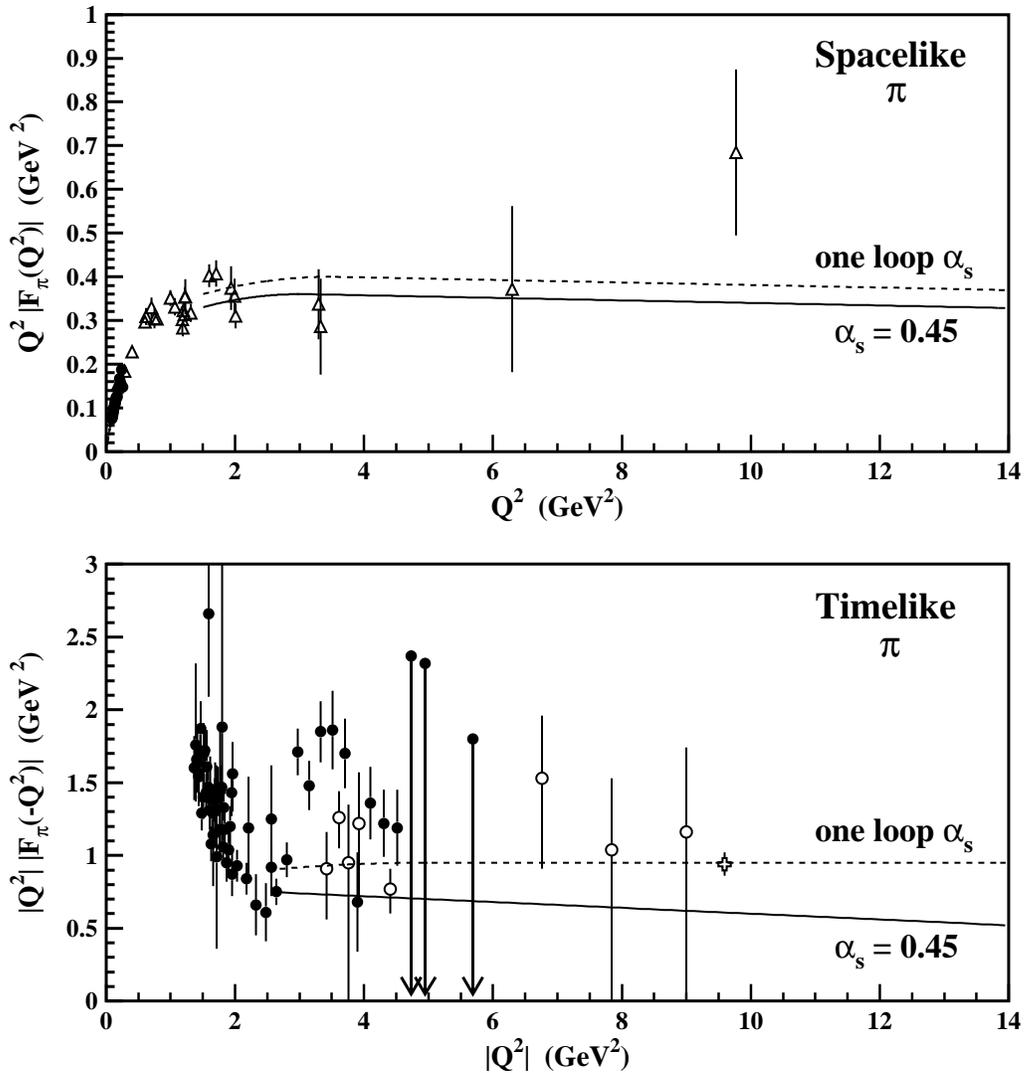}
\caption[Quark Gluon String Model predictions of the spacelike and timelike pion form 
factors.]
{Quark Gluon String Model predictions of the spacelike (top) and timelike (bottom) 
pion form factors.  
The predictions are by Kaidalov $\etal$ \cite{Kaidalovetal_Gluestring}.  
The solid lines are predictions with $\alpha_s(Q^2) = 0.45$, 
and the dashed lines are predictions with the one loop form of $\alpha_s(Q^2)$.  
The solid points are from $\eetopipi$ measurements with pions experimentally identified 
\cite{kpicommontlff_1}-\cite{pitlff_3}.  The open points are from $\eetohh$ 
measurements with the pion fraction of the observed $\hadpair$ determined according to 
a VDM prescription \cite{pitlff_VDM}.  
The value denoted with the plus symbol comes from interpreting the 
$J/\psi \rightarrow \pi^{+}\pi^{-}$ branching ratio as a pion form factor measurement 
as in Ref. \cite{Milanaetal_jpsipipi}.}
\label{fig:pistring}
\end{center}
\end{figure}

The Quark Gluon String Model (QGSM), derived by Kaidalov, Kondratyuk, 
and Tchekin \cite{Kaidalovetal_Gluestring}, is based on parameterizing the interaction 
between the struck quark and the spectator quarks by a 
color gluon string.  The model consists of convoluting two amplitudes: 
the virtual photon coupling to a $\qqbar$ pair and the gluon string between the initial 
$\qqbar$ pair fragmenting into an additional $\qqbar$ pair produced from the vacuum.  
The QGSM model incorporates the Sudakov form factor, which is shown to behave differently 
in the spacelike and timelike regions.  They predict that the ratio of 
timelike-to-spacelike 
form factor at large $|Q^2|$ behaves as \cite{Kaidalovetal_Gluestring}
\begin{equation} 
\frac{F^{tl}_\pi(|Q^2|)}{F^{sl}_\pi(Q^2)} \sim \mathrm{exp}
\left[\frac{8}{27}~\frac{\pi^2}{\mathrm{ln}(|Q^2|/(2.25~\mathrm{GeV}^2))}\right]. 
\label{eq:qgsmtlslratio}
\end{equation}
This prediction leads to a ratio of 1.6 at $|Q^2| = 13.5$ GeV$^2$.  
Figure \ref{fig:pistring} shows the QGSM predictions of the pion form factor in the 
spacelike and timelike regions and are found to be in reasonable agreement with the data.

\newpage
\subsection{Kaon Form Factor}

The charged kaon form factor is generally treated in the same manner as the 
pion form factor.  The difference consists of replacing the down quark 
with a strange quark.  The presence of the strange quark, with its 
larger mass, tends to change the interpretation of the meson into a massive particle 
orbited by the lighter quark, as compared to the near equal mass of the up and down 
quarks in the case of the pion.  The difference in the quark masses leads to mass 
splitting terms and cause theories to acquire slight modifications.

\begin{figure}[!tb]
\begin{center}
\includegraphics[width=14cm]{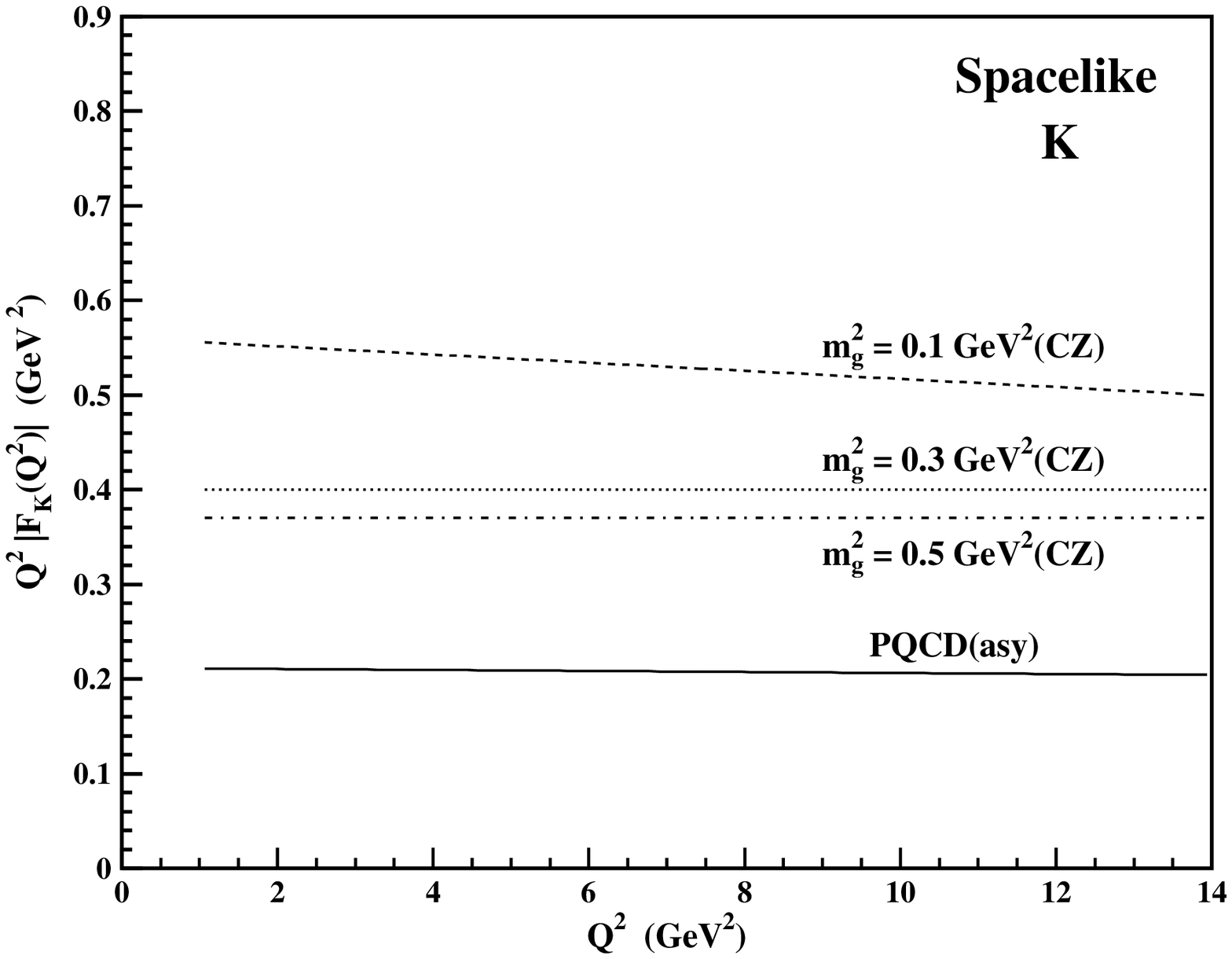}
\caption[PQCD factorization predictions of the spacelike kaon form factor.]
{PQCD factorization predictions of the spacelike kaon form factor.  
The solid line is the PQCD prediction by Lepage and Brodsky \cite{LepageBrodsky_PQCDFF}.  
The other lines are the PQCD predictions using the CZ kaon 
distribution amplitude and the frozen version of the $\alpha_s(Q^2)$ by 
Ji and Amiri \cite{JiAmiri_QCDSR} .  
The dashed, dotted, and dash-dotted lines correspond to 
$m^2_g$ = 0.1, 0.3, and 0.5 GeV$^2$, respectively, and $\Lambda = 0.1$ GeV is assumed.}
\label{fig:kslfact}
\end{center}
\end{figure}

The asymptotic PQCD prediction of the kaon form factor has the same form as the 
pion, only the pion decay constant is replaced by that of the kaon.  
This leads to \cite{LepageBrodsky_PQCDFF}  
\begin{equation}
F_{K}(Q^2) \rightarrow \frac{8\pi~\alpha_s(Q^2)~f^2_{K}}{Q^2}.
\end{equation}  
The spacelike prediction is shown in Figure \ref{fig:kslfact}.  The kaon-to-pion 
form factor ratio from asymptotic PQCD is therefore 
\begin{equation}
\frac{F_{K}(Q^2)}{F_{\pi}(Q^2)} = \frac{f^2_{K}}{f^2_{\pi}} = 1.49\pm0.03 ,
\end{equation}  
using the PDG values \cite{PDG2004} of $f_{K}$ =  159.8 $\pm$ 1.5 MeV and 
$f_{\pi}$ = 130.7 $\pm$ 0.4 MeV. 

\begin{figure}[!tb]
\begin{center}
\includegraphics[width=14cm]{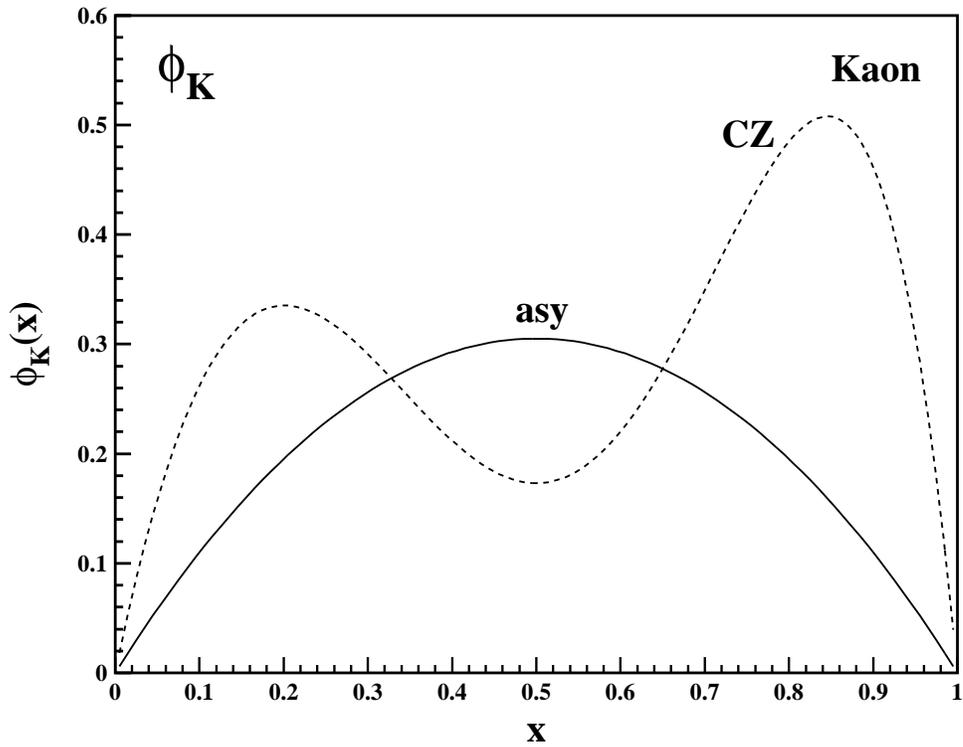}
\caption[Asymptotic and CZ kaon distribution amplitudes as a function of quark 
momentum fraction.]
{Asymptotic and CZ kaon distribution amplitudes as a function of quark 
momentum fraction $x$.  The solid line is the asymptotic form 
($\phi^{asy}_{K}(x) = (f_K\sqrt{3/2})x(1-x)$) 
and the dashed line is the Chernyak-Zhitnitsky form 
($\phi^{CZ}_{~K}(x) = (f_K15/\sqrt{3})~x(1-x)
[0.6(2x-1)^2+~0.25(2x-1)^3+~0.08]$).  
The kaon decay constant is taken to be unity in this figure.}
\label{fig:kda}
\end{center}
\end{figure}

Chernyak and Zhitnitsky determined a distribution amplitude for the kaon based on 
the QCDSR distribution amplitude moment method.  
The CZ distribution amplitude for the kaon is \cite{ChernyakZhitnitsky_QCDSR} 
\begin{equation} 
\phi^{CZ}_{~K}(x) = \frac{15}{\sqrt{3}}~f_K~x(1-x)
[~0.6(2x-1)^2+~0.25(2x-1)^3+~0.08],
\label{eq:czkda}
\end{equation}
Figure \ref{fig:kda} shows the comparison between the CZ and asymptotic 
distributions amplitudes for the 
kaon as a function of quark momentum fraction.  The asymmetric nature of the CZ 
distribution amplitude is caused by $s\overline{s}$ pairs present 
in the vacuum condensate.  The kaon to pion form factor ratio using the CZ distribution 
amplitudes is \cite{ChernyakZhitnitsky_QCDSR}
\begin{equation}
\frac{F_{K}(Q^2)}{F_{\pi}(Q^2)} = \frac{f^2_{K}}{f^2_{\pi}}~\frac{I_{K}}{I_{\pi}} 
= 0.99\pm0.02,
\end{equation}  
where $I_{K}/I_{\pi} = 2/3$ arises from using the CZ distribution amplitudes for 
the pion and kaon.  
Ji and Amiri \cite{JiAmiri_QCDSR} determined 
the spacelike kaon form factor prediction using the CZ kaon distribution amplitude 
in the PQCD factorization scheme.  As with their prediction of the spacelike pion 
form factor, they used a frozen version of $\alpha_s(Q^2)$ 
(see Eqn. \ref{eq:frozenalphas} for the definition of the frozen $\alpha_s(Q^2)$).  
Figure \ref{fig:kslfact} shows that the spacelike form factor prediction of the kaon 
with the CZ distribution 
amplitude is $2-3$ larger than with the asymptotic distribution amplitude.

\begin{figure}[!tb]
\begin{center}
\includegraphics[width=14cm]{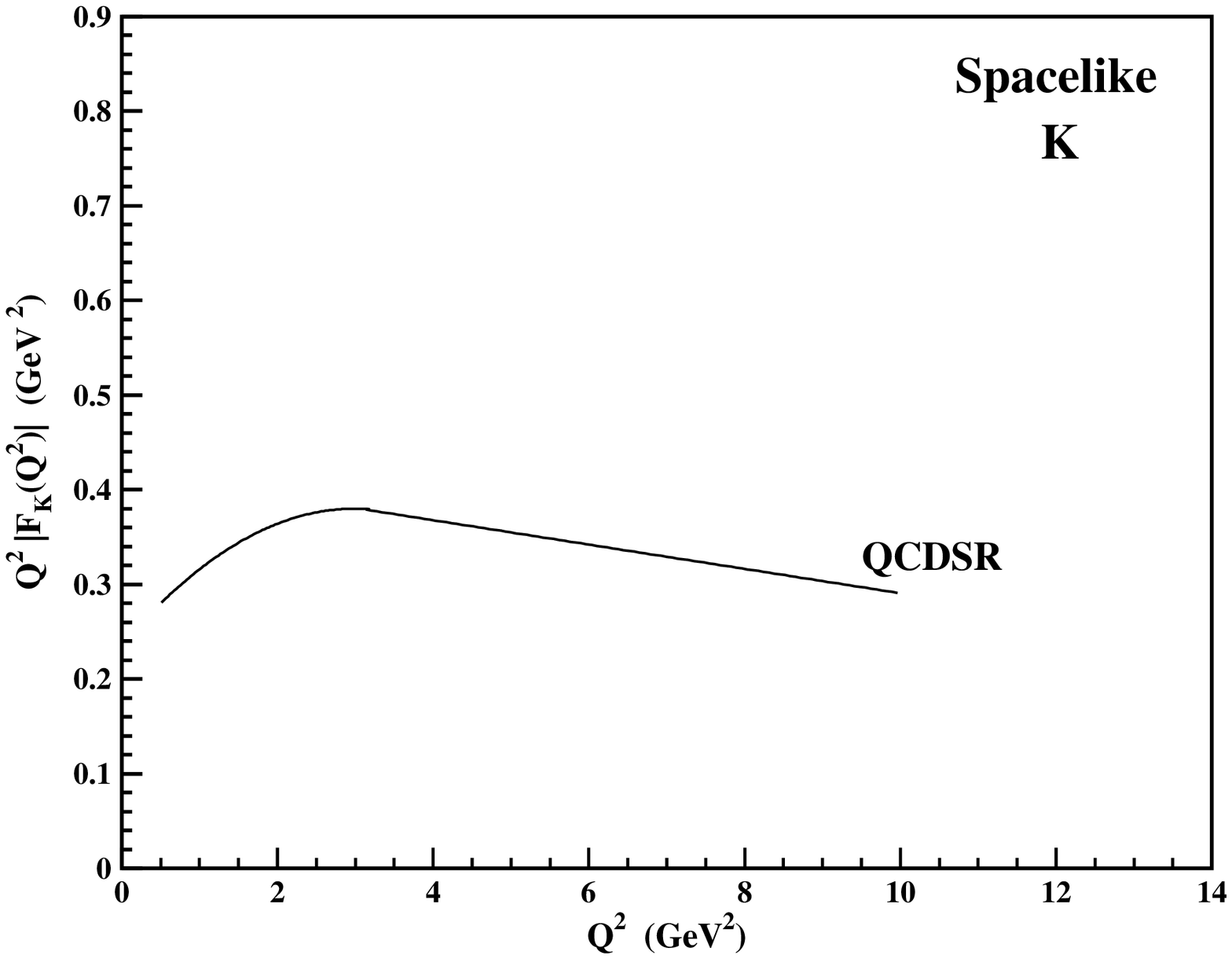}
\caption[Predictions of the spacelike kaon form factor.]
{Predictions of the spacelike kaon form factor.  The line is the 
QCDSR prediction by Bijnens and Khodjamirian \cite{BijnensKhodjamirian_QCDSR}.}
\label{fig:kslqcdsr}
\end{center}
\end{figure}

Bijnens and Khodjamirian \cite{BijnensKhodjamirian_QCDSR} have calculated the 
spacelike kaon form factor using the QCDSR correlator function method 
to twist-6 accuracy.  Their prediction for the spacelike kaon form factor, 
shown in Figure \ref{fig:kslqcdsr}, falls between the PQCD predictions of the kaon 
using the CZ and asymptotic distribution amplitudes.  

Since there is a total absence of experimental data for the spacelike form factor of the 
kaon, and the precision of the existing timelike data is extremely poor, 
it is not possible to determine which of the various theoretical
predictions above describes the nature of the kaon form factor.  
It should also be noted that no explicit calculations exist for the kaon form factor 
with timelike momentum transfers based on PQCD, QCDSR, or Lattice QCD.

\newpage
\clearpage
\section{Proton Form Factors in Theory}


The proton is a spin-1/2 hadron and therefore contains both electric charge and current 
distributions which are described by two distinct form factors.  
An equivalent representation of the form factors is in terms of the helicity 
conserving and helicity changing contributions to the electromagnetic form factors.  
The helicity conserving form factor is given by the Dirac form factor, 
$F^P_1(Q^2)$, and the helicity changing form factor is given by the Pauli 
form factor, $F^P_2(Q^2)$.  The matrix element for \textbf{spacelike} momentum transfers is
\begin{equation}
\langle \mathrm{p}(p_2)|~j^{em}_\mu~|\mathrm{p}(p_1)\rangle = 
\gamma_\mu F_1(Q^2) + \left(\frac{\kappa_p}{2m_p}\right)\sigma_{\mu\nu}q_\nu F_2(Q^2), 
\label{eq:prslmatrixel}
\end{equation}
where $m_p$ and $\kappa_p$ are the mass and anomalous magnetic moment of the proton, 
respectively, and 
$\sigma_{\mu\nu} = \frac{i}{2}(\gamma_\mu\gamma_\nu - \gamma_\nu\gamma_\mu)$.  
For \textbf{timelike} momentum transfer, the matrix element is obtained by replacing 
$\langle \mathrm{p}(p_2)|~j^{em}_\mu~|\mathrm{p}(p_1)\rangle$ by 
$\langle \mathrm{p}(p_2)\overline{\mathrm{p}}(p_1)|~j^{em}_\mu~|0\rangle$. 

The Dirac and Pauli form factors are related to the Sachs 
electric and magnetic form factors, $G^{P}_{E}(Q^2)$ and $G^{P}_{M}(Q^2)$, 
respectively, by
\begin{equation}
G^{P}_{E}(Q^2) = F^{P}_{1}(Q^2) - (\frac{Q^2}{4m^{2}_{p}}) \kappa_{p}F^{P}_{2}(Q^2), 
~~~~~~~~G^{P}_{M}(Q^2) = F^{P}_{1}(Q^2) + \kappa_{p}F^{P}_{2}(Q^2), 
\label{eq:ch2f1f2togegm}
\end{equation}  
where the proton mass $m_p$ = 0.93827 GeV \cite{PDG2004} and 
anomalous magnetic moment of the proton $\kappa_p$ = 1.79 \cite{PDG2004}.  
For protons at rest ($Q^2$ = 0) the form factors are normalized as
\begin{equation}
F^{P}_{1}(0) = 1~~~~~~~~~~F^{P}_{2}(0) = 1 
\end{equation}
and therefore,
\begin{equation}
G^{P}_{E}(0) = 1~~~~~G^{P}_{M}(0) = 1+\kappa_{p} = \mu_p = 2.79 
\end{equation}
where $\mu_p$ is the magnetic moment of the proton in units of the nuclear magneton.  

The proton form factors in the timelike region have an additional relationship. 
At the threshold for $\ppbar$ production ($-Q^2 = 4m^2_p$) 
from Eqn. \ref{eq:ch2f1f2togegm}, 
it follows that the electric and magnetic form factors are equal 
\begin{equation}
G^{P}_{E}(4m^2_p) = G^{P}_{M}(4m^2_p) = F^P_1(4m^2_p) + \kappa_p F^P_2(4m^2_p).
\end{equation}

In the following discussions and figures, the experimental data for spacelike form 
factors are from Refs. \cite{prslff_gmonly1}-\cite{prslff_gmonly6}, with 
$G^{P}_{E}(Q^2) = G^{P}_{M}(Q^2)/\mu_p$ assumed, 
and for timelike form factors are from Refs. \cite{prtlff_2}-\cite{BABARppbarISR}, 
with$|G^{P}_{E}(Q^2)| = |G^{P}_{M}(Q^2)|$ assumed.

\subsection{Perturbative Quantum Chromodynamics}

\begin{figure}[!tb]
\begin{center}
\includegraphics[width=12cm]{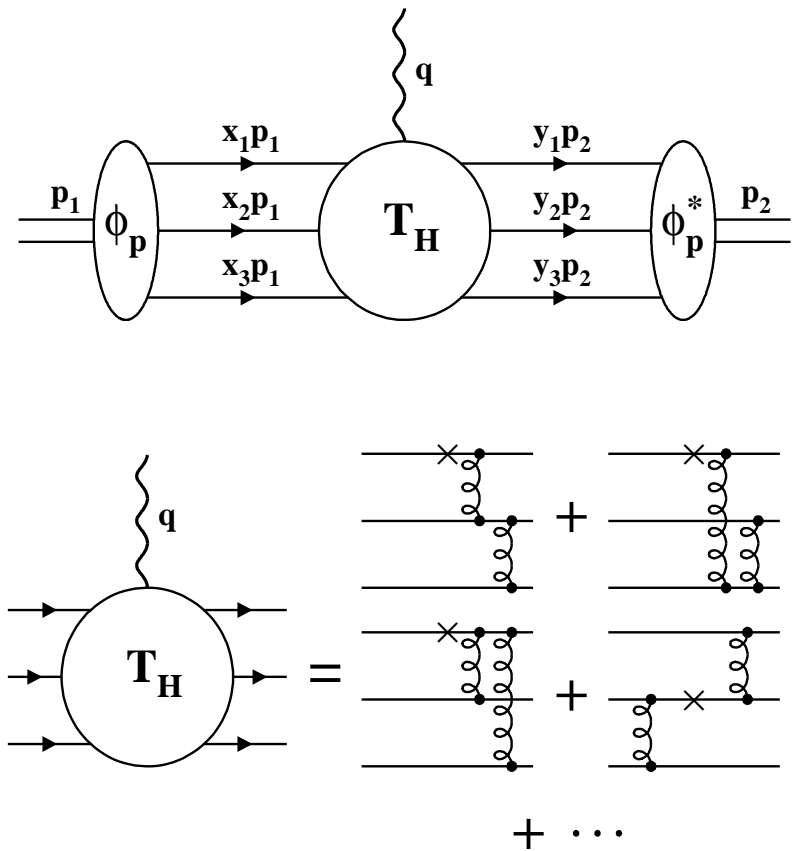}
\caption[Diagrams for the PQCD factorized proton hard scattering process.]
{Diagrams for the PQCD factorized proton hard scattering process.  
The top figure shows the proton approaching from the left in the form of the 
initial proton distribution amplitude ($\phi_{\mathrm{p}}$), 
the virtual photon interacting with the proton (T$_{\mathrm{H}}$), 
and the final proton distribution amplitude ($\phi^{\ast}_{\mathrm{p}}$) 
after the interaction.  The bottom figure shows the leading order terms for 
the hard scattering amplitude  (T$_{\mathrm{H}}$).  The crosses represent 
the quark-photon interaction.}
\label{fig:feyprhard}
\end{center}
\end{figure}

As described in Section 2.1.1, the basic premise of PQCD is the validity of factorization.  
As for the case of mesons, the non-perturbative part contains the proton wave 
function, or the distribution amplitude, and the perturbative part consists of the 
hard scattering amplitude.  
The PQCD factorization diagrams for the proton are shown schematically in 
Figure \ref{fig:feyprhard}.  While the factorization scheme is the same as for the 
mesons, the hard scattering amplitude for the proton is more complicated because of 
the fact that, with three valence quarks, 
two gluons are needed to transfer the momentum from the struck quark to the 
other two.  
This causes the overall form factor to be proportional to $(\alpha_s(Q^2)/Q^2)^2$, 
or $\alpha^2_s(Q^2)/Q^4$, consistent with the behavior predicted from the 
'quark counting rules' \cite{ffscaling1,ffscaling2,ffscaling3}.  
The spin-flip of the quarks in the helicity changing form factor 
is suppressed by $\sim 1/Q^2$ and, at large $Q^2$, 
the Dirac and Pauli form factors are $F^P_1(Q^2) \propto 1/Q^4$ 
and $F^P_2(Q^2) \propto 1/Q^6$, respectively.  
Therefore, $F^P_2(Q^2)$ is neglected in comparison to $F^P_1(Q^2)$ at large $Q^2$, 
and the dominant behavior of the form factors is 
$G^P_M(Q^2) \approx G^P_E(Q^2) \approx F^P_1(Q^2) \propto 1/Q^4$.

The formalism for the proton electromagnetic form factors in the factorization formalism 
was derived by Lepage and Brodsky \cite{LepageBrodsky_PQCDFF,LepageBrodsky_ProtonPQCD}.  
The magnetic form factor of the proton is expressed in the factorized PQCD scheme by 
\begin{equation}
G^P_M(Q^2) = \int^{1}_{0}\int^{1}_{0}[dx][dy]~\phi^{\ast}_p(y_i,Q^2)
~T_H(x_i,y_i,Q^2)~\phi_p(x_i,Q^2),
\label{eq:Gmfact}
\end{equation}
where the hard scattering amplitude is denoted by $T_H(x_i,y_i,Q^2)$, 
the incoming and outgoing proton distribution amplitudes 
are denoted by $\phi_p(x_i,Q^2)$ and $\phi^{\ast}_p(y_i,Q^2)$, respectively.  
The integration variable of the quark momentum fractions is 
$[dx] \equiv dx_1dx_2dx_3~\delta(1 - \Sigma_i x_i)$, and similarly for $[dy]$.  

The hard scattering amplitude $T_H(x_i,y_i,Q^2)$ incorporates the 
short-distance interactions between the constituent quarks inside the proton.  
With the second quark assigned opposite helicity, the lowest order 
contribution from the emission of two hard gluons is given as 
\cite{LepageBrodsky_PQCDFF,LepageBrodsky_ProtonPQCD} 
\begin{equation}
T_H(x_i,y_i,Q^2) = \left[\frac{2~\alpha_s(Q^2)}{3~Q^2}\right]^2
\left[~\sum^{3}_{j=1}e_jT_j(x_i,y_i) + (x_i \leftrightarrow y_i) \right],
\label{eq:prTH}
\end{equation}
where  
\begin{equation}
T_1 = T_3(1 \leftrightarrow 3) = 
\frac{1}{x_2x_3(1-x_3)}\frac{1}{y_2y_3(1-y_3)} - 
\frac{1}{x_3(1-x_1)^2}\frac{1}{y_3(1-y_1)^2}, 
\label{eq:prTHterm1}
\end{equation}
\begin{equation}
T_2 = -\frac{1}{x_1x_3(1-x_1)}\frac{1}{y_1y_3(1-y_3)} 
\label{eq:prTHterm2}
\end{equation}
and $e_j$ is the electric charge of quark $j$. The symbol $(x_i \leftrightarrow y_i)$ 
means to replace $x_i$ with $y_i$ and $y_i$ with $x_i$ in Eqns. \ref{eq:prTHterm1} and 
\ref{eq:prTHterm2}, while the symbol $(1 \leftrightarrow 3)$ in 
Eqn. \ref{eq:prTHterm1} means to replace quark momentum fractions with subscript 1 with 
those of subscript 3 for $T_3$.

\begin{figure}[!tb]
\begin{center}
\includegraphics[width=14cm]{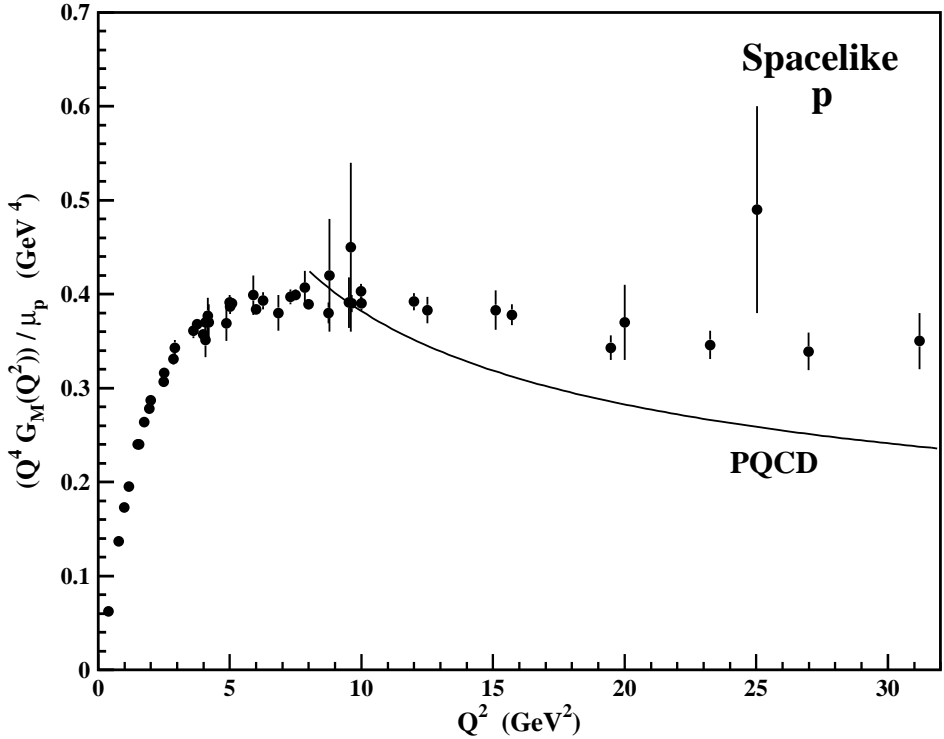}
\caption[PQCD prediction of the proton magnetic form factor in the spacelike region.]
{PQCD prediction of the proton magnetic form factor in the spacelike region.
The line is the PQCD prediction by Lepage and Brodsky 
\cite{LepageBrodsky_PQCDFF} using the simplified distribution 
amplitude $\phi_{p}(x_i,Q^2) = \delta(x_1 - \frac{1}{3})\delta(x_2 - \frac{1}{3})$ and 
is arbitrary normalized at $G^P_M$(8 GeV$^2$) = 0.42.  
The one loop form of $\alpha_s(Q^2)$ is used with 
$n_f = 3$ and $\Lambda = 0.316$ GeV.}  
\label{fig:prslpqcd}
\end{center}
\end{figure}

The proton distribution amplitude in the large $Q^2$, 
or asymptotic, limit ($Q^2 \to \infty$), is 
\cite{LepageBrodsky_PQCDFF,LepageBrodsky_ProtonPQCD} 
\begin{equation}
\phi^{asy}_{~p}(x_i,Q^2) = C~x_1 x_2 x_3
\left(\mathrm{ln}\frac{Q^2}{\Lambda^2}\right)^{-2/3\beta},
\label{eq:prasypida}
\end{equation}
where $\beta = 11 - \frac{2}{3}n_f$, $n_f$ is the number of quark flavors, and 
$C$ is an arbitrary coefficient. The proton does not have an equivalent 
decay constant as in the case of the mesons, so the proton distribution 
amplitude, and therefore the form factor, is not absolutely normalized.  
The magnetic form factor of the proton in the spacelike region at large $Q^2$ is 
\cite{LepageBrodsky_PQCDFF,LepageBrodsky_ProtonPQCD}
\begin{equation}
G^P_M(Q^2) = C^2~\frac{\alpha^2_s(Q^2)}{Q^4}
\left(\mathrm{ln}\frac{Q^2}{\Lambda^2}\right)^{-4/3\beta}.
\label{eq:prpqcd}
\end{equation}
or
\begin{equation}
Q^4~G^P_M(Q^2)/\mu_p 
= (C^2/\mu_p)~\frac{\alpha^2_s(Q^2)}{Q^4}
\left(\mathrm{ln}\frac{Q^2}{\Lambda^2}\right)^{-4/3\beta}~\mathrm{GeV}^4.
\label{eq:q4prpqcd}
\end{equation}
The asymptotic distribution amplitude in Eqn. \ref{eq:prasypida} leads to the bizarre 
PQCD prediction for the magnetic form factor of $G^P_M(Q^2) = 0$ 
for all values of $Q^2$ \cite{LepageBrodsky_PQCDFF}.  
This result points to a proton distribution amplitude having an asymmetric form.  
The magnetic form factor prediction \cite{LepageBrodsky_PQCDFF} as a function of $Q^2$, 
and using a simplified distribution amplitude 
of $\phi_{p}(x_i,Q^2) = \delta(x_1 - \frac{1}{3})\delta(x_2 - \frac{1}{3})$, 
is shown in Figure \ref{fig:prslpqcd}.  
Alternative distribution amplitudes have been derived from QCD Sum Rules 
and are described below.

Various distribution amplitudes for the proton have been developed in the literature.  
They are determined from the QCDSR distribution amplitude moment method.  
The moments of the proton distribution amplitude are defined as 
\cite{ChernyakZhitnitsky_QCDSR}
\begin{equation} 
\langle x^{n_1}_1x^{n_2}_2x^{n_3}_3\rangle 
= \int^{1}_{0} d^{3}x~x^{n_1}_1x^{n_2}_2x^{n_3}_3~\phi(x_i).
\label{eq:prdamoment}
\end{equation}
The distribution amplitudes all have the general form of
\begin{equation}
\phi_p(x_i) = \phi^{asy}_{~p}(x_i)
[Ax^2_1 + Bx^2_2 + Cx^2_3 + Dx_1x_2 + Ex_1 + Fx_2 + Gx_3 + H],
\label{eq:genprda}
\end{equation}
where $\phi^{asy}_{~p}(x_i) = 120f_Nx_1x_2x_3$ and $f_N \approx 5\times10^{-3}$ 
GeV$^2$ is an effective proton decay constant determined by 
the QCDSR \cite{Ioffe_fn,Chungetal_fn}.  
Table \ref{tab:prdacoeff} lists the coefficients 
$A-H$ of the various distribution amplitudes which have been proposed.  
The distribution amplitudes derived by 
Chernyak and Zhitnitsky (CZ) \cite{ChernyakZhitnitsky_QCDSR,ChernyakZhitnitsky_PRDA}, 
King and Sachrajda (KS) \cite{KingSachrajda_PRDA}, and 
Gari and Stefanis (GS) \cite{GariStefanis_PRDA,Stefanis_PRDA} were determined 
from the first two moments ($n \le 2$, where $n = n_1 + n_2 + n_3$ 
and $n_i$ is the moment of the $i$th quark in the proton).  
The distributions derived by Chernyak, Ogloblin, and Zhitnitsky 
(COZ) \cite{Chernyaketal_PRDA} 
used the first three moments ($n \le 3$).  The distribution amplitude derived by 
Stefanis and Bergmann \cite{StefanisBergmann_PRDA} is a hybrid of the 
COZ and GS distribution amplitudes and is called a ``heterotic'' (Het) amplitude.  
The resulting momentum fractions of the $i$th quark are also 
given in Table \ref{tab:prdacoeff}.  
All, except for the asymptotic distribution amplitude, determine the up quark with 
the same helicity as the proton (quark $\#$1) to carry 
$\sim$60$\%$ of the proton's momentum.  For the asymptotic distribution amplitude 
$\langle x_1 \rangle = 33.3\%$

\begin{table}[h]
\caption[Parameters for different proton distribution amplitudes.]
{Parameters for different proton distribution amplitudes.  The variables $A-H$ are 
the coefficients in Eqn. \ref{eq:genprda}.  
The variables $\langle x_i \rangle$ denote the momentum fraction carried by the $i$th 
quark in the proton.}
\begin{center}
\begin{tabular}{|c|c|c|c|c|c|c|}
\hline
 & Asy  & CZ  & KS  & GS  & COZ & Het  \\
 & \cite{LepageBrodsky_PQCDFF,LepageBrodsky_ProtonPQCD}
 & \cite{ChernyakZhitnitsky_QCDSR,ChernyakZhitnitsky_PRDA}
 & \cite{KingSachrajda_PRDA}
 & \cite{GariStefanis_PRDA,Stefanis_PRDA}
 & \cite{Chernyaketal_PRDA}
 & \cite{StefanisBergmann_PRDA}
 \\
\hline
A & 0 &  18.07 &  20.16 &  30.92 &  23.814 & -2.916  \\
B & 0 &  4.63  &  15.12 & -25.28 &  12.978 &  0      \\
C & 0 &  8.82  &  22.68 &  12.94 &  6.174  &  75.25  \\
D & 0 &  0     &  0     &  55.66 &  0      &  16.625 \\
E & 0 &  0     &  1.68  & -23.65 &  0      &  32.756 \\
F & 0 &  0     & -1.68  &  23.65 &  0      &  26.569 \\
G & 0 & -1.68  & -6.72  &  0     &  5.88   & -32.756 \\
H & 1 & -2.94  &  5.04  & -9.92  & -7.098  & -19.773 \\
\hline
$\langle x_1 \rangle$ & 1/3  & 0.63  & 0.55 &  0.63  & 0.579  & 0.572  \\
$\langle x_2 \rangle$ & 1/3  & 0.15  & 0.21 &  0.14  & 0.192  & 0.184  \\
$\langle x_3 \rangle$ & 1/3  & 0.22  & 0.24 &  0.236 & 0.229  & 0.244  \\
\hline
\end{tabular}
\label{tab:prdacoeff}
\end{center}
\end{table}

Isgur and Llewellyn Smith \cite{Isgur_LS_argue1,Isgur_LS_argue2} have argued that 
comparing the proton form factor predictions from the PQCD factorization scheme 
to the existing data is not appropriate because of the endpoint problem.  
They argue that at most $\sim$1$\%$ of the spacelike magnetic form factor 
predicted (with the CZ distribution amplitude) 
can be attributed to the PQCD prediction for $Q^2 < 25$ GeV$^2$.  
They conclude that the major part of the form factors at currently 
accessible energies arises from higher order and nonperturbative effects.  

\begin{figure}[!tb]
\begin{center}
\includegraphics[width=14cm]{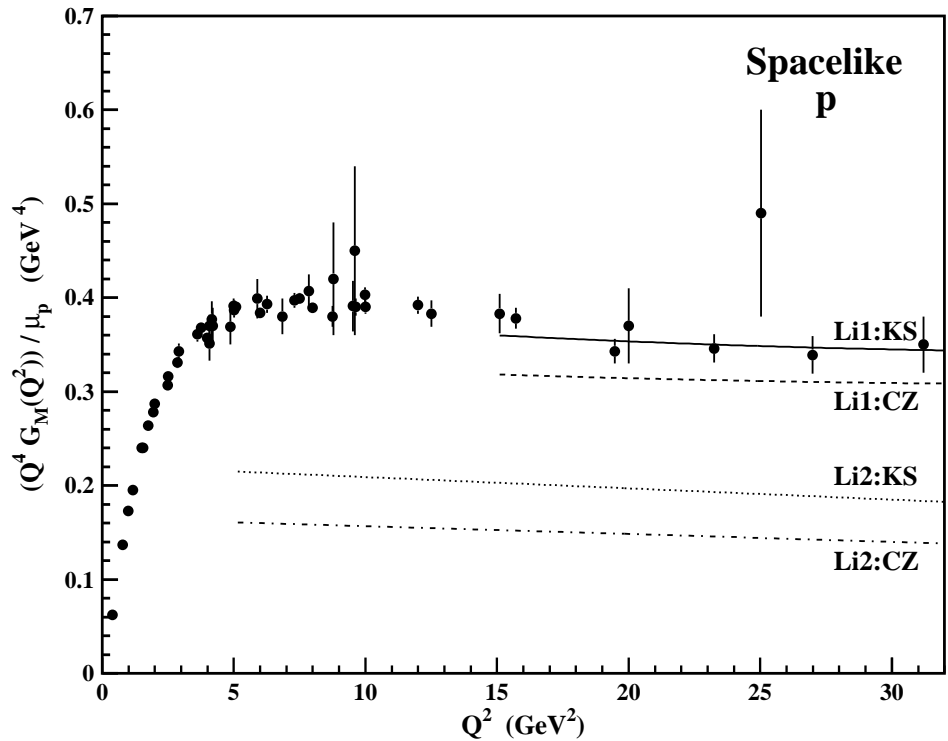}
\caption[Sudakov correction effect on the PQCD predictions of the 
proton magnetic form factor in the spacelike region.]
{Sudakov correction effect on the PQCD predictions of the proton magnetic form factor 
in the spacelike region.  
The solid and dotted lines are the form factor predictions using the KS distribution 
amplitudes with the impact parameter cutoff used by Li \cite{Li_ProtonQCDSR} (Li1:KS) 
and Li and colleagues \cite{Kunduetal_ProtonPQCD} (Li2:KS), respectively.  
The dashed and dash-dotted lines are the form factor predictions 
using the CZ distribution 
amplitudes with the impact parameter cutoff used by Li \cite{Li_ProtonQCDSR} (Li1:CZ) 
and Li and colleagues \cite{Kunduetal_ProtonPQCD} (Li2:CZ), respectively.}  
\label{fig:prslsud}
\end{center}
\end{figure}

To address the endpoint issue, Li \cite{Li_ProtonQCDSR} determined the Sudakov 
correction for the magnetic form factor 
of the proton.  He showed that the Sudakov correction suppresses the 
contributions to the form factor which arise from the endpoint region. 
Figure \ref{fig:prslsud} shows the predictions for the spacelike magnetic form factor 
using the PQCD formalism with the CZ and KS distribution amplitudes.  
The spacelike magnetic form factor prediction using the GS distribution is consistent with 
the CZ and KS predictions \cite{Li_ProtonQCDSR}. 

\begin{figure}[!tb]
\begin{center}
\includegraphics[width=14cm]{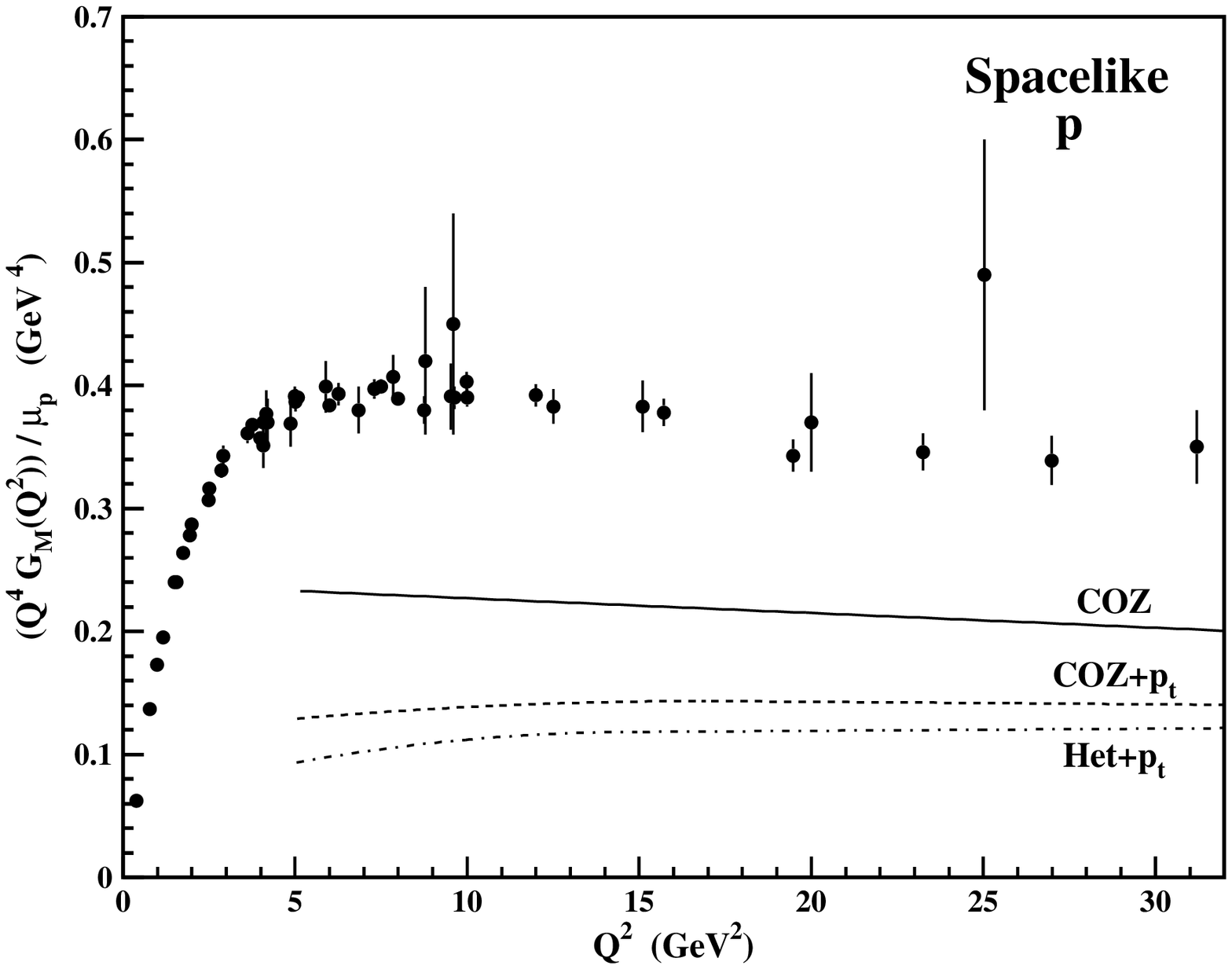}
\caption[Intrinsic transverse momentum effect on the PQCD predictions of the 
proton magnetic form factor in the spacelike region.]
{Intrinsic transverse momentum effect on the PQCD predictions of the 
proton magnetic form factor in the spacelike region.  
The solid and dashed lines are the form factor predictions 
by Bolz $\etal$ \cite{Bolzetal_PrSudArg} using the COZ distribution amplitudes 
without (COZ) and with (COZ+$p_t$) the 
inclusion of the transverse momentum in the proton, respectively.  
The dash-dotted line is the form factor prediction by 
Bolz $\etal$ \cite{Bolzetal_PrSudArg} using the heterotic (Het+$p_t$) distribution amplitude 
with the inclusion of the transverse momentum in the proton.}  
\label{fig:prsltrp}
\end{center}
\end{figure}

Objections were raised to the choice of the impact parameter cutoff used by Li 
\cite{Li_ProtonQCDSR}.  Bolz $\etal$ \cite{Bolzetal_PrSudArg} argued that the 
choice of the impact parameter cutoff used by Li \cite{Li_ProtonQCDSR} for a 
given quark does not suppress the endpoint effects arising from the other quarks.  
They suggest a new cutoff and also study the effect of including the intrinsic 
transverse momentum in the proton.  
Figure \ref{fig:prsltrp} shows the effect of including 
the transverse momentum and the impact parameter cutoff used by 
Bolz $\etal$ \cite{Bolzetal_PrSudArg} on the spacelike magnetic form factor prediction 
using the COZ and heterotic distribution amplitudes.  
Li and colleagues \cite{Kunduetal_ProtonPQCD} 
revisited the impact parameter cutoff used in the analysis of 
Li \cite{Li_ProtonQCDSR} and find the spacelike magnetic form factor prediction is 
decreased by about a factor 2, as shown in Figure \ref{fig:prslsud} (predictions 
marked Li2).

\begin{figure}[!tb]
\begin{center}
\includegraphics[width=14cm]{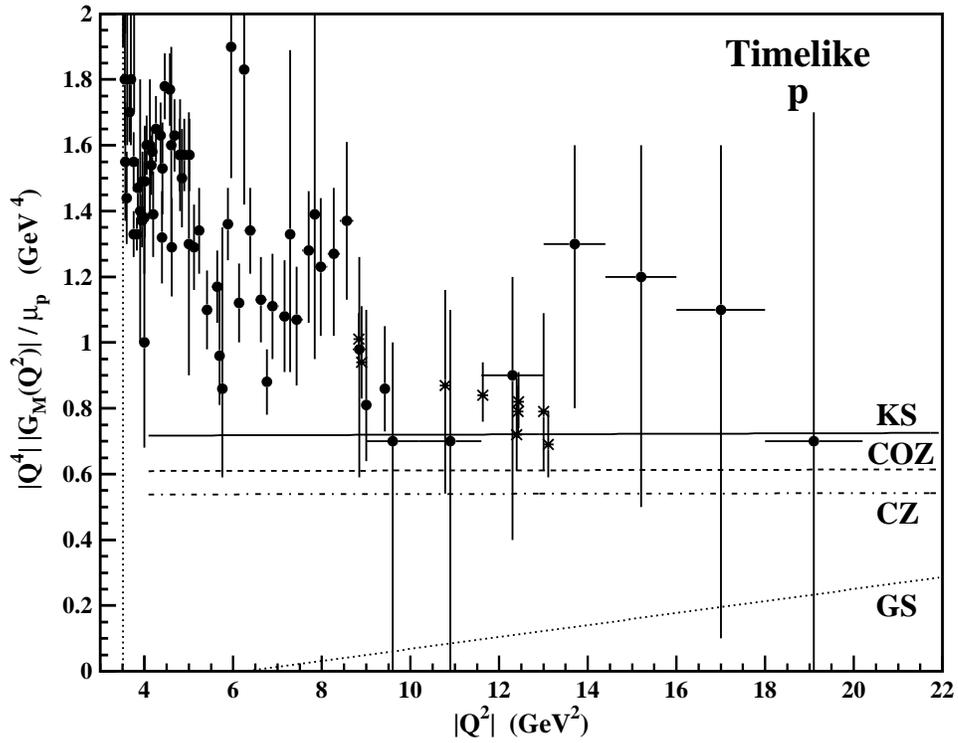}
\caption[PQCD predictions of the proton magnetic form factor in the 
timelike region.]  
{PQCD predictions of the proton magnetic form factor in the 
timelike region. The solid, dashed, dash-dotted, and dotted lines are predictions 
by Hyer \cite{Hyer_tlProtonPQCD} using the KS, COZ, CZ, and GS 
distribution amplitudes, respectively.  The vertical dotted line represents the 
$\ppbar$ production threshold ($|Q^2| = (2m_p)^2 = 3.52$ GeV$^2$).}  
\label{fig:prtlhyer}
\end{center}
\end{figure}

Proton magnetic form factor predictions in the timelike region using the PQCD formalism 
have been made by Hyer \cite{Hyer_tlProtonPQCD}.  He included 
Sudakov corrections and made the form factor predictions using the CZ, KS, 
GS, and COZ distribution amplitudes.  The predictions are shown in \ref{fig:prtlhyer}.  
The predictions have reasonable agreement with the experimental data, with 
the exception of that for the GS distribution amplitude, which is found to be a factor 6 
smaller than the E760/E835 data \cite{e835_1,e835_2,e835_3}.  

\newpage
\clearpage
\subsection{QCD Sum Rules}

\begin{figure}[!tb]
\begin{center}
\includegraphics[width=14cm]{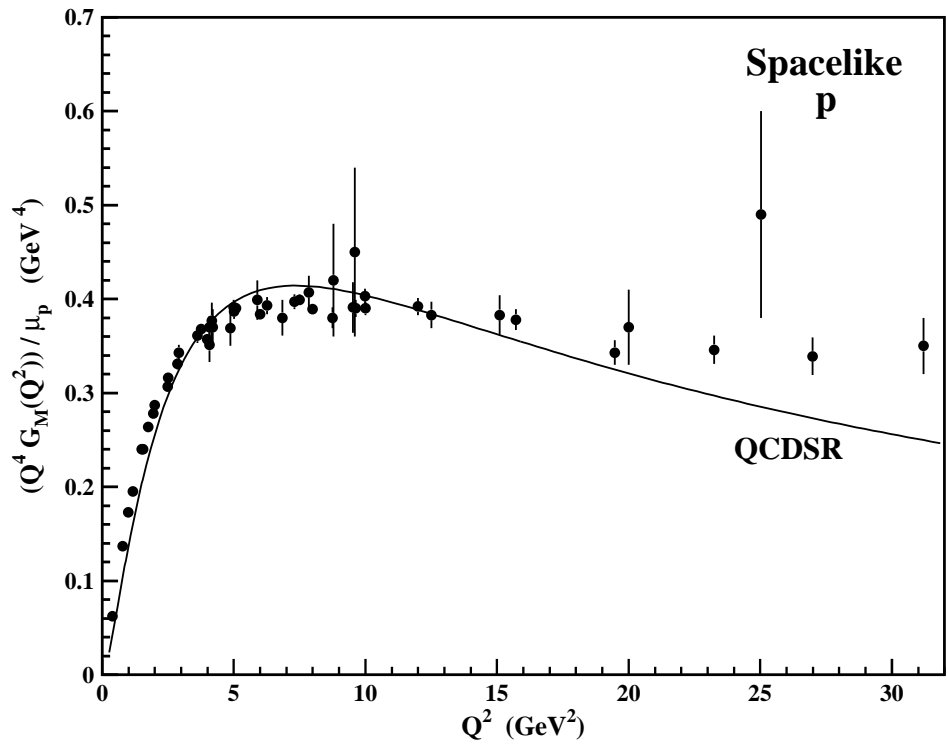}
\caption[Prediction from QCDSR three-point amplitude method with the square representation 
for the proton magnetic form factor in the spacelike region.]  
{Prediction from QCDSR three-point amplitude method with the square representation 
for the proton magnetic form factor in the spacelike region.  
The solid line is the QCDSR local duality prediction 
by Nesterenko and Radyushkin \cite{NesterenkoRadyushkin_PrQCDSR,Radyushkin_QCDSR1}.}  
\label{fig:prslsr1}
\end{center}
\end{figure}

Using the QCDSR three-point amplitude method with the square representation, 
Nesterenko and Radyushkin 
\cite{NesterenkoRadyushkin_PrQCDSR,Radyushkin_QCDSR1} determined the 
magnetic form factor for spacelike momentum transfers to be 
\begin{equation}
G^P_M(Q^2) = \frac{8~\sqrt{T^2-1}}{3~\left\{[4T^2-1][T^2-1]
+T[4T^2-3]\sqrt{T^2-1}\right\}},
\label{eq:prlocaldual}
\end{equation}
where $T = 1 + (Q^2/s_0)$ and $s_0$ = 2.3 GeV$^2$.  The variable $s_0$ is the maximum 
energy of the hadronic current of the quarks to be consistent with forming a proton, 
as described in Section 2.1.2.  
The prediction is shown in Figure \ref{fig:prslsr1} and starts to deviate from the 
experimental data for $Q^2 > 20$ GeV$^2$.

\begin{figure}[!tb]
\begin{center}
\includegraphics[width=14cm]{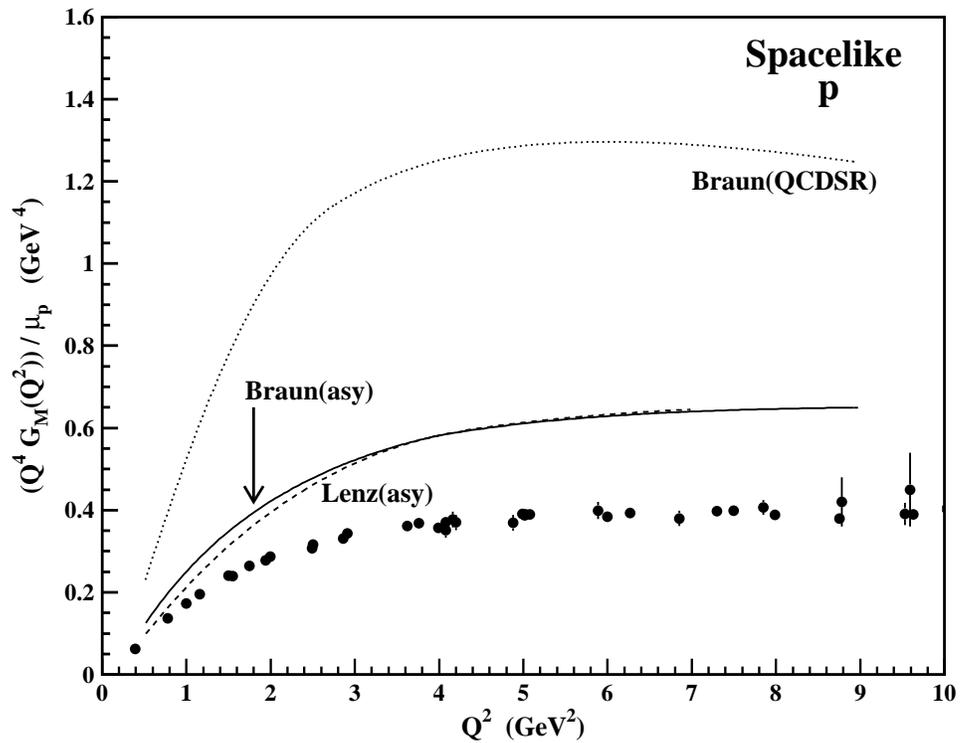}
\caption[QCDSR correlation function predictions of the proton magnetic form factor in the 
spacelike region.]  
{QCDSR correlation function predictions of the proton magnetic form factor in the 
spacelike region.  The solid and dotted lines are the predictions 
by Braun $\etal$ \cite{Braunetal_ProtonQCDSR} using the asymptotic and QCDSR 
correlator function distribution amplitudes, respectively.  The dashed line is the prediction 
by Lenz $\etal$ \cite{Lenzetal_ProtonQCDSR} using the asymptotic distribution amplitude 
with the improved hadronic current for the proton.}  
\label{fig:prslsrcor}
\end{center}
\end{figure}

Predictions for the proton form factors have also been determined using the 
QCDSR correlator function method.  
Braun $\etal$  \cite{Braunetal_ProtonQCDSR} have determined the soft contributions 
of the magnetic form factor using this method.  They found \cite{Braunetal_ProtonQCDSR} 
that the Dirac form factor 
goes as $F^P_1(Q^2) \sim 1/Q^6$ and overestimates the data currently accessible by 
experiment.  They have used two different distribution amplitudes, 
one with the asymptotic form (Eqn. \ref{eq:prasypida}) and one derived to twist-6 
accuracy \cite{Braunetal_ProtonQCDSRDA}.  The predictions using the two 
different distribution amplitudes are shown in Figure \ref{fig:prslsrcor}.

Another prediction of the spacelike magnetic form factor 
using the QCDSR correlator function method has been performed by 
Lenz $\etal$ \cite{Lenzetal_ProtonQCDSR}.  
They suggested an improved hadronic current which explicitly conserves isospin.  
The predictions using this improved hadronic current, and the asymptotic 
distribution amplitude, is shown in Figure \ref{fig:prslsrcor}.  The 
correlator function QCDSR predictions by Braun $\etal$ \cite{Braunetal_ProtonQCDSR} and 
Lenz $\etal$ \cite{Lenzetal_ProtonQCDSR} using the asymptotic distribution amplitudes 
both overestimate the magnetic form factor by $\sim50\%$.  

\newpage
\clearpage
\subsection{Lattice QCD}

As in the case of for pion form factor, the proton form factor predictions 
from Lattice QCD are only for spacelike momentum transfers and 
are in the limited momentum transfer range of $Q^2 < 3$ GeV$^2$.  
The most recent prediction of the electromagnetic form factors of the proton has been 
made by  G\"{o}ckeler $\etal$ \cite{Gockeleretal_PrLattice}.  Their predictions 
are consistent with the VDM dipole form of the form factor, 
\begin{equation}
Q^4 G^P_M(Q^2) = Q^4\left(1+\frac{Q^2}{0.71~\mathrm{GeV}^2}\right)^{-2}.
\label{eq:prdipole}
\end{equation}

\subsection{Other Models}

Other models have been proposed to explain the observed behavior of the proton 
form factors.  They include predictions based on generalized parton distributions, 
meson cloud corrections, a description of the proton as a two-body quark-diquark system, 
and a gluon string tube connecting the valence quarks.

\begin{figure}[!tb]
\begin{center}
\includegraphics[width=14cm]{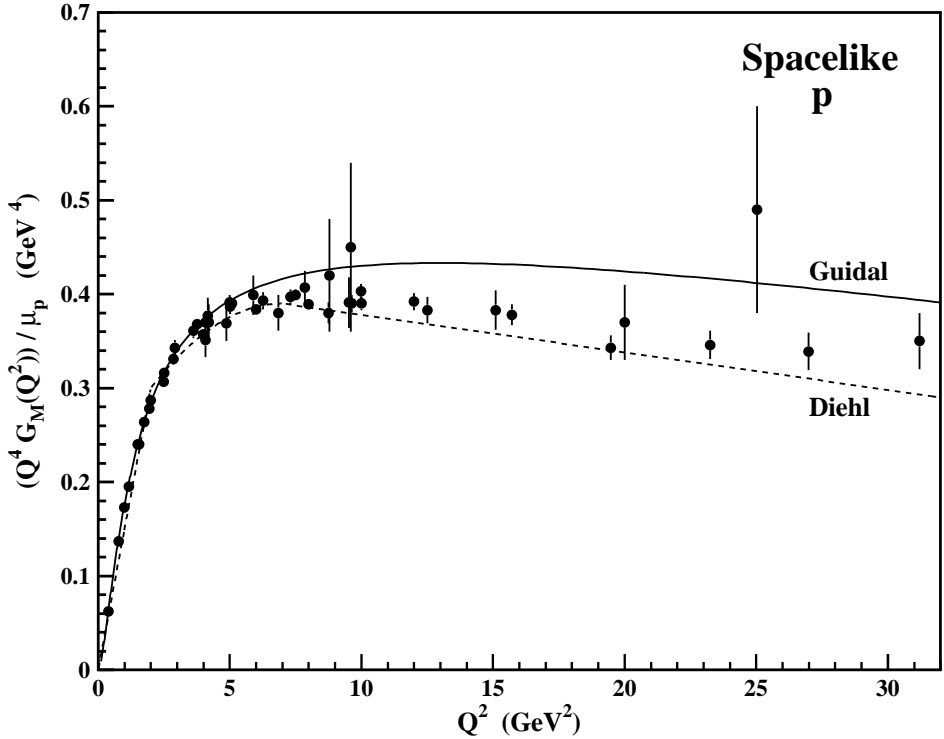}
\caption[Proton magnetic form factor predictions in the 
spacelike region using generalized parton distributions.]  
{Proton magnetic form factor predictions in the 
spacelike region using generalized parton distributions.  
The solid line is the prediction by Guidal $\etal$ \cite{Guidaletal_PrGPD} and 
the dotted line is the prediction by Diehl $\etal$ \cite{Diehletal_PrGPD}.} 
\label{fig:prslgpd}
\end{center}
\end{figure}

Predictions based on generalized parton distributions include the transverse spacial 
distributions of the constituent quarks inside the proton.  The generalized parton 
distribution predictions have been determined by 
Guidal $\etal$ \cite{Guidaletal_PrGPD} and Diehl $\etal$ \cite{Diehletal_PrGPD}.  
Their predictions for the spacelike magnetic form factor are shown in 
Figure \ref{fig:prslgpd} and are found to be consistent with the experimental data.

\begin{figure}[!tb]
\begin{center}
\includegraphics[width=14cm]{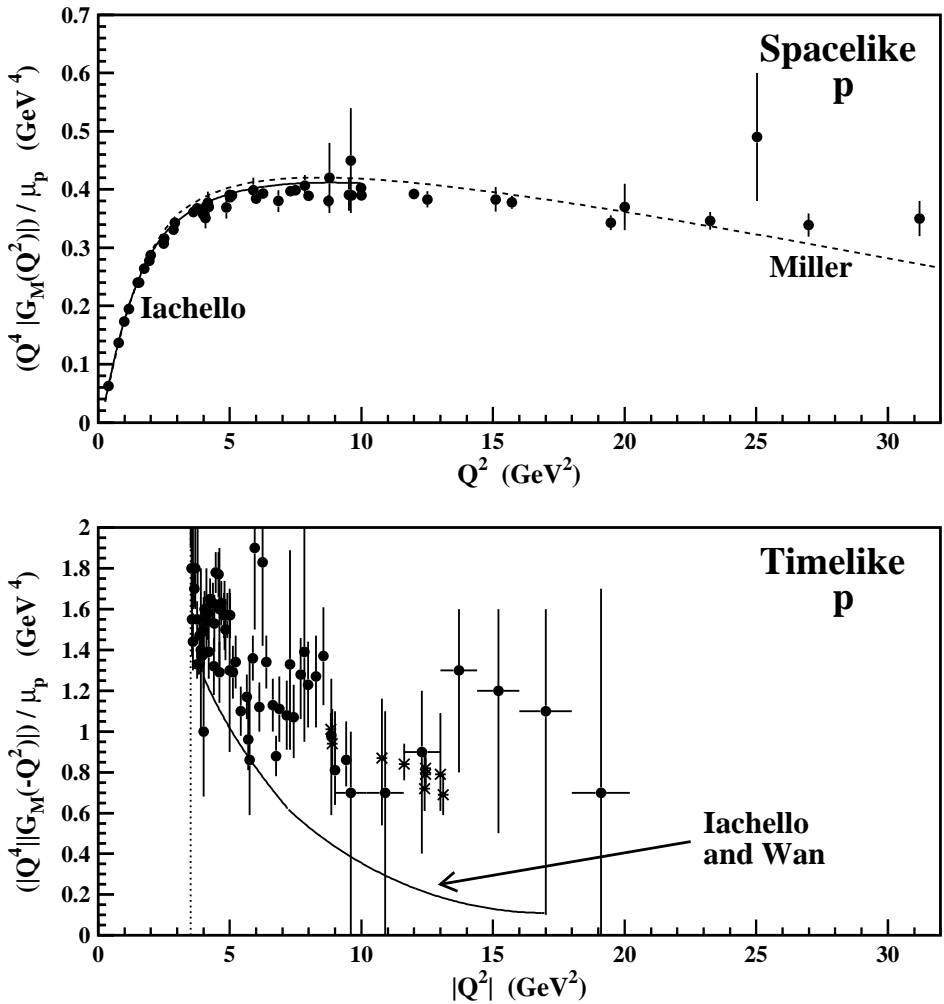}
\caption[Proton magnetic form factor predictions in the spacelike and timelike regions 
from meson cloud models.]  
{Proton magnetic form factor predictions in the spacelike (top) and timelike (bottom) 
regions from meson cloud models.  The dashed line in the top plot is 
by Miller \cite{Miller_SlPrMesonCloud}.  The solid lines in the top and bottom plots are 
by Iachello \cite{Iachello_SlPrMesonCloud} and Iachello and Wan 
\cite{IachelloWan_TlPrMesonCloud}, respectively.}  
\label{fig:prmescloud}
\end{center}
\end{figure}

The meson cloud interpretation consists of describing the proton as a bare core 
of the three valence quarks surrounded by a cloud of virtual particles.  
Two different models have been proposed using different types of particles 
comprising the cloud.  Miller \cite{Miller_SlPrMesonCloud} used a cloud of 
virtual pions.  Iachello, Jackson, and Lande \cite{OriginalIachelloetal} have 
described the cloud as comprised of the isoscalar vector particles $\omega$ and 
$\phi$ and the isovector vector particle $\rho$.   The description of the meson cloud 
in terms of the vector states is very similar to the VDM predictions.  
Iachello \cite{Iachello_SlPrMesonCloud} has updated the spacelike prediction of the 
magnetic form factor and, in collaboration with Wan 
\cite{IachelloWan_TlPrMesonCloud}, has predicted the behavior in the timelike region. 
The meson cloud predictions are shown in Figure \ref{fig:prmescloud}.  Both 
spacelike predictions are consistent with the current 
experimental data, but the timelike prediction by Iachello and 
Wan \cite{IachelloWan_TlPrMesonCloud} underestimate the experimental data.  The 
source for the discrepancy is that the electric form factor in the model by 
Iachello and Wan \cite{IachelloWan_TlPrMesonCloud} increases substantially from 
$G^P_E(Q^2)$ = $G^P_M(Q^2)$ at the $\ppbar$ production threshold 
($|Q^2|$ = 3.52 GeV$^2$) to a maximum of 
$G^P_E(Q^2)$ $\approx$ 8.1 $G^P_M(Q^2)$ at $|Q^2|$ = 16 GeV$^2$.  

\begin{figure}[!tb]
\begin{center}
\includegraphics[width=14cm]{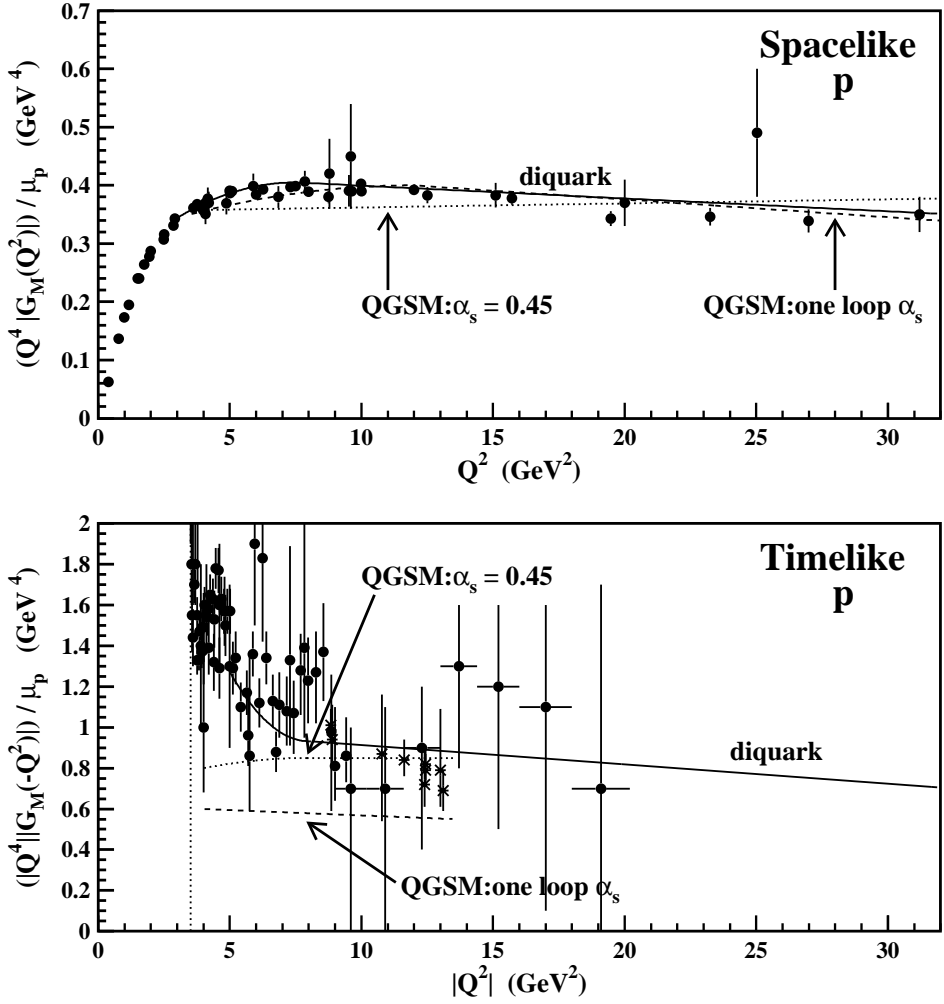}
\caption[Proton magnetic form factor predictions in the spacelike and timelike regions 
from the diquark and Quark Gluon String models.]  
{Proton magnetic form factor predictions in the spacelike (top) and timelike (bottom) 
regions from the diquark and Quark Gluon String models.  The solid line in the top 
and bottom plots are the diquark predictions by 
Kroll $\etal$ \cite{Krolletal_slPrDiquark} 
and Kroll $\etal$ \cite{Krolletal_tlPrDiquark}, respectively.  
The dashed and dotted lines are the predictions by Kaidalov 
$\etal$ \cite{Kaidalovetal_Gluestring}, with the dashed lines using the one loop form 
of $\alpha_s(Q^2)$ and the dotted lines with $\alpha_s(Q^2) = 0.45$.}  
\label{fig:prdiq+string}
\end{center}
\end{figure}

A model has been proposed by Kroll, Sch\"{u}rmann, and Schweiger 
\cite{Krolletal_slPrDiquark} to consider the three valence quarks in the proton as 
an effective 
two-body state consisting of a single quark and a composite diquark.  The 
prediction for the magnetic form factor of the proton with this diquark model 
uses the PQCD factorization scheme with the inclusion of two phenomenological 
diquark form factors, one which treats the diquark in a scalar spin-0 state 
and the other in a vector spin-1 state.  The three valence quark description 
is recovered in the diquark model at large $Q^2$ \cite{Krolletal_slPrDiquark}.  
The diquark model 
has been used to predict the magnetic form factor in both the 
spacelike \cite{Krolletal_slPrDiquark} and 
timelike \cite{Krolletal_tlPrDiquark} regions, with the timelike prediction arising 
from $t$ $\leftrightarrow$ $s$ channel crossing symmetry.  A comparison of the 
diquark model predictions to the existing data is shown in Figure \ref{fig:prdiq+string}.  
With some tuning of the parameters, 
they are found to be consistent with the behavior observed in the data.

The Quark Gluon String Model (QGSM), derived by Kaidalov, Kondratyuk, 
and Tchekin \cite{Kaidalovetal_Gluestring}, is based on parameterizing the interaction 
between the quark (or initial $\qqbar$ pair) struck by the virtual photon 
and the spectator quarks by a color gluon string.  The model for the proton 
is constructed through the convolution of two amplitudes: 
the virtual photon coupling to a $\qqbar$ pair and the gluon string between the initial 
$\qqbar$ pair fragmenting into a diquark-antidiquark pair produced from 
the vacuum, with the diquark in a spin-0 state.  The QGSM model incorporates 
the Sudakov form factor, which are shown to 
behave differently in the spacelike and timelike regions, and predicts that the ratio of 
timelike-to-spacelike magnetic form factors to be 
$G^{P,tl}_M(Q^2)/G^{P,sl}_M(Q^2) \approx$ 1.6 at $|Q^2|$ = 13.5 GeV$^2$.  
Figure \ref{fig:prdiq+string} shows the QGSM predictions of the magnetic form factor 
of the proton in the spacelike and timelike regions and are in reasonable 
agreement with the data.

\newpage
\clearpage
\subsection{The Ratios $\mu_p G^P_E(Q^2)/G^P_M(Q^2)$ and $F_2(Q^2)/F_1(Q^2)$}

\begin{figure}[htbp]
\begin{center}
\includegraphics[width=14.0cm]{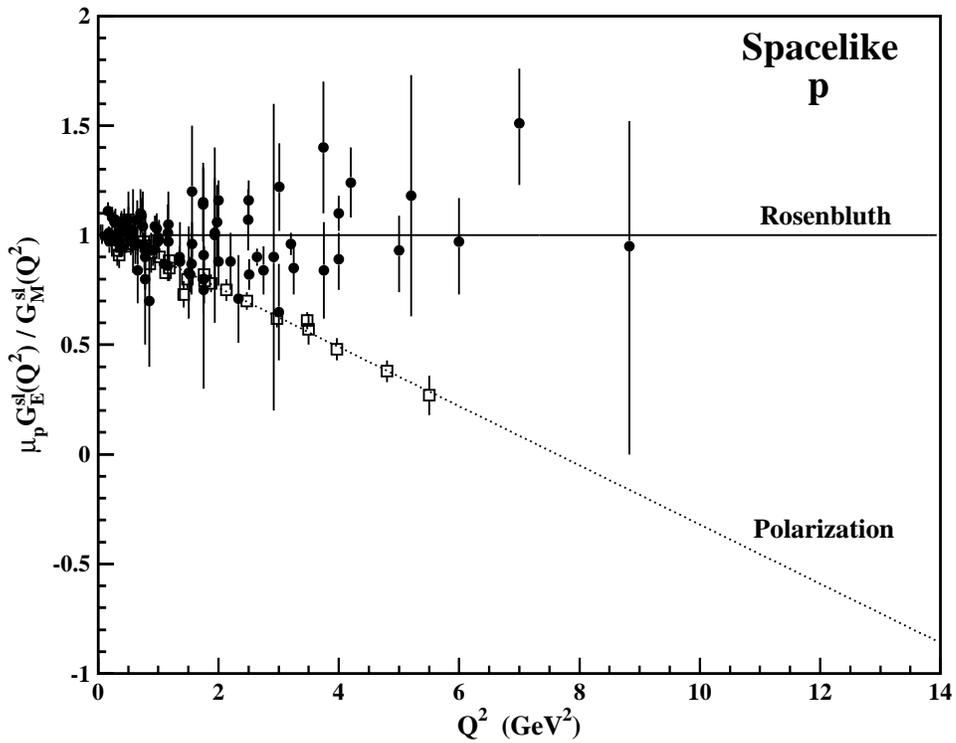}
\caption[The form factor ratio $\mu_p G^P_E(Q^2)/G^P_M(Q^2)$ 
in the spacelike regions.]  
{The form factor ratio $\mu_p G^P_E(Q^2)/G^P_M(Q^2)$ in the spacelike regions.  
The open squares are from the polarization transfer 
\cite{prslff_pol1}-\cite{prslff_pol4} measurements, 
and the solid points are from the Rosenbluth separation 
\cite{prslff_ros1}-\cite{prslff_ros13} measurements.  
The solid line shows the relationship $\mu_p G^P_E(Q^2)/G^P_M(Q^2) = 1$ (Rosenbluth).  
The dotted line is the linear extrapolation of the polarization transfer measurements 
(Polarization).} 
\label{fig:prslgegm}
\end{center}
\end{figure}

For a long time it was believed that $\mu_p G^P_E(Q^2)/G^P_M(Q^2) \sim 1$.  This 
belief was primarily based on the measurements made at SLAC \cite{prslff_ros10,prslff_ros11} 
in which $G^P_E(Q^2)$ and $G^P_M(Q^2)$ were separated 
using the Rosenbluth method \cite{rosenbluthsep}.  For $Q^2 \le 4$ GeV$^2$, the errors 
in $\mu_pG^P_E(Q^2)/G^P_M(Q^2)$ were not large, but became larger in the 
$Q^2 = 4-9$ GeV$^2$ region.    
The measurements of the $\mu_p G^P_E(Q^2)/G^P_M(Q^2)$ ratio in the polarization 
transfer experiments \cite{prslff_pol1}-\cite{prslff_pol4} has caused a reevaluation 
of the behavior of the electric and magnetic form factors.  The polarization transfer 
measurements, in which the product $G^P_E(Q^2)\cdot G^P_M(Q^2)$ is measured, 
have determined that the electric form factor falls faster 
than the magnetic form factor in the range $Q^2 >$ 1 GeV$^2$ with 
$\mu_p G^P_E(Q^2)/G^P_M(Q^2) \sim $ 0.5 at $Q^2$ = 5.5 GeV$^2$.  
A comparison of the Rosenbluth separation and polarization transfer measurements is 
shown in Figure \ref{fig:prslgegm}.

\begin{figure}[htbp]
\begin{center}
\includegraphics[width=14.3cm]{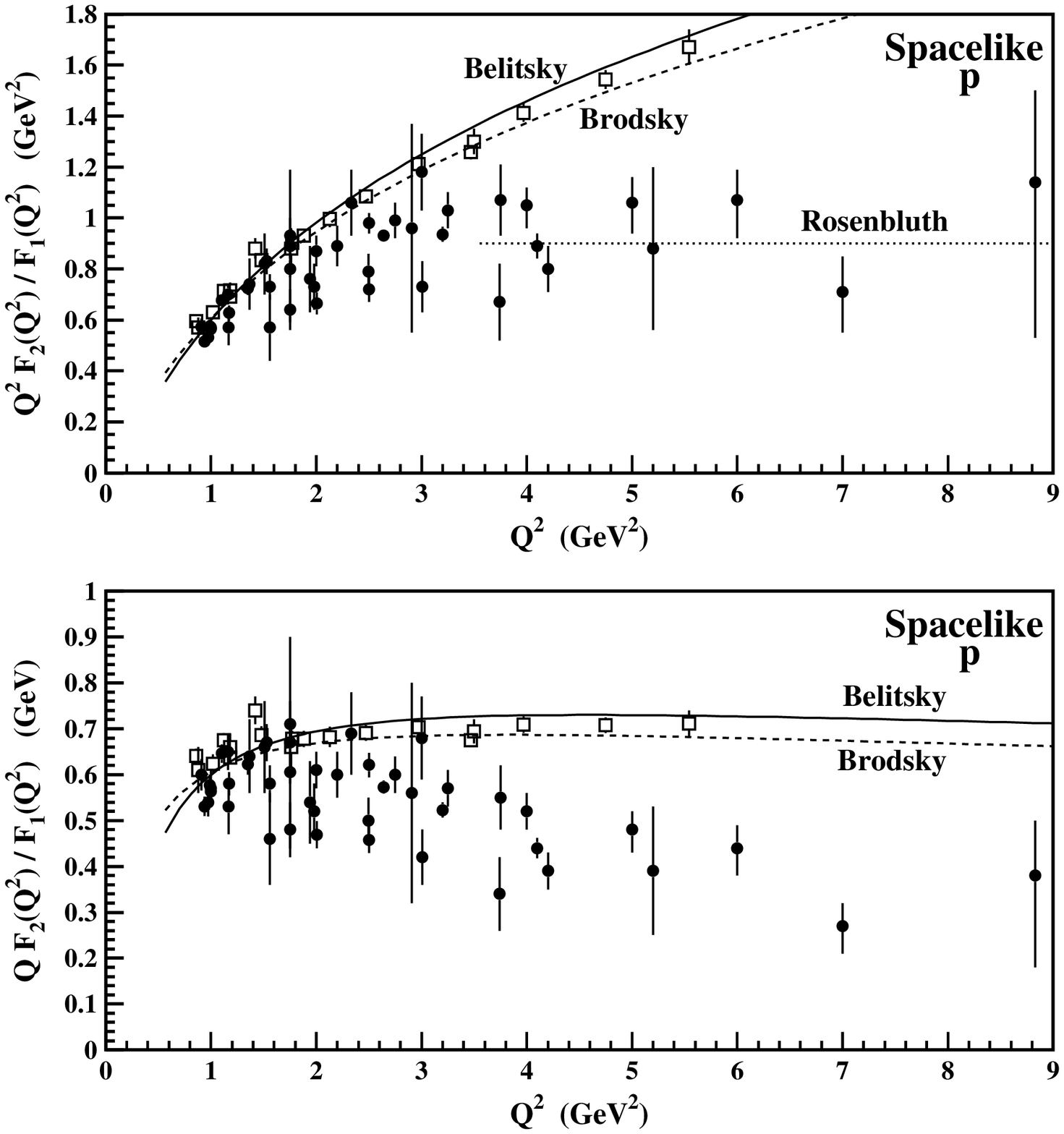}
\caption[The form factor ratios $Q^2\cdot F^P_2(Q^2)/F^P_1(Q^2)$ and 
$Q\cdot F^P_2(Q^2)/F^P_1(Q^2)$ in the spacelike region.]
{The form factor ratios $[Q^2\cdot F^P_2(Q^2)/F^P_1(Q^2)]$ and 
$[Q\cdot F^P_2(Q^2)/F^P_1(Q^2)]$ in the spacelike region.  
The open squares are experimental measurements using polarization transfer 
\cite{prslff_pol2}-\cite{prslff_pol4}, 
and the solid points are experimental measurements using Rosenbluth separation 
\cite{prslff_ros5}-\cite{prslff_ros13}.
The solid and dashed lines are the predictions 
by Belitsky $\etal$ \cite{Belitskyetal_prf1f2} 
and Brodsky $\etal$ \cite{Brodskyetal_prtlf1f2}, respectively.  
The arbitrarily normalized dotted line in the top plot represents the PQCD prediction 
$[Q^2\cdot F^P_2(Q^2)/F^P_1(Q^2)] \approx$ constant.}
\label{fig:prslf2f1}
\end{center}
\end{figure}

The Pauli-to-Dirac form factor ratio predicted by PQCD is 
$[Q^2\cdot F^P_2(Q^2)/F^P_1(Q^2)]
 \approx$ constant 
at large $Q^2$ \cite{LepageBrodsky_PQCDFF}.  
This behavior is supported by the Rosenbluth separation experiments 
for $Q^2 >$ 3 GeV$^2$ (solid points), as shown in Figure \ref{fig:prslf2f1} (top).  
In contrast, the $[Q^2\cdot F^P_2(Q^2)/F^P_1(Q^2)]$ ratio from the polarization transfer 
measurements (open squares) continues to increase.  
On the other hand, polarization transfer 
measurements support the behavior of $[Q\cdot F^P_2(Q^2)/F^P_1(Q^2)] \approx$ 
constant for $Q^2 >$ 3 GeV$^2$, as illustrated by Figure \ref{fig:prslf2f1} (bottom).

Theoretical predictions based on the PQCD factorization scheme have been made to 
address the behavior of the Pauli-to-Dirac form factor ratio.  Belitsky, Ji, and Yuan 
\cite{Belitskyetal_prf1f2} used the distribution amplitudes determined 
by the QCD correlator function method to twist-4 accuracy \cite{Braunetal_ProtonQCDSRDA} 
to account for the higher twist contribution caused by the spin-flip nature 
of the Pauli form factor.  
They find that the form factor ratio has the following form \cite{Belitskyetal_prf1f2} 
\begin{equation}
\frac{F^P_2(Q^2)}{F^P_1(Q^2)} 
= \frac{A}{Q^2}~\mathrm{ln}^2\left(\frac{Q^2}{\Lambda^2}\right),
\label{eq:belitskyf2f1}
\end{equation}
where $A$ is an arbitrary coefficient.  Brodsky $\etal$ \cite{Brodskyetal_prtlf1f2} 
determined the ratio as 
\begin{equation}
\frac{F^P_2(Q^2)}{F^P_1(Q^2)} = \kappa_p~
\frac{\left[1+(Q^2/0.791~\mathrm{GeV}^2)^2
~\mathrm{ln}^{7.1}\left(1+\frac{Q^2}{4m^2_\pi}\right)\right]}
{\left[1+(Q^2/0.380~\mathrm{GeV}^2)^3
~\mathrm{ln}^{5.1}\left(1+\frac{Q^2}{4m^2_\pi}\right)\right]}.
\label{eq:brodskyf2f1}
\end{equation}
They argue \cite{Brodskyetal_prtlf1f2} that the presence of the logarithmic term 
is also not surprising because of the higher twist nature of the Pauli form factor.  
As shown in Figure \ref{fig:prslf2f1} (top), both predictions reproduce the 
$Q^2F^P_2(Q^2)/F^P_1(Q^2)$ behavior from the polarization transfer data quite well.  

\begin{figure}[htbp]
\begin{center}
\includegraphics[width=14.0cm]{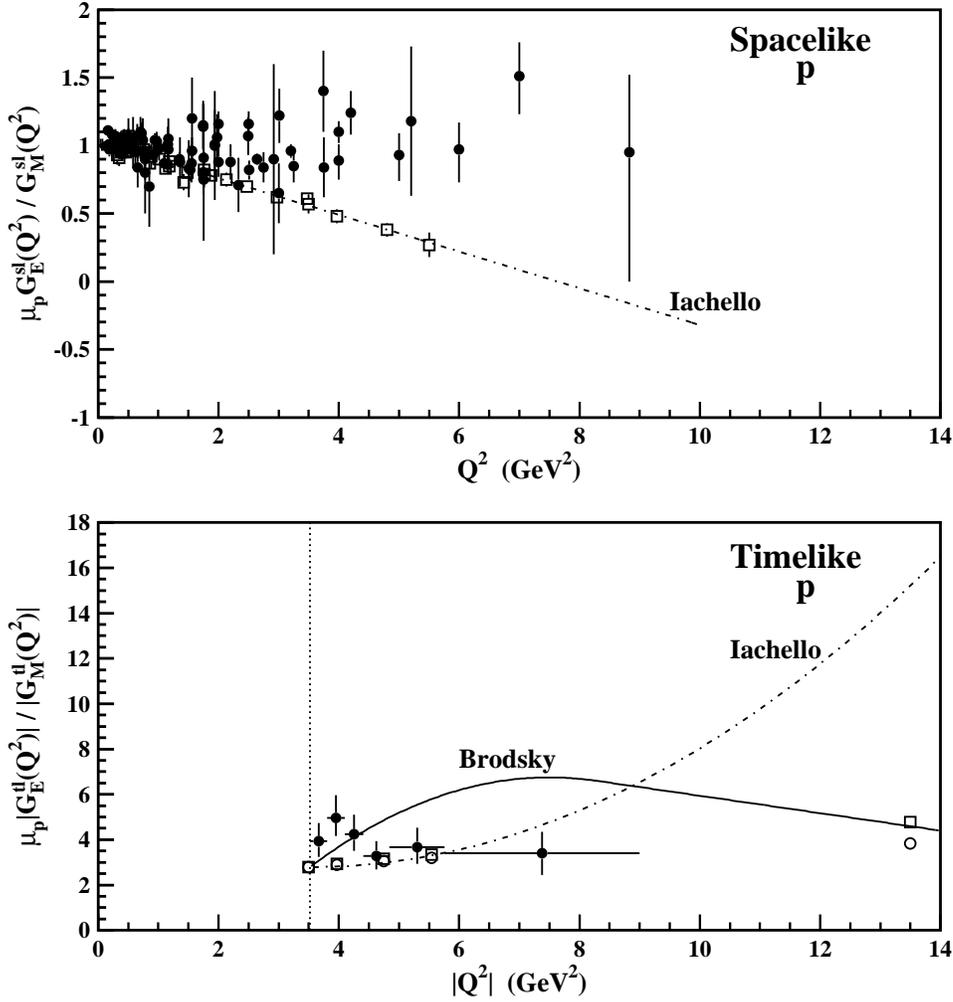}
\caption[The form factor ratio $\mu_p G^P_E(Q^2)/G^P_M(Q^2)$ 
in the spacelike and timelike regions.]  
{The form factor ratio $\mu_p G^P_E(Q^2)/G^P_M(Q^2)$ in the spacelike (top) 
and timelike (bottom) regions.  
The open squares in the top plot are from the polarization transfer 
\cite{prslff_pol1}-\cite{prslff_pol4} measurements, 
and the solid points are from the Rosenbluth 
separation \cite{prslff_ros1}-\cite{prslff_ros13} measurements.  
The dash-dotted line in the top plot is the prediction by 
Iachello \cite{Iachello_SlPrMesonCloud}.  
The vertical dotted line in the bottom plot represents the $\ppbar$ production 
threshold.  The solid and dash-dotted lines in the bottom plot are the predictions 
by Brodsky \cite{Brodsky_prf1f2} 
and Iachello and Wan \cite{IachelloWan_TlPrMesonCloud}, respectively.  
The solid points in the bottom plot are the experimental results 
from Ref. \cite{BABARppbarISR}.  
The open squares and open points in the bottom plot are the extrapolation of the 
behavior of the spacelike data from the polarization transfer and Rosenbluth separation 
experiments into the timelike region, respectively, as described in the text.} 
\label{fig:prsltlgegm}
\end{center}
\end{figure}

The decrease of the electric form factor, with respect to the magnetic form factor, 
with increasing $Q^2$ in the spacelike region 
was predicted by the phenomenological-based meson cloud model of Iachello, 
Jackson, and Lande \cite{OriginalIachelloetal} in 1973.  
It has recently been updated by Iachello \cite{Iachello_SlPrMesonCloud}.
As shown in Figure \ref{fig:prsltlgegm}, the meson cloud predictions agree with the 
$\mu_p G^P_E(Q^2)/G^P_M(Q^2)$ behavior observed by the polarization transfer 
experiments.

Predictions have been made for the behavior of the electric and magnetic form 
factors in the timelike region.  Brodsky \cite{Brodsky_prf1f2} has shown that 
the analytic continuation of a expression of the form of Eqn. 
\ref{eq:brodskyf2f1} causes an enhancement in the electric-to-magnetic form factor 
ratio, as shown by the solid line in Figure \ref{fig:prsltlgegm}. 
Iachello and Wan \cite{IachelloWan_TlPrMesonCloud} have used the meson cloud model 
to determine the electric-to-magnetic form factor in the timelike region.  
The prediction of $|\prelecffqsqtl|$ = $|\prmagffqsqtl|$ by the meson cloud model 
\cite{IachelloWan_TlPrMesonCloud} is compatible with the experimental data 
in the region $|Q^2| < 8$ GeV$^2$ but shows a rapid 
rise with increasing $|Q^2|$.  
On the other hand, the prediction by Brodsky \cite{Brodsky_prf1f2} shows a slow decrease for 
$|Q^2| > 8$ GeV$^2$.

The recent measurements of the electric-to-magnetic form factor ratio in the 
spacelike region can be used to gain insight into the behavior in the timelike region.  
We may assume that the Pauli-to-Dirac form factor ratio $F^P_2(Q^2)/F^P_1(Q^2)$  
for the proton is the same in the spacelike and timelike regions.  If we then assume that 
$G^{P,sl}_E(Q^2) = G^{P,sl}_M(Q^2)/\mu_p$, as indicated by the Rosenbluth measurements in 
the spacelike region, we obtain $|\prelecffqsqtl|$ = 1.38 $|\prmagffqsqtl|$ 
for the timelike form factor.  If we make a linear extrapolation of the 
spacelike results from the polarization transfer measurements, we obtain  
$G^{P,sl}_E$(13.48 GeV$^2$) = $-$0.8 $G^{P,sl}_M$(13.48 GeV$^2$)$/\mu_p$, which leads to 
$|\prelecffqsqtl|$ = 1.75 $|\prmagffqsqtl|$.  
The corresponding ratios $\mu_p |\prelecffqsqtl|/|\prmagffqsqtl|$ for timelike momentum 
transfers of $|Q^2|$ = 13.48 GeV$^2$ are shown in Figure \ref{fig:prsltlgegm}.  
It is interesting to note that the two very different extrapolations lead to such similar 
results.  The results are in reasonable agreement prediction by Brodsky \cite{Brodsky_prf1f2} 
but are a factor $\sim$4 smaller than the prediction by Iachello and Wan 
\cite{IachelloWan_TlPrMesonCloud}.

\baselineskip=24pt
\chapter{Experimental Apparatus}

\section{Cornell Electron Storage Ring}

\begin{figure}[!tb]
\begin{center}
\includegraphics[width=14cm]{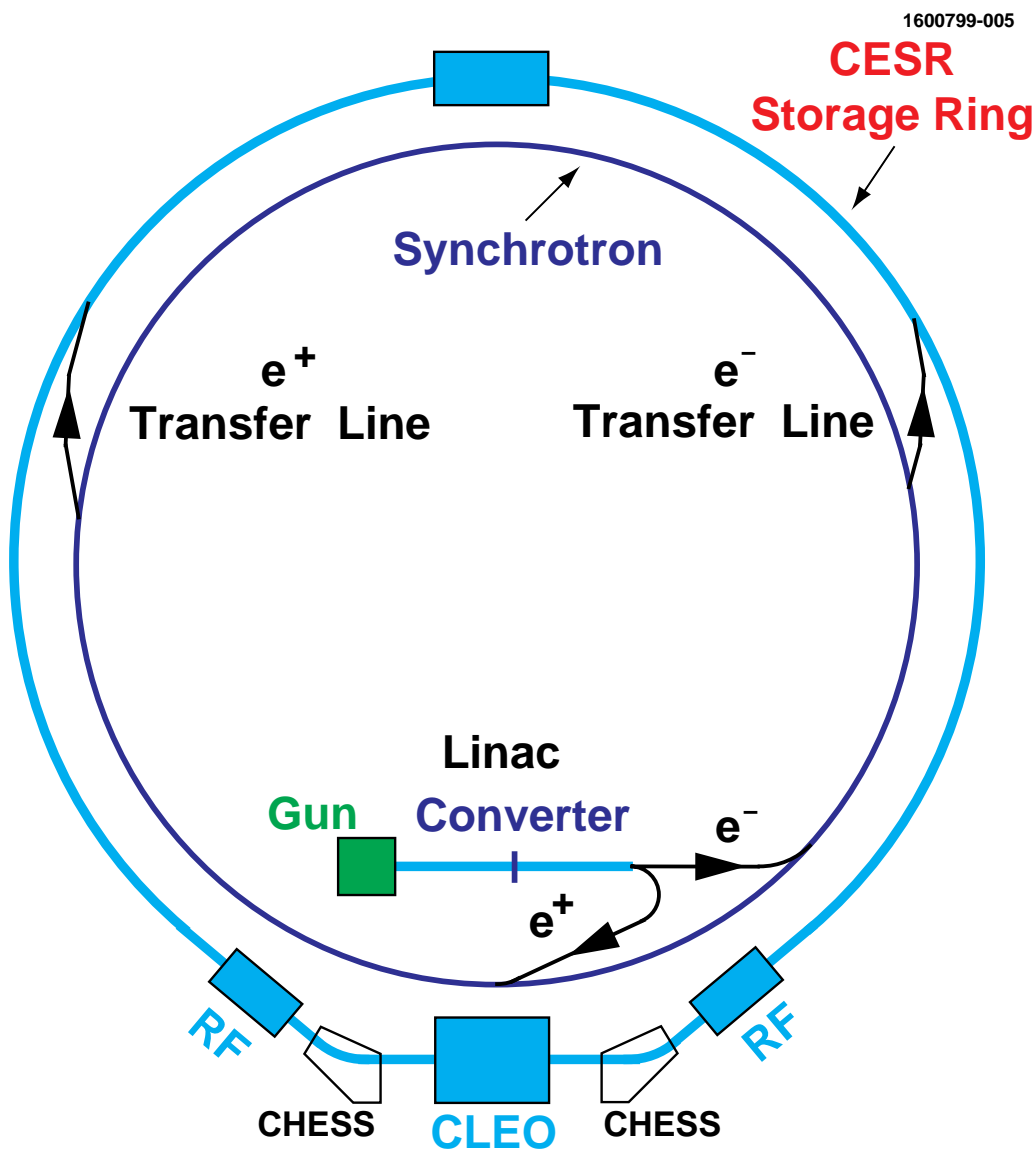}
\caption[The CESR Accelerator Facility.]
{The CESR Accelerator Facility.  The relevant components are the linac (located 
within the ring), synchrotron (located against the inner wall of the tunnel), and storage 
ring (located against the outer wall of the tunnel).  The position of the tungsten target 
used for positron production is shown by the converter.}
\label{fig:CESR}
\end{center}
\end{figure}

The accelerator facility located on the Cornell University campus 
in Ithaca, NY, is commonly referred to as the Cornell Electron Storage Ring, 
or by the acronym CESR. It is actually composed of three parts: 
the linear accelerator or linac, the synchrotron, and the storage ring.  
The synchrotron and storage ring are located in a circular tunnel 768 meters in 
circumference.  The linac in located in the area inside the ring.  The CLEO-c 
detector resides in a 33 by 27 meter room at the south end of the tunnel.  The 
accelerator facility is shown schematically in Figure \ref{fig:CESR}.  

Electrons are generated by heating a filament until they have sufficient energy 
to escape from the filament surface.  
The electrons are then collected by a prebuncher which compresses the 
electrons into packets for acceleration in the linac.  The accelerating 
sections of the linac have oscillating electric fields which are 
synchronized to increase the 
energy of the electrons as they travel through the components.  The electrons are 
accelerated to an energy of about 300 MeV at the end of the linac.  

Positrons are created by colliding an 140 MeV electron beam halfway down the linac 
on a movable tungsten target.  The collisions create showers of 
positrons, electrons, and photons.  The positrons are separated from the electrons 
and accelerated in the rest of the linac to an energy of about 200 MeV.

Bunches of electrons and positrons from the linac are separately fed into 
the synchrotron.  The synchrotron consists of a series of dipole bending magnets and 
four 3-meter long linear accelerators. 
As the energy of the particles is increased by the linear accelerators in the synchrotron, 
the magnetic fields in the dipoles are also increased to keep the particles enclosed in the 
synchrotron.  Once the particles are accelerated to the desired energy of $\sim$2 GeV, they 
are transferred to the storage ring.

The electrons and positrons are steered 
around the storage ring by a series of dipole bending magnets and 
focused by a series of quadrupole and sextupole magnets.  There are also 
final-focusing quadrupole magnets located directly outside of the \\ 
CLEO-c detector.  As the beams go around the 
storage ring, they lose energy by synchrotron radiation.  The energy is replaced 
through the use of superconducting radio frequency (RF) cavities which operate at a 
frequency of 500 MHz.  

The electrons and positrons in the storage ring travel around in 
a single 90 mm $\times$ 50 mm elliptical vacuum pipe  
with its major axis in the horizontal plane.  In optimal running conditions, 
electrons and positrons are separately grouped into nine bunches called trains.  
Each train contains up to five bunches, with each train separated with a 14 ns spacing.  
Electron-positron collisions at non-desirable locations are prevented 
by four electrostatic separators with electric fields in the horizontal plane.  
These separators push the orbit of the electron and positron trains around each other into 
so-called pretzel orbits.  Another interaction point potentially exists at the opposite side 
of the storage ring from the CLEO-c detector, but the beams are separated by 
two vertical electrostatic separators.  
At the interaction point enclosed by the CLEO-c detector, 
the beams do not collide head-on but with a small crossing 
angle of 2.5 mrad ($\approx 1/7~^{\circ}$) into the ring.  This allows for 
bunch-by-bunch collisions of the electron and positron trains.  

The CESR facility operated efficiently at nominal beam energies of $\sim$5.3 GeV 
from 1979 to 2003 for the production of the $\Upsilon(4S)$ resonance.  The change of 
beam energy to $\sim$2 GeV required changes in CESR since the amount of energy emitted 
through synchrotron radiation, which is proportional to $E^4$ \cite{CLEOcCESRc} 
($E$ is the beam energy), decreases.  
Two important beam parameters are affected by the decreased 
amount of synchrotron radiation.
The first parameter is the damping time of the betatron amplitude, which is related to the 
mean value of the transverse momentum of the beam.  At the lower beam energies, 
the damping time, which is proportional to $E^{-3}$ \cite{CLEOcCESRc}, increases.  
The other parameter is the horizontal beam size.  
The total beam size of a particle bunch can be described as a six-dimensional 
phase space envelope (three in position space and three in momentum space) \cite{accbook}.  
At lower beam energies, the horizontal beam size, 
which is proportional to $E^2$ \cite{CLEOcCESRc}, decreases, 
thereby restricting the particle density per bunch.  
These effects decrease the amount of attainable luminosity. 
It can be improved with the insertion of wiggler magnets into CESR.  

A wiggler magnet, sometimes called a Siberian snake, consists of an even number of 
dipole magnets with high magnetic field intensities.  The direction of the magnetic fields 
alternates between the successive magnetic elements of the wiggler.  
The magnetic field configuration 
shakes the beams in the horizontal plane, thereby increasing the amount of synchrotron 
radiation with minimal amounts of overall deviation in the beam path around the ring.  
The emission of the radiation decreases both the mean transverse and longitudinal momenta 
of the beam particles while increasing the dispersion of the momentum components.  
The decrease of the mean transverse momentum decreases the damping time, while the 
increase of the dispersion of the transverse momentum increases the horizontal beam size.  
The mean longitudinal momentum ($E$) is increased in the RF cavities, 
while the dispersion of the longitudinal momentum ($\sigma_{E}$) is decreased.

Twelve wigglers have been installed into CESR for operation at lower beam energies 
(note that only six wigglers magnets had been installed in CESR when the data used in 
this dissertation were taken).  
Each wiggler consists of eight dipole magnets with maximum magnetic field 
strengths of 2.1 Tesla per magnet \cite{CLEOcCESRc}.  The wiggler magnets increase the 
beam size by a factor of 4-8 and decrease the damping time by an order of magnitude, 
as compared to a storage ring without wiggler magnets \cite{CESRminiMAC}.  
The beam energy resolution ($\sigma_{E}/E$) from the wiggler dominated storage ring 
is $\sigma_{E}/E$ = 8.6$\times10^{-4}$, four times larger than without 
wigglers \cite{CESRminiMAC}.  
 
\begin{figure}[!htb]
\begin{center}
\includegraphics[width=14.5cm]{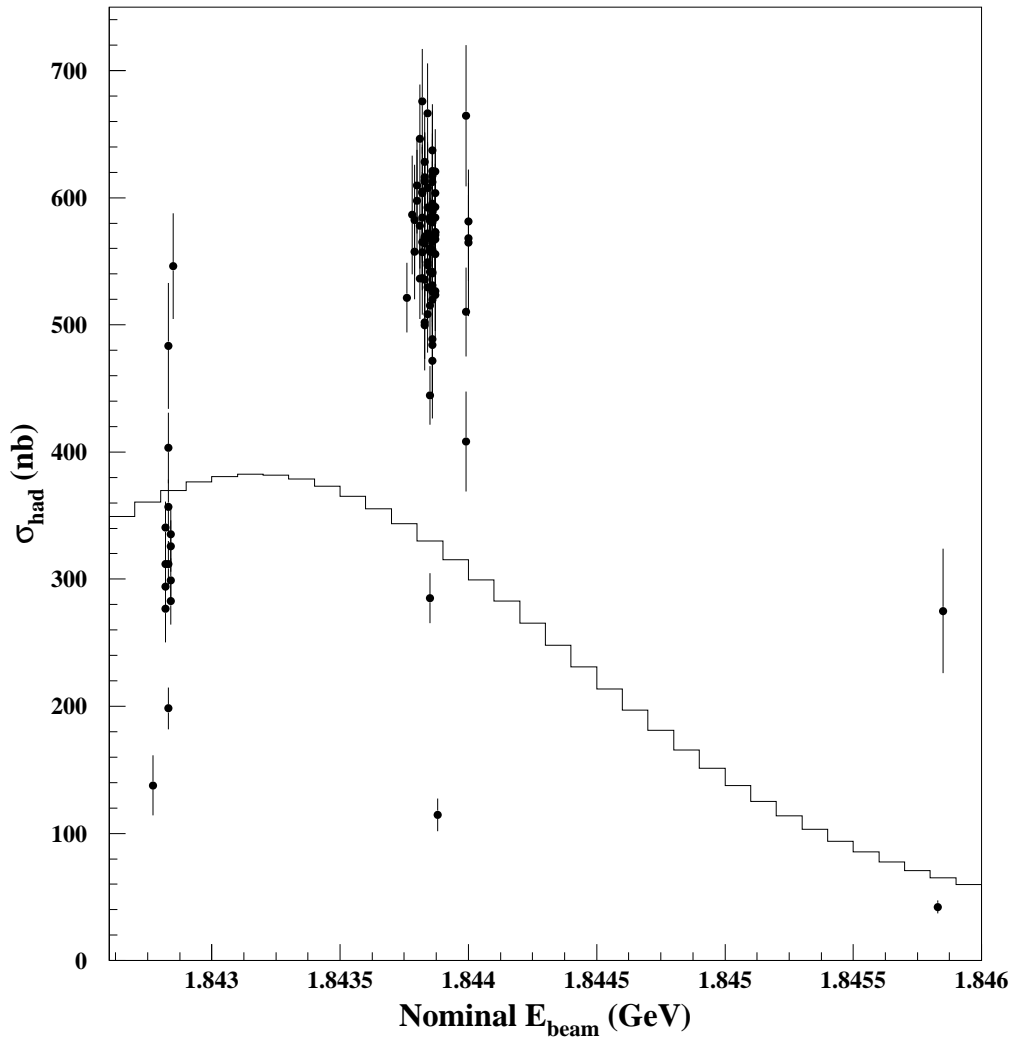}
\caption[Run-by-run hadronic cross section of the CLEO-c $\psip$ data sample 
as a function of beam energy.]
{Run-by-run hadronic cross section of the CLEO-c $\psip$ data sample as a function of 
nominal beam energy.  
The points are individual data runs and the histogram is the result of a 
numerical calculation program which convolutes a non-relativistic Breit-Wigner 
for the $\psip$ resonance, 
the Gaussian beam energy resolution, and a Kureav-Fadin radiative tail.  
The observed cross section for the group of runs near $E_{beam}$ = 1.8438 GeV indicates 
that the $\psip$ peak occurs at a nominal energy which is $\sim$0.7 MeV higher 
than 1.8431 GeV corresponding to the known mass of $\psip$.}
\label{fig:cleocpsi2scs}
\end{center}
\end{figure}

The beam energy is measured by determining the orbit length traveled by the beams 
in the storage ring.  To first order, 
the energy is determined by the bending dipole magnets \cite{BeamEnergy}
\begin{equation}
E_0 = \frac{ec}{2\pi}\sum_{i}|B_i|~\Delta\theta_i~\rho_i,
\label{eq:ebeam}
\end{equation} 
where $e$ is the charge of the electron, $c$ is the speed of light in a vacuum.  
The variables $|B_i|$, $\Delta\theta_i$, and $\rho_i$ are the magnetic field strength, 
bending angle, and radius of 
curvature of the $i$th dipole magnet.  The sum in Eqn. \ref{eq:ebeam} is over all of the 
dipole magnets in the ring.  Additional corrections to the beam energy are caused by 
the RF cavities, electrostatic separators, sextupole and wiggler magnets, 
and the different types, horizontal steering, and hysteresis of the dipole magnets.  
As an example of the uncertainty in the absolute value of the beam energy, 
Figure \ref{fig:cleocpsi2scs} shows the run-by-run hadronic 
cross section of the $\psip$ data sample, collected with the CLEO-c detector in 
December 2003 with CESR in the six wiggler configuration, as a function of 
nominal beam energy.  
Also shown in Figure \ref{fig:cleocpsi2scs} is the result of a numerical calculation 
program \cite{KarlResProgram} which convolutes a non-relativistic Breit-Wigner 
with its mean centered at the $\psip$ mass 
($M(\psip)$ = 3686.093 $\pm$ 0.034 MeV,  $\Gamma(\psip)$ = 281 $\pm$ 17 eV \cite{PDG2004}), 
the Gaussian beam energy resolution, and a Kureav-Fadin radiative tail \cite{KureavFadin}.  
Figure \ref{fig:cleocpsi2scs} shows that the beam energy is $\sim$0.7 MeV larger than 
expected.  This leads to an absolute uncertainty in the center-of-mass energy 
of the $\ee$ collision on the order of 1 MeV.
 
\newpage
\section{The CLEO-c Detector}

The electron and positron beams are focused to collide at a point near the 
midpoint of the CLEO-c detector. Independent of the intermediate states, 
the final result of these collisions 
produces relatively long-lived charged and neutral particles.  There are five different 
types of detected charged particles (and their corresponding antiparticles): 
the electron (denoted by e), muon ($\mu$), pion ($\pi$), kaon (K), and proton (p).  
The most common and easiest neutral particle to detect is the photon, 
while other neutral particles are either very difficult 
(the long lived neutral kaon $K^0_L$ and antineutrons) 
or nearly impossible (neutrons and neutrinos) to observe.  
The rest of this chapter is devoted to describing how we measure the properties 
of particles observed with the CLEO-c detector.
  
\begin{figure}[!htb]
\begin{center}
\includegraphics[width=14.5cm]{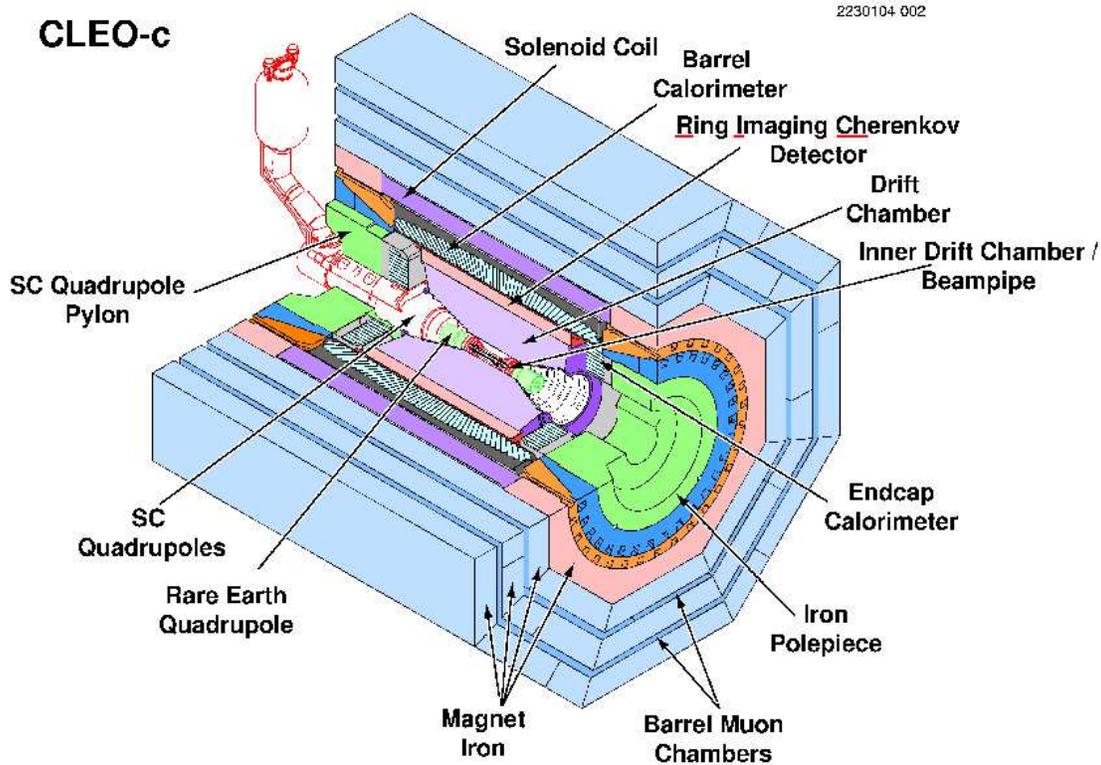}
\caption[The CLEO-c detector.]
{The CLEO-c detector.  The outer and endcap layers of the muon detection chamber are 
omitted.}
\label{fig:CLEOcDet}
\end{center}
\end{figure}

\begin{figure}[!htb]
\begin{center}
\includegraphics[width=15cm]{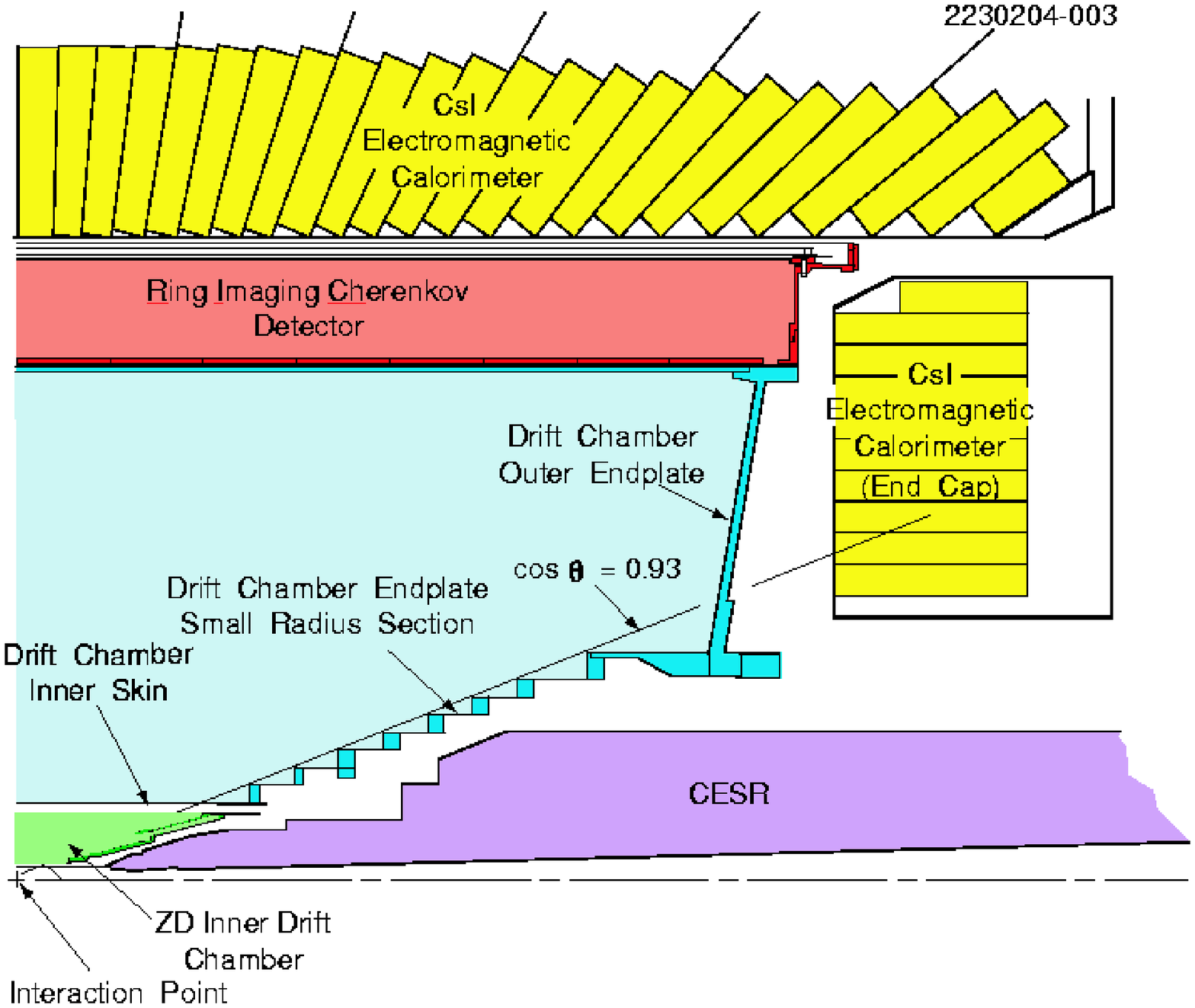}
\caption[Cross section view of the CLEO-c detector in the r-z plane.]
{Cross section of the CLEO-c detector in the r-z plane.}
\label{fig:CLEOcDetCS}
\end{center}
\end{figure}

The CLEO-c detector is a cylindrically symmetric detector with its axis of symmetry 
aligned along the beam axis.  It covers 93$\%$ of solid angle and is thus almost 
completely hermetic.  The main components of the CLEO-c detector are 
the inner drift chamber \cite{CLEOcCESRc}, 
the main drift chamber \cite{CLEOcCESRc,DRIII}, 
the Ring Imaging Cherenkov (RICH) detector \cite{CLEOcCESRc,RICHDet1,RICHDet2}, 
the crystal calorimeter \cite{CLEOcCESRc,CLEOII}, 
and the muon detection chamber \cite{CLEOcCESRc,CLEOII,MuonDet}.  
All of the components, with the exception of the muon detector, 
are operated within a superconducting solenoidal 
coil which produces a uniform 1.0 Tesla magnetic field parallel to the axis of 
symmetry of the detector.  The detector is schematically shown in Figures 
\ref{fig:CLEOcDet} and \ref{fig:CLEOcDetCS}, and its components are described in the 
following sections.  The chapter concludes with a discussion of the trigger and 
data acquisition systems. 

\subsection{The Inner Drift Chamber}

Located immediately outside of the interaction point and beryllium 
beam pipe (radius of 3.5 cm, 0.5 mm thick) is the six-layer inner drift chamber. 
It can detect charged particles with $|$cos$\theta|$ $<$ 0.93, 
where $\theta$ is the angle between the particle and the positron beam, 
and consists of 300 drift cells filled with a helium-propane gas mixture.  
Each cell consists of a sense wire surrounded by eight field wires, forming a nearly 
square cell shape with a half cell size of 5 mm.  
The field wires are shared between neighboring cells, and neighboring 
layers are shifted laterally by one half cell width.  
A potential difference of 1900 V is applied to create an electric field between the 
sense and field wires.  A charged particle ionizes the gas when it passes through a cell.  
The ionized electrons are then attracted toward the sense wire.  
The electric field near the sense wire is strong enough to cause 
the ionized electrons to ionize more atoms, which in turn creates an avalanche of 
electrons on the sense wire.  
The transit time of this electron pulse, synchronized with the timing structure of the 
electron and positron bunches in the storage ring, is converted into a 
distance of closest approach to the sense wire based on the drift velocity 
of the ionized electrons.  The collection of ionized electrons on the 
sense wire constitutes a wire hit and is used in determining the trajectory of the 
charged particle.  

The inner drift chamber covers a radial distance from 4.1 to 11.7 cm.  The inner radial 
wall is made of 1 mm thick aluminum.  The endplates are 16.5 cm machined aluminum plates 
located beyond the 93$\%$ solid angle coverage.  
The outer radial wall is made of 127 $\mu$m thick Mylar.  The helium-propane gas mixture, 
with a radiation length of $\approx$ 330 m, and inner and outer walls constitute a radiation 
length of 1.2$\%$.  The sense and field wires are made of 20 $\mu$m 
diameter gold-plated tungsten and 110 $\mu$m diameter aluminum wires, respectively.  
The wires in each layer are rotated in the $\phi$ direction to determine $z$ information 
of the charged particle.  This rotation creates a so-called stereo angle, with the 
innermost layer being 10.5$^{0}$ and the outermost layer being 15.4$^{0}$ 
(stereo angle is defined as the $\phi$ difference that a wire makes 
between the endplate and the longitudinal center of the detector).  
This detector geometry provides a 680 $\mu$m resolution in the position of the 
charged particle creation along the beam axis ($z_{0}$).  
The inner drift chamber has a momentum resolution of $\sim$0.4$\%$ for charged particles at 
normal incidence (cos$\theta$ = 0) and is the only source of $z$ information for a 
charged particle with transverse momentum $<$ 67 MeV/$c$.

\subsection{The Main Drift Chamber}

Immediately outside of the inner drift chamber is the main drift chamber. 
It covers the radial distance from 13.2 to 82.0 cm and is the primary source of position 
and momentum measurements of charged particles.  The main drift chamber 
consists of 9795 drift cells.  The 
cells are arranged in 47 layers, with the first 16 layers having their field 
and sense wires aligned along the beam axis; the remaining 31 outer layers 
are rotated in a fashion similar to the inner drift chamber.  
The 31 stereo layers are combined into 4-layer ``superlayers'' which have 
alternating stereo angles of about 3$^{0}$.  The wire material, gas mixture, 
cell geometry, endplate material, power supplies, and readout electrons are the same 
as the inner drift chamber, with the exceptions that the half cell size is 7 mm and 
a potential difference of 2100 V is applied in each cell.
The inner wall of the main drift chamber is 2.0 mm thick expanded acrylic with 
20 $\mu$m aluminum skins.  The outer radial wall, made of two layers of 0.8 mm thick 
aluminum cylindrical shells, is lined with 1 cm wide cathode pads.  The cathodes provide a 
longitudinal position measurement at the outer radius and cover 78$\%$ of solid angle.  
The cathodes, when used with the stereo layers, improve the spatial measurement in 
the $z$ direction.  Using a sample of $\Bhabha$ data events, it was shown that the 
resolution in the $z$ direction was improved from 1.5 mm to 1.2 mm with the inclusion 
of information from the cathodes \cite{DRIII}.  
The total radiation length of the main drift chamber is $\sim$2$\%$.  
  
\begin{figure}[!htb]
\begin{center}
\includegraphics[width=12cm]{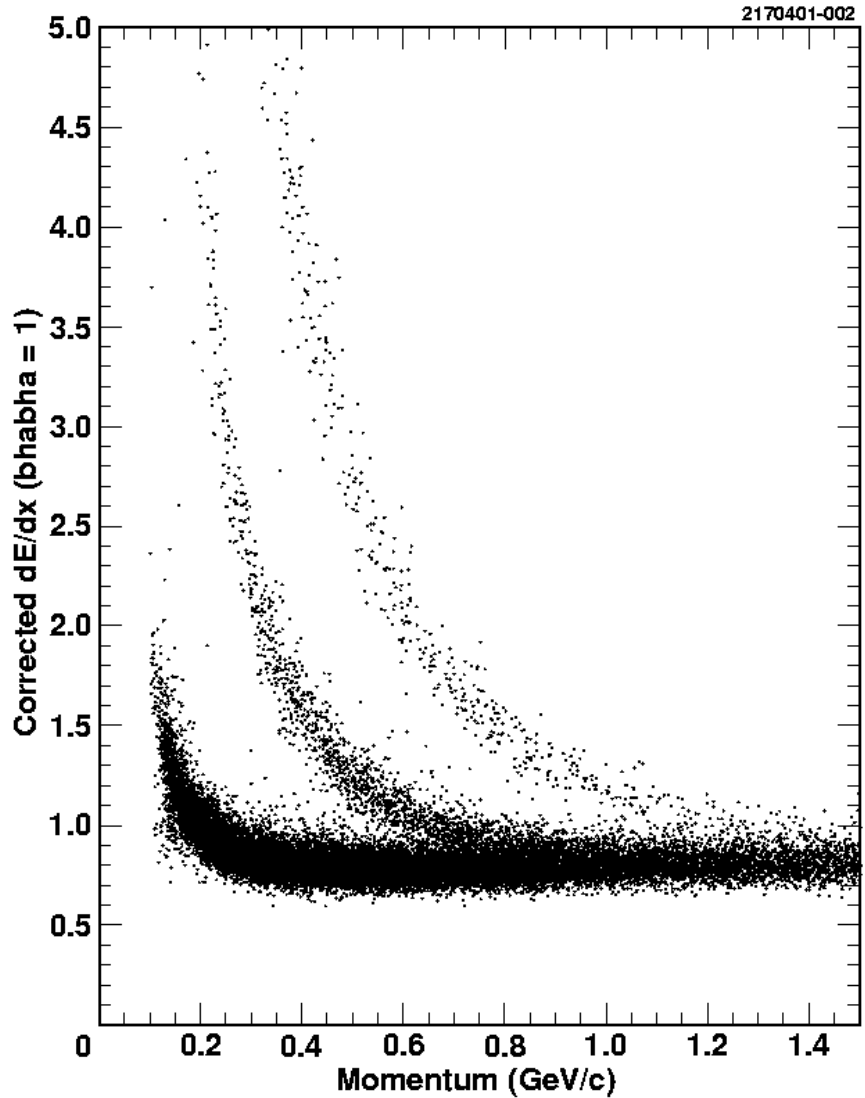}
\caption[Measurement of $dE/dx$ as a function of particle momentum.]
{Measurement of $dE/dx$ as a function of particle momentum. The bands show the 
$dE/dx$ deposited by (from left to right) $\pi$, $K$, and protons, respectively.}
\label{fig:dEdxBands}
\end{center}
\end{figure}

The energy loss of a charged particle due to ionization in the main 
drift chamber can be used for particle identification.  The amount of energy loss 
per unit length ($dE/dx$) is related to the velocity of the particle.
The $\chi^2$-like variable for particle identification relates the measured $dE/dx$ 
to the expected $dE/dx$ for particle hypothesis $i$ ($i$ = e,$\mu$,$\pi$,K, or p) by 
\begin{equation}
S_{i} = \frac{(dE/dx)_{measured} - (dE/dx)_{expected,i}}{\sigma},
\end{equation}
where $\sigma$ is the uncertainty in the $dE/dx$ measurement.  Typical 
$dE/dx$ resolutions are around 6$\%$.  Figure \ref{fig:dEdxBands} 
shows measured $dE/dx$ as a function of momentum.  
It shows that kaons can be well separated from pions with momenta $\lesssim$ 500 MeV/$c$ 
and protons can be well separated from pions and kaons with momenta $\lesssim$ 1 GeV/$c$.

The Kalman fitter procedure \cite{Kalman1,Kalman2} is used to reconstruct the path 
of the charged particle through the beam pipe, inner drift chamber, and main drift chamber. 
A charged particle moving in a vacuum through a solenoidal magnetic field traces out a 
helix, from which the momentum and position of the charged particle can be determined.  
The helix track needs to be corrected for various distortions caused by the material 
that the charged particle traverses.  These distortions are mainly due to 
ionization energy loss and multiple scattering. 
Additional corrections include 
finite signal propagation time along the sense wires, 
flight time corrections of the ionized electrons in the drift cell, 
and the non-uniform magnetic field near the endcaps due to the final-focusing 
quadrupole magnets.  Wire hits from both the inner and main drift chamber are used to 
reconstruct the trajectory of the particle.  The momentum resolution is a function of 
the spatial resolution of individual hits, which is on average 88 $\mu$m.  
The momentum resolution for charged particles 
with 1 GeV/$c$ momenta at normal incidence is $\sim$0.6$\%$.  The resolution is 
worse for charged particles with transverse momenta $<$ 120 MeV/$c$ because they will not 
transverse all 47 cell layers and the outer radius cathodes of the main drift chamber.

\subsection{The RICH Detector}

Outside of the main drift chamber is the Ring Imaging Cherenkov (RICH) detector.  
If the velocity of a charged particle is greater than the speed of light in the medium, 
it emits radiation in the form of Cherenkov photons.  The photons are distributed 
in a conic shape and the apex angle of the cone, called the Cherenkov angle $\Theta$, 
is related to the particle velocity by
\begin{equation}
\mathrm{cos}~\Theta = \frac{1}{\beta n},
\label{eq:cherangle}
\end{equation}
where $\beta$ is the velocity of the particle in units of $c$ and 
$n$ is the index of refraction of the medium.  Equation \ref{eq:cherangle} can be rewritten 
in terms of the momentum and mass of the particle as
\begin{equation}
\mathrm{cos}~\Theta = \frac{1}{n}~\sqrt{1+\frac{m^2}{p^2}}.
\end{equation}
Therefore, the particle can be identified by measuring the Cherenkov angle and its momentum.  

\begin{figure}[!htb]
\begin{center}
\includegraphics[width=13cm]{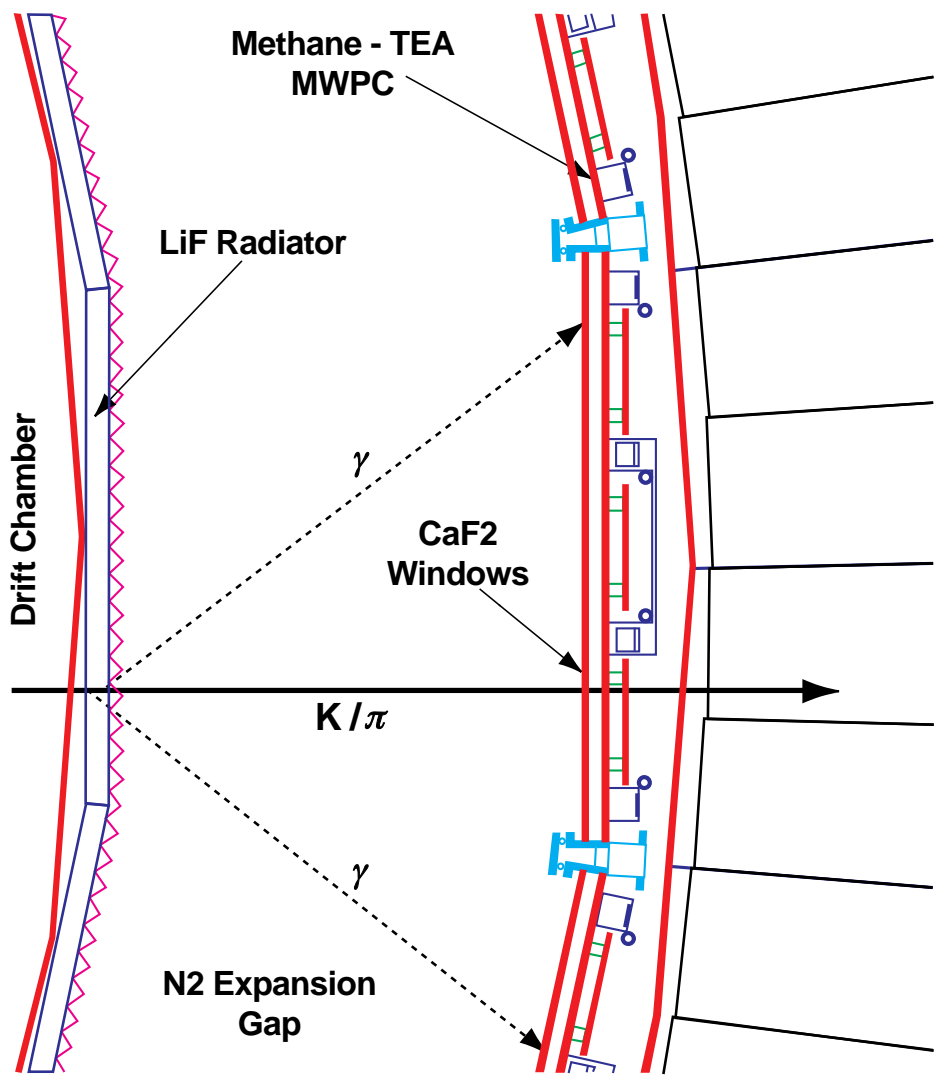}
\caption[Cross section view of the RICH detector.]
{A one-tenth cross section view of the RICH detector in the $r-\phi$ plane.  Also 
shown is the trajectory of a candidate charged track $K$ or $\pi$ and 
its associated Cherenkov photons.}\label{fig:RICHDet}
\end{center}
\end{figure}

The RICH detector is a ``proximity focusing'' Cherenkov detector, which means that the 
Cherenkov photons are not focused and the Cherenkov angle is determined by allowing 
the photons to propagate over a finite space.  The detector is composed 
of the following elements:  
a radiator material where the charged particle radiates Cherenkov photons, 
an expansion volume, and photon detectors.  The radiators are made of lithium 
fluoride (LiF) plates.  The expansion volume is 16 cm long and filled with nitrogen gas.  
The photon detectors are highly segmented 
multiwire proportional chambers (MWPC) filled with a methane-triethylamine 
(TEA) gas behind 2 mm calcium fluoride (CaF$_2$) windows.  The detector covers the 
radial distance from 82 and 101 cm, its radiators cover 83$\%$ of solid angle, and 
constitutes a total radiation length of about 12$\%$.  Figure \ref{fig:RICHDet} 
shows a schematic diagram of a one-tenth view of the RICH detector in the $r-\phi$ plane 
and an example of a charged particle, with its associated Cherenkov photons, 
propagating through its volume.  

The components of the RICH detector are 
optimized for Cherenkov photons with a wavelength of 150 nm, 
for maximum quantum efficiency of the methane-TEA mixture in the photon detector.  
The index of refraction of the LiF radiators is $n$ =  1.50 at 150 nm.  
The nitrogen gas in the expansion volume and the CaF$_2$ windows have 
a transparency of $>$ 99.5$\%$ and $>$ 80$\%$ \cite{RICHDet2}, respectively.

The LiF radiators have a surface area of 17 cm $\times$ 17.5 cm and an average 
thickness of 1.0 cm.  They are arranged in 14 coaxial rings with 30 radiators each.  
Emitted photons from charged particles at normal incidence 
would experience total internal reflection with flat surface radiators.  
The middle four rings, which accept particles with 
$|$cos$\theta|$ $\lesssim$ 0.4, have a ``sawtooth'' outer surface to overcome this problem 
and this radiator geometry is shown in Figure \ref{fig:RICHDet}.  
Radiators at larger $|$cos$\theta|$ which do not have this problem 
have a flat planar outer surface.  

\begin{figure}[!htb]
\begin{center}
\includegraphics[width=15cm]{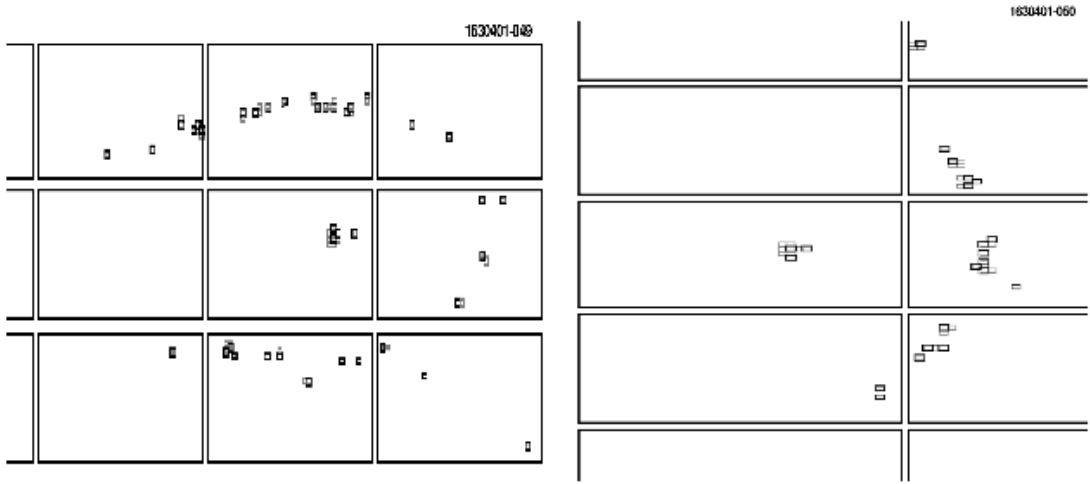}
\caption[Cherenkov rings produced by sawtooth and flat radiators.]
{Cherenkov rings produced by a track transversing the sawtooth (left) 
and flat (right) radiators.  The small rectangles are individual cathode pad hits 
and the large rectangles are the 24 $\times$ 40 cathode cell arrays.  
Only half of the Cherenkov ring is observed from flat radiators (right) 
because the other half experiences total internal reflection in the radiator; 
the ring shape from sawtooth radiators (left) is distorted due to the radiator 
surface geometry.  The cathode hits near the center of the Cherenkov rings 
are from the incident charged particle crossing the cathode plane.}
\label{fig:RICHCHImages}
\end{center}
\end{figure}

The MWPC photon detectors are arranged to cover the same azimuthal angle as the 
radiators.  The photoelectrons produced by a photon are detected by cathode 
pads, which have a surface area of 7.5 mm $\times$ 8.0 mm 
and are located 4.5 mm behind the CaF$_2$ windows.  
The cathode pads are arranged into 24 by 40 pad arrays with a 
total surface area of 30 cm $\times$ 19 cm.  There 
are 8 arrays per MWPC, each separated by a 7 mm spacing.  Axial anode field wires 
are placed 1 mm above 
the cathode pads with a spacing of 2.66 mm with 72 anode wires per MWPC.  The anode 
wires are 20 $\mu$m diameter gold-plated tungsten with a 3$\%$ admixture of rhenium.  
The back plane of the CaF$_2$ windows has 100 $\mu$m wide silver traces 
spaced 2.5 mm apart.  A 2700 V potential is applied between the silver traces 
and the anode wires.  Examples of Cherenkov rings produced in sawtooth and flat radiators 
are shown in Figure \ref{fig:RICHCHImages}.

Precise measurement of the momentum vector of the charged particle 
at the radiator is important for the angular resolution of the Cherenkov cone.  The 
main drift chamber has very good transverse momentum resolution but has larger uncertainty 
in the $z$ direction.  The $z$ momentum component can be improved by as much as 50$\%$ 
if the charged particle can be associated with hits in the cathodes.  The photons 
are traced out from their emitted point to the MWPC while also considering materials 
that it traverses along its path.  
The average number of observed photons for candidate particles is typically 12 and 10 photons 
from sawtooth and flat radiators, respectively.  
The angular resolution of a single photon produced in the sawtooth 
and flat radiators are $\sigma_{\theta}$ = 13.2 and 15.1 mrad, respectively.  
The angular resolution per track is determined by accepting all photons 
which are within $\pm$3$\sigma$ of the expected Cherenkov angle and weighting 
each photon by 1/$\sigma^2_{\theta}$; $\sigma_{\theta}$ = 3.7 and 4.9 mrad for 
sawtooth and flat radiators, respectively.
Possible sources of uncertainties in the angular resolutions are from 
the location of the photon emission point, chromatic dispersion, 
position error of reconstructed photons, and trajectory of the charged track in the radiator.

\begin{figure}[htbp]
\begin{center}
\includegraphics[width=12cm]{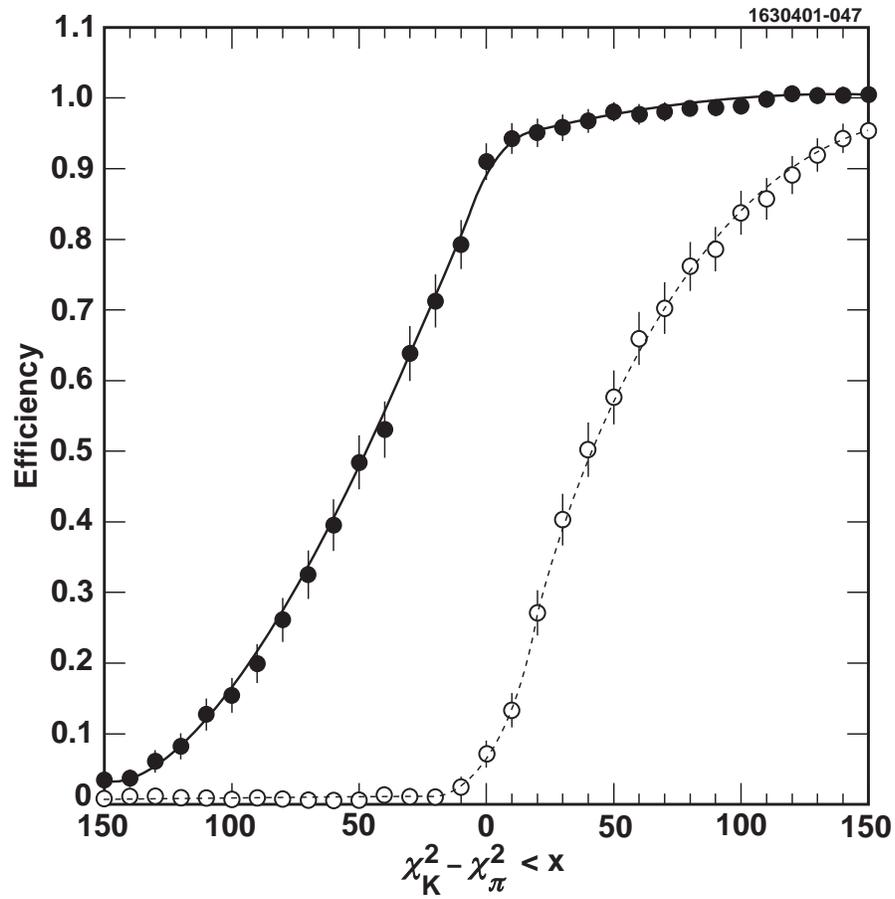}
\caption[Kaon efficiency and pion fake rate determined from RICH Detector.]
{Kaon efficiency (solid points) and pion fake rate (open circles) 
determined from the RICH detector for various cuts on 
$\chi^2_K-\chi^2_\pi$ for kaons and pions with momenta $>$ 0.7 GeV/$c$.  }
\label{fig:piKRICHDiff}
\end{center}
\end{figure}

\begin{figure}[htbp]
\begin{center}
\includegraphics[width=14cm]{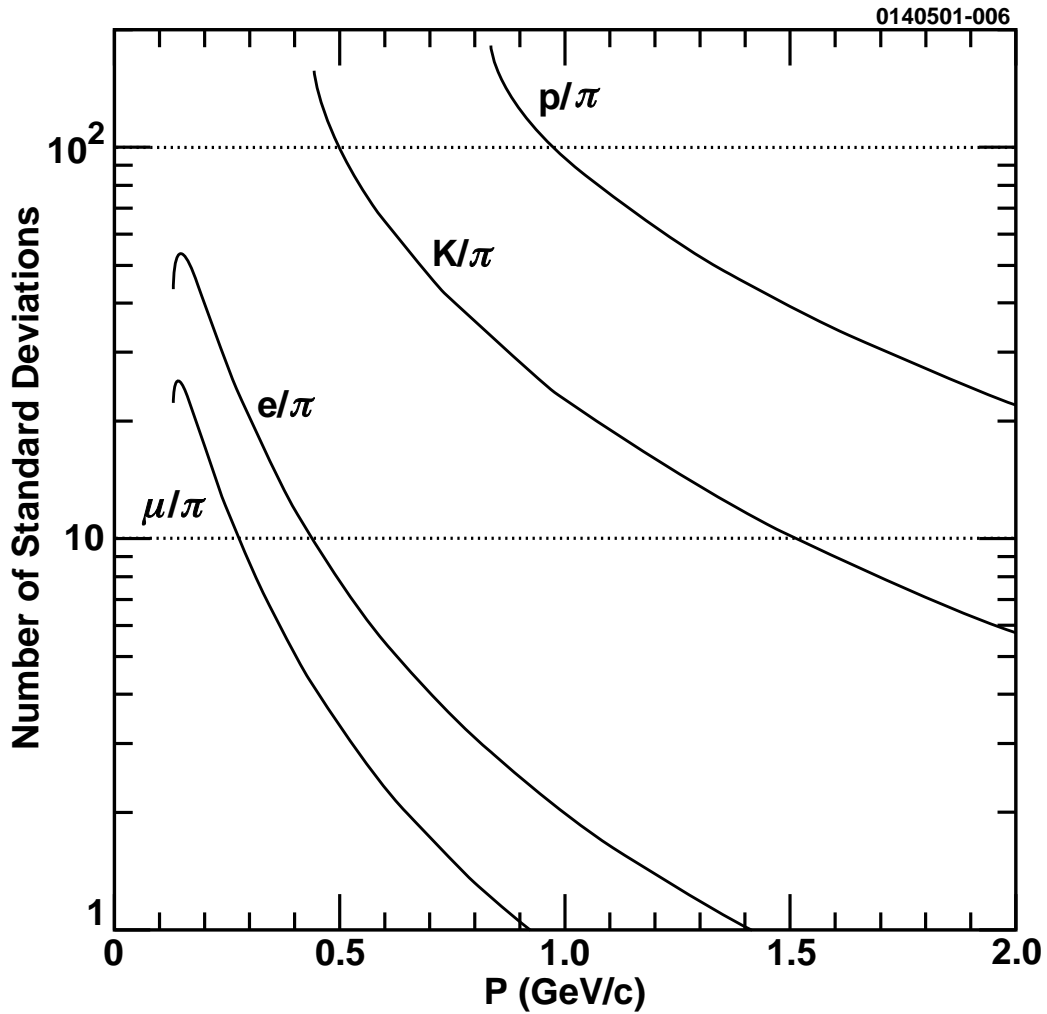}
\caption[Particle separation in the RICH detector.]
{Particle separation in the RICH detector.  The quantity plotted is 
$|\theta_{X}(P) - \theta_{\pi}(P)/\sigma(P)$ versus particle momentum for 
$X = e, \mu, K,$ or $p$, 
where $\sigma(P) = \sigma^{\beta=1}\sqrt{N^{\beta=1}_{\gamma}/N_{\gamma}(P)}$ 
is the rms resolution.}
\label{fig:RICHalltracks}
\end{center}
\end{figure}

Information from the Cherenkov photons is used to derive a likelihood 
for a particular particle hypothesis.  The likelihood weights each possible optical path 
traveled by a photon by considering the length of the radiation path and the refraction 
probabilities from inverse ray tracing.  
A $\chi^2$-like particle identification variable can be obtained by taking the 
difference of the logarithm of two different particle hypotheses, given explicitly as
\begin{equation} 
\chi^2_i - \chi^2_j = -2\mathrm{ln}L_i + 2\mathrm{ln}L_j ,  
\label{eq:RICHchisqdiff}
\end{equation}
where $L_i$ and $L_j$ denote the likelihood for particle hypotheses $i$ and $j$, 
respectively.  Figure \ref{fig:piKRICHDiff} shows an example for the kaon efficiency 
and pion fake rate as a function of different values of $\chi^2_K-\chi^2_\pi$ for 
momenta $>$ 0.7 GeV/$c$.  The kaon efficiency is 92$\%$ and the pion fake rate is 8$\%$ 
when requiring a given particle to be more like a kaon than a pion, 
defined as $\chi^2_K-\chi^2_\pi$ $<$ 0. Figure \ref{fig:RICHalltracks} shows the 
particle separation as function of momentum when both particles are above their 
respective Cherenkov radiation thresholds.

\newpage
\subsection{The Crystal Calorimeter}

Located directly outside of the RICH detector in the radial direction and outside 
of the endcap of the main drift chamber is the electromagnetic calorimeter.  The 
calorimeter consists of 7784 thallium-doped cesium iodide blocks.  
Three different types of reactions can occur in the crystals depending on the incident 
particle \cite{expbook}.  
The first type of reaction is the production of an electromagnetic shower 
produced by incident photons, electrons, and positrons.  Photons produce 
electron-positron pairs, while electrons and positrons emit bremstrahlung radiation.  
The bremstrahlung photons then produce electron-positron pairs.  
This produces a large number of low energy electrons, positrons, and photons.  
The low energy positrons and electrons annihilate to produce pairs of photons, 
and low energy photons begin to ionize the atoms.
The low energy electrons are captured by thallium atoms in the crystal.  The 
deexcitation of thallium emits visible light ($\lambda$ = 560 nm for CsI(Tl) \cite{expbook}), 
which in turn is detected by a silicon photodiode.  
In the second type of reaction, the charged particle directly ionizes atoms in the crystal.  
The liberated electrons are captured by thallium atoms.  This type of reaction occurs 
for all charged particles (with the exception of electrons and positrons).  
The last type of reaction occurs when hadrons interact strongly with an atomic nucleus 
in the crystal.  These strong interactions can produce a large number of neutral 
pions, which decay to pairs of photons and produce electromagnetic 
showers.  

The calorimeter is configured in three sections, a barrel region 
and two endcap regions, as shown in Figure \ref{fig:CLEOcDetCS}.  The barrel region 
accept particles in the $|$cos$\theta|$ range of $<$ 0.8, and 
the endcaps accept 0.85 $< |$cos$\theta| <$ 0.93.  The transition region, 
defined by 0.8 $<$ $|$cos$\theta|$ $<$ 0.85, is generally not used, 
because detector material blocks the interaction point 
and there are shower spillovers between the barrel and the endcap.  
The barrel region covers a radial distance from 102.4 to 142.5 cm 
and has a length of 3.37 meters.  
It contains 6144 blocks arranged in 48 azimuthal rows with 128 blocks per row.  
The front surface of each block is aligned to point back to the interaction point, 
with a minor correction caused by the gaps between blocks pointing a few centimeters 
away to prevent incident particles from passing between adjacent blocks.  
The endcap sections contain 820 blocks and are four-fold symmetric in $\phi$.  
The front surfaces form a plane which is located 124.8 cm away from the interaction point.  
Each crystal block has a front surface area of 5 cm $\times$ 5 cm and a 
length of 30 cm, or 16 radiation lengths.  Each block is wrapped with 
three layers of 0.04 mm white teflon and one layer of 0.01 mm aluminized mylar to 
ensure high internal reflection.  

The light yield from a block is converted into electrical signals 
by four photodiodes attached to the back plane of each block; 
each with an active area of 1 cm $\times$ 1 cm.  
Each photodiode is connected to a separate nearby preamplifier.  The four photodiodes 
for a given block are then summed by a mixer/shaper card.  The output from the mixer/shaper 
card sends a signal to an analog-to-digital converter (ADC) and is also combined with other 
nearby blocks in a trigger tile to be used by the calorimeter trigger, described 
in more detail in the trigger section, Section 3.2.6.  

The ADC values are converted into energies before shower reconstruction begins.  
This requires electronic pedestal subtraction, gain multiplication, and conversion to 
absolute energy units.  Crystal-by-crystal energy calibrations are calculated using 
$\Bhabha$ (Bhabha) data events.

\begin{figure}[htbp]
\begin{center}
\includegraphics[width=14cm]{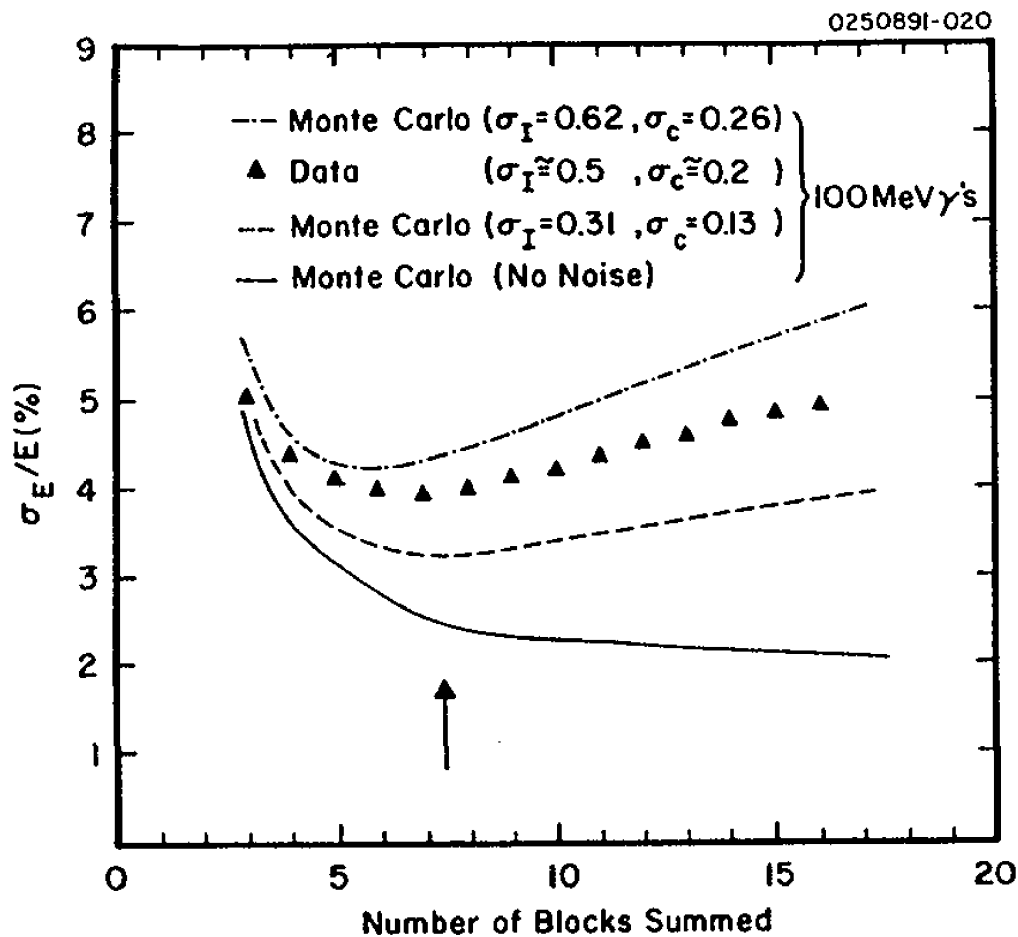}
\caption[Shower energy resolution as a function of the number of summed blocks.]
{Shower energy resolution as a function of the number of summed blocks 
determined for the CLEO II detector.  The calorimeter has remained unchanged since 
the installation of the CLEO II detector.  The curves 
were generated from a Monte Carlo shower simulation of 100 MeV photons with appropriate 
noise included.  The points were measured with experimental data from the 100 MeV photon 
lines from the transitions $\Upsilon(3S) \rightarrow \gamma \chi_{bJ}(2P)$.  The 
arrow indicates the actual number of summed blocks for 100 MeV photons.}
\label{fig:ccshres}
\end{center}
\end{figure}

Energy from neighboring blocks are combined to determine the 
total shower energy from the incident particle, starting with the most energetic block 
if its energy exceeds 10 MeV.  Varying the number of blocks included in determining the 
total energy of a shower as function of shower energy improves the shower energy 
resolution \cite{CLEOII}.  The number of blocks considered in a shower has a logarithmic 
dependence, ranging from 4 blocks at 25 MeV to 13 blocks at 2 GeV .  
The centroid of the shower is determined by an 
energy-weighted average of the block centers used in constructing the shower.  The 
shower energy resolution in the barrel region is 4.0$\%$ at 100 MeV and 
2.2$\%$ at 1 GeV.  Figure \ref{fig:ccshres} shows the shower energy resolution as a function 
of the number of summed blocks.  The angular resolution for barrel showers is 
about 10 mrad.  Showers reconstructed in the endcap regions are of comparable quality.  
  
\newpage
\subsection{The Muon Detector}

\begin{figure}[!tb]
\begin{center}
\includegraphics[width=14cm]{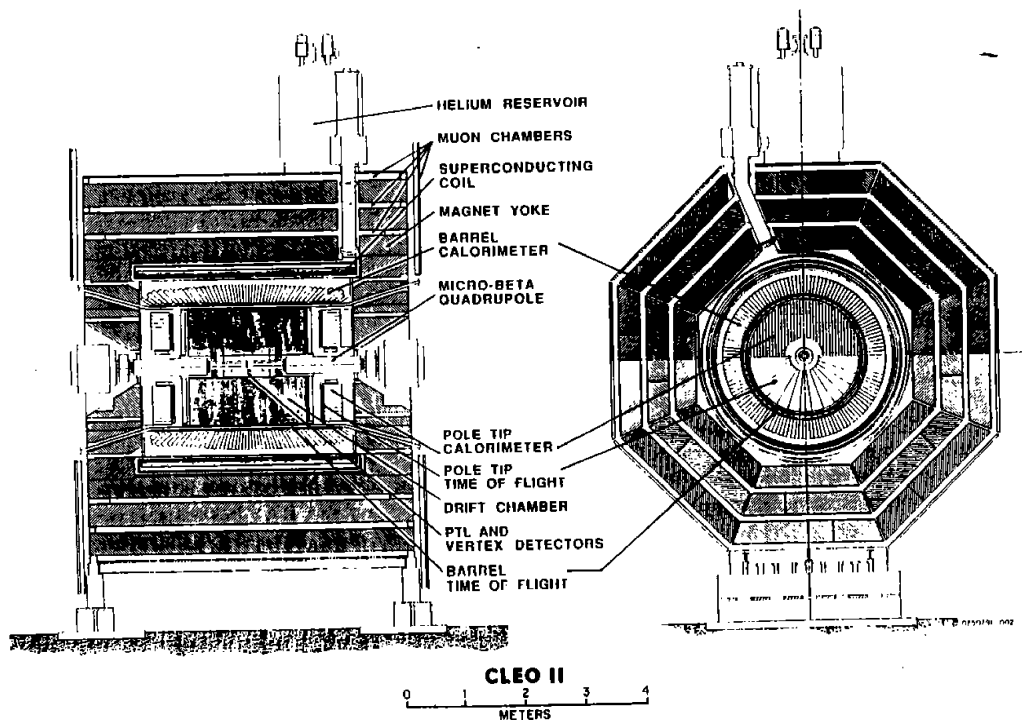}
\caption[The CLEO II detector, showing the barrel and endcap muon chambers.]
{The CLEO II detector, showing the barrel and endcap muon chambers.  The 
muon chambers has remained unchanged since the installation of the CLEO II detector.}
\label{fig:CLEOIIMuonDet}
\end{center}
\end{figure}

The muon detector, as shown in Figure \ref{fig:CLEOIIMuonDet}, consists of a 
system of proportional chambers interspersed in the return iron of the solenoid magnet.  
The muon detector covers 85$\%$ of solid angle.  In the barrel region, 
the chambers are located behind 36, 72, and 108 cm of iron at normal incidence 
and located at respective radial distances of 2.10, 2.46, and 2.82 meters.  
Chambers are also placed outside of the endcap iron at a distance 
of 2.82 meters away from the interaction point.  The total available thickness of 
iron absorber varies from about 7.2 to 10.0 nuclear interaction lengths (n.l.)
depending on their flight direction (1 n.l. = 16.7 cm of iron \cite{PDG2004}).  

\begin{figure}[!tb]
\begin{center}
\includegraphics[width=14cm]{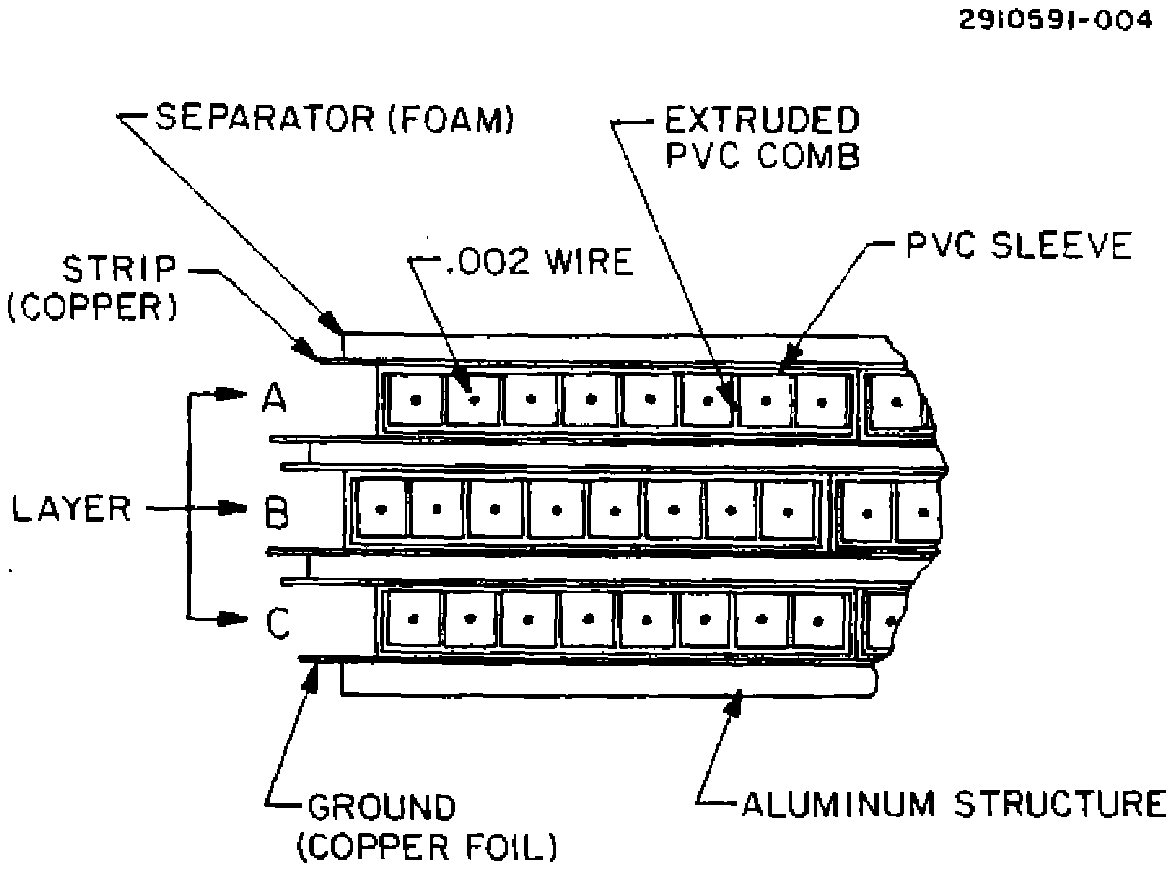}
\caption[Cross section of a muon chamber.]
{Cross section of a muon chamber.  It consists of three layers of 
8-cell proportional counters interspersed with copper pickup strips. 
The middle counter layer is offset by one half cell width to improve geometric 
acceptance.}
\label{fig:mclayer}
\end{center}
\end{figure}

\begin{figure}[!tb]
\begin{center}
\includegraphics[width=14cm]{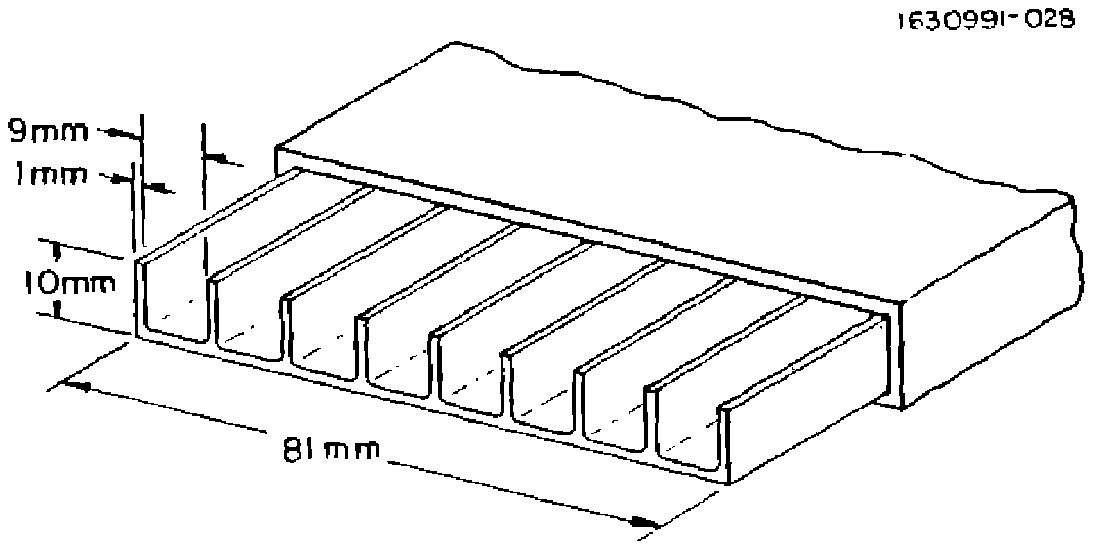}
\caption[Cross section of the proportional counter for the muon detector.]
{Cross section of the proportional counter for the muon detector.}
\label{fig:muoncounter}
\end{center}
\end{figure}

A muon chamber, as shown in Figure \ref{fig:mclayer}, consists of three layers of 
proportional counters interspersed with copper pickup strips.   
The middle counter layer is offset by one half cell width to improve geometric acceptance.  
A proportional counter, as shown in Figure \ref{fig:muoncounter}, is made of 5 m long, 
8.3 cm wide PVC plastic.  Each counter has eight 9 mm $\times$ 9 mm U-shaped rectangle 
cells, all enclosed in a 1 mm thick PVC sleeve.  Each cell has a 50 $\mu$m diameter 
silver-plated copper-beryllium anode wire and is filled with an argon-ethane gas mixture.  
The three sides of a cell are coated with graphite providing the cathode for the cell.  
Each cell is operated in the proportional mode with a potential of 2500 V.  
For position measurements orthogonal to the counter length, 8.3 cm wide 
external copper pickup strips are located on top of each counter.  In total, the 
muon detector consists of 2352 counters and 5472 strips.

To determine if a hit in the muon detector is associated with a charged particle 
track, the track is traced out from the main drift chamber and, 
after taking into account multiple scattering and energy loss, it is projected 
through the muon detector.  A two-dimensional $\chi^2$ fit is performed to test 
if the hit can be associated with the track.  The hit in the muon chamber is identified 
with the track if $\chi^2$ $<$ 16.  
The resolution for anodes located in the three barrel chambers 
(increasing in distance from the interaction point) is 3.7, 4.6, and 5.7 cm, respectively; 
in the endcap chambers, the resolution is 7.2 cm.  
The corresponding resolution for the strips are 5.5, 7.0, 7.5, and 9.0 cm.  
Particles must have momenta in excess of 1.0 GeV/$c$ to be detected in the muon detector.  
In the momentum range 1.5-2.0 GeV/$c$, the efficiency for muons to penetrate 
$>$ 3, $>$ 5, and $>$ 7 n.l. is about 
90$\%$, 85$\%$, and 30$\%$, respectively.  The efficiency is also stable over the 
entire $|$cos$\theta|$ range except for a slightly lower efficiency at the largest 
$|$cos$\theta|$ (near the beampipe).  The fake rate for pions and kaons to transverse 
3 n.l. in the same momentum range is about 4$\%$ and 12$\%$, respectively.

\newpage
\clearpage
\subsection{The Trigger System}

\begin{figure}[!tb]
\begin{center}
\includegraphics[width=14cm]{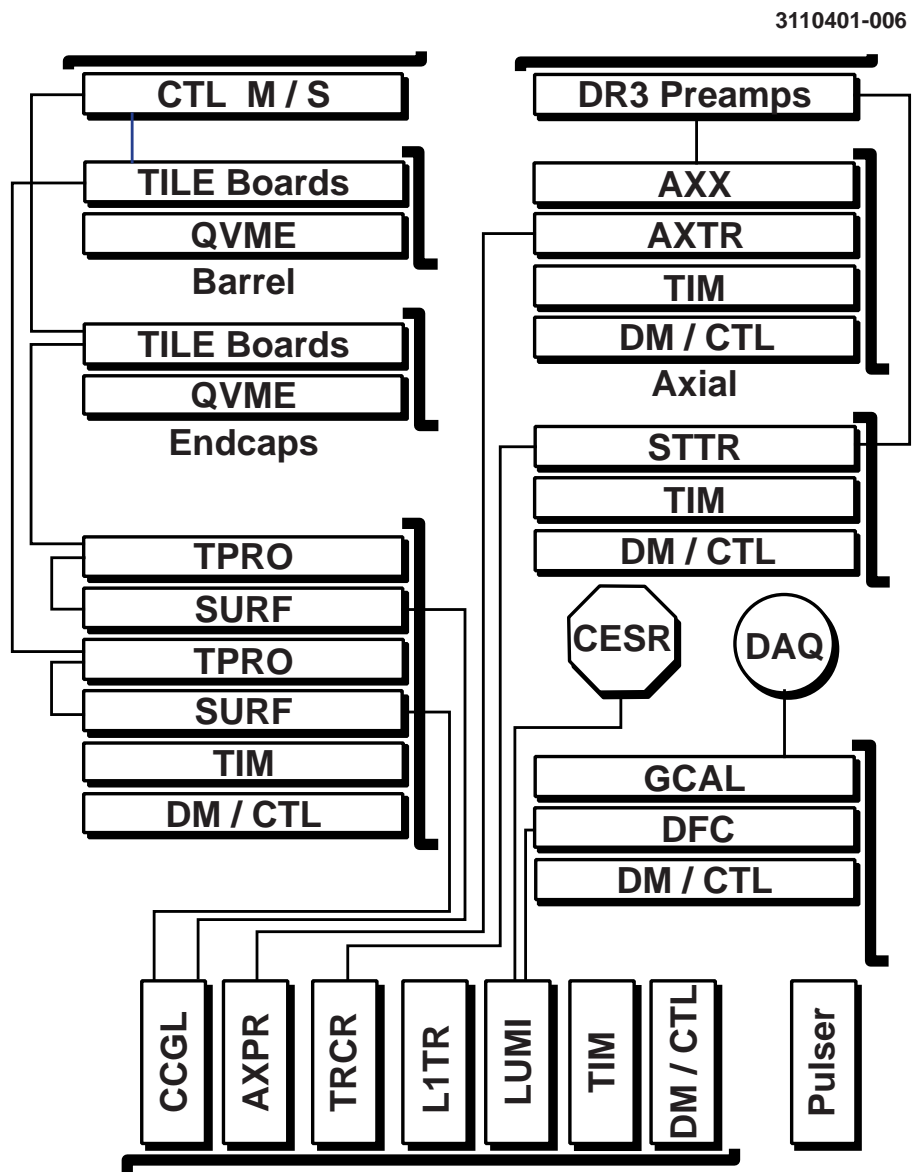}
\caption[Overview of the trigger system.]
{Overview of the trigger system.  For the explanation of symbols, see text.}
\label{fig:trigpicture}
\end{center}
\end{figure}

This description of the trigger is based on Refs. \cite{CLEOcCESRc}, 
\cite{trackingtrigger}, \cite{CCtrigger}, and \cite{GlobalL1trigger}.  
A schematic view of the trigger system \cite{CLEOcCESRc} 
is shown in Figure \ref{fig:trigpicture}.  
Data from the main drift chamber and crystal calorimeter 
are received and processed on separate VME crates 
by appropriate circuit boards to yield basic trigger primitives such as track count 
and topology in the main drift chamber and shower count and topology in the calorimeter.  
The information from both systems is correlated by global trigger circuitry which generates 
an 'pass' strobe every time a valid 
trigger condition is satisfied.  The 'pass' signals are conditionally passed by the 
data flow control (DFC) circuitry to the gating and calibration (GCAL) modules 
for distribution to the data acquisition system.  In addition, online luminosity 
information is determined by the luminosity (LUMI) module from calorimeter information 
and sends it to the CESR accelerator control room via the global trigger.

The trigger system consists of two tracking triggers \cite{trackingtrigger}, 
one using information from the axial layers of the main drift chamber and 
the other using the stereo layers, a calorimeter trigger \cite{CCtrigger}, 
and a decision and gating global trigger system \cite{GlobalL1trigger}.  
Configuration and supervision duties are provided by MVME2304 PowerPC modules, 
which play the dual role of crate controller (CTL) and data mover (DM).  
Also, for clock and 'pass' signal distribution, as well as busy signaling, 
a trigger interface module (TIM) is used.  
The details of the tracking, calorimeter, and global triggers are discussed 
in the following subsections.

\subsubsection{Tracking Triggers}

The limited wire count in the axial portion of the main drift chamber makes it possible 
to build a tracker that examines the complete set of 1696 axial wires in the first 
16 layers of the main drift chamber for possible valid 
patterns caused by tracks having transverse 
momenta $>$ 133 MeV/$c$.  Patterns from tracks missing the central axis of the 
beam pipe by as much as 5 mm are included.  

The axial trigger bins the data into 42 ns wide time slices, three times longer than the 
bunch spacing in a beam train.  This discretization is sufficient for the time 
resolution required by the trigger, namely to determine the time of the interaction, 
with a fixed offset, to within 100 ns.  Pattern recognition is performed by the axial 
tracking (AXTR) boards for the entire axial portion of the main drift chamber.  

The axial processor (AXPR) takes the 112 $\phi$ tracking bits from the AXTR boards and 
produces a 7 bit track count, a 48 bit array which represents 
the azimuthal event topology, and a 2 bit time stamp.  The information from the AXPR 
boards are then passed to the track correlator (TRCR) boards.  

The stereo section of the main drift chamber (layers 17-47) differs from the axial section 
in that the stereo wires are offset with respect to the beam, or $z$, axis; 
the axial wires are almost exactly parallel.  The stereo section is broken up into 
eight superlayers.  The first seven superlayers have four layers of wires each; the last has 
only three.  The odd superlayers are called the U superlayers and have a positive 
$\phi$ tilt with respect to the $z$-axis; 
the even ones are V superlayers and have a negative tilt.

There are too many (8100) wires in the stereo section of the main drift chamber for the 
stereo tracker to examine every wire individually.  Instead, the stereo tracker receives 
1 bit for every 4 by 4 block of wires.  The U and V superlayers are tracked separately.  

The stereo block definitions (which patterns within a 4 by 4 block are defined as track 
segments) and the stereo road definitions (which groups of U or V blocks is considered 
a valid track) were generated from simulated tracks having transverse momenta $>$ 
167 MeV/$c$.  In order to satisfy a stereo block pattern, a hit must be present on at least 
3 out of 4 layers, allowing high track efficiency for realistic wire hit efficiencies.  
Stereo roads, however, do not allow from missing blocks.  

One stereo tracking (STTR) board is responsible for one-sixth of the U or V 
superlayers (12 boards total for the stereo trigger).  To prevent inefficiencies at board 
boundaries, information is shared between neighboring STTRs.  As with the axial trigger, 
pattern recognition is performed every 42 ns.  
The outputs from each STTR are sent to the TRCR boards for further processing.    
   
Topological information from both the axial and stereo tracking trigger hardware is 
received by two TRCR boards.  The TRCRs correlate the U, V, and axial projection bits 
to form a final set of 48 correlated azimuthal 
projection bits as well as high and low momentum track counts.  The derived TRCR output 
information is sent to the Level 1 decision electronics.  

The stereo tracking trigger provides high efficiency per track and good background 
rejection.  The combined axial plus stereo tracking trigger is much less sensitive to 
backgrounds than the axial trigger alone, while maintaining an efficiency 
of $>$ 99$\%$ for single tracks.

\subsubsection{Calorimeter Trigger}
    
The calorimeter trigger incorporates both analog and digital electronics to provide 
\\ pipelined trigger information every 42 ns with a latency of $\sim$2.5 $\mu$s.  Analog 
processing is employed to address the quantization error caused by split energy 
deposition in adjacent calorimeter blocks, and digital field programmable gate arrays are 
used extensively to filter and categorize the calorimeter energy topology.  Timing, 
position, and energy information are all available for use in the calorimeter trigger.  

\begin{figure}[ht]
\begin{center}
\includegraphics[width=14cm]{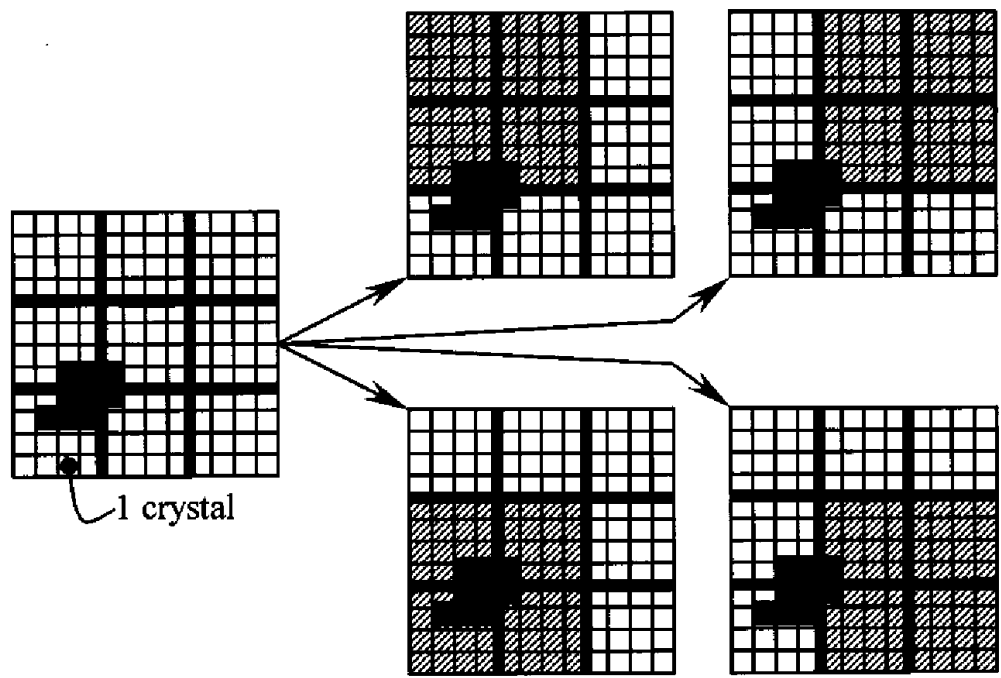}
\caption[An example of calorimeter trigger tiles.]
{An example of calorimeter trigger tiles.  A 12 by 12 block subregion of the calorimeter is 
shown, with a single crystal block represented by an individual box.  An example of a 
photon shower is represented by the cluster of solid boxes, 
and trigger tiles are denoted by the area covered by dashed lines.}
\label{fig:trigtile}
\end{center}
\end{figure}

Complications associated with boundaries in the calorimeter are reduced by 
creating overlapping 'tiles' by forming analog sums of signals from groups of 64 
calorimeter blocks, as shown by the example in Figure \ref{fig:trigtile}.  
A photon striking the calorimeter will deposit nearly all of its 
energy in at least one of the groups of blocks summed in a tile.  Naturally, a signal 
in a single block will appear in four different tiles; it is the task of the 
tile processors (TPRO) to account for this.  

The TPRO boards receive data from as many as 384 active tiles in the 
calorimeter barrel and 120 tiles in the endcaps.  The first task of the TPRO is to filter 
event data so that adjacent or overlapping tiles which contain energy are reduced to a 
single hit.  After filtering the data, the TPRO then determines the number of showers and 
their position in the calorimeter.  The algorithm run by the TPRO boards is a comprise 
between the angular and energy resolutions and the desire to limit the amount of 
information to be processed by the trigger.  The TPROs remove all but the highest threshold 
tile in a group of adjacent or overlapping tiles and project the two-dimensional 
tile information into one-dimensional distributions in $\theta$ and $\phi$.  

Once the individual TPROs have produced their results, one SURF (Sampling Unit for Radio 
Frequency) board combines the four barrel TPRO projections and tile counts.  A second 
SURF board does the same for the endcap TPROs.  The output from the SURF boards are sent 
to the crystal calorimeter global logic (CCGL) for use by the Level 1 
decision electronics.

\subsubsection{Global Trigger}

The global decision and data flow control system produces and distributes 
a trigger decision every 42 ns based on input from the tracking and calorimeter triggers 
described above.  Programmable trigger decision (L1TR) boards monitor this information.  
Tracking and calorimeter information is received and channeled through variable-depth 
pipelines to time-align the data; tracking is available in $\sim$2 $\mu$s while the 
calorimeter requires over 2.5 $\mu$s.  The time-aligned information is presented on a 
shared backplane where several L1TR modules have access to the information for 
performing independent trigger condition evaluation.  

All L1TR boards see the same input information on the Level-1 backplane.  The trigger logic 
section allows the user to define 24 independent trigger ``lines'', each a (potentially 
complex) combinatoric function of the 179 inputs.  Each of the 24 trigger lines 
is routed through a 24 bit prescalar to a 40 bit scalar.  In the present mode of operation 
we typically run with a set of about eight trigger lines.  The definitions of these 
lines are shown in Table  \ref{tab:triggerlines}.  Once the criteria for a trigger line 
is satisfied, the detector information is sent to the data acquisition system. 

\newpage
\begin{table}[ht]
\caption[Definitions of the trigger lines.]{Definitions of the trigger lines.  Note that 
the random trigger only sends out one of 1000 Level 1 decisions to the 
data acquisition system, i.e., prescaled by 1000.}
\medskip
\begin{center}
\begin{tabular}{|l|l|}
\hline
Name & Definition \\
\hline
Hadronic & $(N_{axial}>1)\&(N_{CB~low}>0)$ \\    
Muon Pair & Two back-to-back stereo tracks \\    
Barrel Bhabha & Two back-to-back high showers in barrel \\    
Endcap Bhabha & Two back-to-back high showers in endcaps \\    
Electron track & $(N_{axial}>0)\&(N_{CB~med}>0)$ \\    
Tau & $(N_{stereo}>1)\&(N_{CB~low}>0)$ \\    
Two Track & $N_{axial}>1$ \\    
Random & random 1 kHz source \\    
\hline
\end{tabular}
\label{tab:triggerlines}
\end{center}
\end{table}

\subsection{The Data Acquisition System}

\begin{figure}[htbp]
\begin{center}
\includegraphics[width=13.5cm]{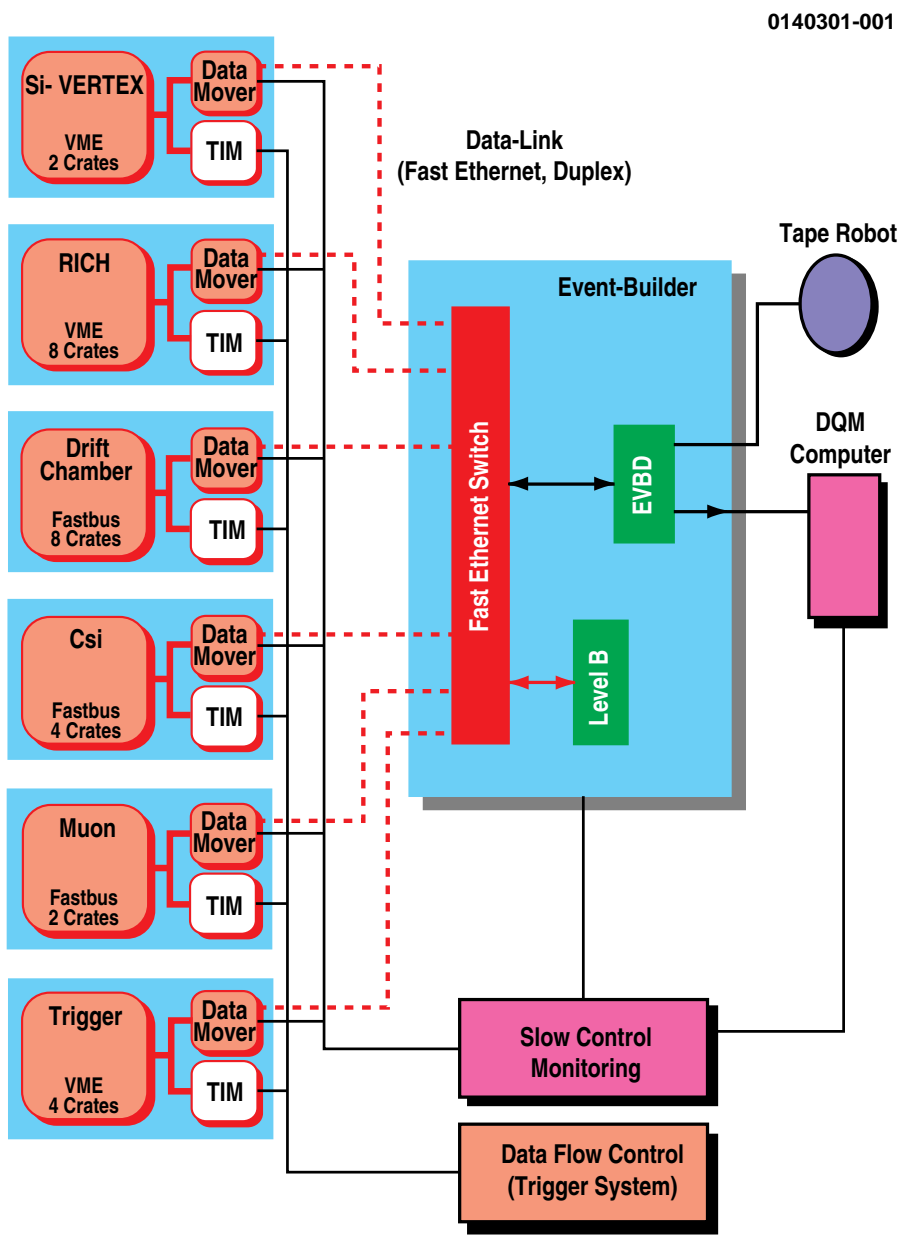}
\caption[Block diagram of the CLEO III data acquisition system.]
{Block diagram of the CLEO III data acquisition system.  The only change between 
the CLEO III and CLEO-c detector, and the corresponding change to the DAQ system, is the 
replacement of the Silicon vertex detector (Si-VERTEX) with the inner drift chamber.}
\label{fig:daq}
\end{center}
\end{figure}

This description of the data acquisition system is based on Ref. \cite{CLEOcCESRc}.  
The data acquisition (DAQ) system \cite{CLEOcCESRc} 
consists of two equally important parts.  The data collection system is responsible 
for the data transfer from the front-end electronics to the mass storage device, while the 
slow control system monitors the data quality and the detector components.  
A block diagram of the DAQ system for the CLEO III detector 
is shown in Figure \ref{fig:daq}.  The only change between the CLEO III and CLEO-c 
detector, and the corresponding change in the DAQ system, is the replacement of 
the Silicon vertex detector (Si-VERTEX) with the inner drift chamber.

For each event accepted by the trigger, approximately 400,000 detector channels have to 
be digitized.  Front-end data conversion is performed in parallel and local buffers 
on each databoard hold the data for later asynchronous readout by the DAQ.  Data 
sparsification is performed directly on the databoards.  The Data Mover, a dedicated 
module in each front-end crate, assures transfer times below 500 $\mu$s and provides 
a second buffer layer.  Using data links based on the Fast Ethernet protocol, the event 
fragments are transmitted from the crates to the Event Builder.  Completely assembled events 
are transferred to mass storage and a fraction of the data is analyzed online 
by the CLEO monitor program (Pass1) to quickly discover problems 
and to ensure the quality of the data written to tape.  

The flow of event data through the data collection system is controlled by a simple control 
protocol.  The basic philosophy is to rearm the experiment to wait for the next trigger 
only when sufficient buffer space is available to receive a new event, i.e., 
a free slot at the databoard.

Independent from the main data path, a slow control system monitors the individual 
detector components.  Run control as well as the initialization of the detector subsystems 
are also part of the slow control system.

The key parameters for the DAQ system are the trigger rate, the acceptable deadtime, 
as well as the average event size.  These quantities constrain the readout and digitization 
time in the front-end electronics as well as the data transfer bandwidth the DAQ system 
has to provide.  The performance parameters of the DAQ system are listed in Figure 
\ref{tab:daqperform}.  The readout time is defined as the time between the trigger 
signal and the end of the digitization process in the front-end electronics.  For each 
event accepted by the trigger, this causes deadtime, and hence it is desirable to 
keep the readout time as short as possible.  
The maximum readout induced detector deadtime is, on average, $<3\%$.  

\begin{table}[h]
\caption[Performance parameters of the DAQ system.]
{Performance parameters of the DAQ system.}
\medskip
\begin{center}
\begin{tabular}{|l|l|}
\hline
Name & Achieved Performance \\
\hline
Maximum Readout Rate & 150 Hz (data taking) \\    
 & 500 Hz (random test trigger) \\    
Average Eventsize & 25 kBytes \\    
Average Readout Time & 30 $\mu$s\\    
Data Transfer Bandwidth & 6 MBytes/s \\    
\hline
\end{tabular}
\label{tab:daqperform}
\end{center}
\end{table}

\baselineskip=24pt
\chapter{Analysis of Data}

\section{Data Sample}
 
Table \ref{tab:cleocdata} lists the continuum and $\psip$ data 
samples used in the present analysis.

\begin{table}[h]
\caption[Continuum and $\psi(2S)$ data samples 
collected with the CLEO-c detector.]{Continuum and $\psi(2S)$ data samples.}
\medskip
\begin{center}
\begin{tabular}{|c|c|c|}
\hline
Data sample & $\sqrt{s}$ (GeV) & $\cal{L}$ (pb$^{-1}$) \\
\hline
Continuum & 3.671 & 20.7 \\ 
$\psip$ & 3.686 & 2.89 \\    
\hline
\end{tabular}
\label{tab:cleocdata}
\end{center}
\end{table}

The luminosity for the continuum data sample is determined by  
comparing the QED processes $\Bhabha$, $\dimuon$, and 
$\gamgamprod$ to their respective calculated cross 
sections \cite{contlumcorr}.  The calculated cross sections and 
the efficiencies for the respective QED final states are determined 
by the Babayaga generator \cite{Babayaga}, which include 
corrections for the initial state radiation and, for the 
$\eetoll$ final states, interference from the J/$\psi$, $\psip$, 
and $\psi(3770)$ resonances.  The luminosity for the $\psip$ data sample 
is determined from only the $\Bhabha$ and 
$\gamgamprod$ processes due to contamination of the $\mumu$ final state from 
the $\psiptomumu$ decays \cite{psiplumcorr}.  The $\psiptomumu$ and 
$\psiptoee$ decays in the $\psip$ sample account for $(43\pm5)\%$ 
and $(2.7\pm0.1)\%$ of the total $\mumu$ and $\ee$ events observed, respectively. 
The errors in the above fractions arise from the current experimental uncertainties 
in the $\psiptomumu$ and $\psiptoee$ branching ratios \cite{PDG2004}.  
The total uncertainties in the luminosities are 1$\%$ \cite{contlumcorr} 
and 3$\%$ \cite{psiplumcorr} for the continuum data sample 
and $\psip$ data samples, respectively.  

The primary data sample for our form factor determinations is the continuum data sample.  
The $\psip$ data sample is used for two purposes: 
(i) it allows the opportunity to test and tune the selection criteria on an 
independent data sample by measuring the $\psip$ branching ratios to 
the final states of interest, and (ii) since the continuum data is only 15 
MeV below the $\psip$ resonance, it allows us to take account of the contribution of 
the tail of the $\psip$ resonance in the continuum data sample.

The number of the produced $\psip$ in the $\psip$ data sample is determined from 
the number of observed hadronic events.  
The number of produced $\psip$ is $1.52\times10^{6}$ with a systematic uncertainty 
of $3\%$ \cite{numofpsip}.

The continuum data sample needs to be corrected for the contamination from 
the tail of the $\psip$ resonance.  A contamination scale factor 
is determined by using the number of observed $\pipiJpsi$ events 
in the continuum and $\psip$ data samples \cite{psipcontamination}.  
The resulting scale factor is 
\begin{equation}
\CpipiJpsi = \frac{N^{\pipiJpsi}_{cont}}{N^{\pipiJpsi}_{\psip}} 
= \frac{221\pm15}{30518\pm175} = 0.0072\pm0.0005 
\label{eq:psipcontamcorr}
\end{equation}
where $N^{\pipiJpsi}_{cont}$ and $N^{\pipiJpsi}_{\psip}$ are the number 
of observed $\pipiJpsi$, $\JPll$ events observed in the continuum and 
$\psip$ data samples, respectively.  This scale factor is found to be in good agreement 
with the one determined from the luminosities and the evaluation of the $\psi(2S)$ 
tail \cite{psipcontamination}.

\section{Monte Carlo Samples}

In order to determine the suitable criteria for event selection and background rejection 
for the $\eetohhbar$ ($h = \pi^+,K^+,p$) processes, Monte Carlo (MC) samples 
are generated for the following processes: $\eetohhbar$ (signal MC), 
$\eetoISRJPhhbar$ (ISR $J/\psi$ MC), and $\eetoll$ (leptonic MC), 
where $l$ = $e,\mu$.  The hadronic final state MC samples 
are generated with the EvtGen generator \cite{EvtGen}, while the leptonic MC 
samples are generated with the Babayaga generator \cite{Babayaga}.  
All MC samples described above incorporate final state radiation emitted from the 
charged particles \cite{PHOTOS}.

The $\eetohhbar$ MC samples are generated with proper angular 
distributions.  The $\eetopipi$ and $\eetoKK$ MC samples are 
generated with a sin$^{2}\theta$ angular distribution, where 
$\theta$ is the angle between the charged hadron and the positron beam, 
as defined in Eqn. \ref{eq:mffdiffcs}.  Two 
different sets of $\eetoppbar$ MC samples are generated based on 
different assumptions of the proton electric form factor.  The two 
different assumptions are $\prelecff$ = 0 and $\prelecff$ = $\prmagff$.  
The proton form factors are related to the differential cross 
section for their pair production at $\sqrt{s}$ as follows
\begin{displaymath}
\frac{d\sigma_{0}(s)}{d\Omega} = \frac{\alpha^{2}}{4s}\beta_{p}
\left[\prmagff^{2}~(1 + \mathrm{cos}^{2}\theta) + 
\left(\frac{4m^{2}_{p}}{s}\right)\prelecff^{2}~(\mathrm{sin}^{2}\theta)\right].
\end{displaymath}
\begin{equation}
= \frac{\alpha^{2}}{4s}\beta_{p}\prmagff^{2}(1+\eta)
\left[1+\left(\frac{1-\eta}{1+\eta}\right)\mathrm{cos}^{2}\theta\right],
\label{eq:tlprad}
\end{equation}
where $\alpha$ is the fine-structure constant, 
$m_{p}$ is the proton mass,
$\beta_{p}$ is the proton velocity (in terms of c) in the laboratory system, 
$\prmagff$ and $\prelecff$ are the magnetic and electric form factor 
of the proton, respectively, 
and 
\begin{equation}
\eta = \frac{4m^{2}_{p}}{s}\frac{\prelecff^{2}}{\prmagff^{2}}.  
\end{equation}
At $\sqrt{s}$ = 3.671 GeV, the angular distributions are 
1 + cos$^{2}\theta$ for $\prelecff$ = 0 and \\ 1 + (0.59)cos$^{2}\theta$ 
for $\prelecff$ = $\prmagff$.

The hadronic final state MC samples consist of 20,000 generated 
events.  Each sample uses the same 10 continuum data run numbers, 
which allows for a sampling of realistic detector effects.

The leptonic MC samples simulate the number of dilepton events in the continuum 
data sample.  
The cross sections listed below are determined by the Babayaga generator 
\cite{Babayaga}.  The cross section for Bhabha events 
($\Bhabha$) at $\sqrt{s}$ = 3.671 GeV and each track having 
$|$cos$\theta|$ $<$ 0.8 is $\sigma_{Bhabha}$ = 126.50 $\pm$ 0.14 nb.  
With a total integrated luminosity of $\cal{L}$ = 20.4 pb$^{-1}$, 
a sample of 2.59$\times$10$^{6}$ Bhabha events is 
generated.  The cross section for dimuon events ($\dimuon$) at 
$\sqrt{s}$ = 3.671 GeV, with the same track $|$cos$\theta|$ requirement, 
is $\sigma_{dimuon}$ = 4.999 $\pm$ 0.015 nb.  This 
corresponds to a sample of 102,500 dimuon events.  The leptonic MC samples 
are generated using all of the continuum data run numbers, and 
the number of events in each run is weighted according to its luminosity.

In order to study other possible background sources, a generic sample of $\psip$ MC 
decays is analyzed.  The sample consists of 40,568,651 events or 
26.7$\times$ the \\ 
CLEO-c $\psip$ data sample.  The $\psiptohhbar$ branching ratios from the 
generic $\psip$ MC sample are not determined 
because the angular distributions of the $\psiptohhbar$ decays are thrown according 
to phase space and no final state radiation is incorporated.

\section{Particle Identification Definitions}

In order to discriminate between the $e$, $\mu$, $\pi$, $K$, and $p$ charged particles, 
particle identification (PID) information from the specific ionization 
($dE/dx$) measured in the main drift chamber and Cherenkov radiation information 
from the RICH detector are used to form a joint $\chi^2$ function. 
For $dE/dx$, we form a quantity $S_i$ ($i=e,\mu,\pi,K,p$), which is 
\begin{equation}
S_i = \frac{(dE/dx)_{meas} - (dE/dx)_{expected,i}}{\sigma}.
\label{eq:dEdxpull}
\end{equation}
The measured $dE/dx$ of the charged track is $(dE/dx)_{meas}$, the expected $dE/dx$ 
for particle hypothesis $i$ is $(dE/dx)_{expected,i}$, and the uncertainty in the 
$dE/dx$ measurement is $\sigma$.  
The information from the RICH detector is given in the form of a likelihood function, 
$-2{\rm log}L$. The joint $\chi^2$ function is
\begin{equation}
\Delta\chi^2 (i-j)= -2{\rm log}L_i + 2{\rm log}L_j + S_i^2 - S_j^2. 
\label{eq:RICHdEdxPID}
\end{equation}
The more negative $\Delta\chi^2$, the higher the likelihood
the particle is of type $i$ compared to type $j$. 

Information from the crystal calorimeter (CC) and the main drift chamber is also used 
for charged particle identification.  A CC shower is ``matched'' 
to a charged particle track if the shower is within 8 cm transverse 
to the position vector of the track at the front of the CC surface and 15 cm along 
the direction of the vector of the track.  The CC energy associated with a track, 
denoted by $E_{tkCC}$, is the sum of all CC crystal energies in the shower associated 
with the track.  The ratio $E_{tkCC}/p$ is also used and is defined as the ratio of the 
CC shower energy associated with the track to the momentum of the track measured 
with the inner and main drift chambers.

\section{Kinematics of Two Track Events}

\begin{figure}[htbp]
\begin{center}
\includegraphics[width=14cm]{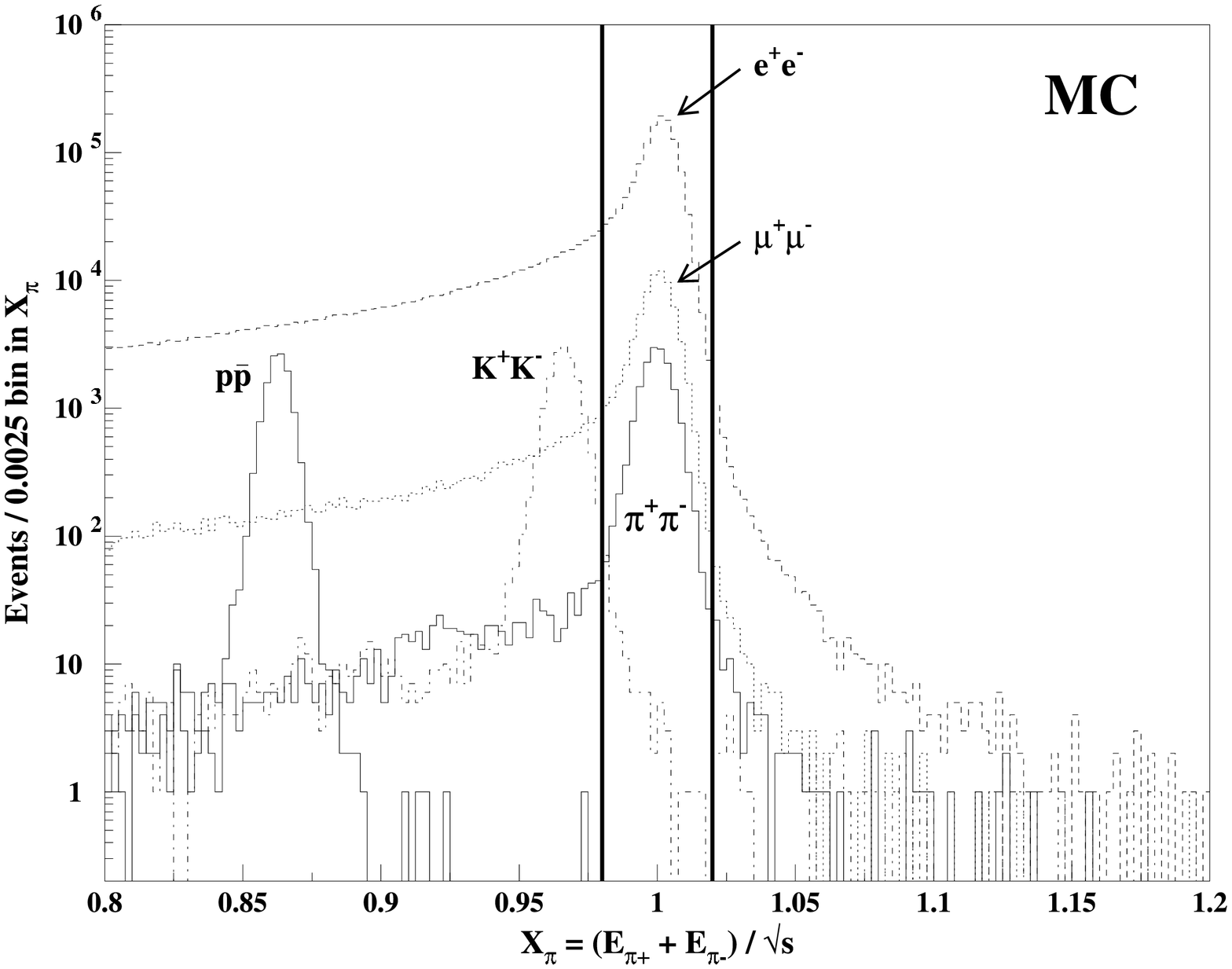}
\caption[MC $X_{\pi}$ distributions with acceptance, trigger, 
and tracking cuts applied.]
{MC $X_{\pi}$ distributions with the acceptance, trigger, and tracking criteria 
applied.  
The distributions are from $\eetopipi$ signal MC (solid histogram at $X_{\pi}$ = 1), 
$\eetoKK$ signal MC (dot-dashed), 
$\eetoppbar$ signal MC obtained with $\prelecff$ = 0 (solid at $X_{\pi}$ = 0.87), 
Bhabha MC (dashed), and dimuon MC (dotted).  The hadronic distributions 
($\pipi$, $\KK$, $\ppbar$) are arbitrary normalized while the dileptonic 
distributions ($\ee$, $\mumu$) correspond to the number of leptonic events at 
$\sqrt{s}$ = 3.671 GeV and a 20.4 pb$^{-1}$ data sample.  
The signal region is defined as 0.98 $<$ $X_{\pi}$ $<$ 1.02 and 
is enclosed between the vertical lines.}
\label{fig:pixwideinit}
\end{center}
\end{figure}

\begin{figure}[htbp]
\begin{center}
\includegraphics[width=14cm]{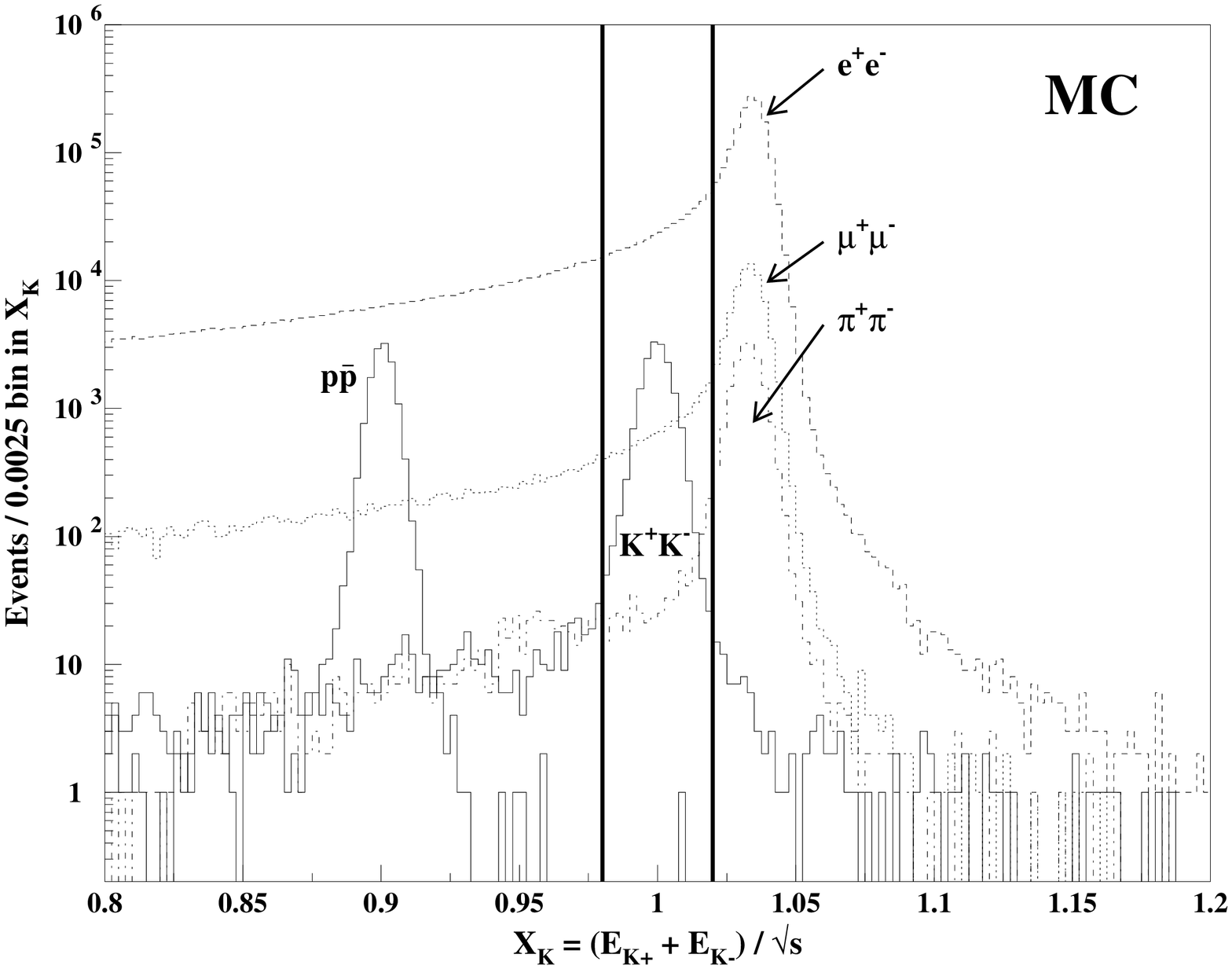}
\caption[MC $X_{K}$ distributions with acceptance, trigger, 
and tracking cuts applied.]
{MC $X_{K}$ distributions with the acceptance, trigger, and tracking criteria 
applied.  
The distributions are from $\eetopipi$ signal MC (dot-dashed), 
$\eetoKK$ signal MC  (solid histogram at $X_{K}$ = 1), 
$\eetoppbar$ signal MC obtained with $\prelecff$ = 0 (solid at $X_{K}$ = 0.90), 
Bhabha MC (dashed), and dimuon MC (dotted).  The hadronic distributions 
($\pipi$, $\KK$, $\ppbar$) are arbitrary normalized while the dileptonic 
distributions ($\ee$, $\mumu$) correspond to the number of leptonic events at 
$\sqrt{s}$ = 3.671 GeV and a 20.4 pb$^{-1}$ data sample.  
The signal region is defined as 0.98 $<$ $X_{K}$ $<$ 1.02 and 
is enclosed between the vertical lines.}
\label{fig:kxwideinit}
\end{center}
\end{figure}

\begin{figure}[htbp]
\begin{center}
\includegraphics[width=14cm]{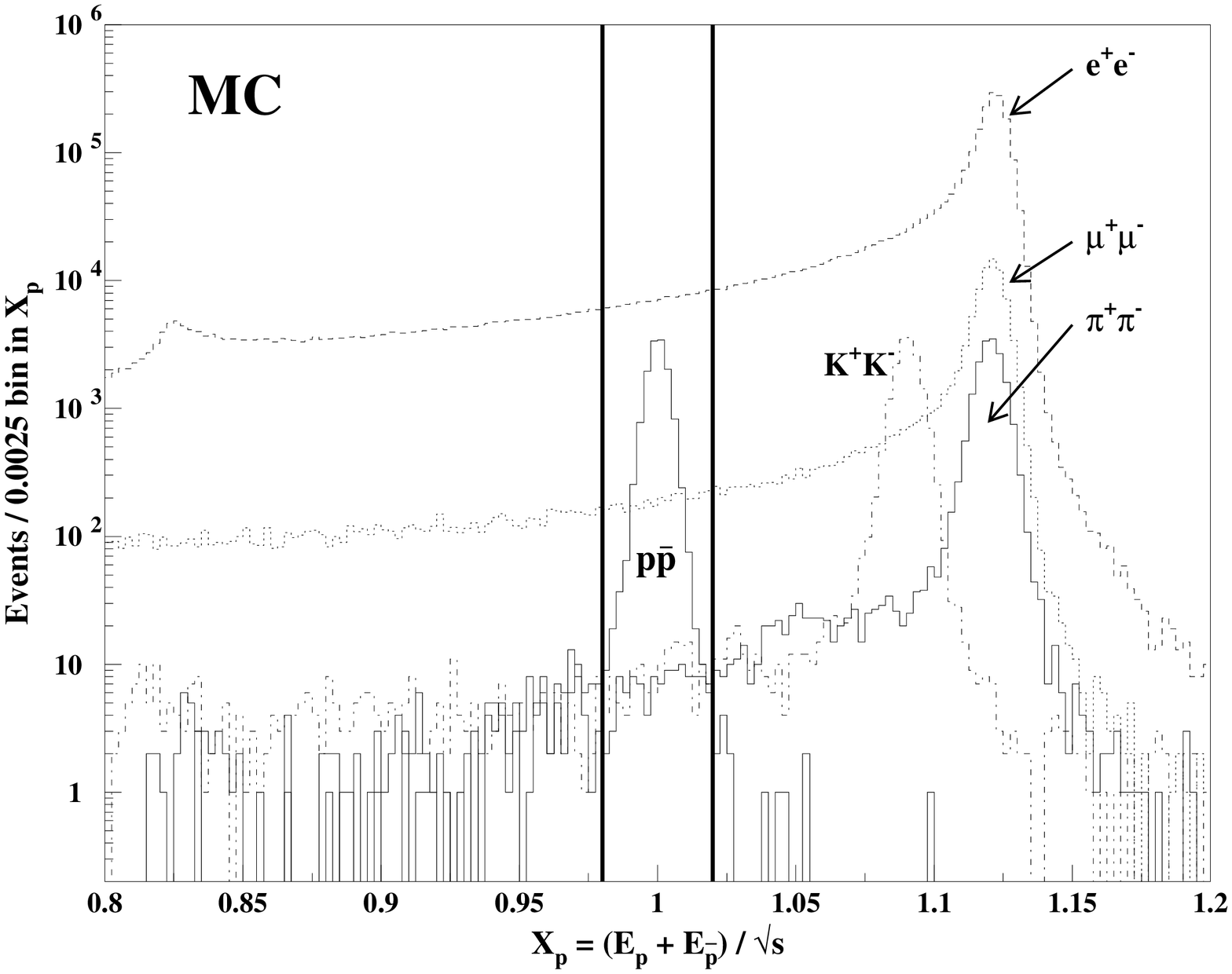}
\caption[MC $X_{p}$ distributions with acceptance, trigger, 
and tracking cuts applied.]
{MC $X_{p}$ distributions with the acceptance, trigger, and tracking criteria 
applied.  
The distributions are from $\eetopipi$ signal MC (solid histogram at $X_{p}$ = 1.12), 
$\eetoKK$ signal MC (dot-dashed), 
$\eetoppbar$ signal MC obtained with $\prelecff$ = 0 (solid at $X_{p}$ = 1), 
Bhabha MC (dashed), and dimuon MC (dotted).  The hadronic distributions 
($\pipi$, $\KK$, $\ppbar$) are arbitrary normalized while the dileptonic 
distributions ($\ee$, $\mumu$) correspond to the number of leptonic events at 
$\sqrt{s}$ = 3.671 GeV and a 20.4 pb$^{-1}$ data sample.  
The signal region is defined as 0.98 $<$ $X_{p}$ $<$ 1.02 and 
is enclosed between the vertical lines.}
\label{fig:prxwideinit}
\end{center}
\end{figure}

Figures \ref{fig:pixwideinit}, \ref{fig:kxwideinit}, and \ref{fig:prxwideinit} 
show the MC distributions for direct production of the two track final states which 
satisfy the acceptance, trigger, and tracking criteria.  The 
variable $X_{h}$ is the total energy 
of the two tracks, assuming the particle hypothesis of interest for each 
track, normalized to $\sqrt{s}$.  The signal regions are defined as 
0.98 $<$ $X_{h}$ $<$ 1.02.  An event passes the acceptance criteria when it has  

\begin{itemize}
  \item{Number of tracks ($N_{trk}$) = 2}
  \item{Net charge ($\Sigma Q$) = 0}
  \item{Each track with $|$cos$\theta$$|$ $<$ 0.8 }
\end{itemize}
The trigger criteria are given in Table \ref{tab:triggerlines}.  
A track passes the tracking criteria when it has 
\begin{itemize}
  \item{$|d_{b}|$ $<$ 5 mm, and $|z_{b}|$ $<$ 5 cm (IP)}
  \item{0.5 $<$ DRHF $<$ 1.2, and $\chi^{2}/dof$ $<$ 10 (Track Quality).}
\end{itemize} 
The interaction point (IP) variables $d_{b}$ and $z_{b}$ are the distances 
between the origin of the helix fit and the position of the $e^+e^-$ 
annihilation in the plane perpendicular to and 
along the axis defined by the positron beam, respectively.  
DRHF is defined as the number of inner and main drift chamber wire ``hits'' observed 
vs. the number of ``hits'' expected from the helix fit.  The reduced 
$\chi^{2}/dof$ is the fit confidence normalized by the number of degrees of freedom 
of the track helix fix.

The $X_{K}$ and $X_{p}$ distributions in Figures \ref{fig:kxwideinit} and 
\ref{fig:prxwideinit}, respectively, show that the 
$\eetoKK$ and $\eetoppbar$ signal regions are sufficiently displaced 
from the dominant $\eetoll$ background region.  
This is not the case for $\eetopipi$ events, shown in Figure \ref{fig:pixwideinit}.  
Since a pion and muon cannot be separated from each other 
using only standard $dE/dx$ or RICH information when they have the expected momenta 
of $\sim$1.83 GeV/$c$, 
muons are rejected in the $\pipi$ final state analysis based on additional information 
from the CC; the criteria are described in Section 4.5.1. 

\begin{figure}[htbp]
\begin{center}
\includegraphics[width=15cm]{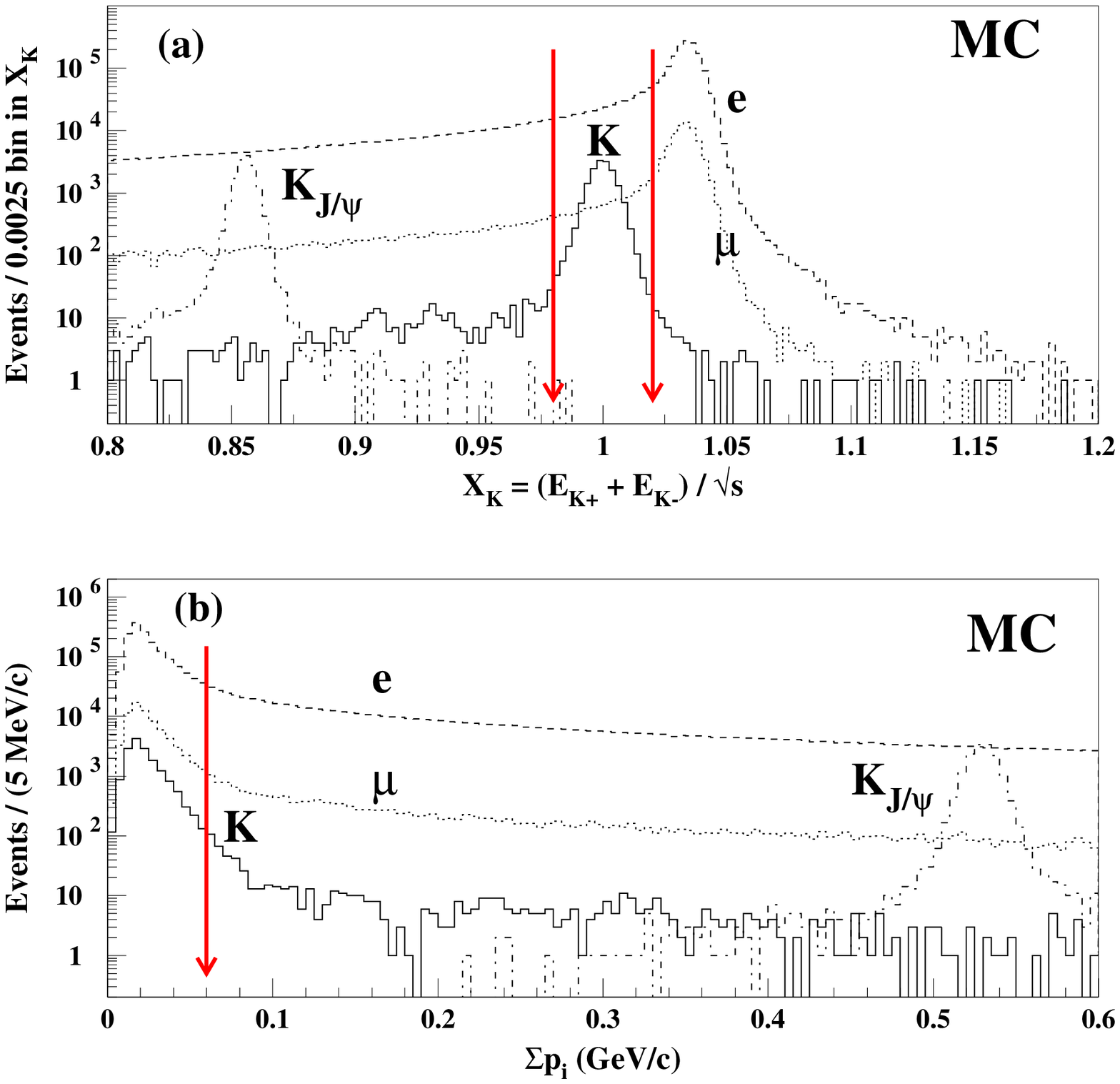}
\caption[MC $X_{K}$ and net momentum distributions with 
acceptance, trigger, and tracking cuts applied.]
{MC $X_{K}$ and net momentum ($\Sigma p_{i}$) distributions with 
acceptance, trigger, and tracking cuts applied.  
The solid histogram is $\eetoKK$ signal MC, the dot-dashed histogram 
is $\eetoISRJPKK$ MC, the dashed histogram is Bhabha MC, 
and the dotted histogram is dimuon MC.  
The signal region is defined as 0.98 $<$ $X_{K}$ $<$ 1.02, designated by 
the arrows in Figure (a), and contains a very small contribution from $\eetoISRJPKK$.  
A cut of $\Sigma p_{i} <$ 60 MeV/$c$ is applied to remove contributions 
from $\eetoISRJPKK$.  
The $\eetoppbar$ and $\eetopipi$ processes have the same characteristics.}
\label{fig:kxwideisrjp}
\end{center}
\end{figure}

Initial state radiation production of $J/\psi$, followed by 
its decay to two charged particles, is an important background to consider.  
Figure \ref{fig:kxwideisrjp} shows the 
$\KK$ signal (solid histogram), ISR $J/\psi \rightarrow \KK$ (dot-dashed), 
Bhabha (dashed), and dimuon (dotted) MC distributions for $X_{K}$ and the 
net momentum of the two tracks.  A cut on the net momentum ($\Sigma p_{i}$) 
of $<$ 60 MeV/$c$ completely removes $\eetoISRJPKK$ events from the $\KK$ final state 
signal region.  The $\eetoppbar$ and $\eetopipi$ 
processes have very similar $X_{h}$ and net momentum characteristics, and similar cuts 
are used.

\section{Event Selection $\&$ Backgrounds}

\subsection{Selection of $\pipi$ Events}

Additional acceptance restrictions are applied to $\pipi$ final state events.  
They are
\begin{itemize}
  \item{Each track must have $|$cos$\theta$$|$ $<$ 0.75 
        (changed from 0.8)}
  \item{Each track must have an associated CC shower.}
\end{itemize} 
These extra requirements are necessary because the RICH detector endplates 
cover the barrel section of the CC for $|$cos$\theta$$|$ $>$ 0.75 and, 
since CC information is very important for the $\pipi$ analysis, 
the pion track must have an associated shower in the CC.

An additional restriction of $\Sigma p_{i} <$ 100 MeV/$c$ is used to 
select $\pipi$ events.  This cut removes possible contamination 
from $J/\psi \rightarrow \pipi$ decays, 
as it did for the $\KK$ final state events 
shown in Figure \ref{fig:kxwideisrjp}b.   

The most important issue for the $\pipi$ analysis is how to distinguish 
a charged pion track from a muon when they have comparable momenta of 
$\sim$1.83 GeV/$c$.  The track momenta are too high for 
using $dE/dx$ and RICH information and too low for the 
muon detector.  The solution is to use the CC to 
distinguish pion tracks which interact hadronically, from muons which do not.   

The muon rejection selection criterion is arbitrarily defined by 
requiring that there be $<$ 0.1 $\mumu$ final state events 
in the $\pipi$ signal region, either from $\dimuon$, or $\psiptomumu$.  
The dimuon MC sample gives 55,361$\pm$235 $\dimuon$ events in the $\pipi$ signal region 
after satisfying the acceptance, trigger, tracking criteria and 
$\Sigma p_{i} <$ 100 MeV/$c$.  Note that the number 
of $\mumu$ events (consisting of both $\dimuon$ and $\psiptomumu$ events) 
in the $\psip$ data sample is expected to be nearly half, or 26,189$\pm$1234. 
The lepton track fake rate efficiency is related to the total 
number of $\leppair$ events by 
\begin{equation}
\epsilon_{l} = \sqrt{\frac{N_{l,bg}}{N_{l,total}}}.
\label{eq:effbgeq}
\end{equation}
For example, a contamination of 0.1 $\mumu$ events in the continuum $\pipi$ signal 
region would require an efficiency of 
$\epsilon_{\mu} = \sqrt{(0.1)/(55,361)} = 1.3\times10^{-3}$. 

The muon detector cannot reject the dimuon background at the desired level of 0.1 events.  
The efficiency of the muon detector for muon tracks with momenta in the range 1.5-2.0 
GeV/$c$ is $\sim$85$\%$ \cite{MuonDet}.  That corresponds to a muon track fake rate 
efficiency of $\epsilon_{\mu}$ = 15$\%$ or, using Eqn. \ref{eq:effbgeq}, 
1245 dimuon events.  It is for this reason that 
more drastic muon rejection is required.

\begin{figure}[htbp]
\begin{center}
\includegraphics[width=12cm]{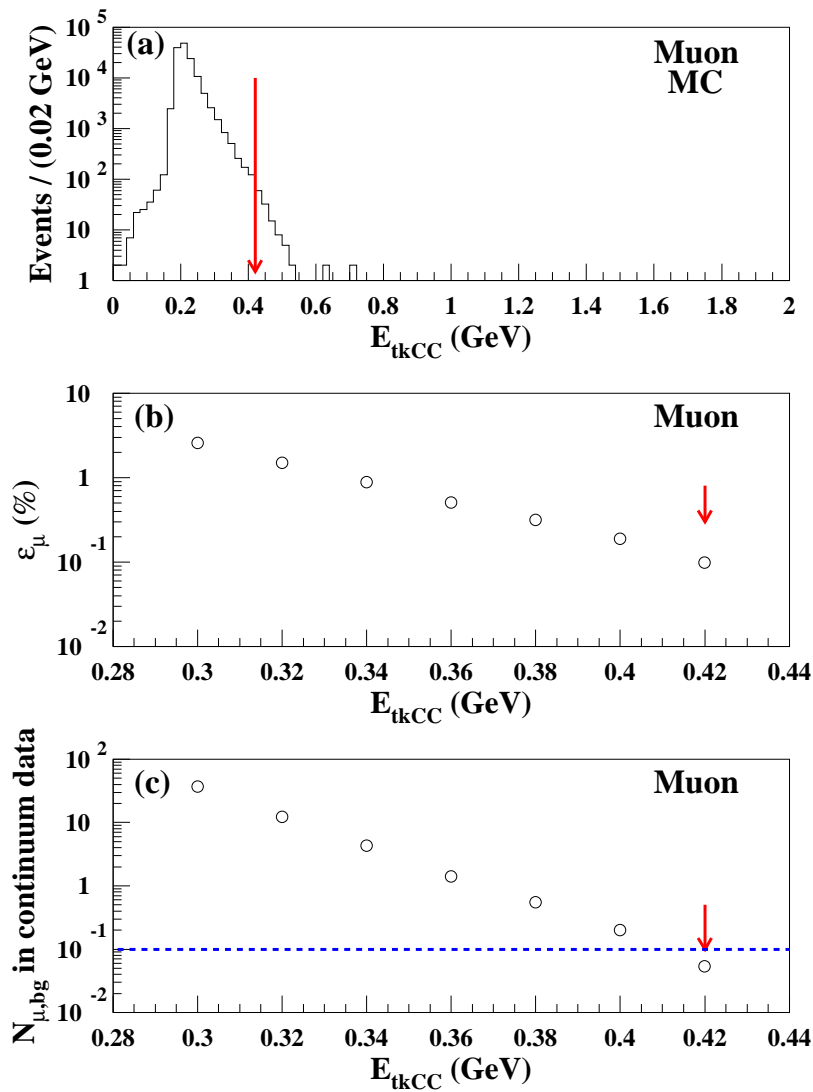}
\caption[CC shower energy distributions for $\mu$ tracks.]
{Figure (a): Shower energy deposited in the CC associated 
with a muon track from the dimuon MC sample.  
Figure (b): Muon track fake rate efficiency as a function of $E_{tkCC}$.  
Figure (c): Number of expected dimuon background events 
in the continuum data sample as a function of $E_{tkCC}$.  The 
values on the abscissa of Figures (b) and (c) imply a cut 
selecting a candidate track with $E_{tkCC}$ greater than the value. 
Requiring the track to have $E_{tkCC}$ $>$ 420 MeV (denoted by the 
arrows) only allows 
0.054$\pm$0.009 dimuon events in the continuum $\pipi$ signal region.}  
\label{fig:mueffbg}
\end{center}
\end{figure}

The muon track fake rate efficiency is determined using the dimuon MC sample.  The 
positive and negative muon tracks are analyzed individually and the efficiencies 
are combined to determine the net efficiency.  A candidate track satisfies the 
acceptance, trigger, and tracking criteria.  The energy deposited in the 
CC from a muon track is shown in Figure \ref{fig:mueffbg}a.  Requiring a muon to 
have $E_{tkCC}$ $>$ 420 MeV has an efficiency of 
$\epsilon_{\mu}$ = (0.099$\pm$0.009)$\%$, which corresponds to 
0.054$\pm$0.009 $\dimuon$ events in the continuum data $\pipi$ signal region.  This 
is a factor two better than our goal.

The behavior of hadronically-interacting pions in the CC is studied 
using data and MC samples with pion tracks having momenta of 
$\sim$1.83 GeV.  Since there is no such sample of pion tracks taken 
with the CLEO-c detector, 
inclusive $\DtoKpi$ decays (charge conjugation is assumed) 
collected with the CLEO III detector are studied.  
The only differences between the CLEO-c and CLEO III detectors are that 
the inner drift chamber replaced a four-layer Silicon vertex detector 
and the solenoidal magnetic field was lowered to 1.0 Tesla from 1.5 Tesla.  
Since the main drift chamber, RICH detector, and CC are the same in 
both detectors, the properties of the pions studied 
using the CLEO III detector are consistent with those 
in the CLEO-c detector.  More information on the CLEO III detector 
can be found in Ref. \cite{CLEOIIIDetector}.  

The sample of inclusive $\DtoKpi$ decays are from a $\sim$3.3 fb$^{-1}$ data sample 
taken at the $\Upsilon(4S)$ resonance ($\sqrt{s}$ $\sim$ 10.58 GeV).  
The $\DtoKpi$ MC samples, which corresponds to four times the $\Upsilon(4S)$ data sample, 
are from $\Upsilon(4S)$ $\rightarrow$ $B\overline{B}$ decays and 
$\ee$ $\rightarrow$ $q\overline{q}$ events ($q$ = $u,d,s,c$).  
The $B$ mesons are decayed according to their branching fractions 
listed in the PDG \cite{PDG2004}.  The $\ee$ $\rightarrow$ $q\overline{q}$ events 
produce hadrons according to the string fragmentation functions 
in the LUND/JETSET model \cite{LUNDJetset1,LUNDJetset2}. 

In order to get a clean sample of $\DtoKpi$ decays with lower background levels, 
the kaon candidate track is required to have $E_{tkCC}/p$ $<$ 0.85, 
$dE/dx$ $|S_K|$ $<$ 3 (see Eqn. \ref{eq:dEdxpull}), 
no associated signal in the muon detector, 
and $\Delta\chi^{2}(K-\pi)$ $<$ 0 (see Eqn. \ref{eq:RICHdEdxPID}).  The pion candidate 
tracks in this CC study are only required to satisfy the tracking criteria.  

\begin{figure}[htbp]
\begin{center}
\includegraphics[width=15cm]{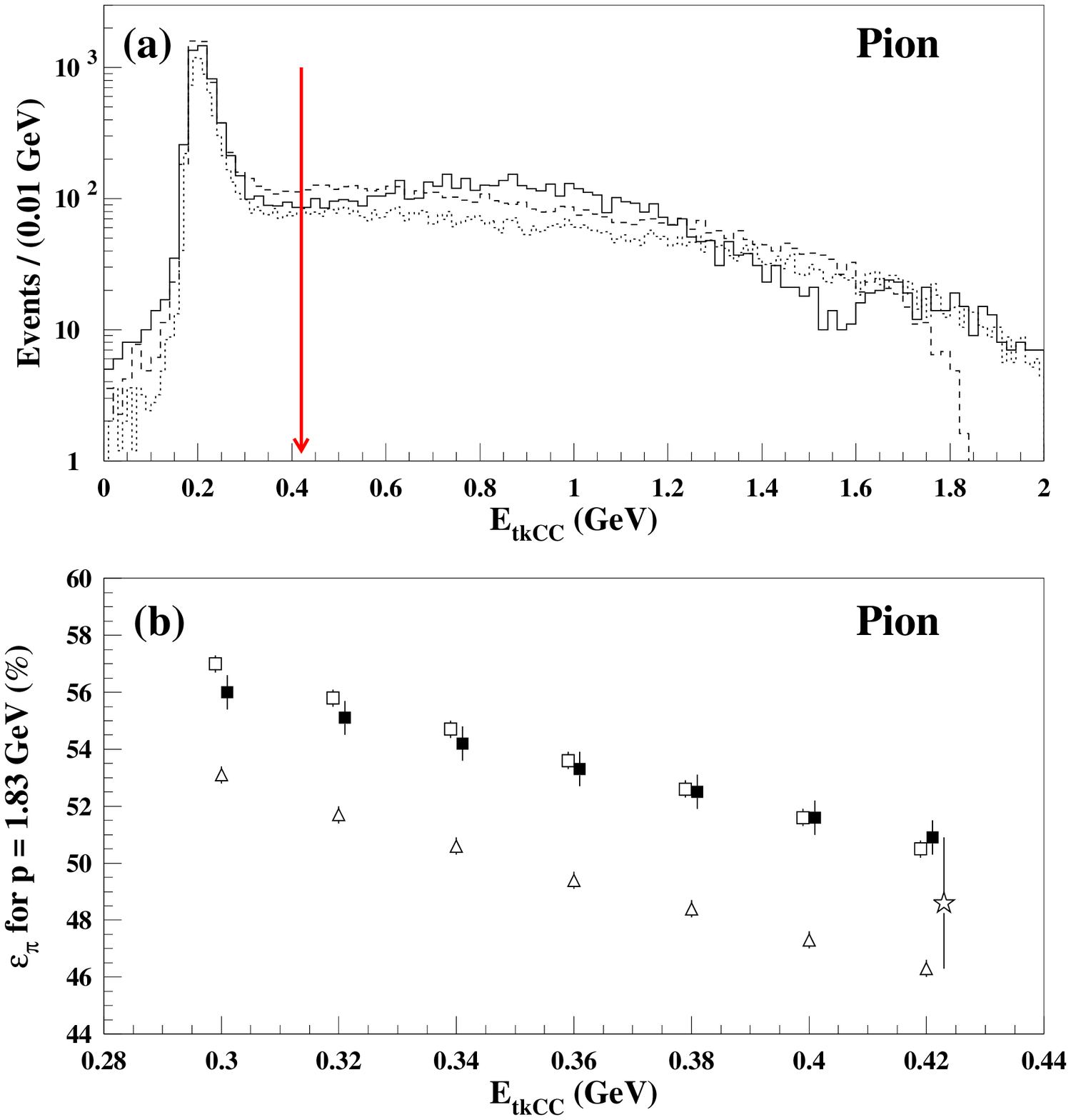}
\caption[CC shower energy distributions for $\pi$ tracks.]
{Figure (a): Pion $E_{tkCC}$ distributions from CLEO III $\DtoKpi$ 
data events (solid histogram), CLEO III $\DtoKpi$ MC events (dotted), 
and $\eetopipi$ signal MC (dashed).  The arrow denotes the $E_{tkCC}$ $>$ 420 MeV cut.  
Figure (b): Pion efficiencies as a function of $E_{tkCC}$ from 
CLEO III $\DtoKpi$ data events (filled squares), 
CLEO III $\DtoKpi$ MC events (open squares), and 
signal MC (open triangles).  The efficiency used in the present analysis 
is the average between the efficiencies 
from the CLEO III $\DtoKpi$ data events ($\blacksquare$)
and the signal MC($\vartriangle$), with the difference between them 
assigned to systematic uncertainty, as shown by the open star.  The star is displaced to 
clearly show its uncertainty.}  
\label{fig:pieff}
\end{center}
\end{figure}

Figure \ref{fig:pieff} shows the pion $E_{tkCC}$ behavior from 
CLEO III data (solid curve, filled squares) and 
MC (dotted curve, open squares) $\DtoKpi$ events.  
They are consistent for pions with momenta
of $\sim$1.83 GeV.  Figure \ref{fig:pieff} also shows the 
$E_{tkCC}$ behavior for pions from the $\eetopipi$ signal MC sample (dashed curve), 
with the efficiency (open triangles) determined in the same manner 
as for the muon fake rate efficiency described above.  The pion efficiency from the 
CLEO III $\DtoKpi$ sample is found to be $\sim$10$\%$ larger 
than that from the signal MC.  The average of these two efficiency 
estimates is used in the present analysis for pions having 
$E_{tkCC}$ $>$ 420 MeV; the difference between them is assigned to 
the systematic uncertainty.

The $\ee$ final state events have unique characteristics which 
allow for easier rejection, but the sheer number of 
$\ee$ events is formidable.  The Bhabha 
MC sample consists of 1,164,559$\pm$1079 $\ee$ events in the $\pipi$ signal region 
after satisfying the acceptance, trigger, and tracking criteria and 
$\Sigma p_{i} <$ 100 MeV/$c$ (there are 162,953$\pm$264 
events in the $\psip$ data sample from both $\Bhabha$ and $\psiptoee$ 
events).  
This requires an electron track fake rate efficiency of $\epsilon_{e}$ $<$ 9$\times10^{-4}$ 
for $<$ 1 $\ee$ event to contaminate the continuum $\pipi$ signal region.  
The easiest way to reject an electron track 
is require it to have an $E_{tkCC}/p$ $<$ 0.85.  
This is normally a sufficient cut, but is not adequate for the present analysis.  
The $E_{tkCC}/p$ $<$ 0.85 cut fails to remove Bhabha events when 
(a) the track goes between two crystals, or 
(b) 'hot' or defective crystals are included in the CC shower.  
Additional rejection of $\ee$ events is performed by using information 
from the $dE/dx$ and the RICH detector.

\begin{figure}[htbp]
\begin{center}
\includegraphics[width=13.9cm]{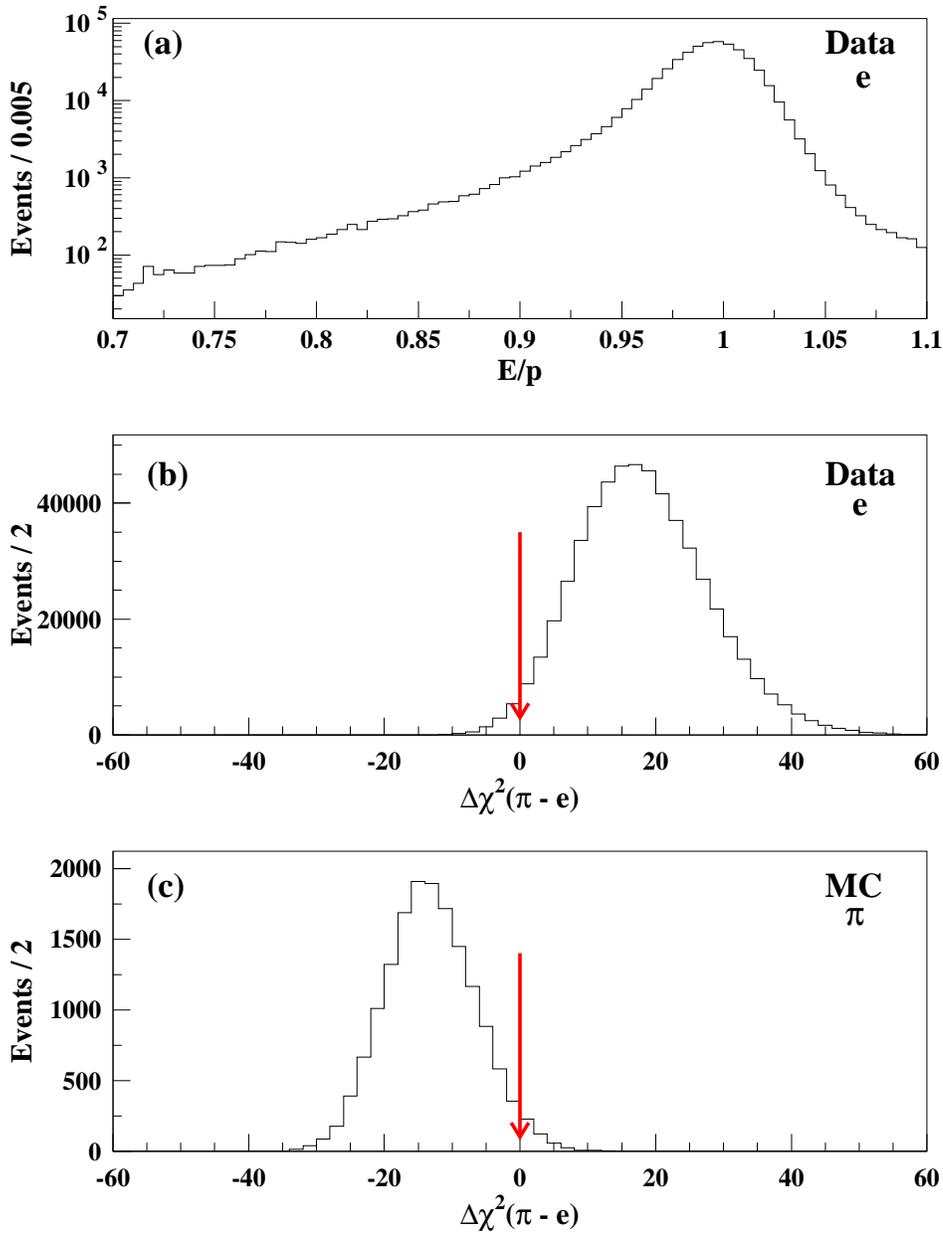}
\caption[Distributions from the continuum data radiative Bhabha 
and $\eetopipi$ signal MC samples.]
{Distributions from the continuum data radiative Bhabha 
and $\eetopipi$ signal MC samples.  
Figure (a): $E_{tkCC}/p$ for the candidate track from the radiative Bhabha sample.  
Figure (b): $\Delta\chi^2(\pi-e)$ for the candidate track from the radiative 
Bhabha sample. 
Figure (c): $\Delta\chi^2(\pi-e)$ for the positive track from the 
$\eetopipi$ signal MC.}
\label{fig:ebg}
\end{center}
\end{figure}

To study the rejection of electrons, radiative Bhabha events in 
the continuum data sample are used.  We require that a candidate radiative Bhabha event 
should have $X_{\pi}$ $<$ 0.975, i.e., below the $\pipi$ signal region.  The 
candidate electron track must have $|$cos$\theta|$ $<$ 0.75, satisfy the track 
quality and IP criteria, $p$ $>$ 1.6 GeV, and 0.7 $<$ $E_{tkCC}/p$ $<$ 1.1, while 
the other track must have 0.85 $<$ $E_{tkCC}/p$ $<$ 1.1, 
and $dE/dx$ $S_{e}$ $>$ -2  (see Eqn. \ref{eq:dEdxpull} for definition of $S_{e}$).  
Figures \ref{fig:ebg}a and Figures \ref{fig:ebg}b show 
the $E_{tkCC}/p$ and $\Delta\chi^2(\pi-e)$ 
distributions for the candidate Bhabha track, respectively.  
Figure \ref{fig:ebg}c shows the $\Delta\chi^2(\pi-e)$ 
distribution for the positive track from the $\eetopipi$ signal MC sample.  
A total number of 556,902 radiative Bhabha events survive the above criteria.
Of these events, 4282$\pm$65 candidate electron tracks 
have an $E_{tkCC}/p$ between 0.7 and 0.85, or $\epsilon_{e}$ = 
(7.69$\pm$0.12)$\times10^{-3}$.  This fake rate efficiency corresponds 
to $\sim$69 $\ee$ events in the $\pipi$ signal region for the 
continuum data sample.  The additional requirement of $\Delta\chi^2(\pi-e) <$ 0
reduces the number of candidate electron tracks to 120$\pm$11 with 
$\epsilon_{e}$ = (2.15$\pm$0.20)$\times10^{-4}$, which gives 
0.054$\pm$0.010 $\ee$ events in the continuum data $\pipi$ signal region.  

By requiring each track in an event to have $E_{tkCC}$ $>$ 420 MeV, 
$E_{tkCC}/p$ $<$ 0.85, and $\Delta\chi^2(\pi-e)$ $<$ 0,  the number of 
$\leppair$ background events in the continuum and $\psip$ data samples are 
$N^{cont}_{\leppair}$ = 0.108$\pm$0.013 and 
$N^{\psip}_{\leppair}$ = 0.021$\pm$0.003, respectively.

From Figure \ref{fig:pixwideinit}, we note that a small number of $\KK$ final state 
events can enter the $\pipi$ signal region.  
From the $\eetoKK$ signal MC, 0.73$\%$ of all $\KK$ events which satisfy the 
acceptance, trigger, and tracking criteria populate the $\pipi$ signal region.  
As will be shown in the Results section (Section 4.8.5), 
approximately 100 observed $\KK$ final state events are found in both 
$\psip$ and continuum data samples.  
Therefore, $\sim$0.73 events in the $\pipi$ signal region 
can be from charged kaon pairs.  
The efficiency for kaon-faking-pion with $\Delta\chi^2(\pi-K)$ $<$ 0 is determined 
to be 2.5$\%$ from the $\eetoKK$ signal MC.  
The $\Delta\chi^2(\pi-K)$ cut reduces the $\KK$ background to 
(0.73)(0.025)$^{2}$ = 0.0005 events and can be safely neglected. 

Table \ref{tab:pipieff} lists the efficiencies for the $\pipi$ 
final state criteria.  The signal MC samples 
are used to determine the efficiency except for the efficiency associated 
with the $E_{tkCC}$ criterion.  
The efficiency for the track $E_{tkCC}$ is determined from an average of the 
$\eetopipi$ signal MC and the CLEO III $\DtoKpi$ data samples.  
The total efficiency for the $\pipi$ final state is $\epsilon_{tot}$ = 0.166 $\pm$ 0.013.  

\begin{table}[h]
\caption[Efficiencies for the $\pipi$ final state.]
{Efficiencies for the $\pipi$ final state.  The individual efficiencies (1-7) are 
determined with respect to the acceptance cuts.}
\begin{center}
\begin{tabular}{|l|c|c|}
\hline
Cuts & Requirement & $\epsilon$ \\    
\hline
\hline
Acceptance & N$_{trk}$ = 2 & \\   
& $\Sigma Q$ = 0  & 0.861 \\   
& $|$cos($\theta_{tk}$)$|$ $<$ 0.75 & \\   
\hline
1. Trigger & L1 Trigger = TRUE & 0.997 \\   
\hline
2. IP & $|d_{b}|$ $<$ 5 mm & 0.999 \\    
& $|z_{b}|$ $<$ 5 cm & \\    
\hline
3. Track Quality & 0.5 $<$ DRHF $<$ 1.2 & 0.9997 \\    
& $\chi^{2}/dof$ $<$ 10 & \\    
\hline
4. $X_{\pi}$ = ($E_{\pi^+}$+$E_{\pi^-}$)/$\sqrt{s}$ 
& 0.98 $<$ $X_{\pi}$ $<$ 1.02 & 0.940 \\    
\hline
5. Net Momentum & $\Sigma p_{i}$ $<$ 100 MeV/$c$ & 0.947 \\
\hline
6. $\pi-K$ Separation & $\Delta\chi^2(\pi - K)$ $<$ 0 & 0.966 \\
\hline
7. $\pi-e$ Separation & $E_{tkCC}/p$ $<$ 0.85 & 0.957 \\
& $\Delta\chi^2(\pi - e)$ $<$ 0 & 0.945 \\
\hline
$\epsilon_{MC}$ & Acc + Cuts 1-7 & 0.702(3) \\
\hline
\hline
$\pi-\mu$ Separation: $\epsilon_{\pi}(E_{tkCC})$ 
& $E_{tkCC}$ $>$ 420 MeV & 0.486(3) \\
\hline	
\hline	
$\epsilon_{tot}$ & $\epsilon_{MC}*(\epsilon_{\pi}(E_{tkCC}))^{2}$ 
& 0.166(2) \\
\hline
\end{tabular}
\label{tab:pipieff}
\end{center}
\end{table}

The generic $\psip$ MC sample is analyzed to test for other 
backgrounds which satisfy the $\pipi$ final state criteria.  
In the 40,568,651 generic $\psip$ decays, there are 3436 $\psiptopipi$ decays.  
After applying the selection criteria, only one background event is found in the 
$\pipi$ signal region.  
The one background event is found to be from a $\psiptomumu$ decay.  
This is consistent with the dimuon rejection determined above.  
The $\psiptopipi$ branching ratio is not determined from this MC sample 
because of the issues discussed at the end of Section 4.2.

\begin{figure}[htbp]
\begin{center}
\includegraphics[width=15cm]{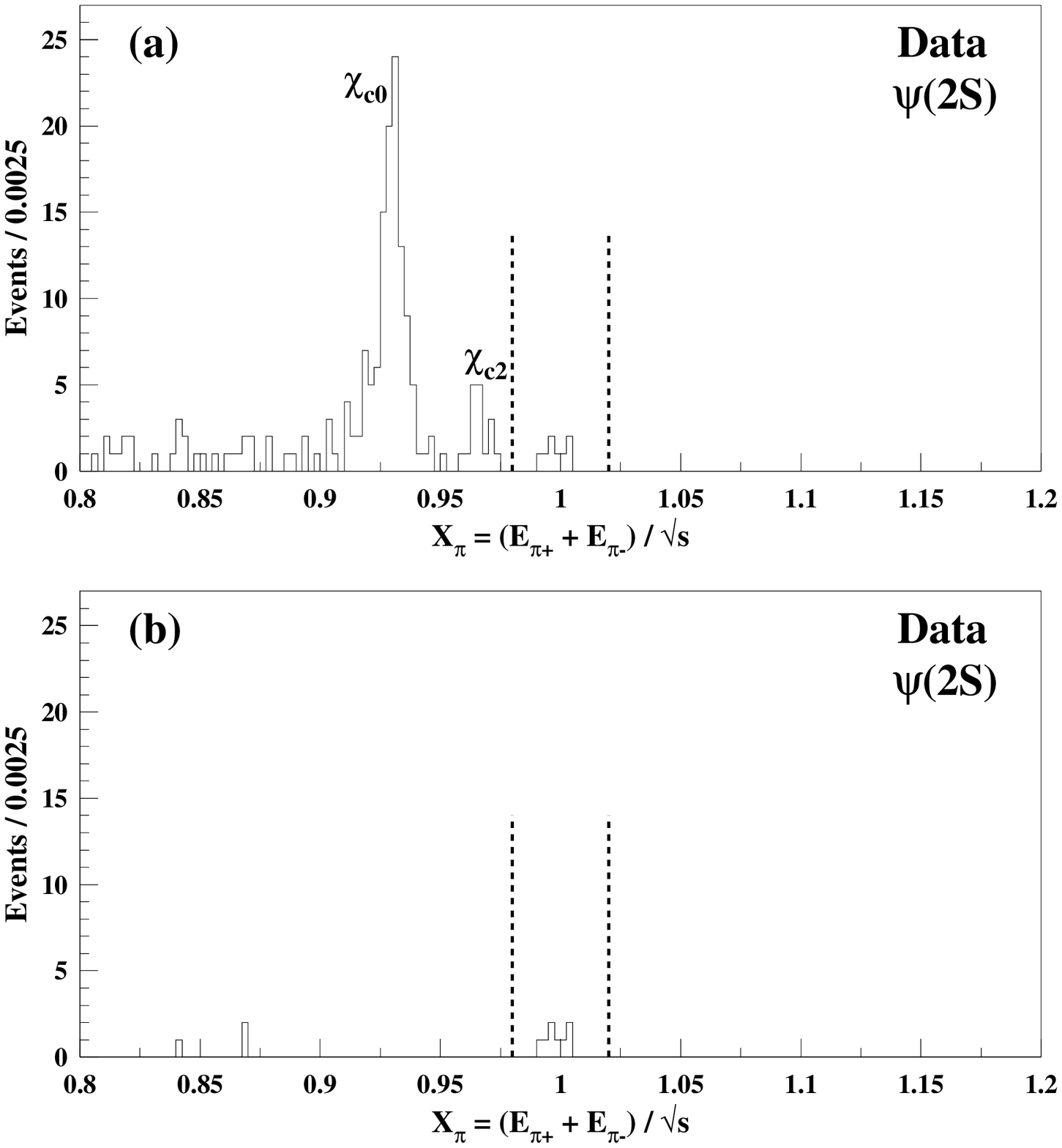}
\caption[Effect of net momentum criterion on the $\psip$ data for the $\pipi$ final state.]
{Effect of net momentum criterion on the $\psip$ data for the $\pipi$ final state.  
Figure (a) shows the $\psip$ data after applying all $\pipi$ final state criteria 
except for the net momentum criterion, while Figure (b) includes it.  The 
$\chi_{c0}$ and $\chi_{c2}$ are clearly seen in Figure (a) at $X_{\pi}$ = 
0.926 and 0.965, respectively, outside of the 0.98 $<$ $X_{\pi}$ $<$ 1.02 
signal region, and are removed in Figure (b) due to the net momentum criterion.}
\label{fig:pi2snetp}
\end{center}
\end{figure}

\begin{figure}[htbp]
\begin{center}
\includegraphics[width=15cm]{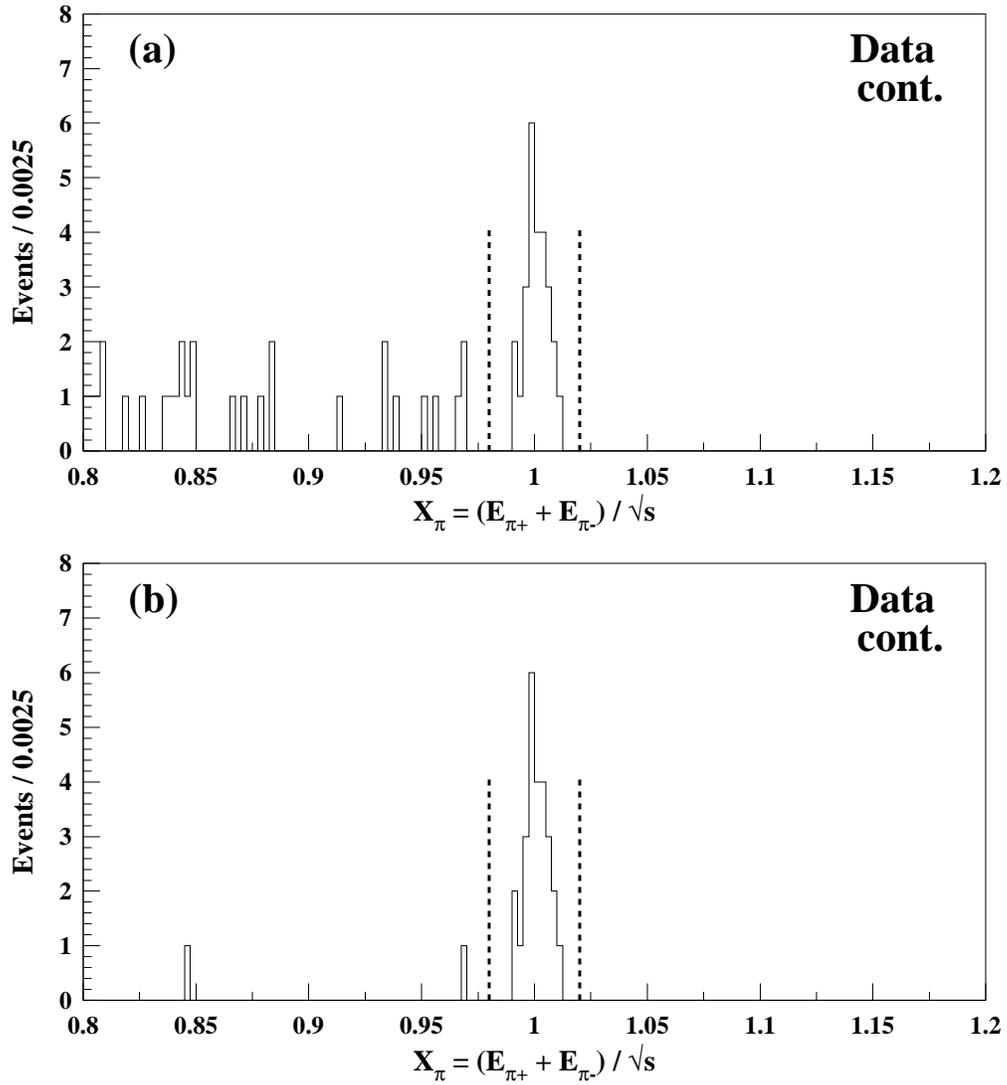}
\caption[Effect of net momentum criterion on the continuum data 
for the $\pipi$ final state.]
{Effect of net momentum criterion on the continuum data for the $\pipi$ final state.  
Figure (a) shows the continuum data after applying all $\pipi$ final state criteria 
except for the net momentum criterion, while Figure (b) includes it.  The 
net momentum criterion removes events below the signal region.}
\label{fig:pictnetp}
\end{center}
\end{figure}

Figures \ref{fig:pi2snetp}a and \ref{fig:pictnetp}a show the $X_{\pi}$ distributions 
for the $\psip$ and continuum data, respectively, after applying 
all $\pipi$ final state criteria except for the net momentum criterion.  
The P-wave spin triplet charmonium 
resonances $\chi_{c0}$ and $\chi_{c2}$ are seen in the $\psip$ data, 
but they are outside of the $X_{\pi}$ signal region. Figure \ref{fig:pi2snetp}b 
shows that only $\psiptopipi$ events survive after applying the net momentum criterion, 
and Figure \ref{fig:pictnetp}b shows that only $\eetopipi$ events 
survive after applying the net momentum criterion.

\begin{figure}[htbp]
\begin{center}
\includegraphics[width=12.8cm]{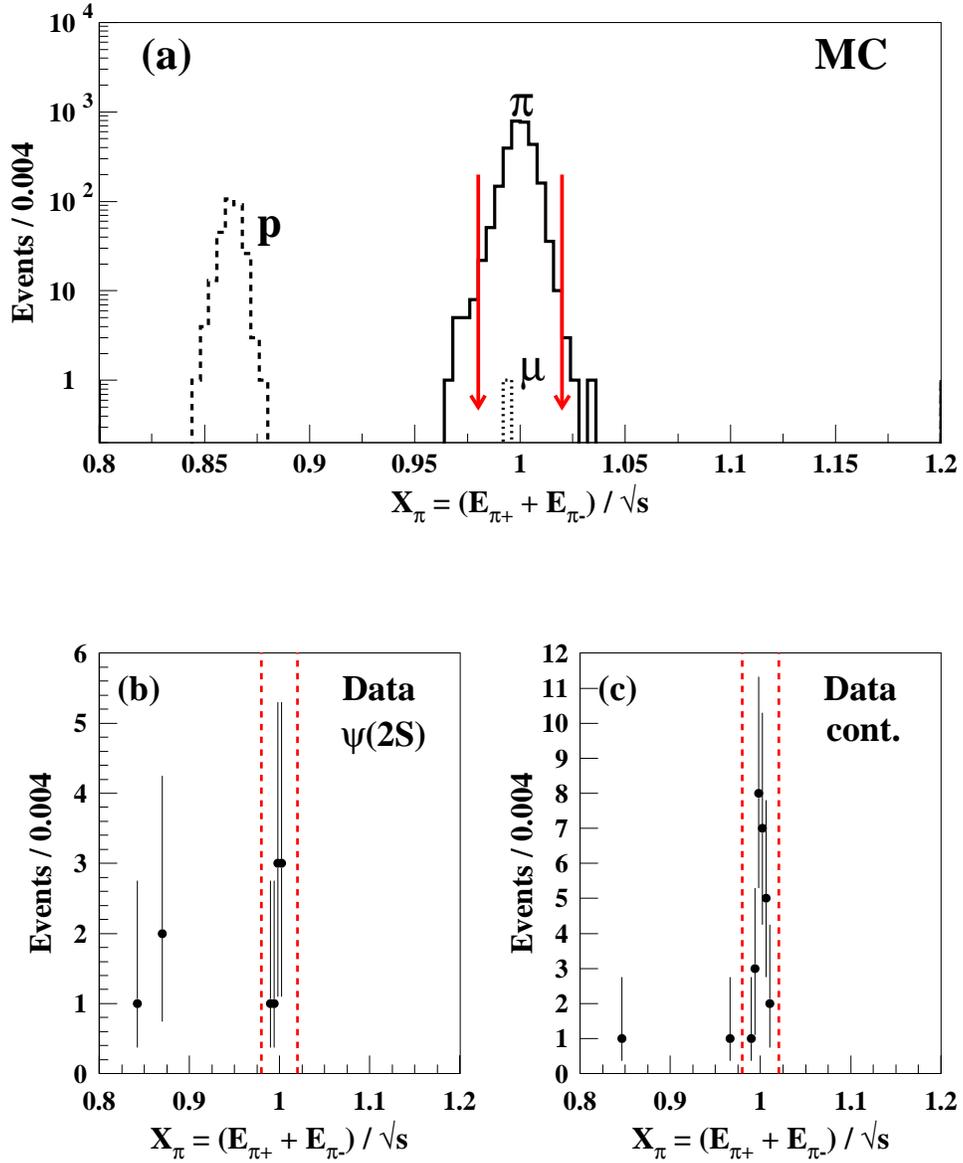}
\caption[MC and data $X_{\pi}$ distributions for $\pipi$ analysis.]
{MC and data $X_{\pi}$ distributions for events 
which satisfy the $\pipi$ final state criteria.  
The arrows in Figure (a) and the dashed lines in Figures (b) and (c) 
denote the signal region of 0.98 $<$ $X_{\pi}$ $<$ 1.02.  
The solid histogram in Figure (a) is 
$\eetopipi$ signal MC, the dotted histogram is dimuon MC, and the dashed histogram is 
$\eetoppbar$ signal MC obtained with $\prelecff$ = 0.  
Figure (b) is for the $\psi(2S)$ data sample.   
Figure (c) is for the continuum data sample.  
There are 8 events in the $\psi(2S)$ signal region and 
26 events in the continuum signal region.}
\label{fig:pixwidemcdata}
\end{center}
\end{figure}

Figure \ref{fig:pixwidemcdata}a shows the $X_{\pi}$ signal region and the 
vicinity after the $\pipi$ final state criteria 
are applied to the $\eetohhbar$ signal and $\eetoll$ MC samples.  Only one 
background event is in MC $X_{\pi}$ signal region, which comes from the $\dimuon$ 
MC sample.  
Figures \ref{fig:pixwidemcdata}b and \ref{fig:pixwidemcdata}c 
show the $X_{\pi}$ distributions after the $\eetopipi$ selection criteria 
have been applied to the $\psi(2S)$ and continuum data samples, respectively.  
In both cases, 1-3 $\psiptoppbar$ events around $X_{\pi}$ = 0.85 are observed, 
far away from the $\pipi$ signal region.  

\newpage
\subsection{Selection of $\KK$ Events}

As can be seen in Figure \ref{fig:kxwideinit}, the $\KK$ final state 
signal region is displaced from the main $\eetoll$ background, 
but it still contains a sizeable amount of leptonic background 
events.  The net momentum constraint and PID information from 
$dE/dx$ and the RICH detector are extremely important 
for suppressing the leptonic background in the $\KK$ final state analysis.  

\begin{figure}[htbp]
\begin{center}
\includegraphics[width=15cm]{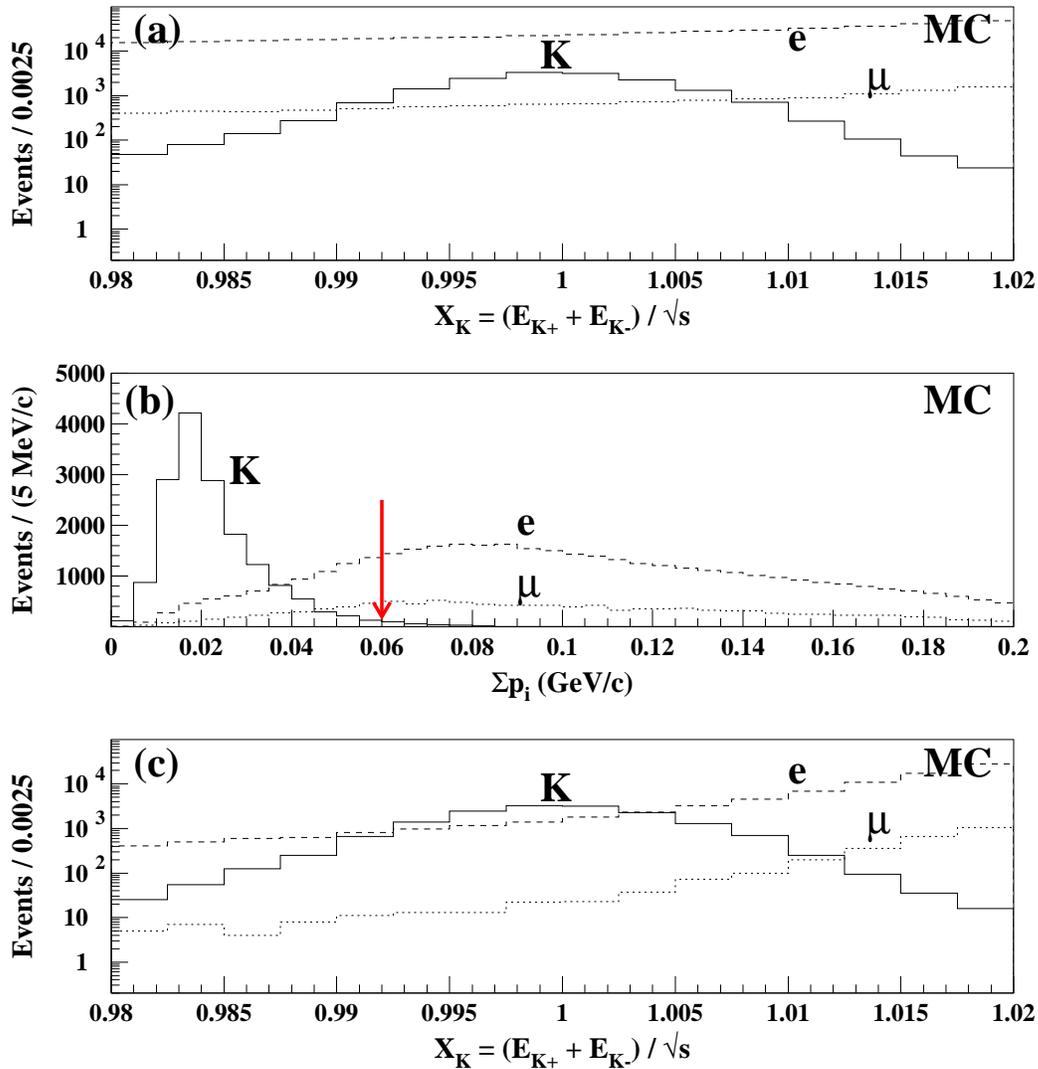}
\caption[MC distributions in the $\KK$ signal region 
with acceptance, trigger, and tracking cuts applied.]
{MC distributions in the $\KK$ signal region 
with acceptance, trigger, and tracking cuts applied.  The solid 
histogram is $\eetoKK$ signal MC, the dashed histogram is Bhabha MC, and 
the dotted histogram is dimuon MC.  
Figure (a) is the $X_{K}$ signal region, 
and Figure (b) is the net monemtum ($\Sigma p_{i}$) of the two tracks.  
The Bhabha MC sample in Figure (b) has been decreased by a factor of 10 
to show more detail.  
A cut is be applied at $\Sigma p_{i}$ $<$ 60 MeV/$c$ to remove 
contamination from the leptonic background.  
Figure (c) is the $X_{K}$ signal region 
after applying the $\Sigma p_{i}$ $<$ 60 MeV/$c$ cut.}
\label{fig:kkincut}
\end{center}
\end{figure}

The net momentum of the $\KK$ pair and the leptonic background in the 
$\KK$ signal region are shown in Figure \ref{fig:kkincut}b.  
The leptonic background peaks around $\Sigma p_{i}\sim$ 0.8 GeV/$c$, so 
a cut is applied requiring $\Sigma p_{i}<$ 0.6 GeV/$c$ and has an 
efficiency of 93$\%$.

\begin{figure}[htbp]
\begin{center}
\includegraphics[width=13cm]{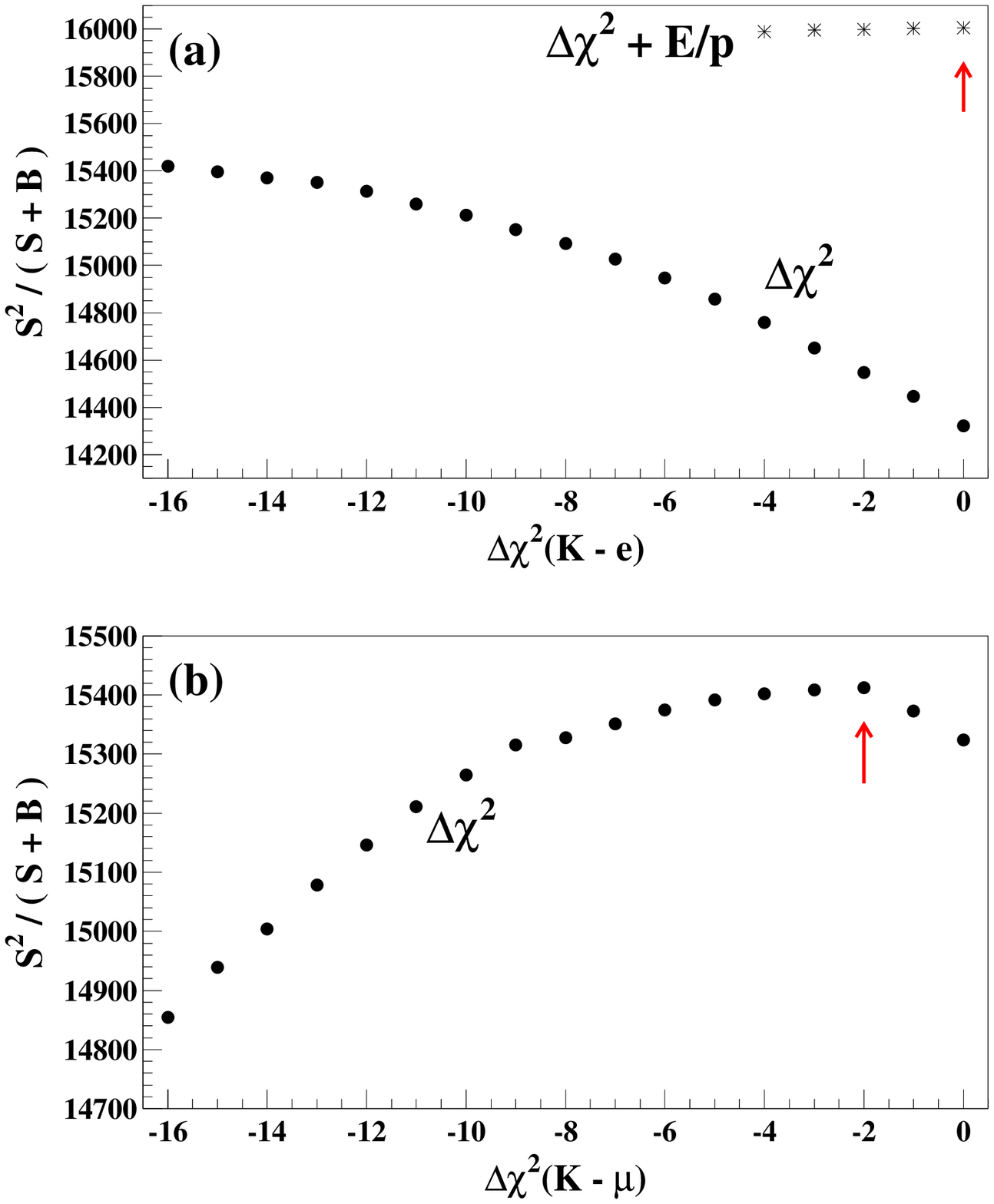}
\caption[Kaon Particle Identification $S^{2}/(S+B)$ Study.]
{Kaon PID $S^{2}/(S+B)$ study.  The values on the abscissa imply a 
cut on $\Delta\chi^2(K-l)$ less than the value.  
The signal (S) used in both figures is from the $\eetoKK$ signal MC.  
The background (B) in Figure (a) is from $\Bhabha$ sideband data.  
The solid points in Figure (a) only have the $\Delta\chi^2(K-e)$ cut applied, 
while the star points have an additional cut of $E_{tkCC}/p$ $<$ 0.85 applied.  
The background (B) in Figure (b) is from $\dimuon$ sideband data.  
The study suggests cuts of $E_{tkCC}/p$ $<$ 0.85 
and $\Delta\chi^2(K-e)$ $<$ 0 (arrow in Figure (a)) and a cut of 
$\Delta\chi^2(K-\mu)$ $<$ -2 (arrow in Figure (b)) on each track.}
\label{fig:ks2spb}
\end{center}
\end{figure}

The PID criteria for the $\KK$ final state is determined from a 
signal squared to signal plus background ($S^{2}/(S+B)$) study.  The signal is from the 
$\eetoKK$ signal MC sample and the background is from 
the continuum data sample.  The background data sample satisfies the 
following criteria:  
1.025 $<$ $X_{K}$ $<$ 1.07, with each track having $E_{tkCC}/p$ $<$ 0.7 for 
the $\dimuon$ sample 
and $E_{tkCC}/p$ $>$ 0.7 for the $\Bhabha$ sample.  Figure \ref{fig:ks2spb} 
shows the $S^{2}/(S+B)$ distributions for different values of 
$\Delta\chi^2(K-l)$ applied (for the definition of $\Delta\chi^2$, see Eqn. 
\ref{eq:RICHdEdxPID}).  
The effect of applying an additional cut on $E_{tkCC}/p$ $<$ 0.85 is also 
considered for rejection of the $\Bhabha$ background.  Applying the $E_{tkCC}/p$ 
cut improves the $S^{2}/(S+B)$ ratio as compared to only 
requiring $\Delta\chi^2(K-e)$.  The $S^{2}/(S+B)$ studies suggest cuts of 
$E_{tkCC}/p$ $<$ 0.85, $\Delta\chi^2(K-e)$ $<$ 0, and $\Delta\chi^2(K-\mu)$ $<$ -2 
for each track.

Table \ref{tab:kkeff} lists the individual and total efficiencies from 
the $\eetoKK$ signal MC sample.  The total efficiency for the $\KK$ 
final state is $\epsilon_{tot}$ = 0.743 $\pm$ 0.003.

\begin{table}[ht]
\caption[Efficiencies for the $\KK$ final state.]
{Efficiencies for the $\KK$ final state.  The individual efficiencies (1-7) 
are determined with respect to the acceptance cuts.} 
\begin{center}
\begin{tabular}{|l|c|c|}
\hline
Cuts & Requirement & $\epsilon$ \\    
\hline
\hline
& N$_{trk}$ = 2 & \\   
Acceptance & $\Sigma Q$ = 0 & 0.868 \\   
& $|$cos$(\theta_{tk})|$ $<$ 0.8 & \\   
\hline
\hline
1. Trigger & L1 Trigger = TRUE &  0.996 \\
\hline
2. $X_{K}$ = ($E_{K^+}$+$E_{K^-}$)/$\sqrt{s}$ 
& 0.98 $<$ $X_{K}$ $<$ 1.02 & 0.948 \\    
\hline
3. IP & $|d_{b}|$ $<$ 5 mm & 0.992 \\    
& $|z_{b}|$ $<$ 5 cm & \\    
\hline
4. Track Quality & 0.5 $<$ DRHF $<$ 1.2 & 0.986 \\    
& $\chi^{2}/dof$ $<$ 10 & \\    
\hline
5. Net Momentum & $\Sigma p_{i}$ $<$ 60 MeV/$c$ & 0.931 \\
\hline
6. $K-e$ Separation &  $E_{tkCC}/p$ $<$ 0.85 & 0.941 \\
& $\Delta\chi^2(K - e)$ $<$ 0 & \\
\hline
7. $K-\mu$ Separation & $\Delta\chi^2(K - \mu)$ $<$ -2 & 0.923 \\
\hline
\hline
$\epsilon_{tot}$ & Acc + Cuts 1-7 & 0.743(3) \\
\hline
\end{tabular}
\label{tab:kkeff}
\end{center}
\end{table}

The generic $\psip$ MC sample is analyzed to test for other 
backgrounds which satisfy the $\KK$ final state criteria.  
In the 40,568,651 generic $\psip$ decays, 
there are only 872 $\psiptoKK$ decays. 
After applying the $\KK$ selection criteria, 486 $\psiptoKK$ events survive.  
The $\psiptoKK$ branching ratio is not determined from this MC sample 
because of the issues discussed at the end of Section 4.2.

\begin{figure}[htbp]
\begin{center}
\includegraphics[width=15cm]{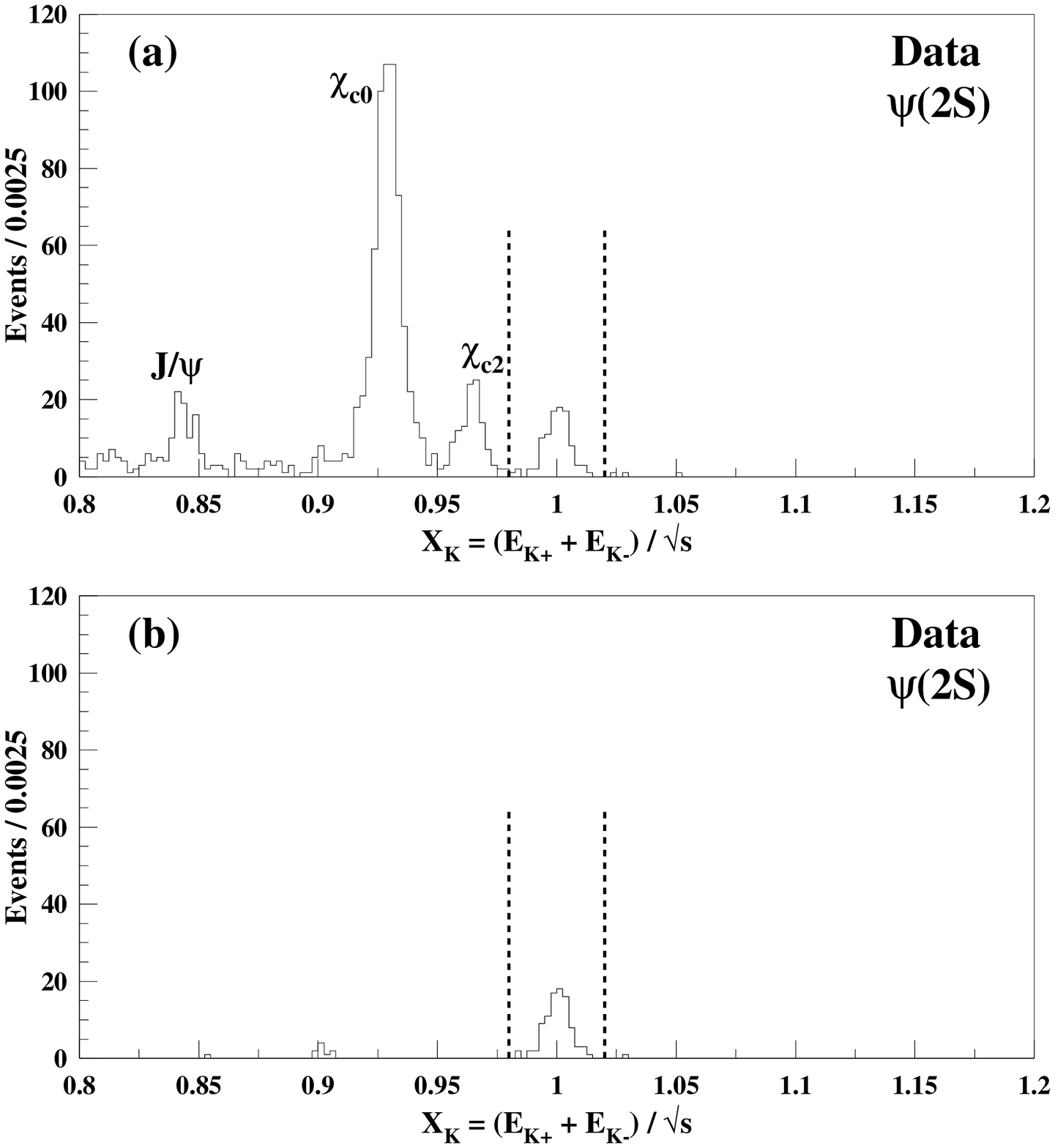}
\caption[Effect of net momentum criterion on the $\psip$ data for the $\KK$ final state.]
{Effect of net momentum criterion on the $\psip$ data for the $\KK$ final state.  
Figure (a) shows the $\psip$ data after applying all $\KK$ final state criteria 
except for the net momentum criterion, while Figure (b) includes it.  The 
J/$\psi$, $\chi_{c0}$, and $\chi_{c2}$ are clearly seen in Figure (a) at $X_{K}$ = 
0.840, 0.926, and 0.965, respectively, outside of the 0.98 $<$ $X_{K}$ $<$ 1.02 
signal region, and are removed in Figure (b) due to the net momentum criterion.}
\label{fig:k2snetp}
\end{center}
\end{figure}

\begin{figure}[htbp]
\begin{center}
\includegraphics[width=15cm]{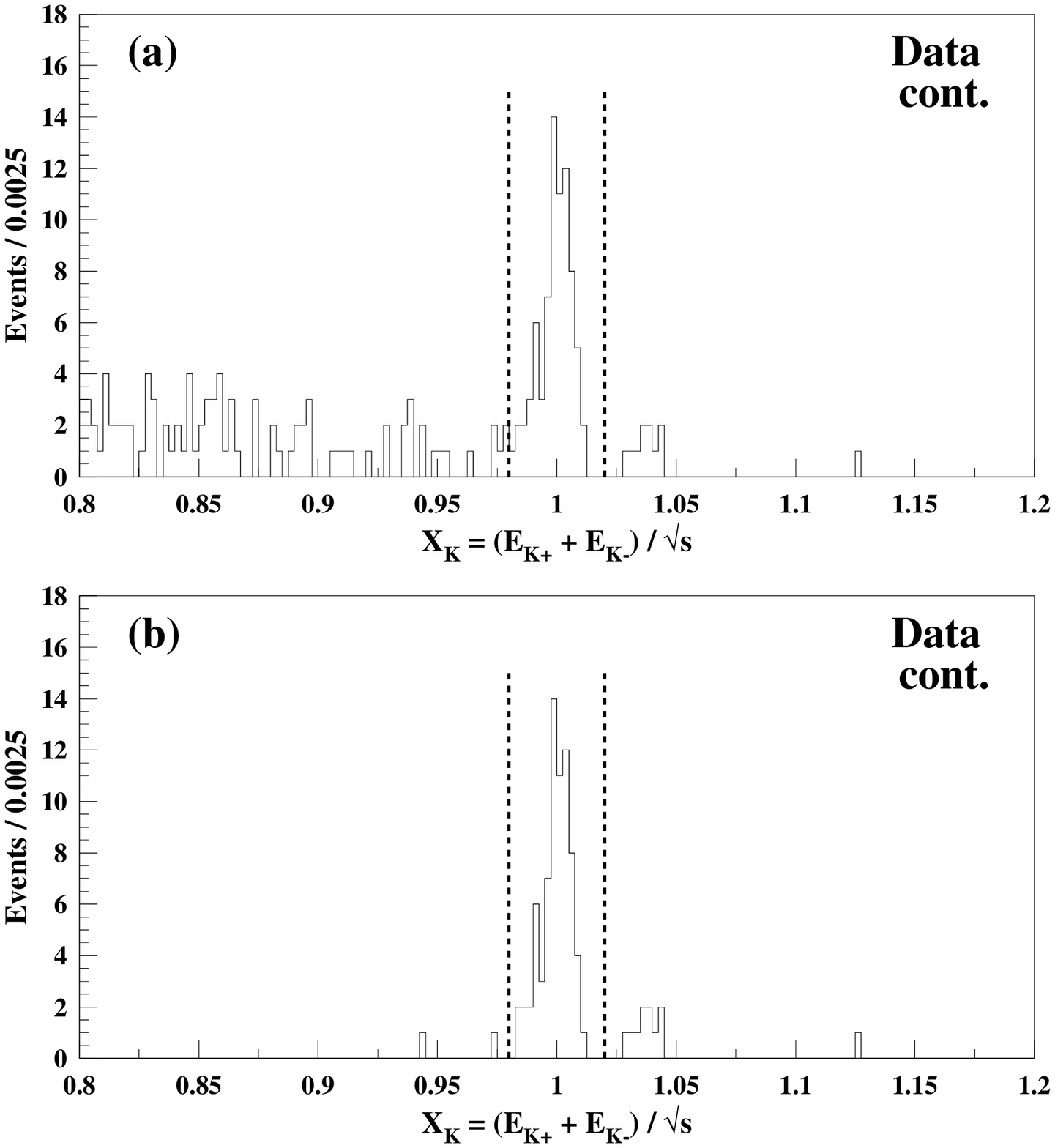}
\caption[Effect of net momentum criterion on the continuum data for the $\KK$ final state.]
{Effect of net momentum criterion on the continuum data for the $\KK$ final state.  
Figure (a) shows the continuum data after applying all $\KK$ final state criteria 
except for the net momentum criterion, while Figure (b) includes it.  The 
net momentum criterion removes events below the 0.98 $<$ $X_{K}$ $<$ 1.02 
signal region but has no effect on the $\eetoll$ events near $X_{K}$ = 1.03.}
\label{fig:kctnetp}
\end{center}
\end{figure}

Figures \ref{fig:k2snetp}a and \ref{fig:kctnetp}a show the $\psip$ and continuum data 
$X_{K}$ distributions, respectively, after applying all $\KK$ final state criteria 
except for the net momentum criterion.  
The charmonium resonances $J/\psi$, $\chi_{c0}$, and $\chi_{c2}$ are seen in 
the $\psip$ data, but they are outside of 
the $X_{K} = 0.98-1.02$ signal region. Figure \ref{fig:k2snetp}b shows that 
only $\psiptoKK$ events survive after applying the net momentum criterion, and 
Figure \ref{fig:kctnetp}b shows that only $\eetoKK$ events 
survive after applying the net momentum criterion.

\begin{figure}[htbp]
\begin{center}
\includegraphics[width=14cm]{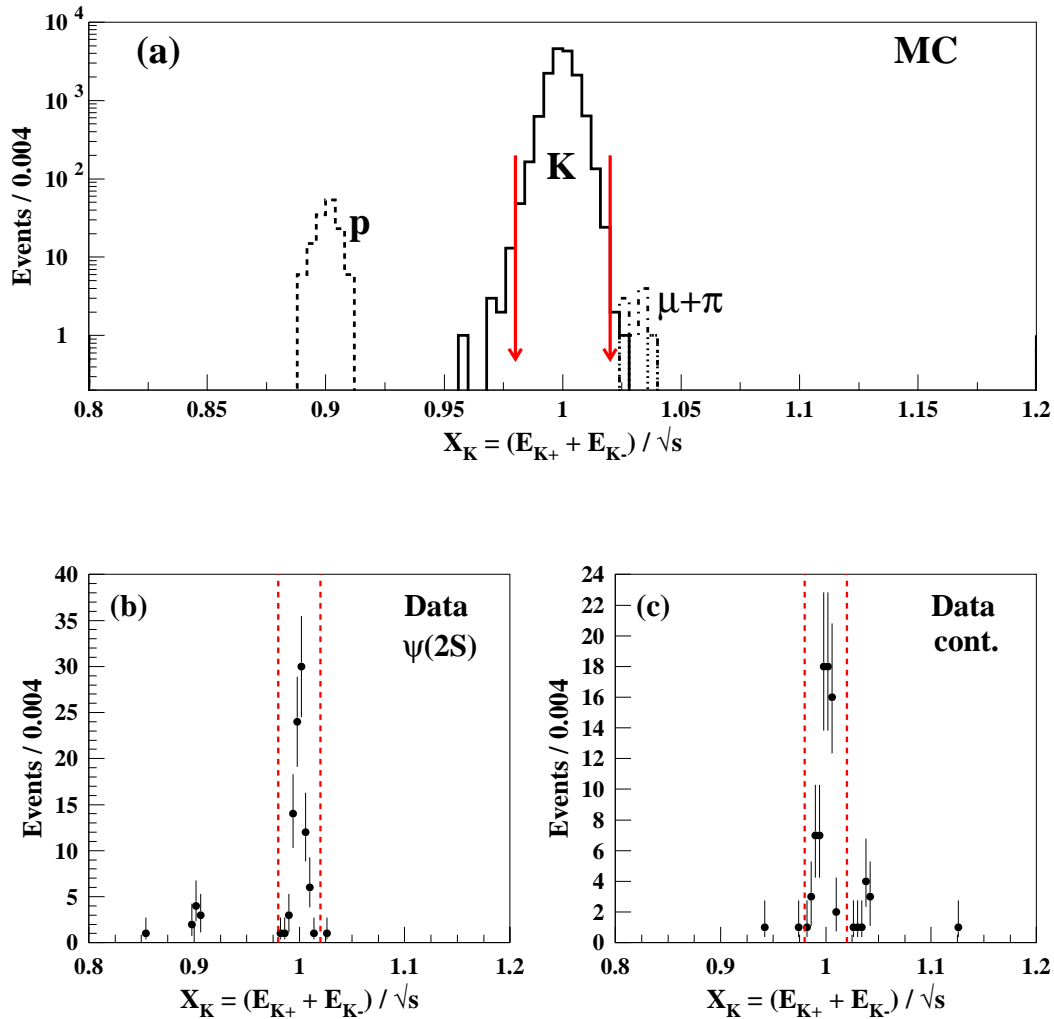}
\caption[MC and data $X_{K}$ distributions for $\KK$ analysis.]
{MC and data $X_{K}$ distributions for events 
which satisfy the $\KK$ final state event selection criteria.  
The arrows in Figure (a) and the dashed lines in Figures (b) and (c) 
denote the signal region of 0.98 $<$ $X_{K}$ $<$ 1.02.  
The solid histogram in Figure (a) is $\eetoKK$ signal MC, 
the dotted histogram is dimuon MC, 
the dot-dashed histogram is the $\eetopipi$ signal MC, and 
the dashed histogram is the $\eetoppbar$ signal MC obtained with $\prelecff$ = 0 .  
Figure (b) is for the $\psi(2S)$ data sample.   
Figure (c) is for the continuum data sample.  
There are 92 events in the $\psi(2S)$ signal region and 
72 events in the continuum signal region.}
\label{fig:kxwidemcdata}
\end{center}
\end{figure}

Figure \ref{fig:kxwidemcdata}a shows the $X_{K}$ signal region and the 
vicinity after the $\KK$ final state criteria 
are applied to the $\eetohhbar$ signal and $\eetoll$ MC samples.  
Only $\eetoKK$ MC events populate the signal region.
Figures \ref{fig:kxwidemcdata}b and \ref{fig:kxwidemcdata}c 
show the $X_{K}$ distributions after the $\eetoKK$ selection criteria 
have been applied to the $\psi(2S)$ and continuum data samples, respectively.  
In both cases, a small number of $\psiptoppbar$ events around $X_{K}$ = 0.9 
are observed, far away from the $\KK$ signal region.  The presence of events 
around $X_{K}$ = 1.03 is consistent with the leptonic background.  The determination 
of the $\leppair$ contamination in the $\KK$ signal region is discussed below.
  
The $\leppair$ contamination is determined by a ratio of the 
number of $\leppair$ events from the leptonic MC samples 
without applying the kaon PID criteria.  
We determine the $\leppair$ events inside the $\KK$ signal region and in the 
peak region of $\eetoll$ production, namely 1.02 $<$ $X_{K}$ $<$ 1.07.  
The ratio of the two gives the leptonic MC scale factor 
\begin{displaymath}
\frac{N^{\leppair}_{MC}(0.98<X_{K}<1.02)}
{N^{\leppair}_{MC}(1.02<X_{K}<1.07)} = 
\frac{84 378\pm290}{1 495 073\pm1223} = 0.0564\pm0.0002.  
\end{displaymath}
We scale the number of events in the $\leppair$ peak region, 1.02 $<$ $X_{K}$ $<$ 1.07, 
in the data samples with this ratio.  
The number of events in the $\leppair$ peak region from the continuum data sample 
is 10.00$^{+3.81}_{-3.22}$.  
Therefore, the leptonic background in the $\KK$ signal region is determined to be 
$N^{cont}_{\leppair}$ = 0.56$^{+0.21}_{-0.18}$.  
The number of events in the $\leppair$ peak region from the $\psip$ data sample is 
1.00$^{+1.75}_{-0.63}$.  
Therefore, the leptonic background in the $\KK$ 
signal region is determined to be 
$N^{\psip}_{\leppair}$ = 0.06$^{+0.10}_{-0.04}$.  

\subsection{Selection of $\ppbar$ Events}

Even though the $\ppbar$ final state signal region is substantially 
displaced from the $\eetoll$ peak region, contamination from the leptonic 
background still exists.  Constraining the net momentum of the $\ppbar$ pair 
and particle identification based on $dE/dx$ and RICH information are 
used to suppress the leptonic background.  

\begin{figure}[htbp]
\begin{center}
\includegraphics[width=15cm]{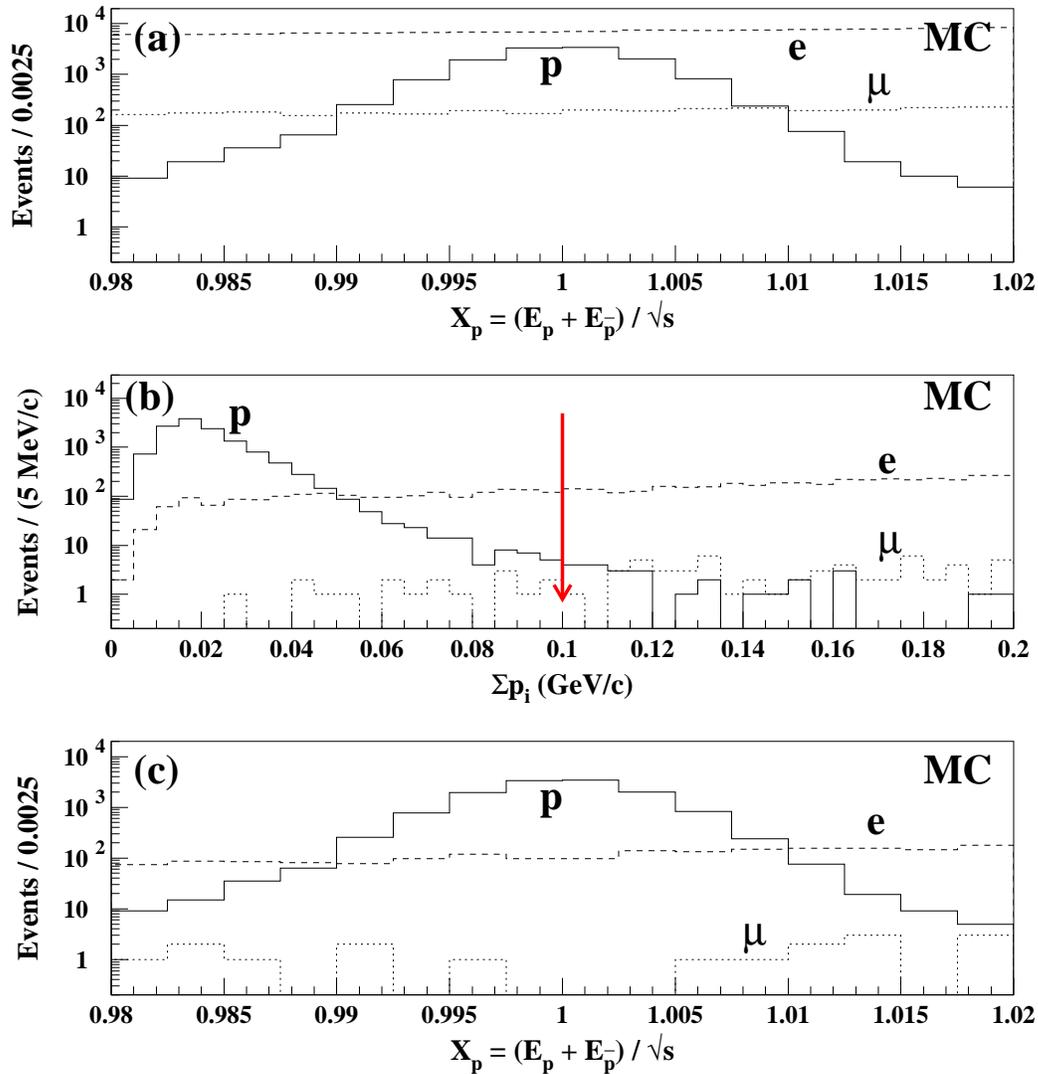}
\caption[MC distributions in the $\ppbar$ signal region 
with acceptance, trigger, and tracking cuts applied.]
{MC distributions in the $\ppbar$ signal region 
with acceptance, trigger, and tracking cuts applied.  The solid 
histogram is the $\eetoppbar$ signal MC obtained with $\prelecff$ = 0, 
the dashed histogram is Bhabha MC, and 
the dotted histogram is dimuon MC.  Figure (a) is the $X_{p}$ signal 
region, and Figure (b) is the net monemtum ($\Sigma p_{i}$) 
of the two tracks.  
A cut is be applied at $\Sigma p_{i}$ $<$ 100 MeV/$c$ to remove 
contamination from the leptonic background.  
Figure (c) is 
the $X_{p}$ signal region after applying the $\Sigma p_{i}$ $<$ 100 MeV/$c$ cut.}
\label{fig:prkincut}
\end{center}
\end{figure}

The net momentum of the $\ppbar$ pair and the leptonic background in the 
$\ppbar$ signal region is shown in Figure \ref{fig:prkincut}.  
Requiring $\Sigma p_{i}<$ 100 MeV/$c$ removed a 
substantial portion of the leptonic background and has an efficiency 
of $>$ 98$\%$.

\begin{figure}[htbp]
\begin{center}
\includegraphics[width=12.5cm]{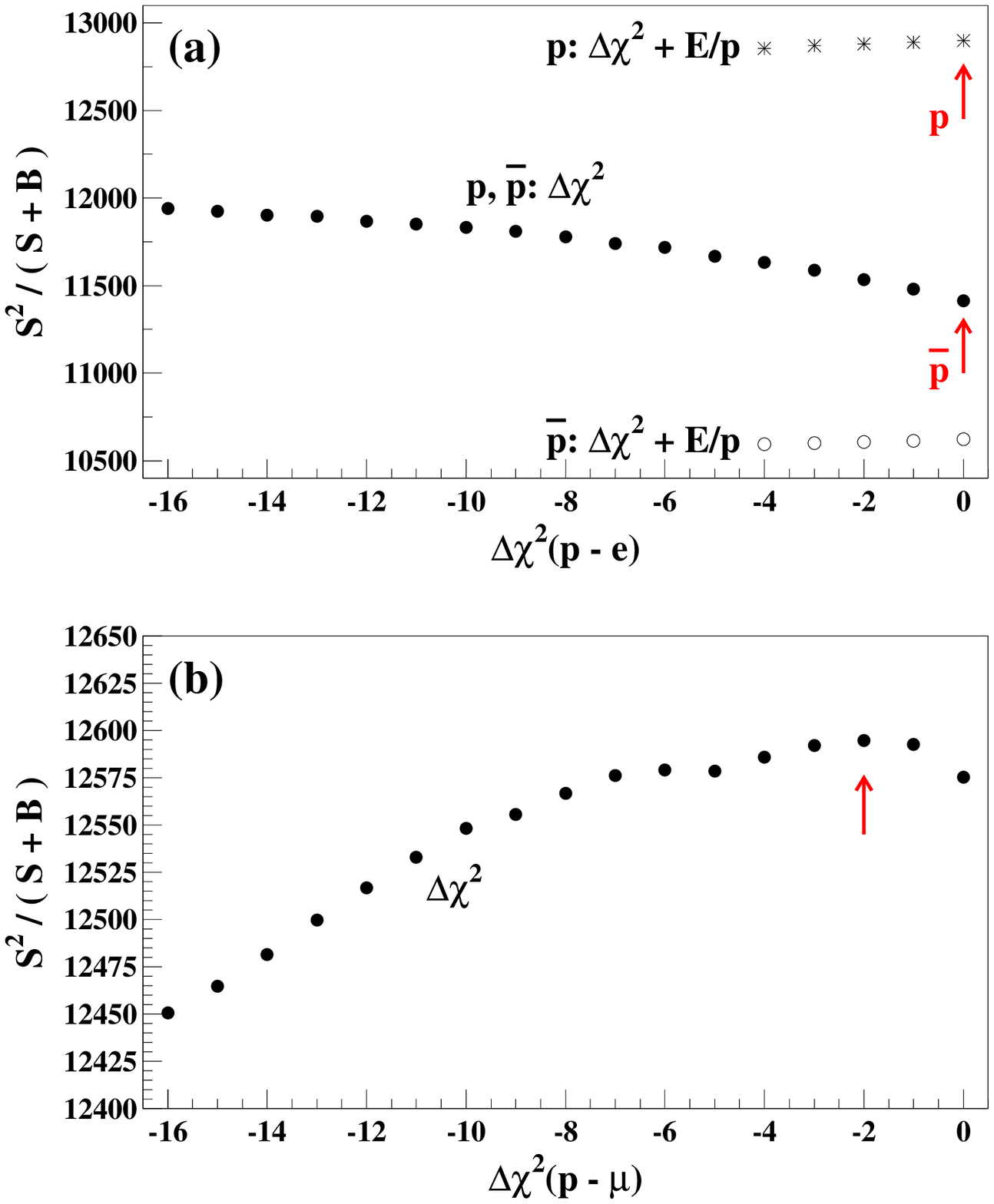}
\caption[Proton Particle Identification $S^{2}/(S+B)$ Study.]  
{Proton PID $S^{2}/(S+B)$ study.  
The values on the abscissa imply a cut on $\Delta\chi^2(p-l)$ less than the value.  
The signal (S) used in both figures is from the $\eetoppbar$ signal MC 
obtained with $\prelecff$ = 0 .  
The background (B) in Figure (a) is from $\Bhabha$ sideband data.  
The solid points in Figure (a) only have the $\Delta\chi^2(p-e)$ cut applied, 
the star points correspond to an additional cut of $E_{tkCC}/p$ $<$ 0.85 
applied to the positive (proton) track, 
and the open points correspond to an additional cut of $E_{tkCC}/p$ $<$ 0.85 
applied to the negative (antiproton) track.  
The background (B) in Figure (b) is from $\dimuon$ sideband data.  
The study suggests cuts of $E_{tkCC}/p$ $<$ 0.85 and $\Delta\chi^2(p-e)$ $<$ 0 for the 
positive track and only $\Delta\chi^2(p-e)$ $<$ 0 for the negative track 
(arrows in Figure (a)).  It also suggests a cut of $\Delta\chi^2(p-\mu)$ $<$ -2 
(arrow in Figure (b)) on each track.}
\label{fig:prs2spb}
\end{center}
\end{figure}

The PID criteria for the $\ppbar$ final state 
is determined from a $S^{2}/(S+B)$ study.  The signal is from the 
$\eetoppbar$ signal MC sample obtained with $\prelecff$ = 0 and the background is from 
the continuum data sample.  The background data sample satisfies the 
following criteria:  1.109 $<$ $X_{p}$ $<$ 1.155, 
with each track having $E_{tkCC}/p$ $<$ 0.7 for the $\dimuon$ sample 
and $E_{tkCC}/p$ $>$ 0.7 for the $\Bhabha$ sample.  Figure \ref{fig:prs2spb} 
shows the $S^{2}/(S+B)$ distributions for different values 
of $\Delta\chi^2(p-l)$ applied (for the definition of $\Delta\chi^2$, see Eqn. 
\ref{eq:RICHdEdxPID}).  
The effect of applying an additional cut of $E_{tkCC}/p$ $<$ 0.85 is also 
considered for rejection of the $\Bhabha$ background.  Applying the $E_{tkCC}/p$ 
cut to the proton candidate track improves the $S^{2}/(S+B)$ ratio but is 
worse for the antiproton candidate, as compared to 
only requiring $\Delta\chi^2(p-e)$.  
The reason is that the antiproton occasionally annihilates in 
the calorimeter.  This has the effect 
of creating a broad bump in the $E_{tkCC}/p$ distribution ranging from 0.4 to 2.0. 
The $S^{2}/(S+B)$ studies suggest cuts of 
$\Delta\chi^2(p-e)$ $<$ 0 and $\Delta\chi^2(p-\mu)$ $<$ -2 for each track, 
with the additional cut of $E_{tkCC}/p$ $<$ 0.85 for the positive track.

Table \ref{tab:ppbareff} lists the individual and total efficiencies 
determined from 
the $\eetoppbar$ signal MC samples.  The total efficiency for the $\ppbar$ 
final state, assuming $\prelecff$ = 0, is $\epsilon_{tot}$ = 0.626 $\pm$ 0.003 
and, assuming $\prelecff$ = $\prmagff$, is $\epsilon_{tot}$ = 0.657 $\pm$ 0.003.

\begin{table}[ht]
\caption[Efficiencies for the $\ppbar$ final state.]
{Efficiencies for the $\ppbar$ final state determined from the $\prelecff$ = 0 
and $\prelecff$ = $\prmagff$ $\eetoppbar$ signal MC samples.  The individual 
efficiencies (1-7) are determined with respect to the acceptance cuts.} 
\begin{center}
\begin{tabular}{|l|c|c|c|}
\hline
Cuts & Requirement & $\epsilon$ & $\epsilon$ \\    
& & $(\frac{\prelecff}{\prmagff} = 0)$ & $(\frac{\prelecff}{\prmagff} = 1)$ \\    
\hline
\hline
& N$_{trk}$ = 2 & & \\   
Acceptance & $\Sigma Q$ = 0 & 0.666 & 0.704 \\   
& $|$cos$(\theta_{tk}|$ $<$ 0.8 & & \\   
\hline
\hline
1. Trigger & L1 Trigger = TRUE & 0.998 & 0.997 \\
\hline
2. $X_{p}$ = ($E_{p}$+$E_{\overline{p}}$)/$\sqrt{s}$ 
& 0.98 $<$ $X_{p}$ $<$ 1.02 & 0.987 & 0.982 \\    
\hline
3. IP & $|d_{b}|$ $<$ 5 mm & 0.999 & 0.998 \\    
& $|z_{b}|$ $<$ 5 cm & & \\    
\hline
4. Track Quality & 0.5 $<$ DRHF $<$ 1.2 & 0.9997 & 0.9995 \\    
& $\chi^{2}/dof$ $<$ 10 & & \\    
\hline
5. Net Momentum & $\Sigma p_{i}$ $<$ 100 MeV/$c$ & 0.986 & 0.981 \\
\hline
6. $p-e$ Separation  & $E_{tkCC}/p$ $<$ 0.85 (+ track) & 0.978 & 0.976 \\
& $\Delta\chi^2(p-e)$ $<$ 0 & & \\
\hline
7. $p-\mu$ Separation  & $\Delta\chi^2(p-\mu)$ $<$ -2 & 0.959 & 0.958 \\
\hline
\hline
$\epsilon_{tot}$ & Acc + Cuts 1-7 & 0.626(3) & 0.657(3) \\
\hline
\end{tabular}
\label{tab:ppbareff}
\end{center}
\end{table}

The generic $\psip$ MC sample is analyzed to test for other 
backgrounds which satisfy the $\ppbar$ final state criteria.  
In the 40,568,651 generic $\psip$ decays, 
there are only 1778 $\psiptoppbar$ decays.  
After applying the $\ppbar$ selection criteria, 
1145 $\psiptoppbar$ events survive.
The $\psiptoppbar$ branching ratio is not determined from this MC sample 
due to the issues discussed at the end of Section 4.2.

\begin{figure}[htbp]
\begin{center}
\includegraphics[width=15cm]{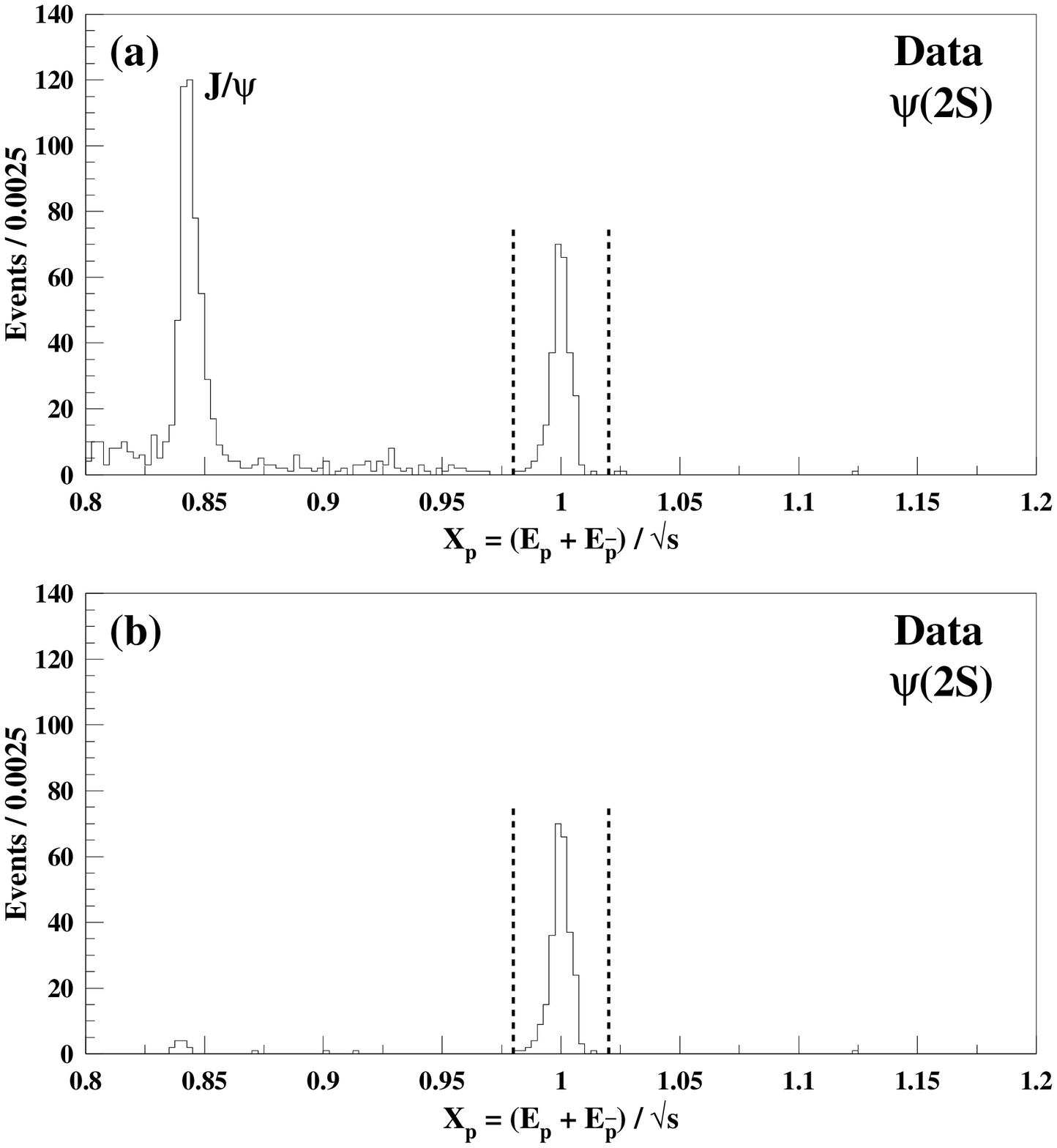}
\caption[Effect of net momentum criterion on the $\psip$ data 
for the $\ppbar$ final state.]
{Effect of net momentum criterion on the $\psip$ data 
for the $\ppbar$ final state.  
Figure (a) shows the $\psip$ data after applying all $\ppbar$ final state criteria 
except for the net momentum criterion, while Figure (b) includes it.  The 
J/$\psi$ is clearly seen in the Figure (a) at $X_{p}$ = 0.840, outside of the 
0.98 $<$ $X_{p}$ $<$ 1.02 signal region, and is removed in Figure (b) 
due to the net momentum criterion.}
\label{fig:pr2snetp}
\end{center}
\end{figure}

\begin{figure}[htbp]
\begin{center}
\includegraphics[width=15cm]{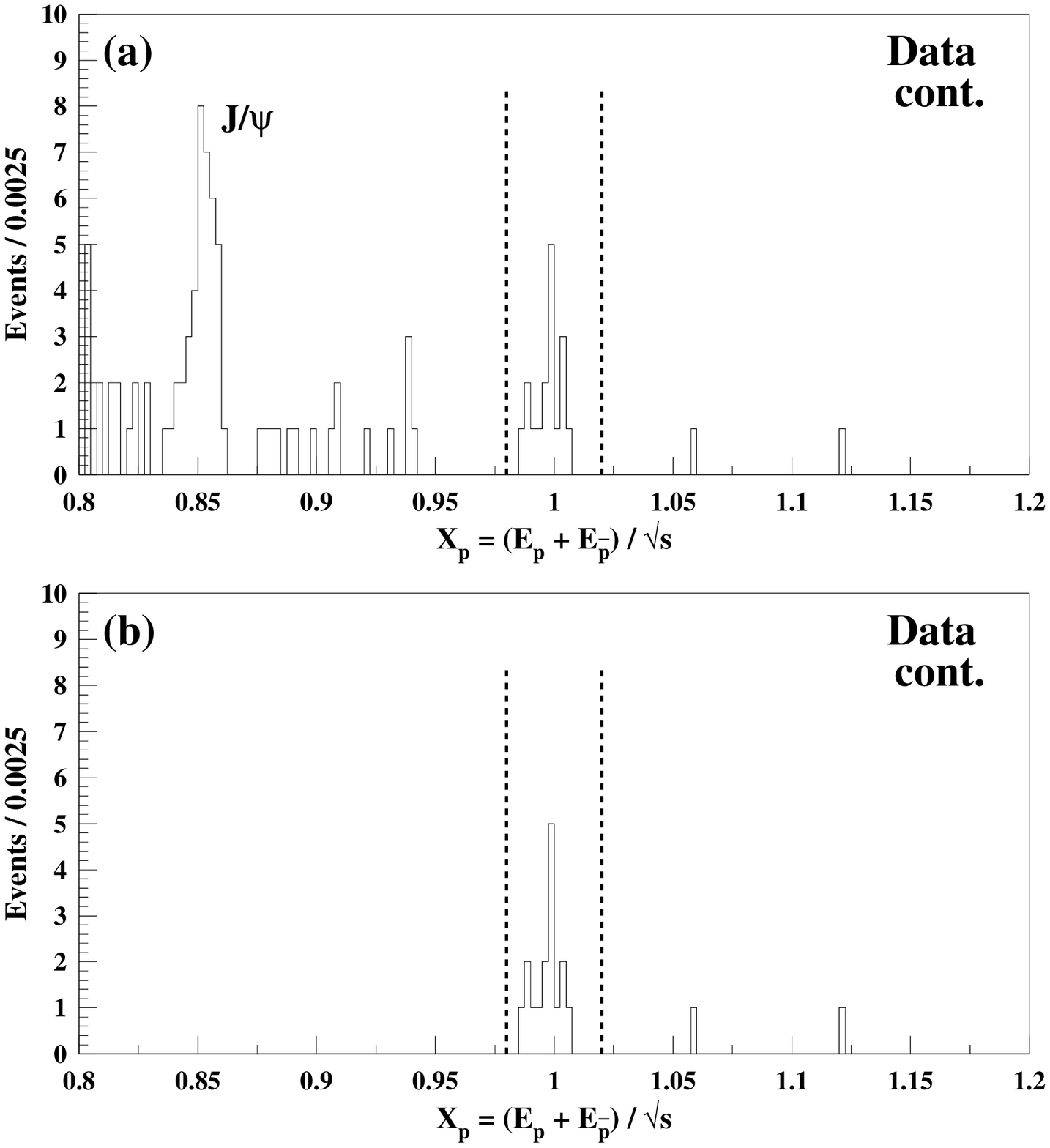}
\caption[Effect of net momentum criterion on the continuum data 
for the $\ppbar$ final state.]
{Effect of net momentum criterion on the continuum data for the $\ppbar$ final state.  
Figure (a) shows the continuum data after applying all $\ppbar$ final state criteria 
except for the net momentum criterion, while Figure (b) includes it.  The 
J/$\psi$ is clearly seen in Figure (a) at $X_{p}$ = 0.840, outside of the 
0.98 $<$ $X_{p}$ $<$ 1.02 signal region, and is removed in Figure (b) 
due to the net momentum criterion.}
\label{fig:prctnetp}
\end{center}
\end{figure}

Figures \ref{fig:pr2snetp}a and \ref{fig:prctnetp}a show the $\psip$ and continuum data 
$X_{p}$ distributions, respectively, after applying all $\ppbar$ final state criteria 
except for the net momentum criterion.  The $J/\psi$ resonance is 
seen in both data samples, but it is outside of 
the $X_{p}$ signal regions. Figure \ref{fig:pr2snetp}b shows that 
only $\psiptoppbar$ events survive after applying the net momentum criterion, 
and Figure \ref{fig:prctnetp}b shows that only 16 $\eetoppbar$ events 
survive after applying the net momentum criterion.

\begin{figure}[htbp]
\begin{center}
\includegraphics[width=15.2cm]{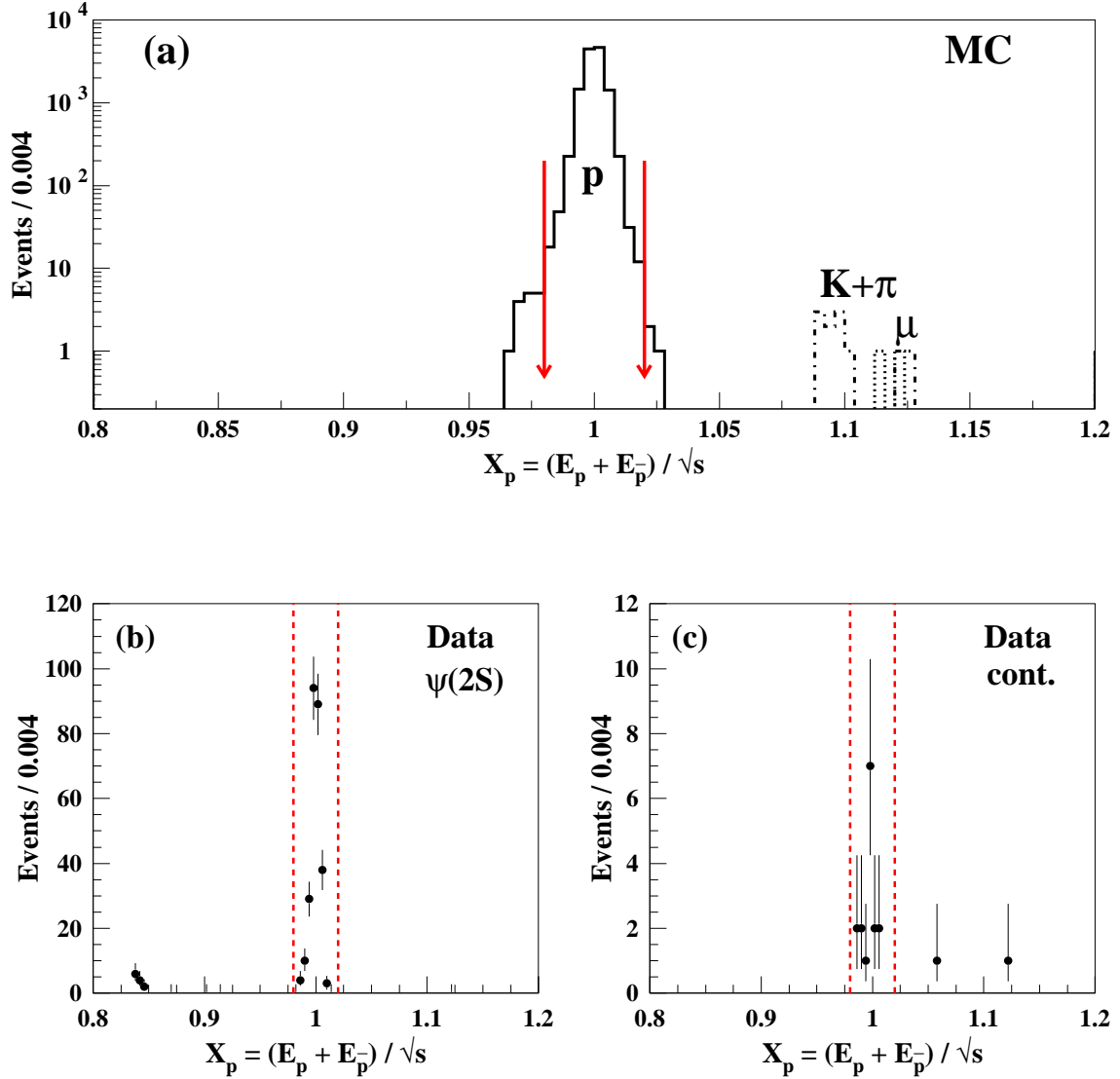}
\caption[MC and data $X_{p}$ distributions for $\ppbar$ analysis.]
{MC and data $X_{p}$ distributions for events 
which satisfy the $\ppbar$ final state event selection criteria.  
The arrows in Figure (a) and the dashed lines in Figures (b) and (c) 
denote the signal region of 0.98 $<$ $X_{p}$ $<$ 1.02.  
The solid histogram in Figure (a) is the $\eetoppbar$ signal MC 
obtained with $\prelecff$ = 0 , 
the dashed histogram is Bhabha MC, 
the dotted histogram is dimuon MC, 
and the dot-dashed histogram is the $\eetopipi$ and $\eetoKK$ signal MC combined.  
Figure (b) is for the $\psi(2S)$ data sample.   
Figure (c) is for the continuum data sample.  
There are 269 events in the $\psi(2S)$ signal region and 16 events 
in the continuum signal region.}
\label{fig:prxwidemcdata}
\end{center}
\end{figure}

Figure \ref{fig:prxwidemcdata}a shows the $X_{p}$ signal region and the 
vicinity after the $\ppbar$ final state event selection criteria 
are applied to the $\eetohhbar$ signal sample and $\eetoll$ MC samples.  
Only $\eetoppbar$ MC events populate the signal region.  
Figures \ref{fig:prxwidemcdata}b and \ref{fig:prxwidemcdata}c 
show the $X_{p}$ distributions after the $\ppbar$ selection criteria have been 
applied to the $\psi(2S)$ and continuum data samples, respectively.  

The same method is used for treating the $\leppair$ background as for the $\KK$ 
final state analysis, as described at the end of Section 4.5.2.  
The peak region of $\eetoll$ production for the $\ppbar$ analysis is 
1.109 $<$ $X_{p}$ $<$ 1.155 and its corresponding leptonic MC scale factor is 
\begin{displaymath}
\frac{N^{\leppair}_{MC}(0.98<X_{p}<1.02)}
{N^{\leppair}_{MC}(1.109<X_{p}<1.155)} = 
\frac{1890\pm43}{1,547,388\pm1244} = (1.22\pm0.03)\times10^{-3}.  
\end{displaymath} 
The number of events in the $\leppair$ peak region in 
both the $\psip$ and continuum data is found to be 1.00$^{+1.75}_{-0.63}$, 
so the $\leppair$ background in the $\ppbar$ signal region 
is $N^{\psip}_{\leppair} = N^{cont}_{\leppair} = (1.22\pm0.03)\times10^{-3}$, 
and is considered negligible in the $\ppbar$ analysis.

\section{Other Backgrounds}

\subsection{Contamination from Charmonium Resonances}

Since the continuum data was taken at $\sqrt{s}$ = 3.671 GeV, contributions from nearby 
charmonium states need to be accounted for.  The resonances considered are the off-mass 
shell production of the charmonium vector states J/$\psi$, $\psip$, $\psi(3770)$, 
$\psi(4040)$, $\psi(4160)$, and $\psi(4415)$, and the P-wave spin triplet states 
$\chi_{c0}$ and $\chi_{c2}$ produced via untagged two-photon fusion.  Since the 
continuum data was only 15 MeV below the $\psip$ resonance, the contribution from the 
$\psip$ is treated directly in the Results section (Section 4.8).  
The resonance parameters of the listed charmonium states are given 
in Table \ref{tab:charmresparam}.

\begin{table}[ht]
\caption[Charmonium resonances parameters.]
{Charmonium resonances parameters.  The full width, $\ee$ partial width, and two-photon 
partial width are denoted by $\Gamma$, $\Gamma_{ee}$, and 
$\Gamma_{\gamma\gamma}$, respectively.  
The parameters for J/$\psi$, $\chi_{c0}$, $\chi_{c2}$, $\psip$, and $\psi(3770)$ 
are from the PDG \cite{PDG2004}.  
The parameters for $\psi(4040)$, $\psi(4160)$, and $\psi(4415)$,  
are from Seth \cite{SethHigherCharmRes}.} 
\begin{center}
\begin{tabular}{|c|c|c|c|c|}
\hline
Resonance & Mass (GeV) & $\Gamma$ (MeV) & $\Gamma_{ee}$ (keV) 
 & (2J+1)$\Gamma_{\gamma\gamma}$ (keV) \\    
\hline
J/$\psi$     & 3.096916(11) & 0.0910(32) & 5.40(17) & ----- \\
$\chi_{c0}$  & 3.41519(34)  & 10.1(8)    & -----    & 2.6(5) \\
$\chi_{c2}$  & 3.55626(11)  & 2.11(16)   & -----    & 2.6(3) \\
$\psip$      & 3.686093(34) & 0.281(17)  & 2.12(12) & ----- \\
$\psi(3770)$ & 3.7700(24)   & 23.6(2.7)  & 0.26(4)  & ----- \\
$\psi(4040)$ & 4.0394(9)    & 88(5)      & 0.89(8)  & ----- \\
$\psi(4160)$ & 4.153(3)     & 107(8)     & 0.83(7)  & ----- \\
$\psi(4415)$ & 4.426(5)     & 119(15)    & 0.71(10) & ----- \\
\hline
\end{tabular}
\label{tab:charmresparam}
\end{center}
\end{table}

\subsubsection{Vector States, $\psi(nS)$}

The contribution from the vector states are determined by convoluting a pure Born-level 
resonance with a Gaussian center-of-mass uncertainty inherent in $\ee$ annihilations.  
The resonance production cross section of a vector state is characterized 
by a non-relativistic Breit-Wigner given as \cite{PDG2004} 
\begin{equation}
\sigma^{BW}_{V}(s) = \frac{3\pi}{s}
\frac{\Gamma_{ee}\Gamma_{\hhbar}}{[(\sqrt{s}-M_{R})^2 + (\Gamma/2)^2]},
\end{equation}
where $s$ is the center-of-mass energy squared of the $\ee$ annihilation, and $M_{R}$, 
$\Gamma$, $\Gamma_{ee}$, $\Gamma_{\hhbar}$ are the mass, full width, $\ee$ partial width, 
and the $\hhbar$ $(\pi^+,K^+,$ or $p$) partial widths of the resonance.  
The $\hhbar$ partial widths of the J$/\psi$ and $\psip$ are 
listed in Table \ref{tab:vectorcharmparwidth}. None are known for the higher vectors.
The distribution of the center-of-mass energy 
of the $\ee$ annihilation is given by the following Gaussian expression 
\begin{equation}
G(s,s^\prime) = \frac{1}{\sqrt{2\pi}~\sigma_{CME}}
~\exp\left(-\frac{(\sqrt{s}-\sqrt{s^\prime})^2}{2~\sigma^2_{CME}}\right),
\end{equation} 
where $s$ is the nominal center-of-mass energy squared, $s^\prime$ is the actual 
center-of-mass energy squared of the $\ee$ annihilation, 
and $\sigma_{CME} = $ 2.3 MeV is the Gaussian uncertainty in the center-of-mass energy.  
The cross section of a resonance at the center-of-mass energy of the continuum 
data ($\sqrt{s}$ = 3.671 GeV) is given by the following convolution expression 
\begin{equation}
\sigma_{V}(s) = 
\int^{\infty}_{0} \sigma^{BW}_{V}(s^\prime)~G(s,s^\prime)~d(s^\prime),
\end{equation} 
where $\sigma_{V}(s)$ is the observed cross section for 
$\ee \rightarrow V \rightarrow \hhbar$.  

\begin{table}[ht]
\caption[J/$\psi$ and $\psip$ partial widths for $\pipi$, $\KK$, and $\ppbar$.]
{J/$\psi$ and $\psip$ partial widths for $\pipi$, $\KK$, and $\ppbar$.  The values 
are from the PDG \cite{PDG2004}.} 
\begin{center}
\begin{tabular}{|c|c|c|c|}
\hline
Resonance & $\Gamma_{\pipi}$ (eV) & $\Gamma_{\KK}$ (eV) & $\Gamma_{\ppbar}$ (eV) \\    
\hline
J/$\psi$ & 13.4$\pm$2.1  & 21.6$\pm$2.9   & 193$\pm$11 \\
$\psip$  & 22.5$\pm$14.1 & 28.1$\pm$19.7  & 58.2$\pm$9.4 \\
\hline
\end{tabular}
\label{tab:vectorcharmparwidth}
\end{center}
\end{table}

The number of expected events in the continuum data sample, 
denoted by $N_{\hhbar}$, are determined by the expression 
\begin{equation}
N_{\hhbar} = \sigma_{V}(s)\cdot\epsilon_{\hhbar}\cdot{\cal L},
\end{equation}
where $\epsilon_{\hhbar}$ = 0.166, 0.743, and 0.657 for the $\pipi$, $\KK$, and $\ppbar$ 
decays, respectively.  Since the $\hhbar$ decay modes 
have not been experimentally observed for 
the $\psi(3770)$, $\psi(4040)$, $\psi(4160)$, and $\psi(4415)$ resonances, 
the J/$\psi$ partial widths for the corresponding final states have been used.  
Table \ref{tab:vectorcharmresult} lists the number of $\hhbar$ events expected from 
the J/$\psi$, $\psi(3770)$, $\psi(4040)$, $\psi(4160)$, and $\psi(4415)$ resonances.  
The total number of $\pipi$, $\KK$, $\ppbar$ events from charmonium vector states, 
excluding $\psip$, are (6.6$\pm$0.8)$\times10^{-4}$, (4.9$\pm$0.6)$\times10^{-3}$, 
and 0.039$\pm$0.003, respectively.

\newpage
\clearpage

\begin{table}[ht]
\caption[Number of expected events from off-mass shell production of the J/$\psi$, 
$\psi(3770)$, $\psi(4040)$, $\psi(4160)$, and $\psi(4415)$ resonances.]
{Number of expected events from off-mass shell production of the J/$\psi$, 
$\psi(3770)$, $\psi(4040)$, $\psi(4160)$, and $\psi(4415)$ resonances.} 
\begin{center}
\begin{tabular}{|c|c|c|c|}
\hline
Resonance & $N_{\pipi}$ & $N_{\KK}$ & $N_{\ppbar}$ \\    
\hline
J/$\psi$     & (20$\pm$3)$\times10^{-5}$ & (15$\pm$2)$\times10^{-4}$ 
 & (117$\pm$7)$\times10^{-4}$  \\
$\psi(3770)$ & (32$\pm$7)$\times10^{-5}$ & (24$\pm$5)$\times10^{-4}$ 
 & (190$\pm$30)$\times10^{-4}$ \\
$\psi(4040)$ & (7.9$\pm$1.4)$\times10^{-5}$ & (6.0$\pm$1.0)$\times10^{-4}$ 
 & (46$\pm$5)$\times10^{-4}$ \\
$\psi(4160)$ & (4.3$\pm$0.8)$\times10^{-5}$ & (3.3$\pm$0.5)$\times10^{-4}$ 
 & (25.3$\pm$2.6)$\times10^{-4}$ \\
$\psi(4415)$ & (1.5$\pm$0.3)$\times10^{-5}$ & (1.14$\pm$0.22)$\times10^{-4}$ 
 & (8.9$\pm$1.3)$\times10^{-4}$ \\
\hline
Total & (6.6$\pm$0.8)$\times10^{-4}$ & (4.9$\pm$0.6)$\times10^{-3}$ 
 & 0.039$\pm$0.003 \\
\hline
\end{tabular}
\label{tab:vectorcharmresult}
\end{center}
\end{table}

\subsubsection{P-wave States, $\chi_{cJ}$}

Untagged two-photon fusion resonance production 
occurs when a photon is emitted by each incident 
electron and positron, the two photons ``fuse'' to form the resonance, 
while the scattered electron and positron are not detected.  Such two-photon events 
have some very distinct characteristics: the total observed 
energy is less than the center-of-mass energy of the two incident beams 
and, because the emitted photons are almost ``real'', 
i.e., they have little transverse momentum, 
the transverse momentum of the produced resonance and hence its decay products is small.  
The resonance produced by the two ``fused'' photons has positive C parity and is 
observed by fully reconstructing its decay into a particular channel. 

The total cross section for producing a resonance, $R$, in two-photon fusion 
is \cite{ggcs}
\begin{displaymath}
\sigma_{\gamma\gamma}(\sqrt{s},m_{R}) = 
\frac{8\alpha^{2}(2J+1)\Gamma_{\gamma\gamma}(R)}{m^{3}_{R}}
\left[\mathrm{f}\left(\frac{m^{2}_{R}}{s}\right)
\left(\ln\frac{s}{m^{2}_{e}}-1\right)^{2} 
- \frac{1}{3}\left(\ln\frac{s}{m^{2}_{R}}\right)^{3}\right],
\end{displaymath}
\begin{equation}
\mathrm{f}\left(\frac{m^{2}_{R}}{s}\right) = 
\left[\left(1 + \frac{m^{2}_{R}}{2s}\right)^{2} \ln\frac{s}{m^{2}_{R}}\right] 
- \left[\frac{1}{2}\left(1 - \frac{m^{2}_{R}}{s}\right)
\left(3 + \frac{m^{2}_{R}}{s}\right)\right]
\end{equation}

\hspace*{-0.75cm}
where $J$, $m_{R}$, and $\Gamma_{\gamma\gamma}(R)$ is the total angular 
momentum, mass, and two-photon partial width of the produced resonance, respectively.  

The number of expected events in the continuum data sample are determined 
by the expression 
\begin{equation}
N_{\hhbar} = \sigma_{\gamma\gamma}(\sqrt{s},m_{R})\cdot\epsilon_{\hhbar}
\cdot{\cal L}\cdot{\cal B}(\chi_{c0,2}\rightarrow\hhbar),
\end{equation}
where $\epsilon_{\hhbar}$ are 0.166, 0.743, and 0.657 for the $\pipi$, $\KK$, and $\ppbar$ 
decays, respectively.  
By using the same efficiencies as for the direct production of the $\hhbar$ final 
states, a conservative estimate of the number of events is determined 
since the efficiency should be lower due to the net momentum and signal region 
selection criteria.  
As is apparent in Figures \ref{fig:pi2snetp}a and \ref{fig:k2snetp}a, 
the $\hhbar$ final states from $\chi_{c0}$ and $\chi_{c2}$ will peak in their 
respective $X_{\pi,K}$ at 0.926 and 0.965, respectively.  
The branching ratios ${\cal B}(\chi_{c0,2}\rightarrow\hhbar)$ are given in 
Table \ref{tab:ggcharmbr}.  
Table \ref{tab:ggcharmresult} lists the results of the number of $\hhbar$ events 
expected from the $\chi_{c0}$ and $\chi_{c2}$ produced in untagged two-photon fusion.  
Under the assumption that all events from the decays of the $\chi_{c0}$ 
and $\chi_{c2}$ were to populate the corresponding signal regions,
the total number of $\pipi$, $\KK$, and $\ppbar$ events 
from $\chi_{c0}$ and $\chi_{c2}$ decays 
produced in untagged two-photon fusion are (2.2$\pm$0.4)$\times10^{-3}$, 
0.011$\pm$0.003, and (3.7$\pm$0.8)$\times10^{-4}$, respectively.

The total number of background events from neighboring charmonium states, 
except from $\psip$, i.e., the sum of the totals in Tables \ref{tab:vectorcharmresult} 
and \ref{tab:ggcharmresult},  
are (2.9$\pm$0.6)$\times10^{-3}$, 0.016$\pm$0.003, and 0.038$\pm$0.009 
for the $\pipi$, $\KK$, and $\ppbar$ final states, respectively, 
These values are considered negligible 
for the determination of the electromagnetic form factors.  
The $\psip$ decays to the $\hhbar$ final states are specifically 
treated in the Results section (Section 4.8).    

\begin{table}[h]
\caption[$\chi_{c0}$ and $\chi_{c2}$ branching ratios for $\pipi$, $\KK$, and $\ppbar$.]
{$\chi_{c0}$ and $\chi_{c2}$ branching ratios for $\pipi$, $\KK$, and $\ppbar$.  The 
values are from the PDG \cite{PDG2004}.} 
\begin{center}
\begin{tabular}{|c|c|c|c|}
\hline
Resonance & ${\cal B}(R\rightarrow\pipi)$ & ${\cal B}(R\rightarrow\KK)$ 
 & ${\cal B}(R\rightarrow\ppbar)$ \\    
\hline
$\chi_{c0}$ & (4.9$\pm$0.5)$\times10^{-3}$ & (6.0$\pm$0.9)$\times10^{-3}$ 
 & (2.24$\pm$0.27)$\times10^{-4}$ \\
$\chi_{c2}$ & (1.77$\pm$0.27)$\times10^{-3}$ & (9.4$\pm$2.1)$\times10^{-4}$ 
 & (6.8$\pm$0.7)$\times10^{-5}$ \\
\hline
\end{tabular}
\label{tab:ggcharmbr}
\end{center}
\end{table}

\begin{table}[h]
\caption[Number of expected events from untagged two-photon fusion production of 
the $\chi_{c0}$ and $\chi_{c2}$ resonances.]
{Number of expected events from untagged two-photon fusion production of 
the $\chi_{c0}$ and $\chi_{c2}$ resonances.} 
\begin{center}
\begin{tabular}{|c|c|c|c|}
\hline
Resonance & $N_{\pipi}$ & $N_{\KK}$ & $N_{\ppbar}$ \\    
\hline
$\chi_{c0}$ & (19$\pm$4)$\times10^{-4}$ & (100$\pm$30)$\times10^{-4}$
 & (33$\pm$8)$\times10^{-5}$ \\
$\chi_{c2}$ & (2.6$\pm$0.5)$\times10^{-4}$ & (6.2$\pm$1.6)$\times10^{-4}$ 
 & (4.0$\pm$0.6)$\times10^{-5}$ \\
\hline
Total & (2.2$\pm$0.4)$\times10^{-3}$ & 0.011$\pm$0.003 
 & (3.7$\pm$0.8)$\times10^{-4}$ \\
\hline
\end{tabular}
\label{tab:ggcharmresult}
\end{center}
\end{table}

\subsection{Common Collider Backgrounds}

Two types of backgrounds are common to all collider experiments.  
The first type of background 
is related to unwanted beam collisions with the residual gas in the 
beam pipe (called beam-gas) and with the beam pipe wall (beam-wall).  
The other type of background is due to 
cosmic rays, typically muons, traversing the detector.  
These backgrounds sources are observed in the IP variables $d_b$ and $z_b$ 
(see Section 4.4 for the definitions of $d_b$ and $z_b$) and the track variable 
$\phi_0$.  The variable $\phi_0$ is the azimuthal angle of the track 
defined in the plane perpendicular to the positron beam (or $z$-axis) 
and with respect to the $x$-axis, which points radially outward from 
the storage ring.  

\begin{figure}[!tb]
\begin{center}
\includegraphics[width=14cm]{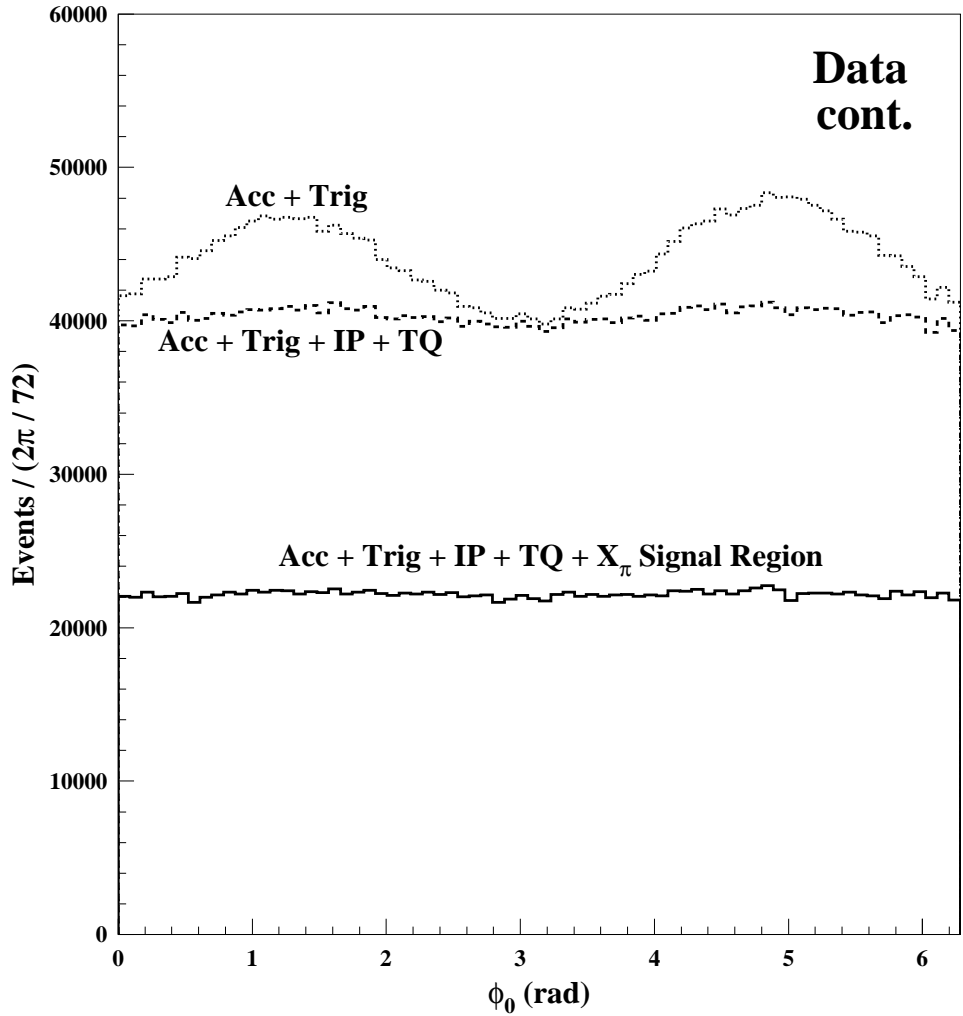}
\caption[Azimuthal angle ($\phi_0$) distribution of positive tracks 
in the continuum data sample with acceptance and trigger selection criteria applied.]
{Azimuthal angle ($\phi_0$) distribution of positive tracks 
in the continuum data sample.  
The dotted histogram are events which satisfy the acceptance (Acc) 
and trigger (Trig) criteria.  
The dashed histogram are events which satisfy the acceptance, trigger, IP, 
and track quality (TQ) criteria.  
The solid histogram are events in the $X_{\pi}$ signal region (0.98 $<$ $X_{p}$ $<$ 1.02) 
which satisfy the acceptance, trigger, IP, and track quality criteria.  Similar 
features are present in the $X_{K}$ and $X_{p}$ signal regions.}
\label{fig:phi0csm}
\end{center}
\end{figure}

Figure \ref{fig:phi0csm} shows the $\phi_0$ distribution for positive tracks
in continuum data events.  A cosmic ray typically traverses the detector 
in the vertical (or $y$) direction.  
This is apparent by the abundance of events at 
$\phi_0$ = $\pi/2$ = 1.57 radians (up) and $\phi_0$ = $3\pi/2$ = 4.71 radians (down).  
A majority of the cosmic ray events is removed by the IP and track quality 
criteria (dashed histogram).  

\begin{figure}[htbp]
\begin{center}
\includegraphics[width=14cm]{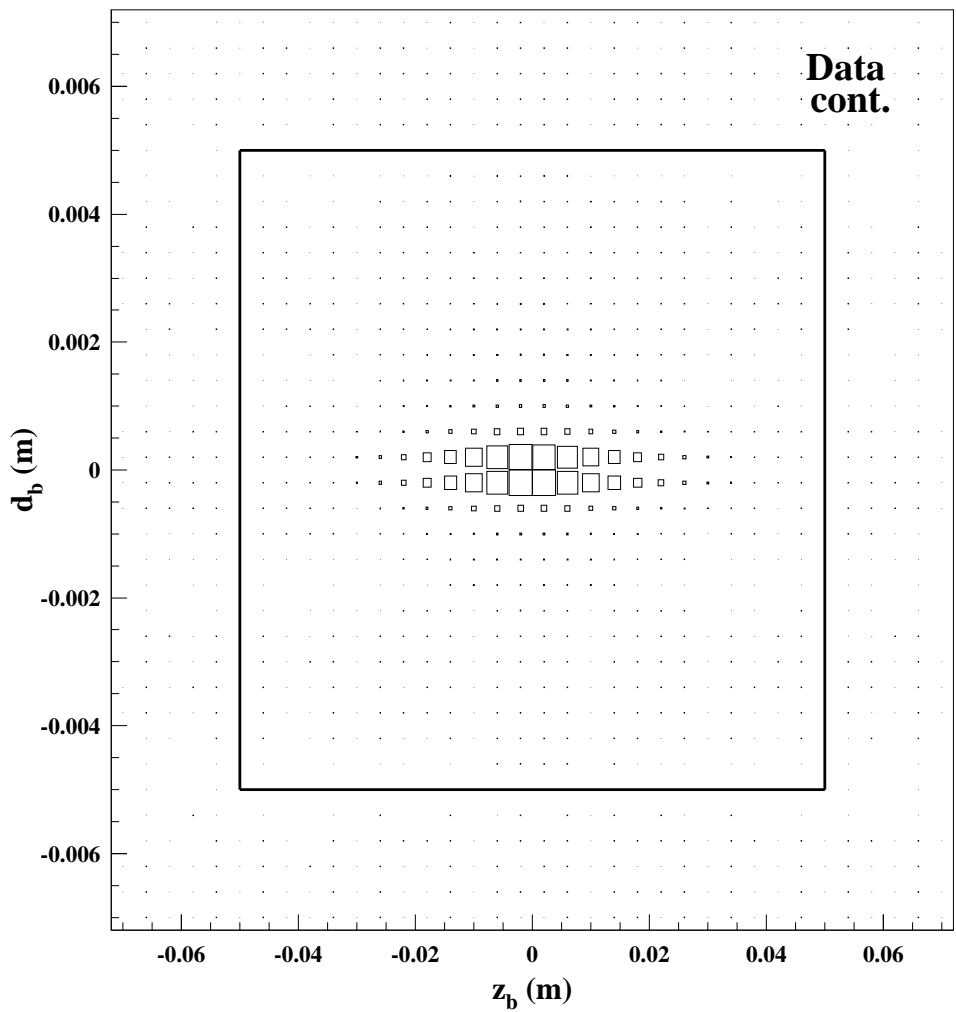}
\caption[Two-dimensional distribution of the IP variables $d_{b}$ and $z_{b}$.]
{Two-dimensional distribution of the IP variables $d_{b}$ and $z_{b}$ 
in the continuum data sample.  Only the acceptance and trigger selection criteria have been applied. 
The IP selection criteria accepts events if the tracks have 
$|d_{b}|$ $<$ 5 mm and $|z_{b}|$ $<$ 5 cm, as shown by the box.  
The outer boundary of the figure corresponds to a total volume which is three times 
larger than is enclosed by the inner box.}
\label{fig:dbzb}
\end{center}
\end{figure}

Figure \ref{fig:dbzb} shows the two-dimensional distribution of the 
IP variables $d_b$ and $z_b$ for events in the continuum data 
sample which pass the acceptance and trigger criteria.  
The background events surrounding the volume defined by the nominal IP region, 
$|d_{b}|$ $<$ 5 mm and $|z_{b}|$ $<$ 5 cm, are analyzed to test for possible 
contamination.   This background volume, which is arbitrarily taken to be 
twice as large as the nominal IP region, is bounded by 
$|d_{b}|$ $<$ 7.2 mm and $|z_{b}|$ $<$ 7.2 cm, 
while excluding the volume defined by the nominal IP region. 
After applying the respective $\hhbar$ final state criteria, 
no events are found in the corresponding $X_h$ signal regions 
originating from this sideband volume.  
Based on these studies, possible contamination from 
beam-gas, beam-wall, and cosmic ray backgrounds are considered negligible.

\newpage
\clearpage
\section{Radiative Corrections}

\begin{figure}[!tb]
\begin{center}
\includegraphics[width=12cm]{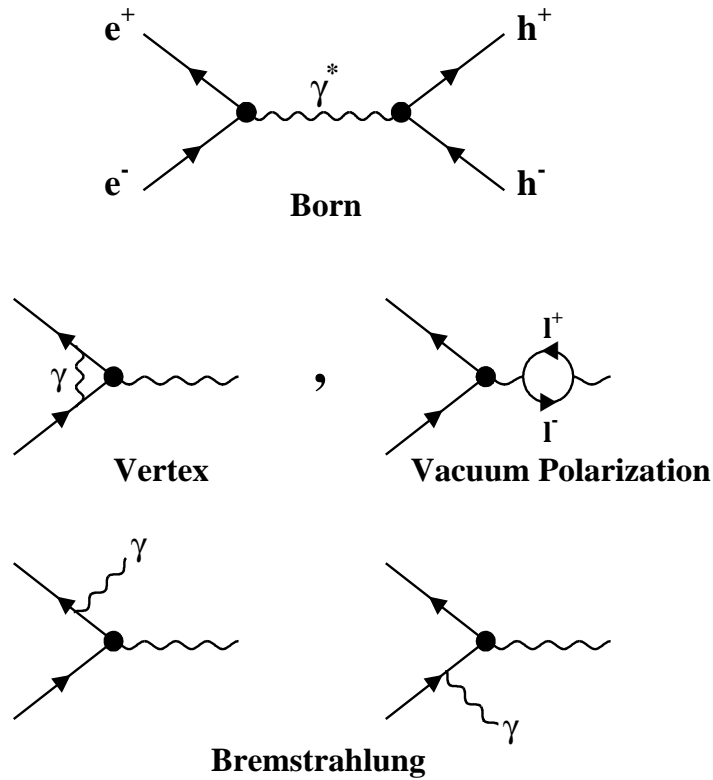}
\caption[Feynman diagrams associated with $\eetohhbar$.]
{Feynman diagrams associated with $\eetohhbar$.  The top figure is the 
Born, or tree-level, diagram.  The middle left figure is the $\ee$ annihilation 
vertex correction, the middle right figure is the vacuum polarization 
correction from lepton pair loops, and the bottom figures are the initial state radiation 
bremstrahlung corrections.}
\label{fig:isrfey} 
\end{center}
\end{figure}

The effect of initial state radiative corrections to the production 
of a virtual photon in $\ee$ annihilation needs to be determined to obtain the Born cross 
section for $\eetohhbar$ $(h = \pi^+,K^+,p)$ 
from the experimentally measured cross section.  Figure \ref{fig:isrfey} (top) 
shows the Born, or tree-level, Feynman diagram for hadron pair production.  
Figure \ref{fig:isrfey} also shows the Feynman diagrams for radiative corrections 
corresponding to the 
$\ee$ annihilation vertex correction, vacuum polarization of the virtual 
photon, and bremstrahlung radiation from the colliding $\ee$ pair.  

The radiative corrections are determined using the method 
of Bonneau and Martin \cite{bonneaumartinrc}, with the addition 
of $\mu$ and $\tau$ pair loops to the vacuum polarization correction.  
This is also the first method suggested by Berends and Kleiss \cite{berendskleissrc}.  
The Born cross section, $\sigma_{0}(s)$, is related to the experimental 
cross section, $\sigma_{exp}(s)$, by 
\begin{equation}
\sigma_{0}(s) = \frac{\sigma_{exp}(s)}{(1+\delta)}
\end{equation}
where $s$ is the center-of-mass energy 
squared of the initial $\ee$ system and $\delta$ is the radiative correction.  
The radiative correction is expressed as \cite{bonneaumartinrc}
\begin{displaymath}
\delta = \frac{2\alpha}{\pi}\left(\frac{3}{4}~\mathrm{ln} \frac{s}{m^{2}_{e}} 
+ \frac{\pi^{2}}{6} - 1\right) 
~+ \sum_{l=e,\mu,\tau} \frac{2\alpha}{\pi}
\left(\frac{1}{3}~\mathrm{ln} \frac{s}{m^{2}_{l}} - \frac{5}{9}\right)
~+~ t(\mathrm{ln} x_{min})
\end{displaymath}
\begin{equation}
+~ t\left[~\int^{x_{max}}_{x_{min}} \frac{dx}{x}\left(1 - x + \frac{x^{2}}{2}\right)
\frac{\sigma_{0}(s(1-x))}{\sigma_{0}(s)}
\frac{\epsilon(x)}{\epsilon(0)}~\right],
\label{eq:fullrc}
\end{equation}
where $\alpha$ is the fine-structure constant, $m_{e}$ is the electron mass, 
$x$ is the ratio of the bremstrahlung photon energy to the beam energy 
defined as $x$ = $\egam/E_{b}$, 
$\epsilon(x)$ is the experimental detection efficiency as a function of $x$, and 
\begin{equation}
t = \frac{2\alpha}{\pi}\left(\mathrm{ln} \frac{s}{m^{2}_{e}} - 1\right).  
\end{equation}
The first term in Eq. \ref{eq:fullrc} is the vertex correction, the second 
term is the sum of the leptonic loop contributions to the vacuum 
polarization correction, and the 
last two terms are associated with bremstrahlung radiation 
from the colliding $\ee$ pair.  Since the integral in the last 
term is infrared divergent, the third term corresponds to the 
low energy cutoff of the bremstrahlung integral.  
In order to evaluate the bremstrahlung integral, the efficiency ratio 
$\epsilon(x)/\epsilon(0)$ is determined from initial state 
radiation (ISR) MC samples and the cross section 
ratio $\sigma_{0}(s(1-x))/\sigma_{0}(s)$ is described in the 
following paragraphs.

As already mentioned in Eqn. \ref{eq:mffdiffcs}, 
the differential cross section for $\eetomm$ ($m$ = $\pi,K$) can be expressed 
as \cite{cabibbogatto} 
\begin{equation}
\frac{d\sigma_{0}(s)}{d\Omega} = \frac{\alpha^{2}}{8s}~\beta^{3}_{m}~
\mff^{2}~\mathrm{sin}^{2}\theta,
\end{equation}
where $\beta_{m}$ and $F_{m}(s)$ is the pseudoscalar meson velocity 
(in terms of c) measured in the laboratory system and 
electromagnetic form factor, respectively. 
Integrating over $\theta$, the total cross section is
\begin{equation}
\sigma_{0}(s) = \frac{\pi\alpha^{2}}{3s}~\beta^{3}_{m}~\mff^{2}.
\label{eq:mmthcs}
\end{equation}
The cross section ratio for $\eetomm$, 
using $\beta_{m}$ = $\sqrt{1 - (4m^{2}_{m}/s)}$, where $m_{m}$ is the mass of the 
meson, and making the PQCD assumption that $\mff \propto s^{-1}$, is 
\begin{equation}
\frac{\sigma_{0}(s(1-x))}{\sigma_{0}(s)} = 
\frac{1}{(1-x)^{9/2}}~\left[\frac{s(1-x) - 4m^{2}_{m}}{s - 4m^{2}_{m}}\right]^{3/2}.
\label{eq:kcrratio}
\end{equation}

As already mentioned in Eqn. \ref{eq:prrldiffcs}, 
the differential cross section for $\eetoppbar$ can be expressed as 
\begin{equation}
\frac{d\sigma_{0}(s)}{d\Omega} = \frac{\alpha^{2}}{4s}\beta_{p}
\left[\prmagff^{2}~(1 + \mathrm{cos}^{2}\theta) + 
\left(\frac{4m^{2}_{p}}{s}\right)\prelecff^{2}~(\mathrm{sin}^{2}\theta)\right].
\end{equation}
Integrating over $\theta$, the total cross section is
\begin{equation}
\sigma_{0}(s) = \frac{4\pi\alpha^{2}}{3s}\beta_{p}~\prmagff^{2}~
\left(1+\frac{2m^{2}_{p}}{s}\cdot r\right),
\label{eq:ppbarthcs}
\end{equation}
where $r = \prelecff^2/\prmagff^2$.  
The cross section ratio $\sigma_{0}(s(1-x))/\sigma_{0}(s)$ 
for $\eetoppbar$, using $\beta_{p}$ = $\sqrt{1 - (4m^{2}_{p}/s)}$ and 
making the PQCD assumption that $\prmagff \propto s^{-2}$, is
\begin{equation}
\frac{\sigma_{0}(s(1-x))}{\sigma_{0}(s)} = 
\frac{1}{(1-x)^{13/2}}~
\left[\frac{s(1-x) - 4m^{2}_{p}}{s - 4m^{2}_{p}}\right]^{1/2}~
\left[\frac{s(1-x) - 2m^{2}_{p}\cdot r}{s - 2m^{2}_{p}\cdot r}\right].
\label{eq:prcrratio}
\end{equation}

The efficiency ratio $\epsilon(x)/\epsilon(0)$ is determined by 
generating ISR MC samples for a particular final state as a function of 
bremstrahlung photon energy, $\egam$.  The ISR MC starts at the continuum center-of-mass 
energy and goes to a generic delta-function ``resonance'' 
and a single ISR photon.  This makes 
the reaction a two-body decay followed by the ``resonance'' decaying to the final state of 
interest.

The ISR MC sample consists of individual subsamples with $\egam$ fixed, 
starting with $\egam$ = 10 keV, and each subsequent subsample with $\egam$ 
increasing by 2 MeV.  There are 10,000 generated events in each individual sample. 
The value of $\epsilon(0)$, the efficiency with no ISR, is determined from a 
10,000 event MC sample generated with the initial center-of-mass energy 
fixed at $\sqrt{s}$ = 3.671 GeV.  
Note that all MC samples have final state radiation incorporated.  

\begin{figure}[htbp]
\begin{center}
\includegraphics[width=15cm]{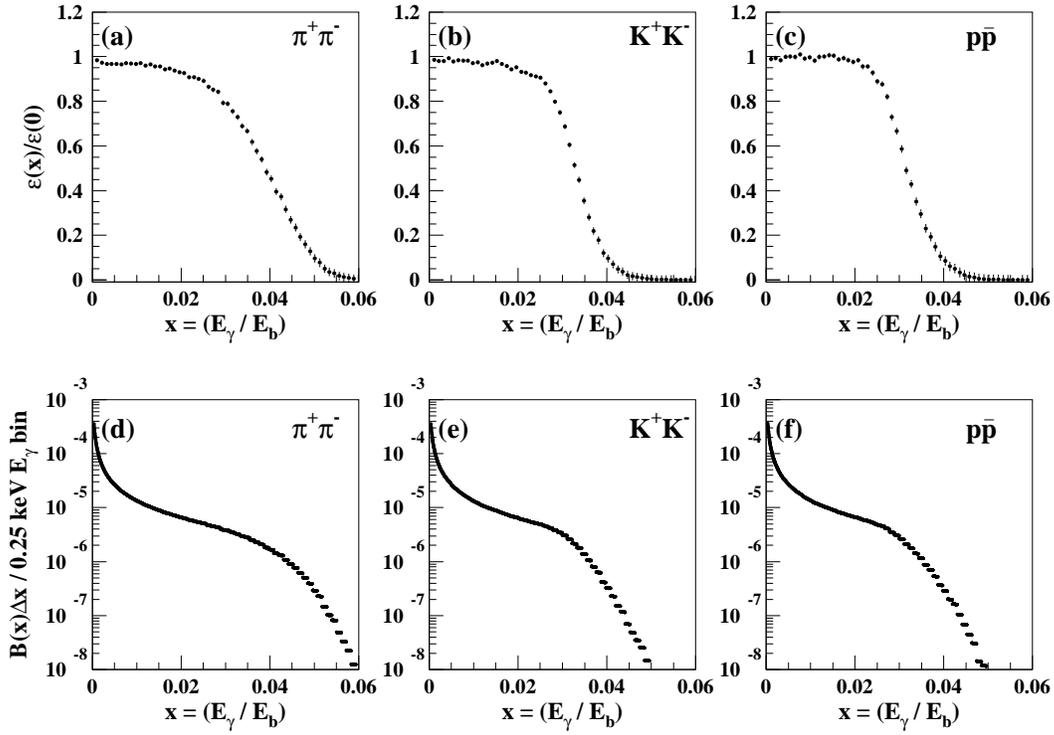}
\caption[Initial state radiative correction efficiency ratio and bremstrahlung 
integral distributions.]
{Efficiency ratio (top row) and bremstrahlung integral (bottom row) distributions.
The left, middle, and right columns are for the $\pipi$, $\KK$, and $\ppbar$ final states, 
respectively.  The bremstrahlung variable $B(x)$ is defined by Eqn. \ref{eq:bremintB}.}
\label{fig:rceffint} 
\end{center}
\end{figure}

Figures \ref{fig:rceffint}a, \ref{fig:rceffint}b, and \ref{fig:rceffint}c (top row) 
show $\epsilon(x)/\epsilon(0)$ as a function of $x$ 
for the $\pipi$, $\KK$, and $\ppbar$ final states, respectively.  
The detection efficiency drops to zero around $x$ = 0.05 ($E_{\gamma}$ = 92 MeV).  
Figures \ref{fig:rceffint}d, \ref{fig:rceffint}e, and \ref{fig:rceffint}f (bottom) 
show how the numerically-evaluated bremstrahlung integral in Eqn. \ref{eq:fullrc} 
varies as a function of $x$.  

\newpage
The radiative correction at $\sqrt{s}$ = 3.671 GeV, with a 
bremstrahlung energy cutoff of $E_{\gamma,min}$ = 10 keV 
($x_{min}$ = $5.448\times10^{-6}$), is 
\begin{displaymath}
1+\delta = 1+(0.0649) + (0.0330) + t(-12.1202) + t\Sigma B(x) \Delta x = 
0.154 + t\Sigma B(x) \Delta x.
\end{displaymath} 
where $t$ = 0.0779 and 
\begin{equation}
B(x) = \frac{1}{x}\left(1-x+\frac{x^{2}}{2}\right)
\frac{\sigma_{0}(s(1-x))}{\sigma_{0}(s)}\frac{\epsilon(x)}{\epsilon(0)}.  
\label{eq:bremintB}
\end{equation}
The radiative corrections to $\sigma_{0}(\eetopipi)$, $\sigma_{0}(\eetoKK)$, 
and $\sigma_{0}(\eetoppbar)$ are listed is Table \ref{tab:rcresults}.  
The radiative corrections have the effect of \textit{increasing} the 
experimental cross sections by 14-19$\%$.  

\begin{table}[h]
\caption[Radiative corrections for $\sigma_{0}(\eetohhbar)$ at $\sqrt{s}$ = 3.671 GeV.]
{Radiative corrections for $\sigma_{0}(\eetopipi)$, $\sigma_{0}(\eetoKK)$, 
and $\sigma_{0}(\eetoppbar)$ at $\sqrt{s}$ = 3.671 GeV.  
The variable $B(x)$ is defined by Eqn. \ref{eq:bremintB}.} 
\begin{center}
\begin{tabular}{|c|c|c|}
\hline
Final State & $\Sigma B(x) \Delta x$ & $1 + \delta$ \\    
\hline
\hline
$\pipi$   & 8.692 & 0.831 \\
\hline
$\KK$     & 8.426 & 0.810 \\
\hline
$\ppbar$  & 8.969 & 0.853 \\
$(\prelecff = 0)$ & & \\
\hline
$\ppbar$  & 9.057 & 0.860 \\
$(\prelecff = \prmagff)$ & & \\
\hline
\end{tabular}
\label{tab:rcresults}
\end{center}
\end{table}

\section{Results for the Form Factors}

In this presentation we use the following convention to specify the number of counts, $N$.  
The $\underline{\mathrm{superscripts}}$ '$\psip$' and 'cont' denote the source of the 
data: the $\psip$ and continuum data samples.  
The $\underline{\mathrm{subscripts}}$ 'obs', '$\leppair$', '$\psip$', and 'cont' denote 
the observed counts, and the contamination counts: 
the $\leppair$ background, the $\psip$, and the continuum contributions, respectively.

\subsection{Determination of ${\cal B}(\psiptohhbar)$}

Form factor measurements are made at energies removed from resonances which can be 
formed or decay into $\ee$, such as J$/\psi$ or $\psip$.  However, 
studying the decays of $\psiptohhbar$ ($h$ = $\pi^+,K^+,p$) 
serve two important purposes: 
they can be used to cross check the analysis procedure for measuring the 
form factors, and they are necessary for determining 
the amount of $\psip$ tail in the continuum data sample for a given 
final state.

The branching ratio for $\psiptohhbar$ is determined by
\begin{equation}
{\cal{B}}(\psiptohhbar) = 
\frac{N(\psiptohhbar)}{\epsilon_{h}N^{\psip}_{prod}},
\label{eq:psipbr}
\end{equation}
where the number of $\psiptohhbar$ signal events is
\begin{equation}
N(\psiptohhbar) = N^{\psip}_{obs} - N^{\psip}_{\leppair} - N^{\psip}_{cont},
\label{eq:psipN}
\end{equation}
and $N^{\psip}_{obs}$ is the number of observed events, 
$N^{\psip}_{\leppair}$ is the expected number of $\leppair$ background events, 
and $N^{\psip}_{cont}$ is the number of $\eetohhbar$ events, 
all for the $\psip$ data sample.  
The $\hhbar$ final state detection efficiency is denoted by $\epsilon_{h}$ 
and $N^{\psip}_{prod}$ = 1.52$\times10^{6}$ is the total number of $\psip$ 
events in the $\psip$ data sample \cite{numofpsip}.  
The number of $\eetohhbar$ events in the $\psip$ data sample 
from the continuum contribution is
\begin{equation}
N^{\psip}_{cont} = A_{n}\left(N^{cont}_{obs} - N^{cont}_{\leppair} - N^{cont}_{\psip}\right),
\label{eq:numcontinpsip}
\end{equation}
where $N^{cont}_{obs}$ is the number of observed events, 
$N^{cont}_{\leppair}$ is the expected number of $\leppair$ background events, 
and $N^{cont}_{\psip}$ is the number of $\psip$ events, all for the continuum data sample.  
The quantity $A_{n}$ in Eqn. \ref{eq:numcontinpsip} is the continuum scaling factor.  
It is defined as 
\begin{equation}
A_{n} = ({\cal{L}}_{\psip}/{\cal{L}}_{cont})\cdot(s_{cont}/s_{\psip})^{n}, 
\label{eq:contscfactor}
\end{equation}
where $n$ denotes the inverse s-dependence of the $\eetohhbar$ cross section, 
i.e., for $\pipi$ and $\KK$, $n$ = 3 ($\sigma_{0}(s) \propto s^{-3}$) gives 
$A_{3}$ = 0.136, and for $\ppbar$, $n$ = 5 ($\sigma_{0}(s) \propto s^{-5}$) 
gives $A_{5}$ = 0.134.  

In turn, the number of $\psip$ events in the continuum data is defined as 
\begin{equation}
N^{cont}_{\psip} = \CpipiJpsi(N^{\psip}_{obs} - N^{\psip}_{\leppair}),
\label{eq:psipincontN}
\end{equation}
where $N^{\psip}_{obs}$ and $N^{\psip}_{\leppair}$ are the number of observed events and 
the expected number of $\leppair$ background events in the $\psip$ data sample, 
respectively. 
The scale factor $\CpipiJpsi$ = 0.0072$\pm$0.0005 is the contamination from $\psip$ in the 
continuum data, as defined in Eqn. \ref{eq:psipcontamcorr}.  
At this level, the continuum content of the $\psip$ data itself is considered negligible.

\subsection{Determination of $\sigma_{0}(\eetohhbar)$}

The Born cross section for $\eetohhbar$ for the continuum data is determined by
\begin{equation}
\sigma_{0}(\eetohhbar) = 
\frac{N^{cont}(\eetohhbar)}{\epsilon_{h}(1+\delta){\cal{L}}_{cont}},
\label{eq:contcs}
\end{equation}

\newpage
\noindent
where the number of $\eetohhbar$ signal events is 
\begin{equation}
N^{cont}(\eetohhbar) = N^{cont}_{obs} - N^{cont}_{\leppair} - N^{cont}_{\psip},
\label{eq:contN}
\end{equation}
and $N^{cont}_{obs}$ is the number of observed events, 
$N^{cont}_{\leppair}$ is the expected number of $\leppair$ background events, 
and $N^{cont}_{\psip}$ is the number of $\psip$ contamination events, 
all for the continuum data sample.  
The $\hhbar$ final state detection efficiency and radiative correction are 
denoted by $\epsilon_{h}$  and $1+\delta$, respectively, 
while ${\cal{L}}_{cont}$ = 20.7 pb$^{-1}$ is the integrated luminosity 
for the continuum data sample.  
The radiative corrections are listed in Table \ref{tab:rcresults}.  
The number of $\psiptohhbar$ events in the continuum data is 
\begin{equation}
N^{cont}_{\psip} = 
\CpipiJpsi(N^{\psip}_{obs} - N^{\psip}_{\leppair} - N^{\psip}_{cont}),
\label{eq:nummeaspsip}
\end{equation}
where $N^{\psip}_{obs}$ is the number of observed events, 
$N^{\psip}_{\leppair}$ is the expected number of $\leppair$ background events, 
and $N^{\psip}_{cont}$ is the number of $\eetohhbar$ events, 
all for the $\psip$ data sample.  
The scale factor $\CpipiJpsi$ is the $\psip$ contamination in the 
continuum data, as defined in Eqn. \ref{eq:psipcontamcorr}.  
The number of $\eetohhbar$ events in the $\psip$ data is defined as 
\begin{equation}
N^{\psip}_{cont} = A_{n}\left(N^{cont}_{obs} - N^{cont}_{\leppair}\right),
\label{eq:continpsipN}
\end{equation}
where $N^{cont}_{obs}$ and $N^{cont}_{\leppair}$ are the number of observed events and 
the expected number of $\leppair$ background events in the continuum data sample, 
respectively.  The quantity $A_{n}$ is the continuum scaling factor, 
as defined in Eqn. \ref{eq:contscfactor}.  
At this level, the $\psip$ content in the continuum data itself is considered negligible.

\subsection{Determination of ${\cal B}(\psiptopipi)$}

\begin{figure}[htbp]
\begin{center}
\includegraphics[width=15.2cm]{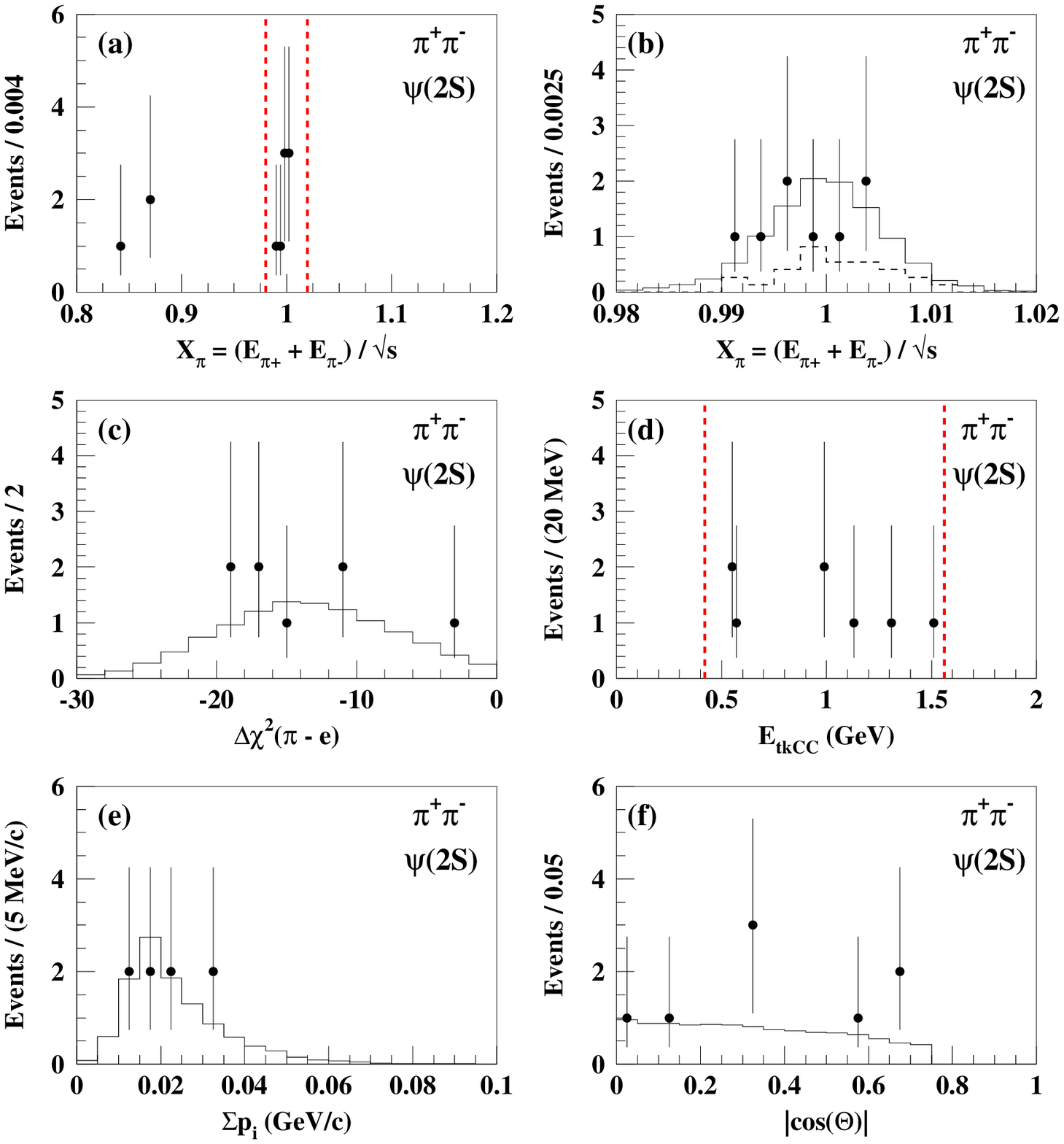}
\caption[The $\psip$ data with the $\pipi$ criteria applied.]
{The $\psip$ data with the $\pipi$ final state criteria applied.  
The points are data events.  The solid histograms are $\eetopipi$ signal MC, 
normalized to the number of observed events in the signal region.
Figure (a): $X_{\pi}$ distribution.  
The $X_{\pi}$ signal region is enclosed by the dashed lines. 
The points near $X_{\pi}$ = 0.85 are associated with $\psiptoppbar$ decays. 
Figure (b): $X_{\pi}$ signal region.  The dashed histogram is the continuum data 
scaled according to Eqn. \ref{eq:contscfactor}.
Figure (c):  $\Delta\chi^2(\pi-e)$ for the positive track.
Figure (d): $E_{tkCC}$ for the positive track.  
Figure (e): Net momentum of the two tracks.  
Figure (f): $|$cos($\theta$)$|$ for the positive track. }
\label{fig:ccpsip}
\end{center}
\end{figure}

Figure \ref{fig:pixwidemcdata} shows both the $X_{\pi}$ signal region and its 
vicinity after the $\pipi$ event selection criteria have been 
applied to the $\psip$ and continuum data samples.  
There are 8 events in the $X_{\pi}$ signal region for the $\psip$ data sample and 
26 events for the continuum data sample.  
Figure \ref{fig:ccpsip} shows that the $\psip$ data events are confined to the regions 
predicted by the signal MC.  

The number of $\eetopipi$ events in the $\psip$ data sample is 
\begin{displaymath}
N^{\psip}_{cont} = A_3[(26.0\pm5.1) - (0.108\pm0.013) - (0.057\pm0.024)] 
= 3.5\pm0.7.
\end{displaymath}  
The number of $\psiptopipi$ signal events is 
\begin{displaymath}
N(\psiptohhbar) = (8.0^{+3.3}_{-2.7}) - (3.5\pm0.7) - (0.021\pm0.002) 
= 4.5^{+3.4}_{-2.8}
\end{displaymath}  
From Eqn. \ref{eq:psipbr}, the $\psiptopipi$ branching ratio is
\begin{displaymath}
{\cal{B}}(\psiptopipi) = \frac{N(\psiptopipi)}{\epsilon_{h}N^{\psip}_{prod}} 
~~~~~~~~~~~~~~~~~~~~~~~~~~~~~~~~~~~~~~~~~~~~~~~~~~~
\end{displaymath}
\begin{displaymath}
= \frac{4.5^{+3.4}_{-2.8}}{(0.166)(1.52\times10^{6})}
= (1.8^{+1.3}_{-1.1}(stat))\times10^{-5}.
\end{displaymath}
This value of ${\cal{B}}(\psiptopipi)$ is a factor $\sim$4 smaller than the 
PDG value of ${\cal{B}}(\psiptopipi)$ = $(8\pm5)\times10^{-5}$ \cite{PDG2004}, 
but consistent with the recent BES I measurement of ${\cal{B}}(\psiptopipi)$ 
= $(0.84\pm0.55^{+0.16}_{-0.35})\times10^{-5}$ \cite{BESIpsi(2S)piK}.  
Our result is also only 1.6$\sigma$ above a null observation.

\subsection{Determination of $\sigma(\eetopipi)$ and $\piff$}

\begin{figure}[htbp]
\begin{center}
\includegraphics[width=15.2cm]{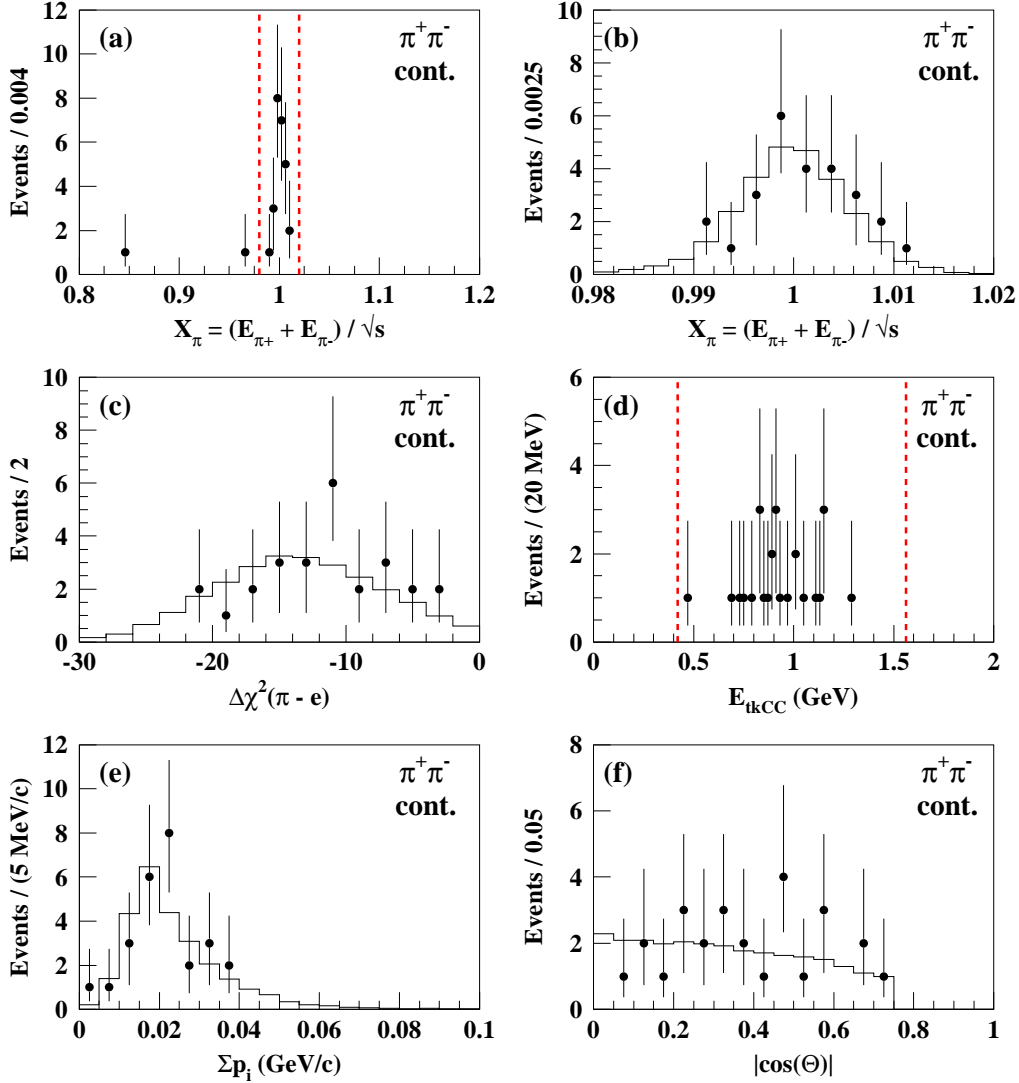}
\caption[The continuum data with the $\pipi$ criteria applied.]
{Continuum data sample with the $\pipi$ final state criteria applied.  
The points are data events.  The solid histogram are $\eetopipi$ signal MC, 
normalized to the number of observed events in the signal region.
Figure (a): $X_{\pi}$ distribution.  
The $X_{\pi}$ signal region is enclosed by the dashed lines. 
Figure (b): $X_{\pi}$ signal region.
Figure (c):  $\Delta\chi^2(\pi-e)$ for the positive track.
Figure (d): $E_{tkCC}$ for the positive track.  
Figure (e): Net momentum of the two tracks.  
Figure (f): $|$cos($\theta$)$|$ for the positive track. }
\label{fig:picont}
\end{center}
\end{figure}

Figure \ref{fig:picont} shows the event distributions for several 
variables for the continuum data and signal MC 
after applying the $\pipi$ event selection criteria.  
There are 26 observed events in the $\pipi$ signal region.  
The data events are distributed in accord with the signal MC predictions.

The values of all measured quantities for the 26 observed events 
are listed in Tables A.1-10 of Appendix A.  Two interesting features are present.  
Table \ref{tab:piapptbl7} lists the 
$dE/dx$+RICH PID variable $\Delta\chi^2(\pi-\mu)$, and it is apparent that it does not 
contain discriminating power between pions and muons.  As seen in 
Table \ref{tab:piapptbl10}, one pion candidate track has an associated signal in the 
muon detector.  A pion track will decay to a muon $\sim$3$\%$ of the time 
according to the $\eetopipi$ signal MC.  
Therefore, one pion decaying to a muon, out of 52 candidates, 
is consistent with the MC prediction.

The number of $\psiptopipi$ events in the continuum data sample is 
\begin{displaymath}
N^{cont}_{\psip} = \CpipiJpsi[(8.0^{3.3}_{2.7}) - (0.021\pm0.002) - (3.5\pm0.7)]  
= 0.03\pm0.02.
\end{displaymath}
The number of $\eetopipi$ signal events is 
\begin{displaymath}
N^{cont}(\eetopipi) = (26.0\pm5.1) - (0.108\pm0.013) - (0.03\pm0.02) 
= 25.9\pm5.1.
\end{displaymath}
Therefore, using Eqn. \ref{eq:contcs}, the cross section for $\eetopipi$ is
\begin{displaymath}
\sigma_{0}(\eetopipi) 
= \frac{N^{cont}(\eetopipi)}{\epsilon_{h}(1+\delta){\cal{L}}_{cont}}
~~~~~~~~~~~~~~~~~~~~~~~~~~~~~~~~~~~~~~~~~~~~~~~~~~~
\end{displaymath}
\begin{displaymath}
= \frac{25.9\pm5.1}{(0.166)(0.832)(20.7~\mathrm{pb^{-1}})} 
= 9.0\pm1.8~\mathrm{pb}.
\end{displaymath}
As stated in Eqn. \ref{eq:mffdiffcs}, the cross section is related to the pion 
electromagnetic form factor by \cite{cabibbogatto}
\begin{equation}
\sigma_{0}(\eetopipi) = \frac{\pi\alpha^{2}\beta^{3}_{\pi}}{3s}\piffsq.
\end{equation}
Inserting the constants in the equation above, 
the pion form factor at $\sqrt{s}$ = 3.671 GeV is
\begin{equation}
|F_{\pi}(13.48~\mathrm{GeV}^{2})| 
= \sqrt{\frac{9.0\pm1.8~\mathrm{pb}}{1598~\mathrm{pb}}} = 0.075\pm0.008
\end{equation}
where the errors are statistical only.

\subsection{Determination of ${\cal B}(\psiptoKK)$}

\begin{figure}[htbp]
\begin{center}
\includegraphics[width=15.2cm]{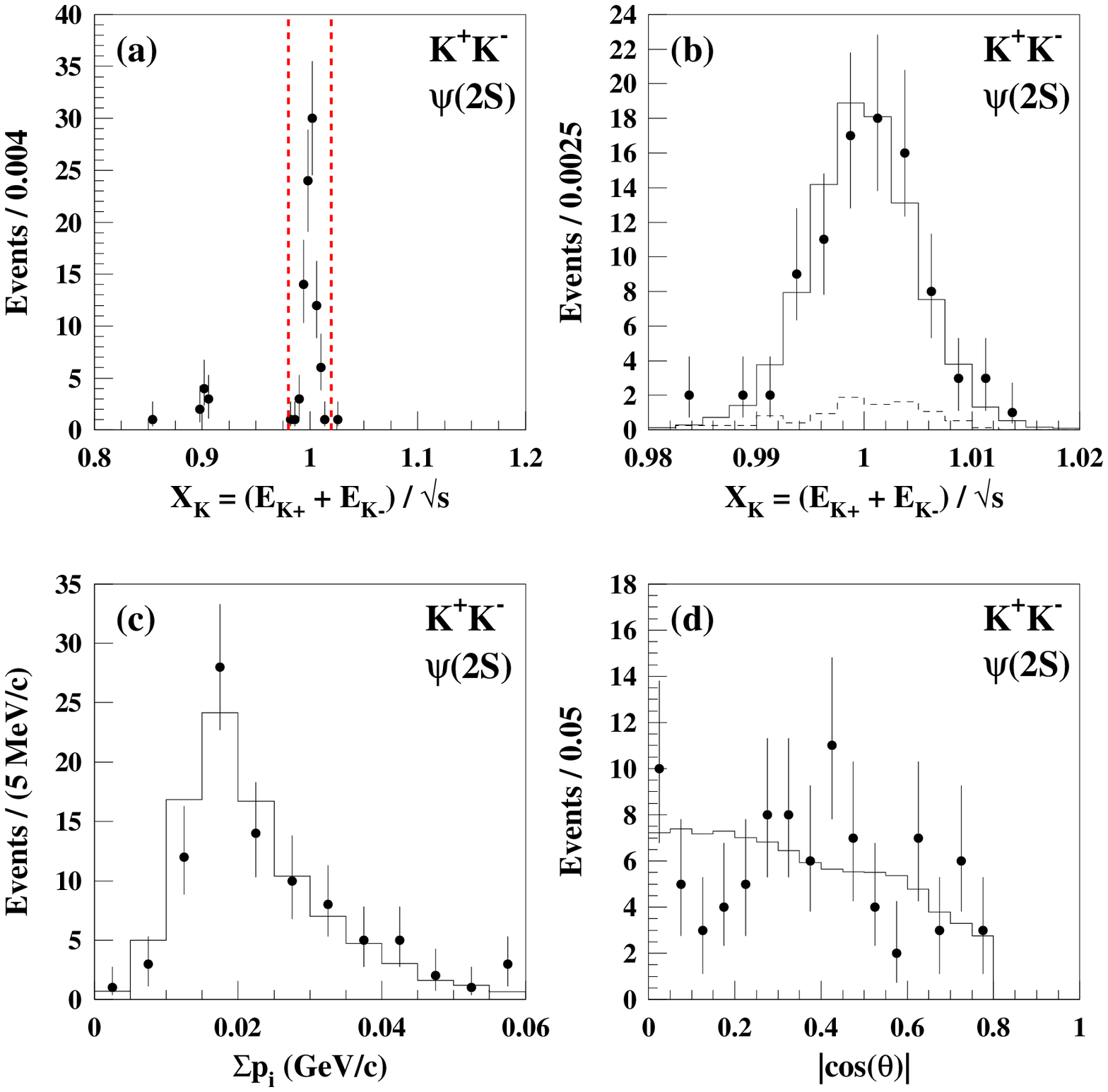}
\caption[The $\psi(2S)$ data with the $\KK$ criteria applied.]
{The $\psip$ data with the $\KK$ final state criteria applied.  
The points are data events.  The solid histograms are the $\eetoKK$ signal MC, 
normalized to the number of observed events in the signal region.
Figure (a): $X_{K}$ distribution.  
The $X_{K}$ signal region is enclosed by the dashed lines. 
The points near $X_{K}$ = 0.90 are associated with $\psiptoppbar$ decays. 
Figure (b): $X_{K}$ signal region.  The dashed histogram is the continuum data 
scaled according to Eqn. \ref{eq:contscfactor}.
Figure (c): Net momentum of the two tracks.  
Figure (d): $|$cos($\theta$)$|$ for the positive track. }
\label{fig:kpsip}
\end{center}
\end{figure}

Figure \ref{fig:kxwidemcdata} shows both the $X_{K}$ signal region and its 
vicinity after the $\KK$ event selection criteria have been 
applied to the $\psi(2S)$ and continuum data sample.  
There are 92 events in the $X_{K}$ signal region for the $\psip$ data sample 
and 72 events for the continuum data sample.  
Figure \ref{fig:kpsip} shows that the distributions of the $\psip$ data events 
are in agreement with the $\eetoKK$ signal MC predictions.  

The number of $\eetoKK$ events in the $\psip$ data sample is 
\begin{displaymath}
N^{\psip}_{cont} = A_3[(72.0\pm8.5) - (0.56^{+0.21}_{-0.18}) - (0.66\pm0.08)] 
= 9.6\pm1.2.
\end{displaymath}  
The number of $\psiptoKK$ signal events is 
\begin{displaymath}
N(\psiptoKK) = (92.0\pm9.6) - (0.06^{+0.10}_{-0.04}) - (9.6\pm1.2) 
= 82.3\pm9.7.
\end{displaymath}
From Eqn. \ref{eq:psipbr}, the $\psiptoKK$ branching ratio is  
\begin{displaymath}
{\cal{B}}(\psiptoKK) = \frac{82.3\pm9.7}{(0.743)(1.52\times10^{6})}
= (7.3\pm0.9(stat))\times10^{-5}.
\end{displaymath}  
This value of ${\cal{B}}(\psiptoKK)$ is consistent with the 
PDG value of ${\cal{B}}(\psiptoKK)$ = $(10\pm7)\times10^{-5}$ \cite{PDG2004} and 
the BES I measurement of ${\cal{B}}(\psiptoKK)$ 
= $(6.1\pm1.4^{+1.5}_{-1.3})\times10^{-5}$  \cite{BESIpsi(2S)piK}.

\subsection{Determination of $\sigma(\eetoKK)$ and $\kff$}

\begin{figure}[htbp]
\begin{center}
\includegraphics[width=15.2cm]{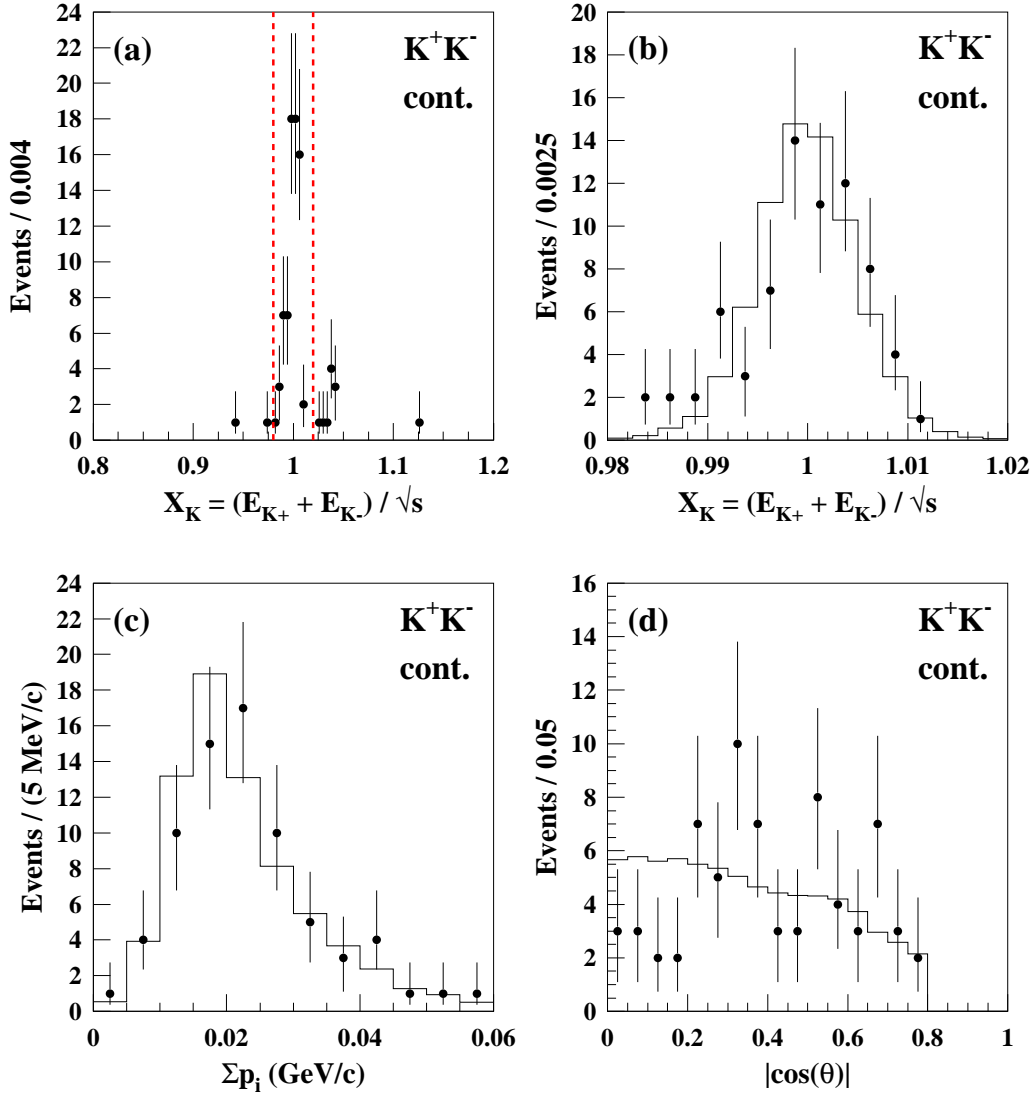}
\caption[The continuum data with the $\KK$ criteria applied.]
{Continuum data sample with the $\KK$ final state criteria applied.  
The points are data events.  The solid histograms are $\eetoKK$ signal MC, 
normalized to the number of observed events in the signal region.
Figure (a): $X_{K}$ distribution.  
The $X_{K}$ signal region is enclosed by the dashed lines. 
The points near $X_{K}$ = 1.03 are associated with $\eetoll$ events. 
Figure (b): $X_{K}$ signal region.
Figure (c): Net momentum of the two tracks.  
Figure (d): $|$cos($\theta$)$|$ for the positive track. }
\label{fig:kcont}
\end{center}
\end{figure}

Figure \ref{fig:kcont} shows the event distributions for several 
variables for the continuum data and signal MC 
after applying the $\KK$ event selection criteria.  
There are 72 observed events in the $\KK$ signal region.  The 
distributions of data events are in agreement with the signal MC predictions.

The number of $\psiptoKK$ events in the continuum data sample is 
\begin{displaymath}
N^{cont}_{\psip} = \CpipiJpsi[(92.0\pm9.6) - (0.06^{+0.10}_{-0.04}) - (9.7\pm1.2)]  
= 0.59\pm0.08.
\end{displaymath}
The number of $\eetoKK$ signal events is 
\begin{displaymath}
N^{cont}(\eetoKK) = (72.0\pm8.5) - (0.56^{+0.21}_{-0.18}) - (0.59\pm0.08) 
= 70.9\pm8.5.
\end{displaymath}
Therefore, using Eqn. \ref{eq:contcs}, the cross section for $\eetoKK$ is
\begin{displaymath}
\sigma_{0}(\eetoKK) 
= \frac{N^{cont}(\eetoKK)}{\epsilon_{h}(1+\delta){\cal{L}}_{cont}}
~~~~~~~~~~~~~~~~~~~~~~~~~~~~~~~~~~~~~~~~~~~~~~~~~~~
\end{displaymath}
\begin{displaymath}
= \frac{70.9\pm8.5}{(0.743)(0.810)(20.7~\mathrm{pb^{-1}})} 
= 5.7\pm0.7~\mathrm{pb}.
\end{displaymath}
As stated in Eqn. \ref{eq:mffdiffcs}, the cross section is related to the kaon 
electromagnetic form factor by \cite{cabibbogatto}
\begin{equation}
\sigma_{0}(\eetoKK) = \frac{\pi\alpha^{2}\beta^{3}_{K}}{3s}\kffsq.
\end{equation}
Inserting the constants in the equation above, 
the kaon form factor at $\sqrt{s}$ = 3.671 GeV is
\begin{equation}
|F_{K}(13.48~\mathrm{GeV}^{2})| 
= \sqrt{\frac{5.7\pm0.7~\mathrm{pb}}{1440~\mathrm{pb}}} = 0.063\pm0.004
\end{equation}
where the errors are statistical only.

\subsection{Determination of ${\cal B}(\psiptoppbar)$}

\begin{figure}[htbp]
\begin{center}
\includegraphics[width=15.2cm]{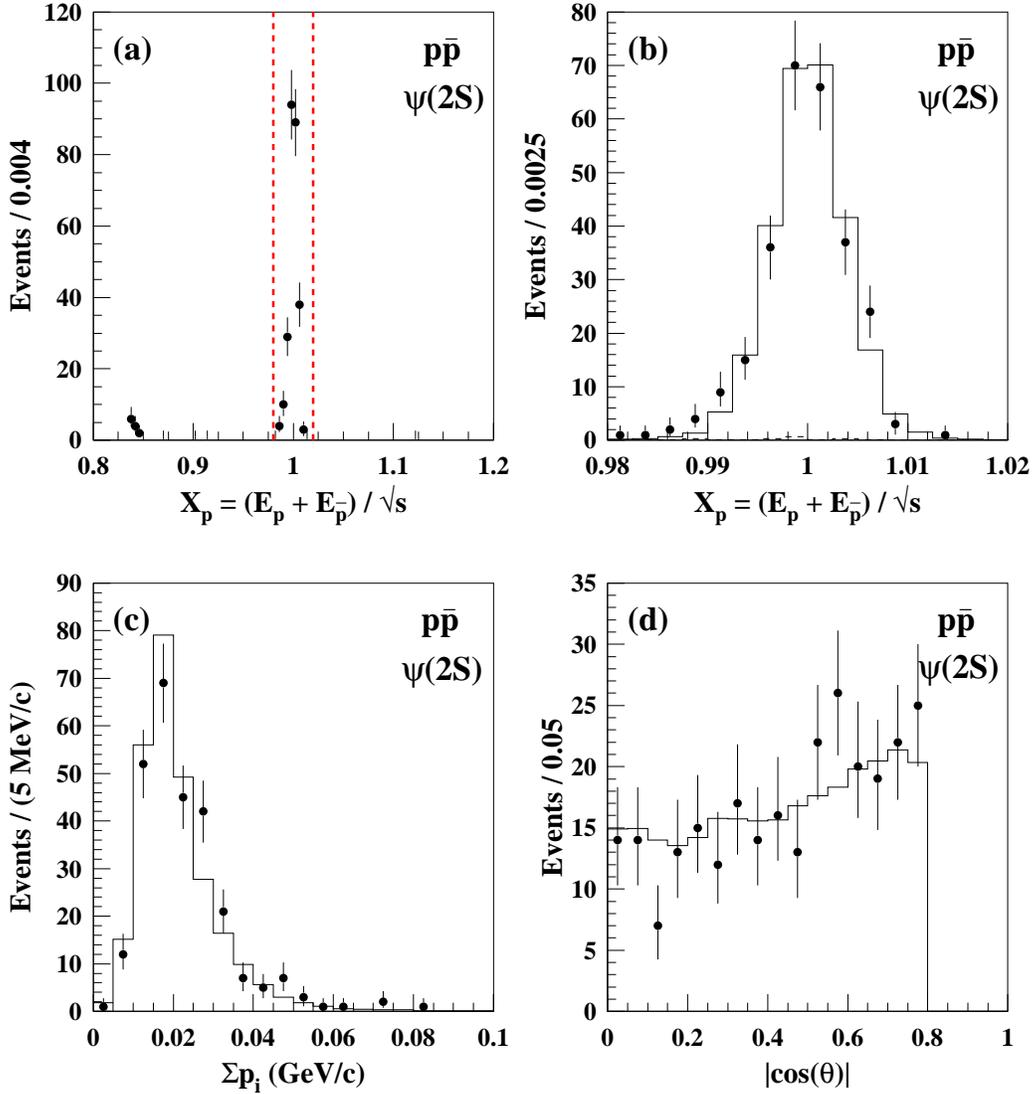}
\caption[The $\psi(2S)$ data with the $\ppbar$ criteria applied.]
{The $\psip$ data with the $\ppbar$ final state criteria applied.  
The points are data events.  The solid histograms are $\eetoppbar$ signal MC, 
normalized to the number of observed events in the signal region.
Figure (a): $X_{p}$ distribution.  
The $X_{p}$ signal region is enclosed by the dashed lines. 
Figure (b): $X_{p}$ signal region.  The dashed histogram is the continuum data 
scaled according to Eqn. \ref{eq:contscfactor}.
Figure (c): Net momentum of the two tracks.  
Figure (d): $|$cos($\theta$)$|$ for the positive track. }
\label{fig:prpsip}
\end{center}
\end{figure}

Figure \ref{fig:prxwidemcdata} shows both the $X_{p}$ signal region and its 
vicinity after the $\ppbar$ event selection criteria have been 
applied to the $\psi(2S)$ and continuum data samples. 
There are 269 events in the $X_{p}$ signal region for the $\psip$ data 
sample and 16 events for the continuum data sample.  
Figure \ref{fig:prpsip} compares the distributions of $\psip$ data events 
with the $\eetoppbar$ signal MC prediction calculated with $\prelecff$ = 0.  
There is good agreement between the data and MC predictions.

The number of $\eetoppbar$ events in the $\psip$ data sample is 
\begin{displaymath}
N^{\psip}_{cont} = A_5[(16.0^{+4.8}_{-3.7}) - (1.94\pm0.18)] 
= 1.9\pm0.6.
\end{displaymath}
Note that the $\leppair$ contamination is negligible in both the continuum and 
$\psip$ data samples, as discussed at the end of Section 4.5.3.  
The number of $\psiptoppbar$ signal events is 
\begin{displaymath}
N(\psiptoppbar) = (269.0\pm16.4) - (1.9\pm0.6) 
= 267.1\pm16.4.
\end{displaymath}
From Eqn. \ref{eq:psipbr}, the $\psiptoppbar$ branching ratio is  
\begin{displaymath}
{\cal{B}}(\psiptoppbar) = \frac{267.1\pm16.4}{(0.626)(1.52\times10^{6})}
= (2.81\pm0.17(stat))\times10^{-4}.
\end{displaymath}  
This value of ${\cal{B}}(\psiptoppbar)$ is 2.4$\sigma$ larger than the PDG 
value of ${\cal{B}}(\psiptoppbar)$ = $(2.07\pm0.31)\times10^{-4}$ \cite{PDG2004} but is 
in good agreement with the E760 result 
${\cal{B}}(\psiptoppbar)$ = $(2.61^{+0.31}_{-0.21}\pm0.17\pm0.17)\times10^{-4}$ 
\cite{e835_psipppbar}.  Assuming $\prelecff$ = $\prmagff$ decreases the branching ratio 
by 0.04$\%$.

\subsection{Determination of $\sigma(\eetoppbar)$ and $\prmagff$}

\begin{figure}[htbp]
\begin{center}
\includegraphics[width=15.2cm]{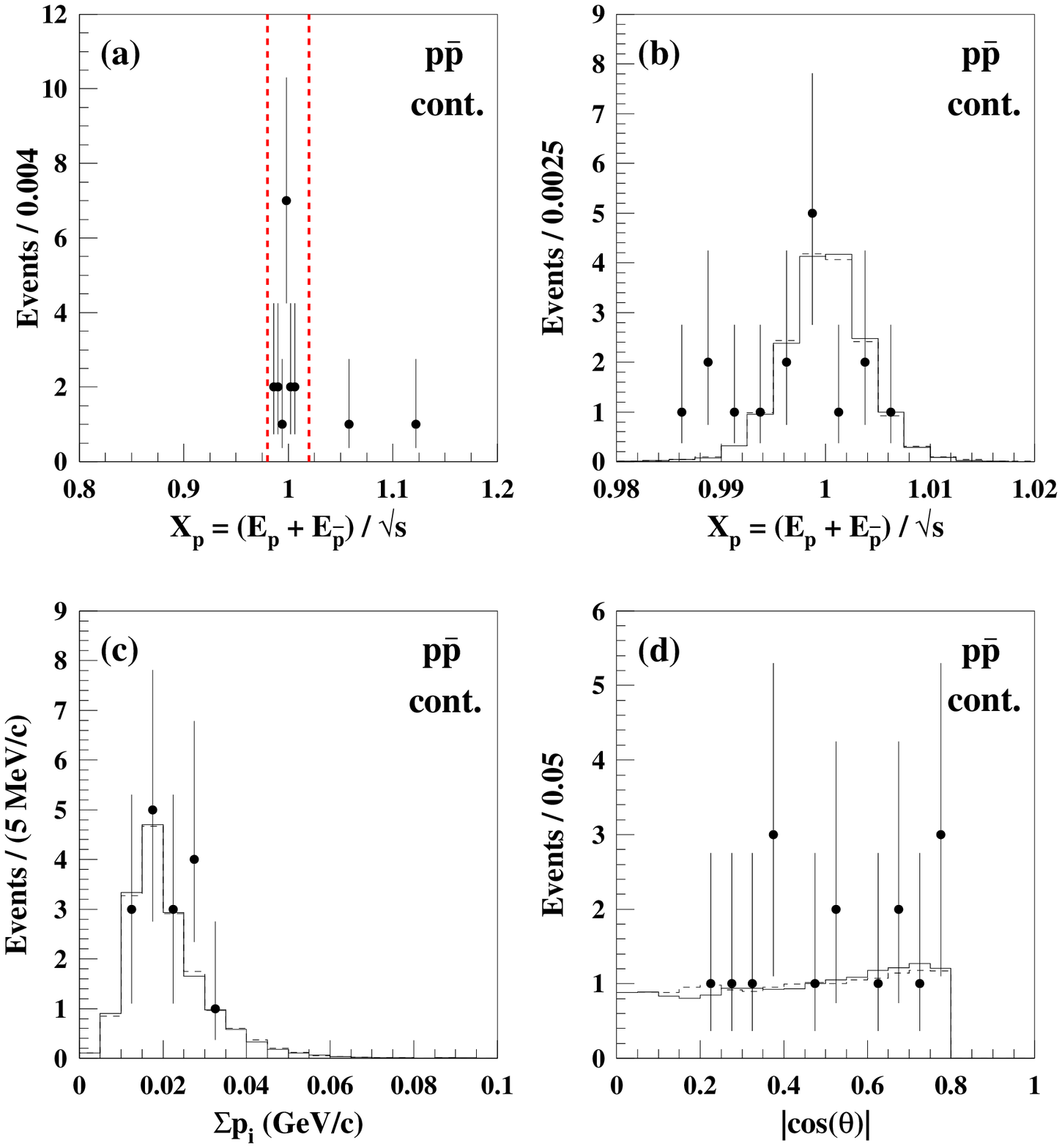}
\caption[The continuum data with the $\ppbar$ criteria applied.]
{Continuum data sample with the $\ppbar$ final state criteria applied.  
The points are data events. 
The solid histograms are the $\eetoppbar$ signal MC obtained with $\prelecff$ = 0, 
and the dashed histograms are the $\eetoppbar$ signal MC obtained with 
$\prelecff$ = $\prmagff$ .  The signal MC samples are normalized to the number of 
observed events in the signal region.
Figure (a): $X_{p}$ distribution.  
The $X_{p}$ signal region is enclosed by the dashed lines. 
Figure (b): $X_{p}$ signal region.
Figure (c): Net momentum of the two tracks.  
Figure (d): $|$cos($\theta$)$|$ for the positive track. }
\label{fig:prcont}
\end{center}
\end{figure}

Figure \ref{fig:prcont} shows the event distributions for several 
variables for the continuum data and signal MC after applying the 
$\ppbar$ event selection criteria under the two assumptions $\prelecff$ = 0 
and $\prelecff$ = $\prmagff$.  There are 16 observed events in the $\ppbar$ signal region.  
The distributions of the data events are in agreement with both MC predictions, 
which differ very little.

The values of all measured quantities for the 16 observed events 
are listed in Tables A.11-20 of Appendix A.  
The $E_{tkCC}/p$ values for the antiproton candidates are listed in 
Table \ref{tab:prapptbl10}.  The $E_{tkCC}/p$ $<$ 0.85 requirement is applied 
only to the protons, and not to the antiprotons because of 
their possible annihilation in the CC.  Table \ref{tab:prapptbl10} shows that 
5 of the 16 antiproton candidates have an $E_{tkCC}/p$ $>$ 0.85, 
consistent with the expectation of annihilation.

The number of $\psiptoppbar$ events in the continuum data sample is 
\begin{displaymath}
N^{cont}_{\psip} = \CpipiJpsi[(269.0\pm16.4) - (2.1^{+0.6}_{-0.5})]  
= 1.92\pm0.18.
\end{displaymath}  
Note that the $\leppair$ contamination is negligible in both the continuum and 
$\psip$ data samples, as discussed at the end of Section 4.5.3.  
The number of $\eetoppbar$ signal events is 
\begin{displaymath}
N^{cont}(\eetoppbar) = (16.0^{+4.8}_{-3.7}) - (1.92\pm0.18) 
= 14.1^{+4.8}_{-3.7}.
\end{displaymath}
Therefore, using Eqn. \ref{eq:contcs}, the cross section for $\eetoppbar$ is 
\begin{displaymath}
\sigma_{0}(\eetoppbar) = 
\frac{14.1^{+4.8}_{-3.7}}
{(0.626)(0.853)(20.7~\mathrm{pb^{-1}})} = 
1.27^{+0.43}_{-0.33}~\mathrm{pb},~~~~~~~~~~~~~(~\prelecff = 0~)
\end{displaymath}
and
\begin{displaymath}
\sigma_{0}(\eetoppbar) = 
\frac{14.1^{+4.8}_{-3.7}}
{(0.657)(0.860)(20.7~\mathrm{pb^{-1}})} = 
1.20^{+0.41}_{-0.31}~\mathrm{pb},~~~~(~\prelecff = \prmagff~).
\end{displaymath}
Solving Eqn. \ref{eq:ppbarthcs} for $\prmagff$ gives 
\begin{equation}
\prmagff = \sqrt{\frac{3s}{4\pi\alpha^{2}}
\frac{\sigma_{0}(\eetoppbar)}{\beta_{p}(1+\frac{2m^{2}_{p}}{s}
\frac{\prelecff^{2}}{\prmagff^{2}})}}.
\end{equation}  
The magnetic form factor of the proton is therefore 
\begin{displaymath}
|G^{P}_{M}(13.48~\mathrm{GeV}^{2})| = 
\sqrt{\frac{1.27^{+0.43}_{-0.33}~\mathrm{pb}}{5542~\mathrm{pb}}} 
= 0.0152^{+0.0026}_{-0.0020},~~~~~~~~~~\mathrm{assuming}~(~\prelecff = 0~),
\end{displaymath}
and
\begin{displaymath}
|G^{P}_{M}(13.48~\mathrm{GeV}^{2})| = 
\sqrt{\frac{1.20^{+0.41}_{-0.31}~\mathrm{pb}}{6266~\mathrm{pb}}} 
= 0.0139^{+0.0024}_{-0.0018},~\mathrm{assuming}~(~\prelecff = \prmagff~).
\end{displaymath}
where all errors are statistical only.

Due to the low statistics, a definitive measurement of the 
$\prelecff/\prmagff$ ratio is not possible from our data.  A rough measurement can be 
done by calculating the ratio of the number of events with 
$|$cos$\theta|<0.5$ to the events with $0.5<|$cos$\theta|<0.8$ from the continuum data 
and comparing it to the ratio obtained from the 
$\prelecff$ = 0 and $\prelecff$ = $\prmagff$ MC samples.  
The experimental ratio is found to be 0.78$^{+0.49}_{-0.38}$.  The MC results are 
1.282$\pm$0.023 for $\prelecff$ = 0, and 1.416$\pm$0.025 for $\prelecff$ = $\prmagff$.  
It is clear that our data does not have the statistical precision to distinguish between 
the two.

\section{Systematic Uncertainties}

Various sources of systematic uncertainty arise from possible biases in 
the event selection criteria, the treatment of the leptonic background, 
the treatment $\psip$ contamination in the continuum data sample, the 
statistical uncertainty in the detection efficiency, the bremstrahlung energy 
cutoff in the radiative correction, and the uncertainty in the 
luminosity.

The trigger uncertainty was studied in Ref. \cite{psipcontamination} 
for events with only two hard tracks and determined to be at most 
0.62$\%$.  A conservative estimate of the trigger uncertainty 
is taken to be 1$\%$.  
  
Tracking uncertainties was studied in Ref. \cite{dhadsyst} and, 
again for the case of two hard tracks, in Ref. \cite{psipcontamination}.  
A tracking uncertainty of 0.7$\%$ was found for charged kaons, 
but a study of hard $\mu$ tracks in $\psip$ $\rightarrow$ $X$ $J/\psi$, 
$J/\psi$ $\rightarrow$ $\mu^{+}\mu^{-}$ decays determined an uncertainty 
of 1.0$\%$.  The tracking uncertainty per track is taken to be 1$\%$.  

The biases imposed by the signal region, net momentum, and $E_{tkCC}$ 
criteria are studied by varying the cuts, recalculating the efficiency, 
leptonic background, and radiative correction, and stating the difference 
in the cross section as a systematic uncertainty.  
Tables \ref{tab:pionkinsyst} and \ref{tab:kaonkinsyst} list their effect on 
$\sigma_{0}(\eetopipi)$ and $\sigma_{0}(\eetoKK)$, respectively.  For 
$\sigma_{0}(\eetoppbar)$,  Table \ref{tab:prge0kinsyst} lists their effect on 
$\eetoppbar$ events with the assumption $\prelecff$ = 0 and 
Table \ref{tab:prgegmkinsyst} lists their effect on 
$\eetoppbar$ events with the assumption $\prelecff$ = $\prmagff$.  
The final results are listed in Table \ref{tab:systematics}.

\begin{table}[h]
\caption[Systematic uncertainties in $\sigma_{0}(\eetopipi)$ from signal region, 
net momentum, and $E_{tkCC}$ variation.]
{Systematic uncertainties in $\sigma_{0}(\eetopipi)$ from signal region, 
net momentum, and $E_{tkCC}$ variation.  The larger variation is taken as the systematic 
uncertainty.} 
\begin{center}
\begin{tabular}{|c|c|c|}
\hline
$X_{\pi}$ Signal Region & $\sigma_{0}(\eetopipi)$ (pb) & Change ($\%$) \\    
\hline
$0.98 < X_{\pi} < 1.02$ (nominal) & 9.045 & --- \\
$0.985 < X_{\pi} < 1.015$         & 9.416 & $+4.1$ \\
$0.975 < X_{\pi} < 1.025$         & 8.846 & $-2.2$ \\ 
\hline
\hline
Net momentum (MeV/$c$) &  &  \\    
\hline
$\Sigma p_{i} <$ 100 (nominal) & 9.045 & --- \\
$\Sigma p_{i} <$ 60            & 9.479 & $+4.8$ \\
$\Sigma p_{i} <$ 150           & 9.009 & $-0.4$ \\
\hline
\hline
$E_{tkCC}$ (MeV) &  &  \\    
\hline
$E_{tkCC} <$ 420 (nominal) & 9.045  & --- \\
$E_{tkCC} <$ 540           & 10.013 & $+10.7$ \\
$E_{tkCC} <$ 380           & 8.737  & $-3.4$ \\
\hline
\end{tabular}
\label{tab:pionkinsyst}
\end{center}
\end{table}

\begin{table}[h]
\caption[Systematic uncertainties in $\sigma_{0}(\eetoKK)$ from signal region and 
net momentum variation.]
{Systematic uncertainties in $\sigma_{0}(\eetoKK)$ from signal region 
and net momentum variation.  
The larger variation is taken as the systematic uncertainty.} 
\begin{center}
\begin{tabular}{|c|c|c|}
\hline
$X_{K}$ Signal Region & $\sigma_{0}(\eetoKK)$ (pb) & Change ($\%$) \\    
\hline
$0.98 < X_{K} < 1.02$ (nominal) & 5.687 & --- \\
$0.985 < X_{K} < 1.015$         & 5.657 & $-0.5$ \\
$0.975 < X_{K} < 1.025$         & 5.666 & $-0.4$ \\ 
\hline
\hline
Net momentum (MeV/$c$) &  &  \\    
\hline
$\Sigma p_{i} <$ 60 (nominal) & 5.687 & --- \\
$\Sigma p_{i} <$ 50           & 5.770 & $+1.5$ \\
$\Sigma p_{i} <$ 80           & 5.541 & $-2.6$ \\
\hline
\end{tabular}
\label{tab:kaonkinsyst}
\end{center}
\end{table}

\begin{table}[h]
\caption[Systematic uncertainties in $\sigma_{0}(\eetoppbar)$ for $\prelecff$ = 0 
from signal region and net momentum variation.]
{Systematic uncertainties in $\sigma_{0}(\eetoppbar)$ for $\prelecff$ = 0 
from signal region and net momentum variation.  
The larger variation is taken as the systematic uncertainty.} 
\begin{center}
\begin{tabular}{|c|c|c|}
\hline
$X_{p}$ Signal Region & $\sigma_{0}(\eetoppbar)$ (pb) & Change ($\%$) \\    
\hline
$0.98 < X_{p} < 1.02$ (nominal) & 1.274 & --- \\
$0.985 < X_{p} < 1.015$         & 1.321 & $+3.7$ \\
$0.975 < X_{p} < 1.025$         & 1.248 & $-2.0$ \\ 
\hline
\hline
Net momentum (MeV/$c$) &  &  \\    
\hline
$\Sigma p_{i} <$ 100 (nominal) & 1.274 & --- \\
$\Sigma p_{i} <$ 60            & 1.326 & $+4.1$ \\
$\Sigma p_{i} <$ 150           & 1.360 & $+6.8$ \\
\hline
\end{tabular}
\label{tab:prge0kinsyst}
\end{center}
\end{table}

\begin{table}[h]
\caption[Systematic uncertainties in $\sigma_{0}(\eetoppbar)$ for $\prelecff$ = $\prmagff$ 
from signal region and net momentum variation.]
{Systematic uncertainties in $\sigma_{0}(\eetoppbar)$ for $\prelecff$ = $\prmagff$ 
from signal region and net momentum variation.  
The larger variation is taken as the systematic uncertainty.} 
\begin{center}
\begin{tabular}{|c|c|c|}
\hline
$X_{p}$ Signal Region & $\sigma_{0}(\eetoppbar)$ (pb) & Change ($\%$) \\    
\hline
$0.98 < X_{p} < 1.02$ (nominal) & 1.204 & --- \\
$0.985 < X_{p} < 1.015$         & 1.249 & $+3.7$ \\
$0.975 < X_{p} < 1.025$         & 1.179 & $-2.0$ \\ 
\hline
\hline
Net momentum (MeV/$c$) &  &  \\    
\hline
$\Sigma p_{i} <$ 100 (nominal) & 1.204 & --- \\
$\Sigma p_{i} <$ 60            & 1.252 & $+4.0$ \\
$\Sigma p_{i} <$ 150           & 1.287 & $+6.9$ \\
\hline
\end{tabular}
\label{tab:prgegmkinsyst}
\end{center}
\end{table}

The uncertainty in the $dE/dx$+RICH PID criteria is determined by studying a substantial 
statistical sample of pion, kaon, and proton tracks at the desired momenta.  The 
only source of such samples are from $D^{0}$ $\rightarrow$ $K^{-}\pi^{+}$ 
decays for charged pions and kaons and 
$\Lambda$ $\rightarrow$ $p\pi$ decays for protons 
with data taken at $\sqrt{s}$ = 10.58 GeV with the CLEO III detector 
(note that charge conjugation is implied for $D^{0}$ and $\Lambda$ and 
their decays).  The PID uncertainty is studied by calculating the 
efficiencies for finding $D^{0}$s and $\Lambda$s using the method and 
resources described in Ref. \cite{CLEOIIIPIDsyst}.  
Either the mean value, or the uncertainty, of the difference between 
the CLEO III data and MC efficiency for a particular particle type is taken 
as the systematic uncertainty per track.  
For pions with track momenta of $\sim$1.83 GeV/c, the difference is $-2.7\pm1.9\%$; 
for kaons with track momenta of $\sim$1.77 GeV/c, the difference is $+0.2\pm1.2\%$; and
for protons with track momenta of $\sim$1.58 GeV/c, the difference is $-1.2\pm1.6\%$.  
Therefore, the pion, kaon, and proton PID criteria is assigned a 
2.7$\%$, 1.2$\%$, and 1.6$\%$ uncertainty per track, respectively.

The uncertainty in the pion identification efficiency from the 
$E_{tkCC}$ criteria is 2.3$\%$ per track and is described in Section 4.5.1.  

Uncertainties from the $\psip$ contamination, leptonic background, and 
statistical uncertainty in efficiency determination due to finite 
MC sample sizes is determined by individually varying the mean values 
by $\pm1\sigma$.    The uncertainty in the bremstrahlung energy cutoff in 
the radiative correction is determined by varying the cutoff energy by a
factor of two, i.e., changing $E_{\gamma,min}$ to 5 keV and 20 keV. The 
results are listed in Table \ref{tab:systematics}.

The determination of the continuum data luminosity 
systematic uncertainty is documented in Ref. \cite{contlumcorr} and 
is determined to be 1$\%$.  

The individual and total systematic uncertainties for 
$\sigma_{0}(\eetohhbar)$ are summarized in Table \ref{tab:systematics}.  
The total uncertainty is determined by a sum in quadrature of all 
individual contributions.  
The total systematic uncertainty on $\sigma_{0}(\eetopipi)$ is 14.6$\%$, 
in $\sigma_{0}(\eetoKK)$ it is 4.4$\%$, in $\sigma_{0}(\eetoppbar)$ 
for $\prelecff$ = 0 it is $8.8\%$ and for $\prelecff$ = $\prmagff$ it is $8.9\%$.  

\begin{table}[h]
\caption[Sources of systematic uncertainty for $\sigma_{0}(\eetohhbar)$.]
{Sources of systematic uncertainty for $\sigma_{0}(\eetohhbar)$.  
The values are listed as percentages.  The total uncertainty is determined by a sum 
in quadrature of all individual contributions.} 
\begin{center}
\begin{tabular}{|c|c|c|c|c|}
\hline
\hline
Source & $\pipi$ & $\KK$ & $\ppbar$ & $\ppbar$ \\ 
& & & $(\frac{\prelecff}{\prmagff} = 0)$ & $(\frac{\prelecff}{\prmagff} = 1)$ \\    
\hline
\hline
Trigger & 1.0 & 1.0 & 1.0 & 1.0 \\
\hline
Tracking & 2$\times$1.0 & 2$\times$1.0 & 2$\times$1.0 & 2$\times$1.0 \\
\hline
$X_{h}$ Signal Region & 4.1 & 0.5 & 3.7 & 3.7 \\
\hline
Net Momentum & 4.8 & 2.6 & 6.8 & 6.9 \\
\hline
$E_{tkCC}$ & 10.7 & --- & --- & --- \\
\hline
$E_{tkCC}/p$ & $>$0.1 & $>$0.1 & $>$0.1 & $>$0.1  \\
\hline
$dE/dx$+RICH PID & 2$\times$2.7 & 2$\times$1.2 & 2$\times$1.6 & 2$\times$1.6 \\
\hline
$\epsilon_{\pi}(E_{tkCC})$ & 2$\times$2.3 & --- & --- & --- \\
\hline
$\psip$ Contam & $>$0.1 & 0.1 & 0.9 & 1.0 \\
\hline
Leptonic BG & $>$0.1 & 0.3 & $>$0.1 & $>$0.1 \\
\hline
MC statistics & 1.3 & 0.4 & 0.5 & 0.5 \\
\hline
$x_{min}$ & 0.2 & 0.2 & 0.2 & 0.2 \\
\hline
$\cal{L}$ & 1.0 & 1.0 & 1.0 & 1.0 \\
\hline
\hline
Total ($\%$) & 14.6 & 4.4 & 8.8 & 8.9 \\
\hline
\hline
\end{tabular}
\label{tab:systematics}
\end{center}
\end{table}

\baselineskip=24pt
\chapter{Conclusions}

\begin{figure}[htbp]
\begin{center}
\includegraphics[width=15cm]{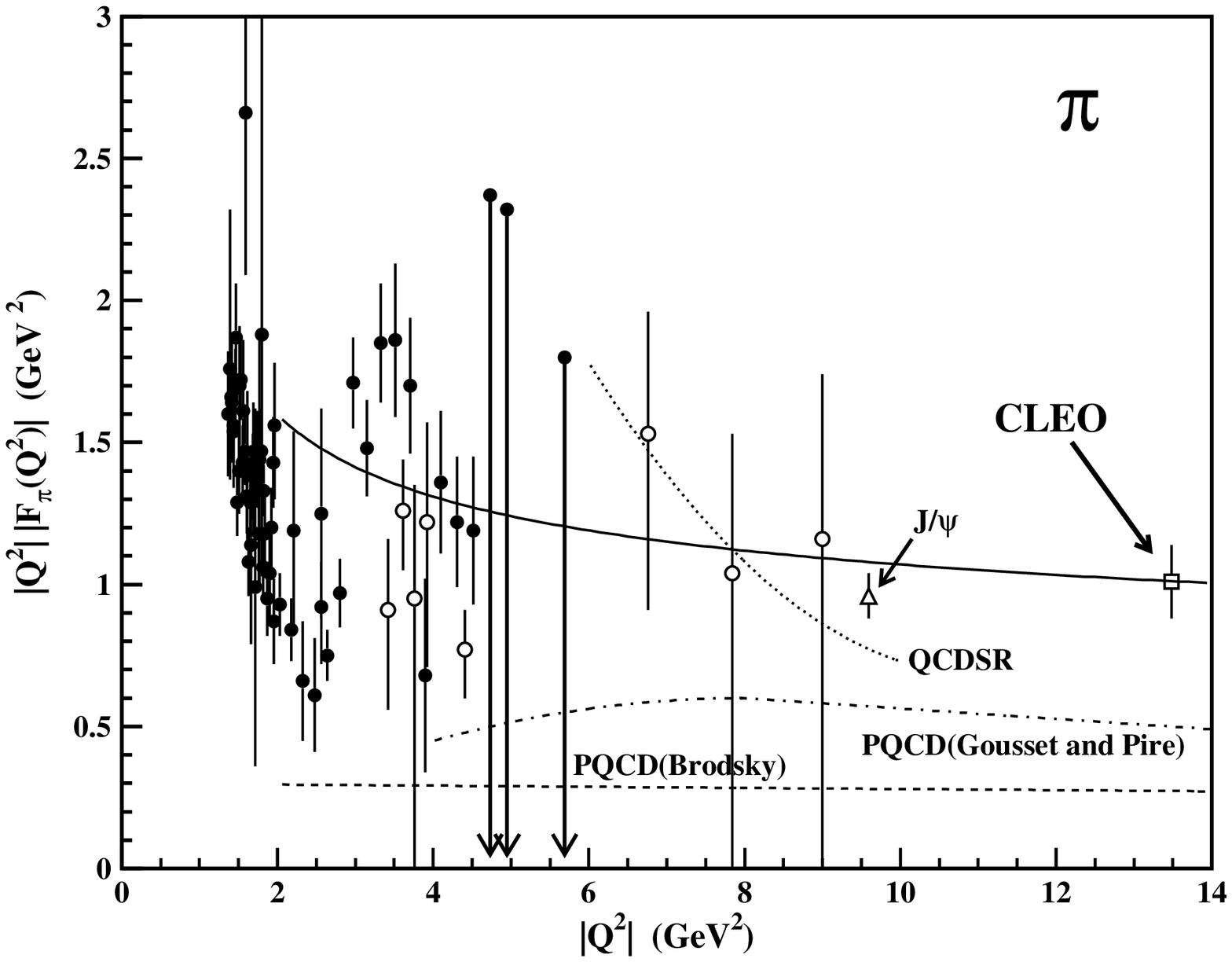}
\caption[Experimental status of the pion form factors with timelike momentum transfer.]
{Experimental status of the pion form factors with timelike momentum transfer.  
The solid points are from $\eetopipi$ measurements with pions experimentally identified 
\cite{kpicommontlff_1}-\cite{pitlff_3}.  The open points are from $\eetohh$ 
measurements with the pion fraction of the observed $\hadpair$ determined according to 
a VDM prescription \cite{pitlff_VDM}. 
The value denoted with the triangle comes from interpreting the 
$J/\psi \rightarrow \pi^{+}\pi^{-}$ branching ratio as a pion form factor measurement 
as in Ref. \cite{Milanaetal_jpsipipi}.  The result from the present analysis is denoted 
by the open square.  The arbitrarily normalized solid line shows the variation of 
$\alpha_{s}(|Q^{2}|)$ using its two-loop form with $n_{f}$ = 4 and $\Lambda$ = 0.322 GeV.  
The dashed and dash-dotted lines are the timelike PQCD predictions by 
Brodsky $\etal$ \cite{Brodskyetal_tlPQCD} and 
Gousset and Pire \cite{GoussetPire_tlPQCD}, respectively.  
The dotted line is the timelike QCD Sum Rules (QCDSR) 
prediction by Bakulev $\etal$ \cite{Bakulevetal_tlQCDSR}.}
\label{fig:pifinaltl} 
\end{center}
\end{figure}
\begin{figure}[htbp]
\begin{center}
\includegraphics[width=15cm]{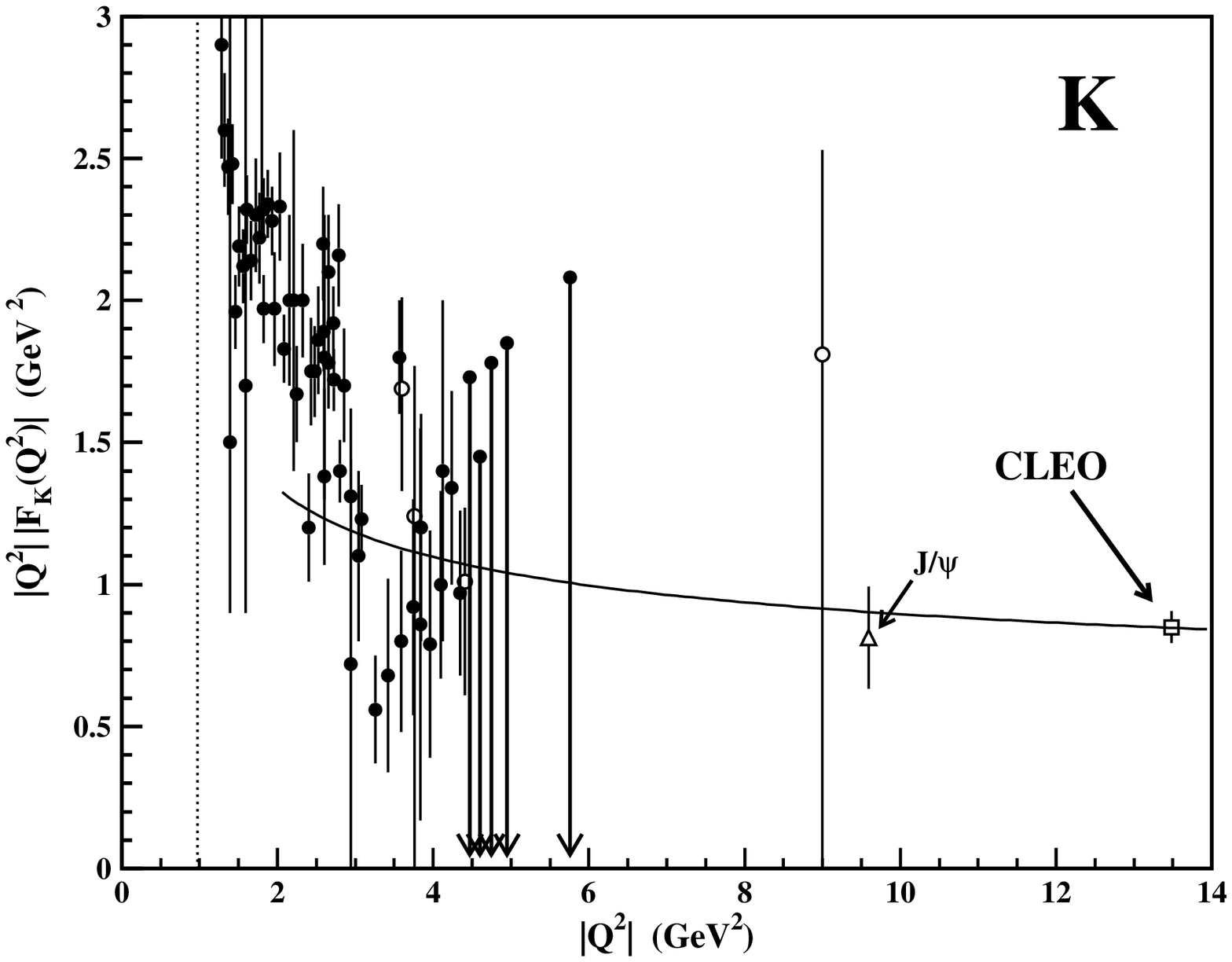}
\caption[Experimental status of the kaon form factors with timelike momentum transfer.]
{Experimental status of the kaon form factors with timelike momentum transfer.  
The solid points are from $\eetoKK$ measurements with kaons experimentally identified 
\cite{kpicommontlff_1}-\cite{kpicommontlff_3}, 
\cite{ktlff_1}-\cite{ktlff_4}. The open points are from $\eetohh$ 
measurements with the kaon fraction of the observed $\hadpair$ determined according to 
a VDM prescription \cite{ktlff_VDM}.
The value denoted with the triangle comes from interpreting the 
$J/\psi \rightarrow \KK$ decay via a virtual photon  
as a kaon form factor measurement, as described in the text.  
The result from the present analysis is denoted by the open square.  
The vertical dotted line specifies the threshold for $\KK$ production, 
i.e., $|Q^{2}|$ = $(2m_{K})^{2}$ = 0.975 GeV$^{2}$.  
The arbitrarily normalized solid line shows the variation of $\alpha_{s}(|Q^{2}|)$ 
using its two-loop form with $n_{f}$ = 4 and $\Lambda$ = 0.322 GeV.}
\label{fig:kfinaltl} 
\end{center}
\end{figure}
\begin{figure}[htbp]
\begin{center}
\includegraphics[width=15cm]{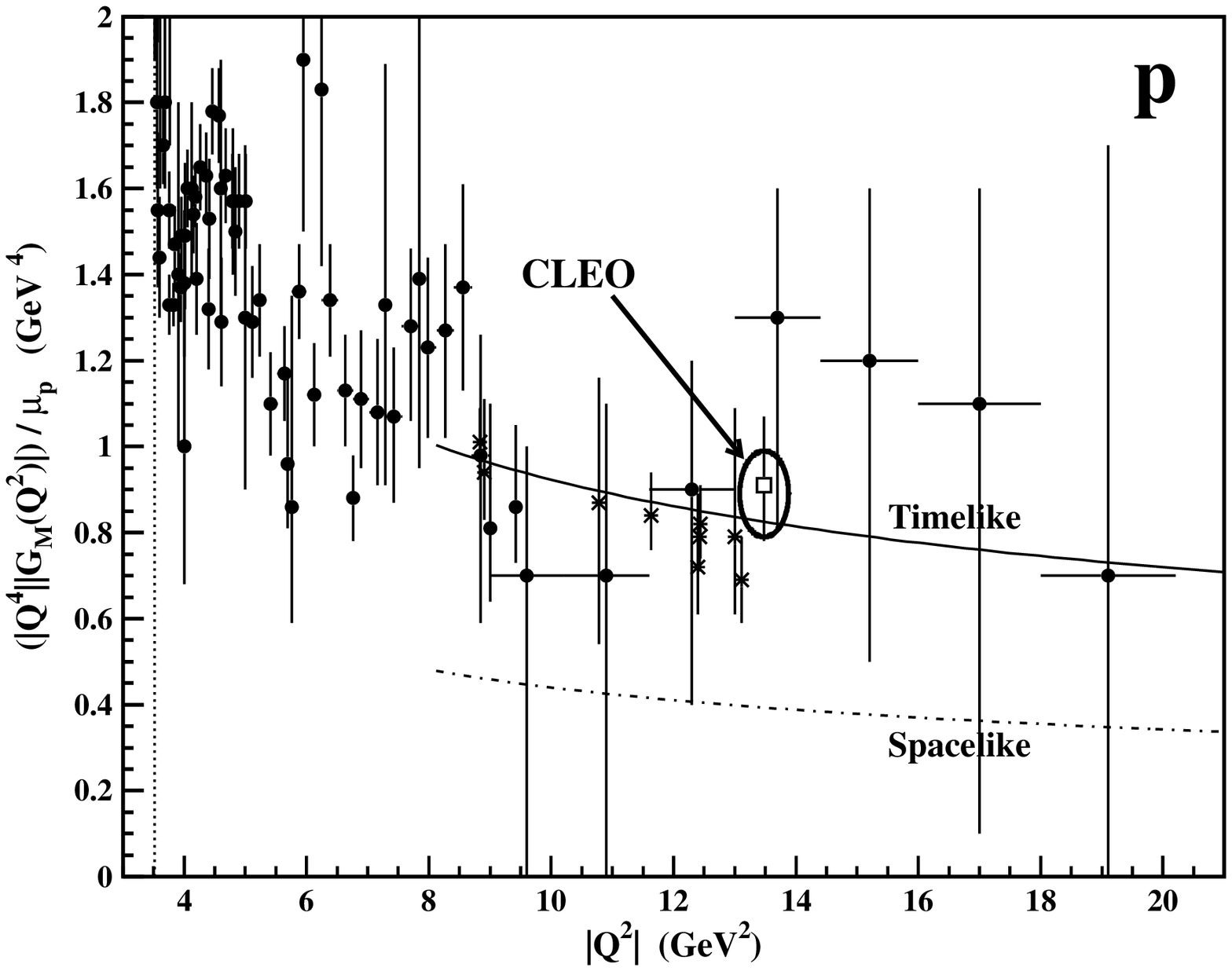}
\caption[Experimental status of the proton magnetic form factor with timelike 
momentum transfer assuming $\prelecff$ = $\prmagff$.]
{Experimental status of the proton magnetic form factor with timelike 
momentum transfer assuming $|G^P_E(Q^2)|$ = $|G^P_M(Q^2)|$.
The solid points are from $\ppbar$ final state measurements 
\cite{prtlff_2}-\cite{BABARppbarISR}, 
while the stars are the E760/E835 measurements from $\ppbartoee$ events 
\cite{e835_1}-\cite{e835_3}.
The result from the present analysis is denoted by the open square.  The ellipse 
represents the variation of the central value with the different assumptions 
of $|G^P_E(Q^2)|$ listed in Table \ref{tab:geass}.  
The vertical dotted line specifies the threshold for $\ppbar$ production, 
i.e., $|Q^{2}|$ = $(2m_{p})^{2}$ = 3.52 GeV$^{2}$.  
The arbitrarily normalized solid line shows the variation of $\alpha_{s}^2(|Q^{2}|)$ 
using its two-loop form with $n_{f}$ = 4 and $\Lambda$ = 0.322 GeV.  
The dashed line is the fit result of the variation of $\alpha_{s}^2(|Q^{2}|)$ for 
the spacelike momentum transfer data.}
\label{fig:prfinaltl} 
\end{center}
\end{figure}

Using the 20.7 pb$^{-1}$ of $e^{+}e^{-}$ data collected at $\sqrt{s}$ = 3.671 GeV, 
the electromagnetic form factors of the pion, kaon, and proton have been measured 
at a timelike momentum transfer of $|Q^{2}| = s$ = 13.48 GeV$^{2}$.  The results are 
\begin{displaymath}
|F_\pi(13.48~\mathrm{GeV}^2)| = 0.075\pm0.008(stat)\pm0.005(syst),
\end{displaymath}
\begin{displaymath}
|F_K(13.48~\mathrm{GeV}^2)| = 0.063\pm0.004(stat)\pm0.001(syst),
\end{displaymath}
and, assuming $|G^P_E(Q^2)|$ = $|G^P_M(Q^2)|$,
\begin{displaymath}
|G^P_M(13.48~\mathrm{GeV}^2)| = 0.0139^{+0.0024}_{-0.0018}(stat)\pm0.0006(syst).
\end{displaymath}
In order to facilitate comparison with PQCD predictions, equivalently       
\begin{equation}
|Q^2||F_\pi(13.48~\mathrm{GeV}^2)| = (1.01\pm0.11(stat)\pm0.07(syst))~\mathrm{GeV}^2,
\end{equation}
\begin{equation}
|Q^2||F_K(13.48~\mathrm{GeV}^2)| = (0.85\pm0.05(stat)\pm0.02(syst))~\mathrm{GeV}^2,
\end{equation}
and
\begin{equation}
|Q^4||G^P_M(13.48~\mathrm{GeV}^2)|/\mu_p 
= (0.91^{+0.16}_{-0.12}(stat)\pm0.04(syst))~\mathrm{GeV}^4.
\end{equation}
The results are displayed in Figures \ref{fig:pifinaltl}, \ref{fig:kfinaltl}, 
and \ref{fig:prfinaltl} as $|Q^2||F_\pi|$, $|Q^2||F_K|$, and \\ 
$|Q^4||G_M^p|/\mu_p$, together with the existing world data for the same.

The precision result for $|Q^2||F_\pi(13.48~\mathrm{GeV}^2)|$ is the first such directly 
measured result.  Figure \ref{fig:prfinaltl} shows that the PQCD predictions 
by Gousset and Pire \cite{GoussetPire_tlPQCD} and Brodsky $\etal$ \cite{Brodskyetal_tlPQCD}  
are factor two and three smaller than our result, respectively.  
The prediction by Bakulev $\etal$ \cite{Bakulevetal_tlQCDSR}, 
based on the QCD Sum Rules, is also lower than our result.  
Our result also provides empirical validity for 
$|Q^2||F_\pi(9.6\;\mathrm{GeV}^2)|=(0.94\pm0.06)$ GeV$^2$ obtained by interpreting 
$\Gamma(J/\psi\to\pi^+\pi^-)/\Gamma(J/\psi\to e^+e^-)$ as a measure of the form 
factor \cite{Milanaetal_jpsipipi}.   Together, the two appear to support the PQCD 
prediction of $\alpha_s(|Q^2|)/|Q^2|$ variation of the form factor at large $Q^2$.  
Bebek $\etal$ \cite{pislff_elprod4} have reported 
$Q^2F_\pi(9.77~\mathrm{GeV}^2)$ = 0.69$\pm$0.19 GeV$^2$ for the spacelike form 
factor.  Within errors this is consistent with being nearly factor two smaller 
than the timelike form factors for $Q^2 > 9$ GeV$^2$, as found for protons.

The measurement of the kaon form factor stands alone at present.  
While no explicit predictions exist for the kaon form factor in the timelike region, 
the $\alpha_s(|Q^2|)$ and $|Q^2|$ independent PQCD prediction \cite{LepageBrodsky_PQCDFF}
$|F_K(Q^2)|/|F_\pi(Q^2)|$ = $f_K^2/f_\pi^2$ = 1.49$\pm$0.03 is in disagreement 
with our result 
$|F_K(13.48\;\mathrm{GeV}^2)|/|F_\pi(13.48~\mathrm{GeV}^2)|$ = 0.84$\pm$0.11.  
Chernyak and Zhitnitsky \cite{ChernyakZhitnitsky_QCDSR} use the PQCD factorization 
scheme, but with two-humped distribution amplitudes obtained by the 
QCD Sum Rules.  
They obtain a factor 2/3 multiplying $f_K^2/f_\pi^2$ and predict 
$|F_K(Q^2)|/|F_\pi(Q^2)|$ = 0.99$\pm$0.02, which is consistent with our result 
within errors.
However, the discrepancy between our experimental results and the 
PQCD based theoretical predictions has to be considered as unexplained.

Seth \cite{SethJpsiKK} has pointed out that because the $J/\psi \rightarrow \KK$ 
decay can proceed through a virtual photon 
as well as three gluons, and the two decays have been shown to be orthogonal  
\cite{SuzukiJpsi,RosnerJpsi}, the measured ${\cal B}(J/\psi \rightarrow \KK)$ can be 
related to the kaon form factor after removing the contribution from the three 
gluon decay.  The $J/\psi \rightarrow \KK$ branching fraction can be written in terms 
of the amplitudes for the one photon $A_{\gamma}$ 
and three gluon $A_{ggg}$ decay as \cite{RosnerJpsi}
\begin{equation}
{\cal B}(J/\psi \rightarrow \KK) = |A_{\gamma} + A_{ggg}|^2 = 
|A_{\gamma}|^2 + |A_{ggg}|^2 = {\cal B}_{~~\gamma}^{\KK} + {\cal B}_{~ggg}^{\KK}, 
\end{equation}
where ${\cal B}_{~~\gamma}^{\KK}$ and ${\cal B}_{~ggg}^{\KK}$ are the 
branching fractions for the one photon and three gluon components of the 
$J/\psi \rightarrow \KK$ decay.  Using the assumption that the decay 
$J/\psi \rightarrow K^0_{S}K^0_{L}$ only occurs through the three gluon channel, 
its branching fraction can be used to determine ${\cal B}_{~ggg}^{\KK}$.  
The one photon component is related to the kaon form factor by \cite{WangetalJpsi}
\begin{equation}
{\cal B}_{~~\gamma}^{\KK} = 2{\cal B}(J/\psi \rightarrow \ee)
\left(\frac{p_K}{M_{J/\psi}}\right)^3|F_K(M^2_{J/\psi})|^2,
\end{equation}
where $(p_K/M_{J/\psi})^3 = 0.106$ is the phase space factor, 
$p_K$ is the kaon momentum in the $J/\psi$ rest frame, 
and $M_{J/\psi}$ and ${\cal B}(J/\psi \rightarrow \ee)$ are the mass and the 
$\ee$ branching fraction of $J/\psi$, respectively.  
Using the PDG values \cite{PDG2004} of 
${\cal B}(J/\psi \rightarrow \KK) = (2.37\pm0.31)^{-4}$,  
${\cal B}_{~ggg}^{\KK} = {\cal B}(J/\psi \rightarrow K^0_{S}K^0_{L}) = (1.46\pm0.26)^{-4}$,  
${\cal B}(J/\psi \rightarrow \ee) = 0.0593\pm0.0010$,  and 
$M_{J/\psi} = 3096.916\pm0.011$ MeV, 
the kaon form factor at $|Q^2| = M^2_{J/\psi} = 9.6$ GeV$^2$ is 
\begin{displaymath}
|Q^2||F_K(9.6~\mathrm{GeV}^2)| = (0.81\pm0.18)~\mathrm{GeV}^2. 
\end{displaymath}
This result is consistent with our result, 
\begin{displaymath}
|Q^2||F_K(13.48~\mathrm{GeV}^2)| = (0.85\pm0.05(stat)\pm0.02(syst))~\mathrm{GeV}^2,
\end{displaymath}
as shown in Figure \ref{fig:kfinaltl}.  Using 
$|Q^2||F_\pi(9.6\;\mathrm{GeV}^2)|=(0.94\pm0.06)$ GeV$^2$ \cite{Milanaetal_jpsipipi}, 
the form factor ratio 
$|F_K(9.6~\mathrm{GeV}^2)|/|F_\pi(9.6~\mathrm{GeV}^2)|$ = 0.86$\pm$0.20 
is also consistent with our experimental result, 0.84$\pm$0.11, 
at $|Q^2| = 13.48$ GeV$^2$ 
and smaller than the theoretical predictions discussed above.  
Together, the two also appear to support the PQCD prediction 
of $\alpha_s(|Q^2|)/|Q^2|$ variation of the form factor at large $Q^2$. 

The result for $|G^P_M(13.48~\mathrm{GeV}^2)|$ is in excellent agreement with the 
results of the Fermilab E760/E835 experiments in which the reverse reaction 
$p\bar{p}\to e^+e^-$ was measured \cite{e835_1,e835_2,e835_3}.  The result is also 
consistent with the magnetic form factor in the timelike region being twice as 
large as in the spacelike region.

The extraction of the experimental value of the proton magnetic form factor 
is dependent on the particular assumption of the electric form 
factor.  The magnetic form factor has already been determined using two different 
assumptions: $|\prelecffqsqtl|$ = 0 and 
$|\prelecffqsqtl|$ = $|\prmagffqsqtl|$, where the superscript $tl$ refers to timelike 
values.  The recent measurements of the 
electric-to-magnetic form factor ratio in the spacelike region 
can be used to determine the same ratio in the timelike region by assuming that 
the Pauli-to-Dirac form factor ratio 
for the proton is the same in the spacelike and timelike regions.  
The electric-to-magnetic form factor ratio from the Rosenbluth measurements 
in the spacelike region is $G^{P,sl}_E(Q^2) = G^{P,sl}_M(Q^2)/\mu_p$, where the 
superscript $sl$ refers to spacelike values.  This corresponds 
to $F^P_2(Q^2)/F^P_1(Q^2) = 0.0855$, 
and leads to $|\prelecffqsqtl|$ = 1.38 $|\prmagffqsqtl|$.  
A linear extrapolation of the spacelike results from the polarization transfer measurements 
($G^{P,sl}_E$(13.48 GeV$^2$) = $-$0.8 $G^{P,sl}_M$(13.48 GeV$^2$)$/\mu_p$) 
corresponds to $F^P_2(Q^2)/F^P_1(Q^2) = 0.2027$, 
and leads $|\prelecffqsqtl|$ = 1.75 $|\prmagffqsqtl|$.  
The different assumptions of the 
electric form factor affect the differential and total cross sections and therefore the 
value of the magnetic form factor. 

As mentioned earlier (Section 4.2), the differential cross section for $\eetoppbar$ 
is given by Eqn. \ref{eq:tlprad} as 
\begin{equation}
\frac{d\sigma_{0}}{d\Omega} \propto 1 + B~\mathrm{cos}^{2}\theta,
\label{eq:prfinaldiffcs}
\end{equation}
where $B = (1-\eta)/(1+\eta)$ and
\begin{equation}
\eta = \frac{4m^{2}_{p}}{|Q^2|}\frac{|\prelecffqsqtl|^{2}}{|\prmagffqsqtl|^{2}}.
\end{equation}
The corresponding total cross section is
\begin{equation}
\sigma_{0} = \frac{\pi\alpha^{2}}{3|Q^2|}~\beta_{p}
(1+\eta)~[3+B]\times|\prmagffqsqtl|^{2}~ 
\equiv \sigma_{kin}~|\prmagffqsqtl|^{2},
\label{eq:prfinalcs}
\end{equation}
where $\sigma_0 = N/[\epsilon(1+\delta){\cal L}]$, $N = 14.1^{+4.8}_{-3.7}$, 
$(1+\delta) = 0.856$, ${\cal L}$ = 20.7 pb$^{-1}$, and the 
efficiencies as listed in Table \ref{tab:geass}.  
Table \ref{tab:geass} shows the results of the four different assumptions of 
$|\prelecffqsqtl|/|\prmagffqsqtl|$ on the present results for the 
$|\prmagffqsqtl|$ measurement.  
It is interesting to note that with all these possible variations of 
$G^{P,sl}_E(Q^2)/G^{P,sl}_M(Q^2)$, $|Q^4||\prmagffqsqtl|/\mu_p$ varies by only 
$\pm$0.1 GeV$^4$, which is well within the errors of our measurement.

\begin{table}[h]
\caption[Effect of different assumptions of $|\prelecffqsqtl|/|\prmagffqsqtl|$ on the 
extracted value of $|\prmagffqsqtl|$.]
{Effect of different assumptions of $|\prelecffqsqtl|/|\prmagffqsqtl|$  on the 
extracted value of $|\prmagffqsqtl|$. 
The variable $B$ is the coefficient of $\mathrm{cos}^{2}\theta$ 
in the differential cross section given by Eqn. \ref{eq:prfinaldiffcs}.  
The $\sigma_{kin}$ and $|\prmagffqsqtl|$ terms are 
determined from Eqn. \ref{eq:prfinalcs}.  The errors are the 
statistical and systematic uncertainties summed in quadrature.}
\medskip
\begin{center}
\begin{tabular}{|c|c|c|c|c|}
\hline
 & \multicolumn{4}{|c|}{$|\prelecffqsqtl|/|\prmagffqsqtl|$ } \\
\hline
 & 0 & 1 & 1.38 & 1.75 \\
\hline
\hline
$B = (1-\eta)/(1+\eta)$ & 1 & 0.59 & 0.34 & 0.11 \\
Efficiency: $\epsilon$ & 0.626 & 0.657 & 0.668 & 0.685 \\
$\sigma_0$ (pb) & 1.27$^{+0.44}_{-0.35}$ & 1.21$^{+0.42}_{-0.33}$ 
 & 1.19$^{+0.42}_{-0.33}$ & 1.16$^{+0.41}_{-0.32}$ \\
$\sigma_{kin}$ (pb) & 5541 & 6265 & 6919 & 7757 \\
$|\prmagffqsqtl|$ & 0.0152$^{+0.0027}_{-0.0021}$ & 0.0139$^{+0.0025}_{-0.0019}$ 
 & 0.0131$^{+0.0023}_{-0.0018}$ & 0.0122$^{+0.0022}_{-0.0017}$ \\
\hline
$|Q^4||\prmagffqsqtl|/\mu_p$ & 0.99$^{+0.18}_{-0.14}$ & 0.91$^{+0.16}_{-0.13}$ 
 & 0.85$^{+0.15}_{-0.12}$ & 0.79$^{+0.14}_{-0.11}$ \\
(GeV$^4$) & & & & \\
\hline
\end{tabular}
\label{tab:geass}
\end{center}
\end{table}

\addcontentsline{toc}{chapter}{References}
\renewcommand\bibname{References}

\appendix
\baselineskip=24pt
\chapter{Event Lists for $\eetopipi$ and $\eetoppbar$ 
at $\sqrt{s}$ =  3.671 GeV}

Tables A.1-10 list the properties of all 26 events included in the 
final event selection for the channel $\eetopipi$ as described in Chapter 4.  
Tables A.11-20 list the properties of all 16 events included in the 
final event selection for the channel $\eetoppbar$ as described in Chapter 4. 

\newpage
\clearpage

\begin{table}[htbp]
\caption[Properties of the events for the $\eetopipi$ 
final selection (Part 1).]
{Properties of the events for the $\eetopipi$ final selection (Part 1).  
The variable $X_{\pi}$ is defined as 
$X_{\pi} \equiv (E(+)+E(-))/\sqrt{s}$, 
where $E(\pm)$ is $E(\pm) = \sqrt{|p(\pm)|^2 + m^2_\pi}$. 
The variable $\Sigma p_{i}$ is the net momentum of the two tracks.  
Acolin is the acolinearity between the two tracks.  
The variables $p(+)$ and $p(-)$ are the momenta of the positive 
and negative tracks, respectively.}
\begin{center}
\begin{tabular}{|c|c|c|c|c|c|c|}
\hline
Run & Event & $X_{\pi}$ & $\Sigma p_{i}$ & Acolin & $p(+)$ & $p(-)$ \\
 & & & (GeV/c) & (deg.) & (GeV/c) & (GeV/c) \\
\hline
203165 &  39815 &  0.999 &  0.039 &  0.407 &  1.810 &  1.848 \\    
203182 &  45425 &  0.992 &  0.023 &  0.391 &  1.806 &  1.826 \\    
203203 &  10540 &  0.996 &  0.025 &  0.427 &  1.832 &  1.811 \\    
203218 &  27556 &  1.002 &  0.020 &  0.558 &  1.830 &  1.839 \\    
203232 &  12545 &  1.003 &  0.036 &  0.306 &  1.852 &  1.818 \\    
203247 & 101424 &  1.008 &  0.017 &  0.531 &  1.845 &  1.845 \\   
203280 &  41768 &  1.007 &  0.013 &  0.383 &  1.845 &  1.841 \\    
203328 &   7923 &  0.991 &  0.022 &  0.583 &  1.819 &  1.808 \\    
203363 &  52988 &  1.008 &  0.035 &  0.508 &  1.860 &  1.829 \\    
203949 &  64635 &  1.003 &  0.013 &  0.159 &  1.841 &  1.830 \\    
203972 &  42766 &  1.003 &  0.016 &  0.484 &  1.837 &  1.834 \\    
204001 &  56773 &  1.012 &  0.019 &  0.161 &  1.861 &  1.843 \\    
204002 &  21854 &  0.998 &  0.019 &  0.585 &  1.827 &  1.826 \\    
204003 &  87811 &  0.998 &  0.012 &  0.284 &  1.830 &  1.822 \\    
204014 &  27853 &  1.000 &  0.005 &  0.155 &  1.831 &  1.831 \\    
204020 &  11274 &  0.999 &  0.016 &  0.492 &  1.826 &  1.829 \\    
204064 &  45495 &  1.006 &  0.020 &  0.419 &  1.848 &  1.833 \\    
204083 &  53808 &  0.999 &  0.031 &  0.166 &  1.844 &  1.813 \\   
204117 &  56203 &  0.998 &  0.009 &  0.287 &  1.828 &  1.826 \\    
204134 &  39364 &  0.997 &  0.023 &  0.501 &  1.817 &  1.833 \\    
204160 &   8433 &  0.993 &  0.022 &  0.479 &  1.810 &  1.825 \\    
204195 &   1949 &  0.996 &  0.031 &  0.578 &  1.811 &  1.836 \\    
204197 &  39119 &  1.007 &  0.027 &  0.673 &  1.851 &  1.835 \\    
204213 &   2886 &  1.005 &  0.021 &  0.291 &  1.848 &  1.830 \\    
204297 &  32047 &  1.002 &  0.021 &  0.626 &  1.837 &  1.831 \\    
204327 &  82333 &  1.001 &  0.020 &  0.532 &  1.827 &  1.837 \\    
\hline
\hline
\multicolumn{2}{|c|}{Requirement} 
 & 0.98$-$1.02 & $<$ 0.100 & --- & \multicolumn{2}{|c|}{---} \\
\hline
\end{tabular}
\end{center}
\end{table}

\newpage
\clearpage

\begin{table}[htbp]
\caption[Properties of the events for the $\eetopipi$ 
final selection (Part 2).]
{Properties of the events for the $\eetopipi$ final selection (Part 2).  
The variables cos$\theta$, $\phi_{0}$, and $\chi^{2}/dof$ are the 
cosine of the angle between the charged track and the positron beam, 
the angle of the charged track in the plane perpendicular 
to the positron beam, and the 
reduced $\chi^2$ of the helix fit of the charged track, respectively.  
The $+$ and $-$ specify the charge of the track.}
\begin{center}
\begin{tabular}{|c|c|c|c|c|c|c|c|}
\hline
Run & Event & cos$\theta(+)$ & cos$\theta(-)$ 
& $\phi_{0}(+)$ & $\phi_{0}(-)$ 
& $\chi^{2}/dof(+)$ & $\chi^{2}/dof(-)$ \\
 & & & & (deg.) & (deg.) & & \\
\hline
203165 &  39815 &  -0.238 &   0.241 &  285.7 &  106.1 &  0.818 &  0.734 \\
203182 &  45425 &   0.499 &  -0.494 &   37.4 &  217.1 &  0.861 &  0.880 \\
203203 &  10540 &   0.063 &  -0.059 &  127.0 &  306.7 &  0.956 &  0.498 \\
203218 &  27556 &  -0.250 &   0.258 &  226.0 &   46.3 &  0.831 &  1.208 \\
203232 &  12545 &   0.585 &  -0.589 &  187.8 &    8.0 &  1.506 &  1.014 \\
203247 & 101424 &   0.445 &  -0.436 &   10.6 &  190.5 &  1.032 &  0.775 \\
203280 &  41768 &   0.595 &  -0.593 &  300.3 &  120.8 &  1.126 &  1.105 \\
203328 &   7923 &  -0.315 &   0.324 &  169.4 &  349.1 &  0.887 &  1.334 \\
203363 &  52988 &   0.378 &  -0.384 &   25.9 &  205.5 &  0.830 &  0.628 \\
203949 &  64635 &   0.121 &  -0.120 &  200.4 &   20.7 &  0.783 &  1.098 \\
203972 &  42766 &  -0.533 &   0.534 &   92.3 &  271.7 &  0.839 &  0.986 \\
204001 &  56773 &   0.247 &  -0.245 &  357.5 &  177.4 &  1.352 &  1.395 \\
204002 &  21854 &   0.346 &  -0.352 &  248.3 &   68.8 &  1.377 &  0.961 \\
204003 &  87811 &   0.658 &  -0.655 &  174.8 &  354.7 &  1.046 &  0.941 \\
204014 &  27853 &   0.152 &  -0.154 &  359.7 &  179.8 &  0.637 &  1.081 \\
204020 &  11274 &   0.682 &  -0.677 &   57.6 &  237.1 &  0.689 &  0.953 \\
204064 &  45495 &   0.584 &  -0.579 &  340.2 &  160.4 &  0.739 &  0.947 \\
204083 &  53808 &  -0.486 &   0.486 &  208.9 &   29.1 &  0.791 &  1.185 \\
204117 &  56203 &  -0.203 &   0.201 &   53.1 &  232.8 &  0.631 &  0.996 \\
204134 &  39364 &   0.487 &  -0.488 &  332.3 &  152.9 &  1.273 &  1.542 \\
204160 &   8433 &   0.266 &  -0.269 &  285.2 &  105.7 &  0.836 &  0.551 \\
204195 &   1949 &  -0.345 &   0.343 &  275.3 &   95.9 &  0.971 &  0.832 \\
204197 &  39119 &   0.397 &  -0.404 &  128.3 &  307.7 &  1.862 &  1.157 \\
204213 &   2886 &  -0.469 &   0.473 &  191.0 &   11.1 &  1.333 &  0.694 \\
204297 &  32047 &   0.119 &  -0.112 &   70.4 &  249.9 &  1.327 &  0.540 \\
204327 &  82333 &  -0.735 &   0.729 &  350.4 &  170.5 &  1.072 &  0.858 \\
\hline
\hline
\multicolumn{2}{|c|}{Requirement} 
 & \multicolumn{2}{|c|}{$|$cos$\theta(\pm)|$ $<$ 0.75} 
 & \multicolumn{2}{|c|}{---} 
 & \multicolumn{2}{|c|}{$\chi^{2}/dof(\pm)$ $<$ 10} \\
\hline
\end{tabular}
\end{center}
\end{table}

\newpage
\clearpage

\begin{table}[htbp]
\caption[Properties of the events for the $\eetopipi$ 
final selection (Part 3).]
{Properties of the events for the $\eetopipi$ final selection (Part 3).
The variables $d_{b}$ and $z_{b}$ are the distances 
between the origin of the helix fit and the position of the $e^+e^-$ 
annihilation in the plane perpendicular to and 
along the axis defined by the positron beam, respectively.  
DRHF is defined as the ratio of the number of Drift Chamber wire ``hits'' observed 
to the number of ``hits'' expected from the helix fit.  
The $+$ and $-$ specify the charge of the track.}
\begin{center}
\begin{tabular}{|c|c|c|c|c|c|c|c|}
\hline
Run & Event & $d_{b}(+)$ & $d_{b}(-)$ & $z_{b}(+)$ & $z_{b}(-)$ 
& DRHF(+) & DRHF(-) \\
 & & (mm) & (mm) & (cm) & (cm) & & \\
\hline
203165 &  39815 &   0.224 &  -0.105 &   0.330 &   0.205 &  0.907 &  0.907 \\
203182 &  45425 &  -0.118 &   0.075 &   0.192 &   0.340 &  0.963 &  0.944 \\
203203 &  10540 &   0.357 &  -0.132 &  -1.373 &  -1.344 &  0.963 &  0.963 \\
203218 &  27556 &  -0.092 &   0.409 &   0.414 &   0.524 &  0.852 &  0.963 \\
203232 &  12545 &  -0.090 &  -0.019 &  -1.335 &  -1.329 &  0.944 &  0.944 \\
203247 & 101424 &  -0.083 &   0.013 &  -1.132 &  -0.917 &  0.944 &  0.981 \\
203280 &  41768 &   0.101 &   0.048 &  -0.628 &  -0.618 &  0.944 &  0.963 \\
203328 &   7923 &   0.012 &  -0.008 &  -0.314 &   0.206 &  0.870 &  0.926 \\
203363 &  52988 &  -0.407 &   0.154 &   0.893 &   0.547 &  0.926 &  0.926 \\
203949 &  64635 &   0.012 &   0.025 &  -0.880 &  -0.642 &  0.963 &  0.944 \\
203972 &  42766 &   0.252 &   0.019 &   0.626 &   0.540 &  1.000 &  0.907 \\
204001 &  56773 &  -0.291 &  -0.065 &  -0.628 &  -0.906 &  0.833 &  0.926 \\
204002 &  21854 &   0.005 &  -0.302 &  -1.629 &  -1.691 &  0.963 &  0.926 \\
204003 &  87811 &  -0.181 &   0.115 &  -0.390 &  -0.306 &  0.963 &  0.944 \\
204014 &  27853 &  -0.048 &  -0.038 &  -1.045 &  -1.253 &  0.944 &  0.926 \\
204020 &  11274 &   0.315 &  -0.496 &   1.179 &   1.293 &  0.963 &  0.963 \\
204064 &  45495 &  -0.184 &   0.071 &   0.379 &   0.459 &  0.963 &  0.926 \\
204083 &  53808 &   0.052 &  -0.127 &   0.797 &   0.559 &  0.907 &  0.944 \\
204117 &  56203 &  -0.087 &  -0.008 &   0.320 &   0.388 &  0.889 &  0.963 \\
204134 &  39364 &   0.176 &  -0.046 &   0.755 &   0.389 &  0.963 &  0.907 \\
204160 &   8433 &   0.217 &  -0.268 &  -0.295 &  -0.157 &  0.944 &  0.926 \\
204195 &   1949 &  -0.078 &   0.016 &   1.503 &   1.401 &  0.926 &  0.963 \\
204197 &  39119 &   0.066 &   0.108 &  -0.148 &  -0.333 &  0.926 &  0.870 \\
204213 &   2886 &   0.202 &  -0.035 &  -0.194 &   0.257 &  0.944 &  0.981 \\
204297 &  32047 &   0.243 &  -0.186 &   1.100 &   1.024 &  0.926 &  0.907 \\
204327 &  82333 &  -0.074 &   0.067 &   1.170 &   0.895 &  0.907 &  0.944 \\
\hline
\hline
\multicolumn{2}{|c|}{Requirement} 
 & \multicolumn{2}{|c|}{$|d_{b}(\pm)|$ $<$ 5} 
 & \multicolumn{2}{|c|}{$|z_{b}(\pm)|$ $<$ 5} 
 & \multicolumn{2}{|c|}{0.5$-$1.2} \\
\hline
\end{tabular}
\end{center}
\end{table}

\newpage
\clearpage

\begin{table}[htbp]
\caption[Properties of the events for the $\eetopipi$ 
final selection (Part 4).]
{Properties of the events for the $\eetopipi$ final selection (Part 4). 
The variables $S(i)$ [$i = e,\mu,\pi,K,p$] are the pulls based on the ionization 
energy loss ($dE/dx$) of the charged track in the Drift Chamber and 
are defined as 
$S(i) \equiv ((dE/dx)_{\mathrm{meas}}-(dE/dx)_{\mathrm{expected,}i})/\sigma$. 
The measured $dE/dx$ of the charged track is $(dE/dx)_{\mathrm{meas}}$, 
the expected $dE/dx$ for particle hypothesis $i$ 
is $(dE/dx)_{\mathrm{expected,}i}$, 
and the uncertainty in the $dE/dx$ measurement is $\sigma$.  
The values below are for positive tracks only, 
the values for the negative tracks 
are given in Table \ref{tab:piapptbl5}. }
\begin{center}
\begin{tabular}{|c|c|c|c|c|c|c|}
\hline
Run & Event & $S(e^+)$ & $S(\mu^+)$ 
& $S(\pi^+)$ & $S(K^+)$ & $S(p)$ \\
\hline
203165 &  39815 &  -3.4158 &  -0.5323 &  -0.0014 &  1.4992 &  -0.0489 \\
203182 &  45425 &  -4.8347 &  -1.9824 &  -1.4359 &  0.1406 &  -1.5020 \\  
203203 &  10540 &  -2.6720 &   0.2610 &   0.7919 &  2.2988 &   0.8366 \\  
203218 &  27556 &  -3.5736 &  -0.8540 &  -0.3457 &  1.1227 &  -0.3138 \\  
203232 &  12545 &  -3.5419 &  -0.5487 &   0.0076 &  1.6331 &   0.1378 \\  
203247 & 101424 &  -3.4473 &  -0.4866 &   0.0620 &  1.6538 &   0.1586 \\ 
203280 &  41768 &  -3.7438 &  -0.7175 &  -0.1535 &  1.4903 &  -0.0502 \\
203328 &   7923 &  -2.6994 &   0.1601 &   0.6775 &  2.1333 &   0.6690 \\  
203363 &  52988 &  -3.7028 &  -0.8622 &  -0.3285 &  1.2495 &  -0.1732 \\  
203949 &  64635 &  -4.8308 &  -1.9759 &  -1.4164 &  0.2250 &  -1.2822 \\  
203972 &  42766 &  -3.6171 &  -0.4889 &   0.0993 &  1.7808 &   0.2210 \\  
204001 &  56773 &  -2.1967 &   0.6125 &   1.1320 &  2.6073 &   1.3276 \\  
204002 &  21854 &  -3.2839 &  -0.3697 &   0.1738 &  1.7161 &   0.2373 \\  
204003 &  87811 &  -4.5225 &  -1.4148 &  -0.8195 &  0.8990 &  -0.7298 \\  
204014 &  27853 &  -2.6898 &   0.3253 &   0.8792 &  2.4377 &   0.9667 \\  
204020 &  11274 &  -3.8742 &  -0.7395 &  -0.1498 &  1.5327 &  -0.0774 \\
204064 &  45495 &  -4.0482 &  -1.0278 &  -0.4488 &  1.2306 &  -0.2776 \\  
204083 &  53808 &  -2.7069 &   0.4102 &   0.9854 &  2.6120 &   1.1295 \\  
204117 &  56203 &  -2.7890 &   0.1844 &   0.7319 &  2.2740 &   0.8088 \\  
204134 &  39364 &  -2.6972 &   0.3675 &   0.9260 &  2.4852 &   0.9547 \\  
204160 &   8433 &  -3.6618 &  -0.7608 &  -0.2183 &  1.3183 &  -0.2196 \\  
204195 &   1949 &  -2.6643 &   0.3413 &   0.8878 &  2.4098 &   0.8935 \\  
204197 &  39119 &  -3.8755 &  -1.0414 &  -0.4966 &  1.0867 &  -0.3290 \\  
204213 &   2886 &  -4.4551 &  -1.4863 &  -0.9107 &  0.7699 &  -0.7440 \\  
204297 &  32047 &  -3.3744 &  -0.6618 &  -0.1488 &  1.3229 &  -0.0479 \\
204327 &  82333 &  -2.7469 &   0.6104 &   1.2223 &  2.9337 &   1.2984 \\
\hline
\end{tabular}
\label{tab:piapptbl4}
\end{center}
\end{table}

\newpage
\clearpage

\begin{table}[htbp]
\caption[Properties of the events for the $\eetopipi$ 
final selection (Part 5).]
{Properties of the events for the $\eetopipi$ final selection (Part 5). 
The variables $S(i)$ [$i = e,\mu,\pi,K,p$] are the pulls based on the ionization 
energy loss ($dE/dx$) of the charged track in the Drift Chamber 
and are defined as 
$S(i) \equiv ((dE/dx)_{\mathrm{meas}}-(dE/dx)_{\mathrm{expected,}i})/\sigma$. 
The measured $dE/dx$ of the charged track is $(dE/dx)_{\mathrm{meas}}$, 
the expected $dE/dx$ for particle hypothesis $i$ 
is $(dE/dx)_{\mathrm{expected,}i}$, 
and the uncertainty in the $dE/dx$ measurement is $\sigma$.  
The values below are for negative tracks only, 
the values for the positive tracks 
are given in Table \ref{tab:piapptbl4}. }
\begin{center}
\begin{tabular}{|c|c|c|c|c|c|c|}
\hline
Run & Event & $S(e^-)$ & $S(\mu^-)$ 
& $S(\pi^-)$ & $S(K^-)$ & $S(\overline{p})$ \\
\hline
203165 &  39815 &  -3.6418 &  -0.8653 &  -0.3446 &  1.1810 &  -0.2407 \\    
203182 &  45425 &  -3.6141 &  -0.7130 &  -0.1742 &  1.3722 &  -0.1571 \\    
203203 &  10540 &  -4.4347 &  -1.5859 &  -1.0454 &  0.5104 &  -1.0899 \\    
203218 &  27556 &  -2.3698 &   0.5825 &   1.1130 &  2.6179 &   1.1840 \\    
203232 &  12545 &  -4.4078 &  -1.4868 &  -0.9339 &  0.6622 &  -0.9521 \\    
203247 & 101424 &  -3.3184 &  -0.2232 &   0.3461 &  1.9919 &   0.4501 \\    
203280 &  41768 &  -2.5923 &   0.6279 &   1.2069 &  2.8513 &   1.2911 \\    
203328 &   7923 &  -4.4856 &  -1.7139 &  -1.1859 &  0.3337 &  -1.2445 \\    
203363 &  52988 &  -2.9475 &   0.0653 &   0.6133 &  2.1694 &   0.6445 \\   
203949 &  64635 &  -2.0894 &   0.9977 &   1.5549 &  3.1043 &   1.6347 \\    
203972 &  42766 &  -2.8245 &   0.3217 &   0.9009 &  2.5340 &   1.0073 \\    
204001 &  56773 &  -3.4900 &  -0.6101 &  -0.0647 &  1.5029 &   0.0720 \\
204002 &  21854 &  -3.5622 &  -0.5561 &   0.0071 &  1.6099 &   0.0742 \\
204003 &  87811 &  -4.6514 &  -1.5548 &  -0.9618 &  0.7469 &  -0.9070 \\   
204014 &  27853 &  -4.4539 &  -1.6971 &  -1.1624 &  0.3934 &  -1.0748 \\    
204020 &  11274 &  -2.7461 &   0.6925 &   1.3191 &  3.0726 &   1.4103 \\    
204064 &  45495 &  -4.6621 &  -1.6189 &  -1.0318 &  0.6718 &  -0.9284 \\    
204083 &  53808 &  -4.5409 &  -1.5315 &  -0.9573 &  0.6904 &  -0.9424 \\    
204117 &  56203 &  -3.7167 &  -0.8560 &  -0.3151 &  1.2328 &  -0.2468 \\    
204134 &  39364 &  -3.1438 &  -0.1954 &   0.3541 &  1.9148 &   0.4497 \\    
204160 &   8433 &  -2.5479 &   0.5153 &   1.0738 &  2.6363 &   1.1388 \\    
204195 &   1949 &  -3.3511 &  -0.2538 &   0.3250 &  1.9726 &   0.4387 \\    
204197 &  39119 &  -2.3580 &   0.6411 &   1.1886 &  2.7242 &   1.2912 \\    
204213 &   2886 &  -5.0126 &  -2.0889 &  -1.5181 &  0.1484 &  -1.4308 \\    
204297 &  32047 &  -3.6854 &  -0.7180 &  -0.1581 &  1.4437 &  -0.0687 \\
204327 &  82333 &  -4.0414 &  -0.8221 &  -0.2124 &  1.5379 &  -0.0889 \\
\hline
\end{tabular}
\label{tab:piapptbl5}
\end{center}
\end{table}

\newpage
\clearpage

\begin{table}[htbp]
\caption[Properties of the events for the $\eetopipi$ 
final selection (Part 6).]
{Properties of the events for the $\eetopipi$ final selection (Part 6).  
This table deals with separating pions from electrons based on $dE/dx$ and 
RICH information.  Columns 1 and 3 are $\Delta S^2(\pi$$-$$e) 
= S^2(\pi) - S^2(e)$ for positive and negative tracks, respectively.  
Columns 2 and 4 are $\Delta \chi^2_{RICH}(\pi$$-$$e) 
= -2$ log$L(\pi)+2$ log$L(e)$ for positive and negative tracks, respectively. 
Column 1+2 (3+4) is the sum of columns 1 and 2 (3 and 4).  
A track is more likely to be a pion than an electron if the value is $<$0.}
\begin{center}
\begin{tabular}{|c|c|c|c|c|c|c|c|}
\hline
Run & Event & 1 & 2 & 1+2 & 3 & 4 & 3+4 \\
\hline
203165 &  39815 &  -11.668 &   0.311 &  -11.356 &  -13.144
 &   0.256 &  -12.887 \\    
203182 &  45425 &  -21.313 &  -0.040 &  -21.353 &  -13.031
 &   0.345 &  -12.686 \\    
203203 &  10540 &   -6.512 &   0.577 &   -5.935 &  -18.573
 &  -1.626 &  -20.199 \\    
203218 &  27556 &  -12.651 &   0.666 &  -11.985 &   -4.377
 &  -0.242 &   -4.620 \\    
203232 &  12545 &  -12.545 &   0.899 &  -11.646 &  -18.556
 &  -2.235 &  -20.792 \\    
203247 & 101424 &  -11.880 &   0.690 &  -11.190 &  -10.892
 &  -1.909 &  -12.801 \\    
203280 &  41768 &  -13.992 &  -2.672 &  -16.665 &   -5.264
 &   0.099 &   -5.164 \\    
203328 &   7923 &  -6.8277 &  -2.911 &   -9.739 &  -18.714
 &  -0.372 &  -19.086 \\    
203363 &  52988 &  -13.603 &   0.208 &  -13.395 &   -8.312
 &  -1.065 &   -9.377 \\    
203949 &  64635 &  -21.330 &   4.142 &  -17.189 &   -1.948
 &   1.341 &   -0.607 \\    
203972 &  42766 &  -13.074 &  -0.245 &  -13.319 &   -7.166
 &  -0.501 &   -7.667 \\    
204001 &  56773 &   -3.544 &  -0.442 &   -3.986 &  -12.176
 &  -1.522 &  -13.699 \\    
204002 &  21854 &  -10.754 &   0.334 &  -10.420 &  -12.689
 &  -1.952 &  -14.641 \\    
204003 &  87811 &  -19.781 &  -0.577 &  -20.358 &  -20.710
 &   1.470 &  -19.241 \\    
204014 &  27853 &   -6.462 &  -2.439 &   -8.901 &  -18.486
 &   6.159 &  -12.327 \\    
204020 &  11274 &  -14.987 &   0.450 &  -14.537 &   -5.801
 &   0.216 &   -5.585 \\    
204064 &  45495 &  -16.187 &   2.926 &  -13.261 &  -20.671
 &   0.084 &  -20.587 \\    
204083 &  53808 &   -6.356 &  -1.062 &   -7.418 &  -19.703
 &   0.329 &  -19.374 \\    
204117 &  56203 &   -7.243 &   4.108 &   -3.135 &  -13.715
 &  -2.932 &  -16.647 \\    
204134 &  39364 &  -6.4176 &   1.879 &   -4.538 &   -9.758
 &  -0.296 &  -10.054 \\    
204160 &   8433 &  -13.361 &  -0.723 &  -14.084 &   -5.339
 &  -1.246 &   -6.585 \\    
204195 &   1949 &   -6.310 &   0.027 &   -6.284 &  -11.124
 &   1.420 &   -9.704 \\    
204197 &  39119 &  -14.773 &  -0.135 &  -14.907 &   -4.147
 &   0.744 &   -3.403 \\    
204213 &   2886 &  -19.018 &  -0.740 &  -19.758 &  -22.822
 &  -0.885 &  -23.707 \\    
204297 &  32047 &  -11.364 &   0.209 &  -11.156 &  -13.557
 &  -1.601 &  -15.158 \\    
204327 &  82333 &  -6.0512 &  -0.534 &   -6.585 &  -16.288
 &  -1.120 &  -17.407 \\    
\hline
\hline
\multicolumn{2}{|c|}{Requirement} 
 & --- & --- & $<$ 0 
 & --- & --- & $<$ 0 \\
\hline
\end{tabular}
\end{center}
\end{table}

\newpage
\clearpage

\begin{table}[htbp]
\caption[Properties of the events for the $\eetopipi$ 
final selection (Part 7).]
{Properties of the events for the $\eetopipi$ final selection (Part 7).  
This table deals with separating pions from muons based on $dE/dx$ and 
RICH information.  Columns 1 and 3 are $\Delta S^2(\pi$$-$$\mu) 
= S^2(\pi) - S^2(\mu)$ for positive and negative tracks, respectively.  
Columns 2 and 4 are $\Delta \chi^2_{RICH}(\pi$$-$$\mu) 
= -2$ log$L(\pi)+2$ log$L(\mu)$ for positive and negative tracks, respectively. 
Column 1+2 (3+4) is the sum of columns 1 and 2 (3 and 4).  A track 
is more likely to be a pion than an muon if the value is $<$0.}
\begin{center}
\begin{tabular}{|c|c|c|c|c|c|c|c|}
\hline
Run & Event & 1 & 2 & 1+2 & 3 & 4 & 3+4 \\
\hline
203165 &  39815 &  -0.283 &   0.201 &  -0.082 &  -0.630
 &   0.120 &  -0.510 \\    
203182 &  45425 &  -1.868 &   0.038 &  -1.831 &  -0.478
 &   0.203 &  -0.275 \\    
203203 &  10540 &   0.559 &   0.316 &   0.875 &  -1.422
 &  -0.588 &  -2.010 \\    
203218 &  27556 &  -0.610 &   0.373 &  -0.237 &   0.900
 &  -0.023 &   0.876 \\    
203232 &  12545 &  -0.301 &   0.448 &   0.148 &  -1.338
 &  -0.868 &  -2.207 \\    
203247 & 101424 &  -0.233 &   0.370 &   0.137 &   0.070
 &  -0.736 &  -0.666 \\    
203280 &  41768 &  -0.491 &  -1.032 &  -1.523 &   1.062
 &   0.110 &   1.173 \\    
203328 &   7923 &   0.433 &  -1.136 &  -0.703 &  -1.531
 &  -0.072 &  -1.603 \\    
203363 &  52988 &  -0.635 &   0.152 &  -0.484 &   0.372
 &  -0.416 &  -0.045 \\
203949 &  64635 &  -1.898 &   3.450 &   1.552 &   1.422
 &   0.681 &   2.103 \\    
203972 &  42766 &  -0.229 &  -0.038 &  -0.267 &   0.708
 &  -0.156 &   0.553 \\    
204001 &  56773 &   0.906 &  -0.097 &   0.810 &  -0.368
 &  -0.498 &  -0.866 \\    
204002 &  21854 &  -0.107 &   0.277 &   0.170 &  -0.309
 &  -0.703 &  -1.012 \\    
204003 &  87811 &  -1.330 &  -0.170 &  -1.500 &  -1.492
 &   0.678 &  -0.815 \\    
204014 &  27853 &   0.667 &  -1.015 &  -0.348 &  -1.529
 &   4.920 &   3.391 \\    
204020 &  11274 &  -0.524 &   0.234 &  -0.291 &   1.261
 &   0.120 &   1.380 \\    
204064 &  45495 &  -0.855 &  -0.400 &  -1.255 &  -1.556
 &   0.086 &  -1.471 \\    
204083 &  53808 &   0.803 &  -0.357 &   0.446 &  -1.429
 &   0.216 &  -1.213 \\    
204117 &  56203 &   0.502 &   0.080 &   0.581 &  -0.633
 &  -1.099 &  -1.733 \\    
204134 &  39364 &   0.722 &   0.878 &   1.601 &   0.087
 &  -0.041 &   0.047 \\
204160 &   8433 &  -0.531 &  -0.239 &  -0.770 &   0.888
 &  -0.405 &   0.483 \\    
204195 &   1949 &   0.672 &   0.124 &   0.796 &   0.041
 &   0.680 &   0.721 \\    
204197 &  39119 &  -0.838 &  -0.009 &  -0.846 &   1.002
 &   0.346 &   1.348 \\    
204213 &   2886 &  -1.380 &  -0.209 &  -1.589 &  -2.059
 &  -0.310 &  -2.369 \\    
204297 &  32047 &  -0.416 &   0.180 &  -0.236 &  -0.490
 &  -0.620 &  -1.110 \\    
204327 &  82333 &   1.121 &  -0.169 &   0.952 &  -0.631
 &  -0.447 &  -1.077 \\    
\hline
\end{tabular}
\label{tab:piapptbl7}
\end{center}
\end{table}

\newpage
\clearpage

\begin{table}[htbp]
\caption[Properties of the events for the $\eetopipi$ 
final selection (Part 8).]
{Properties of the events for the $\eetopipi$ final selection (Part 8).  
This table deals with separating pions from kaons based on $dE/dx$ and 
RICH information.  Columns 1 and 3 are $\Delta S^2(\pi$$-$$K) 
= S^2(\pi) - S^2(K)$ for positive and negative tracks, respectively.  
Columns 2 and 4 are $\Delta \chi^2_{RICH}(\pi$$-$$K) 
= -2$ log$L(\pi)+2$ log$L(K)$ for positive and negative tracks, respectively. 
Column 1+2 (3+4) is the sum of columns 1 and 2 (3 and 4).  A track 
is more likely to be a pion than an kaon if the value is $<$0.}
\begin{center}
\begin{tabular}{|c|c|c|c|c|c|c|c|}
\hline
Run & Event & 1 & 2 & 1+2 & 3 & 4 & 3+4 \\
\hline
203165 &  39815 &  -2.248 &  -28.836 &  -31.083 &  -1.276
 &   -7.619 &   -8.895 \\   
203182 &  45425 &   2.042 &  -23.463 &  -21.421 &  -1.853
 &  -29.295 &  -31.147 \\    
203203 &  10540 &  -4.657 &  -38.716 &  -43.373 &   0.832
 &  -15.533 &  -14.700 \\    
203218 &  27556 &  -1.141 &  -45.013 &  -46.154 &  -5.615
 &  -24.845 &  -30.460 \\   
203232 &  12545 &  -2.667 &  -38.662 &  -41.329 &   0.434
 &  -11.827 &  -11.393 \\    
203247 & 101424 &  -2.731 &  -31.920 &  -34.651 &  -3.848
 &  -17.838 &  -21.686 \\    
203280 &  41768 &  -2.197 &  -22.150 &  -24.347 &  -6.674
 &  -28.714 &  -35.388 \\    
203328 &   7923 &  -4.092 &  -16.930 &  -21.021 &   1.295
 &  -23.840 &  -22.545 \\    
203363 &  52988 &  -1.453 &  -24.021 &  -25.475 &  -4.330
 &   -7.901 &  -12.231 \\    
203949 &  64635 &   1.956 &  -61.203 &  -59.247 &  -7.219
 &  -50.836 &  -58.054 \\    
203972 &  42766 &  -3.162 &  -28.373 &  -31.535 &  -5.609
 &  -16.671 &  -22.280 \\    
204001 &  56773 &  -5.516 &  -28.234 &  -33.751 &  -2.254
 &  -57.159 &  -59.413 \\    
204002 &  21854 &  -2.915 &  -42.901 &  -45.816 &  -2.592
 &  -33.268 &  -35.860 \\    
204003 &  87811 &  -0.137 &  -25.477 &  -25.614 &   0.367
 &  -33.886 &  -33.519 \\    
204014 &  27853 &  -5.169 &   -0.595 &   -5.765 &   1.196
 &  -74.474 &  -73.278 \\    
204020 &  11274 &  -2.327 &  -19.427 &  -21.754 &  -7.701
 &  -16.342 &  -24.043 \\    
204064 &  45495 &  -1.312 &  -29.268 &  -30.581 &   0.613
 &  -20.971 &  -20.358 \\    
204083 &  53808 &  -5.852 &  -30.594 &  -36.445 &   0.440
 &  -35.001 &  -34.562 \\    
204117 &  56203 &  -4.635 &  -38.807 &  -43.443 &  -1.421
 &  -28.982 &  -30.402 \\    
204134 &  39364 &  -5.319 &  -44.112 &  -49.431 &  -3.541
 &  -35.119 &  -38.660 \\    
204160 &   8433 &  -1.690 &  -24.125 &  -25.815 &  -5.797
 &  -36.961 &  -42.758 \\    
204195 &   1949 &  -5.019 &  -48.936 &  -53.955 &  -3.786
 &  -43.839 &  -47.624 \\    
204197 &  39119 &  -0.934 &  -18.757 &  -19.691 &  -6.009
 &  -21.239 &  -27.247 \\    
204213 &   2886 &   0.237 &  -35.883 &  -35.646 &   2.283
 &  -18.696 &  -16.414 \\    
204297 &  32047 &  -1.728 &  -38.732 &  -40.460 &  -2.059
 &   -9.927 &  -11.986 \\    
204327 &  82333 &  -7.113 &  -22.856 &  -29.969 &  -2.320
 &   -3.089 &   -5.409 \\   
\hline
\hline
\multicolumn{2}{|c|}{Requirement} 
 & --- & --- & $<$ 0 
 & --- & --- & $<$ 0 \\
\hline
\end{tabular}
\end{center}
\end{table}

\newpage
\clearpage

\begin{table}[htbp]
\caption[Properties of the events for the $\eetopipi$ 
final selection (Part 9).]
{Properties of the events for the $\eetopipi$ final selection (Part 9).  
This table deals with separating pions from protons based on $dE/dx$ and 
RICH information.  Columns 1 and 3 are $\Delta S^2(\pi$$-$$p) 
= S^2(\pi) - S^2(p)$ for positive and negative tracks, respectively.  
Columns 2 and 4 are $\Delta \chi^2_{RICH}(\pi$$-$$p) 
= -2$ log$L(\pi)+2$ log$L(p)$ for positive and negative tracks, respectively. 
Column 1+2 (3+4) is the sum of columns 1 and 2 (3 and 4).  A track 
is more likely to be a pion than an proton if the value is $<$0.}
\begin{center}
\begin{tabular}{|c|c|c|c|c|c|c|c|}
\hline
Run & Event & 1 & 2 & 1+2 & 3 & 4 & 3+4 \\
\hline
203165 &  39815 &  -0.002 &   -70.294 &   -70.297 &   0.061
 &   -21.822 &   -21.761 \\   
203182 &  45425 &  -0.194 &  -102.678 &  -102.872 &   0.006
 &  -100.294 &  -100.288 \\    
203203 &  10540 &  -0.073 &  -115.476 &  -115.548 &  -0.095
 &  -146.877 &  -146.972 \\    
203218 &  27556 &   0.021 &  -110.937 &  -110.916 &  -0.163
 &   -82.639 &   -82.803 \\    
203232 &  12545 &  -0.019 &  -149.792 &  -149.811 &  -0.034
 &  -128.522 &  -128.556 \\    
203247 & 101424 &  -0.021 &  -135.227 &  -135.249 &  -0.083
 &  -129.667 &  -129.750 \\    
203280 &  41768 &   0.021 &  -195.545 &  -195.524 &  -0.210
 &  -115.735 &  -115.945 \\    
203328 &   7923 &   0.012 &  -127.004 &  -126.993 &  -0.142
 &   -80.387 &   -80.529 \\    
203363 &  52988 &   0.078 &   -85.402 &   -85.324 &  -0.039
 &   -55.816 &   -55.856 \\    
203949 &  64635 &   0.362 &  -148.569 &  -148.207 &  -0.254
 &  -129.155 &  -129.410 \\    
203972 &  42766 &  -0.039 &  -123.886 &  -123.925 &  -0.203
 &   -91.949 &   -92.152 \\    
204001 &  56773 &  -0.481 &  -143.892 &  -144.373 &  -0.001
 &  -141.017 &  -141.018 \\    
204002 &  21854 &  -0.026 &  -111.513 &  -111.539 &  -0.005
 &  -117.623 &  -117.629 \\    
204003 &  87811 &   0.139 &  -110.106 &  -109.967 &   0.102
 &  -116.477 &  -116.375 \\    
204014 &  27853 &  -0.161 &   -91.503 &   -91.665 &   0.196
 &  -175.258 &  -175.062 \\    
204020 &  11274 &   0.016 &   -79.879 &   -79.863 &  -0.249
 &   -64.357 &   -64.605 \\    
204064 &  45495 &   0.124 &  -154.176 &  -154.052 &   0.203
 &   -85.709 &   -85.506 \\    
204083 &  53808 &  -0.305 &  -160.211 &  -160.516 &   0.028
 &  -139.807 &  -139.779 \\    
204117 &  56203 &  -0.118 &  -110.080 &  -110.199 &   0.038
 &  -139.011 &  -138.973 \\    
204134 &  39364 &  -0.054 &  -155.698 &  -155.752 &  -0.077
 &  -165.091 &  -165.168 \\    
204160 &   8433 &  -0.001 &  -106.285 &  -106.286 &  -0.144
 &  -172.356 &  -172.500 \\    
204195 &   1949 &  -0.010 &  -151.338 &  -151.348 &  -0.087
 &  -103.506 &  -103.593 \\    
204197 &  39119 &   0.138 &   -88.223 &   -88.084 &  -0.254
 &   -75.811 &   -76.065 \\    
204213 &   2886 &   0.276 &  -179.510 &  -179.234 &   0.257
 &  -106.701 &  -106.444 \\    
204297 &  32047 &   0.020 &  -145.397 &  -145.377 &   0.020
 &   -59.201 &   -59.180 \\    
204327 &  82333 &  -0.192 &  -135.996 &  -136.188 &   0.037
 &   -58.977 &   -58.940 \\    
\hline
\end{tabular}
\end{center}
\end{table}

\newpage
\clearpage

\begin{table}[htbp]
\caption[Properties of the events for the $\eetopipi$ 
final selection (Part 10).]
{Properties of the events for the $\eetopipi$ final selection (Part 10).  
Columns 1 and 4 are $E_{tkCC}/p$ for positive and negative tracks, respectively.  
Columns 2 and 5 are $E_{tkCC}$ for positive and negative tracks, respectively.  
Columns 3 and 6 are the distances transversed in the Muon Chamber (in terms of 
nuclear interaction lengths) by positive and negative tracks, respectively. }
\begin{center}
\begin{tabular}{|c|c|c|c|c|c|c|c|}
\hline
Run & Event & 1 & 2 & 3 & 4 & 5 & 6 \\
& &  & (GeV) & & & (GeV) &  \\
\hline
203165 &  39815 &  0.632 &  1.144 &  0.00 &  0.378 &  0.698 &  0.00 \\
203182 &  45425 &  0.496 &  0.896 &  0.00 &  0.512 &  0.935 &  0.00 \\
203203 &  10540 &  0.632 &  1.158 &  0.00 &  0.722 &  1.308 &  0.00 \\
203218 &  27556 &  0.382 &  0.698 &  0.00 &  0.464 &  0.853 &  0.00 \\
203232 &  12545 &  0.445 &  0.824 &  0.00 &  0.392 &  0.713 &  0.00 \\
203247 & 101424 &  0.485 &  0.895 &  0.00 &  0.596 &  1.100 &  0.00 \\
203280 &  41768 &  0.396 &  0.731 &  0.00 &  0.795 &  1.463 &  0.00 \\
203328 &   7923 &  0.500 &  0.910 &  0.00 &  0.529 &  0.955 &  0.00 \\
203363 &  52988 &  0.445 &  0.827 &  0.00 &  0.404 &  0.739 &  0.00 \\
203949 &  64635 &  0.446 &  0.822 &  0.00 &  0.505 &  0.923 &  0.00 \\
203972 &  42766 &  0.464 &  0.853 &  0.00 &  0.317 &  0.581 &  0.00 \\
204001 &  56773 &  0.591 &  1.101 &  0.00 &  0.273 &  0.502 &  0.00 \\
204002 &  21854 &  0.406 &  0.742 &  0.00 &  0.389 &  0.710 &  0.00 \\
204003 &  87811 &  0.492 &  0.901 &  0.00 &  0.402 &  0.732 &  0.00 \\
204014 &  27853 &  0.617 &  1.130 &  0.00 &  0.319 &  0.584 &  0.00 \\
204020 &  11274 &  0.553 &  1.010 &  0.00 &  0.415 &  0.760 &  0.05 \\
204064 &  45495 &  0.253 &  0.468 &  3.95 &  0.385 &  0.706 &  0.00 \\
204083 &  53808 &  0.695 &  1.282 &  0.00 &  0.357 &  0.648 &  0.00 \\
204117 &  56203 &  0.572 &  1.045 &  0.00 &  0.495 &  0.904 &  0.00 \\
204134 &  39364 &  0.516 &  0.938 &  0.00 &  0.466 &  0.855 &  0.00 \\
204160 &   8433 &  0.537 &  0.972 &  0.00 &  0.337 &  0.615 &  0.00 \\
204195 &   1949 &  0.553 &  1.001 &  0.00 &  0.498 &  0.915 &  0.00 \\
204197 &  39119 &  0.493 &  0.913 &  0.00 &  0.571 &  1.047 &  0.00 \\
204213 &   2886 &  0.471 &  0.870 &  0.00 &  0.538 &  0.984 &  0.00 \\
204297 &  32047 &  0.623 &  1.144 &  0.00 &  0.451 &  0.826 &  0.00 \\
204327 &  82333 &  0.431 &  0.788 &  0.00 &  0.385 &  0.708 &  0.00 \\
\hline
\hline
\multicolumn{2}{|c|}{Requirement} 
 & $<$ 0.85 & $>$ 0.420 & --- 
 & $<$ 0.85 & $>$ 0.420 & --- \\
\hline
\end{tabular}
\label{tab:piapptbl10}
\end{center}
\end{table}

\newpage
\clearpage

\begin{table}[htbp]
\caption[Properties of the events for the $\eetoppbar$ 
final selection (Part 1).]
{Properties of the events for the $\eetoppbar$ final selection (Part 1).  
The variable $X_{p}$ is defined as 
$X_{p} \equiv (E(+)+E(-))/\sqrt{s}$, 
where $E(\pm)$ is $E(\pm) = \sqrt{|p(\pm)|^2 + m^2_{p}}$. 
The variable $\Sigma p_{i}$ is the net momentum of the two tracks.  
Acolin is the acolinearity between the two tracks.  
The variables $p(+)$ and $p(-)$ are the momenta of the positive 
and negative tracks, respectively.}
\begin{center}
\begin{tabular}{|c|c|c|c|c|c|c|}
\hline
Run & Event & $X_{p}$ & $\Sigma p_{i}$ & Acolin & $p(+)$ & $p(-)$ \\
 & & & (GeV/c) & (deg.) & (GeV/c) & (GeV/c) \\
\hline
203131 &  37307 &  0.997 &  0.013 &  0.485 &  1.571 &  1.571 \\
203224 &  11782 &  1.003 &  0.011 &  0.398 &  1.583 &  1.585 \\
203240 &  39099 &  0.987 &  0.025 &  0.854 &  1.545 &  1.555 \\
203267 &  26988 &  0.997 &  0.026 &  0.406 &  1.582 &  1.559 \\
203267 &  61798 &  1.007 &  0.016 &  0.489 &  1.597 &  1.587 \\
203273 &  35020 &  1.000 &  0.019 &  0.606 &  1.582 &  1.572 \\
203274 &  18555 &  0.999 &  0.019 &  0.284 &  1.585 &  1.567 \\
203352 &  26890 &  0.999 &  0.020 &  0.490 &  1.568 &  1.584 \\
203359 &  41016 &  0.990 &  0.032 &  0.912 &  1.546 &  1.567 \\
203375 &  74490 &  0.993 &  0.030 &  0.477 &  1.575 &  1.548 \\
203403 &  44417 &  1.000 &  0.022 &  0.236 &  1.567 &  1.588 \\
203418 &   6948 &  1.002 &  0.017 &  0.567 &  1.586 &  1.578 \\
204014 &  15143 &  1.004 &  0.023 &  0.355 &  1.597 &  1.576 \\
204064 &  24504 &  1.000 &  0.013 &  0.446 &  1.578 &  1.575 \\
204069 &  53356 &  0.988 &  0.025 &  0.753 &  1.544 &  1.558 \\
204277 &  19757 &  0.990 &  0.016 &  0.216 &  1.549 &  1.563 \\
\hline
\hline
\multicolumn{2}{|c|}{Requirement} 
 & 0.98$-$1.02 & $<$ 0.100 & --- & \multicolumn{2}{|c|}{---} \\
\hline
\end{tabular}
\end{center}
\end{table}

\newpage
\clearpage

\begin{table}[htbp]
\caption[Properties of the events for the $\eetoppbar$ 
final selection (Part 2).]
{Properties of the events for the $\eetoppbar$ final selection (Part 2).  
The variables cos$\theta$, $\phi_{0}$, and $\chi^{2}/dof$ are the 
cosine of the angle between the charged track and the positron beam, 
the angle of the charged track in the plane perpendicular 
to the positron beam, and the 
reduced $\chi^2$ of the helix fit of the charged track, respectively.  
The $+$ and $-$ specify the charge of the track.}
\begin{center}
\begin{tabular}{|c|c|c|c|c|c|c|c|}
\hline
Run & Event & cos$\theta(+)$ & cos$\theta(-)$ 
& $\phi_{0}(+)$ & $\phi_{0}(-)$ 
& $\chi^{2}/dof(+)$ & $\chi^{2}/dof(-)$ \\
 & & & & (deg.) & (deg.) & & \\
\hline
203131 &  37307 &  -0.680 &   0.677 &  294.5 &  115.1 &  1.043 &  1.034 \\
203224 &  11782 &  -0.386 &   0.380 &   10.0 &  189.8 &  1.013 &  1.226 \\
203240 &  39099 &   0.604 &  -0.614 &  327.9 &  148.4 &  0.444 &  1.019 \\
203267 &  26988 &   0.544 &  -0.550 &  206.0 &   26.2 &  0.789 &  1.775 \\
203267 &  61798 &   0.453 &  -0.458 &  238.6 &   59.0 &  0.684 &  0.997 \\ 
203273 &  35020 &   0.540 &  -0.548 &  222.9 &   43.2 &  1.092 &  0.971 \\ 
203274 &  18555 &  -0.770 &   0.772 &  191.8 &   12.1 &  0.789 &  0.615 \\ 
203352 &  26890 &  -0.396 &   0.393 &  321.8 &  142.2 &  0.673 &  1.035 \\
203359 &  41016 &   0.301 &  -0.315 &  191.6 &   11.3 &  0.557 &  0.990 \\ 
203375 &  74490 &   0.285 &  -0.283 &  101.7 &  281.2 &  0.835 &  1.825 \\
203403 &  44417 &   0.356 &  -0.353 &   39.9 &  219.7 &  0.779 &  1.114 \\
203418 &   6948 &   0.744 &  -0.744 &  249.7 &   70.5 &  1.012 &  0.898 \\ 
204014 &  15143 &   0.232 &  -0.232 &  231.9 &   52.2 &  1.131 &  1.004 \\
204064 &  24504 &  -0.681 &   0.687 &  159.1 &  339.0 &  0.922 &  0.870 \\ 
204069 &  53356 &   0.751 &  -0.759 &   67.8 &  247.3 &  1.303 &  1.304 \\
204277 &  19757 &   0.768 &  -0.768 &  289.0 &  109.4 &  1.051 &  1.358 \\
\hline
\hline
\multicolumn{2}{|c|}{Requirement} 
 & \multicolumn{2}{|c|}{$|$cos$\theta(\pm)|$ $<$ 0.80} 
 & \multicolumn{2}{|c|}{---} 
 & \multicolumn{2}{|c|}{$\chi^{2}/dof(\pm)$ $<$ 10} \\
\hline
\end{tabular}
\end{center}
\end{table}

\newpage
\clearpage

\begin{table}[htbp]
\caption[Properties of the events for the $\eetoppbar$ 
final selection (Part 3).]
{Properties of the events for the $\eetoppbar$ final selection (Part 3).
The variables $d_{b}$ and $z_{b}$ are the distances 
between the origin of the helix fit and the position of the $e^+e^-$ 
annihilation in the plane perpendicular to and 
along the axis defined by the positron beam, respectively.  
DRHF is defined as the ratio of the number of Drift Chamber wire ``hits'' observed 
to the number of ``hits'' expected from the helix fit.  
The $+$ and $-$ specify the charge of the track.}
\begin{center}
\begin{tabular}{|c|c|c|c|c|c|c|c|}
\hline
Run & Event & $d_{b}(+)$ & $d_{b}(-)$ & $z_{b}(+)$ & $z_{b}(-)$ 
& DRHF(+) & DRHF(-) \\
 & & (mm) & (mm) & (cm) & (cm) & & \\
\hline
203131 &  37307 &  -0.100 &   0.084 &  -1.208 &  -1.311 &  0.944 &  0.981 \\
203224 &  11782 &  -0.010 &   0.090 &   0.249 &  -0.058 &  0.944 &  0.944 \\
203240 &  39099 &   0.155 &  -0.029 &  -0.610 &  -0.624 &  0.926 &  0.926 \\
203267 &  26988 &   0.166 &  -0.035 &   0.852 &   1.011 &  0.926 &  0.963 \\
203267 &  61798 &  -0.073 &  -0.100 &   0.489 &   0.284 &  0.963 &  0.926 \\
203273 &  35020 &   0.008 &   0.049 &   0.052 &   0.185 &  0.870 &  0.926 \\
203274 &  18555 &   0.011 &  -0.261 &  -0.864 &  -0.798 &  0.981 &  0.963 \\
203352 &  26890 &   0.079 &  -0.184 &   0.542 &   0.568 &  0.944 &  0.963 \\
203359 &  41016 &  -0.066 &   0.037 &  -0.689 &  -0.783 &  0.944 &  0.944 \\
203375 &  74490 &  -0.197 &   0.173 &   1.022 &   0.876 &  0.944 &  0.926 \\
203403 &  44417 &   0.128 &   0.215 &   1.379 &   1.670 &  0.981 &  0.963 \\
203418 &   6948 &  -0.171 &  -0.181 &  -0.019 &  -0.018 &  0.981 &  0.944 \\
204014 &  15143 &  -0.548 &   0.302 &   1.534 &   1.520 &  0.889 &  0.963 \\
204064 &  24504 &  -0.177 &   0.028 &   1.271 &   1.140 &  0.907 &  0.944 \\
204069 &  53356 &   0.061 &   0.322 &   0.666 &   0.443 &  0.963 &  0.963 \\
204277 &  19757 &  -0.119 &   0.112 &  -0.744 &  -0.973 &  0.926 &  0.981 \\
\hline
\hline
\multicolumn{2}{|c|}{Requirement} 
 & \multicolumn{2}{|c|}{$|d_{b}(\pm)|$ $<$ 5} 
 & \multicolumn{2}{|c|}{$|z_{b}(\pm)|$ $<$ 5} 
 & \multicolumn{2}{|c|}{0.5$-$1.2} \\
\hline
\end{tabular}
\end{center}
\end{table}

\newpage
\clearpage

\begin{table}[htbp]
\caption[Properties of the events for the $\eetoppbar$ 
final selection (Part 4).]
{Properties of the events for the $\eetoppbar$ final selection (Part 4). 
The variables $S(i)$ [$i = e,\mu,\pi,K,p$] are the pulls based on the 
ionization energy loss ($dE/dx$) of the charged track in the Drift Chamber 
and are defined as 
$S(i) \equiv ((dE/dx)_{\mathrm{meas}}-(dE/dx)_{\mathrm{expected,}i})/\sigma$. 
The measured $dE/dx$ of the charged track is $(dE/dx)_{\mathrm{meas}}$, 
the expected $dE/dx$ for particle hypothesis $i$ 
is $(dE/dx)_{\mathrm{expected,}i}$, 
and the uncertainty in the $dE/dx$ measurement is $\sigma$.  
The values below are for positive tracks only, 
the values for the negative tracks 
are given in Table \ref{tab:prapptbl5}. }
\begin{center}
\begin{tabular}{|c|c|c|c|c|c|c|}
\hline
Run & Event & $S(e^+)$ & $S(\mu^+)$ 
& $S(\pi^+)$ & $S(K^+)$ & $S(p)$ \\
\hline
203131 &  37307 &  -2.7385 &   0.8449 &   1.4250 &  2.6909 &   0.0163 \\
203224 &  11782 &  -1.7344 &   1.7652 &   2.3205 &  3.5351 &   1.0517 \\
203240 &  39099 &  -2.2952 &   1.3861 &   1.9648 &  3.1733 &   0.3509 \\ 
203267 &  26988 &  -2.5557 &   0.9376 &   1.5042 &  2.7544 &   0.1961 \\ 
203267 &  61798 &  -2.6684 &   0.8160 &   1.3874 &  2.6730 &   0.1687 \\ 
203273 &  35020 &  -2.2696 &   1.0713 &   1.6103 &  2.7953 &   0.3605 \\ 
203274 &  18555 &  -2.9767 &   0.7211 &   1.3259 &  2.6687 &  -0.0532 \\
203352 &  26890 &  -4.1479 &  -0.8983 &  -0.3467 &  0.8794 &  -1.6993 \\
203359 &  41016 &  -1.1329 &   2.2322 &   2.7464 &  3.8073 &   1.3165 \\
203375 &  74490 &  -2.3373 &   0.8277 &   1.3395 &  2.4593 &   0.1143 \\ 
203403 &  44417 &  -1.3077 &   2.2583 &   2.8114 &  3.9868 &   1.4239 \\
203418 &   6948 &  -0.6430 &   3.3571 &   3.9706 &  5.2886 &   2.5629 \\
204014 &  15143 &  -2.5385 &   0.6083 &   1.1123 &  2.3125 &   0.0675 \\
204064 &  24504 &  -3.4627 &  -0.0157 &   0.5409 &  1.8571 &  -0.7183 \\ 
204069 &  53356 &  -2.8516 &   0.9016 &   1.4799 &  2.7763 &  -0.1252 \\ 
204277 &  19757 &  -4.0103 &  -0.4269 &   0.1459 &  1.4579 &  -1.3926 \\
\hline
\end{tabular}
\label{tab:prapptbl4}
\end{center}
\end{table}

\newpage
\clearpage

\begin{table}[htbp]
\caption[Properties of the events for the $\eetoppbar$ 
final selection (Part 5).]
{Properties of the events for the $\eetoppbar$ final selection (Part 5). 
The variables $S(i)$ [$i = e,\mu,\pi,K,p$] are the pulls based on the 
ionization energy loss ($dE/dx$) of the charged track in the Drift Chamber 
and are defined as 
$S(i) \equiv ((dE/dx)_{\mathrm{meas}}-(dE/dx)_{\mathrm{expected,}i})/\sigma$. 
The measured $dE/dx$ of the charged track is $(dE/dx)_{\mathrm{meas}}$, 
the expected $dE/dx$ for particle hypothesis $i$ 
is $(dE/dx)_{\mathrm{expected,}i}$, 
and the uncertainty in the $dE/dx$ measurement is $\sigma$.  
The values below are for negative tracks only, 
the values for the positive tracks 
are given in Table \ref{tab:prapptbl4}. }
\begin{center}
\begin{tabular}{|c|c|c|c|c|c|c|}
\hline
Run & Event & $S(e^-)$ & $S(\mu^-)$ 
& $S(\pi^-)$ & $S(K^-)$ & $S(\overline{p})$ \\
\hline
203131 &  37307 &  -2.738930 &   0.8908 &   1.4773 &  2.7539 &   0.0408 \\
203224 &  11782 &  -2.383696 &   0.9368 &   1.4754 &  2.6670 &   0.2493 \\
203240 &  39099 &  -3.469368 &   0.0042 &   0.5745 &  1.8056 &  -0.9217 \\
203267 &  26988 &  -1.005250 &   2.6353 &   3.1923 &  4.3585 &   1.7381 \\
203267 &  61798 &  -3.954325 &  -0.7604 &  -0.2151 &  1.0255 &  -1.4281 \\
203273 &  35020 &  -3.558450 &  -0.2287 &   0.3266 &  1.5558 &  -1.0138 \\
203274 &  18555 &  -3.641736 &   0.0977 &   0.7138 &  2.0622 &  -0.8179 \\
203352 &  26890 &  -2.845031 &   0.4209 &   0.9584 &  2.1551 &  -0.2683 \\
203359 &  41016 &  -2.949881 &   0.2458 &   0.7700 &  1.9154 &  -0.5294 \\
203375 &  74490 &  -2.197144 &   1.2037 &   1.7407 &  2.8696 &   0.2714 \\
203403 &  44417 &  -4.254406 &  -1.1532 &  -0.6168 &  0.6119 &  -1.8017 \\
203418 &   6948 &  -4.259047 &  -0.5353 &   0.0917 &  1.4939 &  -1.3811 \\
204014 &  15143 &  -2.228664 &   1.1651 &   1.6944 &  2.9200 &   0.4585 \\
204064 &  24504 &  -3.520055 &   0.0353 &   0.6063 &  1.9467 &  -0.7323 \\
204069 &  53356 &  -3.062407 &   0.7606 &   1.3575 &  2.7208 &  -0.1809 \\
204277 &  19757 &  -2.461900 &   1.4513 &   2.0550 &  3.4318 &   0.5428 \\
\hline
\end{tabular}
\label{tab:prapptbl5}
\end{center}
\end{table}

\newpage
\clearpage

\begin{table}[htbp]
\caption[Properties of the events for the $\eetoppbar$ 
final selection (Part 6).]
{Properties of the events for the $\eetoppbar$ final selection (Part 6).  
This table deals with separating protons from electrons based on $dE/dx$ and 
RICH information.  Columns 1 and 3 are $\Delta S^2(p$$-$$e) 
= S^2(p) - S^2(e)$ for positive and negative tracks, respectively.  
Columns 2 and 4 are $\Delta \chi^2_{RICH}(p$$-$$e) 
= -2$ log$L(p)+2$ log$L(e)$ for positive and negative tracks, respectively. 
Column 1+2 (3+4) is the sum of columns 1 and 2 (3 and 4).  
A track is more likely to be a proton than an electron if the value is $<$0.}
\begin{center}
\begin{tabular}{|c|c|c|c|c|c|c|c|}
\hline
Run & Event & 1 & 2 & 1+2 & 3 & 4 & 3+4 \\
\hline
203131 &  37307 &   -7.499 &   -90.334 &   -97.833 &   -7.500 &   -24.676 &   -32.176 \\
203224 &  11782 &   -1.902 &   -54.274 &   -56.176 &   -5.619 &   -15.512 &   -21.132 \\
203240 &  39099 &   -5.145 &   -62.874 &   -68.019 &  -11.186 &  -106.279 &  -117.465 \\
203267 &  26988 &   -6.493 &   -91.694 &   -98.187 &    2.011 &   -50.344 &   -48.333 \\
203267 &  61798 &   -7.092 &   -72.475 &   -79.567 &  -13.597 &   -46.258 &   -59.855 \\
203273 &  35020 &   -5.021 &  -120.827 &  -125.848 &  -11.634 &   -51.287 &   -62.921 \\
203274 &  18555 &   -8.858 &   -41.762 &   -50.619 &  -12.593 &   -94.390 &  -106.983 \\
203352 &  26890 &  -14.317 &  -107.075 &  -121.393 &   -8.022 &   -57.359 &   -65.381 \\
203359 &  41016 &    0.450 &  -151.575 &  -151.125 &   -8.421 &   -12.197 &   -20.618 \\
203375 &  74490 &   -5.450 &   -62.220 &   -67.670 &   -4.754 &   -78.443 &   -83.197 \\
203403 &  44417 &    0.318 &   -63.508 &   -63.190 &  -14.854 &  -223.358 &  -238.211 \\
203418 &   6948 &    6.155 &  -100.519 &   -94.363 &  -16.232 &   -61.634 &   -77.866 \\
204014 &  15143 &   -6.439 &   -92.645 &   -99.084 &   -4.756 &   -89.057 &   -93.813 \\
204064 &  24504 &  -11.475 &  -115.501 &  -126.976 &  -11.854 &  -113.530 &  -125.384 \\
204069 &  53356 &   -8.116 &   -82.173 &   -90.289 &   -9.346 &   -44.003 &   -53.349 \\
204277 &  19757 &  -14.143 &   -78.231 &   -92.374 &   -5.766 &   -21.996 &   -27.762 \\
\hline
\hline
\multicolumn{2}{|c|}{Requirement} 
 & --- & --- & $<$ 0 
 & --- & --- & $<$ 0 \\
\hline
\end{tabular}
\end{center}
\end{table}

\newpage
\clearpage

\begin{table}[htbp]
\caption[Properties of the events for the $\eetoppbar$ 
final selection (Part 7).]
{Properties of the events for the $\eetoppbar$ final selection (Part 7).  
This table deals with separating protons from muons based on $dE/dx$ and 
RICH information.  Columns 1 and 3 are $\Delta S^2(p$$-$$\mu) 
= S^2(p) - S^2(\mu)$ for positive and negative tracks, respectively.  
Columns 2 and 4 are $\Delta \chi^2_{RICH}(p$$-$$\mu) 
= -2$ log$L(p)+2$ log$L(\mu)$ for positive and negative tracks, respectively. 
Column 1+2 (3+4) is the sum of columns 1 and 2 (3 and 4).  A track 
is more likely to be a proton than an muon if the value is $<$0.}
\begin{center}
\begin{tabular}{|c|c|c|c|c|c|c|c|}
\hline
Run & Event & 1 & 2 & 1+2 & 3 & 4 & 3+4 \\
\hline
203131 &  37307 &  -0.714 &   -90.334 &   -91.048 &  -0.792 &   -24.081 &   -24.873 \\
203224 &  11782 &  -2.010 &   -52.893 &   -54.903 &  -0.815 &   -16.121 &   -16.937 \\
203240 &  39099 &  -1.798 &   -62.874 &   -64.672 &   0.849 &  -105.826 &  -104.976 \\
203267 &  26988 &  -0.841 &   -91.694 &   -92.534 &  -3.924 &   -50.308 &   -54.232 \\
203267 &  61798 &  -0.637 &   -72.254 &   -72.892 &   1.461 &   -46.395 &   -44.934 \\
203273 &  35020 &  -1.018 &  -120.815 &  -121.833 &   0.976 &   -51.543 &   -50.567 \\
203274 &  18555 &  -0.517 &   -41.755 &   -42.272 &   0.659 &   -94.390 &   -93.730 \\
203352 &  26890 &   2.081 &  -107.075 &  -104.995 &  -0.105 &   -57.217 &   -57.323 \\
203359 &  41016 &  -3.250 &  -149.820 &  -153.070 &   0.220 &   -11.361 &   -11.141 \\
203375 &  74490 &  -0.672 &   -62.220 &   -62.892 &  -1.375 &   -77.204 &   -78.579 \\
203403 &  44417 &  -3.073 &   -62.663 &   -65.735 &   1.916 &  -222.964 &  -221.048 \\
203418 &   6948 &  -4.702 &  -100.519 &  -105.220 &   1.621 &   -61.010 &   -59.389 \\
204014 &  15143 &  -0.365 &   -92.259 &   -92.625 &  -1.147 &   -89.631 &   -90.778 \\
204064 &  24504 &   0.516 &  -115.501 &  -114.985 &   0.535 &  -113.959 &  -113.424 \\
204069 &  53356 &  -0.797 &   -82.173 &   -82.970 &  -0.546 &   -44.094 &   -44.640 \\
204277 &  19757 &   1.757 &   -78.219 &   -76.462 &  -1.812 &   -21.996 &   -23.807 \\
\hline
\hline
\multicolumn{2}{|c|}{Requirement} 
 & --- & --- & $<$ -2 
 & --- & --- & $<$ -2 \\
\hline
\end{tabular}
\end{center}
\end{table}

\newpage
\clearpage

\begin{table}[htbp]
\caption[Properties of the events for the $\eetoppbar$ 
final selection (Part 8).]
{Properties of the events for the $\eetoppbar$ final selection (Part 8).  
This table deals with separating protons from pions based on $dE/dx$ and 
RICH information.  Columns 1 and 3 are $\Delta S^2(p$$-$$\pi) 
= S^2(p) - S^2(\pi)$ for positive and negative tracks, respectively.  
Columns 2 and 4 are $\Delta \chi^2_{RICH}(p$$-$$\pi) 
= -2$ log$L(p)+2$ log$L(\pi)$ for positive and negative tracks, respectively. 
Column 1+2 (3+4) is the sum of columns 1 and 2 (3 and 4).  A track 
is more likely to be a proton than an pion if the value is $<$0.}
\begin{center}
\begin{tabular}{|c|c|c|c|c|c|c|c|}
\hline
Run & Event & 1 & 2 & 1+2 & 3 & 4 & 3+4 \\
\hline
203131 &  37307 &  -2.030 &   -90.334 &   -92.364 &  -2.181 &   -23.738 &   -25.919 \\
203224 &  11782 &  -4.278 &   -51.788 &   -56.067 &  -2.115 &   -16.573 &   -18.687 \\
203240 &  39099 &  -3.737 &   -62.874 &   -66.612 &   0.519 &  -105.540 &  -105.021 \\
203267 &  26988 &  -2.224 &   -91.694 &   -93.918 &  -7.170 &   -50.272 &   -57.441 \\
203267 &  61798 &  -1.896 &   -72.111 &   -74.007 &   1.993 &   -46.511 &   -44.518 \\
203273 &  35020 &  -2.463 &  -120.801 &  -123.264 &   0.921 &   -51.733 &   -50.812 \\
203274 &  18555 &  -1.755 &   -41.773 &   -43.528 &   0.159 &   -94.390 &   -94.230 \\
203352 &  26890 &   2.768 &  -107.075 &  -104.308 &  -0.847 &   -57.075 &   -57.922 \\
203359 &  41016 &  -5.809 &  -148.655 &  -154.465 &  -0.313 &   -10.803 &   -11.116 \\
203375 &  74490 &  -1.781 &   -62.220 &   -64.001 &  -2.956 &   -76.269 &   -79.226 \\
203403 &  44417 &  -5.877 &   -62.071 &   -67.947 &   2.866 &  -222.613 &  -219.748 \\
203418 &   6948 &  -9.197 &  -100.519 &  -109.716 &   1.899 &   -60.544 &   -58.645 \\
204014 &  15143 &  -1.233 &   -91.993 &   -93.226 &  -2.661 &   -94.247 &   -96.908 \\
204064 &  24504 &   0.223 &  -115.501 &  -115.278 &   0.169 &  -114.277 &  -114.108 \\
204069 &  53356 &  -2.175 &   -82.173 &   -84.347 &  -1.810 &   -44.107 &   -45.917 \\
204277 &  19757 &   1.918 &   -78.205 &   -76.287 &  -3.928 &   -21.996 &   -25.924 \\
\hline
\end{tabular}
\end{center}
\end{table}

\newpage
\clearpage

\begin{table}[htbp]
\caption[Properties of the events for the $\eetoppbar$ 
final selection (Part 9).]
{Properties of the events for the $\eetoppbar$ final selection (Part 9).  
This table deals with separating protons from kaons based on $dE/dx$ and 
RICH information.  Columns 1 and 3 are $\Delta S^2(p$$-$$K) 
= S^2(p) - S^2(K)$ for positive and negative tracks, respectively.  
Columns 2 and 4 are $\Delta \chi^2_{RICH}(p$$-$$K) 
= -2$ log$L(p)+2$ log$L(K)$ for positive and negative tracks, respectively. 
Column 1+2 (3+4) is the sum of columns 1 and 2 (3 and 4).  A track 
is more likely to be a proton than an kaon if the value is $<$0.}
\begin{center}
\begin{tabular}{|c|c|c|c|c|c|c|c|}
\hline
Run & Event & 1 & 2 & 1+2 & 3 & 4 & 3+4 \\
\hline
203131 &  37307 &   -7.241 &   -90.289 &   -97.530 &   -7.582 &   -36.224 &   -43.807 \\
203224 &  11782 &  -11.391 &   -38.167 &   -49.558 &   -7.051 &   -27.987 &   -35.038 \\
203240 &  39099 &   -9.947 &   -62.874 &   -72.821 &   -2.411 &  -107.370 &  -109.781 \\
203267 &  26988 &   -7.548 &   -91.694 &   -99.242 &  -15.975 &   -33.068 &   -49.043 \\
203267 &  61798 &   -7.117 &   -74.137 &   -81.253 &    0.988 &   -43.297 &   -42.309 \\
203273 &  35020 &   -7.684 &  -107.202 &  -114.886 &   -1.393 &   -56.538 &   -57.931 \\
203274 &  18555 &   -7.119 &   -46.069 &   -53.188 &   -3.584 &   -94.381 &   -97.965 \\
203352 &  26890 &    2.114 &  -107.035 &  -104.920 &   -4.572 &   -47.864 &   -52.437 \\
203359 &  41016 &  -12.763 &  -153.117 &  -165.880 &   -3.388 &   -12.560 &   -15.949 \\
203375 &  74490 &   -6.035 &   -62.220 &   -68.255 &   -8.161 &   -43.880 &   -52.040 \\
203403 &  44417 &  -13.867 &   -70.242 &   -84.109 &    2.872 &  -206.506 &  -203.634 \\
203418 &   6948 &  -21.401 &  -100.519 &  -121.919 &   -0.324 &   -45.542 &   -45.866 \\
204014 &  15143 &   -5.343 &   -93.899 &   -99.242 &   -8.316 &  -101.827 &  -110.143 \\
204064 &  24504 &   -2.933 &  -115.456 &  -118.389 &   -3.253 &  -125.285 &  -128.539 \\
204069 &  53356 &   -7.692 &   -81.952 &   -89.644 &   -7.370 &   -38.665 &   -46.035 \\
204277 &  19757 &   -0.186 &   -68.257 &   -68.444 &  -11.482 &   -21.996 &   -33.478 \\
\hline
\end{tabular}
\end{center}
\end{table}

\newpage
\clearpage

\begin{table}[htbp]
\caption[Properties of the events for the $\eetoppbar$ 
final selection (Part 10).]
{Properties of the events for the $\eetoppbar$ final selection (Part 10).  
Columns 1 and 3 are $E_{CC}/p$ for positive and negative tracks, respectively.  
Columns 2 and 4 are the distances transversed in the Muon Chamber (in terms of 
nuclear interaction lengths) by positive and negative tracks, respectively. }
\begin{center}
\begin{tabular}{|c|c|c|c|c|c|}
\hline
Run & Event & 1 & 2 & 3 & 4 \\
\hline
203131 &  37307 &  0.129 &  0.00 &  0.237 &  0.00 \\
203224 &  11782 &  0.268 &  0.00 &  1.028 &  0.00 \\
203240 &  39099 &  0.193 &  0.00 &  0.713 &  0.00 \\
203267 &  26988 &  0.129 &  0.00 &  0.137 &  0.00 \\
203267 &  61798 &  0.143 &  0.00 &  0.123 &  0.00 \\
203273 &  35020 &  0.143 &  0.00 &  0.852 &  0.00 \\
203274 &  18555 &  0.127 &  0.00 &  0.603 &  0.00 \\
203352 &  26890 &  0.471 &  0.05 &  0.882 &  0.00 \\
203359 &  41016 &  0.260 &  0.00 &  1.085 &  0.00 \\
203375 &  74490 &  0.135 &  0.00 &  0.232 &  0.00 \\
203403 &  44417 &  0.128 &  0.00 &  0.751 &  0.00 \\
203418 &   6948 &  0.135 &  0.00 &  0.723 &  0.00 \\
204014 &  15143 &  0.113 &  0.00 &  0.181 &  0.00 \\
204064 &  24504 &  0.181 &  0.00 &  0.694 &  0.00 \\
204069 &  53356 &  0.176 &  0.00 &  0.442 &  0.00 \\
204277 &  19757 &  0.197 &  0.00 &  0.990 &  0.00 \\
\hline
\hline
\multicolumn{2}{|c|}{Requirement} 
 & $<$ 0.85 & --- & --- & --- \\
\hline
\end{tabular}
\label{tab:prapptbl10}
\end{center}
\end{table}

\baselineskip=24pt
\chapter{Results from Other Experiments}

This appendix summerizes the electromagnetic form factor results from other experiments.  
Tables B.1-3 contain measurements of the pion form factor in the timelike region.  
Tables B.4-7 contain measurements of the pion form factor in the spacelike region from $\pi$-$e$ 
scattering experiments.  
Table B.8 contains measurements of the pion form factor 
in the spacelike region from electroproduction experiments.
Tables B.9,10 contain measurements of the kaon form factor in the timelike region.
Table B.11 contains measurements of the kaon form factor in the spacelike region from $K$-$e$ 
scattering experiments.  
Tables B.12,13 contain measurements of the proton magentic form factor in the timelike region, 
assuming $|G^{P}_{E}(Q^2)|$ = $|G^{P}_{M}(Q^2)|$. 
Tables B.14,15 contain measurements of the proton magentic form factor in the timelike region 
from the Babar experiment. 
Tables B.16,17 contain measurements of the proton magentic form factor in the spacelike region, 
assuming $G^{P}_{E}(Q^2)$ = $G^{P}_{M}(Q^2)/\mu_p$. 
Table B.18 contains measurements of proton form factor ratios in the timelike region 
from the Babar experiment. 
Table B.19 contains measurements of proton form factor ratios in the spacelike region 
from polarization transfer experiments. 
Tables B.19-23 contain measurements of proton form factor ratios in the spacelike region from 
Rosenbluth separation experiments.  
All errors listed in the tables have statistical and systematic uncertainties 
summed in quadrature.

\newpage
\clearpage

\begin{table}
\caption[Pion electromagnetic form factor in the timelike region (Part 1).]
{Pion electromagnetic form factor in the timelike region (Part 1).}
\begin{center}
\begin{tabular}{|c|l|l|c|}
\hline
$|Q^{2}|$ (GeV$^{2}$) & $|F_{\pi}|$ & $|Q^{2}||F_{\pi}|$ (GeV$^{2}$) & Ref. \\
\hline
1.369 & 1.17$\pm$0.16 & 1.60$\pm$0.22 & \cite{pitlff_1} \\
1.385 & 1.15$\pm$0.08 & 1.60$\pm$0.11 & \cite{pitlff_2} \\
1.39 & 1.26$^{+0.40}_{-0.28}$ & 1.76$^{+0.56}_{-0.39}$ &\cite{kpicommontlff_1} \\
1.409 & 1.17$\pm$0.09 & 1.66$\pm$0.13 & \cite{pitlff_2} \\
1.416 & 1.16$\pm$0.15 & 1.64$\pm$0.21 & \cite{pitlff_1} \\
1.433 & 1.07$\pm$0.07 & 1.54$\pm$0.11 & \cite{pitlff_2} \\
1.44 & 1.086$^{+0.132}_{-0.151}$ & 1.56$^{+0.19}_{-0.22}$ & \cite{pitlff_VDM} \\
1.457 & 1.16$\pm$0.10 & 1.69$\pm$0.14 & \cite{pitlff_2} \\
1.464 & 1.28$\pm$0.13 & 1.87$\pm$0.19 & \cite{pitlff_1} \\
1.481 & 0.87$\pm$0.08 & 1.29$\pm$0.12 & \cite{pitlff_2} \\
1.506 & 0.93$\pm$0.10 & 1.40$\pm$0.15 & \cite{pitlff_2} \\
1.513 & 1.12$\pm$0.14 & 1.70$\pm$0.21 & \cite{pitlff_1} \\
1.530 & 1.12$\pm$0.09 & 1.72$\pm$0.14 & \cite{pitlff_2} \\
1.555 & 0.92$\pm$0.08 & 1.43$\pm$0.13 & \cite{pitlff_2} \\
1.563 & 1.03$\pm$0.16 & 1.61$\pm$0.25 & \cite{pitlff_1} \\
1.580 & 0.93$\pm$0.08 & 1.47$\pm$0.12 & \cite{pitlff_2} \\
1.59 & 1.67$^{+0.57}_{-0.36}$ & 2.66$^{+0.91}_{-0.57}$ & \cite{kpicommontlff_1} \\
1.605 & 0.82$\pm$0.08 & 1.31$\pm$0.13 & \cite{pitlff_2} \\
1.613 & 0.87$\pm$0.17 & 1.41$\pm$0.27 & \cite{pitlff_1} \\
1.631 & 0.66$\pm$0.08 & 1.08$\pm$0.12 & \cite{pitlff_2} \\
1.656 & 0.78$\pm$0.07 & 1.29$\pm$0.12 & \cite{pitlff_2} \\
1.664 & 0.69$\pm$0.21 & 1.14$\pm$0.35 & \cite{pitlff_1} \\
1.682 & 0.82$\pm$0.09 & 1.38$\pm$0.15 & \cite{pitlff_2} \\
1.69 & 0.846$^{+0.125}_{-0.146}$ & 1.43$^{+0.21}_{-0.25}$ & \cite{pitlff_VDM} \\
\hline
\end{tabular}
\end{center}
\end{table}

\begin{table}
\caption[Pion electromagnetic form factor in the timelike region (Part 2).]
{Pion electromagnetic form factor in the timelike region (Part 2).}
\begin{center}
\begin{tabular}{|c|l|l|c|}
\hline
$|Q^{2}|$ (GeV$^{2}$) & $|F_{\pi}|$ & $|Q^{2}||F_{\pi}|$ (GeV$^{2}$) & Ref. \\
\hline
1.708 & 0.82$\pm$0.08 & 1.40$\pm$0.14 & \cite{pitlff_2} \\
1.716 & 0.57$\pm$0.37 & 0.99$\pm$0.63 & \cite{pitlff_1} \\
1.734 & 0.83$\pm$0.10 & 1.44$\pm$0.17 & \cite{pitlff_2} \\
1.761 & 0.67$\pm$0.10 & 1.18$\pm$0.17 & \cite{pitlff_2} \\
1.769 & 0.81$\pm$0.26 & 1.44$\pm$0.46 & \cite{pitlff_1} \\
1.788 & 0.82$\pm$0.08 & 1.47$\pm$0.15 & \cite{pitlff_2} \\
1.80 & 1.05$^{+1.58}_{-0.48}$ & 1.88$^{+2.82}_{-0.85}$ & \cite{kpicommontlff_1} \\
1.814 & 0.58$\pm$0.07 & 1.06$\pm$0.12 & \cite{pitlff_2} \\
1.823 & 0.73$\pm$0.05 & 1.33$\pm$0.09 & \cite{pitlff_3} \\
1.841 & 0.64$\pm$0.08 & 1.18$\pm$0.14 & \cite{pitlff_2} \\
1.869 & 0.51$\pm$0.07 & 0.95$\pm$0.13 & \cite{pitlff_2} \\
1.896 & 0.55$\pm$0.08 & 1.04$\pm$0.16 & \cite{pitlff_2} \\
1.924 & 0.62$\pm$0.07 & 1.20$\pm$0.14 & \cite{pitlff_2} \\
1.946 & 0.73$\pm$0.08 & 1.43$\pm$0.16 & \cite{pitlff_3} \\
1.952 & 0.45$\pm$0.08 & 0.87$\pm$0.15 & \cite{pitlff_2} \\
1.96 & 0.798$^{+0.112}_{-0.131}$ & 1.56$^{+0.22}_{-0.26}$ & \cite{pitlff_VDM} \\
2.031 & 0.46$\pm$0.05 & 0.93$\pm$0.11 & \cite{pitlff_3} \\
2.176 & 0.39$\pm$0.05 & 0.84$\pm$0.11 & \cite{pitlff_3} \\
$\langle$2.21$\rangle$ & 0.54$\pm$0.16 & 1.19$\pm$0.35 &\cite{ kpicommontlff_3} \\
2.326 & 0.28$\pm$0.09 & 0.66$\pm$0.21 & \cite{pitlff_3} \\
2.481 & 0.25$\pm$0.08 & 0.61$\pm$0.20 & \cite{pitlff_3} \\
$\langle$2.56$\rangle$ & 0.361$^{+0.064}_{-0.078}$ & 0.92$^{+0.16}_{-0.20}$ & \cite{pitlff_VDM} \\
2.56 & 0.49$\pm$0.14 & 1.25$\pm$0.37 & \cite{kpicommontlff_2} \\
2.641 & 0.28$\pm$0.04 & 0.75$\pm$0.09 & \cite{pitlff_3} \\
2.806 & 0.35$\pm$0.04 & 0.97$\pm$0.12 & \cite{pitlff_3} \\
2.976 & 0.57$\pm$0.05 & 1.71$\pm$0.16 & \cite{pitlff_3} \\
\hline
\end{tabular}
\end{center}
\end{table}

\begin{table}
\caption[Pion electromagnetic form factor in the timelike region (Part 3).]
{Pion electromagnetic form factor in the timelike region (Part 3).}
\begin{center}
\begin{tabular}{|c|l|l|c|}
\hline
$|Q^{2}|$ (GeV$^{2}$) & $|F_{\pi}|$ & $|Q^{2}||F_{\pi}|$ (GeV$^{2}$) & Ref. \\
\hline
3.151 & 0.47$\pm$0.05 & 1.48$\pm$0.17 & \cite{pitlff_3} \\
3.331 & 0.56$\pm$0.06 & 1.85$\pm$0.21 & \cite{pitlff_3} \\
3.42 & 0.267$^{+0.072}_{-0.101}$ & 0.91$^{+0.25}_{-0.35}$ & \cite{pitlff_VDM} \\
3.516 & 0.53$\pm$0.08 & 1.86$\pm$0.27 & \cite{pitlff_3} \\
3.61 & 0.349$^{+0.049}_{-0.057}$ & 1.26$^{+0.18}_{-0.21}$ & \cite{pitlff_VDM} \\
3.706 & 0.46$\pm$0.07 & 1.70$\pm$0.24 & \cite{pitlff_3} \\
3.76 & 0.253$^{+0.105}_{-0.253}$ & 0.95$^{+0.40}_{-0.95}$ & \cite{pitlff_VDM} \\
3.901 & 0.17$\pm$0.09 & 0.68$\pm$0.34 & \cite{pitlff_3} \\
3.92 & 0.312$^{+0.090}_{-0.129}$ & 1.22$^{+0.35}_{-0.51}$ & \cite{pitlff_VDM} \\
4.101 & 0.33$\pm$0.06 & 1.36$\pm$0.25 & \cite{pitlff_3} \\
4.306 & 0.28$\pm$0.05 & 1.22$\pm$0.23 & \cite{pitlff_3} \\
4.41 & 0.174$^{+0.031}_{-0.038}$ & 0.77$^{+0.14}_{-0.17}$ & \cite{pitlff_VDM} \\
4.516 & 0.26$\pm$0.06 & 1.19$\pm$0.26 & \cite{pitlff_3} \\
4.731 & $<$0.50 & $<$2.37 & \cite{pitlff_3} \\
4.951 & $<$0.47 & $<$2.32 & \cite{pitlff_3} \\
5.688 & $<$0.32 & $<$1.80 & \cite{pitlff_3} \\
5.76 & $<$0.160 & $<$0.92 & \cite{pitlff_VDM} \\
6.76 & 0.227$^{+0.064}_{-0.091}$ & 1.53$^{+0.43}_{-0.62}$ & \cite{pitlff_VDM} \\
7.84 & 0.133$^{+0.063}_{-0.133}$ & 1.04$^{+0.49}_{-1.04}$ & \cite{pitlff_VDM} \\
9.00 & 0.129$^{+0.064}_{-0.129}$ & 1.16$^{+0.58}_{-0.16}$ & \cite{pitlff_VDM} \\ 
\hline
\end{tabular}
\end{center}
\end{table}

\begin{table}
\caption[Pion electromagnetic form factor in the spacelike region from 
$\pi$-$e$ scattering experiments (Part 1).]
{Pion electromagnetic form factor in the spacelike region from 
$\pi$-$e$ scattering experiments (Part 1).}  
\begin{center}
\begin{tabular}{|c|l|l|c|}
\hline
$Q^{2}$ (GeV$^2$) & $F_{\pi}$ & $Q^2~F_{\pi}$ (GeV$^2$) & Ref. \\
\hline
0.0138 & 1.00$\pm$0.05 & 0.0138$\pm$0.0007 & \cite{pislff_sc1} \\
0.0149 & 1.00$\pm$0.05 & 0.0149$\pm$0.0007 & \cite{pislff_sc1} \\
0.0150 & 0.972$\pm$0.004 & 0.01457$\pm$0.00005 & \cite{pislff_sc4} \\
0.0159 & 0.99$\pm$0.04 & 0.0158$\pm$0.0007 & \cite{pislff_sc1} \\
0.0169 & 0.99$\pm$0.05 & 0.0168$\pm$0.0008 & \cite{pislff_sc1} \\
0.0170 & 0.960$\pm$0.003 & 0.01631$\pm$0.00005 & \cite{pislff_sc4} \\
0.0179 & 0.99$\pm$0.05 & 0.0178$\pm$0.0008 & \cite{pislff_sc1} \\
0.0190 & 0.961$\pm$0.023 & 0.0183$\pm$0.0004 & \cite{pislff_sc1} \\
0.0190 & 0.966$\pm$0.003 & 0.01835$\pm$0.00006 & \cite{pislff_sc4} \\
0.0200 & 0.957$\pm$0.022 & 0.0191$\pm$0.0004 & \cite{pislff_sc1} \\
0.0210 & 0.973$\pm$0.023 & 0.0204$\pm$0.0005 & \cite{pislff_sc1} \\
0.0210 & 0.962$\pm$0.003 & 0.02021$\pm$0.00007 & \cite{pislff_sc4} \\
0.0220 & 0.974$\pm$0.025 & 0.0214$\pm$0.0005 & \cite{pislff_sc1} \\
0.0230 & 0.956$\pm$0.004 & 0.02199$\pm$0.00008 & \cite{pislff_sc4} \\
0.0231 & 0.958$\pm$0.024 & 0.0221$\pm$0.0006 & \cite{pislff_sc1} \\
0.0241 & 0.947$\pm$0.025 & 0.0228$\pm$0.0006 & \cite{pislff_sc1} \\
0.0250 & 0.951$\pm$0.004 & 0.02378$\pm$0.00009 & \cite{pislff_sc4} \\
0.0251 & 0.951$\pm$0.026 & 0.0239$\pm$0.0006 & \cite{pislff_sc1} \\
0.0261 & 0.947$\pm$0.027 & 0.0247$\pm$0.0007 & \cite{pislff_sc1} \\
0.0270 & 0.948$\pm$0.004 & 0.02559$\pm$0.00011 & \cite{pislff_sc4} \\
0.0272 & 0.95$\pm$0.03 & 0.0257$\pm$0.0008 & \cite{pislff_sc1} \\
0.0282 & 0.95$\pm$0.03 & 0.0269$\pm$0.0009 & \cite{pislff_sc1} \\
0.0290 & 0.940$\pm$0.004 & 0.02727$\pm$0.00012 & \cite{pislff_sc4} \\
0.0292 & 0.920$\pm$0.027 & 0.0269$\pm$0.0008 & \cite{pislff_sc1} \\
\hline
\end{tabular}
\end{center}
\end{table}

\begin{table}
\caption[Pion electromagnetic form factor in the spacelike region from 
$\pi$-$e$ scattering experiments (Part 2).]
{Pion electromagnetic form factor in the spacelike region from 
$\pi$-$e$ scattering experiments (Part 2).}  
\begin{center}
\begin{tabular}{|c|l|l|c|}
\hline
$Q^{2}$ (GeV$^2$) & $F_{\pi}$ & $Q^2~F_{\pi}$ (GeV$^2$) & Ref. \\
\hline
0.0302 & 0.93$\pm$0.03 & 0.0282$\pm$0.0009 & \cite{pislff_sc1} \\
0.0310 & 0.940$\pm$0.005 & 0.02915$\pm$0.00015 & \cite{pislff_sc4} \\
0.0312 & 0.95$\pm$0.03 & 0.0298$\pm$0.0010 & \cite{pislff_sc1} \\
0.0317 & 0.950$\pm$0.014 & 0.0301$\pm$0.0004 & \cite{pislff_sc2} \\
0.0323 & 0.93$\pm$0.03 & 0.0299$\pm$0.0010 & \cite{pislff_sc1} \\
0.0330 & 0.943$\pm$0.005 & 0.03113$\pm$0.00016 & \cite{pislff_sc4} \\
0.0333 & 0.86$\pm$0.04 & 0.0287$\pm$0.0013 & \cite{pislff_sc1} \\
0.0337 & 0.954$\pm$0.014 & 0.0321$\pm$0.0005 & \cite{pislff_sc2} \\
0.0343 & 0.94$\pm$0.04 & 0.0323$\pm$0.0015 & \cite{pislff_sc1} \\
0.0350 & 0.931$\pm$0.005 & 0.03257$\pm$0.00019 & \cite{pislff_sc4} \\
0.0353 & 0.92$\pm$0.06 & 0.0325$\pm$0.0020 & \cite{pislff_sc1} \\
0.0358 & 0.963$\pm$0.016 & 0.0345$\pm$0.0006 & \cite{pislff_sc2} \\
0.0370 & 0.936$\pm$0.006 & 0.03463$\pm$0.00022 & \cite{pislff_sc4} \\
0.0378 & 0.994$\pm$0.017 & 0.0376$\pm$0.0006 & \cite{pislff_sc2} \\
0.039 & 0.925$\pm$0.010 & 0.0361$\pm$0.0004 & \cite{pislff_sc3} \\
0.0390 & 0.926$\pm$0.006 & 0.03610$\pm$0.00023 & \cite{pislff_sc4} \\
0.0399 & 0.954$\pm$0.018 & 0.0381$\pm$0.0007 & \cite{pislff_sc2} \\
0.0419 & 0.964$\pm$0.020 & 0.0404$\pm$0.0008 & \cite{pislff_sc2} \\
0.0420 & 0.921$\pm$0.005 & 0.03870$\pm$0.00021 & \cite{pislff_sc4} \\
0.043 & 0.930$\pm$0.011 & 0.0400$\pm$0.0005 & \cite{pislff_sc3} \\
0.0439 & 0.938$\pm$0.021 & 0.0412$\pm$0.0009 & \cite{pislff_sc2} \\
0.0460 & 0.939$\pm$0.021 & 0.0432$\pm$0.0010 & \cite{pislff_sc2} \\
0.0460 & 0.915$\pm$0.005 & 0.04208$\pm$0.00023 & \cite{pislff_sc4} \\
0.047 & 0.906$\pm$0.012 & 0.0426$\pm$0.0006 & \cite{pislff_sc3} \\
0.0480 & 0.938$\pm$0.023 & 0.0450$\pm$0.0011 & \cite{pislff_sc2} \\
0.0500 & 0.911$\pm$0.005 & 0.04555$\pm$0.00027 & \cite{pislff_sc4} \\
0.0501 & 0.963$\pm$0.024 & 0.0482$\pm$0.0012 & \cite{pislff_sc2} \\
0.051 & 0.917$\pm$0.013 & 0.0467$\pm$0.0007 & \cite{pislff_sc3} \\
0.0521 & 0.985$\pm$0.026 & 0.0513$\pm$0.0014 & \cite{pislff_sc2} \\
0.0540 & 0.895$\pm$0.006 & 0.0483$\pm$0.0003 & \cite{pislff_sc4} \\
0.0542 & 0.957$\pm$0.028 & 0.0518$\pm$0.0015 & \cite{pislff_sc2} \\
\hline
\end{tabular}
\end{center}
\end{table}

\begin{table}
\caption[Pion electromagnetic form factor in the spacelike region from 
$\pi$-$e$ scattering experiments (Part 3).]
{Pion electromagnetic form factor in the spacelike region from 
$\pi$-$e$ scattering experiments (Part 3).}  
\begin{center}
\begin{tabular}{|c|l|l|c|}
\hline
$Q^{2}$ (GeV$^2$) & $F_{\pi}$ & $Q^2~F_{\pi}$ (GeV$^2$) & Ref. \\
\hline
0.055 & 0.912$\pm$0.014 & 0.0501$\pm$0.0008 & \cite{pislff_sc3} \\
0.0562 & 0.86$\pm$0.03 & 0.0482$\pm$0.0017 & \cite{pislff_sc2} \\
0.0580 & 0.894$\pm$0.007 & 0.0519$\pm$0.0004 & \cite{pislff_sc4} \\
0.0583 & 0.88$\pm$0.03 & 0.0515$\pm$0.0019 & \cite{pislff_sc2} \\
0.059 & 0.876$\pm$0.015 & 0.0517$\pm$0.0009 & \cite{pislff_sc3} \\
0.0603 & 0.92$\pm$0.04 & 0.0557$\pm$0.0021 & \cite{pislff_sc2} \\
0.0620 & 0.899$\pm$0.007 & 0.0558$\pm$0.0004 & \cite{pislff_sc4} \\
0.0623 & 0.93$\pm$0.04 & 0.0581$\pm$0.0024 & \cite{pislff_sc2} \\
0.063 & 0.915$\pm$0.016 & 0.0577$\pm$0.0010 & \cite{pislff_sc3} \\
0.0644 & 0.90$\pm$0.04 & 0.0582$\pm$0.0027 & \cite{pislff_sc2} \\
0.0660 & 0.887$\pm$0.008 & 0.0585$\pm$0.0005 & \cite{pislff_sc4} \\
0.0664 & 0.90$\pm$0.04 & 0.060$\pm$0.003 & \cite{pislff_sc2} \\
0.067 & 0.870$\pm$0.018 & 0.0583$\pm$0.0012 & \cite{pislff_sc3} \\
0.0685 & 0.89$\pm$0.05 & 0.061$\pm$0.003 & \cite{pislff_sc2} \\
0.0700 & 0.886$\pm$0.008 & 0.0620$\pm$0.0006 & \cite{pislff_sc4} \\
0.0705 & 0.88$\pm$0.05 & 0.062$\pm$0.004 & \cite{pislff_sc2} \\
0.072 & 0.889$\pm$0.020 & 0.0640$\pm$0.0014 & \cite{pislff_sc3} \\
0.0740 & 0.881$\pm$0.009 & 0.0652$\pm$0.0007 & \cite{pislff_sc4} \\
0.076 & 0.873$\pm$0.022 & 0.0663$\pm$0.0017 & \cite{pislff_sc3} \\
0.0780 & 0.877$\pm$0.010 & 0.0684$\pm$0.0008 & \cite{pislff_sc4} \\
0.080 & 0.875$\pm$0.023 & 0.0700$\pm$0.0019 & \cite{pislff_sc3} \\
0.0830 & 0.870$\pm$0.009 & 0.0722$\pm$0.0007 & \cite{pislff_sc4} \\
0.084 & 0.896$\pm$0.025 & 0.0752$\pm$0.0021 & \cite{pislff_sc3} \\
0.088 & 0.849$\pm$0.028 & 0.0747$\pm$0.0024 & \cite{pislff_sc3} \\
0.0890 & 0.846$\pm$0.009 & 0.0753$\pm$0.0008 & \cite{pislff_sc4} \\
0.092 & 0.853$\pm$0.029 & 0.0785$\pm$0.0027 & \cite{pislff_sc3} \\
0.0950 & 0.851$\pm$0.011 & 0.0808$\pm$0.0010 & \cite{pislff_sc4} \\
0.1010 & 0.825$\pm$0.010 & 0.0833$\pm$0.0010 & \cite{pislff_sc4} \\
0.1070 & 0.834$\pm$0.011 & 0.0893$\pm$0.0012 & \cite{pislff_sc4} \\
0.1130 & 0.829$\pm$0.012 & 0.0937$\pm$0.0014 & \cite{pislff_sc4} \\
0.1190 & 0.822$\pm$0.013 & 0.0978$\pm$0.0015 & \cite{pislff_sc4} \\
\hline
\end{tabular}
\end{center}
\end{table}

\begin{table}
\caption[Pion electromagnetic form factor in the spacelike region from 
$\pi$-$e$ scattering experiments (Part 4).]
{Pion electromagnetic form factor in the spacelike region from 
$\pi$-$e$ scattering experiments (Part 4).}  
\begin{center}
\begin{tabular}{|c|l|l|c|}
\hline
$Q^{2}$ (GeV$^2$) & $F_{\pi}$ & $Q^2~F_{\pi}$ (GeV$^2$) & Ref. \\
\hline
0.1250 & 0.815$\pm$0.014 & 0.1019$\pm$0.0018 & \cite{pislff_sc4} \\
0.1310 & 0.807$\pm$0.015 & 0.1057$\pm$0.0019 & \cite{pislff_sc4} \\
0.1370 & 0.804$\pm$0.017 & 0.1101$\pm$0.0023 & \cite{pislff_sc4} \\
0.1440 & 0.785$\pm$0.015 & 0.1130$\pm$0.0021 & \cite{pislff_sc4} \\
0.1530 & 0.809$\pm$0.014 & 0.1237$\pm$0.0022 & \cite{pislff_sc4} \\
0.1630 & 0.750$\pm$0.016 & 0.1223$\pm$0.0026 & \cite{pislff_sc4} \\
0.1730 & 0.731$\pm$0.021 & 0.126$\pm$0.004 & \cite{pislff_sc4} \\
0.1830 & 0.766$\pm$0.022 & 0.140$\pm$0.004 & \cite{pislff_sc4} \\
0.1930 & 0.738$\pm$0.024 & 0.142$\pm$0.005 & \cite{pislff_sc4} \\
0.2030 & 0.727$\pm$0.027 & 0.148$\pm$0.006 & \cite{pislff_sc4} \\
0.2130 & 0.78$\pm$0.03 & 0.167$\pm$0.007 & \cite{pislff_sc4} \\
0.2230 & 0.70$\pm$0.04 & 0.156$\pm$0.008 & \cite{pislff_sc4} \\
0.2330 & 0.65$\pm$0.04 & 0.150$\pm$0.010 & \cite{pislff_sc4} \\
0.2430 & 0.77$\pm$0.05 & 0.187$\pm$0.012 & \cite{pislff_sc4} \\
0.2530 & 0.58$\pm$0.06 & 0.147$\pm$0.016 & \cite{pislff_sc4} \\
\hline
\end{tabular}
\end{center}
\end{table}

\begin{table}
\caption[Pion electromagnetic form factor in the spacelike region from 
electroproduction experiments.]
{Pion electromagnetic form factor in the spacelike region from 
electroproduction experiments.  The values listed as Ref. \cite{pislff_elprod4} 
are from the reevaluation in Ref. \cite{pislff_elprod6}.}  
\begin{center}
\begin{tabular}{|c|l|l|c|}
\hline
$Q^{2}$ (GeV$^2$) & $F_{\pi}$ & $Q^2~F_{\pi}$ (GeV$^2$) & Ref. \\
\hline
0.18 & 0.850$\pm$0.044 & 0.153$\pm$0.008 & \cite{pislff_elprod1} \\
0.29 & 0.634$\pm$0.029 & 0.184$\pm$0.008 & \cite{pislff_elprod1} \\
0.40 & 0.570$\pm$0.016 & 0.228$\pm$0.006 & \cite{pislff_elprod1} \\
0.60 & 0.493$\pm$0.022 & 0.296$\pm$0.013 & \cite{pislff_elprod6} \\
0.62 & 0.445$\pm$0.016 & 0.309$\pm$0.019 & \cite{pislff_elprod2} \\
0.70 & 0.471$\pm$0.032 & 0.330$\pm$0.022 & \cite{pislff_elprod6} \\
0.75 & 0.407$\pm$0.031 & 0.305$\pm$0.023 & \cite{pislff_elprod6} \\
0.79 & 0.384$\pm$0.014 & 0.303$\pm$0.011 & \cite{pislff_elprod1} \\
1.00 & 0.351$\pm$0.018 & 0.351$\pm$0.018 & \cite{pislff_elprod6} \\
1.07 & 0.309$\pm$0.019 & 0.331$\pm$0.020 & \cite{pislff_elprod2} \\
1.18 & 0.256$\pm$0.026 & 0.302$\pm$0.031 & \cite{pislff_elprod4} \\
1.19 & 0.238$\pm$0.017 & 0.283$\pm$0.020 & \cite{pislff_elprod1} \\
1.20 & 0.269$\pm$0.012 & 0.323$\pm$0.014 & \cite{pislff_elprod2} \\
1.20 & 0.262$\pm$0.014 & 0.314$\pm$0.017 & \cite{pislff_elprod2} \\
1.20 & 0.294$\pm$0.019 & 0.353$\pm$0.023 & \cite{pislff_elprod3} \\
1.22 & 0.290$\pm$0.030 & 0.354$\pm$0.037 & \cite{pislff_elprod3} \\
1.31 & 0.242$\pm$0.015 & 0.317$\pm$0.020 & \cite{pislff_elprod2} \\
1.60 & 0.251$\pm$0.016 & 0.402$\pm$0.026 & \cite{pislff_elprod6} \\
1.71 & 0.238$\pm$0.020 & 0.407$\pm$0.034 & \cite{pislff_elprod3} \\
1.94 & 0.193$\pm$0.025 & 0.374$\pm$0.049 & \cite{pislff_elprod4} \\
1.99 & 0.179$\pm$0.021 & 0.356$\pm$0.042 & \cite{pislff_elprod3} \\
2.01 & 0.154$\pm$0.014 & 0.310$\pm$0.028 & \cite{pislff_elprod2} \\
3.30 & 0.102$\pm$0.023 & 0.337$\pm$0.076 & \cite{pislff_elprod3} \\
3.33 & 0.086$\pm$0.033 & 0.286$\pm$0.110 & \cite{pislff_elprod4} \\
6.30 & 0.059$\pm$0.030 & 0.372$\pm$0.189 & \cite{pislff_elprod4} \\
9.77 & 0.070$\pm$0.019 & 0.684$\pm$0.186 & \cite{pislff_elprod4} \\
\hline
\end{tabular}
\end{center}
\end{table}

\begin{table}
\caption[Kaon electromagnetic form factor in the timelike region (Part 1).]
{Kaon electromagnetic form factor in the timelike region (Part 1).}
\begin{center}
\begin{tabular}{|c|l|l|c|}
\hline
$|Q^{2}|$ (GeV$^{2}$) & $|F_{K}|$ & $|Q^{2}||F_{K}|$ (GeV$^{2}$) & Ref. \\
\hline
$\langle$1.28$\rangle$ & 2.3$\pm$0.3 & 2.9$\pm$0.4 & \cite{ktlff_3} \\
$\langle$1.32$\rangle$ & 1.97$\pm$0.16 & 2.6$\pm$0.2 & \cite{ktlff_3} \\
$\langle$1.37$\rangle$ & 1.81$\pm$0.12 & 2.47$\pm$0.17 & \cite{ktlff_3} \\
1.39 & 1.1$^{+1.3}_{-0.5}$ & 1.5$^{+1.8}_{-0.6}$ & \cite{kpicommontlff_1} \\
$\langle$1.42$\rangle$ & 1.75$\pm$0.10 & 2.48$\pm$0.14 & \cite{ktlff_3} \\
$\langle$1.46$\rangle$ & 1.34$\pm$0.09 & 1.96$\pm$0.13 & \cite{ktlff_3} \\
$\langle$1.51$\rangle$ & 1.45$\pm$0.09 & 2.19$\pm$0.14 & \cite{ktlff_3} \\
$\langle$1.56$\rangle$ & 1.36$\pm$0.08 & 2.12$\pm$0.13 & \cite{ktlff_3} \\
1.59 & 1.0$^{+1.2}_{-0.5}$ & 1.7$^{+1.9}_{-0.8}$ & \cite{kpicommontlff_1} \\
$\langle$1.61$\rangle$ & 1.44$\pm$0.08 & 2.32$\pm$0.12 & \cite{ktlff_3} \\
$\langle$1.66$\rangle$ & 1.29$\pm$0.08 & 2.14$\pm$0.14 & \cite{ktlff_3} \\
$\langle$1.72$\rangle$ & 1.35$\pm$0.12 & 2.3$\pm$0.2 & \cite{ktlff_3} \\
$\langle$1.77$\rangle$ & 1.25$\pm$0.09 & 2.22$\pm$0.16 & \cite{ktlff_3} \\
1.80 & 1.8$^{+1.2}_{-0.6}$ & 3.3$^{+2.1}_{-1.1}$ & \cite{kpicommontlff_1} \\
$\langle$1.82$\rangle$ & 1.27$\pm$0.06 & 2.32$\pm$0.11 & \cite{ktlff_3} \\
$\langle$1.821$\rangle$ & 1.08$\pm$0.06 & 1.97$\pm$0.12 & \cite{ktlff_4} \\
$\langle$1.88$\rangle$ & 1.24$\pm$0.06 & 2.34$\pm$0.12 & \cite{ktlff_3} \\
$\langle$1.93$\rangle$ & 1.18$\pm$0.06 & 2.28$\pm$0.12 & \cite{ktlff_3} \\
$\langle$1.959$\rangle$ & 1.00$\pm$0.10 & 1.97$\pm$0.20 & \cite{ktlff_4} \\
$\langle$2.029$\rangle$ & 1.15$\pm$0.10 & 2.33$\pm$0.19 & \cite{ktlff_2} \\
$\langle$2.087$\rangle$ & 0.88$\pm$0.06 & 1.83$\pm$0.12 & \cite{ktlff_4} \\
$\langle$2.154$\rangle$ & 0.94$\pm$0.13 & 2.0$\pm$0.3 & \cite{ktlff_2} \\
$\langle$2.21$\rangle$ & 0.90$\pm$0.26 & 2.0$\pm$0.6 & \cite{kpicommontlff_3} \\
$\langle$2.249$\rangle$ & 0.74$\pm$0.07 & 1.67$\pm$0.17 & \cite{ktlff_4} \\
$\langle$2.324$\rangle$ & 0.88$\pm$0.09 & 2.0$\pm$0.2 & \cite{ktlff_2} \\
2.40 & 0.50$\pm$0.08 & 1.20$\pm$0.19 & \cite{ktlff_1} \\
$\langle$2.432$\rangle$ & 0.72$\pm$0.08 & 1.75$\pm$0.19 & \cite{ktlff_4} \\
$\langle$2.479$\rangle$ & 0.71$\pm$0.09 & 1.75$\pm$0.16 & \cite{ktlff_2} \\
$\langle$2.527$\rangle$ & 0.73$\pm$0.07 & 1.86$\pm$0.19 & \cite{ktlff_4} \\
$\langle$2.590$\rangle$ & 0.87$\pm$0.09 & 2.2$\pm$0.2 & \cite{ktlff_2} \\
$\langle$2.590$\rangle$ & 0.73$\pm$0.05 & 1.89$\pm$0.14 & \cite{ktlff_4} \\
$\langle$2.6$\rangle$ & 0.53$\pm$0.12 & 1.38$\pm$0.31 & \cite{ktlff_VDM} \\
2.6 & 0.68$\pm$0.19 & 1.8$\pm$0.5 & \cite{kpicommontlff_2} \\
\hline
\end{tabular}
\end{center}
\end{table}

\begin{table}
\caption[Kaon electromagnetic form factor in the timelike region (Part 2).]
{Kaon electromagnetic form factor in the timelike region (Part 2).}
\begin{center}
\begin{tabular}{|c|l|l|c|}
\hline
$|Q^{2}|$ (GeV$^{2}$) & $|F_{K}|$ & $|Q^{2}||F_{K}|$ (GeV$^{2}$) & Ref. \\
\hline
$\langle$2.655$\rangle$ & 0.67$\pm$0.06 & 1.78$\pm$0.16 & \cite{ktlff_4} \\
$\langle$2.655$\rangle$ & 0.81$\pm$0.09 & 2.1$\pm$0.2 & \cite{ktlff_2} \\
$\langle$2.721$\rangle$ & 0.71$\pm$0.05 & 1.92$\pm$0.13 & \cite{ktlff_2} \\
$\langle$2.721$\rangle$ & 0.63$\pm$0.04 & 1.72$\pm$0.11 & \cite{ktlff_4} \\
$\langle$2.787$\rangle$ & 0.77$\pm$0.06 & 2.16$\pm$0.18 & \cite{ktlff_2} \\
$\langle$2.804$\rangle$ & 0.50$\pm$0.04 & 1.40$\pm$0.11 & \cite{ktlff_4} \\
$\langle$2.854$\rangle$ & 0.59$\pm$0.08 & 1.7$\pm$0.2 & \cite{ktlff_2} \\
$\langle$2.940$\rangle$ & 0.45$\pm$0.04 & 1.31$\pm$0.13 & \cite{ktlff_4} \\
$\langle$2.940$\rangle$ & 0.24$\pm$0.31 & 0.72$\pm$0.90 & \cite{ktlff_2} \\
$\langle$3.043$\rangle$ & 0.37$\pm$0.09 & 1.1$\pm$0.3 & \cite{ktlff_2} \\
$\langle$3.078$\rangle$ & 0.40$\pm$0.04 & 1.23$\pm$0.12 & \cite{ktlff_4} \\
$\langle$3.256$\rangle$ & 0.17$\pm$0.06 & 0.56$\pm$0.19 & \cite{ktlff_4} \\
$\langle$3.421$\rangle$ & 0.20$\pm$0.10 & 0.68$\pm$0.34 & \cite{ktlff_4} \\
$\langle$3.570$\rangle$ & 0.50$\pm$0.07 & 1.8$\pm$0.2 & \cite{ktlff_2} \\
$\langle$3.589$\rangle$ & 0.22$\pm$0.09 & 0.80$\pm$0.32 & \cite{ktlff_4} \\
3.6 & 0.47$^{+0.09}_{-0.10}$ & 1.7$^{+0.3}_{-0.4}$ & \cite{ktlff_VDM} \\
$\langle$3.742$\rangle$ & 0.24$\pm$0.10 & 0.92$\pm$0.38 & \cite{ktlff_4} \\
3.76 & 0.33$^{+0.14}_{-0.33}$ & 1.2$^{+0.5}_{-1.2}$ & \cite{ktlff_VDM} \\
$\langle$3.840$\rangle$ & 0.22$\pm$0.18 & 0.86$\pm$0.69 & \cite{ktlff_4} \\
$\langle$3.847$\rangle$ & 0.30$\pm$0.12 & 1.2$\pm$0.4 & \cite{ktlff_2} \\
$\langle$3.958$\rangle$ & 0.20$\pm$0.10 & 0.79$\pm$0.40 & \cite{ktlff_4} \\
$\langle$4.099$\rangle$ & 0.24$\pm$0.08 & 1.00$\pm$0.33 & \cite{ktlff_4} \\
$\langle$4.121$\rangle$ & 0.33$\pm$0.14 & 1.4$\pm$0.6 & \cite{ktlff_2} \\
$\langle$4.239$\rangle$ & 0.32$\pm$0.08 & 1.34$\pm$0.34 & \cite{ktlff_4} \\
$\langle$4.345$\rangle$ & 0.22$\pm$0.07 & 0.97$\pm$0.29 & \cite{ktlff_4} \\
4.41 & 0.23$^{+0.06}_{-0.09}$ & 1.0$^{+0.3}_{-0.4}$ & \cite{ktlff_VDM} \\
$\langle$4.471$\rangle$ & $<$0.39 & $<$1.73 & \cite{ktlff_4} \\
$\langle$4.599$\rangle$ & $<$0.32 & $<$1.45 & \cite{ktlff_4} \\
$\langle$4.750$\rangle$ & $<$0.37 & $<$1.78 & \cite{ktlff_4} \\
$\langle$4.948$\rangle$ & $<$0.37 & $<$1.85 & \cite{ktlff_4} \\
$\langle$5.758$\rangle$ & $<$0.36 & $<$2.08 & \cite{ktlff_4} \\
5.76 & $<$0.14 & $<$0.81 & \cite{ktlff_VDM} \\
9.0 & 0.10$^{+0.08}_{-0.20}$ & 0.9$^{+0.7}_{-1.8}$ & \cite{ktlff_VDM} \\
\hline
\end{tabular}
\end{center}
\end{table}

\begin{table}
\caption[Kaon electromagnetic form factor in the spacelike region from 
$K$-$e$ scattering experiments.]
{Kaon electromagnetic form factor in the spacelike region from 
$K$-$e$ scattering experiments.}  
\begin{center}
\begin{tabular}{|c|l|l|c|}
\hline
$Q^{2}$ (GeV$^{2}$) & $F_{K}$ & $Q^{2}~F_{K}$ (GeV$^{2}$) & Ref. \\
\hline
0.0175 & 0.982$\pm$0.012 & 0.0172$\pm$0.0002 & \cite{kslff_sc2} \\
0.0225 & 0.949$\pm$0.013 & 0.0213$\pm$0.0003 & \cite{kslff_sc2} \\
0.0275 & 0.986$\pm$0.016 & 0.0271$\pm$0.0004 & \cite{kslff_sc2} \\
0.0325 & 0.981$\pm$0.019 & 0.0319$\pm$0.0006 & \cite{kslff_sc2} \\
0.0375 & 0.958$\pm$0.022 & 0.0359$\pm$0.0008 & \cite{kslff_sc2} \\
0.0409 & 0.964$\pm$0.016 & 0.0394$\pm$0.0006 & \cite{kslff_sc1} \\
0.0425 & 0.943$\pm$0.026 & 0.0401$\pm$0.0011 & \cite{kslff_sc2} \\
0.0475 & 0.936$\pm$0.029 & 0.0445$\pm$0.0014 & \cite{kslff_sc2} \\
0.0491 & 0.949$\pm$0.021 & 0.0466$\pm$0.0010 & \cite{kslff_sc1} \\
0.0525 & 0.90$\pm$0.03 & 0.0473$\pm$0.0017 & \cite{kslff_sc2} \\
0.0572 & 0.943$\pm$0.026 & 0.0540$\pm$0.0015 & \cite{kslff_sc1} \\
0.0575 & 0.94$\pm$0.04 & 0.0539$\pm$0.0022 & \cite{kslff_sc2} \\
0.0625 & 0.97$\pm$0.04 & 0.0605$\pm$0.0027 & \cite{kslff_sc2} \\
0.0654 & 0.922$\pm$0.033 & 0.0603$\pm$0.0021 & \cite{kslff_sc1} \\
0.0675 & 0.86$\pm$0.05 & 0.059$\pm$0.003 & \cite{kslff_sc2} \\
0.0725 & 0.86$\pm$0.06 & 0.062$\pm$0.004 & \cite{kslff_sc2} \\
0.0736 & 0.88$\pm$0.04 & 0.0654$\pm$0.0029 & \cite{kslff_sc1} \\
0.0775 & 0.85$\pm$0.07 & 0.066$\pm$0.005 & \cite{kslff_sc2} \\
0.0818 & 0.95$\pm$0.05 & 0.078$\pm$0.004 & \cite{kslff_sc1} \\
0.0850 & 0.85$\pm$0.06 & 0.073$\pm$0.005 & \cite{kslff_sc2} \\
0.0899 & 0.82$\pm$0.06 & 0.074$\pm$0.005 & \cite{kslff_sc1} \\
0.0950 & 0.89$\pm$0.08 & 0.085$\pm$0.007 & \cite{kslff_sc2} \\
0.0981 & 0.88$\pm$0.07 & 0.086$\pm$0.007 & \cite{kslff_sc1} \\
0.1063 & 0.87$\pm$0.09 & 0.093$\pm$0.010 & \cite{kslff_sc1} \\
0.1145 & 0.95$\pm$0.12 & 0.109$\pm$0.014 & \cite{kslff_sc1} \\
\hline
\end{tabular}
\end{center}
\end{table}

\begin{table}
\caption[Proton magnetic form factor in the timelike region (Part 1).]
{Proton magnetic form factor in the timelike region (Part 1).  
$|G^{P}_{E}(Q^2)|$ = $|G^{P}_{M}(Q^2)|$ is assumed.}
\begin{center}
\begin{tabular}{|c|l|l|c|}
\hline
$|Q^{2}|$ (GeV$^{2}$) & $|G^{P}_{M}|$ 
& $|Q^{4}||G^{P}_{M}|/\mu_{p}$ (GeV$^{4}$) & 
Ref. \\
\hline
3.52 & 0.51$\pm$0.08 & 2.3$\pm$0.4 & \cite{prtlff_3} \\
3.523 & 0.53$^{+0.06}_{-0.04}$ & 2.36$^{+0.27}_{-0.18}$ & \cite{prtlff_6} \\
3.553 & 0.39$\pm$0.05 & 1.8$\pm$0.2 & \cite{prtlff_6} \\
3.572 & 0.34$\pm$0.04 & 1.55$\pm$0.18 & \cite{prtlff_6} \\
3.599 & 0.31$\pm$0.03 & 1.44$\pm$0.14 & \cite{prtlff_6} \\
3.61 & 0.42$^{+0.14}_{-0.08}$ & 2.0$^{+0.7}_{-0.4}$ & \cite{prtlff_3}\\
3.69 & 0.36$\pm$0.05 & 1.8$\pm$0.2 & \cite{prtlff_8} \\
$\langle$3.76$\rangle$ & 0.262$\pm$0.014 & 1.33$\pm$0.07 & \cite{prtlff_7} \\
$\langle$3.76$\rangle$ & 0.39$\pm$0.06 & 2.0$\pm$0.3 & \cite{prtlff_2} \\
$\langle$3.83$\rangle$ & 0.253$\pm$0.010 & 1.33$\pm$0.05 & \cite{prtlff_7} \\
$\langle$3.90$\rangle$ & 0.25$\pm$0.08 & 1.4$\pm$0.4 & \cite{prtlff_2} \\
$\langle$3.94$\rangle$ & 0.247$\pm$0.014 & 1.37$\pm$0.08 & \cite{prtlff_7} \\
4.00 & 0.24$\pm$0.03 & 1.38$\pm$0.17 & \cite{prtlff_8} \\
4.00 & 0.175$^{+0.067}_{-0.055}$ & 1.00$^{+0.38}_{-0.32}$ & \cite{prtlff_9} \\
$\langle$4.0$\rangle$ & 0.26$\pm$0.03 & 1.49$\pm$0.17 & \cite{prtlff_4} \\
$\langle$4.12$\rangle$ & 0.26$\pm$0.03 & 1.6$\pm$0.2 & \cite{prtlff_2} \\
$\langle$4.18$\rangle$ & 0.252$\pm$0.011 & 1.58$\pm$0.07 & \cite{prtlff_7} \\
$\langle$4.2$\rangle$ & 0.22$\pm$0.02 & 1.39$\pm$0.13 & \cite{prtlff_4} \\
$\langle$4.4$\rangle$ & 0.19$\pm$0.02 & 1.32$\pm$0.14 & \cite{prtlff_4} \\
4.41 & 0.22$\pm$0.02 & 1.53$\pm$0.14 & \cite{prtlff_8} \\
$\langle$4.60$\rangle$ & 0.21$\pm$0.04 & 1.6$\pm$0.3 & \cite{prtlff_2} \\
$\langle$4.6$\rangle$ & 0.17$\pm$0.02 & 1.29$\pm$0.15 & \cite{prtlff_4} \\
$\langle$4.8$\rangle$ & 0.19$\pm$0.02 & 1.57$\pm$0.17 & \cite{prtlff_4} \\
4.84 & 0.179$\pm$0.018 & 1.50$\pm$0.15 & \cite{prtlff_9} \\
$\langle$5.0$\rangle$ & 0.14$\pm$0.04 & 1.3$\pm$0.4 & \cite{prtlff_4} \\
5.693 & 0.083$^{+0.017}_{-0.013}$ & 0.96$^{+0.20}_{-0.15}$ & \cite{prtlff_5} \\
5.76 & 0.072$^{+0.041}_{-0.023}$ & 0.86$^{+0.49}_{-0.27}$ & \cite{prtlff_9} \\
5.95 & 0.15$\pm$0.03 & 1.9$\pm$0.4 & \cite{prtlff_8} \\
\hline
\end{tabular}
\end{center}
\end{table}

\begin{table}
\caption[Proton magnetic form factor in the timelike region (Part 2).]
{Proton magnetic form factor in the timelike region (Part 2).  
$|G^{P}_{E}(Q^2)|$ = $|G^{P}_{M}(Q^2)|$ is assumed.}
\begin{center}
\begin{tabular}{|c|l|l|c|}
\hline
$|Q^{2}|$ (GeV$^{2}$) & $|G^{P}_{M}|$ 
& $|Q^{4}||G^{P}_{M}|/\mu_{p}$ (GeV$^{4}$) & 
Ref. \\
\hline
6.25 & 0.131$^{+0.037}_{-0.029}$ & 1.83$^{+0.52}_{-0.41}$ & \cite{prtlff_9} \\
6.76 & 0.054$\pm$0.006 & 0.88$\pm$0.10 & \cite{prtlff_9} \\
7.29 & 0.070$^{+0.039}_{-0.022}$ & 1.33$^{+0.56}_{-0.42}$ & \cite{prtlff_9} \\
7.84 & 0.063$^{+0.036}_{-0.020}$ & 1.39$^{+0.79}_{-0.44}$ & \cite{prtlff_9} \\
8.41 & $<$0.073 & $<$1.85 & \cite{prtlff_9} \\
8.84 & 0.0359$\pm$0.0030 & 1.01$\pm$0.08 & \cite{e835_2} \\
8.9 & 0.033$^{+0.006}_{-0.004}$ & 0.94$^{+0.17}_{-0.11}$ & \cite{e835_1} \\
9.00 & 0.028$^{+0.010}_{-0.006}$ & 0.81$^{+0.29}_{-0.17}$ & \cite{prtlff_9} \\
9.42 & 0.027$^{+0.006}_{-0.004}$ & 0.86$^{+0.19}_{-0.13}$ & \cite{prtlff_9} \\
10.78 & 0.021$^{+0.007}_{-0.008}$ & 0.87$^{+0.29}_{-0.33}$ & \cite{e835_2} \\
11.63 & 0.0174$^{+0.0021}_{-0.0017}$ & 0.84$^{+0.10}_{-0.08}$ & \cite{e835_3} \\
12.4 & 0.013$^{+0.003}_{-0.002}$ & 0.72$^{+0.17}_{-0.11}$ & \cite{e835_1} \\
12.43 & 0.0143$\pm$0.0015 & 0.79$\pm$0.08 & \cite{e835_2} \\
12.43 & 0.0148$^{+0.0017}_{-0.0014}$ & 0.82$^{+0.09}_{-0.08}$ & \cite{e835_3} \\
13.0 & 0.013$^{+0.005}_{-0.003}$ & 0.79$^{+0.30}_{-0.18}$ & \cite{e835_1} \\
13.11 & 0.0112$\pm$0.0017 & 0.69$\pm$0.10 & \cite{e835_2} \\
\hline
\end{tabular}
\end{center}
\end{table}

\begin{table}[htbp]
\caption[Proton magnetic form factor in the timelike region from the Babar experiment (Part 1).]
{Proton magnetic form factor in the timelike region from the Babar experiment (Part 1).  
The measurements are from $\ee$ annihilations after initial state radiation 
\cite{BABARppbarISR}.  $|G^{P}_{E}(Q^2)|$ = $|G^{P}_{M}(Q^2)|$ is assumed.}
\begin{center}
\begin{tabular}{|c|l|l|}
\hline
$|Q^{2}|$ (GeV$^{2}$) & $|F_p|$ = $|G^{P}_{M}|$ 
& $|Q^{4}||G^{P}_{M}|/\mu_{p}$ (GeV$^{4}$) \\
\hline
3.57$\pm$0.05 & 0.453$^{+0.023}_{-0.025}$ &  2.07$^{+0.11}_{-0.13}$ \\
3.66$\pm$0.05 & 0.354$\pm$0.017 & 1.70$\pm$0.09 \\
3.76$\pm$0.05 & 0.305$\pm$0.015 & 1.55$\pm$0.09 \\
3.85$\pm$0.05 & 0.276$\pm$0.015 & 1.47$\pm$0.09 \\
3.95$\pm$0.05 & 0.266$\pm$0.015 & 1.49$\pm$0.09 \\
4.05$\pm$0.05 & 0.273$\pm$0.014 & 1.60$\pm$0.09 \\
4.15$\pm$0.05 & 0.250$\pm$0.014 & 1.54$\pm$0.09 \\
4.26$\pm$0.05 & 0.254$\pm$0.014 & 1.65$\pm$0.10 \\
4.36$\pm$0.05 & 0.239$\pm$0.013 & 1.63$\pm$0.10 \\
4.46$\pm$0.06 & 0.250$\pm$0.013 & 1.78$\pm$0.10 \\
4.57$\pm$0.06 & 0.237$\pm$0.013 & 1.77$\pm$0.11 \\
4.68$\pm$0.06 & 0.207$\pm$0.013 & 1.63$\pm$0.11 \\
4.79$\pm$0.06 & 0.191$\pm$0.013 & 1.57$\pm$0.11 \\
4.90$\pm$0.06 & 0.183$\pm$0.013 & 1.57$\pm$0.11 \\
5.01$\pm$0.06 & 0.174$\pm$0.012 & 1.57$\pm$0.11 \\
5.12$\pm$0.06 & 0.137$\pm$0.013 & 1.29$\pm$0.13 \\
5.23$\pm$0.06 & 0.137$\pm$0.013 & 1.34$\pm$0.13 \\
\hline
\end{tabular}
\end{center}
\end{table}

\begin{table}[htbp]
\caption[Proton magnetic form factor in the timelike region from the Babar experiment (Part 2).]
{Proton magnetic form factor in the timelike region from the Babar experiment (Part 2).  
The measurements are from $\ee$ annihilations after initial state radiation 
\cite{BABARppbarISR}.  $|G^{P}_{E}(Q^2)|$ = $|G^{P}_{M}(Q^2)|$ is assumed.}
\begin{center}
\begin{tabular}{|c|l|l|}
\hline
$|Q^{2}|$ (GeV$^{2}$) & $|F_p|$ = $|G^{P}_{M}|$ 
& $|Q^{4}||G^{P}_{M}|/\mu_{p}$ (GeV$^{4}$) \\
\hline
5.41$\pm$0.12 & 0.105$\pm$0.010 & 1.10$\pm$0.12 \\
5.64$\pm$0.12 & 0.103$\pm$0.009 & 1.17$\pm$0.11 \\
5.88$\pm$0.12 & 0.110$\pm$0.008 & 1.36$\pm$0.11 \\
6.13$\pm$0.12 & 0.083$\pm$0.008 & 1.12$\pm$0.12 \\
6.38$\pm$0.13 & 0.092$\pm$0.008 & 1.34$\pm$0.13 \\
6.63$\pm$0.13 & 0.072$\pm$0.008 & 1.13$\pm$0.13 \\
6.89$\pm$0.13 & 0.065$\pm$0.009 & 1.11$\pm$0.16 \\
7.16$\pm$0.13 & 0.059$\pm$0.009 & 1.08$\pm$0.17 \\
7.43$\pm$0.14 & 0.054$^{+0.008}_{-0.010}$ & 1.07$^{+0.16}_{-0.20}$ \\
7.70$\pm$0.14 & 0.060$^{+0.008}_{-0.010}$ & 1.28$^{+0.18}_{-0.22}$ \\
7.98$\pm$0.14 & 0.054$\pm$0.009 & 1.23$\pm$0.21 \\
8.27$\pm$0.14 & 0.052$^{+0.008}_{-0.010}$ & 1.27$^{+0.20}_{-0.25}$ \\
8.56$\pm$0.15 & 0.052$\pm$0.009 & 1.37$\pm$0.24 \\
8.85$\pm$0.15 & 0.035$^{+0.010}_{-0.014}$ & 0.98$^{+0.28}_{-0.39}$ \\
9.6$\pm$0.6  & 0.021$^{+0.009}_{-0.021}$ &  0.7$^{+0.3}_{-0.7}$ \\
10.9$\pm$0.7 & 0.017$^{+0.008}_{-0.017}$ &  0.7$^{+0.4}_{-0.7}$ \\
12.3$\pm$0.7 & 0.016$^{+0.005}_{-0.009}$ &  0.9$^{+0.3}_{-0.5}$ \\
13.7$\pm$0.7 & 0.019$^{+0.005}_{-0.008}$ &  1.3$^{+0.3}_{-0.5}$ \\
15.2$\pm$0.8 & 0.015$^{+0.005}_{-0.009}$ &  1.2$^{+0.4}_{-0.7}$ \\
17.0$\pm$1.0 & 0.011$^{+0.005}_{-0.010}$ &  1.1$^{+0.5}_{-1.0}$ \\
19.1$\pm$1.1 & 0.005$^{+0.008}_{-0.005}$ &  0.7$^{+1.0}_{-0.7}$ \\
\hline
\end{tabular}
\end{center}
\end{table}


\begin{table}
\caption[Proton magnetic form factor in the spacelike region (Part 1).]
{Proton magnetic form factor in the spacelike region (Part 1).  
Results determined by Rosenbluth separation, 
and $G^P_E(Q^2)$ = $G^P_M(Q^2)/\mu_p$ is assumed.}  
\begin{center}
\begin{tabular}{|c|l|l|c|}
\hline
$Q^{2}$ (GeV) & $G^P_M/\mu_p$ & $Q^4G^P_M/\mu_p$ (GeV$^4$) & Ref. \\
\hline
0.39 & 0.410$\pm$0.008 & 0.0624$\pm$0.0012 & \cite{prslff_gmonly1} \\
0.78 & 0.226$\pm$0.005 & 0.137$\pm$0.003 & \cite{prslff_gmonly1} \\
0.999 & 0.1733$\pm$0.0016 & 0.1730$\pm$0.0016 & \cite{prslff_gmonly4} \\
1.16 & 0.145$\pm$0.003 & 0.195$\pm$0.004 & \cite{prslff_gmonly1} \\
1.498 & 0.1071$\pm$0.0010 & 0.2403$\pm$0.0022 & \cite{prslff_gmonly4} \\
1.55 & 0.100$\pm$0.002 & 0.240$\pm$0.005 & \cite{prslff_gmonly1} \\
1.75 & 0.0861$\pm$0.0017 & 0.264$\pm$0.005 & \cite{prslff_gmonly1} \\
1.94 & 0.0739$\pm$0.0015 & 0.278$\pm$0.006 & \cite{prslff_gmonly1} \\
1.999 & 0.0718$\pm$0.0007 & 0.2869$\pm$0.0028 & \cite{prslff_gmonly4} \\
2.495 & 0.04932$\pm$0.00016 & 0.307$\pm$0.001 & \cite{prslff_gmonly5} \\
2.502 & 0.0505$\pm$0.0004 & 0.3161$\pm$0.0025 & \cite{prslff_gmonly4} \\
2.862 & 0.0404$\pm$0.0007 & 0.331$\pm$0.006 & \cite{prslff_gmonly6} \\
2.91 & 0.0405$\pm$0.0010 & 0.343$\pm$0.008 & \cite{prslff_gmonly1} \\
3.621 & 0.0275$\pm$0.0006 & 0.361$\pm$0.008 & \cite{prslff_gmonly6} \\
3.759 & 0.02606$\pm$0.00026 & 0.368$\pm$0.004 & \cite{prslff_gmonly4} \\
3.990 & 0.02242$\pm$0.00013 & 0.357$\pm$0.002 & \cite{prslff_gmonly5} \\
4.08 & 0.0222$\pm$0.0007 & 0.370$\pm$0.012 & \cite{prslff_gmonly1} \\
4.08 & 0.0211$\pm$0.0011 & 0.351$\pm$0.018 & \cite{prslff_gmonly2} \\
4.16 & 0.0218$\pm$0.0011 & 0.377$\pm$0.019 & \cite{prslff_gmonly2} \\
4.20 & 0.0210$\pm$0.0011 & 0.370$\pm$0.019 & \cite{prslff_gmonly2} \\
4.88 & 0.0155$\pm$0.0008 & 0.369$\pm$0.019 & \cite{prslff_gmonly2} \\
4.991 & 0.0157$\pm$0.0003 & 0.391$\pm$0.008 & \cite{prslff_gmonly6} \\
5.017 & 0.0154$\pm$0.0003 & 0.387$\pm$0.008 & \cite{prslff_gmonly6} \\
5.027 & 0.01543$\pm$0.00028 & 0.390$\pm$0.007 & \cite{prslff_gmonly6} \\
5.075 & 0.01513$\pm$0.00016 & 0.390$\pm$0.004 & \cite{prslff_gmonly4} \\
5.89 & 0.0115$\pm$0.0006 & 0.399$\pm$0.021 & \cite{prslff_gmonly2} \\
5.996 & 0.01068$\pm$0.00008 & 0.384$\pm$0.003 & \cite{prslff_gmonly5} \\
\hline
\end{tabular}
\end{center}
\end{table}

\begin{table}[htp]
\caption[Proton magnetic form factor in the spacelike region (Part 2).]
{Proton magnetic form factor in the spacelike region (Part 2).  
Results determined by Rosenbluth separation, 
and $G^P_E(Q^2)$ = $G^P_M(Q^2)/\mu_p$ is assumed.}  
\begin{center}
\begin{tabular}{|c|l|l|c|}
\hline
$Q^{2}$ (GeV) & $G^P_M/\mu_p$ & $Q^4G^P_M/\mu_p$ (GeV$^4$) & Ref. \\
\hline
6.270 & 0.00999$\pm$0.00023 & 0.393$\pm$0.009 & \cite{prslff_gmonly4} \\
6.85 & 0.0081$\pm$0.0004 & 0.380$\pm$0.019 & \cite{prslff_gmonly2} \\
7.300 & 0.00745$\pm$0.00015 & 0.397$\pm$0.008 & \cite{prslff_gmonly6} \\
7.498 & 0.00709$\pm$0.00011 & 0.399$\pm$0.006 & \cite{prslff_gmonly4} \\
7.85 & 0.0066$\pm$0.0003 & 0.407$\pm$0.018 & \cite{prslff_gmonly2} \\
7.988 & 0.00610$\pm$0.00006 & 0.389$\pm$0.004 & \cite{prslff_gmonly5} \\
8.752 & 0.00496$\pm$0.00015 & 0.380$\pm$0.011 & \cite{prslff_gmonly4} \\
8.78 & 0.0054$\pm$0.0008 & 0.42$\pm$0.06 & \cite{prslff_gmonly2} \\
9.53 & 0.0043$\pm$0.0003 & 0.391$\pm$0.027 & \cite{prslff_gmonly3} \\
9.59 & 0.0049$\pm$0.0010 & 0.45$\pm$0.09 & \cite{prslff_gmonly2} \\
9.629 & 0.00421$\pm$0.00010 & 0.390$\pm$0.009 & \cite{prslff_gmonly6} \\
9.982 & 0.00404$\pm$0.00008 & 0.403$\pm$0.008 & \cite{prslff_gmonly4} \\
10.004 & 0.00390$\pm$0.00007 & 0.390$\pm$0.007 & \cite{prslff_gmonly5} \\
11.99 & 0.00273$\pm$0.00006 & 0.392$\pm$0.009 & \cite{prslff_gmonly6} \\
12.50 & 0.00245$\pm$0.00009 & 0.383$\pm$0.014 & \cite{prslff_gmonly4} \\
15.10 & 0.00168$\pm$0.00009 & 0.383$\pm$0.021 & \cite{prslff_gmonly4} \\
15.72 & 0.00153$\pm$0.00004 & 0.378$\pm$0.011 & \cite{prslff_gmonly6} \\
19.47 & 0.00090$\pm$0.00003 & 0.343$\pm$0.013 & \cite{prslff_gmonly6} \\
20.00 & 0.00093$\pm$0.00009 & 0.37$\pm$0.04 & \cite{prslff_gmonly4} \\
23.24 & 0.000641$\pm$0.000028 & 0.346$\pm$0.015 & \cite{prslff_gmonly6} \\
25.03 & 0.00078$\pm$0.00017 & 0.49$\pm$0.11 & \cite{prslff_gmonly4} \\
26.99 & 0.000465$\pm$0.000027 & 0.339$\pm$0.020 & \cite{prslff_gmonly6} \\
31.20 & 0.00036$\pm$0.00003 & 0.35$\pm$0.03 & \cite{prslff_gmonly6} \\
\hline
\end{tabular}
\end{center}
\end{table}

\begin{table}[htbp]
\caption[Proton electromagnetic form factor ratios in the timelike region from the Babar 
experiment.]
{Proton electromagnetic form factor ratios in the timelike region from the Babar 
experiment.  The measurements are from $\ee$ annihilations after initial state radiation 
\cite{BABARppbarISR}.}
\begin{center}
\begin{tabular}{|c|l|l|}
\hline
$|Q^{2}|$ (GeV$^{2}$) & $|G^{P}_{E}|/|G^{P}_{M}|$ & $\mu_p|G^{P}_{E}|/|G^{P}_{M}|$ \\
\hline
3.663$\pm$0.140 & 1.41$^{+0.29}_{-0.25}$ &  3.93$^{+0.81}_{-0.70}$ \\
3.952$\pm$0.149 & 1.78$^{+0.36}_{-0.29}$ &  4.97$^{+1.00}_{-0.81}$ \\
4.256$\pm$0.154 & 1.52$^{+0.31}_{-0.26}$ &  4.24$^{+0.86}_{-0.73}$ \\
4.625$\pm$0.215 & 1.18$^{+0.23}_{-0.22}$ &  3.29$^{+0.64}_{-0.61}$ \\
5.30$\pm$0.46   & 1.32$^{+0.31}_{-0.27}$ &  3.68$^{+0.86}_{-0.75}$ \\
7.38$\pm$1.62   & 1.22$^{+0.34}_{-0.34}$ &  3.40$^{+0.95}_{-0.95}$ \\
\hline
\end{tabular}
\end{center}
\end{table}

\newpage
\clearpage

\begin{table}
\caption[Proton electromagnetic form factor ratios in the spacelike region from 
polarization transfer experiments.]
{Proton electromagnetic form factor ratios in the spacelike region from 
polarization transfer experiments.}
\begin{center}
\begin{tabular}{|c|l|l|l|l|c|}
\hline
$Q^{2}$ (GeV) & $\mu_pG^P_E/G^P_M$ & $F_2/F_1$ & $Q~F_2/F_1$ 
& $Q^2~F_2/F_1$ & Ref. \\
\hline
0.32 & 0.93$\pm$0.07 & & & & \cite{prslff_pol2} \\
0.35 & 0.91$\pm$0.06 & & & & \cite{prslff_pol2} \\
0.38 & 0.95$\pm$0.05 & & & & \cite{prslff_pol1} \\
0.38 & 1.00$\pm$0.10 & & & & \cite{prslff_pol1} \\
0.39 & 0.96$\pm$0.03 & & & & \cite{prslff_pol2} \\
0.46 & 0.95$\pm$0.03 & & & & \cite{prslff_pol2} \\
0.49 & 0.979$\pm$0.017 & & & & \cite{prslff_pol4} \\
0.50 & 1.02$\pm$0.05 & & & & \cite{prslff_pol1} \\
0.50 & 1.07$\pm$0.06 & & & & \cite{prslff_pol1} \\
0.57 & 0.96$\pm$0.04 & & & & \cite{prslff_pol2} \\
0.76 & 0.97$\pm$0.04 & & & & \cite{prslff_pol2} \\
0.79 & 0.951$\pm$0.016 & & & & \cite{prslff_pol4} \\
0.86 & 0.87$\pm$0.03 & 0.691$\pm$0.017 & 0.641$\pm$0.016 & 0.595$\pm$0.015 & \cite{prslff_pol2} \\
0.88 & 0.92$\pm$0.09 & 0.646$\pm$0.048 & 0.606$\pm$0.045 & 0.568$\pm$0.042 & \cite{prslff_pol2} \\
1.02 & 0.90$\pm$0.04 & 0.618$\pm$0.020 & 0.624$\pm$0.020 & 0.630$\pm$0.020 & \cite{prslff_pol2} \\
1.12 & 0.83$\pm$0.03 & 0.638$\pm$0.015 & 0.675$\pm$0.016 & 0.714$\pm$0.017 & \cite{prslff_pol2} \\
1.18 & 0.85$\pm$0.06 & 0.607$\pm$0.028 & 0.660$\pm$0.030 & 0.716$\pm$0.033 & \cite{prslff_pol2} \\
1.18 & 0.883$\pm$0.022 & 0.586$\pm$0.010 & 0.637$\pm$0.011 & 0.692$\pm$0.012 & \cite{prslff_pol4} \\
1.42 & 0.73$\pm$0.06 & 0.620$\pm$0.027 & 0.739$\pm$0.032 & 0.881$\pm$0.038 & \cite{prslff_pol2} \\
1.48 & 0.80$\pm$0.04 & 0.564$\pm$0.016 & 0.686$\pm$0.020 & 0.834$\pm$0.024 & \cite{prslff_pol4} \\
1.76 & 0.82$\pm$0.13 & 0.497$\pm$0.044 & 0.659$\pm$0.058 & 0.875$\pm$0.077 & \cite{prslff_pol2} \\
1.77 & 0.79$\pm$0.04 & 0.510$\pm$0.014 & 0.678$\pm$0.018 & 0.902$\pm$0.024 & \cite{prslff_pol4} \\
1.88 & 0.78$\pm$0.04 & 0.495$\pm$0.013 & 0.678$\pm$0.018 & 0.930$\pm$0.025 & \cite{prslff_pol4} \\
2.13 & 0.75$\pm$0.05 & 0.468$\pm$0.015 & 0.682$\pm$0.022 & 0.996$\pm$0.032 & \cite{prslff_pol4} \\
2.47 & 0.70$\pm$0.04 & 0.439$\pm$0.010 & 0.691$\pm$0.017 & 1.085$\pm$0.026 & \cite{prslff_pol4} \\
2.97 & 0.62$\pm$0.04 & 0.408$\pm$0.009 & 0.703$\pm$0.016 & 1.211$\pm$0.028 & \cite{prslff_pol4} \\
3.47 & 0.61$\pm$0.04 & 0.363$\pm$0.008 & 0.675$\pm$0.015 & 1.258$\pm$0.028 & \cite{prslff_pol4} \\
3.50 & 0.57$\pm$0.07 & 0.371$\pm$0.014 & 0.694$\pm$0.026 & 1.298$\pm$0.049 & \cite{prslff_pol3} \\
3.97 & 0.48$\pm$0.05 & 0.356$\pm$0.009 & 0.709$\pm$0.018 & 1.413$\pm$0.036 & \cite{prslff_pol3} \\
4.75 & 0.38$\pm$0.05 & 0.325$\pm$0.008 & 0.708$\pm$0.017 & 1.543$\pm$0.037 & \cite{prslff_pol3} \\
5.54 & 0.27$\pm$0.09 & 0.302$\pm$0.012 & 0.711$\pm$0.029 & 1.674$\pm$0.068 & \cite{prslff_pol3} \\
\hline
\end{tabular}
\end{center}
\end{table}

\begin{table}
\caption[Proton electromagnetic form factor ratios in the spacelike region from 
Rosenbluth separation experiments (Part 1).]
{Proton electromagnetic form factor ratios in the spacelike region from 
Rosenbluth separation experiments (Part 1).}
\begin{center}
\begin{tabular}{|c|l|c|}
\hline
$Q^{2}$ (GeV) & $\mu_pG^P_E/G^P_M$ & Ref. \\
\hline
0.038 & 1.02$\pm$0.03 & \cite{prslff_ros1} \\
0.049 & 1.00$\pm$0.02 & \cite{prslff_ros1} \\
0.061 & 0.98$\pm$0.02 & \cite{prslff_ros1} \\
0.068 & 1.01$\pm$0.04 & \cite{prslff_ros1} \\
0.130 & 0.997$\pm$0.020 & \cite{prslff_ros8} \\
0.16 & 1.11$\pm$0.04 & \cite{prslff_ros2} \\
0.18 & 1.01$\pm$0.03 & \cite{prslff_ros2} \\
0.19 & 0.97$\pm$0.05 & \cite{prslff_ros2} \\
0.190 & 0.999$\pm$0.015 & \cite{prslff_ros8} \\
0.23 & 1.08$\pm$0.05 & \cite{prslff_ros2} \\
0.27 & 1.06$\pm$0.05 & \cite{prslff_ros2} \\
0.270 & 0.992$\pm$0.023 & \cite{prslff_ros8} \\
0.29 & 1.07$\pm$0.05 & \cite{prslff_ros2} \\
0.31 & 0.97$\pm$0.05 & \cite{prslff_ros2} \\
0.330 & 0.980$\pm$0.016 & \cite{prslff_ros8} \\
0.35 & 0.97$\pm$0.07 & \cite{prslff_ros2} \\
0.389 & 0.95$\pm$0.04 & \cite{prslff_ros7} \\
0.39 & 1.06$\pm$0.05 & \cite{prslff_ros2} \\
0.39 & 1.03$\pm$0.04 & \cite{prslff_ros3} \\
0.390 & 1.020$\pm$0.021 & \cite{prslff_ros8} \\
0.43 & 1.05$\pm$0.07 & \cite{prslff_ros2} \\
0.450 & 1.02$\pm$0.03 & \cite{prslff_ros8} \\
0.47 & 1.02$\pm$0.06 & \cite{prslff_ros2} \\
\hline
\end{tabular}
\end{center}
\end{table}

\begin{table}[htp]
\caption[Proton electromagnetic form factor ratios in the spacelike region from 
Rosenbluth separation experiments (Part 2).]
{Proton electromagnetic form factor ratios in the spacelike region from 
Rosenbluth separation experiments (Part 2).}
\begin{center}
\begin{tabular}{|c|l|l|l|l|c|}
\hline
$Q^{2}$ (GeV) & $\mu_pG^p_E/G^p_M$ & $F_2/F_1$ & $Q~F_2/F_1$ & $Q^2~F_2/F_1$ & Ref. \\
\hline
0.50 & 1.07$\pm$0.13 & & & & \cite{prslff_ros2} \\
0.530 & 1.01$\pm$0.05 & & & & \cite{prslff_ros8} \\
0.54 & 1.00$\pm$0.09 & & & & \cite{prslff_ros2} \\
0.58 & 1.02$\pm$0.19 & & & & \cite{prslff_ros2} \\
0.580 & 0.969$\pm$0.020 & & & & \cite{prslff_ros8} \\
0.584 & 0.98$\pm$0.03 & & & & \cite{prslff_ros7} \\
0.62 & 0.96$\pm$0.09 & & & & \cite{prslff_ros2} \\
0.650 & 0.96$\pm$0.05 & & & & \cite{prslff_ros8} \\
0.65 & 1.07$\pm$0.09 & & & & \cite{prslff_ros12} \\
0.66 & 0.84$\pm$0.15 & & & & \cite{prslff_ros2} \\
0.70 & 1.10$\pm$0.11 & & & & \cite{prslff_ros2} \\
0.720 & 1.09$\pm$0.08 & & & & \cite{prslff_ros8} \\
0.74 & 1.04$\pm$0.16 & & & & \cite{prslff_ros2} \\
0.779 & 0.95$\pm$0.05 & & & & \cite{prslff_ros7} \\
0.78 & 0.8$\pm$0.3 & & & & \cite{prslff_ros2} \\
0.78 & 0.90$\pm$0.05 & & & & \cite{prslff_ros3} \\
0.78 & 0.93$\pm$0.18 & & & & \cite{prslff_ros4} \\
0.780 & 0.94$\pm$0.04 & & & & \cite{prslff_ros8} \\
0.85 & 0.7$\pm$0.3 & 0.85$\pm$0.22 & 0.78$\pm$0.21 & 0.72$\pm$0.19 & \cite{prslff_ros2} \\
0.91 & 0.93$\pm$0.07 & 0.629$\pm$0.036 & 0.600$\pm$0.034 & 0.573$\pm$0.032 & \cite{prslff_ros12} \\
0.940 & 1.04$\pm$0.05 & 0.548$\pm$0.022 & 0.531$\pm$0.021 & 0.515$\pm$0.021 & \cite{prslff_ros8} \\
0.973 & 1.03$\pm$0.07 & 0.546$\pm$0.030 & 0.539$\pm$0.030 & 0.531$\pm$0.030 & \cite{prslff_ros7} \\
0.99 & 0.97$\pm$0.05 & 0.580$\pm$0.023 & 0.577$\pm$0.023 & 0.574$\pm$0.023 & \cite{prslff_ros6} \\
\hline
\end{tabular}
\end{center}
\end{table}

\begin{table}[htp]
\caption[Proton electromagnetic form factor ratios in the spacelike region from 
Rosenbluth separation experiments (Part 3).]
{Proton electromagnetic form factor ratios in the spacelike region from 
Rosenbluth separation experiments (Part 3).}
\begin{center}
\begin{tabular}{|c|l|l|l|l|c|}
\hline
$Q^{2}$ (GeV) & $\mu_pG^p_E/G^p_M$  & $F_2/F_1$ & $Q~F_2/F_1$ & $Q^2~F_2/F_1$ & Ref. \\
\hline
1.00 & 0.99$\pm$0.05 & 0.564$\pm$0.022 & 0.564$\pm$0.022 & 0.564$\pm$0.022 & \cite{prslff_ros9} \\
1.000 & 0.98$\pm$0.09 & 0.571$\pm$0.041 & 0.571$\pm$0.041 & 0.571$\pm$0.041 & \cite{prslff_ros10} \\
1.100 & 0.87$\pm$0.04 & 0.616$\pm$0.019 & 0.646$\pm$0.020 & 0.678$\pm$0.021 & \cite{prslff_ros8} \\
1.16 & 1.01$\pm$0.13 & 0.515$\pm$0.051 & 0.555$\pm$0.055 & 0.598$\pm$0.059 & \cite{prslff_ros3} \\
1.16 & 0.86$\pm$0.07 & 0.606$\pm$0.032 & 0.653$\pm$0.035 & 0.703$\pm$0.038 & \cite{prslff_ros4} \\
1.168 & 1.05$\pm$0.15 & 0.49$\pm$0.06 & 0.53$\pm$0.06 & 0.57$\pm$0.07 & \cite{prslff_ros7} \\
1.17 & 0.97$\pm$0.06 & 0.536$\pm$0.024 & 0.580$\pm$0.026 & 0.627$\pm$0.029 & \cite{prslff_ros9} \\
1.350 & 0.90$\pm$0.05 & 0.536$\pm$0.020 & 0.623$\pm$0.023 & 0.724$\pm$0.027 & \cite{prslff_ros8} \\
1.363 & 0.88$\pm$0.18 & 0.54$\pm$0.07 & 0.64$\pm$0.08 & 0.74$\pm$0.10 & \cite{prslff_ros7} \\
1.51 & 0.83$\pm$0.21 & 0.54$\pm$0.08 & 0.66$\pm$0.10 & 0.82$\pm$0.12 & \cite{prslff_ros4} \\
1.53 & 0.82$\pm$0.08 & 0.542$\pm$0.031 & 0.670$\pm$0.038 & 0.829$\pm$0.047 & \cite{prslff_ros6} \\
1.55 & 0.87$\pm$0.14 & 0.51$\pm$0.05 & 0.64$\pm$0.06 & 0.79$\pm$0.08 & \cite{prslff_ros3} \\
1.557 & 1.2$\pm$0.3 & 0.37$\pm$0.08 & 0.46$\pm$0.10 & 0.57$\pm$0.13 & \cite{prslff_ros7} \\
1.56 & 0.96$\pm$0.10 & 0.47$\pm$0.03 & 0.58$\pm$0.04 & 0.73$\pm$0.05 & \cite{prslff_ros9} \\
1.75 & 1.15$\pm$0.18 & 0.36$\pm$0.05 & 0.48$\pm$0.06 & 0.63$\pm$0.08 & \cite{prslff_ros3} \\
1.75 & 1.14$\pm$0.17 & 0.36$\pm$0.04 & 0.48$\pm$0.06 & 0.64$\pm$0.08 & \cite{prslff_ros4} \\
1.750 & 0.80$\pm$0.08 & 0.508$\pm$0.027 & 0.673$\pm$0.037 & 0.890$\pm$0.048 & \cite{prslff_ros8} \\
1.75 & 0.75$\pm$0.11 & 0.53$\pm$0.04 & 0.71$\pm$0.05 & 0.93$\pm$0.07 & \cite{prslff_ros9} \\
1.75 & 0.91$\pm$0.06 & 0.457$\pm$0.019 & 0.605$\pm$0.025 & 0.800$\pm$0.033 & \cite{prslff_ros11} \\
1.752 & 0.8$\pm$0.5 & 0.51$\pm$0.17 & 0.67$\pm$0.23 & 0.89$\pm$0.30 & \cite{prslff_ros7} \\
1.94 & 1.0$\pm$0.4 & 0.39$\pm$0.11 & 0.55$\pm$0.15 & 0.76$\pm$0.21 & \cite{prslff_ros3} \\
1.94 & 1.01$\pm$0.25 & 0.39$\pm$0.07 & 0.54$\pm$0.09 & 0.76$\pm$0.13 & \cite{prslff_ros5} \\
1.98 & 1.06$\pm$0.17 & 0.37$\pm$0.04 & 0.52$\pm$0.06 & 0.73$\pm$0.09 & \cite{prslff_ros6} \\
2.00 & 0.88$\pm$0.11 & 0.43$\pm$0.03 & 0.61$\pm$0.04 & 0.87$\pm$0.06 & \cite{prslff_ros9} \\
2.003 & 1.16$\pm$0.09 & 0.331$\pm$0.021 & 0.469$\pm$0.030 & 0.664$\pm$0.043 & \cite{prslff_ros10} \\
2.20 & 0.88$\pm$0.13 & 0.41$\pm$0.03 & 0.60$\pm$0.05 & 0.89$\pm$0.08 & \cite{prslff_ros12} \\
2.33 & 0.71$\pm$0.20 & 0.45$\pm$0.06 & 0.69$\pm$0.09 & 1.06$\pm$0.13 & \cite{prslff_ros9} \\
2.497 & 1.07$\pm$0.14 & 0.315$\pm$0.029 & 0.50$\pm$0.05 & 0.79$\pm$0.07 & \cite{prslff_ros10} \\
2.50 & 1.16$\pm$0.09 & 0.290$\pm$0.018 & 0.458$\pm$0.029 & 0.72$\pm$0.05 & \cite{prslff_ros6} \\
2.50 & 0.82$\pm$0.07 & 0.393$\pm$0.017 & 0.621$\pm$0.027 & 0.98$\pm$0.04 & \cite{prslff_ros11} \\
2.64 & 0.90$\pm$0.04 & 0.353$\pm$0.009 & 0.573$\pm$0.014 & 0.932$\pm$0.023 & \cite{prslff_ros13} \\
2.75 & 0.84$\pm$0.11 & 0.361$\pm$0.024 & 0.60$\pm$0.04 & 0.99$\pm$0.07 & \cite{prslff_ros12} \\
2.91 & 2.7$\pm$0.7 & 0.01$\pm$0.08 & 0.02$\pm$0.13 & 0.03$\pm$0.23 & \cite{prslff_ros3} \\
2.91 & 0.9$\pm$0.7 & 0.33$\pm$0.14 & 0.56$\pm$0.24 & 0.96$\pm$0.41 & \cite{prslff_ros5} \\
\hline
\end{tabular}
\end{center}
\end{table}

\begin{table}[htp]
\caption[Proton electromagnetic form factor ratios in the spacelike region from 
Rosenbluth separation experiments (Part 4).]
{Proton electromagnetic form factor ratios in the spacelike region from 
Rosenbluth separation experiments (Part 4).}
\begin{center}
\begin{tabular}{|c|l|l|l|l|c|}
\hline
$Q^{2}$ (GeV) & $\mu_pG^p_E/G^p_M$  & $F_2/F_1$ & $Q~F_2/F_1$ 
& $Q^2~F_2/F_1$ & Ref. \\
\hline
3.00 & 0.65$\pm$0.22 & 0.39$\pm$0.05 & 0.68$\pm$0.09 & 1.18$\pm$0.15 & \cite{prslff_ros9} \\
3.007 & 1.22$\pm$0.20 & 0.24$\pm$0.03 & 0.42$\pm$0.06 & 0.73$\pm$0.10 & \cite{prslff_ros10} \\
3.25 & 0.85$^{+0.11}_{-0.12}$ & 0.316$^{+0.021}_{-0.022}$ & 0.57$\pm$0.04 & 1.03$\pm$ & \cite{prslff_ros11} \\
3.20 & 0.96$\pm$0.05 & 0.292$\pm$0.009 & 0.523$\pm$0.016 & 0.936$\pm$0.029 & \cite{prslff_ros13} \\
3.74 & 1.4$\pm$0.3 & 0.18$\pm$0.04 & 0.34$\pm$0.08 & 0.67$\pm$0.15 & \cite{prslff_ros6} \\
3.75 & 0.84$\pm$0.22 & 0.29$\pm$0.04 & 0.55$\pm$0.07 & 1.07$\pm$0.14 & \cite{prslff_ros12} \\
4.00 & 0.89$^{+0.12}_{-0.14}$ & 0.261$^{+0.018}_{-0.021}$ & 0.52$\pm$0.04 & 1.05$^{+0.07}_{-0.09}$ & \cite{prslff_ros11} \\
4.07 & 3.2$\pm$1.3 & -0.04$\pm$0.11 & -0.07$\pm$0.23 & -0.15$\pm$0.46 & \cite{prslff_ros3} \\
4.10 & 1.10$\pm$0.08 & 0.217$\pm$0.011 & 0.440$\pm$0.022 & 0.89$\pm$0.05 & \cite{prslff_ros13} \\
4.20 & 1.24$\pm$0.16 & 0.190$\pm$0.021 & 0.39$\pm$0.04 & 0.80$\pm$0.09 & \cite{prslff_ros12} \\
5.00 & 0.93$^{+0.16}_{-0.19}$ & 0.212$^{+0.020}_{-0.023}$ & 0.48$^{+0.04}_{-0.05}$ & 1.06$^{+0.10}_{-0.12}$ & \cite{prslff_ros11} \\
5.20 & 1.18$\pm$0.55 & 0.17$\pm$0.06 & 0.39$\pm$0.14 & 0.88$\pm$0.32 & \cite{prslff_ros12} \\
6.00 & 0.97$^{+0.20}_{-0.24}$ & 0.178$^{+0.020}_{-0.025}$ & 0.44$^{+0.05}_{-0.06}$ & 1.07$^{+0.12}_{-0.15}$ & \cite{prslff_ros11} \\
7.00 & 1.51$^{+0.25}_{-0.28}$ & 0.101$^{+0.020}_{-0.023}$ & 0.27$^{+0.05}_{-0.06}$ & 0.71$^{+0.14}_{-0.16}$ & \cite{prslff_ros11} \\
8.83 & 0.95$^{+0.57}_{-0.95}$ & 0.13$^{+0.04}_{-0.07}$ & 0.38$^{+0.12}_{-0.20}$ & 1.14$^{+0.36}_{-0.61}$ & \cite{prslff_ros11} \\
\hline
\end{tabular}
\end{center}
\end{table}

\addcontentsline{toc}{chapter}{Curriculum Vitae}
\renewcommand{\baselinestretch}{1}

\begin{center}
\bf
\Large
Peter Zweber\\

\medskip

\large
\bf Curriculum Vitae
\normalsize

\medskip

\rm

Northwestern University \\
Department of Physics and Astronomy \\
Evanston, IL 60208-3112\\
email address: pete@handel.phys.northwestern.edu\\
Tel: (847) 491 8630  
\end{center}

\renewcommand{\baselinestretch}{1.2}

\rm
\begin{tabbing}
{\bf Personal Information} \\
Name: \= \hspace{1.0in} Peter Karl Zweber \\
Date of Birth: \= \hspace{0.45in} January 15, 1975 \\ 
Place of Birth: \= \hspace{0.40in} St. Paul, MN \\
Sex: \= \hspace{1.15in} Male \\
Citizenship: \=  \hspace{0.60in} United States \\
Marital Status: \= \hspace{0.35in} Unmarried \\
\end{tabbing}

\begin{tabbing}
{\bf Education} \\
2006: \= \hspace{1.06in} Ph.D. in Physics, Northwestern University\\
1999: \= \hspace{1.06in} M.S. in Physics, Northwestern University\\
1998: \= \hspace{1.06in} B.S. in Physics, University of Minnesota, Twin Cities\\
1993: \= \hspace{1.06in} High School Diploma, Simley High School,\\
      \= \hspace{1.50in} Inver Grove Heights, MN \\
\end{tabbing}

\begin{tabbing}
{\bf Employment} \\
1999-2006: \= \hspace{0.68in} Graduate Student Research Assistant, Northwestern University\\
           \= \hspace{1.50in} Advisor: Prof. K. K. Seth\\
2001-2005: \= \hspace{0.68in} Visiting Fellow, Cornell University\\
1998-1999: \= \hspace{0.68in} Graduate Student Research Assistant, Northwestern University\\
           \= \hspace{1.50in} Supervisor: Prof. B. Gobbi\\
1997:      \= \hspace{1.05in} Undergraduate Research Opportunities Program Participant,\\
           \= \hspace{1.50in} University of Minnesota, Twin Cities\\
           \= \hspace{1.50in} Supervisor: Prof. E. A. Peterson\\
\end{tabbing}
\bigskip

\noindent
{\bf Award}

\rm
\noindent
Selected by the U.S. Department of Energy to attend 
``The 50th Meeting of the Nobel Laureates'', Lindau, Germany, June 2000.
\medskip
\medskip

\noindent
{\bf Invited Talk}

\rm
\noindent
``Search for $X(3872)$ in $\gamma\gamma$ Fusion and ISR at CLEO'', 
presented by P. Zweber for the CLEO Collaboration, 
1st Meeting of the APS Topical Group on Hadronic Physics, 
\\ Fermilab, Oct. 24-26, 2004.
\medskip
\medskip

\noindent
{\bf Conference Presentations}

\rm
\noindent
``Electromagnetic Form Factors of the Pion, Kaon, and 
Proton at Q$^{2}$ = 13.48 GeV$^{2}$'', presented by 
P. Zweber for the CLEO Collaboration, Particles and Nuclei 
International Conference (PANIC 05), Santa Fe, New Mexico, 
Oct. 24-28, 2005.

\medskip
\rm
\noindent
``Ferreting Through $X(3872)$ Alternatives'', 
presented by P. Zweber for the CLEO Collaboration, 
APS April Meeting, 
Denver, Colorado, May 1-4, 2004.
\medskip
\medskip

\noindent
{\bf Conference Poster}

\rm
\noindent
``The First Precision Measurement of the Charged Kaon Form Factor'', 
P. Zweber and K. K. Seth for the CLEO Collaboration, 
Kaon 2005 International Workshop,
Northwestern University, Evanston, Illinois, June 13-17, 2005.  
\medskip
\medskip

\noindent
{\bf Seminar Presentations}

\rm
\noindent
``$X(3872)$ and Electromagnetic Form Factors at CLEO'', 
Northwestern University, \\ Evanston, Illinois, Oct. 17, 2005.

\medskip
\rm
\noindent
``Selected Results from CLEO'', University of Minnesota, Twin Cities, 
\\ Minneapolis, Minnesota, Sept. 27, 2005.
\medskip
\medskip

\noindent
{\bf Publication List as Primary Author}

\rm
\noindent
``Measurement of Interference between Electromagnetic and Strong Amplitudes 
in $\psi(2S)$ Decays to Two Pseudoscalar Mesons'', 
CLEO Collaboration, S. Dobbs {\itshape{et al.}}, 
submitted to Phys. Rev. Lett., e-Print Archive: hep-ex/0603020,

\medskip
\rm
\noindent
``Precision Measurements of the Timelike Electromagnetic Form Factors 
of Pion, Kaon, and Proton'', 
CLEO Collaboration, T. K. Pedlar {\itshape{et al.}}, 
Phys. Rev. Lett. {\bf 95}, 261803 (2005).

\medskip
\rm
\noindent
``Precision Measurements of the Charged Pion, Charged Kaon, and Proton 
Electromagnetic Form Factors at $s$ = 13.48 $GeV^2$'', 
P. Zweber for the CLEO Collaboration, contributed to Particles and 
Nuclei International Conference (PANIC 05), Santa Fe, New Mexico, 
Oct. 24-28,  2005, e-Print Archive: hep-ex/0512050.

\medskip
\rm
\noindent
``Proton and Charged Kaon Timelike Form Factors at $\sqrt{s}$ = 3.67 GeV'', 
CLEO Collaboration, S. Dobbs {\itshape{et al.}}, 
CLEO CONF 05-09, LP2005-441, submitted to the XXII International 
Symposium On Lepton and Photon Interactions At High Energies, 
Uppsala, Sweden, June 30 to July 5, 2005.

\medskip
\rm
\noindent
``Search for $X(3872)$ in $\gamma\gamma$ Fusion and Radiative Production 
at CLEO'', CLEO Collaboration, S. Dobbs {\itshape{et al.}}, 
Phys. Rev. Lett. {\bf 94}, 032004 (2005).

\medskip
\rm
\noindent
``Search for $X(3872)$ in $\gamma\gamma$ Fusion and ISR at CLEO'', 
P. Zweber for the CLEO Collaboration, 
J. Phys. Conf. Ser. {\bf 9}, 75 (2005).

\medskip
\rm
\noindent
``Search for $X(3872)$ in Untagged $\gamma\gamma$ Fusion and 
Initial State Radiation with CLEO III'', 
CLEO Collaboration, Z. Metreveli {\itshape{et al.}}, 
CLEO CONF 04-07, ICHEP04 ABS10-0768, 
submitted to the 32nd International Conference on High Energy Physics, 
Beijing, China, August 16-22, 2004, e-Print Archive: hep-ex/0408057.
\medskip
\medskip

\noindent
{\bf Publication List with significant contribution}

\rm
\noindent
``Experimental Study of $\chi_{b}(2P) \rightarrow \pi \pi~\chi_{b}(1P)$'', 
CLEO Collaboration, C. Cawlfield {\itshape{et al.}},
Phys. Rev. {\bf D73}, 012003 (2006).

\medskip
\rm
\noindent
``Measurement of the Resonance Parameters of the 
$\chi_{1}(1^{3}$P$_{1})$ and $\chi_{2}(1^{3}$P$_{2})$ 
States of Charmonium Formed in Antiproton-Proton 
Annihilations'', FNAL E835 Collaboration, M. Andreotti {\itshape{et al.}}, 
Nucl. Phys. {\bf B717}, 34 (2005).


\begin{thebibliography}{xxx}


\bibitem{perkinsbook} D. H. Perkins, ``Introduction to High Energy Physics'', 
Third Edition, Addison-Wesley, Reading, MA (1987). 

\bibitem{PDG2004} S. Eidelman $\etal$, Phys. Lett. \textbf{B592}, 1 (2004). 

\bibitem{greenbergcolor}O. W. Greenberg, Phys. Rev. Lett. \textbf{13}, 598 (1964).

\bibitem{Cornellpotential}E. Eichten $\etal$, Phys. Rev. \textbf{D17}, 3090 (1978) 
[Erratum-ibid. \textbf{D21}, 313 (1980)].


\bibitem{OriginalLattice}K. G. Wilson, 
Phys. Rev. \textbf{D10}, 2445 (1974).

\bibitem{asyfree1}D. J. Gross and F. Wilczek, Phys. Rev. Lett. \textbf{30}, 1343 (1973).

\bibitem{asyfree2}H. D. Politzer, Phys. Rev. Lett. \textbf{30}, 1346 (1973).



\bibitem{Hofstadter1}R. Hofstadter, Rev. Mod. Phys. \textbf{28}, 214 (1956).

\bibitem{Hofstadter2}R. Hofstadter, Ann. Rev. Nucl. Part. Sci \textbf{7}, 231 (1957).

\bibitem{Hofstadter3}R. Hofstadter and R. W. McAllister, 
Phys. Rev. \textbf{98}, 217 (1955).

\bibitem{Hofstadter4}R. Hofstadter and R. Herman, 
Phys. Rev. Lett. \textbf{6}, 293 (1961).

\bibitem{Wilson1}R. M. Littauer, H. F. Schopper, and R. R. Wilson, 
Phys. Rev. Lett. \textbf{7}, 141 (1961).

\bibitem{cabibbogatto}N. Cabibbo and R. Gatto, 
Phys. Rev. \textbf{124}, 1577 (1961).





\bibitem{GounarisSakurai}G. J. Gounaris and J. J. Sakurai, 
Phys. Rev. Lett. \textbf{21}, 244 (1968).

\bibitem{Dubnickaetal1}S. Dubnicka, 
Nuovo Cimento \textbf{A100}, 1 (1988) [Erratum-ibid. \textbf{A103}, 469 (1990)]. 

\bibitem{Dubnickaetal2}M. E. Biagini, S. Dubnicka, E. Etim, and P. Kolar, 
Nuovo Cimento \textbf{A104}, 363 (1991).

\bibitem{deMeloetal}J. P. B. C. de Melo, T. Frederico, E. Pace, and G. Salme, 
Phys. Lett. \textbf{B581}, 75 (2004).

\bibitem{Bruchetal}C. Bruch, A. Khodjamirian, and J. H. Kuhn, 
Eur. Phys. J. \textbf{C39}, 41 (2005).

\bibitem{ktlff_VDM}BCF Collaboration, M. Bernardini $\etal$, 
Phys. Lett. \textbf{B46}, 261 (1973).



\bibitem{ffscaling1}S. J. Brodsky and G. R. Farrar, 
Phys. Rev. Lett. \textbf{31}, 1153 (1973).

\bibitem{ffscaling2}S. J. Brodsky and G. R. Farrar, 
Phys. Rev. \textbf{D11}, 1309 (1975).

\bibitem{ffscaling3}V. A. Matveev, R. M. Muradyan, and A. N. Tavkhelidze, 
Lett. Nuovo Cimento \textbf{7}, 719 (1973).

\bibitem{FarrarJackson_PionPQCD}G. R. Farrar and D. J. Jackson, 
Phys. Rev. Lett. \textbf{43}, 246 (1979).

\bibitem{LepageBrodsky_PQCDFF}G. P. Lepage and S. J. Brodsky, 
Phys. Rev. \textbf{D22}, 2157 (1980).

\bibitem{Isgur_LS_argue1}N. Isgur and C. H. Llewellyn Smith, 
Phys. Rev. Lett. \textbf{52}, 1080 (1984).

\bibitem{Isgur_LS_argue2}N. Isgur and C. H. Llewellyn Smith, 
Nucl. Phys. \textbf{B317}, 526 (1989).

\bibitem{Rad_argue}A. V. Radyushkin, Nucl. Phys. \textbf{A532}, 141 (1991).


\bibitem{kpicommontlff_1}VEPP-2 Collaboration, V. E. Balakin $\etal$, 
Phys. Lett. \textbf{B41}, 205 (1972). 
\bibitem{kpicommontlff_2}MEA Collaboration, B. Esposito $\etal$, 
Phys. Lett. \textbf{B67}, 239 (1977). 
\bibitem{kpicommontlff_3}MEA Collaboration, B. Esposito $\etal$, 
Lett. Nuovo Cim. \textbf{28}, 337 (1980).

\bibitem{pitlff_1}OLYA Collaboration, A. D. Bukin $\etal$, 
Phys. Lett. \textbf{B73}, 226 (1978).
\bibitem{pitlff_2}OLYA and CMD Collaborations, L. M. Barkov $\etal$, 
Nucl. Phys. \textbf{B256}, 365 (1985). 
\bibitem{pitlff_3}DM2 Collaboration, D. Bisello $\etal$, 
Phys. Lett. \textbf{B220}, 321 (1989).
\bibitem{pitlff_VDM}BCF Collaboration, D. Bollini $\etal$, 
Lett. Nuovo Cim. \textbf{14}, 418 (1975).

\bibitem{pislff_sc1}G. T. Adylov $\etal$, Nucl. Phys. \textbf{B128}, 461 (1977).
\bibitem{pislff_sc2}E. B. Dally $\etal$, Phys. Rev. \textbf{D24}, 1718 (1981).
\bibitem{pislff_sc3}E. B. Dally $\etal$, Phys. Rev. Lett. \textbf{48}, 375 (1982).
\bibitem{pislff_sc4}S. R. Amendolia $\etal$, Nucl. Phys. \textbf{B277}, 168 (1986).

\bibitem{pislff_elprod1}C. N. Brown $\etal$, Phys. Rev. \textbf{D8}, 92 (1973).
\bibitem{pislff_elprod2}C. J. Bebek $\etal$, Phys. Rev. \textbf{D9}, 1229 (1974).
\bibitem{pislff_elprod3}C. J. Bebek $\etal$, Phys. Rev. \textbf{D13}, 25 (1976).
\bibitem{pislff_elprod4}C. J. Bebek $\etal$, Phys. Rev. \textbf{D17}, 1693 (1978).
\bibitem{pislff_elprod6}J. Volmer $\etal$, Phys. Rev. Lett. \textbf{86}, 1713 (2001). 

\bibitem{Brodskyetal_tlPQCD}S. J. Brodsky, C.-R. Ji, A. Pang, and D. G. Robertson, 
Phys. Rev. \textbf{D57}, 245 (1998).

\bibitem{Frazer}W. R. Frazer, Phys. Rev. \textbf{115}, 1763 (1959). 

\bibitem{StermanStoler}G. Sterman and P. Stoler, 
Ann. Rev. Nucl. Part. Sci. \textbf{47}, 193 (1997). 

\bibitem{carlsonmilana_elprod}C. E. Carlson and J. Milana, Phys. Rev. Lett. \textbf{65}, 
1717 (1990).




\bibitem{ktlff_1}BCF Collaboration, M. Bernardini $\etal$, 
Phys. Lett. \textbf{B44}, 393 (1973). 
\bibitem{ktlff_2}DM1 Collaboration, B. Delcourt $\etal$, 
Phys. Lett. \textbf{B99}, 257 (1981). 
\bibitem{ktlff_3}OLYA Collaboration, P. M. Ivanov $\etal$, 
Phys. Lett. \textbf{B107}, 297 (1981). 
\bibitem{ktlff_4}DM2 Collaboration, D. Bisello $\etal$, 
Z Phys. \textbf{C39}, 13 (1988).

\bibitem{kslff_sc1}E. B. Dally $\etal$, Phys. Rev. Lett. \textbf{45}, 232 (1980).
\bibitem{kslff_sc2}S. R. Amendolia $\etal$, Phys. Lett. \textbf{B178}, 435 (1986).


\bibitem{prtlff_2}DM1 Collaboration, B. Delcourt $\etal$, 
Phys. Lett. \textbf{B86}, 395 (1979). 
\bibitem{prtlff_3}Mulhouse-Strasbourg-Torino Collaboration, 
G. Bassompierre $\etal$, Nuovo Cimento \textbf{A73}, 347 (1983). 
\bibitem{prtlff_4}DM2 Collaboration, D. Bisello $\etal$, 
Nucl. Phys. {\bf B224}, 379 (1983).
\bibitem{prtlff_5}DM2 Collaboration, D. Bisello $\etal$, 
Z. Phys. \textbf{C48}, 23 (1990). 
\bibitem{prtlff_6}PS170 Collaboration, G. Bardin $\etal$, 
Phys. Lett. \textbf{B255}, 149 (1991).
\bibitem{prtlff_7}PS170 Collaboration, G. Bardin $\etal$, 
Phys. Lett. \textbf{B257}, 514 (1991).
\bibitem{prtlff_8}FENICE Collaboration, A. Antonelli $\etal$, 
Phys. Lett. \textbf{B334}, 431 (1994).
\bibitem{prtlff_9}BES Collaboration, M. Ablikim $\etal$, 
Phys. Lett. \textbf{B630}, 14 (2003).
\bibitem{BABARppbarISR}BaBar Collaboration, B. Aubert $\etal$, 
Phys. Rev. \textbf{D73}, 012005 (2006).
\bibitem{e835_1}E760 Collaboration, T. A. Armstrong $\etal$, 
Phys. Rev. Lett. \textbf{70}, 121 (1993). 
\bibitem{e835_2}E835 Collaboration, M. Ambrogiani $\etal$, 
Phys. Rev. \textbf{D60}, 032002 (1999). 
\bibitem{e835_3}E835 Collaboration, M. Andreotti $\etal$, 
Phys. Lett. \textbf{B559}, 20 (2003).

\bibitem{prslff_gmonly1}W. Bartel $\etal$, Phys. Rev. Lett. \textbf{17}, 608 (1966).
\bibitem{prslff_gmonly2}W. Albrecht $\etal$, Phys. Rev. Lett.\textbf{17}, 1192 (1966).
\bibitem{prslff_gmonly3}W. Albrecht $\etal$, Phys. Rev. Lett. \textbf{18}, 1014 (1967).
\bibitem{prslff_gmonly4}P. N. Kirk $\etal$, Phys. Rev. \textbf{D8}, 63 (1973).
\bibitem{prslff_gmonly5}S. Rock $\etal$, Phys. Rev. \textbf{D46}, 24 (1992).
\bibitem{prslff_gmonly6}A. F. Sill $\etal$, Phys. Rev. \textbf{D48}, 29 (1993).  


\bibitem{rosenbluthsep}M. N. Rosenbluth, Phys. Rev. \textbf{79}, 615 (1950). 

\bibitem{poltrans_th1}A. I. Akhiezer and M. P. Rekalo, Sov. J. Part. Nucl. \textbf{3}, 
277 (1974). 
\bibitem{poltrans_th2}R. Arnold, C. Carlson, and F. Gross, Phys. Rev. \textbf{C23}, 
363 (1981). 

\bibitem{prslff_ros1}D. Fr\`{e}rejacque $\etal$, Phys. Rev. \textbf{141}, 1308 (1966). 
\bibitem{prslff_ros2}T. Janssens $\etal$, Phys. Rev. \textbf{142}, 922 (1966).
\bibitem{prslff_ros3}W. Bartel $\etal$, Phys. Rev. Lett. \textbf{17}, 608 (1966).
\bibitem{prslff_ros4}W. Albrecht $\etal$, Phys. Rev. Lett. \textbf{17}, 1192 (1966).
\bibitem{prslff_ros5}W. Albrecht $\etal$, Phys. Rev. Lett. \textbf{18}, 1014 (1967).
\bibitem{prslff_ros6}J. Litt $\etal$, Phys. Lett. \textbf{B31}, 40 (1970).
\bibitem{prslff_ros7}C. Berger $\etal$, Phys. Lett. \textbf{B35}, 87 (1971). 
\bibitem{prslff_ros8}L. E. Price $\etal$, Phys. Rev. \textbf{D4}, 45 (1971).
\bibitem{prslff_ros9}W. Bartel $\etal$, Nucl. Phys. \textbf{B58}, 429 (1973).
\bibitem{prslff_ros10}R. C. Walker $\etal$, Phys. Rev. \textbf{D49}, 5671 (1994).
\bibitem{prslff_ros11}L. Andivahis $\etal$, Phys. Rev. \textbf{D50}, 5491 (1994).
\bibitem{prslff_ros12}M. E. Christy $\etal$, Phys. Rev. \textbf{C70}, 015206 (2004).
\bibitem{prslff_ros13}I. A. Qattan $\etal$, Phys. Rev. Lett. \textbf{94}, 142301 (2005).  

\bibitem{prslff_pol1}B. D. Milbrath $\etal$, Phys. Rev. Lett. \textbf{80}, 452 (1998)
[Erratum-ibid. \textbf{82}, 2221 (1999)].
\bibitem{prslff_pol2}O. Gayou $\etal$, Phys. Rev. \textbf{C64}, 038202 (2001).
\bibitem{prslff_pol3}O. Gayou $\etal$, Phys. Rev. Lett. \textbf{88}, 092301 (2002).
\bibitem{prslff_pol4}V. Punjabi $\etal$, Phys. Rev. \textbf{C71}, 055202 (2005).



\bibitem{ParisiPetronzio_FrozenAlphas}G. Parisi and R. Petronzio, 
Phys. Lett. \textbf{B94}, 51 (1980).

\bibitem{Cornwall_FrozenAlphas}J. M. Cornwall, 
Phys. Rev. \textbf{D26}, 1453 (1982).


\bibitem{QCDSRTheory}M. A. Shifman, A. I. Vainshtein, and V. I. Zakharov, 
Nucl. Phys. \textbf{B147}, 385 (1979).


\bibitem{ChernyakZhitnitsky_QCDSR}V. L. Chernyak and A. R. Zhitnitsky, 
Phys. Rep. \textbf{112}, 173 (1984).

\bibitem{NesterenkoRadyushkin_QCDSR}V. A. Nesterenko and A. V. Radyushkin, 
Phys. Lett. \textbf{B115}, 410 (1982).

\bibitem{IoffeSmilga_QCDSR}B. L. Ioffe and A. V. Smilga, 
Phys. Lett. \textbf{B114}, 353 (1982).

\bibitem{BraunHalperin_QCDSR}V. M. Braun and I. E. Halperin, 
Phys. Lett. \textbf{B328}, 457 (1994).

\bibitem{Lattice1}T. Draper, R. M. Woloshyn, W. Wilcox, and K.-F. Liu, 
Nucl. Phys. \textbf{B318}, 319 (1989).



\bibitem{EfremovRadyushkin_PionPQCD}A. V. Efremov and A. V. Radyushkin, 
Phys. Lett. \textbf{B94}, 245 (1980).

\bibitem{LepageBrodsky_PionPQCD}G. P. Lepage and S. J. Brodsky, 
Phys. Lett. \textbf{B87}, 359 (1979).


\bibitem{JiAmiri_QCDSR}C.-R. Ji and F. Amiri, 
Phys. Rev. \textbf{D42}, 3764 (1990).


\bibitem{LiSterman_PionSud}H.-N. Li and G. Sterman, 
Nucl. Phys. \textbf{B381}, 129 (1992).

\bibitem{JakobKroll_Ptrans}R. Jakob and P. Kroll, 
Phys. Lett. \textbf{B315}, 463 (1993) [Erratum-ibid. \textbf{B315}, 545 (1993)].

\bibitem{Fieldetal_NLOPQCD}R. D. Field, R. Gupta, S. Otto, and L. Chang, 
Nucl. Phys. \textbf{B186}, 429 (1981).

\bibitem{DittesRadyushkin_NLOPQCD}F.-M. Dittes and A. V. Radyushkin, 
Sov. J. Nucl. Phys. \textbf{34}, 293 (1981).

\bibitem{BraatenTse_NLOPQCD}E. Braaten and S.-M. Tse, 
Phys. Rev. \textbf{D35}, 2255 (1987).

\bibitem{MSbar_RenormScheme}W. A. Bardeen, A. J. Buras, D. W. Duke, and T. Muta, 
Phys. Rev. \textbf{D18}, 3998 (1978).

\bibitem{MOM_RenormScheme}W. Celmaster and R. J. Gonsalves, 
Phys. Rev. \textbf{D20}, 1420 (1979).


\bibitem{Melicetal_PQCD}B. Meli\'{c}, B. Ni\v{z}i\'{c}, and K. Passek, 
Phys. Rev. \textbf{D60}, 074004 (1999).

\bibitem{Huangetal_PQCDkt}Y. Huang, X.-G. Wu, and X.-H. Wu, 
Phys. Rev. \textbf{D70}, 053007 (2004).

\bibitem{HuangWu_QCDSRT3}T. Huang and X.-G. Wu, 
Phys. Rev. \textbf{D70}, 093013 (2004).


\bibitem{GoussetPire_tlPQCD}T. Gousset and B. Pire, 
Phys. Rev. \textbf{D51}, 15 (1995).


\bibitem{Milanaetal_jpsipipi}J. Milana, S. Nussinov, and M. G. Olsson, 
Phys. Rev. Lett. \textbf{71}, 2533 (1993).

\bibitem{Bakulevetal_tlQCDSR}A. P. Bakulev, A. V. Radyushkin, and N. G. Stefanis, 
Phys. Rev. \textbf{D62}, 113001 (2000).



\bibitem{Braunetal_QCDSR}V. M. Braun, A. Khodjamirian, and M. Maul, 
Phys. Rev. \textbf{D61}, 073004 (2000).


\bibitem{Agaev5}S. S. Agaev, 
Phys. Rev. \textbf{D72}, 074020 (2005).

\bibitem{renormalon}H. Contopanagos and G. Sterman,
Nucl. Phys. \textbf{B419}, 77 (1994).



\bibitem{Bakulevetal_QCDSR1}A. P. Bakulev, K. Passek-Kumericki, W. Schroers 
and N. G. Stefanis, 
Phys. Rev. \textbf{D70}, 033014 (2004) [Erratum-ibid. \textbf{D70}, 079906 (2004)].


\bibitem{APT1}D. V. Shirkov and I. L. Solovtsov, 
Phys. Rev. Lett. \textbf{79}, 1209 (1997).

\bibitem{APT2}D. V. Shirkov, 
Theor. Math. Phys. \textbf{119}, 438 (1999).

\bibitem{APT3}I. L. Solovtsov and D. V. Shirkov, 
Theor. Math. Phys. \textbf{120}, 1220 (1999).



\bibitem{Lattice2}J. van der Heide, M. Lutterot, J. H. Koch, and E. Laermann, 
Phys. Lett. \textbf{B566}, 131 (2003).

\bibitem{Lattice3}F. D. R. Bonnet, R. G. Edwards, G. T. Fleming, R. Lewis, 
and D. G. Richards, Nucl. Phys. Proc. Suppl. \textbf{129}, 299 (2004).

\bibitem{Lattice4}A. M. Abdel-Rehim and R. Lewis, 
Nucl. Phys. Proc. Suppl. \textbf{140}, 299 (2005).

\bibitem{Lattice5}G. T. Fleming, F. D. R. Bonnet, R. G. Edwards, R. Lewis, 
and D. G. Richards, Nucl. Phys. Proc. Suppl. \textbf{140}, 302 (2005).

\bibitem{Lattice6}S. Capitani, C. Gattringer, and C. B. Lang, 
hep-lat/0511040.

\bibitem{Lattice7}F. D. R. Bonnet, R. G. Edwards, G. T. Fleming, R. Lewis, 
and D. G. Richards, Phys. Rev. \textbf{D72}, 054506 (2005).



\bibitem{Facciolietal_Instanton}P. Faccioli, A. Schwenk, and E. V. Shuryak, 
Phys. Rev. \textbf{D67}, 113009 (2003).

\bibitem{Carvalhoetal_MesonCloud}F. Carvalho, F. O. Dur\~{a}es, F. S. Navarra, 
and M. Nielsen, 
Nucl. Phys. \textbf{C69}, 065202 (2004).

\bibitem{Kaidalovetal_Gluestring}A. B. Kaidalov, L. A. Kondratyuk, D. V. Tchekin, 
Phys. Atom. Nucl. \textbf{63}, 1395 (2000).

\bibitem{BijnensKhodjamirian_QCDSR}J. Bijnens and A. Khodjamirian, 
Eur. Phys. J. \textbf{C26}, 67 (2002).



\bibitem{LepageBrodsky_ProtonPQCD}G. P. Lepage and S. J. Brodsky, 
Phys. Rev. Lett. \textbf{43}, 545 (1979).

\bibitem{Ioffe_fn}B. L. Ioffe, 
Nucl. Phys. \textbf{B188}, 317 (1981) [Erratum-ibid. \textbf{B191}, 591 (1981)].

\bibitem{Chungetal_fn}Y. Chung, H. G. Dosch, M. Kremer, and D. Schall, 
Nucl. Phys. \textbf{B197}, 55 (1982).

\bibitem{ChernyakZhitnitsky_PRDA}V. L. Chernyak and I. R. Zhitnitsky, 
Nucl. Phys. \textbf{B246}, 52 (1984).

\bibitem{KingSachrajda_PRDA}I. D. King and C. T. Sachrajda, 
Nucl. Phys. \textbf{B279}, 785 (1987).

\bibitem{GariStefanis_PRDA}M. Gari and N. G. Stefanis, 
Phys. Rev. \textbf{D35}, 1074 (1987).

\bibitem{Stefanis_PRDA}N. G. Stefanis, 
Phys. Rev. \textbf{D40}, 2305 (1989) [Erratum-ibid. \textbf{D44}, 1616 (1991)].

\bibitem{Chernyaketal_PRDA}V. L. Chernyak, A. A. Ogloblin, and I. R. Zhitnitsky, 
Z. Phys. \textbf{C42}, 569 (1989).

\bibitem{StefanisBergmann_PRDA}N. G. Stefanis and M. Bergmann, 
Phys. Rev. \textbf{D47}, 3685 (1993).

\bibitem{Li_ProtonQCDSR}H.-N. Li, 
Phys. Rev. \textbf{D48}, 4243 (1993).


\bibitem{Bolzetal_PrSudArg}J. Bolz, R. Jakob, P. Kroll, M. Bergmann, and N. G. Stefanis, 
Z. Phys. \textbf{C66}, 267 (1995).

\bibitem{Kunduetal_ProtonPQCD}B. Kundu, H.-N. Li, J. Samuelson, and P. Jain, 
Eur. Phys. J. \textbf{C8}, 637 (1999).

\bibitem{Hyer_tlProtonPQCD}T. Hyer, 
Phys. Rev. \textbf{D47}, 3875 (1993).


\bibitem{NesterenkoRadyushkin_PrQCDSR}V. A. Nesterenko and A. V. Radyushkin, 
Phys. Lett. \textbf{B128}, 439 (1983).

\bibitem{Radyushkin_QCDSR1}A. V. Radyushkin, 
Nucl. Phys. \textbf{A532}, 141 (1991).


\bibitem{Braunetal_ProtonQCDSR}V. M. Braun, A. Lenz, N. Mahnke, and E. Stein, 
Phys. Rev. \textbf{D65}, 074011 (2002).

\bibitem{Braunetal_ProtonQCDSRDA}V. M. Braun, R. J. Fries, N. Mahnke, and E. Stein, 
Nucl. Phys. \textbf{B589}, 381 (2000) [Erratum-ibid. \textbf{B607}, 433 (2001)].

\bibitem{Lenzetal_ProtonQCDSR}A. Lenz, M. Wittmann, and E. Stein, 
Phys. Lett. \textbf{B581}, 199 (2004).



\bibitem{Gockeleretal_PrLattice}M. G\"{o}ckeler, T. R. Hemmert, R. Horsley, 
P. E. L. Rakow, A. Sch\"{a}fer, and G. Schierholz, 
Phys. Rev. \textbf{D71}, 034508 (2005).



\bibitem{Guidaletal_PrGPD}M. Guidal, M. V. Polyakov, A. V. Radyushkin, 
and M. Vanderhaeghen, 
Phys. Rev. \textbf{D72}, 054013 (2005).


\bibitem{Diehletal_PrGPD}M. Diehl, T. Feldmann, R. Jakob, and P. Kroll, 
Eur. Phys. J. \textbf{C39}, 1 (2005).





\bibitem{Miller_SlPrMesonCloud}G. A. Miller, 
Phys. Rev. \textbf{C66}, 032201 (2004).

\bibitem{Iachello_SlPrMesonCloud}F. Iachello, 
Eur. Phys. J. \textbf{A19}, 29 (2004).

\bibitem{IachelloWan_TlPrMesonCloud}F. Iachello and Q. Wan, 
Phys. Rev. \textbf{C69}, 055204 (2004).



\bibitem{Krolletal_slPrDiquark}P. Kroll, M. Sch\"{u}rmann, and W. Schweiger, 
Int. J. Mod. Phys. \textbf{A6}, 4107 (1991).

\bibitem{Krolletal_tlPrDiquark}P. Kroll, Th. Pilsner, M. Sch\"{u}rmann, 
and W. Schweiger, 
Phys. Lett. \textbf{B316}, 546 (1993).




\bibitem{Belitskyetal_prf1f2}A. V. Belitsky, X. Ji, and F. Yuan, 
Phys. Rev. Lett. \textbf{91}, 092003 (2003).

\bibitem{Brodskyetal_prtlf1f2}S. J. Brodsky, C. E. Carlson, J. R. Hiller, and D. S. Hwang, 
Phys. Rev. \textbf{D69}, 054022 (2004).

\bibitem{OriginalIachelloetal}F. Iachello, A. D. Jackson, and A. Lande, 
Phys. Lett. \textbf{B43}, 191 (1973).

\bibitem{Brodsky_prf1f2}S. J. Brodsky, hep-ph/0208158.



\bibitem{CLEOcCESRc}R. A. Briere $\etal$, CLEO-c/CESR-c Taskforces and CLEO Collaboration, 
Cornell University LEPP Report No. CLNS 01/1742 (2001), unpublished.

\bibitem{CESRminiMAC}D. Rice (for the CESR staff), ``CESR-c Configuration and Performance'', 
presented at miniMAC, Cornell University, Ithaca, NY, USA, July 22-23, 2005.

\bibitem{accbook} H. Wiedemann, ``Particle Accelerator Physics'', 
Springer-Verlag, New York, NY (1993).

\bibitem{BeamEnergy}CLEO Collaboration internal document, 
V\'{e}ronique Boisvert, CBX 00-70.

\bibitem{KarlResProgram}CLEO Collaboration internal document, 
Karl Berkelman, CBX 03-12.

\bibitem{KureavFadin}E. A. Kureav and V. S. Fadin, 
Sov. J. Nucl. Phys. \textbf{41}, 466 (1985).

\bibitem{DRIII}D. Peterson $\etal$, Nucl. Instrum. Meth. \textbf{A478}, 142 (2002).

\bibitem{RICHDet1}M. Artuso $\etal$, Nucl. Instrum. Meth. \textbf{A441}, 374 (2000).

\bibitem{RICHDet2}M. Artuso $\etal$, Nucl. Instrum. Meth. \textbf{A554}, 147 (2005).

\bibitem{CLEOII}Y. Kubota $\etal$, Nucl. Instrum. Meth. \textbf{A320}, 66 (1992).

\bibitem{MuonDet}D. Bortoletto $\etal$, Nucl. Instrum. Meth. \textbf{A320}, 114 (1992).

\bibitem{Kalman1}CLEO Collaboration internal document, 
Robert Kutschke and Anders Ryd, CBX 96-20.

\bibitem{Kalman2}P. Billoir, Nucl. Instrum. Meth. \textbf{255}, 352 (1984). 

\bibitem{expbook}R. Fernow, ``Introduction to Experimental Particle Physics'', 
Cambridge, New York, NY (1986). 

\bibitem{trackingtrigger}R. M. Hans, C. L. Plager, M. A. Selen, and M. J. Haney, 
IEEE Trans. Nucl. Sci. \textbf{48}, 552 (2001). 

\bibitem{CCtrigger}G. D. Gollin, J. A. Ernst, J. B. Williams, R. M. Hans, and M. J. Haney, 
IEEE Trans. Nucl. Sci. \textbf{48}, 547 (2001). 

\bibitem{GlobalL1trigger}M. A. Selen, R. M. Hans, and M. J. Haney, 
IEEE Trans. Nucl. Sci. \textbf{48}, 562 (2001). 




\bibitem{Babayaga}C. M. Carlomi Calame {\itshape{et al.}}, hep-ph/0312014, 
and in Proceedings of the Workshop on Hadronic Cross-Section at Low Energy 
(SIGHAD03), 8-10 October 2003, Pisa, Italy, ed. M. Incagli and 
G. Graziano, 258 (Elsevier, Amsterday, 2004).

\bibitem{contlumcorr}CLEO Collaboration internal document, 
S. Mehrabyan and B. K. Heltsley, CBX 05-10.  

\bibitem{psiplumcorr}CLEO Collaboration internal document, 
B. K. Heltsley, CBX 04-11. 

\bibitem{numofpsip}CLEO Collaboration internal document, 
H. Muramatsu and T. Skwarnicki, CBX 04-35.

\bibitem{psipcontamination}CLEO Collaboration internal document, 
B. K. Heltsley and H. Mahlke-Krueger, CBX 04-47.  

\bibitem{EvtGen}David J. Lange,  Nucl. Instrum. Meth. \textbf{A462}, 
152 (2001).

\bibitem{PHOTOS}E. Barberio and Z. W\c{a}s, 
Comput. Phys. Commun. {\bf 79}, 291 (1994).

\bibitem{ggcs}V. M. Budnev, I. F. Ginzburg, G. V. Meledin, and V. G. Serbo, 
Phys. Rep. {\bf 15C}, 181 (1975). 

\bibitem{bonneaumartinrc}G. Bonneau and F. Martin, 
Nucl. Phys. \textbf{B27}, 381 (1971).

\bibitem{berendskleissrc}F. A. Berends and R. Kleiss, 
Nucl. Phys. \textbf{B178}, 141 (1981).

\bibitem{bkhhmrc}CLEO Collaboration internal document, 
B. K. Heltsley and H. Mahlke-Krueger, CBX 04-16.

\bibitem{cleopsip2body}CLEO Collaboration internal document, 
S. B. Athar, L. Breva-Newell, and J. Yelton, CBX 05-11.

\bibitem{dhadsyst}CLEO Collaboration internal document, 
D. Cassel \textit{et al.}, CBX 05-07.
 
\bibitem{CLEOIIIDetector}G. Viehhauser, Nucl. Instrum. Methods A {\bf 462}, 146 (2001).

\bibitem{LUNDJetset1}T.~Sj\"ostrand \textit{et al.}, 
Computer Physics Commun. {\bf 135}, 238 (2001).

\bibitem{LUNDJetset2}T.~Sj\"ostrand,  L.\ L\"onnblad, S. Mrenna, and P. Skans,
\textit{PYTHIA 6.3 Physics and Manual}, hep-ph/0308153.

\bibitem{SethHigherCharmRes}K. K. Seth, Phys. Rev. \textbf{D72}, 017501 (2005).

\bibitem{BESIpsi(2S)piK}S. W. Ye, ``Study of some VP and PP modes of $\psip$ 
decays'', Ph.D. thesis, Univerisity of Science and Technology of China, 1997 
(in Chinese).
 
\bibitem{e835_psipppbar}E760 Collaboration, T. A. Armstrong $\etal$, 
Phys. Rev. \textbf{D47}, 772 (1993). 

\bibitem{CLEOIIIPIDsyst}CLEO Collaboration internal document, 
R. A. Briere, G. P. Chen, A. J. Sadoff, and G. Tatishvili, CBX 03-37.

\bibitem{SethJpsiKK}K. K. Seth, to be published. 

\bibitem{SuzukiJpsi}M. Suzuki, Phys. Rev. {\bf D60}, 051501(R) (1999). 

\bibitem{RosnerJpsi}J. L. Rosner, Phys. Rev. {\bf D60}, 074029 (1999). 
 
\bibitem{WangetalJpsi}P. Wang, C. Z. Yuan, X. H. Mo, 
Phys. Rev. {\bf D69}, 057502 (2004). 
 

\end{thebibliography}
\end{document}